\ifx\pdfminorversion\undefined\else\pdfminorversion=7\fi
\ifx\pdfsuppresswarningpagegroup\undefined\else\pdfsuppresswarningpagegroup=1\fi
\ifx\texorpdfstring\undefined\newcommand\texorpdfstring[2]{#1}\fi
\PassOptionsToPackage{hyperfootnotes=true,unicode,colorlinks}{hyperref}
\documentclass[aps,prd,11pt,onecolumn,superscriptaddress,tightenlines,nofootinbib,preprintnumbers,floatfix,altaffilletter]{revtex4-2}
\usepackage[utf8]{inputenc}
\usepackage{silence}
\usepackage[caption=false]{subfig}
\ifnum\paperheight=0\relax\paperheight=11in\relax\fi

\ActivateFilters[unknownclass]
\usepackage{subfig,graphics,graphicx,xcolor,cancel,slashed,multirow,siunitx,tikz,latexsym,mathtools,amsmath,amssymb,amsbsy,amsfonts,array,bm,mathrsfs,setspace,orcidlink,physics}
\DeactivateFilters[unknownclass]
\allowdisplaybreaks
\usepackage[normalem]{ulem}
\usepackage[hyperfootnotes=true,unicode,colorlinks]{hyperref}
\usepackage[capitalise]{cleveref}
\usepackage{booktabs}

\newcommand{\targetsens}{$10^{-29}~ e \cdot \textnormal{cm}$ }

\let\origfootnote\footnote
\def\footnote{\unskip\origfootnote}
\graphicspath{{FIGS/}}

\begin{document}
\title{Electric dipole moments and the search for new physics}
\author{Ricardo Alarcon} \affiliation{Arizona State University, Tempe, AZ 85287, USA}
\author{Jim Alexander}\affiliation{
  Cornell University, Ithaca, New York, USA}
\author{Vassilis Anastassopoulos}\affiliation{
  University of Patras, Dept. of Physics, Patras-Rio, Greece}
  \author{	Takatoshi Aoki	}	\affiliation{	The University of Tokyo, Meguro-ku, Tokyo, Japan	}

\author{Rick Baartman}\affiliation{
  TRIUMF, Vancouver, British Columbia, Canada}
\author{Stefan Bae\ss{}ler}\affiliation{University of Virginia, 382 McCormick Road, Charlottesville, VA 22903, USA}\affiliation{Oak Ridge National Laboratory, 1 Bethel Valley Road, Oak Ridge, TN 
    37830, USA}
\author{ Larry Bartoszek} \affiliation{Bartoszek Engineering, Aurora, IL 60506, USA.}
\author{Douglas H. Beck} \affiliation{University of Illinois at Urbana-Champaign,
Urbana, IL 61801, USA}
\author{Franco Bedeschi}\affiliation{
  National Institute for Nuclear Physics (INFN-Pisa), Pisa, Italy}
  \author{	Robert Berger	}	\affiliation{	Philipps-Universitaet Marburg, Fachbereich Chemie, Hans-Meerwein-Str. 4, 35032 Marburg, Germany	}
% ...
  
\newpage\begin{abstract}
Static electric dipole moments of nondegenerate systems probe mass scales for physics beyond the Standard Model well beyond those reached directly at high energy colliders. Discrimination between different physics models, however, requires complementary searches in atomic-molecular-and-optical, nuclear and particle physics. In this report, we discuss the current status and prospects in the near future for a compelling suite of such experiments, along with developments needed in the encompassing theoretical framework.
\end{abstract}

\date{March 2022}
\maketitle

\newpage
\vspace{-5mm}
\section{Executive Summary}

\vspace{-5mm}

Observation of an electric dipole moment (EDM) in any experimental system (electron, neutron, proton, atom, molecule) at current or near-future sensitivity would yield exciting new physics. Assuming maximal breaking of CP symmetry, EDMs probe beyond the Standard Model (BSM) mass scales well beyond those directly probed at high energy colliders. However, a single discovery alone can not by itself discriminate between the many viable BSM theories, rule out baryogenesis scenarios for the observed matter-antimatter asymmetry of the Universe, or tell whether CP symmetry is spontaneously or explicitly broken in Nature. To accomplish such goals a well coordinated program of complementary EDM searches in atomic/molecular, nuclear, and particle physics experiments is needed. In this report a compelling suite of such experiments and an encompassing theoretical framework are proposed to discover and establish the next fundamental theory of physics.

Searches for fundamental EDMs have a long history, beginning with the neutron EDM (nEDM, \cref{sec:neutron}) beam experiments of Purcell and Ramsey in 1949. The nEDM is sensitive to $\theta$QCD, quark electric dipole and chromo-electric moments, the gluon chromo-electric moment, and CP-violating four-fermi interactions at the leading order in an effective field theory. Modern experiments use ultracold neutrons (UCN) that can be polarized and stored in room-temperature bottles for hundreds of seconds, leading to very precise measurements. These techniques have been developed around the world over decades, with a best limit of $d_n < 1.8\times 10^{-26}$~$e\cdot$cm (90\% C.L.) reported by the Paul Scherrer Institut in 2020. Several UCN experiments are being developed around the world with the goal of $10^{-27}$ within the next 5--10 years and $10^{-28}$ in 10--15 years.

Like neutrons, atoms and molecules (\cref{sec:amo}) have been sensitive probes of EDMs for decades and currently set the best limits on the electron EDM, semileptonic CP-violating four-fermi interactions,
and quark chromo-EDMs; they also are competitive with the nEDM for sensitivity to quark EDMs
and $\theta$QCD, providing an excellent check on both types of experiments. Atom and molecule-based searches for fundamental symmetry violations have advanced rapidly in recent
years, and have excellent prospects for further advances.
Improvements in sensitivity of one, two-three, and four-six orders of magnitude appear to be realistic on the few, 5--10, and 15--20 year time scales, respectively, by leveraging major advancements made using quantum science techniques and the increasing availability of exotic species with extreme sensitivity. These gains open an exciting pathway to probe PeV-scale physics using “tabletop” scale experiments.

Finally, a whole new class of EDM experiments is possible in storage rings (\cref{sec:proton}). A ring at BNL or Fermilab, using technology off the shelf and a design based on the successful muon $g-2$ experiments can be operational and produce first results for the proton in less than 10 years. The proton storage ring is expected to reach $10^{-29}e\cdot\rm cm$ sensitivity in 10 years for the proton, and the same level in an additional five years for the deuteron. The exceptional sensitivity is due to several features designed to suppress systematics. Furthermore, the storage ring experiment is sensitive to dark matter, dark energy models, and axions. New electron and muon EDM storage ring measurements are proposed at Jlab, PSI and Fermilab. The PSI and Fermilab experiments will use the so-called ``frozen spin" technique. The latter is an improvement of the current muon g-2 experiment at Fermilab while the former is a whole new experiment at PSI and will improve the sensitivity to the muon EDM by four orders of magnitude.

Fermion EDMs originate at a high mass scale through new complex CP-violating phases ($e.g.$, through new Yukawa couplings) and feed down to lower energy scales via dimension four and higher operators in a Standard Model effective theory. These elementary particle EDMs then manifest in bound states like the proton and neutron, and even atoms and molecules. At the quark-nucleon level, lattice QCD plays a crucial role in this matching between energy scales. At lower energies still, nucleon chiral perturbation theory, and finally, nuclear and atomic theories are needed. The reverse holds as well: if a nucleon, atomic, or molecular EDM is measured, the underlying BSM physics is tested and can be diagnosed. This remarkable effective theory framework, encompassing tens-of-orders of magnitude in energy, is in place today (\cref{sect:th}). Crucially, the low-energy theory is continually improved as new high energy models are invented.


In this white paper, we discuss the motivation to search for EDMs, the theoretical framework necessary to understand and interpret them, and current and planned experiments to find them.

\newpage
\tableofcontents

\newpage
\section{Introduction}


The observation of a nonzero value of the permanent electric dipole moment (EDM) of non-degenerate systems such as elementary particles, atoms, or molecules would signal the existence of new interactions violating both time-reversal (T) and parity (P) invariance, 
or equivalently CP~\cite{Luders:1954zz}, the combination of charge conjugation and parity. 

CP violation exists in the Standard Model (SM) of particle 
interactions due to the strong CP phase $\bar \theta$, 
a phase in the Cabibbo-Kobayashi-Maskawa (CKM) quark
mixing matrix~\cite{Kobayashi:1973fv}, and possibly  a similar phase
in the Pontecorvo-Maki–Nakagawa–Sakata (PMNS) leptonic mixing matrix~\cite{Pontecorvo:1957qd,Maki:1962mu,Nunokawa:2007qh}. 
Physical manifestations of the CKM and PMNS  phases require the interplay of three fermion families  and therefore their impact on flavor-diagonal 
CP probes such as EDMs is  very small, many orders of magnitude below the current and future experimental sensitivity 
(see \cite{Pospelov:2005pr,Yamaguchi2021,Ema2022} and references therein).
Given this background, EDMs play a prominent role in searches for  new physics, for at least three reasons:

(1) Due to the smallness of the CKM and PMNS contributions, 
EDMs are a very clean  probe of new sources of CP
violation arising from physics beyond the Standard Model (BSM)~\cite{Pospelov:2005pr,Chupp:2017rkp}, 
providing at the same time very strong constraints on the SM $\bar \theta$-term  ($\bar \theta \sim 10^{-10}$).  
Should a positive signal emerge in the future, 
a combination of EDM measurements and appropriate theory  input will be able to disentangle $\bar \theta$ from BSM sources of CPV, 
either way providing deep insight on the nature of CP in a fundamental theory.

(2) 
At the current and future sensitivity level, assuming maximal breaking of CP, 
EDMs  probe BSM mass scales that can be considerably higher than the ones directly probed at high energy colliders. 

(3) EDMs  can  shed light on one of the great puzzles of modern physics, i.e. 
the origin of the  cosmic matter-antimatter asymmetry,  quantified by the number of baryons minus antibaryons per black body photon, 
\(6.104\pm0.058\times 10^{-10}\) \cite{Bennett:2003bz,Planck:2015fie,Planck:2018vyg,Fields:2019pfx}. 
Such an asymmetry is difficult to include as an initial condition in  an inflationary
cosmological scenario~\cite{Coppi:2004za} and therefore requires dynamical generation, baryogenesis. 
According  to Sakharov's conditions~\cite{Sakharov:1967dj} CP violation is one of three 
key necessary conditions for baryogenesis.  
The strength of the CPV in the CKM matrix is, however, too small to explain baryogenesis~\cite{Shaposhnikov:1987tw,Farrar:1993sp,Gavela:1993ts,Huet:1994jb}. 
Similarly, CP-violation due to the $\bar \theta$ term is unlikely to lead to appreciable baryon asymmetry~\cite{Dolgov:1991fr,Gross:1980br}. 
Thus, BSM CP violation is expected to have played a major role in baryogenesis and EDMs provide a strong tool to probe it. 
In particular, EDMs are major tools to test  low-scale baryogenesis mechanisms, 
such as electroweak baryogenesis (see \cite{Morrissey:2012db} and references therein).

The discovery of an EDM in a single system will be  paradigm-shifting but  will  not be sufficient to 
qualitatively nor quantitatively discriminate among BSM models and rule out baryogenesis scenarios. To achieve these goals, it is necessary to observe or bound EDMs in \textit{complementary} systems, so that  the main features of CP violation at low energy can be identified  and  systematically  connected to the underlying BSM scenarios. 
This, thus, opens up a clear target for a joint experiment-theory program to elucidate novel sources of CP violation and their implications 
for fundamental interactions and cosmology.

\newpage 
\section{Theory} \label{sect:th}



\subsection{General considerations}
\label{sect:th1}
The Standard Model admits two CP-violating phases in dimension 4 operators, the strong CP phase $\bar \theta$ and the CKM phase $\delta_\mathrm{CKM}$. The former is known to be extraordinarily tiny, whereas the latter is order one. There is also one CP-violating phase in the PMNS matrix, or three phases in the case of Majorana neutrinos. These are not yet measured. Given such limited information, the fundamental status of CP symmetry in our universe is unknown. The nonzero CKM phase leaves open the possibility that CP is a fundamental symmetry of nature, but is spontaneously broken. In this case, the nature of the spontaneous breaking could lead to substantial variation in the sizes of CP phases arising in different couplings; for example, the breaking of CP could be correlated with the breaking of flavor symmetries~\cite{Nir:1996am}.  On the other hand, if CP is simply not a symmetry of nature at any level, then we expect all CP phases to be order-one numbers (with possible exceptions like $\bar \theta$ that could be dynamically relaxed to zero).

Models beyond the Standard Model typically admit a plethora of new CP violating phases. For instance, any scenario with new vector-like fermions that also have Yukawa couplings to the Higgs boson allows for CP violation, and will generate EDMs of SM particles through two-loop diagrams~\cite{Barr:1990vd}. Simple dimensional analysis tells us that new physics at a mass scale $M_\mathrm{NP}$, with a CP-violating phase $\delta_\mathrm{CPV}$, contributing to an EDM at the $\ell$-loop order will generate an EDM of order
\begin{equation}
d_f \sim e q_f  \sin(\delta_\mathrm{CPV}) \left(\frac{g^2}{16\pi^2}\right)^\ell \xi_\mathrm{FV} \frac{m_f}{M_\mathrm{NP}^2}.
\label{eq:crudeestimate}
\end{equation}
Here $d_f$ is the EDM of the SM fermion $f$ (e.g., the electron); $m_f$ and $q_f$ are the mass and charge of that fermion; $g$ is the typical size of a coupling appearing in the calculation (generally $O(1)$); and $\xi_\mathrm{FV}$ is a possible enhancement factor due to flavor violation. Setting this factor to 1 is conservative in terms of estimating the mass reach $M_\mathrm{NP}$, although models in which it is 1 (``Minimal Flavor Violation'' or MFV) tend to be rather special. This approximate estimate is confirmed by many concrete calculations in specific models (e.g., the selected early references~\cite{Ellis:1982tk, Chia:1982gp, Polchinski:1983zd, GavelaLegazpi:1982ud, delAguila:1983dfr,Barr:1990vd, Leigh:1990kf}). To give a sense of the numbers, if we consider the electron EDM and assume that $\delta_\mathrm{CPV}$ is maximal, $\xi_\mathrm{FV} = 1$, and $g \approx 0.6$ is of order the weak coupling constant, we find that an experimental constraint $|d_e| \leq d_e^\mathrm{max}$ translates into
\begin{equation}
M_\mathrm{NP} \gtrsim \sqrt{\frac{10^{-29}\,e\,\mathrm{cm}}{d_e^\mathrm{max}}}  \times \begin{cases} 50\,\mathrm{TeV}, & \ell =1 \\ 2\,\mathrm{TeV},& \ell = 2 \end{cases}.
\label{eq:massreach}
\end{equation}
While this estimate is heuristic, it gives us the immediate qualitative lesson that current precision searches for CP violation are already sensitive to mass scales above those accessible to the LHC, even for effects arising at the two-loop order. Note that every two orders of magnitude improvement in the EDM bound translates into one order of magnitude improvement  in the mass reach. On the other hand, if we focus on a particular mass range---say, new physics at the TeV or 10 TeV scale, accessible to a current or future collider experiment---then every order of magnitude improvement in the EDM constraint translates to an order of magnitude stronger constraint on the CP phase $\delta_\mathrm{CPV}$. If new physics exists but has tiny CP-violating phases, this could be a strong clue that CP may be a spontaneously broken fundamental symmetry of nature.

Finally, in certain BSM scenarios EDMs of composite systems (like atoms, molecules, or nuclei) can be induced via four fermion operators arising at tree level, avoiding the loop suppression in Eq.~\eqref{eq:massreach}. An example is a scalar lepto-quark model \cite{Barr:1992cm,Fuyuto:2018scm,Dekens:2018bci} where tree-level lepto-quark exchange leads to CP-odd electron-quark operators that, in turn, induce EDMs of paramagnetic systems. However, such couplings intrinsically involve new sources of flavor violation beyond the Standard Model Yukawa couplings, so one must be cautions in comparing estimated mass reach with the MFV assumption in Eq.~\eqref{eq:massreach}. Another example is the minimal left-right symmetric model \cite{Mohapatra:1974hk,Senjanovic:1975rk} where $W_L$-$W_R$ mixing leads to CP-odd four-quark operators that induce a neutron and diamagnetic EDM without loop suppression. However, in these models additional suppression can arise from small dimensionless quantities such as ratios of fermion masses. An explicit example of such suppression is that $W_L$-$W_R$ mixing is at least suppressed by $m_b/m_t$ \cite{Maiezza:2014ala} to be consistent with measured quark masses. 

If CP violation is discovered, our first task is to assess whether it arises from the strong CP phase $\bar \theta$ or from a higher-dimension operator in the Standard Model EFT. We expect that $\bar\theta$ dominantly contributes to hadronic CP violation, but it also feeds into the electron EDM. Further theoretical work is needed to fully characterize the imprint of $\bar \theta$ in the full range of experimental observables. Only by measuring CP violation in multiple, complementary systems could we make the case that we are observing the effect of $\bar \theta$ rather than, say, a quark chromo-electric dipole moment or a CP-violating four fermion interaction. A discovery of nonzero $\bar \theta$ would be significant. For example, in axion solutions to the strong CP problem,  $\bar \theta$ is dynamically relaxed~\cite{Peccei:1977ur, Peccei:1977hh, Weinberg:1977ma, Wilczek:1977pj}, but generally not to exactly zero, due to effects violating the Peccei-Quinn symmetry~\cite{Barr:1992qq,kamionkowski:1992mf, holman:1992us, Ghigna:1992iv}. Measurement of nonzero $\bar \theta$ would be an important clue about the symmetry structure of the high-energy theory. More practically, a nonzero $\bar \theta$ would allow an axion to mediate long-range forces rather than simply spin-dependent ones, an effect that can be searched for \href{https://hep.ph.liv.ac.uk/ariadne/}{experimentally}~\cite{Arvanitaki:2014dfa}.

If a new CP violating effect is observed that, after correlating the results from multiple systems, cannot be attributed to $\bar \theta$, the implications are even more profound. This would point to new physics in higher dimension operators in the Standard Model, suppressed by a heavy mass scale. As in the example estimate~\eqref{eq:massreach} for the EDM, knowledge of such an operator would immediately point to a maximum mass scale at which the effect might be generated. This would immediately and dramatically strengthen the case for building further energy-frontier experiments, such as collider experiments that could directly produce the particles responsible for generating  the operator. Precision experiments themselves can begin to unravel the nature of the new physics by carrying out complementary, redundant tests of many systems. These could clarify which operators are producing the effect and, ideally, measure the coefficients of multiple operators in the SM EFT. These measurements could serve as a ``fingerprint'' of a given model. For instance, some models would produce a signal in polar molecule EDM experiments that arises dominantly from the electron EDM, while others would produce a signal dominantly through a CP-violating electron-nucleon interaction. Only by measuring CP violation in multiple molecules could one assess which is being observed, and begin to home in on the underlying new physics models.

\subsection{Theoretical framework: status and challenges}
Considering that EDMs are very low-energy measurements, the above discussion can be neatly streamlined in an effective field theory (EFT) framework. Assuming that new CP-violating sources are associated to energy scales well above the electroweak scale 
(the case of light and weakly coupled new physics is considered 
for example in Refs~\cite{LeDall:2015ptt,Fuyuto:2019vfe}), 
their low-energy effects can be captured by local higher-dimensional operators in the Standard Model Effective Field Theory (see \citet{Engel:2013lsa} for a systematic review). The first relevant CP-odd operators appear at dimension six---suppressed by two powers of the new physics mass scale---and consist of lepton and quark electric dipole moments, quark and gluon chromo-electric dipole moments, four-fermion operators, and interactions among Higgs and gauge bosons. 
The CP-odd operators can be evolved to lower energy scales (and in the process `mix' with each other) and matched to an EFT that only contain the light SM degrees of freedom: light quarks and leptons, photons, and gluons. 

So at the hadronic scale of order $1$~GeV,   CP-violating interactions from the SM and beyond can be captured by 
a handful of CPV operators, ordered according to their dimension. 
First, at dimension four one has a CPV quark mass term  or equivalently the QCD topological charge, \(\overline\Theta\)   (the two are related by the singlet axial anomaly). 
The leading BSM CPV  dimension-six  operators from the high-scale manifest themselves   
as dimension-five and dimension-six operators that can be grouped in three  classes: 
\begin{itemize}
    \item EDMs of the elementary fermions (light quarks and electron), coupling the spin of the fermion to the electric field. 
    \item Chromo-Electric Dipole Moments (cEDMs) of quarks and gluon, coupling the spin of the quarks and gluons to the chromo-electric field.   The gluon cEDM operator is also known in the literature as the CPV Weinberg three-gluon operator~\cite{Weinberg:1989dx}.
    \item Various CPV four-fermion operators. They are divided in two groups. On one hand, semi-leptonic operators involving a quark and a lepton bilinear, that contribute to atomic and molecular EDMs. On the other hand,  four-quark operators that contribute primarily to CP violation in hadronic systems (such as nucleon EDM and nuclear moments via  CPV pion-nucleon couplings).
\end{itemize}
The leading four-electron operators  in the  Standard Model that are not hermitian, and thus admit CP-violating phases, arise at dimension eight (of the form $(h^\dagger L {\bar e})^2$), so their effects are expected to be smaller by at least a factor of $v^2/M_\mathrm{NP}^2$ than those of dimension-six operators.

In a next step, the resulting effective operators can be matched to CP-violating interactions among the relevant low-energy degrees of freedom: leptons, pions, nucleons, and photons. Chiral EFT has proven to be a powerful tool to organize the various CP-odd hadronic interactions based on the chiral symmetry properties of the underlying sources of CP violation \cite{Mereghetti:2010tp,deVries:2012ab}. The associated power counting predicts that for any CP-odd dimension-six operator, EDMs can be expressed in terms of a handful of CP-violating hadronic and (semi-)leptonic interactions. Chiral EFT also allows for a systematic derivation of CP-violating nuclear forces \cite{deVries:2020iea} that 
can then be used in combination with few- and many-body techniques to compute EDMs of light nuclei and diamagnetic atoms and molecules.  
Both the matching from quark-gluon operators to chiral EFT and the calculation of EDMs of nuclei and diamagnetic atoms require non-perturbative techniques that  suffer at the moment from large theoretical uncertainties. As discussed below, 
these uncertainties dilute the physics reach of very sensitive experimental searches.  Therefore, 
a systematic approach to reduce theoretical uncertainties is 
an integral part of the EDM physics program for the next decade.

The matching from quark-gluon operators to chiral EFT involves non-perturbative QCD because the hadronic coupling constants (typically low-energy constants or LECs) associated to CP-violating interactions are, in most cases, not determined by symmetry considerations (a rare exception is the QCD $\bar \theta$ term, where certain hadronic couplings are connected to known CP-even matrix elements). 
Calculations of the LECs have been performed in the past mostly within QCD sum rules or other hadronic models, 
leading to results whose uncertainties are hard to quantify. 
In recent years it has become clear that lattice QCD (LQCD) is the most promising tool to obtain QCD-based results with quantified 
uncertainties, albeit with many challenges. 
While most effort in LQCD so far has focused on matrix elements associated to the QCD $\bar \theta$ term, the higher-dimensional operators such as the quark and gluon chromo-EDMs are becoming the target of investigation (see \citet{Shindler:2021bcx} for a recent review). 
The current situation is well exemplified by looking at the  status of the neutron EDM expressed in terms of 
SM ($\bar \theta$) and 
BSM sources of CP violation ($d_q$ and $\tilde{d}_q$ denote the quark EDMs and cEDMs, respectively,
and $\tilde{d}_G$, the gluon cEDM). 
Working at a renormalization scale of $\mu=2$~GeV in the $\overline{\rm{MS}}$ scheme, and putting together input form 
lattice QCD and QCD sum rules, we have
\begin{eqnarray}
d_n&=&    - (1.5 \pm 0.7) \cdot 10^{-3}  \   { \bar{\theta} } \,  e\,  {\rm fm}  \nonumber \\
& &-(0.20\pm0.01)  { d_u}   \, + \, (0.78\pm0.03 )  {d_d}   \, + \, (0.0027\pm0.016)  { d_s}  \nonumber \\
&&-(0.55\pm0.28) e{ \tilde{d}_u} \,  -\, (1.1\pm0.55) e { \tilde{d}_d}  \, + \,  (50\pm40) \, {\rm MeV}  e\, { \tilde{d}_G}   ~.
\label{eq:nEDM}
\end{eqnarray}
The coefficient of the $\bar \theta$ term has been computed in LQCD in \citet{Dragos:2019oxn}, 
while  other lattice calculations point out that larger systematic effects might be possible~\cite{Bhattacharya:2021lol}.
The matrix elements relating $d_n$ to $d_q$ have been precisely calculated in LQCD~\cite{Gupta:2018lvp,Bhattacharya:2015esa}. 
The quoted matrix elements of the quark cEDM operators are obtained via QCD sum rules~\cite{Pospelov_qCEDM,Pospelov_deuteron,Hisano1}. For the Weinberg operator we quote a range covering the  QCD sum rules~\cite{Pospelov_Weinberg,Haisch:2019bml} (lower  value) 
and  Naive Dimensional Analysis (NDA)~\cite{Weinberg:1989dx} (higher  value). 
Finally, the neutron EDM dependence on BSM operators whose matrix elements are even less known (such as four-quark operators) is not included in Eq.~(\ref{eq:nEDM}).

A quick look at Eq.~(\ref{eq:nEDM}) 
illustrates  several lessons: 
(i) First,  a single EDM is not sufficient to disentangle 
the many possible sources of CP violation from BSM physics. 
Even though  the low-energy couplings ($d_q$, $\tilde{d}_q$, ...) 
are correlated in a given underlying model, 
if we want to probe the origin of CP violation, the search for EDMs in  multiple systems is essential. 
(ii)   Even in the unrealistic case in which 
only one source of CPV is active at low-energy, with the exception of $d_q$ the hadronic  uncertainties greatly  dilute the nominal  constraining and diagnosing power of EDM searches (i.e. the 
one obtained by using central values for all matrix elements, ignoring their uncertainty). This of course applies to all hadronic and nuclear EDMs. 
In a realistic situation in which multiple CPV operators are relevant at the hadronic scale, the situation is even worse. 
For example,  \citet{Chien:2015xha} studied the case in which the underlying CP violation originates in the couplings of quarks and gluons to the Higgs boson.  
The  dilution effect comes about  because a given high-energy coupling  generates via renormalization evolution  and threshold corrections  a number of  operators at low-energy, whose contribution can cancel each other due to the poorly known matrix elements. 
The study in \citet{Chien:2015xha} concluded  that 
once matrix elements are known at the  10-25\% level,  
room for cancellations is much reduced  
and one essentially exploits the full power of experimental constraints. 
Therefore 25\% represents a minimal  target uncertainty for hadronic matrix elements 
relating the strange quark EDM and quark / gluon cEDMs to nucleon EDM and CP-violating pion-nucleon couplings. 
This uncertainty might be within reach in the next decade, as discussed below.



\subsection{Lattice QCD input at the hadronic scale}

As described above, effective field theory allows us to express the low-energy effects of the physics beyond the standard model in terms of the matrix elements of a series of operators composed of quark and gluon fields and factors that can be determined reliably in perturbation theory. The estimation of these matrix elements, themselves, cannot be carried out using perturbation theory in the strong coupling since the latter is \(O(1)\) or larger at the relevant scales. Na\"\i{}ve dimensional analysis can provide an order of magnitude estimate for them. For many of these matrix elements, general analyticity arguments can provide `sum rules' involving them, and one can try to use phenomenological analyses and assumptions about resonance contributions to estimate them with somewhat greater precision.  In many cases, moreover, chiral effective field theory can estimate some of the leading contributions to the matrix elements in terms of low-energy constants. The fundamental issue with all these estimates is that the approximations involved are usually uncontrolled and the series involved are not well behaved; resulting in unreliable error estimates and an inability to compare the different methods.

In recent years, LQCD has been able to estimate these matrix elements with far greater control and reliable estimates of the uncertainty.  At heart, LQCD carries out a discretized version of the Feynman path integral numerically using importance sampling through Markov chain Monte Carlo over fields in a small finite universe with possibly unphysical values of the four relevant QCD parameters: the strong coupling, and the up, down, and strange quark masses\footnote{The charm quark is sometimes included, the heavier quarks can be integrated out at the level of accuracy of the nucleon matrix elements we are interested here.}. The results then need to be extrapolated to the continuum and physical values of the QCD parameters. 

To calculate the nucleon matrix elements, one proceeds by putting in interpolating fields creating and destroying nucleons out of the vacuum, propagating them in Euclidean time to remove contaminations from other, possibly multihadron, states coupling to the same interpolating fields, and inserting a discretized version of the operator in the Green's function.  The lattice being a hard cutoff, this operator, regularized through discretization, in general has mixing with operators of the same and lower dimensions that diverge as the continuum limit is taken.  A non-perturbative renormalization procedure is generically necessary to subtract these divergences. Alternatively a smearing called `gradient-flow' can be applied to the discretized matrix elements, which introduces its own hard cutoff scale, but that lets us obtain matrix elements in the continuum. Finally, these continuum quantities have to be converted to the \(\overline{\rm MS}\)-scheme used in phenomenology perturbatively~\cite{Hasenfratz:2022wll,Mereghetti:2021nkt,Kim:2021qae,Cirigliano:2020msr,Bhattacharya:2015rsa,Kniehl:2020sgo,Rizik:2020naq}. This last step forces us to work at a fine enough discretization that one has a window between the scale at which perturbation theory becomes reliable and the scale where discretization (or smearing) artifacts dominate.

One additional systematic arises from the fact that the signal-to-noise for a nucleon state degrades exponentially with the Euclidean time evolution, especially when the quark masses are light.  As a result, the excited state contamination can often not be satisfactorily removed merely by the Euclidean time evolution alone.  Instead, one needs to carry out multi-state fits to remove the remaining contamination and extract the matrix elements. It has, however, been discovered recently that the contamination from the light \(N\pi\) and \(N\pi\pi\) states are often not discernible from these fits alone, leading to additional systematics that are not yet under control.

Fortunately, the quark EDMs  contribution to nucleon EDM is quantified by the tensor charges of the nucleon at leading order, and these seem insensitive to assumptions about the contributing excited state spectrum. Moreover, the renormalization of the tensor charges is multiplicative, thus simplified with respect to higher dimensional operators. As a result, lattice calculations of the light quark tensor charge of the nucleons give results\footnote{We will quote matrix elements of the neutron renormalized in the \(\overline{\rm MS}\)-scheme at \(2~\rm GeV\) at the isospin symmetric limit, unless stated otherwise. The quoted errors represent nominal \(1\sigma\) uncertainties.} with about 5\% precision, and there are bounds on the strange quark contribution~\cite{Aoki:2021kgd}:
\begin{equation}
    g_T^{u-d} = 0.989(34) \qquad g_T^u = 0.784(28)(10)\qquad g_T^d = - 0.204(11)(10)\qquad g_T^s= -0.0027(16)
\end{equation}

There is a long history of lattice calculations of the nucleon EDM due to the QCD topological charge~\cite{Shintani:2005xg,Berruto:2005hg,Shindler:2014oha,Guo:2015tla,Shindler:2015aqa,Alexandrou:2015spa,Shintani:2015vsx}, but they often did not reach statistical precision necessary to obtain a nonzero signal. Recent calculations that extrapolate from heavier quark masses could control the statistical errors giving~\cite{Dragos:2019oxn} \(d_N = -0.00152(71)\) and \(d_P=0.0011(10)\) in units of \(\bar\theta e\;\rm fm\). Other recent calculations~\cite{Alexandrou:2020mds,Bhattacharya:2021lol} with quark masses closer to the physical point have statistical uncertainties larger by an order of magnitude, but have pointed out that the \(N\pi\) contamination may not be under control at these quark masses. Similar situation holds for the calculation of the nEDM due to the isovector quark chromo-EDM operator~\cite{Abramczyk:2017oxr,Kim:2018rce,TanmoyUnpublished}.  The isoscalar quark chromo-EDM operator, which has the additional complication of power divergent mixing with the CPV mass-term, has, so far, not been studied to the same extent, though the coefficient of the  power divergence for the qCEDM operator has become available recently~\cite{Kim:2021qae}.

The techniques for carrying out the matrix element of the CPV Weinberg three-gluon operator have been reported~\cite{Dragos:2017wms,TanmoyUnpublished}. The main issue here is that this operator has mixing with lower-dimensional operators that need to be subtracted.  The power divergent coefficient for the Weinberg operator has been calculated in perturbation theory~\cite{Rizik:2020naq}, but the nonperturbative renormalization of this operator is also complex. The mixing with lower dimensional operators needs to be subtracted nonperturbatively~\cite{Maiani:1991az}. The gradient flow can be used to implement the subtraction and obtain a gauge invariant continuum limit, which can then be connected to the \(\overline{\rm MS}\) scheme perturbatively. 
The full renormalization of the qCEDM and Weinberg operators is still ongoing by several LQCD groups. It is also important to remark, that beside the renormalization pattern, the calculation of the relevant nucleon matrix elements is possibly contaminated by light excited states, a systematic uncertainty that needs to be fully understood for a robust nucleon EDM calculation. The LQCD calculation of the nucleon EDM due to the CPV four-quark operators has not yet been attempted.

To obtain the electric dipole moments of nuclei and atoms, in addition to the calculation of the nucleon EDMs, one needs the matrix elements of the various quark bilinears within the nucleons and the pion-nucleon CPV matrix elements. The former can be used to calculate the various semi-leptonic CPV four-fermion operators, and many lattice groups have calculated these quantities. The FLAG collaboration has calculated the current world averages based on these, and we summarize their findings~\cite{Aoki:2021kgd} here. The tensor matrix element, as discussed above, is known to about 5\% for the light quarks. For the axial charge, the lattice results are at or below the 5\% precision for the light quarks:
\begin{equation}
g_A^{u-d} = 1.246(28)\qquad g_A^u= 0.777(25)(30) \qquad g_A^d = -0.438(18)(30) \qquad g_A^s = -0.053(8)\,.
\end{equation}
For the isovector charge, the phenomenological determination \(g_A^{u-d} = 1.2724(23)\) is an order of magnitude more precise, but the lattice results are consistent with it. The scalar matrix elements are currently known to only about 10\% precision: \(g_S^{u-d}=1.02(10)\).  The individual quark contributions are known only to about 20\% and are usually given by quoting\footnote{Only preliminary calculations are available for the \(u\) and \(d\) constributions separately.} \(\sigma_{\pi N} \equiv m_u g_S^u + m_d g_S^d = 64.9(1.5)(13.2)~\rm MeV\) and \(\sigma_s \equiv m_s g_S^s = 41.0(8.8)\).

The CPV pion-nucleon coupling can also be calculated on the lattice for all these sources of CPV, but no calculations with full control of systematics are available yet. Whereas the latter are given by matrix elements of the electromagnetic current (in presence of the CPV interaction) at zero four-momentum transfer, the CPV pion-nucleon coupling can be extracted from the pion-pole contribution to the matrix element of the axial or pseudoscalar currents.  The axial and pseudoscalar calculations are related by the PCAC relation, and recent lattice developments have resolved the apparent violation of PCAC in the CP-conserving sector~\cite{Jang:2019vkm} as being due to unresolved excited state contaminations.

In summary, LQCD provides a controlled determination for these matrix elements. The matrix elements of the light quark EDM operator and the semi-leptonic four-Fermi operators involving light quark are already sufficient to start discriminating between the various models. For the operators involving strange quarks, for the QCD topological charge operator and for the other BSM operators, however, the precision on the matrix elements extrapolated to the physical point need to be increased by about an order of magnitude, or about 100x in statistics in a brute-force approach, to make full use of the experimental constraints.  In addition, the contamination from light excited states needs to be controlled: some improvement in this is expected if the statistical precision is improved, but currently the known ways of attacking this problem directly involve including multihadron interpolation operators as sources and sinks, increasing the computational cost by another order of magnitude. Alternatively, since the signal-to-noise problems in nucleon two-point functions comes from large phase fluctuations, methods like contour deformations that reduce these fluctuations~\cite{Detmold:2021ulb,Detmold:2020ncp} may ultimately prove to be useful in giving us reliable signal up to Euclidean time separations where the excited state effects can be measured and removed.

\subsection{Nuclear, atomic, and molecular systems}    
    The EDMs of light nuclei are interesting for both theoretical and experimental reasons. Proposals have existed for some time now, to measure the EDMs of light atomic nuclei in storage rings \cite{edm_proposal,CPEDM:2019nwp}. This is an exciting prospect as the nuclear EDMs, unlike atomic systems, do not suffer from Schiff screening. On the theoretical side, EDMs of light nuclei are attractive as they can be tackled with firm theoretical tools. Nowadays EDMs of light nuclei are fully calculated with CP-even and CP-violating chiral EFT nucleon-nucleon potentials \cite{deVries:2011an,Bsaisou:2014zwa,Yang:2020ges,Froese:2021civ}. In particular, the contributions from the constituent nucleon EDMs  and from CP-odd pion-exchange nuclear forces are known to good accuracy. These computations already show the discriminating power of EDMs of different systems. For instance, the ratio of the deuteron-to-neutron EDM could already exclude the QCD $\bar \theta$ term as the dominant source of CP violation \cite{Lebedev:2004va,Dekens:2014jka}. Open questions remain regarding the role of short-distance CP-odd nucleon-nucleon operators whose role is not well understood. While not the target of experimental investigations, EDMs of intermediate mass nuclei could provide interesting theoretical laboratories and form a bridge between ab initio chiral EFT computations and many-body methods suitable for heavier nuclei. The proton and neutron EDM are discussed in detail in  \cref{sec:proton,sec:neutron}, respectively.

In the arena of atomic and molecular EDMs, 
current experimental efforts focus on systems with large atomic number, which complicates 
the nuclear physics input. 
In fact, a big outstanding question concerns the computation of EDMs of nuclei, atoms, and molecules in terms of the effective CP-odd hadronic and (semi-)leptonic interactions. This step is of key importance in the program of connecting observables to the underlying CP-violating parameters of BSM scenarios. 
For paramagnetic systems, the largest contributions arise from the electron EDM and CP-violating electron-nucleon interactions 
(which can also be induced by high-scale purely hadronic CP-violating interactions~\cite{Flambaum:2019ejc,Flambaum2020Hadronic2,Flambaum2020Internucleon}). The associated atomic and molecular matrix elements can be computed with many-body QED methods and are relatively well under control. 

The computation of the nuclear Schiff moments that can give rise to EDMs in atoms and molecules that contain heavy nuclei is another story. Here the strong nuclear many-body problem must be solved, at least approximately.  Two basic approaches have been taken so far: the nuclear shell model \cite{Caurier05} and nuclear density functional theory (DFT) \cite{Schunck19}.  The first method, which requires that the nucleus be not too strongly deformed, has been applied several times to the Schiff moment of ${}^199$Hg, most recently in \citet{Yanase20}, and the second has been applied both to that isotope and light actinides such as ${}^{225}$Ra \cite{Dobaczewski2005,Dobaczewski2018}.  The results exhibit fairly large discrepancies, pointing to a problem with these phenomenological approaches: there is no way to reliably estimate uncertainty.  The dependence of results on model parameters can be examined, as can correlations of the Schiff moment with other observables but ``systematic" uncertainty associated with the model itself is hard to assess.

For this reason, attention in recent years as turned to ab initio methods, in which, nominally at least, one can estimate error at every step.  Two developments led to the
quick growth of ab initio theory: the arrival  of 
chiral (EFT) and the creation
of several nonperurbtive many body methods for addressing nuclear structure that 
allow  controlled approximations despite the absence of small parameters.
Chiral EFT provides a complete set of operators 
at each order in $p/\Lambda$ or $m_\pi/\Lambda$, where $p$ is a nucleon momentum
and $\Lambda$ is the breakdown scale of the theory. 
All nuclear operator can then be expressed as a sum of
operators in the complete set, with a finite number of coefficients
(at each order) multiplying them.  The power counting is guaranteed to
work only in perturbation theory, but it does quite well in practice in
non-perturbative many-body computations. 

Two ab initio methods that make use of operators from chiral EFT have been developed to the point that they will be useful for the computation of Schiff moments:  coupled-clusters theory and the in-medium similarity renormalization group
Coupled clusters theory is based on a form for the ground
state wave function of a nucleus in which one- and two-body operators repeatedly move nucleons from occupied to
empty orbitals.   The method can be extended to three- and more-body operators, leading to a limit in which it is exact.  Estimates of the effects of the higher-body operators allow make uncertainty analysis possible. 
The method has been used mainly to compute nuclear
spectra and transition rates but also is being applied to other phenomena, e.g., for photo-excitation reactions
\cite{Hagen14,Bonaiti21}.  Recently, it was used in a calculation of the
the neutrinoless double-beta decay matrix element for $^{48}$Ca \cite{Novario21}, an observable that has features in common with Schiff moments 

The IMSRG works by including correlations into 
effective Hamiltonians and other operators rather than into nuclear states.
To do so, the IMSRG solves flow equations that gradually transform the operators so that to work with  pre-specified and simple ``reference state" (a Slater determinant, for example) in place of the actual ground state. 
The flow
equations drive the Hamiltonian into a from that decouples the reference state from others, so that the reference state
is the ground state of a transformed Hamiltonian.  The method is more flexible than coupled clusters because one can choose the reference state independently; it doesn't have to be a Slater determinant and can even be an ensemble of states.  Like coupled clusters theory, the IMSRG has been used to compute to many observables, including binding and excitation energies, transition rates, and nuclear radii, in medium mass nuclei \cite{Hergert16,Hergert17,Stroberg19}, and it has also been applied to double-beta decay \cite{Yao20,Belley21}.)  

One version of the IMSRG, the In-Medium Generator Coordinate Method (IM-GCM), builds the reference state from 
a deformed mean-field state, projected onto good
angular momentum.  The sophisticated reference state makes the method particularly able to describe collective properties such as E2 transitions \cite{Yao20} and will be useful in computing the enhanced Schiff moments of octupole-deformed nuclei such as ${}^{225}$Ra.  Another variant, the valence-space IMSRG, uses valence single-particle spaces to construct reference ensembles, leading to shell-model-like calculations with rigorously derived effective operators.  This approach will be useful in nuclei such as $^{199}$Hg that are complicated but not too deformed. 

The main problem in applying any of these methods right away to calculate Schiff moments in heavy nuclei is the need for large amounts of computing time and memory. 
The practitioners of coupled-clusters theory have already taken steps to rewrite their codes to exploit accelerators such as GPUs, but the researchers using the IMSRG have yet to take those steps.  The availability of convenient computing resources will thus be crucial in advancing the computation of Schiff moments in the next few years.

Atomic and molecular EDMs are discussed in detail in \cref{sec:amo}.


\subsection{Discovery and diagnosing power: 
the importance of multiple probes}     

Ultimately, the goal of the EDM search program is to discover new physics and learn about the underlying sources of CP violation, 
and whether or not they are related to the baryon asymmetry of the universe.   To achieve this goal, the search for EDMs in multiple systems and the associated theoretical developments are crucial. 

{\it Discovery potential} -- 
Since for different underlying sources of CP violation the 
hierarchy and pattern  of EDMs can be quite different, 
to maximize the discovery potential it is mandatory to 
push the experimental sensitivity in multiple systems. 

The discovery potential in EDM searches can be roughly quantified by the reach in mass scale, assuming maximal CP violation. 
This was already discussed in \cref{sect:th1} using the 
electron EDM as an example. 
\Cref{tab:scales}   gives a crude estimate of the mass  reach of different operators (concrete models will have additional $O(1)$ factors modifying these estimates, if not larger factors due to, e.g., flavor violation or suppressed CP phases). The EDMs follow Eq.~\eqref{eq:crudeestimate}, with $g = 0.6$ (of order a weak coupling) for the electron but $g = 1.2$ (approximately $g_s(m_Z)$) for quarks. In each case we assumed maximal CP violation and no additional flavor factors.  The running up and down quark masses at the weak scale are estimated as $1.5\,\mathrm{MeV}$ and $3\,\mathrm{MeV}$ respectively, so they have equal $q_f m_f$ and appear in one entry for the EDM. The quark CEDM is estimated with a color factor of $4/3$ and otherwise similar scaling to~\eqref{eq:crudeestimate}. The gluon CEDM (Weinberg operator) does not arise from 1-loop renormalizable gluon couplings, so we assume it arises at two loops, scaling as $\frac{g_s^5}{(16\pi^2)^2} \frac{m_f}{M_\mathrm{NP}^3}$. We show two possibilities, one where $m_f$ is the top quark mass (this  arises in SUSY from a top/stop/gluino diagram) and one where $m_f = M_\mathrm{NP}$. Note that the former is the only entry in the table for which the reach scales with the {\em cube} root of the operator coefficient.  For $d_e$, $d_q$, and $\tilde{d}_q$ 
we have taken a target  sensitivity of $10^{-29}\,\mathrm{cm}$. For the Weinberg operator we further divide by $100\,\mathrm{MeV}$ (by crude dimensional analysis, this is roughly how the neutron EDM scales relative to the gluon CEDM).
These choices assume that the 
physical EDM sensitivities will reach  $10^{-29}\,e\,\mathrm{cm}$ 
and that the  hadronic matrix elements 
connecting the hadronic EDMs to  $d_q$, $\tilde{d}_q$, $\tilde{d}_G$  are of O(1). 

In all cases we see that the mass reach is very high -- EDMs are exploring uncharted territory. 
Alternatively, if we insist that the scale of new physics is close to the electroweak scale, EDMs probe very small CP-violating couplings, still providing invaluable information for model building and understanding  the nature of CP symmetry and its breaking.

\begin{table}
\begin{center}
\begin{tabular}{ c|c|c } 
 Operator & Loop order & Mass reach \\ 
 \hline
 Electron EDM & 1 & $48\,\mathrm{TeV} \sqrt{10^{-29}\,e\,\mathrm{cm} / d_e^\mathrm{max}}$ \\ 
  & 2 & $2\,\mathrm{TeV} \sqrt{10^{-29}\,e\,\mathrm{cm} / d_e^\mathrm{max}}$ \\ 
 Up/down quark EDM & 1 & $130\,\mathrm{TeV} \sqrt{10^{-29}\,e\,\mathrm{cm} / d_q^\mathrm{max}}$ \\ 
  & 2 & $13\,\mathrm{TeV} \sqrt{10^{-29}\,e\,\mathrm{cm} / d_q^\mathrm{max}}$ \\ 
 Up-quark CEDM & 1 & $210\,\mathrm{TeV} \sqrt{10^{-29}\,\mathrm{cm} / {\tilde d}_u^\mathrm{max}}$ \\ 
  & 2 & $20\,\mathrm{TeV} \sqrt{10^{-29}\,\mathrm{cm} / {\tilde d}_u^\mathrm{max}}$ \\   
 Down-quark CEDM & 1 & $290\,\mathrm{TeV} \sqrt{10^{-29}\,\mathrm{cm} / {\tilde d}_d^\mathrm{max}}$ \\ 
  & 2 & $28\,\mathrm{TeV} \sqrt{10^{-29}\,\mathrm{cm} / {\tilde d}_d^\mathrm{max}}$ \\  
 Gluon CEDM & 2 ($\propto m_t$) & $22\,\mathrm{TeV} \sqrt[3]{10^{-29}\,\mathrm{cm}/(100\,\mathrm{MeV}) / {\tilde d}_G^\mathrm{max}}$ \\ 
 & 2 & $260\,\mathrm{TeV} \sqrt{10^{-29}\,\mathrm{cm}/(100\,\mathrm{MeV})/ {\tilde d}_G^\mathrm{max}}$ \\
 \hline
\end{tabular}
\end{center}
\caption{Crude estimate of the mass  reach of different operators. 
See text for explanation of the notation and assumptions used in deriving the estimates.
\label{tab:scales}}
\end{table}

\begin{figure}
    \centering
  \includegraphics[scale=0.4]{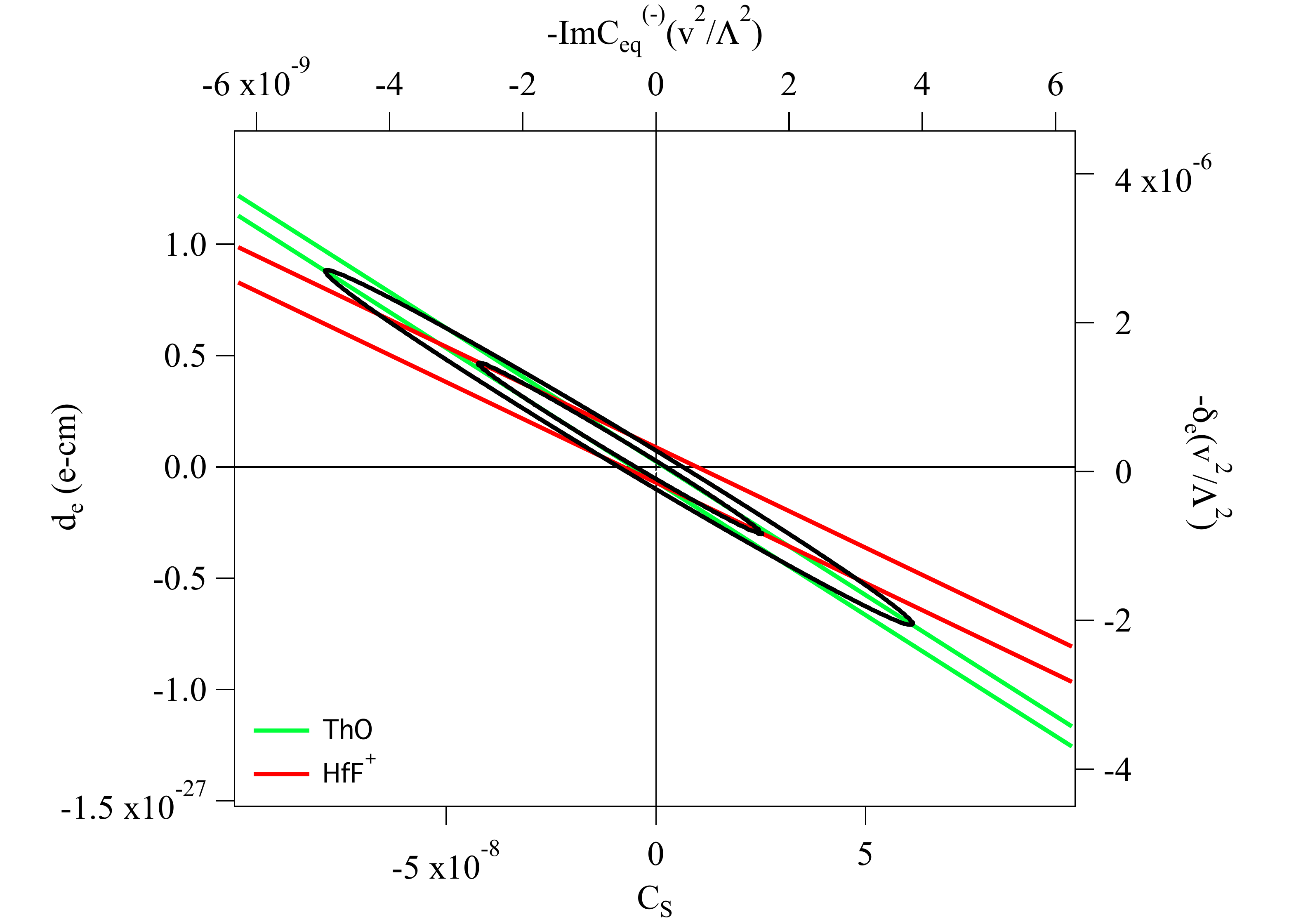}
    \caption{Experimental constraints on the electron EDM 
    $d_e$ and the semi-leptonic CP-violating coupling  $C_S$ from the experimental results in ThO and HfF$^+$ molecules,  
    with 1$\sigma$ experimental error bars. Also shown are the  68\% and 95\% $\chi^2$ contours for all paramagnetic systems,  including Cs, Tl, YbF.  The top and right axes show the  dimensionless Wilson coefficients $\delta_e$ and $\mathrm{Im}\, C_{eq}^{(-)}$ normalized to the squared scale ratio $(v/\Lambda)^2$, as defined in \citet{Chupp:2017rkp}. 
    The figure is taken from \citet{Chupp:2017rkp}.     }
    \label{fig:paramagnetic}
\end{figure}

\begin{figure}
    \centering
  \includegraphics[scale=0.33]{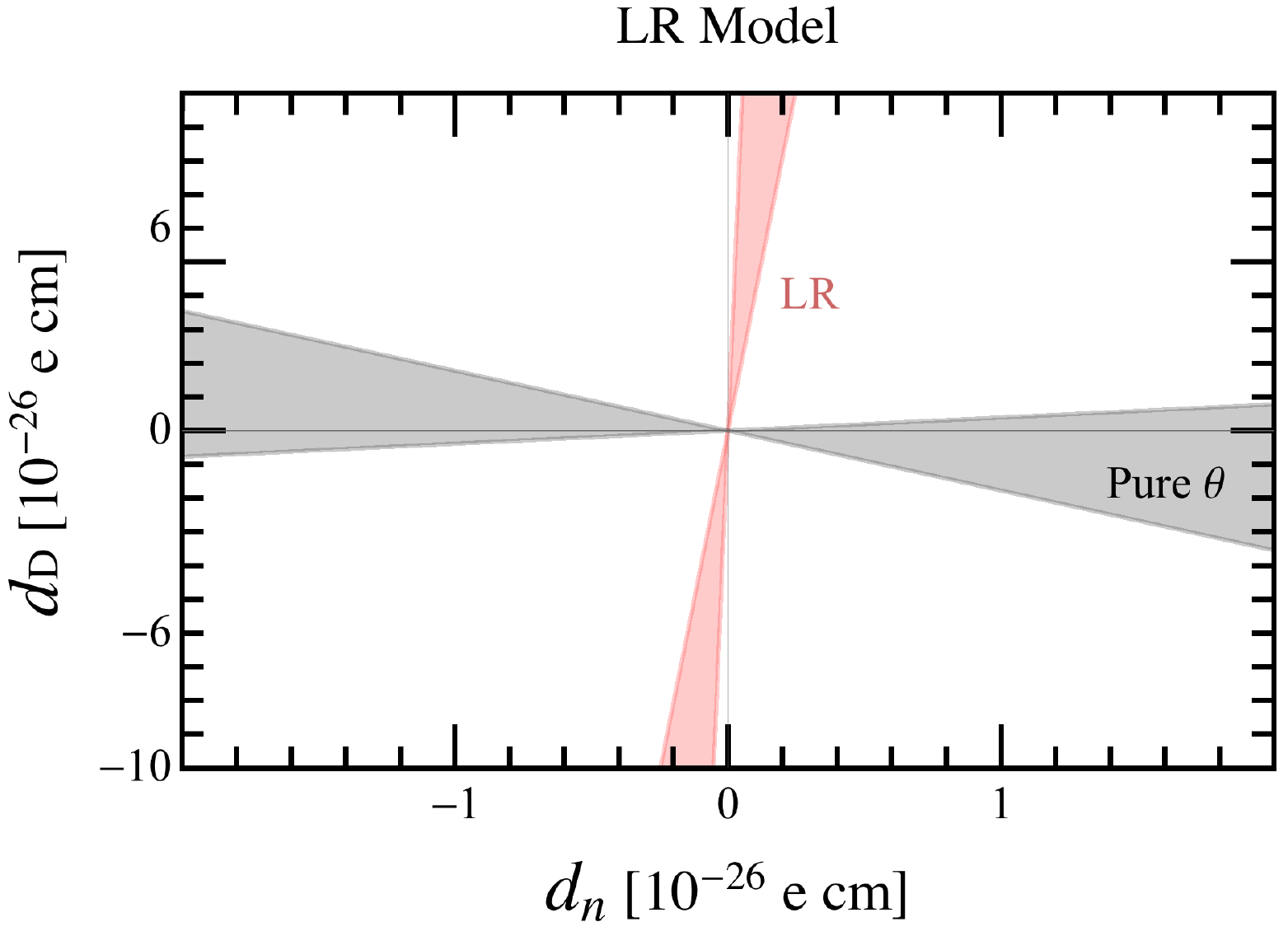}
\includegraphics[scale=0.33]{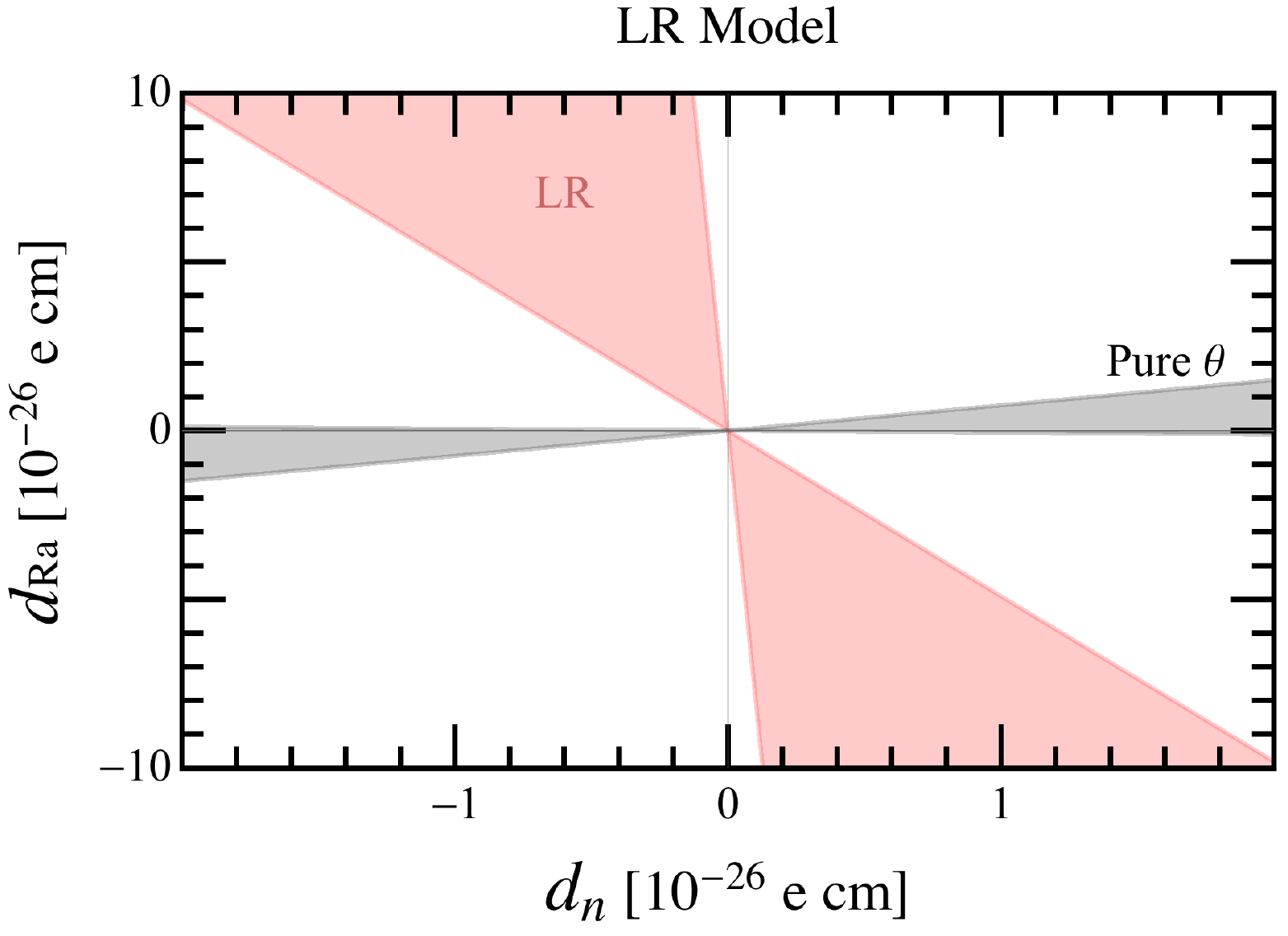}
\includegraphics[scale=0.33]{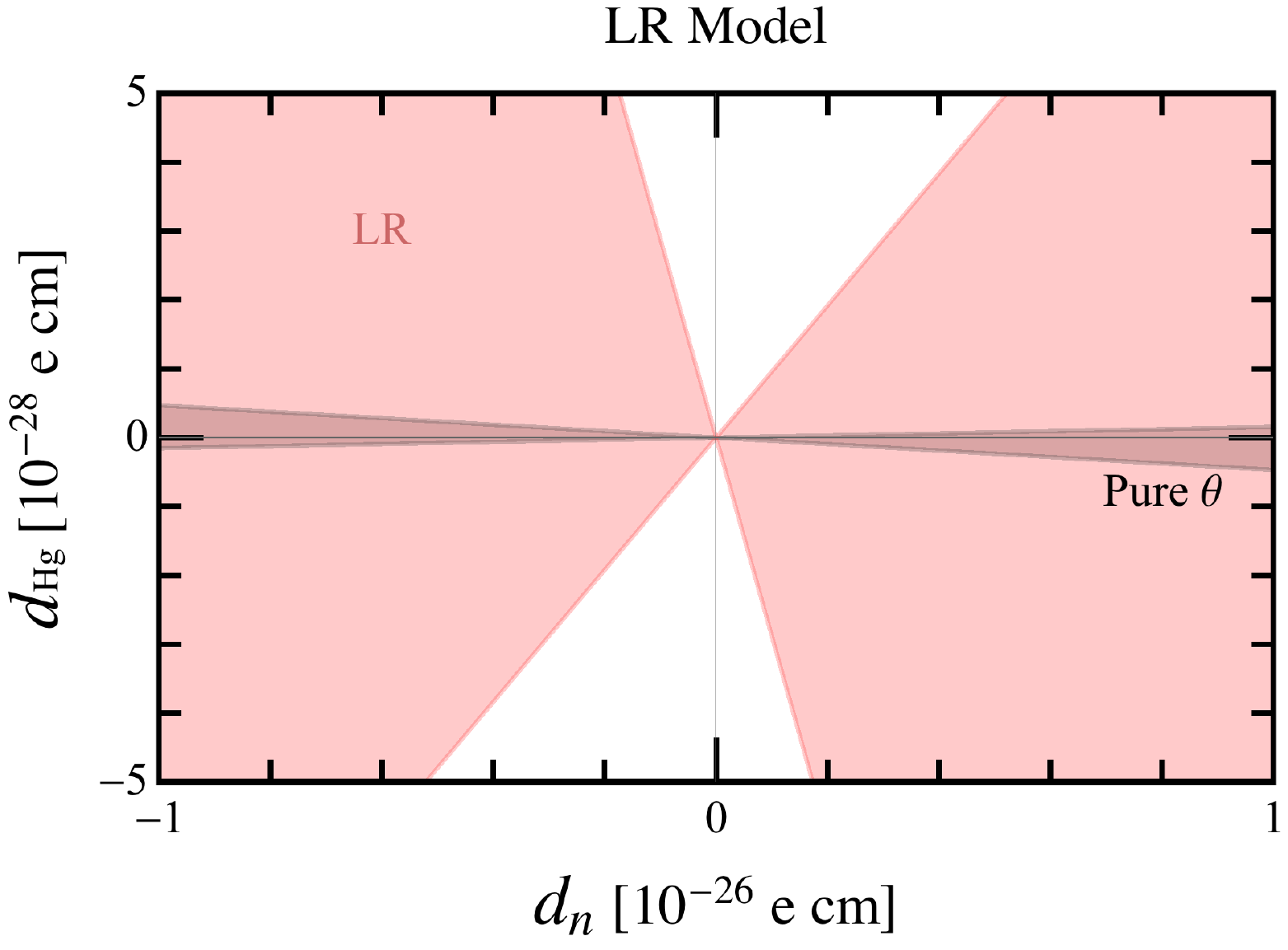}
    \caption{Correlation between various EDMs and the neutron EDM for a pure-$\bar \theta$ scenario (gray) and the mLRSM (red). The bands indicate the uncertainty in the ratios arising from hadronic and nuclear matrix elements.}
    \label{fig:EDMShifts}
\end{figure}

{\it Diagnosing power} -- 
While the  discovery of a non-zero EDM in a single system will be 
paradigm-shifting, it will not be sufficient to learn about the underlying theory of CP violation. 
To achieve this one needs to study the correlation among measurements in different systems. 
We illustrate this with two examples: 

\begin{itemize}
    \item EDMs of paramagnetic systems are sensitive not only to the electron EDM $d_e$ but also to a CP-violating  four-fermion operator coupling the electron spin to the quark scalar density, with coefficient $C_S$. 
    Disentangling between the two requires at least two measurements. 
    \Cref{fig:paramagnetic}, taken from \citet{Chupp:2017rkp}, 
    illustrates the experimental constraints in the $d_e$-$C_S$ plane. Observation of a non-zero result in at least two molecular systems will be needed to determine the electron EDM $d_e$ and $C_S$ separately.

    \item Hadronic and nuclear EDMs: 
\Cref{fig:EDMShifts} \cite{deVries:2021sxz} shows the correlation 
among hadronic EDMs in  a pure $\bar \theta$ scenario and a specific BSM scenario, the minimal left-right symmetric model  (LRSM). The left panel shows  the prediction of the deuteron-versus-neutron EDM, while the middle  and 
right panel shows the neutron vs Ra and Hg, respectively,  
where the uncertainties are larger and the diagnosing power largely diluted. 
Measurements of EDM ratios outside the gray bands would indicate a non-$\bar \theta$ source of CP violation, showing the power of multiple measurements.

\end{itemize}

\section{Neutron EDM}
\label{sec:neutron}
Neutron is a convenient probe for EDM. Because of its neutral charge, neutrons can be subject to large electric fields (in a beam or in a bottle), without impacting their motions, to measure the EDM.
The full strength of the applied electric field is experienced by the neutrons without shielding from the polarizable electron cloud, an effect that limits the EDM sensitivity in light atoms.
Neutrons have a spin of 1/2 and can be readily spin polarized using polarizing mirrors (for neutron beam) or superconducting magnets via the 
${\bm{\mu}}\cdot {\bm{B}}$ 
potential (for ultracold neutrons). Using polarized neutrons, many important studies on the parity-violating phenomena in formulating the electroweak theory have been carried out, including the search for the neutron EDM (nEDM)~\cite{EPurcell1950}.
Although the initial attempt on nEDM failed to reveal the anticipated parity non-conservation, the non-observation of the nEDM severely suppresses the QCD $\bar \theta$ term, giving rise to the strong CP problem. Various BSM models are being constrained by comparing to the ever-increasing precision of nEDM measurements.
Today, motivated by the need to find new sources of CP violation for baryogenesis, the community of fundamental neutron physics has been stepping up the effort worldwide to search for nEDM.

\subsection{Brief history}
\subsubsection{Beam experiments}
Searches for the neutron EDM (nEDM) began in 1949 by Purcell and Ramsey~\cite{Smith1957} at the Oak Ridge reactor. A beam of neutrons, polarized using a magnetized iron mirror, passed through a region of magnetic and electric fields; upon exit, the spin state of the neutrons was analyzed by reflecting off another magnetized iron mirror. The nEDM was measured through the nuclear magnetic resonance (NMR): EDM interactions induced a shift in the Larmour precession frequency when the neutron was subject to a {\it strong} external electric field in parallel with a {\it weak} magnetic field.  In the attempt to improve the NMR signal, Ramsey developed the well-known method of separated oscillatory fields~\cite{Ramsey1950}: by separating the oscillating field that rotated the neutron spin into two regions, a much narrower resonance was achieved. The resulting interferogram contained narrow fringes, the width of which scaled inversely with the duration of free spin precession. The precision in the precession frequency measurements could be enhanced by increasing the length of the region of the interaction fields and by using slower neutrons.

Continued beam experiments in 1969 in the Brookhaven high flux beam reactor revealed a systematic effect associated with the motional field~\cite{Cohen69, Baird69}. 
As polarized neutrons moved through the interaction region of static electric and magnetic fields, they experienced an additional magnetic field, $\frac{v}{c} \times E$, arising from the relativistic effect, first pointed out in the context of an atomic EDM search~\cite{Sandars1964}. Coupled with a small, but inevitable misalignment (of a few mrad) between the $E$ and $B$ fields, the strength of the total magnetic field experienced by the neutrons varied upon the field reversals. This effect gave rise to a systematic shift to the precession frequency that limited the EDM sensitivity to $10^{-21}\, e\cdot\rm cm$. Subsequent improvements involved routinely rotating the whole apparatus to reverse the direction of the neutron beam through the apparatus; the best nEDM sensitivity using neutron beams was carried out at the ILL high flux reactor in 1977 and achieved $3\times 10^{-24}\, e\cdot\rm cm$~\cite{Dress77}.

\subsubsection{UCN experiments}
Ultracold neutrons (UCN) have velocities up to 5~m/s, wavelengths of 500~$\text{\AA}$, and an effective temperature of order 3~mK. These neutrons can be easily polarized and stored inside room-temperature material bottles for several hundreds of seconds.
Ramsey's method of separated oscillatory fields, applied to a bottle of stored UCN in a precisely controlled magnetic field, resulted in superb interferograms with extremely narrow fringes, giving orders of magnitude enhancement in the precision of resonant frequency measurements. 
To measure the nEDM, an additional large electric field ($E$) is applied parallel to the magnetic field ($B$);  
the neutron precession speeds up (or slows down) in proportion to the strength of both the applied electric field and the size of the nEDM.  The precession frequency is $\nu = -2 (\mu \cdot B + d \cdot E)/h.$ The change in the Larmor precession frequency between the two configurations with the $B$ and $E$ fields parallel and anti-parallel measures the nEDM. 
The nEDM sensitivity per cycle is 
\begin{equation}
\sigma_{d_n} \approx \frac{\hbar}{2 \alpha E T_{fp} \sqrt{N}},
\label{eq:EDMsensitivity}
\end{equation}
\noindent where $\hbar$ is Planck's constant, $\alpha$ is the visibility of the Ramsey fringes, $E$ is the strength of the electric field, $T_{fp}$ is the time allowed for neutron free precession, and $N$ is the number of neutrons per cycle counted by the UCN detector in the spin analyzer. Because the velocity of the UCN in a bottle averages to 0, i.e., $\langle v \rangle =0$, the dominant systematic effect of motional field associated with nonparallel $E$ and $B$ fields is greatly suppressed.

The technique of UCN storage has been developed for nEDM measurements for decades at the Institut Laue-Langevin (ILL)~\cite{SMITH1990, Baker2006} and the Petersburg Nuclear Physics Institute (PNPI)~\cite{Serebrov2014}. 
The best experimental limit was reached recently by the group at the Paul Scherrer Institut (PSI). Using the inherited Sussex-Rutherford-ILL apparatus with substaintial hardware improvements, including modernizing the magnetometry needed to control the next dominant systematic effect of the geometric phase~\cite{Pendlebury2004}, they reported in 2020 a new nEDM result of $(0.0 \pm 1.1_{\rm stat} \pm 0.2_{\rm sys})\times 10^{-26}\, e\cdot\rm cm$~\cite{Abel2020}, corresponding to a limit of $|d_n| < 1.8\times 10^{-26}$~$e\cdot$cm (90\% C.L.).

Achieving this sensitivity required exquisite knowledge of the magnetic fields. It was only realized with the introduction of a comagnetometer, first implemented in the ILL experiment~\cite{Pendlebury1992}. Polarized $^{199}$Hg (in a small fraction of its vapor pressure) was introduced to co-habit the neutron precession chambers and monitor the temporal variations of the ambient magnetic fields. The $^{199}$Hg atoms are insensitive to the electric field, as indicated by its null EDM~\cite{BGraner2016}. They track the same variations of the magnetic field experienced by the neutrons. Fluctuations of the neutron precession frequency resulted from varying magnetic fields can thus be removed by what is inferred using the precession signal of Hg atoms.

A magnetic field gradient of $\sim$1 pT/cm (or 0.1~nT/m) could cause a false-EDM in the UCN of magnitude $|d_{\rm false}|\sim 10^{-28}\, e\cdot\rm cm$~\cite{Lamoreaux2005}. Currently, the room-temperature nEDM experiments use laser-polarized $^{199}$Hg as comagnetometer. 
However, the Hg atoms also suffers a  false-EDM, that is larger than that on neutrons because of their higher average thermal speed. The gradient has to be controlled and measured precisely to allow for corrections of the Hg false-EDM. 
In addition, the center of mass of the UCN in the cell is somewhat lower than that of the $^{199}$Hg comagnetometer, a temporal change in the vertical static gradient between measurement cycles with field reversals could not be fully tracked by the comagnetometer and corrected in the analysis. 
The PSI group measured the UCN height spectrum and residual gradients in a storage cell by using spin-echo spectrometry~\cite{PSI3}. Reversing the $B_0$ field and the associated gradients assesses the size of the vertical displacement and the Hg geometric phase to first order~\cite{Pendlebury2015}. Higher order effects might emerge as the sensitivity improves.

Despite the many advantages of UCN to improve nEDM measurements, the main bottleneck limiting the progress of nEDM experiments worldwide is the density of UCN.  
In the US, its first UCN facility was developed in 2000 and became operational in 2005 at the Los Alamos Neutron Science Center (LANSCE). Other UCN-producing facilities hosting active physics measurement programs are the ILL (in France) and PSI (in Switzerland). Another UCN facility coming online is at TRIUMF (in Canada).

\subsection{Experimental considerations}
\subsubsection{Statistical}

As discussed earlier, most of modern nEDM experiments are performed using UCN (see, however, \citet{Piegsa2013}). As seen in Eq.~\ref{eq:EDMsensitivity}, the number of neutrons counted per cycle is an important factor contributing to the statistical sensitivity of an nEDM experiment. 
For decades the UCN source at the Institut Laue Langevin (ILL)~\cite{Steyerl1986} has provided UCN to various UCN based experiments, including nEDM experiments (e.g. Refs.~\cite{SMITH1990,Baker2006,Serebrov2015, Pendlebury2015}). This source produces UCN by slowing down vertically-extracted cold neutrons using receding blades. However, in this case, the achievable UCN density is limited by Liouville's theorem. By employing the so-called superthermal method~\cite{Golub1975}, it is possible to surmount this limit. In this method, a UCN converter, which has a temperature much higher than that of UCN ($\sim$mK) but lower than that of typical cold neutrons (several 10's of kelvin), converts cold neutrons to ultracold neutrons. Incident neutrons are downscattered with their kinetic energies dissipated into the bulk medium to create phonons.
The reverse process of neutron upscattering (i.e., gaining energy from coupling to the thermal bath of phonons) is heavily reduced as the population number of phonons is suppressed by the Boltzman factor $\exp(-\Delta/k_B T)$, where $\Delta$ is the energy difference between cold and ultracold neutrons and $T$ is the temperature of the converter. 

There are many new UCN sources around the world currently operational, being developed, or proposed~\cite{Zimmer2016, Anghel2009, Ito2018, SAhmed2019,  Korobkina2014, Kahlenberg2017, FRMII, Serebrov_2017, Leung2019, Shin2021} based on the principles of the superthermal process.
Superfluid liquid helium (LHe) and solid deuterium (SD$_2$) have been successfully applied as UCN converters. 
In both cases, the interaction between cold neutrons and phonons in the converter converts cold neutrons to UCN. The cross section for converting cold neutrons to UCN integrated over the cold neutron energies for SD$_2$ is approximately 10 times larger than that for superfluid liquid helium (see e.g Refs.~\cite{Yoshiki_2003, Frei2010}). However, the lifetime of UCN in SD$_2$ is limited to $\approx$ 40~ms resulted from the nuclear capture, which is absent in isotopically pure LHe.  
In natural LHe, the UCN lifetime can be as long as several hundred seconds.
The choice between SD$_2$ and LHe as UCN converter is made based on the desired characteristics for the applications. \Cref{tab:UCNsources} lists selected UCN sources that are either operational or under development that are hosting an nEDM experiment. 
\begin{table}
    \centering
    \begin{tabular}{l|l|l|l|l|l} \hline\hline
    UCN source & Location & Converter & Neutron source & Status & Ref.\\ \hline
    ILL (Turbine) & ILL & Receding blades & Reactor & Operational & \cite{Steyerl1986} \\
    SuperSUN & ILL & LHe & Reactor & Commissioning & \cite{Zimmer2016} \\
    PSI & PSI & SD$_2$ & Spallation & Operational & \cite{Anghel2009,Becker2015,Bison2020}\\
    LANL & LANL & SD$_2$ & Spallation & Operational & \cite{Ito2018}\\
    TRIUMF & TRIUMF & LHe & Spallation & Under development & \cite{SAhmed2019} \\
    TUM & TUM & SD$_2$ & Reactor & Under development & \cite{FRMII} \\ \hline \hline
    \end{tabular}
    \caption{``General-purpose" UCN sources that are either operational or under development hosting nEDM experiments}
    \label{tab:UCNsources}
\end{table}

For experiments performed in room temperature vacuum, the other two factors affecting the statistical sensitivity in Eq.~\ref{eq:EDMsensitivity}, namely $E$ and $T_{fp}$, cannot be improved dramatically. The strength of the electric field is typically limited to $12-15$~kV/cm. One of the limiting factors is the field emission at the so-called triple junction, where the electrodes, the vacuum, and the insulator (needed to confine UCN between the electrodes) meet. The free precession time $T_{fp}$ is limited to $130-200$~s due to various mechanisms that causes loss of UCN when they come in contact with the wall confining UCN. 

One way to improve both $E$ and $T_{fp}$ is to perform the experiment in superfluid helium. This is the approach taken by the CryoEDM~\cite{Baker2010} and the nEDM@SNS experiment~\cite{Ahmed2019}.  Superfluid LHe was expected to be a better electric insulator than vacuum, thereby making it possible to apply much larger $E$. The design goal for nEDM@SNS is to achieve 75~kV/cm. Promising results have been obtained from their currently ongoing R\&D on generating large electric fields in superfluid liquid helium~\cite{Ito2016,Phan2021}. Performing an nEDM experiment in a cryogenic environment (such as in a bath of superfluid liquid helium) also allows $T_{fp}$ to be increased as some of the mechanisms responsible for UCN loss due to material interactions are suppressed at low temperatures. In the case of the nEDM@SNS experiment, a significant improvement in the neutron statistics $N$ is expected from the in-situ production of UCN directly inside the measurement cells using the superthermal process with superfluid liquid helium.

The idea of beam experiments has also been revisited: the spurious effect of the motional field could be utilized to extract the EDM signal in conjunction with the pulsed structure of the cold neutron beam at a spallation source~\cite{Piegsa2013, Piegsa2019}. Modular units of vacuum chambers containing precision field plates could be assembled together to lengthen the free precession region to tens and hundreds of meters. In this case there is also no need for UCN confining walls, which should allows for a substantial increase in $E$. The European Spallation Neutron Source (ESS) could provide the high-intensity beam of cold neutrons needed to realize these ideas for future nEDM experiments.

\subsubsection{Systematics}
Improved statistical sensitivity calls for a better control of possible systematic effects. The history of the nEDM search has been a history of uncovering and controlling new systematic effects as well as that of continuously improving statistical sensitivity. 

Given the much stronger response to a magnetic field than that to an electric field (e.g. the precession frequency is 30~Hz for a 1~$\mu$T field while an EDM of $d_n = 10^{-27}\, e\cdot\rm cm$ in a 15~kV/cm electric field causes a precession of 7~nHz), it is of paramount importance to monitor the stability and uniformity of the magnetic field. For example, change in the strength of the magnetic field correlated with the direction of the electric field (such as the case for leakage current) causes an effect that mimics an nEDM signal. For this reason, many of the modern nEDM experiments employ a comagnetometer, a magnetometer that consists of a spin polarized species that occupy the same volume as the stored UCN, thus providing the spatial and temporal average of the magnetic field experienced by the stored UCN. $^{199}$Hg has been the choice for many experiments. $^{129}$Xe has been considered for the Tucan experiment at TRIUMF. The nEDM@SNS experiment will use $^3$He. 

Currently the largest known systematic effect is the so-called geometric phase effect. This effect, which gives a shift in precession frequency linear in $E$, arises as a result of interplay between nonuniformity of the magnetic field and the motional magnetic field. It was first pointed out in the context of EDM experiments using atomic beams~\cite{Commins1991,Regan2002}. The first observation for confined particles, which was made in an nEDM experiment~\cite{Baker2006}, was reported in \citet{Pendlebury2004}. The geometric phase effect affects both neutrons and the comagnetometer. Its size depends on the particle's trajectory, which is characterized by the position-velocity correlation functions. Typically, the effect is much bigger for the comagnetomter as the effect is proportional to the velocity of the particle. Since this is currently a dominant source of systematic effect, intensive studies have been performed by several collaborations~\cite{Lamoreaux2005,Barabanov2006,Harris2006,Clayton2011,Pignol2012,Pignol2015,Swank2016,Pignol2019}.

\subsection{Experiments being developed}

\begin{table}[htbp]
    \centering
    \begin{tabular}{p{0.15\linewidth} | p{0.1\linewidth} | p{0.20\linewidth} | p{0.4\linewidth} |  p{0.05\linewidth}} \hline\hline
    Experiment & Location & UCN source & Features &  Ref. \\ \hline
    n2EDM & PSI & Spallation, SD$_2$ & Ramsey method, double cell, $^{199}$Hg comagnetometer &  \cite{PSI_apparatus_2021}\\ \hline
    
   PanEDM & ILL & Reactor, LHe & Ramsey method, double cell, $^{199}$Hg comagnetometer & \cite{Wurm2019} \\ \hline  
    
   LANL nEDM & LANL & Spallation, SD$_2$ & Ramsey method, double cell, $^{199}$Hg comagnetometer &  \cite{Ito2018}\\ \hline

 Tucan & TRIUMF & Spallation, LHe & Ramsey method, double cell, $^{129}$Xe comagnetometer &   \cite{Martin2020} \\ \hline
   
    nEDM@SNS & ORNL & In-situ production in LHe & Cryogenic, double cell, $^3$He comagnetometer, $^3$He as the spin analyzer & \cite{Ahmed2019} \\ \hline\hline
    \end{tabular}
    \caption{A list of the nEDM experiments that are being developed}
    \label{tab:nEDMexperiments}
\end{table}

Several nEDM experiments are being developed around the world. Tab.~\ref{tab:nEDMexperiments} lists the ongoing efforts.
The PSI collaboration inherited the ILL apparatus in 2006 and subsequently made technical improvements to nearly all elements of the system, in particular adding the capabilities of field trimming, atomic magnetometry, and modern NMR techniques including spin echo~\cite{PSI2, PSI3}.  
In 2020, the PSI group published a new nEDM measurement of $(0.0 \pm 1.1_{\rm stat} \pm 0.2_{\rm sys})\times 10^{-26}\, e\cdot\rm cm$~\cite{Abel2020}. 
This result includes a moderate statistical improvement and a factor of 5 reduction in systematic errors. 
Built on the experience in the past decade, the collaboration is putting together the n2EDM apparatus, which incorporates the design of a double chamber for spin precession---an idea first implemented in the PNPI experiment~\cite{Serebrov2015}. 
In the double-chamber geometry, high voltage is applied to the central electrode to create electric fields in opposite directions in the two cells, and a common magnetic field is applied parallel/anti-parallel to the electric fields.  The spin precession frequencies for UCN in the two chambers thus differ only by the interaction of EDM coupled to the applied $E$ fields. This significantly reduces the impact of temporal drift of the ambient magnetic fields.
The double-cell configuration also increases the overall neutron counting statistics.
The majority of the current nEDM experiments use a ``room temperature" apparatus, applying the Ramsey's separated oscillatory fields method, with external neutron polarimetry. 
In addition to n2EDM at PSI, other ongoing efforts include the double-chamber nEDM experiment led by the PNPI group using the ILL turbine source~\cite{Serebrov2015}, the panEDM experiment using the new SuperSUN source at ILL~\cite{Wurm2019}, the TUCAN experiment using the superfluid helium UCN source being developed at TRIUMF~\cite{Martin2020}, and the LANL nEDM experiment~\cite{Ito2018}.
Many of the contemporary nEDM experiments also make use of large-scale magnetically shielded rooms (MSR): multi-layers of nested shells with alloys of high magnetic susceptibilities are used to suppress the ambient fields by a factor of 100,000 to a million~\cite{Altarev:2015fra}. The MSRs bring the stabilities of the magnetic field to the level needed to reveal and mitigate subtle systematic effects associated with residual field gradients.

At LANSCE, the newly-completed upgrade of the UCN facility~\cite{Ito2018}
provides the necessary UCN density to meet the demand of a nEDM experiment with tenfold sensitivity improvement. A factor of 5--6 increase in the UCN output has already been achieved (as measured both in the UCN$\tau$ experiment~\cite{Gonzalez2021} and in a nEDM test apparatus~\cite{Ito2018}). 
The LANL nEDM experiment takes the same Ramsey approach by using a room-temperature apparatus coupled to the newly-upgraded, solid deuterium-based UCN source. 
The apparatus operates in vacuum and uses the Ramsey's method of separated oscillatory fields, which is a mature technology developed in prior nEDM experiments~\cite{Pendlebury2000, Baker2006}. 
The low-risk technology together with the high-yield UCN source at LANL opens up a timely opportunity to substantially increase the nEDM sensitivity before the nEDM@SNS experiment becomes fully operational.

\begin{figure}
    \centering
    \includegraphics[width=0.9\textwidth]{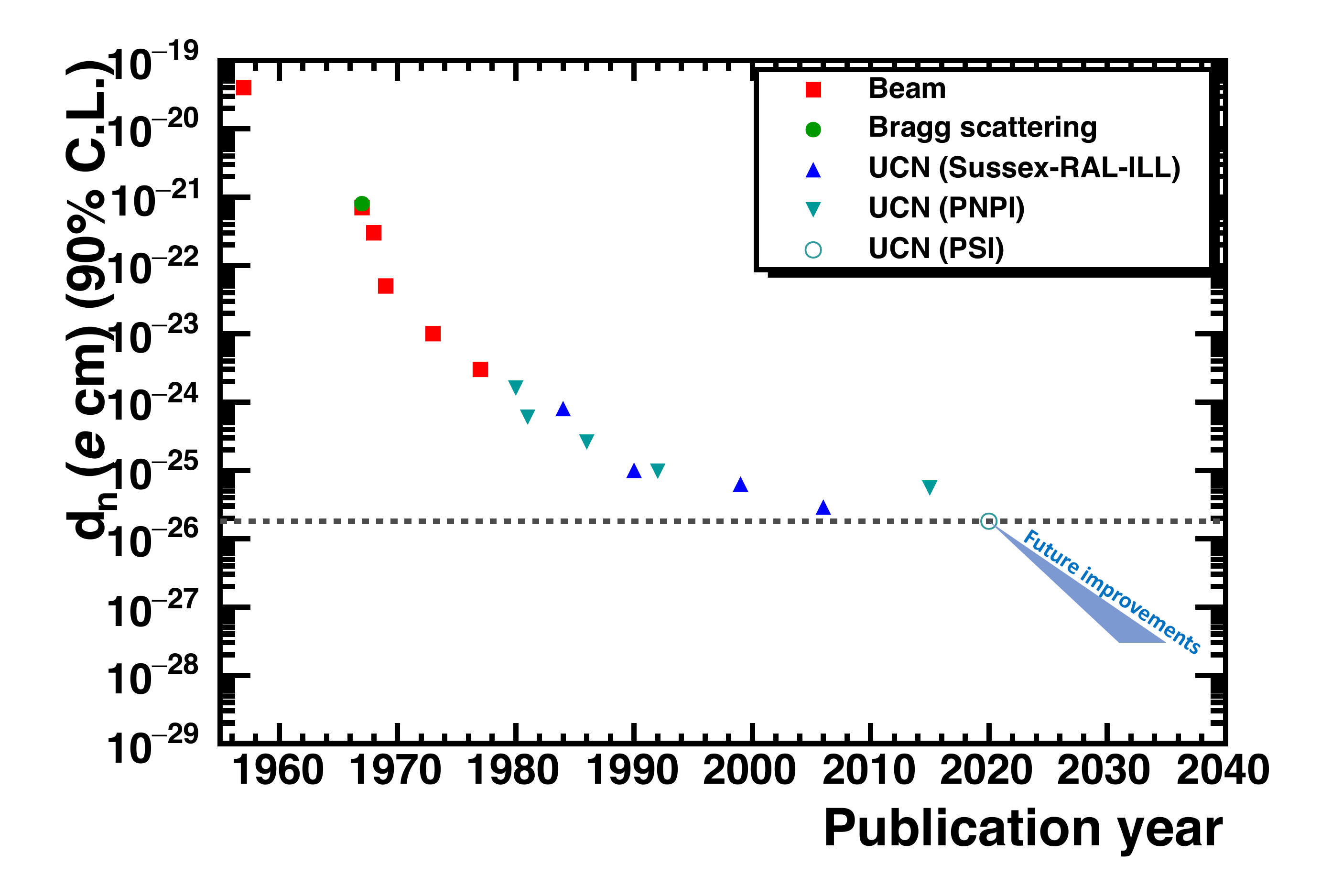}
    \caption{Evolution of the nEDM results along with projected future results}
    \label{fig:nEDM_evolution}
\end{figure}

The nEDM@SNS experiment has been under development for the past two decades as it involves many technological innovations to enable nEDM breakthroughs.
In 1994, Golub and Lamoreaux proposed a new method~\cite{GOLUB1994} to improve EDM measurements. It calls for the innovative use of superfluid helium as the UCN production target as well as a noble-liquid detector to measure the neutron precession. Performing an experiment immersed in a bath of superfluid helium, a significant improvement in all of $E$, $N$, and $T_{fp}$ is expected, with a goal sensitivity of $\delta d_n = 3 \times 10^{-28}\, e\cdot\rm cm$. 
The nEDM@SNS apparatus also has double chambers, with the $E$ fields pointing in opposite directions.
In this (sub-Kelvin) cryogenic environment, $^3$He atoms replace $^{199}$Hg as the comagnetometer. One major advantage of using $^3$He dissolved in superfluid LHe as a comagnetometer is its ability to study the systematic effects due to the geometric phase effect (the known leading systematic effect) by varying the temperature: changing the temperature changes the phonon density in LHe, thereby changing the mean free path of $^3$He atoms, which in turn changes the aforementioned correlation function. The spin-dependent cross-sections for neutron absorption on $^3$He is used to analyze the neutron spin precession. Scintillation in liquid He produced when the reaction products of the neutron capture on $^3$He travel in the medium signals the time at which the spins of the neutrons and $^{3}$He are anti-parallel.  

The use of $^3$He dissolved in $^4$He as a comagnetometer will allow the study of the ``geometric phase" systematic error in great detail.
The nEDM@SNS apparatus will be capable of two different readout techniques: the dressed-spin method originally proposed by Golub and Lamoreaux~\cite{GOLUB1994}, and a free-precession method without spin dressing in which the $^3$He magnetization is directly detected by SQUID magnetometers. In the dressed-spin method, a strong ($\sim$0.1 mT), ~1 kHz magnetic field (“dressing field”) is applied transverse to the static electric and magnetic fields, with the dressing field parameters set such that the neutron and $^3$He have the same effective gyromagnetic ratio (“critical dressing”). At critical dressing, the difference between neutron and $^3$He effective precession rates are insensitive to changes in the (near-static) magnetic field. Dressing field parameters are varied around this critical dressing condition and a feedback-null technique is designed to be sensitive to a neutron EDM.  The two possible modes of operation, i.e. traditional free precession and spin dressing,  will allow an independent check of the results. 

The nEDM@SNS experiment uses challenging technologies involving large-scale cryogenic systems (for the current experimental design, see \citet{Ahmed2019}). Considerable results have been obtained from the R\&D efforts in all areas, including: polarized $^3$He~\cite{Lamoreaux2002,Ye2008,Ye2009,SEckel2012,Baym2013,Baym2015,Baym2015b}, 
dressed spins~\cite{Esler2007, Chu2011, Chu2015},
magnetic field monitoring~\cite{Nouri2015}, non-magnetic feedthrough\cite{Cianciolo2018}, SQUID based magnetometers~\cite{YJKim2015}, noble gas scintillation and its detection~\cite{Ito2012,Ito2013,Gehman2013,Seidel2014,Phan2020,Loomis2021}, electrical breakdown in LHe~\cite{Ito2016,Phan2021}, HV generation in superfluid helium~\cite{Clayton2018}, cryogenic magnetic field studies~\cite{Slutsky2017}, possible systematic effects~\cite{Lamoreaux2005,Barabanov2006,Schmid2008,Clayton2011,Steyerl2014,Pignol2015,Golub2015,Golub2015b,Swank2016}, and apparatus for studying spin dressing and systematic effects~\cite{Korobkina2014}.  
The SNS project is the only project making use of a superconducting shield, meaning it will not be affected by intrinsic fluctuations in the room temperature ferromagnetic shields, a technology which has not been tested at sensitivities below
$10^{-26}\, e\cdot\rm cm$, and is relied on by all the other nEDM projects discussed above. 
In contrast to all the other projects discussed above the SNS project produces UCN in its heart so that its sensitivity can be increased by moving to a more intense cold neutron beam, e.g. the NIST center for neutron research; an external UCN source is not required.

\subsection{Conclusions}
The nEDM limit has been improved steadily, as shown in Fig.~\ref{fig:nEDM_evolution}, with substantial efforts in developing techniques in spin manipulation, magnetometry, and precision field controls to mitigate various systematic effects.
Ongoing efforts will improve the current nEDM limit by one to two orders of magnitude over the coming $\sim$ 10 years.
Improving the nEDM sensitivity beyond $10^{-28}\, e\cdot\rm cm$ would require a large-scale apparatus with built-in magnetometry and a high-intensity neutron source. A high-current spallation target, coupled to super-cooled helium, is technically feasible to significantly enhance the UCN yields beyond the capabilities of current UCN sources and increase the density of UCN by several orders of magnitude to reach several thousands per cubic centimeters~\cite{Leung2019}. To keep pushing the nEDM sensitivity, a new neutron facility (both UCN and cold neutrons) in the US is needed.

A successful measurement of the nEDM will provide the most cleanly-interpretable information on P and T violation in the light quark sector.
One of the leading dark matter candidates (axions) was originally proposed to help explain the mysterious absence of T violation in QCD, and the present upper bound on the EDM strongly constrains theories beyond the Standard Model. If no nEDM is discovered, the nEDM experiments,  in combination with ongoing EDM searches in atomic and molecular systems, will push the limits on the mass scale for new T violation physics above 100 TeV with direct sensitivity to the Higgs sector, conclusively test the minimal supersymmetric model for electroweak baryogenesis, and tightly constrain model-independent analyses of P and T violation.  Discovery of a nonzero EDM at this level, on the other hand, would reveal a completely new source of T (and thus CP) violation, that is needed to advance the development towards a unified theory of fundamental forces of nature. 
Given the high scientific impact of the nEDM physics, continued support of nEDM research and the investment on a high-intensity neutron facility should be the highest priorities in this field.

\section{Atomic and Molecular EDMs}
\label{sec:amo}
\newcommand{\Esca}{\mathcal{E}}
\newcommand{\Evec}{\vec{\mathcal{E}}}

Atoms and molecules have been sensitive platforms for precision measurements of symmetry violations for many decades, including CP-violation (CPV) through EDMs~\cite{Safronova2018,Chupp:2017rkp}.  These experiments currently set the best limits on the electron EDM, semileptonic CPV interactions, and quark chromo-EDMs; they also are competitive with nEDM for sensitivity to quark EDMs and $\theta_{QCD}$~\cite{ACME2018,Graner2016}.  These searches are generally referred to as ``AMO" searches since they use atomic, molecular, and optical techniques to achieve their goals, and 
leverage the very advanced AMO techniques for quantum control.

In all atoms and molecules, CPV effects are revealed by measuring energy shifts when an electric field, $\Evec$, is applied to the system. Its effect can be described by a term in the atomic/molecular Hamiltonian of the form 
\begin{equation}
    H_{\rm CPV} \propto \mathbf{J}\cdot\mathbf{n}. \label{eq:AMOHamiltonian}
\end{equation}
Here, the axial vector $\mathbf{J}$ is an intra-atomic/molecular angular momentum (e.g. the electron spin $\mathbf{S}$, when searching for for the electron EDM). The polar vector $\mathbf{n}$ is parallel to $\Evec$, which acts to electrically polarize the atom or molecule along $\mathbf{n}$. Generically, we can write $\mathbf{n}=\mathcal{P}(\mathcal{E})\hat{\Esca}$, where $\hat{\Esca}$ is a unit vector along the applied field, and $\mathcal{P}(\mathcal{E})$, which takes values $-1\le\mathcal{P}\le 1$, is the dimensionless electrical polarization of the atom or molecule.

\subsection{Observable effects of CP-violation in atoms and molecules} \label{sec:AMO:ObservableEffects}

Because of the compositeness of atoms and molecules, and of the nuclei within them, AMO searches for CPV physics do not directly measure any single underlying CPV parameter. Moreover, even the concept of searching for EDMs using neutral atoms or molecules can be confusing: since the electric field acting on non-relativistic, point-like constituents in a bound, neutral system interacting only electrostatically must vanish~\cite{Schiff1963}, it might appear that EDMs of electrons and nuclei are unobservable in atoms and molecules.  This effect, known as ``Schiff screeening'', arises because, as an external $\Esca$-field polarizes the atom or molecule, the internal field on each charged constituent will exactly cancel the external field---i.e., the external $\Esca$-field is fully screened.   

Here we discuss the numerous loopholes in this argument, and how they are exploited to study CPV effects in atoms and molecules. Most simply, semi-leptonic electron-nucleon four-fermion couplings are not subject to any screening effect. In addition, electrons moving near heavy (high-$Z$) nuclei are highly relativistic; the electron EDM then may be understood to interact with an effective internal electric field, $\Esca_{\rm ef\!f}$. Remarkably and counter-intuitively, this effective field can be orders of magnitude stronger than the applied external $\Esca$-field \cite{Sandars1964a,Sandars1965}. 

The EDMs of nuclei in neutral atoms and molecules, which are deeply non-relativistic, indeed experience severe screening of an applied $\Esca$-field, and are practically unobservable. Instead, here the lowest-order CPV effects that can be measured arise from higher-order electromagnetic moments of a finite-size nucleus. In most experiments to date, observable CPV arising from hadronic 
sources arises from the nuclear Schiff moment (NSM). The NSM, $\mathcal{S}$, is
equivalent to a static charge distribution on the surface of the nucleus that produces a uniform, CPV-induced $\Esca$-field inside the nucleus parallel to the nuclear spin $\mathbf{I}$~\cite{Flambaum2002,Ginges2004}. This intra-nuclear electric field interacts with the charge of a penetrating electron to produce energy shifts in the atom/molecule. This Schiff screening is also sidestepped by magnetic effects~\cite{Porsev2011}, imperfect screening of oscillating fields~\cite{Flambaum2018,Tan2019,Flambaum2020Solids}. In addition, nuclei with spin $I\ge 1$ obtain a CPV-induced magnetic quadrupole moment (MQM) that will interact with an unpaired electron spin. The NSM and MQM can be induced by all purely hadronic underlying CPV interactions: quark EDMs and chromo-EDMs, the Weinberg 3-gluon operator, and the QCD $\bar{\theta}$ term (see \cref{sect:th}).

Experiments searching for CP violation with atoms and molecules can be usefully divided into two classes, according to whether the atom/molecule is paramagnetic (with unpaired electron spins) or diamagnetic (with closed electron shells, but nonzero nuclear spin).  In diamagnetic systems, the observables most sensitive to underlying CPV physics are the nuclear Schiff moment and a tensor-pseudotensor (T-PT) semileptonic interaction. Paramagnetic systems are most sensitive to the electron EDM and a scalar-pseudoscalar (S-PS) nucleon-electron coupling. When the nuclear spin is $I\ge 1$ in a paramagnetic system, the MQM also can contribute substantially.   

In the atomic/molecular Hamiltonian of Eq.~\ref{eq:AMOHamiltonian}, the angular momentum $\mathbf{J}$ corresponds to the electron spin $\mathbf{S}$ for the electron EDM and S-PS coupling, the nuclear spin $\mathbf{I}$ for the NSM or T-PT coupling, or their hyperfine-coupled vector sum, $\mathbf{F} = \mathbf{S}+\mathbf{I}$, for the MQM. Because of the similarities of these experiments, all are referred to as ``EDM experiments" even if they are searching for semi-leptonic couplings, MQMs, or NSMs.

\begin{figure}
    \centering
    \includegraphics{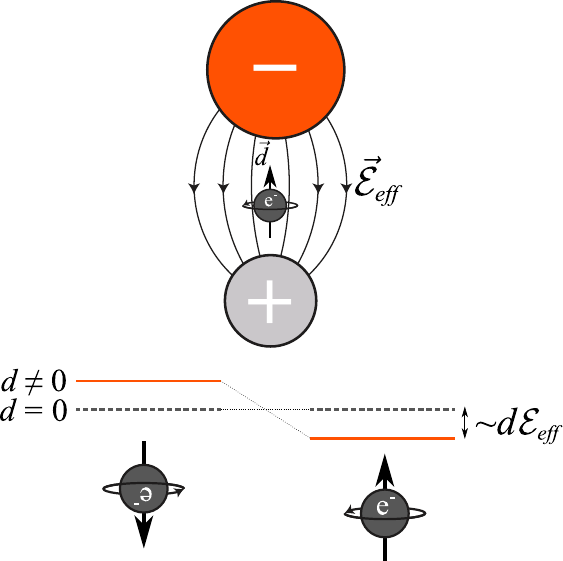}
    \caption{Intuitive picture of molecular EDM searches. The Hamiltonian $H_{\rm CPV}$ leads to energy shifts between states with spin angular momentum $\vec{J}$ oriented along and against the axis of internal polarization, $\hat{\vec{n}}$.
    The figure shows the specific example of an unpaired valence electron with an EDM along its spin axis, which will experience CP-violating energy shifts correlated with the relative orientation of the molecular effective internal field and electron spin. The molecule depicted here is fully polarized ($\mathcal{P} = 1$), as required for maximal energy shifts. 
    The approach for other CP-violating moments, such as nuclear Schiff moments or magnetic quadrupole moments, is entirely analogous. The idea for atomic EDM searches is also similar, though here the degree of atomic polarization (caused by dispacement of the valence electron wavefunction relative to the nucleus and core electrons) is usally much smaller, i.e. $\mathcal{P}\ll 1$.}
    \label{fig:EDMShifts-amo}
\end{figure}

A simple picture of how 
the energy shifts associated with $H_{\rm CPV}$ can be measured is shown in  \cref{fig:EDMShifts-amo}, for the easy to understand case of the electron EDM.
Here, an unpaired valence electron in a polar molecule will experience a large effective internal electric field $\Evec_{ef\!f}$, leading to an energy splitting $\pm\vec{d}_e\cdot\Evec_{ef\!f}$.  The polarization axis $\hat{\mathbf{n}}$, and hence $\Evec_{ef\!f}$, is parallel to the externally applied field $\Evec_{lab}$; for the fully polarized molecule shown here, the polarization is $\mathcal{P} = 1$.  Since $\vec{d}_e\propto\vec{S}$, where $\vec{S}$ is the electron spin, the CP-violating Hamiltonian is $H_{\rm CPV} \propto\vec{S}\cdot\hat{\Esca}_{lab}$. 

This interaction is fully analogous to the Hamiltonian relevant to e.g.~neutron EDM searches.  
Also similar is the experimental measurement method:
maximum sensitivity is achieved by creating a superposition of states with opposite signs of $S_z$, the projection of electron spin relative to $\Evec_{lab}$ (and therefore also $\Evec_{ef\!f}$), and measuring the subsequent spin precession. The spin precession frequency is equal to the energy shift between spin up and spin down states, divided by $\hbar$. Note that this basic idea is the same for all the various sources of CP-violation in atoms and molecules, whether the coupling is to the electron spin, nuclear spin, or both.

    \subsection{Relation between atom/molecule CPV energy shifts and underlying CPV parameters} \label{sec:AMO:ShiftsFromParameters}
    
The relation between underlying CPV interactions and atomic/molecular energy shifts includes several conversion factors associated with phenomena at different scales. Quite generally, the CPV energy shift $\Delta E_{\rm CPV}$ between spin up and spin down states, due to an underlying CPV parameter $C_{\rm CPV}$ (such as an electron EDM $d_e$, a quark chromo-EDM $\tilde{d}_q$, a four-fermion S-PS coupling constant $C_S$, etc.) in an electric field $\mathbf{\mathcal{E}}$ can be written in the form
\begin{equation}
    \Delta E_{\rm CPV} = Q(C_{\rm CPV}) \times \mathcal{P}(\mathcal{E}).
\end{equation}
Here, again $\mathcal{P}(\mathcal{E})$ is the dimensionless electrical polarization of the system; the factor $Q(C_{\rm CPV})$ encodes the intrinsic sensitivity of the particular atomic or molecular state to $C_{\rm CPV}$. We discuss these two terms separately.

The electric polarization $\mathcal{P}(\mathcal{E})$ is qualitatively different in atoms than in polar molecules. The polarization arises due to Stark-induced mixing of opposite parity energy levels, $(|+\rangle $and$ |-\rangle)$. Full polarization $|\mathcal{P}|=1$ corresponds to complete mixing of these states. From quantum mechanical static perturbation theory, the size of this mixing corresponds to the ratio of the dipole matrix element, $e\mathcal{E}\langle-|r|+\rangle$, to the energy splitting, $E_+-E_-$. For atoms with valence electron(s) in an $s$-orbital, this arises from $s$-$p$ mixing. Since $s$ and $p$ orbitals are typically split by $\sim$~eV~$\sim 0.1 e^2/a_0$ (where $a_0$ is the Bohr radius) and dipole matrix elements are typically $\sim e\mathcal{E}a_0$, to reach full polarization would require $\mathcal{E} \sim 0.1 e/a_0^2 \sim 1$ GV/cm. 
The largest fields that can be applied continuously in the lab are $\mathcal{E}_{\rm max} \sim 300$ kV/cm.  Hence, in atoms, full polarization is never achieved; instead, in this regime typically $\mathcal{P}(\mathcal{E}) \sim 10^{-4}[\mathcal{E}/(100$ kV/cm)].

In polar molecules, the internal field from the negative ion on the positive ion naturally has magnitude $\sim e/a_0^2$, so an $s$-orbital of the positive ion can be fully polarized~\cite{Sushkov1985,Khriplovich1997}.  However, in the absence of any externally applied field, the molecular eigenstates are also eigenstates of angular momentum, where there is no preferred orientation of one ion relative to the other.  Here, applying an external field mixes \textit{rotational} states of opposite parity (e.g. $|+\rangle = |J=0\rangle$ and $|-\rangle=|J=1\rangle$.  By contrast with atoms, here a sufficiently strong applied $\mathcal{E}$-field can polarize the rotational motion of the molecule---and hence orient the molecular axis (and the internal $\mathcal{E}$-field of the molecule) along it. The energy splitting between molecular rotational states is typically $10^{-4} e$V or even smaller, so that in molecules external applied fields of $\mathcal{E} \sim 100$ kV/cm (or much smaller) is sufficient to reach $|\mathcal{P}|\approx 1$.  This is the reason that molecular systems typically have CPV-induced energy shifts $10^3-10^4$ times larger than those in atoms.  Note that in molecules with complex hyperfine or rovibrational structure, $\mathcal{P}$ may be not a
monotonic function of the laboratory field and corresponding molecular calculations are required~\cite{Petrov2018,Kurchavov2020,Baturo2021,Petrov2021,Kurchavov2021}.   In experiments with \textit{rotating} magnetic and electric fields used to trap ions, $\mathcal{P}$ is function of both rotating electric and magnetic fields, which  must be sufficiently large to maximize sensitivity~\cite{ Cairncross2017, Petrov2018b, Kurchavov2020, Kurchavov2021}.

Next we consider the intrinsic sensitivity of the system, $Q(C_{\rm CPV})$. This can be written as a product of four factors:
\begin{equation}
    Q(C_{\rm CPV}) = Q_{\rm at}(C_{\rm CPV})\times Q_{\rm nuc}(C_{\rm CPV}) = S_{\rm at}^{C_{\rm CPV}}(Z) q_{\rm at} \times S_{\rm nuc}^{C_{\rm CPV}}(Z)  q_{\rm nuc}.
\end{equation}
Here, the subscript `at' corresponds to effects associated with atomic/molecular/ionic structure, and `nuc' to effects of nuclear structure.  The factors $S^{C_{\rm CPV}}(Z)$ correspond to generic, dimensionless scaling factors for atoms/ions/molecules with atomic number $Z$, while the dimensionful factors $q$ depend on the detailed structure of the electronic or nuclear state in the system. Hence, the overall intrinsic sensitivity has the form $Q = S(Z)q$ for atoms/ions and for the nuclei within them. Because the atomic scaling factors $S_{\rm at}^{C_{\rm CPV}}(Z)$ are associated with relativistic effects or very short range electron-nucleon interactions, they depend on the electron wavefunctions near the nucleus of charge $Z$, which grow rapidly with $Z$. The nuclear scaling factors $S_{\rm nuc}(Z)$ are typically proportional to the nuclear surface area or volume, and hence grow roughly linearly with $Z$. Hence, sensitive atomic/molecular EDM searches all use systems with a heavy (high-$Z$) atom/ion/molecule. The factors $q$ depend on the detailed structure of the electronic or nuclear state in the system, and have no systematic variation with $Z$.

Paramagnetic systems are, again, sensitive primarily to underlying CPV parameters that couple to electron spin and are independent of nuclear spin (such as $d_e$ or $C_S$). Hence, in these systems we set $Q_{\rm nuc}(C_{\rm CPV})=1$.  Here, the scaling of sensitivity is roughly $S_{\rm at}^{C_{\rm CPV}}(Z) \sim Z^3$.  Diamagnetic systems are, again, sensitive mainly to effects that couple only to nuclear spin~\cite{Ginges2004}, such as the NSM $(\mathbf{\mathcal{S}})$ (which, again, can be induced by any underlying purely hadronic CPV interaction). For the NSM, the atom/ion scaling factors roughl satisfy $S_{\rm at}^{C_{\rm CPV}}(Z) \propto Z^2$, while the NSM itself scales roughly as $\mathcal{S} \propto Z^{2/3}$. For the case of the MQM in paramagnetic systems, again $S_{\rm at}^{C_{\rm CPV}}(Z) \propto Z^2$.  For spherical nuclei, the MQM is roughly independent of $Z$, while for quadrupole-deformed nuclei there is a collective enhancement~\cite{Flambaum1994,Flambaum2014,Lackenby2018} which scales roughly as $\mathcal{M} \propto Z^{2/3}$.

The structure-dependent factors $q$ require detailed calculations of electron or nuclear wavefunctions in order to evaluate~\cite{Safronova2018,Kudashov2014,Gaul2017,Gaul2020,Skripnikov2016,Sudip:2016b,Abe:2018,Zhang:2021,Skripnikov2015ThO,Skripnikov2016,HaaEliIli20,HaaDoeBoe21,Abe2014,Prasannaa2020,Zakharova2021RaOH,Zakharova2021YbOH,Oleynichenko2022,Fleig2017HfF,Skripnikov2017}.  For electrons in atoms/ions, $q_{\rm at}$ is largest for systems where $s$ and $p$ orbitals are the main components of the states mixed by the external (for atoms) or internal (for ions in a molecule) $\mathcal{E}$-field.  The values of $q_{\rm at}$ have been calculated using sophisticated relativistic ab-initio computational methods for all experimental systems now of interest.  These calculations can be benchmarked by comparing to many types of atomic/molecular experimental data, including energy levels, dipole matrix elements, and especially, for paramagnetic systems, hyperfine structure~\cite{Kozlov1997,Skripnikov2020BW} (which is sensitive to wavefunctions near the nucleus, much as are the CPV structure factors). From these benchmarks, calculations have demonstrated accuracy for $q_{\rm at}$ below $10\%$ for most molecular systems, and typically a few percent or less for atomic systems. Input from these AMO-theoretical calculations is also critical to help select new species for future experimental studies.  These computational methods have proposed improvements via quantum algorithms for molecular EDM searches in near-term quantum devices~\cite{Peruzzo2014,Villela2021}.

The nuclear structure factors $q_{\rm nuc}$ have much larger uncertainties~\cite{Engel2013}.  For ordinary spherical nuclei, calculations of $q_{\rm nuc}$ associated with the NSM and MQM typically have uncertainties estimated to be of order $100\%$ or more. Hence, though measurements with these systems can detect nonzero CPV, for now they will not be useful in determining accurate values of underlying CPV parameters. In certain highly-deformed nuclei, these structure factors are enhanced relative to the case of spherical nuclei. 
For example, MQMs are enhanced (by a factor of typically 10-20) in quadrupole-deformed nuclei~\cite{Flambaum1994}. In addition, NSMs are dramatically enhanced (by a factor of typically $\sim300$, or possibly much more~\cite{Auerbach1996,Dobaczewski2005}) in octupole-deformed nuclei. In both cases, the same deformation that creates the enhancement is expected to enable calculations of nuclear structure factors $q_{\rm nuc}$ with smaller uncertainties, by benchmarking calculations agains nuclear energy levels, magnetic moments, shape deformations, and transition matrix elements.  Even if large uncertainties in the nuclear structure factors persist, the sensitivity of atomic and molecular measurements on these systems provides a powerful means for discovery of nonzero flavor-conserving hadronic CPV effects.

\begin{figure}
    \centering
    \includegraphics[width=0.7\textwidth]{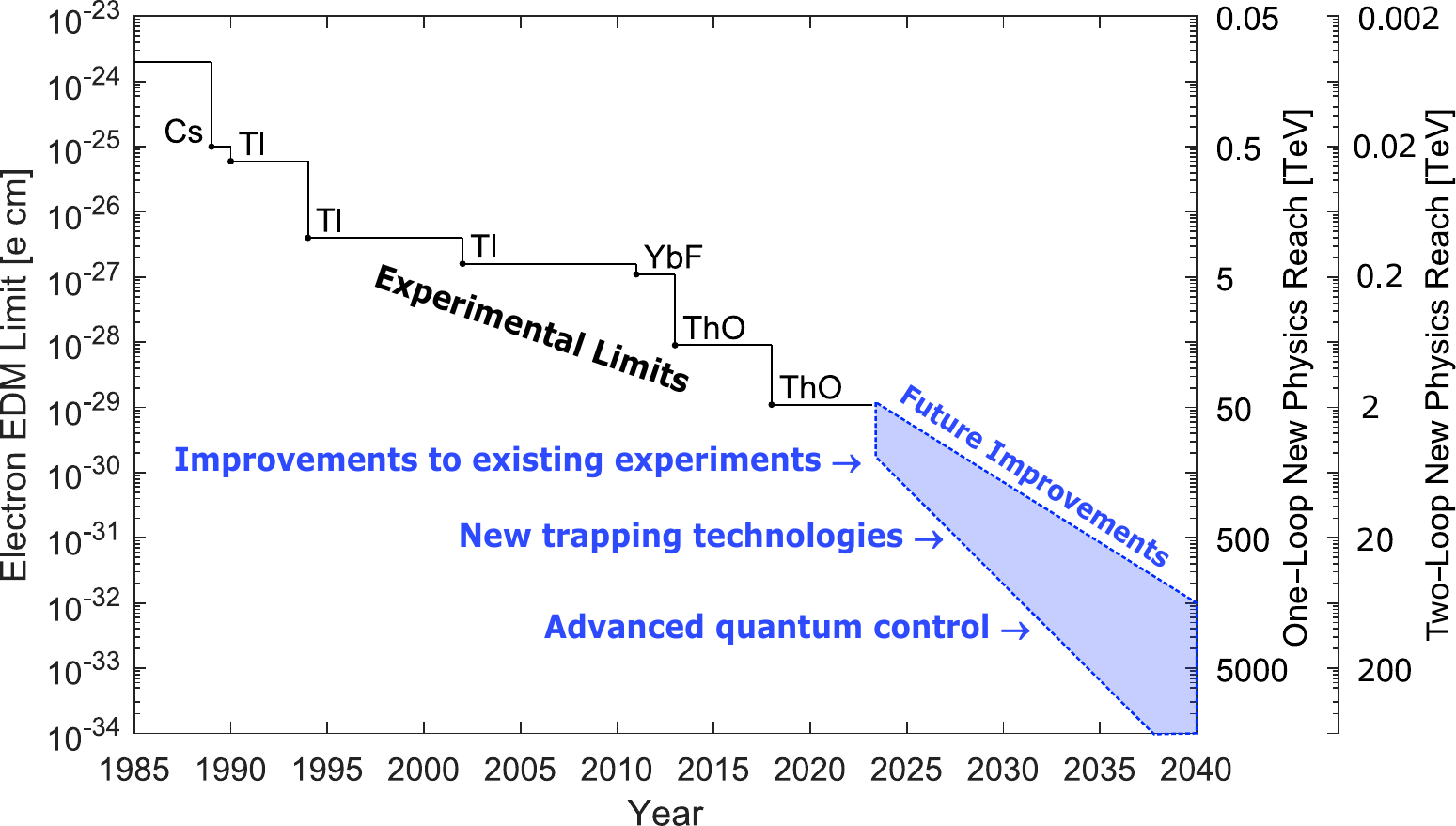}
    \caption{Electron EDM limits versus time, along with new physics reach for one-loop and two-loop effects (see Eq.~\ref{eq:massreach}).  All electron EDM experiments to date use AMO techniques. The solid line indicates the most sensitive experimental limit, including the species used.  The shaded area indicates potential future improvements discussed in the text.  Improvements in the next few years are driven largely by improvements to existing experiments and are quite likely, though as we go more into the future the projection becomes increasingly speculative and uncertain.}
    \label{fig:EDMReachVsTime}
\end{figure}

As discussed in  \cref{sect:th}, these searches are already probing energy scales beyond the reach of direct collider searches.  Even more excitingly, AMO searches for fundamental symmetry violations offer realistic pathways to orders of magnitude of improvement in the not-too-distant future.  First, modern molecular experiments are still relatively new, and recent advances such as laser cooling and trapping at ultracold temperatures~\cite{Moses2017,Isaev2018,Hutzler2020Review,Fitch2021Review}, matrix isolation~\cite{Vutha2018Atoms,Vutha2018PRA,Singh2019,Upadhyay2020}, and advanced control in molecular ion traps~\cite{Zhou2020,Fan2021} offer orders of magnitude increases in sensitivity through improvements in coherence time and count rates of existing approaches.  Second, heavy nuclei with octupole deformations can lead to enhanced sensitivity to hadronic CP-violation, up to a thousandfold larger than in spherical nuclei~\cite{Auerbach1996,Dobaczewski2005,Parker2015,Dobaczewski2018,Flambaum2019Schiff}. Combined with enhancement from large molecular fields, molecular species with deformed nuclei can be up to $\sim\!10^7$ times more intrinsically sensitive~\cite{Sushkov1985,Flambaum2019Schiff} than the current state of the art~\cite{Graner2016}. 

\subsection{Atomic Searches}

Though EDM searches using atoms are intrinsically several orders of magnitude less sensitive than analogous searches using polar molecules, atom-based experiments have many advantages. It is typically much easier to measure far larger numbers of atoms than molecules and hence increase statistical sensitivity; moreover, atomic spin coherence times can be much longer than has been demonstrated with molecules, leading to better resolution of small energy shifts.  Indeed, until 2011 the best limits on the electron EDM and S-PS semi-leptonic CPV interactions came from experiments using paramagnetic atoms, most recently using Tl $(Z=81)$ atoms~\cite{Regan2002}. Still today, the best limits on quark chromo-EDMs and T-PT semi-leptonic interactions come from experiments using diamagnetic $^{199}$Hg $(Z=80$) atoms~\cite{Graner2016,Swallows2013,Sahoo2018}.  In these extraordinary experiments, performed at the University of Washington, vapor cells containing $\sim 10^{15}$ $^{199}$Hg atoms were subjected to electric fields of $\Esca\sim 10$ kV/cm (corresponding to polarization $\mathcal{P} \sim 10^{-4}$), with over 100~s coherence time of spins oriented and then probed with nearly 100\% efficiency using lasers. A new generation of this experiment is underway, with projected improvement in sensitivity by a factor of a few.  Related experiments using vapor cells of $^{129}$Xe $(Z=54)$ atoms have also reported recent new results~\cite{Allmendinger2019,Sachdeva2019,Sakurai2019}. Here, the product of the scaling factors and intrinsic sensitivities are a factor of ~6 smaller than in $^{199}$Hg, and to date these experiments use $\sim 3$ times smaller $\mathcal{E}$-fields and measure spin precession with uncertainty $\sim 60$ times less precise (albeit with much shorter total integration time). Nevertheless, because of the large number of underlying hadronic and semileptonic CPV effects that can lead to a measurable signal in these systems~\cite{Engel2013}, these $^{129}$Xe experiments, along with all other searches for hadronic CPV, are complementary and important for setting robust limits with a global analysis~\cite{Chupp2015,Chupp2019}. 

\subsubsection{Pathways to improved sensitivity with atoms}

In the meantime, several new atomic EDM searches are in preparation, and some have already achieved preliminary results. Each of these new experiments seeks to take advantage of the extremely powerful techniques of laser cooling and trapping for atoms. These ultracold and (sometimes) trapped atoms can yield long spin coherence times, and access to new species with enhanced sensitivity factors. Typically spin preparation and readout can be done very efficiently in these systems using lasers. 

For example, experiments are underway with ultracold paramagnetic atoms such as Cs ($Z=55$) and Fr ($Z=87$).  The Cs EDM experiment~\cite{Zhu2013,Tang2018} being developed at Pennsylvania State University will use large numbers of both Cs and Rb ($Z=37$) atoms trapped in an optical lattice. Because the Rb atoms have much lower scaling factor than Cs, and hence less sensitivity to CPV effects such as the electron EDM, they can be used to monitor magnetic field variations as well as many effects leading to possible systematic errors. Experiments planning to use Fr atoms~\cite{Wundt2012,Inoue2014,Feinberg2018,Aoki2021,Shitara2021,FrLOI} are in preparation: an optical lattice at Tohoku University/The University of Tokyo/RIKEN, and a fountain configuration at LBNL/TRIUMF.  In both systems, projected sensitivities are comparable with the best current limits from molecular experiments.

New experiments with atoms, primarily sensitive to underlying hadronic CPV, are also underway. A particularly promising approach is to take advantage of the large structural enhancement of the NSM in the octupole-deformed nuclei $^{225}$Ra or $^{223}$Ra $(Z=88)$~\cite{Parker2015,Bishof2016,Auerbach1996,Dobaczewski2005,Spevak1997}. Here, the Ra atoms are laser-cooled and optically trapped, then subjected to potentially very large electric fields. Long spin coherence times are expected. A proof-of-principle measurement with $\sim 500$ $^{225}$Ra atoms achieved 20~s spin coherence time with an electric field of $\mathcal{E} = 67$ kV/cm. Numerous upgrades to increase the electric field~\cite{Ready2021}, the number of trapped atoms~\cite{Booth2020}, the spin coherence time, and the spin detection efficiency~\cite{Rabga2020} are in progress, with projections of several orders of magnitude improved sensitivity. This would be sufficient to match or surpass typical limits from the $^{199}$Hg experiment.

\subsection{Molecular Searches}

As mentioned earlier, CP-violating EDM-like signals in molecules can be $\sim\! 10^3-10^4$ times larger than in atoms due to the ability to reach the strong field, order-unity polarization limit, where the intra-molecular fields can be efficiently aligned with the lab frame. This applies to nearly all underlying CPV phenomena: electron EDM, and both semileptonic and hadronic CPV interactions~\cite{Safronova2018,DeMille2017,Chupp2019,Cairncross2019,Hutzler2020Review}.  Though the complexity of molecules makes them challenging to control, their large intrinsic sensitivity combined with modern experimental advances have resulted in major advances.  Molecules are now the most sensitive probe of the eEDM~\cite{Hudson2011,Baron2014,Cairncross2017,ACME2018}, having overtaken the most sensitive experiment with atomic Tl~\cite{Regan2002} for the first time in 2011~\cite{Hudson2011} and improved their sensitivity by two orders of magnitude since that time.  Although molecular amplification of CP-violating moments has been known for decades, only recently were experimental techniques developed which could offer quantum state control of molecules with sufficient sophistication for significant statistical sensitivity and robust suppression of systematic errors.

The strongest constraints on hadronic CPV still come from EDM measurements on $^{199}$Hg~\cite{Graner2016} and free neutrons~\cite{Abel2020}. However, molecules have similar advantages in this area and many experiments are now underway to exploit their intrinsically high sensitivity.  Furthermore, as discussed in \cref{sect:th1}, the complex parameter space of new sources of CP-violation, especially in the hadronic sector, provides strong motivation to study multiple different systems sensitive to such effects: a single positive result will not be sufficient to completely isolate the underlying source(s) of new physics~\cite{Chupp2015,Chupp2019,Fleig2018,Gaul2019}.  Furthermore, interpreting experimental results as limits on CP-violating parameters requires critical input from a wide range of theoretical methods to understand molecular and nuclear structure, which is especially challenging for the heavy species used in these experiments as discussed earlier~\cite{Engel2013,Safronova2018}.

Three molecular experiments, so far, have been able to set an eEDM limit that surpasses the sensitivity of any previous atomic experiment.  The ACME experiment~\cite{ACME2018} uses a cryogenic molecular beam of ThO in a metastable electronic state, and currently sets the most sensitive eEDM limit of $|d_e|< 1.1\times10^{-29}~e$~cm.  ACME is currently developing a third-generation experiment that will improve coherence time with a longer beam line, increase count rates through enhancements to molecular flux and detection efficiency, and suppress systematic errors through better control of experimental imperfections~\cite{Panda2019,Wu2020,Masuda2021}.  The JILA HfF$^+$ experiment uses trapped molecular ions, which enables considerably longer coherence time than can be achieved by beam experiments.  The first generation result~\cite{Cairncross2017} set a limit of $|d_e|< 1.3\times10^{-28}~e$~cm, and is being upgraded via extended coherence times~\cite{Zhou2020}, a new apparatus to afford higher count rates, and use of a molecular species with an even larger intrinsic sensitivity, ThF$^+$~\cite{Gresh2016,Ng2022}.  The Imperial College YbF experiment~\cite{Hudson2011}, which was the first molecular experiment to surpass the limits set by atoms, has implemented several improvements to both molecular flux and preparation/readout efficiency, including leveraging laser cooling and optical forces~\cite{Ho2020,Alauze2021}.

Experiments are also under underway to search for hadronic CP-violation by taking advantage of the intrinsically enhanced sensitivity of molecules, relative to atoms.  Nuclear Schiff moments and nuclear magnetic quadrupole moments give access to sources of hadronic parameters such as those discussed in \cref{sect:th1}~\cite{Ginges2004,Engel2013,Safronova2018}.  These experiments include the CENTReX nuclear Schiff moment search utilizing TlF~\cite{Hunter2012,Grasdijk2021}, nuclear magnetic quadrupole moment searches with $^{173}$YbOH~\cite{Kozyryev2017PolyEDM,Maison2019,Denis2020,Pilgram2021} and TaO$^+$~\cite{Fleig2017TaO,Chung2021}, and experiments with short-lived radioactive species such as RaF~\cite{GarciaRuiz2020,Udrescu2021} and RaOCH$_3^+$\cite{Fan2021,Yu2021}, which are discussed in a later section.   The accidental near-degeneracy of opposite parity states in $^{207}$PbF provides a further experimental approach, especially in excited vibrational states~\cite{Baturo2021}.

In addition to static CP-violation, there is motivation to search for oscillating CP-violating observables as well, in particular with hadronic sources.  The axion (and axion-like fields) can induce, for example, an oscillating value of $\theta_{QCD}$, which in turn results in oscillating CP-violating observables~\cite{Graham2011,Graham2013,Stadnik2014,Budker2014,Flambaum2020SpinRotation,Flambaum2020SpinRotation,Arvanitaki2021}.  These experiments generally benefit from the same enhancements as static EDMs in terms of heavy species, but offer a variety of distinct experimental approaches.  Note that static EDM searches can also be used to probe the effect induced by the exchange of axion-like particles between electrons and nucleons~\cite{Stadnik2018,Maison2021,Maison2021Axion}.

\subsection{Pathways to improved sensitivity with molecules}

These searches rely on coherent precession of electron or nuclear spin in an internal molecular field, with figure of merit given roughly by (coherence time)$\times$(system sensitivity)$\times$(count rate)$^{1/2}$. We discuss how each of these areas has significant untapped experimental potential in molecular systems.  

\subsubsection{Advanced molecular cooling methods}

Laser cooling and the quantum control afforded by ultracold temperatures is a main driver of quantum science advances with atoms, for example the atomic clocks which now reach unprecedented $<10^{-20}$ uncertainty~\cite{Bothwell2022}.  Implementing these techniques in molecules suitable for use in searches for CPV effects would result in orders of magnitude of improvement in sensitivity.  Laser cooling relies on the application of forces via repeated cycles of photon excitation followed by spontaneous decay.  Laser cooling of molecules is a challenge since the spontaneous decay step can result in excitation of internal vibrational modes, thereby populating states which are not addressed by the laser~\cite{DiRosa2004,Isaev2016Poly,Hutzler2020Review,Fitch2021Review}.  Since the first laser cooling of a molecule in 2010~\cite{Shuman2010}, the field has advanced rapidly and has resulted in several groups having directly cooled and trapped molecules at ultracold temperatures~\cite{McCarron2018SrF,Caldwell2019,Anderegg2019,Ding2020}, including loading into optical tweezers~\cite{Anderegg2019} and lattices~\cite{Wu2021}.  While no molecular symmetry violation search has yet utilized these techniques, there are many promising approaches with laser-coolable species, many of which are underway, including the diatomic species YbF~\cite{Tarbutt2013,Lim2018,Fitch2021YbF,Alauze2021}, BaF~\cite{Aggarwal2018}, RaF~\cite{Isaev2010,Kudashov2014,Gaul2019,Gaul2017,Gaul2020,Zhang:2021,Petrov2020}, HgF~\cite{Prasannaa2015,Yang2019}, and TlF~\cite{Cho1991,Hunter2012,Grasdijk2021}, along with polyatomic $^{174}$YbOH,~\cite{Kozyryev2017PolyEDM,Denis2019,Prasannaa2019,Gaul2020,Augenbraun2020YbOH,Zakharova2021YbOH,Petrov2021}, RaOH~\cite{Isaev2017RaOH,Kozyryev2017PolyEDM,Gaul2020,Zakharova2021RaOH}, BaOH~\cite{Denis2019,Gaul2020}, and SrOH~\cite{Kozyryev2017SrOH,Gaul2020}.  These experiments can benefit from increased coherence times due to trapping or use of a fountain~\cite{Tarbutt2013,Cheng2016}.  Coherence time can also be increased via molecular beam brightening and/or slowing using a combination of optical~\cite{Ho2020,Alauze2021}, electric~\cite{Aggarwal2021}, and magnetic~\cite{Augenbraun2021ZS} methods.

Molecules can also be coherently assembled from ultracold atoms without heating, thereby creating them in a trap at ultracold temperatures directly~\cite{Ni2008}.  This technique is considerably more advanced than the methods for direct cooling of molecules, but can be applied to only the small subset of molecular species built from laser-coolable atoms.  Until recently, no candidate species of this type, with high sensitivity to CPV effects, had been identified~\cite{Meyer2009}. In the past few years, several molecules that likely can be assembled from ultracold atoms, with intrinsic sensitivities comparable to that of the best other species, have been identified~\cite{Fleig2021, Klos2022,Sunaga2019Heavy,Sunaga2019AMD}. This includes species sensitive both to electron EDM such as RaAg~\cite{Fleig2021,Sunaga2019Heavy} and to  hadronic CPV such as $^{223}$FrAg~\cite{Klos2022,Fleig2021}.

\subsubsection{Matrix isolation}

The experiments discussed thus far have all relied on atoms or molecules being in the gas phase to obtain maximum coherence.  However, the ability to engineer appreciable coherence in the solid state would result in major gains in statistical sensitivity, since the number density could be many orders of magnitude higher than in any gas phase experiment~\cite{Pryor1987,Arndt1993,Kozlov2006EDM}.  
Previous experimental approaches used solids such as nickel-zinc ferrite~\cite{VK78}, GdIG~\cite{Hei2005},  ferroelectrics~\cite{Eckel2012} , and GGG~\cite{Kim2015} by searching for bulk linear magnetoelectric effects~\cite{Shapiro68,Ignat69}.
	
A new approach currently being pursued is to isolate atoms or molecules with high CPV sensitivity within a solid matrix, such as a frozen gas at cryogenic temperatures~\cite{Vutha2018Atoms,Vutha2018PRA,Singh2019,Upadhyay2020}, and to manipulate them using lasers akin to molecular beam experiments.
If these systems can be engineered to have high enough dopant density while maintaining sufficient coherence, they could improve statistical sensitivity to a wide range of CP-violating physics by many orders of magnitude.  Work toward these goals is ongoing, and coherence in matrix-isolated atoms has been achieved for 0.1~s~\cite{Upadhyay2020}, providing an exciting motivation for continued research and development.

\subsubsection{Radioactive species}

Heavy nuclei with octupole $(\beta_3)$ deformations, such as some isotopes of Fr, Ra, Th, Pa, and others, can have  sensitivities to CP-violation enhanced by up to a thousand-fold compared to spherical nuclei~\cite{Auerbach1996,Dobaczewski2005,Parker2015,Dobaczewski2018,Flambaum2019Schiff,Flambaum2020Schiff,Flambaum2020SchiffStable}. Combined with molecular enhancements, heavy molecular species with deformed nuclei can be up to $10^7$ times more intrinsically sensitive~\cite{Sushkov1985,Flambaum2019Schiff} than the current most sensitive experiment with atomic Hg~\cite{Graner2016}.  An example species, which is perhaps the subject of the most study to date, is radium: it has a nuclear deformation which has been carefully characterized~\cite{Gaffney2013,Butler2020}, and the atom, atomic ion, and many radium-containing molecules can be laser cooled~\cite{Parker2015,Isaev2010,Kudashov2014,Isaev2017RaOH,Fan2019}. As mentioned previously, the diamagnetic atom $^{225}$Ra is the subject of an EDM experiment at Argonne National Lab~\cite{Parker2015,Bishof2016,Auerbach1996,Dobaczewski2005,Spevak1997}. The RaF molecule was recently  studied spectroscopically~\cite{GarciaRuiz2020,Udrescu2021}, and several Ra-containing polyatomic molecular ions~\cite{Yu2021} were recently synthesized, trapped and cooled in an ion trap~\cite{Fan2021}.  Due to the combination of nuclear and molecular enhancements, these Ra-containing molecules are so sensitive that even the ability to trap one at a time with second-scale coherence would enable probes of new physics at the frontiers of hadronic CP violation~\cite{Isaev2010,Yu2021,Fan2021}.

Molecules containing other deformed nuclei are also of experimental and theoretical interest.  For example, isotopes of many heavy nuclei such as Eu, Ac, Th, and others have longer radioactive half-lives than the $^{225}$Ra isotope ($\tau\approx$2~weeks) needed for a hadronic CPV search, yet have comparable sensitivity~\cite{Flambaum2019Schiff,Skripnikov2020Act,Flambaum2020Schiff,Flambaum2020SchiffStable}.  
The case of $^{229}$Pa is curious, as it is purported to possess an extremely small
splitting between opposite parity nuclear states~\cite{Ahmad1982,Ahmad2015} which would result in a factor of 30 to 3000 further enhancement compared to $^{225}$Ra~\cite{Flambaum2008,Singh2019,Flambaum2020Schiff,Dobaczewski2018}.  The degree of enhancement depends on the assumptions made about the nuclear structure of $^{229}$Pa, which still has considerable uncertainty.  Molecules containing superheavy elements have also been explored due to their very high sensitivity to CP-violating effects~\cite{Gaul2019,Zhang:2021,Mitra2021Superheavy}.
Experimental facilities such as FRIB, TRIUMF, and ISOLDE offer exciting opportunities to realize a wide range of physics with short-lived and exotic species.

\subsubsection{Advanced quantum control}

The measurements discussed here rely on creation of quantum superpositions which are closely analogous to those used in quantum information science (QIS). While the techniques used are typically ``primitive'' compared to those used in modern QIS experiments, advances in the QIS field are a resource for future dramatic improvements in sensitivity to EDM-like signals~\cite{Cloet2019}.  In particular, techniques for entanglement-based spin squeezing have already been demonstrated to significantly surpass the standard quantum limit of sensitivity in atomic clock systems~\cite{Hosten2016}, which are very similar to the systems used to search for CPV in atoms and molecules. Applying these techniques to experiments aimed at detecting EDM-like signals could provide significant gains in sensitivity for both atoms~\cite{Aoki2021} and molecules, up to a factor of $\sim\sqrt{N}$ (where $N$ is the number of particles in a single measurement). 

This rapidly developing QIS technology provides a far-future pathway for continued gains in sensitivity to CPV signals.  
New techniques using ultracold, trapped atoms and molecules, as discussed above, are especially attractive for realizing such gains from entanglement. Here, the trapped ensembles of atoms or molecules have sufficiently high density to engineer the particle-particle interactions needed to generate entanglement.  Realizing these ideas will require significant advances in experimental capabilities and theoretical development of suitable protocols, but the advances being made currently in these areas show great promise for the future.

\subsection{Summary and Outlook} 

AMO-based searches for fundamental symmetry violations have advanced rapidly in recent years, and have excellent prospects for further advances in the near, medium, and long terms.  Improvements in sensitivity of one, two-three, and four-six orders of magnitude appear to be realistic on few, 5-10, and 15-20 year time scales, respectively, by leveraging major advancements made using quantum science techniques and the increasing availability of exotic species with extreme sensitivity.  This is an exciting pathway to probe PeV-scale physics using ``tabletop'' scale experiments.

Given the complementary approaches, both in terms of experimental methods and sensitivity to new physics parameter space, supporting and pursuing many efforts, both experimental and theoretical, is critical.  Measurements from multiple experiments are required to determine the actual source of the new physics, and significantly differing methods will give considerable robustness against systematic errors.   Furthermore, there are many opportunities for a wide range of synergies; methods to create, cool, control, and understand the increasingly complex systems used for these searches will have an extremely wide range of applications outside of fundamental symmetry violations, from quantum sensing to quantum many-body physics and beyond.  Similarly, a wide range of advances in other fields such as sensitive photon detection, integrated photonics, atomic and molecular spectroscopy, advanced photon sources, \textit{ab initio} methods, advanced computational infrastructure, and facilities to create and handle exotic species will directly benefit the field.

 Given the low relative cost of these experiments in terms of both funding ($\lesssim\$$10 M, and often $\lesssim$1 M) and personnel (typically $\lesssim$ 10 people), pursuing many simultaneously is feasible.  However, advancing to the next generation will require increases in scale and complexity. Many of the new approaches discussed here require sustained R\&D budgets, theory support, and access to facilities when working with exotic nuclei, continued over several experimental generations to fully realize their projected gains. The field is moving very rapidly and requires risk tolerance, but it has proven that it can deliver results from a variety of novel approaches.

\section{Storage Ring EDMs}
\label{sec:proton}
The EDM of charged particles has been directly probed only for particles in a storage ring.
There are two different approaches pursued globally for the proton EDM. One is U.S.-based by the storage ring EDM collaboration (srEDM), relying on presently available technology and on symmetries to effectively eliminate the systematic error sources. The other is a European effort (JEDI) focused on smaller rings, requiring further technological and conceptual developments.

The JEDI (Jülich Electric Dipole moment Investigations) collaboration is pursuing the project of a precision storage ring to measure the EDMs of polarized proton and deuteron beams with unprecedented sensitivity. Besides R\&D covering, e.g., prototype development and spin dynamics simulations, a precursor experiment with polarized deuterons has been conducted at the Cooler Synchrotron COSY of Forschungszentrum Jülich (Germany), and is currently being analyzed. 
The next step of the research will require a new class of hitherto not existing precision storage rings. This encompasses the complete chain from design and building to operation of the storage ring and includes instrumentation for control/feedback of the beam(s) and its polarization. Such a ring requires a large electric field ($\approx 10$MV/m) and a proton momentum of 707 MeV/c, resulting in a storage ring of $\approx 500$~m circumference. Recently, it was realized that, before starting the construction of a such a ring, an intermediate step, a so-called prototype with approximately 100~m circumference, is needed, which demonstrates the functionality of all components and allows for a first direct measurement of the proton EDM as well as searches for axions/axion-like-particles~\cite{SeungPyo19,pretz2020statistical,kim_new_2021}. The current state-of-the-art of the development is summarized in a recent CERN Yellow Report~\cite{CPEDM:2019nwp}.

The U.S. has the option of putting the ring inside the 805~m AGS ring-tunnel at Brookhaven National Laboratory (BNL), saving the cost of the tunnel. This will result in a comfortable electric field of 4.4~MV/m, which does not require the time and expense of a prototype ring with a 10~MV/m electric field, beyond the present state of the art.

Finally, we note that recently a direct measurement of the electron EDM in a low-energy storage ring has been proposed~\cite{Suleiman:2021whz}. This new method uses a small spin-transparent Figure-8 ring~\cite{PhysRevLett.124.194801} with two-energy levels at different sections. In addition, there is a small-effort proposal to modify the current muon $(g-2)$ ring at Fermilab to demonstrate the dedicated ``frozen'' spin method  with muons~\cite{Semertzidis:1999kv,JPARCmedmLOI,EDM:2003olr,Miller2004,Adelmann:2010zz,Adelmann:2021udj}, while taking muon EDM data with better sensitivity than the current parasitic EDM goals. We note that an ambitious muon EDM program is being planned at PSI that will take advantage of a proposed high intensity muon beam (HIMB)~\cite{Crivellin:2018qmi,Aiba:2021bxe,Adelmann:2021udj}.

The following subsections detail the US-based srEDM proposal and end with a brief description of the PSI muon EDM proposal.

\subsection{Summary of the Storage Ring Proton EDM Experiment}
\begin{itemize}
\item Proton EDM sensitivity \targetsens\!\!. 
\item Improves the sensitivity to QCD CP-violation ($\theta_{\textnormal{QCD}}$) by three orders of magnitude, currently set by the neutron EDM experimental limits. 
\item New Physics reach at $10^3$~TeV mass scale.
\item Probes CP-violation in the Higgs sector with best sensitivity~\cite{edmtheory}.
\item Highly symmetric, magic momentum storage ring lattice in order to control systematics.
    \begin{itemize}
        \item Proton magic momentum =$\SI{0.7007}{GeV/c}$.
        \item Proton polarimetry peak sensitivity at the magic momentum.
        \item Optimal electric bending and magnetic focusing.
        \item $\num{2e10}$ polarized protons per fill. One fill every twenty minutes.
        \item Stores simultaneously clockwise (CW) and counterclockwise (CCW) bunches.
        \item Stores simultaneously longitudinally and radially polarized bunches with positive and negative helicities.
        \item 24-fold symmetric storage ring lattice.
        \item Changes sign of the focusing/defocusing quadrupoles within 0.1\% of ideal current setting per flip.
        \item Keeps the vertical spin precession rate low when the beam planarity is within $\SI{0.1}{mm}$ over the whole circumference and the maximum split between the counter-rotating (CR) beams is $<\SI{0.01}{mm}$.
        \item Closed orbit automatically compensates spin precession from radial magnetic fields.
        \item Circumference = $\SI{800}{m}$ with $E=\SI{4.4}{MV/m}$, a conservative electric field strength. 
    \end{itemize}
\item 3 -- 5 years of construction and 2 -- 3 years to collect  required statistics to first physics publication. 
\item Sensitive to vector dark matter/dark energy (DM/DE) models~\cite{graham_paper}.
   DM/DE signal proportional to $\beta=v/c$. Magic momentum pEDM ring with $\beta=0.6$.
\item pEDM is highly complementary with atomic and molecular (AMO) EDM experiments~\cite{edmtheory2}.
   AMO: many different effects, “sole source analysis”, unknown cancellations~\cite{Chupp:2017rkp}.
\item After proton EDM, add magnetic bending for deuteron/$^3$He EDM measurements.
   Deuteron and $^3$He EDM measurements complementary physics to proton EDM.
\end{itemize}

\subsection{History}
 The proposed method has its origins in the measurements of the anomalous magnetic moment of the muon in the 1950-70s at CERN. The CERN I experiment~\cite{cern_report} was limited by statistics. The sensitivity breakthrough was to go to a magnetic storage ring. The CERN II result was then limited by the systematics of knowing the magnetic field seen by the muons in the quadrupole magnet. The CERN III experiment~\cite{cern_report,cern3} used an ingenious method to overcome this. It was realized that an electric field at the so-called ``magic'' momentum does not influence the particle $(g-2)$ precession. Rather, the electric field precesses the momentum and the spin at exactly the same rate, so the difference is zero. The fact that all electric fields have this feature, opened up  the possibility of using electric quadrupoles in the ring to focus the beam, while the magnetic field is kept uniform. 
 
 The precession rate of the longitudinal component of the spin in a storage ring with electric and magnetic fields is given by:
\begin{equation}
    \dv{\bm{\beta} \cdot \bm{s}}{t} = -\frac{e}{m} \bm{s}_\perp  \cdot \qty[\qty(\frac{g-2}{2}) \hat{\beta} \times \bm{B} + \qty(\frac{g\beta}{2} - \frac{1}{\beta})\frac{\bm{E}}{c}].
    \label{eq:omega}
\end{equation}

The CERN III experiment used a bending magnetic field with electric quadrupoles for focusing at the ``magic'' momentum, given by $\beta^2=2/g$; see the electric field term in \Cref{eq:omega}. The CERN III experiment, and the BNL E821~\cite{bennett_final_2006} were limited by statistics not systematics. The recent announcement of the $(g-2)$ experimental results~\cite{fnal1} from Fermilab at $\SI{460}{ppb}$ has confirmed the BNL results, with similar statistical and smaller systematic errors. We believe that the FNAL E989 final results, at about $\SI{140}{ppb}$, will have equal statistical and systematic errors. The storage ring/magic momentum breakthrough gained a factor of \num{2e3} in systematic error. 

BNL E821 set a ``parasitic'' limit on the EDM of the muon: $d_{\mu} < \num{1.9e-19}~ e \cdot \textnormal{cm}$~\cite{bnl_edm}. For FNAL E989, we expect this result to improve by up to two orders of magnitude. The statistical and systematic errors will then be roughly equal. The dominant systematic error effect is due to radial magnetic fields. For the pEDM experiment, we plan to use a storage ring at the proton magic momentum with electric bending and magnetic focusing. This will have a negligible radial magnetic field  systematic effect (see below), while the main systematic errors will drop out with simultaneous clockwise and counterclockwise storage.

For both BNL E821 and FNAL E989, new systematic effects were discovered that were not in the original proposals. These small effects have been mitigated, so they are not limiting factors. For the pEDM experiment, e.g., we can get  $10^{11}$ polarized protons per fill from the BNL LINAC/Booster system, and we use symmetries to handle the systematics down to the level of sensitivity. We expect that at that level we perhaps will also discover new small systematics effects, similarly to the $(g-2)$ experiments.

\subsection{The storage ring EDM method}

\begin{figure}
    \centering
    \includegraphics[width=.9\textwidth]{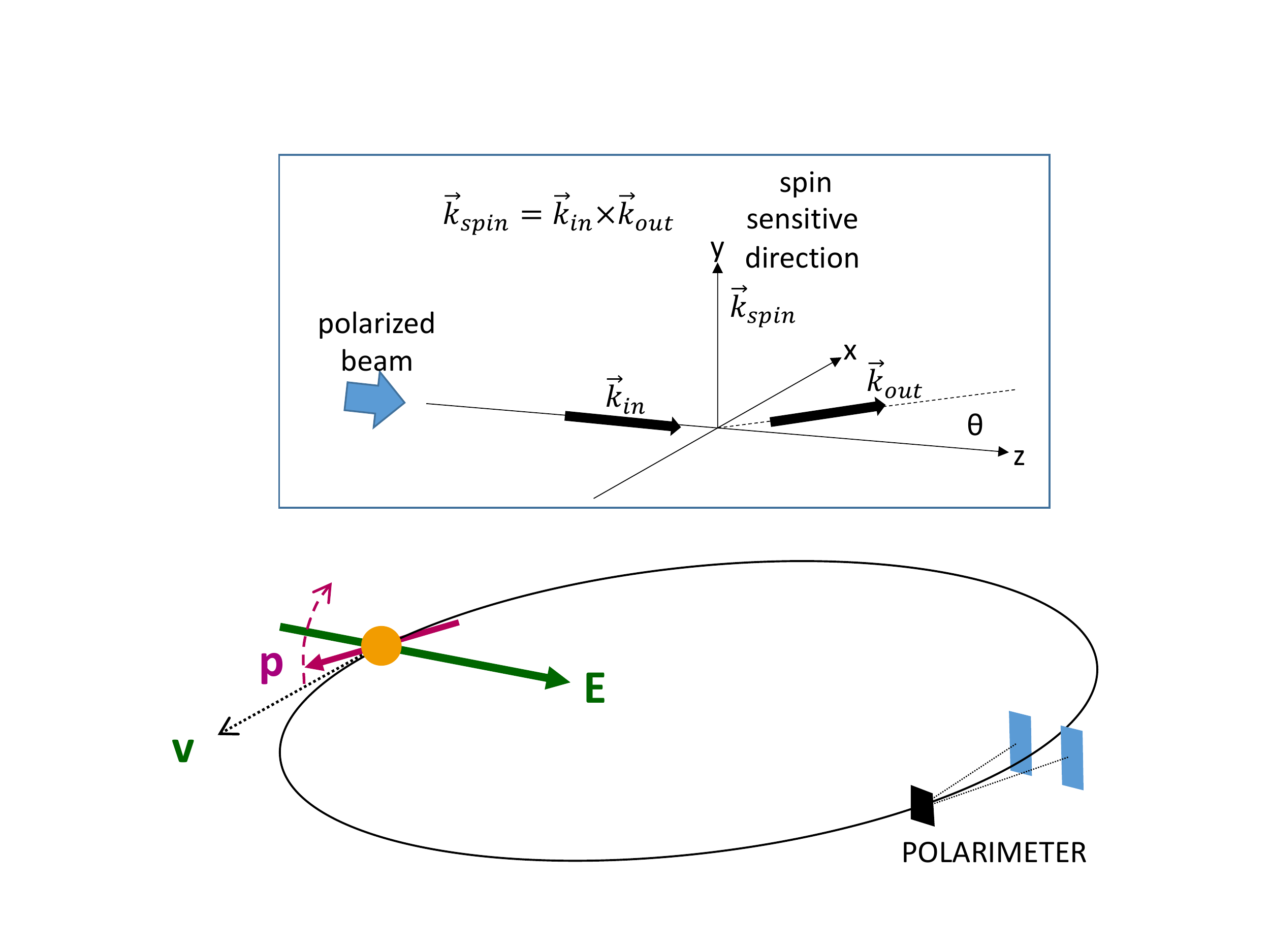}
    \caption{Diagram of storage ring EDM concept, with the horizontal spin precession locked to the momentum precession rate (``frozen'' spin). The radial electric field acts on the particle EDM for the duration of the storage time.  Positive and negative helicity bunches are stored, as well as bunches with their polarization pointing in the radial direction for systematic error cancellations. In addition, simultaneous clockwise and counterclockwise storage is used to cancel the main systematic errors. The ring circumference is about 800~m. The top inset shows the cross section geometry that is enhanced in parity-conserving Coulomb and nuclear scattering as the EDM signal increases over time.}
    \label{fig:EJS_ring}
\end{figure}

The concept of the storage ring EDM experiment is illustrated in \cref{fig:EJS_ring}. There are three starting requirements. (1) The proton beam must be highly polarized in the ring plane. (2) The momentum of the beam must match the magic value of $p=0.7007$~GeV/c where the ring plane precession is ``frozen.'' (3) The polarization is initially along the axis of the beam velocity.

The electric field acts along the radial direction toward the center of the ring (E). It is perpendicular to the spin axis (p) and therefore perpendicular to the axis of the EDM. In this situation the spin will precess in the vertical plane as shown. The appearance of a vertical polarization component with time is the signal for a non-vanishing EDM. This signal is measured at the polarimeter where a sample of the beam is continuously brought to a carbon target. Elastic proton scattering is measured by two downstream detectors (blue). Any asymmetry in the rates into these two detectors is a manifestation of the vertical polarization component and hence of the EDM.

A limited number of sensitive storage ring EDM experimental methods have been developed with various degrees of sensitivity and levels of systematic error, see \Cref{tab:lattices}~\cite{farley_new_2004,symmetric}. Here we only address the method based on the hybrid-symmetric ring lattice, which has been studied extensively and shown to perform well, applying
presently available technologies. The other methods, although promising, are outside the scope of this document, requiring additional studies and further technical developments.

The hybrid-symmetric ring method is built on the all-electric ring method, improving it in a number of critical ways that make it practical with present technology. It replaces the electric focusing with alternating gradient magnetic focusing, still allowing simultaneous CW and CCW storage and eliminating the main systematic error source by design. A major improvement in this design is the enhanced ring-lattice symmetry, eliminating the next most-important systematic error source, that of the average vertical beam velocity within the bending sections~\cite{symmetric}.

Symmetries in the hybrid-symmetric ring with \targetsens sensitivity:

\begin{enumerate}
    \item CW and CCW beam storage simultaneously.
    \item Positive and negative helicity of longitudinal and radial beam polarizations.
    \item Current flip of the magnetic quadrupoles.
    \item Beam planarity to better than \SI{0.1}{mm} and beam splitting of the counter-rotating (CR) beams to $<\SI{0.01}{mm}$.
\end{enumerate}

\begin{table*}
    \footnotesize
  \centering
  \caption{Storage ring Electric Dipole Moment experiment options}\label{tab:lattices}  
  \begin{tabular}[t]{p{0.20\linewidth} p{0.01\linewidth} p{0.15\linewidth} p{0.02\linewidth} p{0.25\linewidth} p{0.01\linewidth} p{0.3\linewidth}}
    \toprule
    Fields & & Example & & EDM signal term & & Comments \\ \midrule
    
    Dipole magnetic field $\bm{B}$ (Parasitic).
    & & Muon $(g-2)$ experiment.
    & & Tilt of the spin precession plane. (Limited statistical sensitivity due to spin precession.) 
    & & Eventually limited by geometrical alignment. Requires consecutive CW and CCW injection to eliminate systematic errors. \vspace{1cm}\\
    
    Combination of electric and magnetic fields ($\vb{E}, \vb{B}$) (Combined lattice).
    & & Deuteron, $^3\textnormal{He},$ proton.
    & & $\dv{\bm{s}}{t} \approx \bm{d}\times\qty(\bm{v}\times\bm{B})$
    & & High statistical sensitivity. Requires consecutive CW and CCW injection, with main fields flipping sign to eliminate systematic errors.\vspace{0.5cm}\\
    
    Radial electric field ($\vb{E}$) and electric focusing ($\vb{E}$) (All-electric lattice).
    & & Proton.
    & & $\dv{\bm{s}}{t} = \bm{d}\times\bm{E}$
    & & Allows simultaneous CW and CCW storage. Requires demonstration of adequate sensitivity to radial $\bm{B}$-field systematic error source.\vspace{0.5cm}\\
    
    Radial electric field ($\vb{E}$) and magnetic focusing ($\vb{B}$)
    (Hybrid, symmetric lattice).
    & & Proton.
    & & $\dv{\bm{s}}{t} = \bm{d}\times\bm{E}$
    & & Allows simultaneous CW and CCW storage. Only lattice to achieve direct cancellation of the main systematic error sources (its own ``co-magnetometer''). \\
    \bottomrule
\end{tabular}
\end{table*}

\subsection{Highly symmetric lattice}
A highly symmetric lattice is necessary to limit the dark matter/dark energy systematics, see~\cite{graham_paper,symmetric}. The 24-fold symmetric ring parameters are given in \Cref{tab:specs} and some of the beam parameters in \Cref{tab:magicgamma}. The \Cref{tab:specs} ring circumference is \SI{800}{m}, with bending electric field \SI{4.4}{MV/m}. This circumference is close to that of the BNL AGS tunnel, which saves tunnel construction costs. $E=\SI{4.4}{MV/m}$ is conservative. A ``green field'' pEDM experiment could have up to $E=\SI{5}{MV/m}$ without R\&D progress, see~\cite{electrode1,electrode2,electrode3}.

\begin{table}[tbp]
  \centering
  \caption{Ring and beam parameters for the hybrid-symmetric ring design. The beam planarity refers to the average vertical orbit of the counter-rotating (CR) beams with respect to gravity around the ring.}
  \begin{tabular}[t]{ll}
    \toprule
    Quantity                                                      & Value                        \\
    \midrule
    Bending Radius $R_{0}$                                        & \SI{95.49}{m}                \\
    Number of periods                                             & 24                           \\
    Electrode spacing                                             & \SI{4}{cm}                   \\
    Electrode height                                              & \SI{20}{cm}                  \\
    Deflector shape                                               & cylindrical                  \\
    Radial bending $E$-field                                      & \SI{4.4}{MV/m}               \\
    Straight section length                                       & \SI{4.16}{m}                 \\
    Quadrupole length                                             & \SI{0.4}{m}                  \\
    Quadrupole strength                                           & \SI{\pm 0.21}{T/m}           \\
    Bending section length                                        & \SI{12.5}{m}                 \\
    Bending section circumference \,\,\,                                & \SI{600}{m}                  \\
    Total circumference                                           & \SI{800}{m}               \\
    Cyclotron frequency                                           & \SI{224}{kHz}                \\
        Tunes, $Q_{x}, ~ Q_{y}$                                       & 2.699, 2.245                 \\
    Particles per bunch                                           & \num{1.17e8}                 \\
    RF voltage                                                    & \SI{1.89}{k V}               \\
    Harmonic number, $h$                                          & 80                           \\
    Synchrotron tune, $Q_{s}$                                     & \num{3.81e-3}                \\
    Beam planarity better than                                                & \SI{0.1}{mm}               \\
    CR-beam splitting less than                                             & \SI{0.01}{mm}                \\
    \bottomrule
  \end{tabular}
  \label{tab:specs}
\end{table}%
\begin{table}[tbp]
  \centering
  \caption{``Magic'' parameters for protons, values obtained from \citet{mooser_direct_2014}.}
  \begin{tabular}{lllll}
    \toprule
    $G$           & $\beta$         & $\gamma$        & $p$                  & $KE$
    \\
    \midrule
    $\num{1.793}\qq{}$ & $\num{0.598}\qq{}$ & $\num{1.248}\qq{}$ & $\SI{700.7}{MeV/c}\qq{}$ & $\SI{233}{MeV}$\\
    \bottomrule
  \end{tabular}
  \label{tab:magicgamma}
\end{table}

\subsection{Polarimetry}
 
Tests with beams and polarimeters at several laboratories (BNL, KVI, COSY) have consistently demonstrated over more than a decade that the requirements of storage ring EDM search are within reach~\cite{brantjes2012correcting,ref1,ref2,ref3,hempelmann-phase-2017,jedi_collaboration_how_2016,ref6,ref7}. Of particular importance, it has been shown that polarimeters based on forward elastic scattering offer a way to calibrate and correct geometrical and counting rate systematic errors in real time. Sextupole field adjustments along with electron cooling yield long lifetimes for a ring-plane polarization whose direction may be controlled using polarimeter-based feedback. Given the extensive model-based studies demonstrating that ring designs using the symmetries described above can control EDM systematics at the \targetsens level~\cite{symmetric}, the optimum path forward is to continue these developments on a full-scale hybrid, symmetric-lattice machine. 

The features of the forward-angle elastic scattering polarimeter are listed here.

\begin{figure}
\centering
\includegraphics[width=0.5\textwidth]{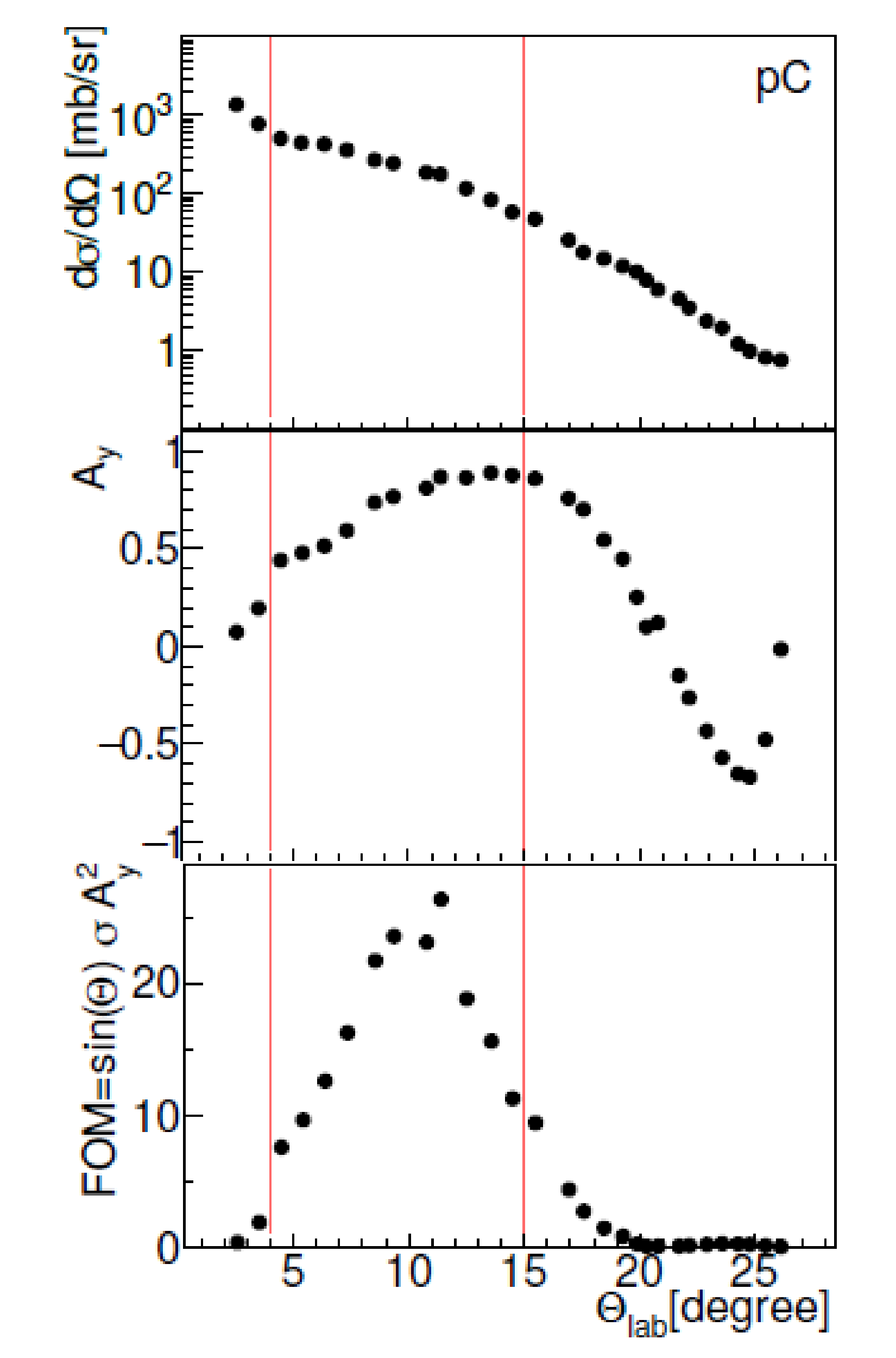}
\caption{Angular distributions of p+C elastic scattering differential cross section, analyzing power, and modified figure of merit ($FOM=(\sin{\theta}) \sigma A^2_y$). The red lines show typical boundaries for data collection in a polarimeter.}\label{fig:FoM}
\end{figure}

\begin{itemize}
    \item Carbon target, observing elastic scattering between \ang{5} and \ang{15}. Target thickness: \SIrange{2}{4}{cm}. Angular distributions are shown in \Cref{fig:FoM} from \citet{hom88}.
    \item CW and CCW polarimeters share target in middle. Calibrate using vertical polarization and by allowing the horizontal spin to fully precess.
    \item Detector: position sensitive $\Delta E$, segmented calorimeter.
    \item Efficiency: $\sim 1$\% using particles removed from beam, which become part of the useful data stream.
    \item Analyzing power = 0.6, under Monte-Carlo (MC) estimation.
    \item Signal accumulation rate at \targetsens $\SI{e-9}{rad/s}$.
    \item Full azimuthal coverage and forward/backward polarization allow first-order systematic error monitoring. Corrections to signal made to second order. Tested successfully at $10^{-5}$~\cite{brantjes2012correcting}.
\end{itemize}

\subsection{Statistics}
The statistical sensitivity of a single measurement, similar to the neutron EDM case, is inversely proportional to the beam polarization, the analyzing power, the spin coherence time (SCT) and the square root of the number of detected events. One advantage of the storage ring method over using neutrons is that high-intensity, highly polarized beams with small values in the relevant phase-space parameters are readily available. As a consequence, it is possible to achieve long SCT with horizontally polarized beams, as was calculated analytically and demonstrated at COSY~\cite{jedi_collaboration_how_2016,hempelmann-phase-2017}.

Under optimized running conditions, where the beam storage duration is for half the SCT, the EDM statistical sensitivity of the method is given by~\cite{kim_new_2021},
\begin{equation}
    \sigma_d = \frac{2.33\hbar}{P_0 A E \sqrt{k N_{cyc} T_{exp}\tau_p}},
\end{equation}
where  $P_0$ ($\sim 0.8$) is the horizontal beam polarization, $A$ ($\sim0.6$) is the asymmetry, $E$ ($\SI{3.3}{MV/m} = \SI{4.4}{MV/m} \times \SI{600}{m}/\SI{800}{m}$) is the average radial electric field integrated around the ring, $k$ (1\%) is the polarimeter detector efficiency, $N_{cyc}$ ($\sim \num{2e10}$) is the stored particles per cycle, $T_{\textnormal{ exp}}$  (\SI{1e8}{s}) is the total duration of the experiment and $\tau_p$ (\SI{2e3}{s}) is the in-plane (horizontal) beam polarization lifetime (equivalent to SCT). The SCT of \SI{2e3}{s}, i.e., an optimum storage time of \SI{e3}{s}, is assumed here in order to achieve a statistical sensitivity at \targetsens~level, while assuming the total experiment duration is \SI{80}{million} seconds (in practice, corresponding to roughly five calendar years). Such a beam storage might require stochastic cooling due to IBS and beam-gas interactions. The SCT of the beam itself (without stochastic cooling) is estimated to be greater than \SI{2e3}{s}, as indicated by preliminary results with high-precision beam/spin-dynamics simulations and is limited by the simulation speed. 

\subsection{The muon EDM storage ring experiment at PSI}

The muEDM initiative at the Paul Scherrer Institute, Switzerland, prepares a search for a muon EDM using the frozen-spin method\cite{Farley2004} in two steps.
During the initial phase, crucial instrumentation for the frozen-spin technique will be demonstrated\cite{Adelmann:2010zz,Adelmann:2021udj}, using an existing compact solenoid with a magnetic field strength of 3T. 
By using surface muons with a momentum of about 28MeV/c and a polarization of at least 95\% we expect a sensitivity of better than $3\times10^{-21} e\cdot {\rm cm}$. 
The muons will be injected along the main field direction similar as in the concept for the (g-2) experiment at JPARC. 
A short magnetic-field pulse in the central plane of the solenoid, triggered by an entrance scintillator, will twist the muons into a stable orbit within the weakly focusing field area. 
The electric field, $E\approx aBc\beta\gamma^2 \approx 3{\rm kV/cm}$, for the frozen-spin condition will be applied by a concentric electrode system made of thin graphite foils to reduce multiple scattering of decay positrons and eddy currents from the magnetic kick used for injection.
A positron detection system outside the storage zone records longitudinal positron asymmetry, along the magnet field direction. The signal for an EDM is the increase in positron asymmetry with time after injection.
The final experiment will use a dedicated 3T magnet with a small magnetic-field gradient between injection area and the storage zone increasing significantly the injection efficiency. By using muons with a momentum larger than 125 MeV/c and an accordingly higher electric field of about 20kV/cm a sensitivity of better than $6\times10^{-23} e\cdot {\rm cm}$ is anticipated within a year of data-taking.

\subsection{Conclusions}
\begin{figure}
    \centering
    \includegraphics[width=0.85\textwidth]{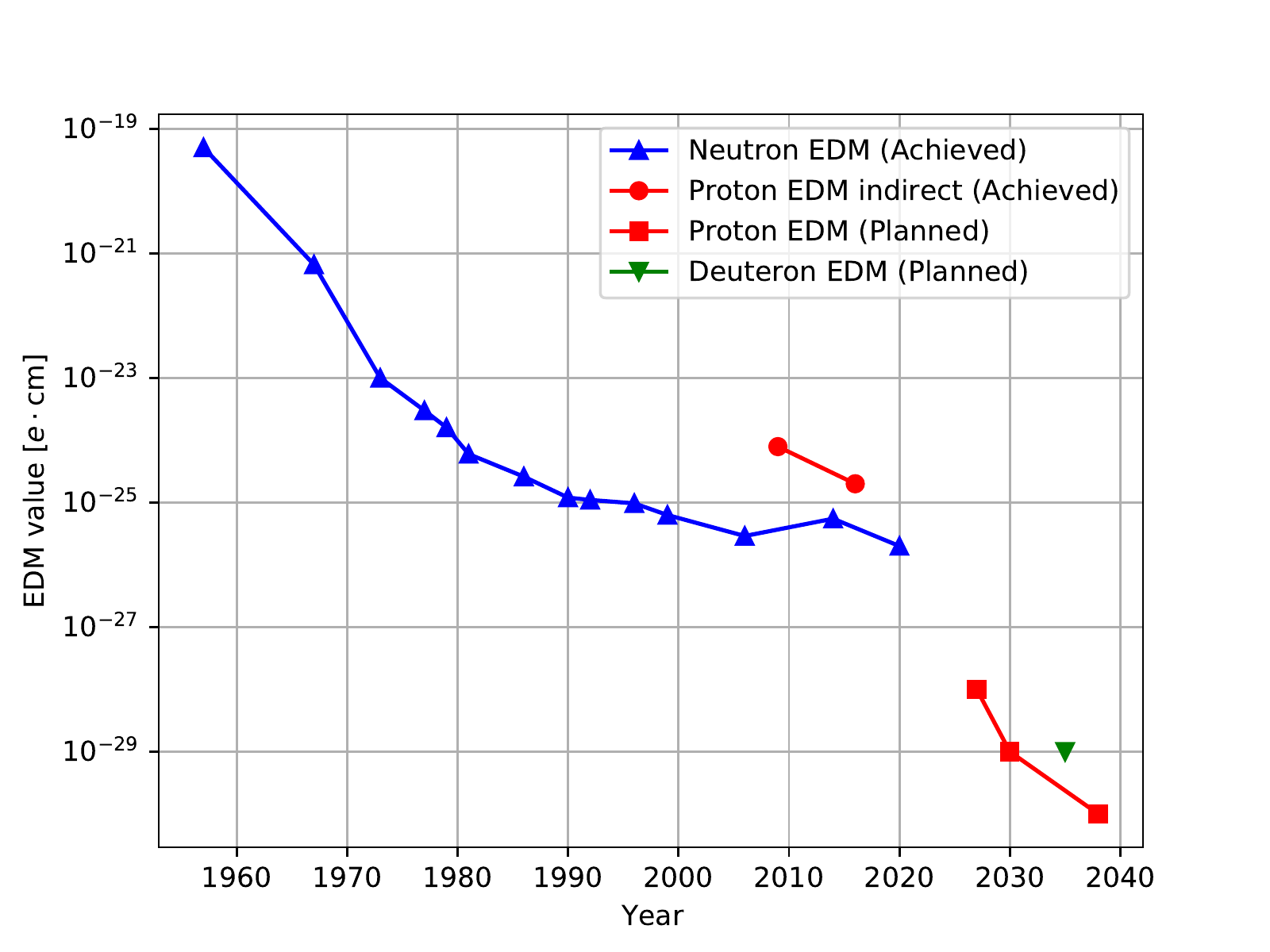}
    \caption{The neutron and proton (indirect) EDM limits relative to publication year are shown here. The storage ring EDM projected sensitivities for the proton and deuteron nuclei are also shown as a function of year. The $^3$He nucleus storage ring EDM is projected to be similar to that of the deuteron.}
    \label{fig:edmplot}
\end{figure}

A storage ring proton EDM experiment offering an unprecedented statistical sensitivity to \targetsens level, can be built based on present technology. At this level, it would be the best EDM experiment on one of the simplest hadrons. The method, based on the hybrid-symmetric ring lattice, is the only one today that can eliminate the main sources of EDM systematic error. The experiment would also make it possible to study the deuteron/$^3\textnormal{He}$ EDM with a sensitivity about an order of magnitude less than the one stated above.  We expect construction to take 3 -- 5 years and another 2 -- 3 years to collect the required statistics to first physics publication.    \Cref{fig:edmplot} shows the neutron EDM limits as well as the proton indirect limits (from $^{199}$Hg) as a function of publication year. We also show the projected EDM sensitivity of the storage ring EDM method for the proton and deuteron nuclei. The storage ring $^3$He nucleus EDM (equivalent to a neutron EDM) sensitivity is also estimated to be similar to that of the deuteron. A proposed experiment at PSI, using the frozen spin technique, will improve the sensitivity of the muon EDM by roughly 3-4 orders of magnitude over the current limit.

\section{Summary and Outlook}

An EDM observed in a single system (electron, proton, neutron, deuteron, atom, molecule) will very likely signal the discovery of new physics beyond the Standard Model of Particle Physics, with far-reaching impact, but it will not be sufficient to discriminate among the many viable BSM models or rule out baryogenesis scenarios. To achieve these goals it is necessary to observe or bound EDMs in complementary systems, so that the main features of CP violation at low energy can be identified and systematically connected to the underlying BSM models. A promising experiment-theory program between particle, nuclear, and AMO disciplines has been described in this white paper. In the next 10 years gains in sensitivity by several orders of magnitude over current bounds are possible and even likely for electrons, nucleons, atoms and molecules, with the very real chance of discovery. Some of these experiments also offer opportunities for dark matter, dark energy, and axion searches. Likewise, improvements in theory, including lattice QCD, low energy effective field theories, and theories of nuclear and atomic systems will dramatically improve the interpretation of experimental results.

\section{Endorsements}

In addition to the listed authors, the following people endorse and support this white paper:
\begin{itemize}
\setlength\itemsep{-0.4em}
\item Marcel Demarteau, Oak Ridge National Laboratory, Oak Ridge, TN 37831, USA.
\item Rajan Gupta, Los Alamos National Laboratory, Los Alamos, NM 87545, USA.
\item Steve K. Lamoreaux, Yale University, New Haven, CT 06520, USA.
\item Keh-Fei Liu, University of Kentucky, Lexington, KY 40506, USA.
\item Zheng-Tian Lu, University of Science and Technology of China, Hefei, China
\item Konstantinos Orginos, Jefferson Laboratory, Newport News, VA 23606, USA.
\item Anthony Palladino, Boston University, Boston, MA, 02215 USA.
\item Robert P. Redwine, Massachusetts Institute of Technology, Cambridge, MA 02139, USA.
\item W. Michael Snow, Indiana University, Bloomington, IN 47405, USA.
\item Wim Ubachs, Vrije Universiteit Amsterdam, The Netherlands
\item Liang Yang,  University of California San Diego, La Jolla, CA 92093, USA.
\item Jun Ye, JILA, NIST, and University of Colorado, Boulder, CO 80309, USA.
\end{itemize}
 %


\bibliographystyle{apsrev4-2}
\bibliography{Theory,neutron,amo1,amo2,Hadron,Misc,SREDM,SREDM_new}

\begin{thebibliography}{396}%
\makeatletter
\providecommand \@ifxundefined [1]{%
 \@ifx{#1\undefined}
}%
\providecommand \@ifnum [1]{%
 \ifnum #1\expandafter \@firstoftwo
 \else \expandafter \@secondoftwo
 \fi
}%
\providecommand \@ifx [1]{%
 \ifx #1\expandafter \@firstoftwo
 \else \expandafter \@secondoftwo
 \fi
}%
\providecommand \natexlab [1]{#1}%
\providecommand \enquote  [1]{``#1''}%
\providecommand \bibnamefont  [1]{#1}%
\providecommand \bibfnamefont [1]{#1}%
\providecommand \citenamefont [1]{#1}%
\providecommand \href@noop [0]{\@secondoftwo}%
\providecommand \href [0]{\begingroup \@sanitize@url \@href}%
\providecommand \@href[1]{\@@startlink{#1}\@@href}%
\providecommand \@@href[1]{\endgroup#1\@@endlink}%
\providecommand \@sanitize@url [0]{\catcode `\\12\catcode `\$12\catcode
  `\&12\catcode `\#12\catcode `\^12\catcode `\_12\catcode `\%12\relax}%
\providecommand \@@startlink[1]{}%
\providecommand \@@endlink[0]{}%
\providecommand \url  [0]{\begingroup\@sanitize@url \@url }%
\providecommand \@url [1]{\endgroup\@href {#1}{\urlprefix }}%
\providecommand \urlprefix  [0]{URL }%
\providecommand \Eprint [0]{\href }%
\providecommand \doibase [0]{https://doi.org/}%
\providecommand \selectlanguage [0]{\@gobble}%
\providecommand \bibinfo  [0]{\@secondoftwo}%
\providecommand \bibfield  [0]{\@secondoftwo}%
\providecommand \translation [1]{[#1]}%
\providecommand \BibitemOpen [0]{}%
\providecommand \bibitemStop [0]{}%
\providecommand \bibitemNoStop [0]{.\EOS\space}%
\providecommand \EOS [0]{\spacefactor3000\relax}%
\providecommand \BibitemShut  [1]{\csname bibitem#1\endcsname}%
\let\auto@bib@innerbib\@empty
\bibitem [{\citenamefont {Luders}(1954)}]{Luders:1954zz}%
  \BibitemOpen
  \bibfield  {author} {\bibinfo {author} {\bibfnamefont {G.}~\bibnamefont
  {Luders}},\ }\href@noop {} {\bibfield  {journal} {\bibinfo  {journal} {Kong.\
  Dan.\ Vid.\ Sel.\ Mat.\ Fys.\ Med.}\ }\textbf {\bibinfo {volume} {28N5}},\
  \bibinfo {pages} {1} (\bibinfo {year} {1954})}\BibitemShut {NoStop}%
\bibitem [{\citenamefont {Kobayashi}\ and\ \citenamefont
  {Maskawa}(1973)}]{Kobayashi:1973fv}%
  \BibitemOpen
  \bibfield  {author} {\bibinfo {author} {\bibfnamefont {M.}~\bibnamefont
  {Kobayashi}}\ and\ \bibinfo {author} {\bibfnamefont {T.}~\bibnamefont
  {Maskawa}},\ }\href {https://doi.org/10.1143/PTP.49.652} {\bibfield
  {journal} {\bibinfo  {journal} {Prog.Theor.Phys.}\ }\textbf {\bibinfo
  {volume} {49}},\ \bibinfo {pages} {652} (\bibinfo {year} {1973})}\BibitemShut
  {NoStop}%
\bibitem [{\citenamefont {Pontecorvo}(1957)}]{Pontecorvo:1957qd}%
  \BibitemOpen
  \bibfield  {author} {\bibinfo {author} {\bibfnamefont {B.}~\bibnamefont
  {Pontecorvo}},\ }\href@noop {} {\bibfield  {journal} {\bibinfo  {journal}
  {Zh. Eksp. Teor. Fiz.}\ }\textbf {\bibinfo {volume} {34}},\ \bibinfo {pages}
  {247} (\bibinfo {year} {1957})}\BibitemShut {NoStop}%
\bibitem [{\citenamefont {Maki}\ \emph {et~al.}(1962)\citenamefont {Maki},
  \citenamefont {Nakagawa},\ and\ \citenamefont {Sakata}}]{Maki:1962mu}%
  \BibitemOpen
  \bibfield  {author} {\bibinfo {author} {\bibfnamefont {Z.}~\bibnamefont
  {Maki}}, \bibinfo {author} {\bibfnamefont {M.}~\bibnamefont {Nakagawa}},\
  and\ \bibinfo {author} {\bibfnamefont {S.}~\bibnamefont {Sakata}},\ }\href
  {https://doi.org/10.1143/PTP.28.870} {\bibfield  {journal} {\bibinfo
  {journal} {Prog.\ Theor.\ Phys.}\ }\textbf {\bibinfo {volume} {28}},\
  \bibinfo {pages} {870} (\bibinfo {year} {1962})}\BibitemShut {NoStop}%
\bibitem [{\citenamefont {Nunokawa}\ \emph {et~al.}(2008)\citenamefont
  {Nunokawa}, \citenamefont {Parke},\ and\ \citenamefont
  {Valle}}]{Nunokawa:2007qh}%
  \BibitemOpen
  \bibfield  {author} {\bibinfo {author} {\bibfnamefont {H.}~\bibnamefont
  {Nunokawa}}, \bibinfo {author} {\bibfnamefont {S.~J.}\ \bibnamefont
  {Parke}},\ and\ \bibinfo {author} {\bibfnamefont {J.~W.}\ \bibnamefont
  {Valle}},\ }\href {https://doi.org/10.1016/j.ppnp.2007.10.001} {\bibfield
  {journal} {\bibinfo  {journal} {Prog.Part.Nucl.Phys.}\ }\textbf {\bibinfo
  {volume} {60}},\ \bibinfo {pages} {338} (\bibinfo {year} {2008})},\ \Eprint
  {https://arxiv.org/abs/0710.0554} {arXiv:0710.0554 [hep-ph]} \BibitemShut
  {NoStop}%
\bibitem [{\citenamefont {Pospelov}\ and\ \citenamefont
  {Ritz}(2005)}]{Pospelov:2005pr}%
  \BibitemOpen
  \bibfield  {author} {\bibinfo {author} {\bibfnamefont {M.}~\bibnamefont
  {Pospelov}}\ and\ \bibinfo {author} {\bibfnamefont {A.}~\bibnamefont
  {Ritz}},\ }\href {https://doi.org/10.1016/j.aop.2005.04.002} {\bibfield
  {journal} {\bibinfo  {journal} {Annals Phys.}\ }\textbf {\bibinfo {volume}
  {318}},\ \bibinfo {pages} {119} (\bibinfo {year} {2005})},\ \Eprint
  {https://arxiv.org/abs/hep-ph/0504231} {arXiv:hep-ph/0504231 [hep-ph]}
  \BibitemShut {NoStop}%
\bibitem [{\citenamefont {Yamaguchi}\ and\ \citenamefont
  {Yamanaka}(2021)}]{Yamaguchi2021}%
  \BibitemOpen
  \bibfield  {author} {\bibinfo {author} {\bibfnamefont {Y.}~\bibnamefont
  {Yamaguchi}}\ and\ \bibinfo {author} {\bibfnamefont {N.}~\bibnamefont
  {Yamanaka}},\ }\bibfield  {journal} {\bibinfo  {journal} {Physical Review D}\
  }\textbf {\bibinfo {volume} {103}},\ \href
  {https://doi.org/10.1103/physrevd.103.013001} {10.1103/physrevd.103.013001}
  (\bibinfo {year} {2021})\BibitemShut {NoStop}%
\bibitem [{\citenamefont {Ema}\ \emph {et~al.}(2022)\citenamefont {Ema},
  \citenamefont {Gao},\ and\ \citenamefont {Pospelov}}]{Ema2022}%
  \BibitemOpen
  \bibfield  {author} {\bibinfo {author} {\bibfnamefont {Y.}~\bibnamefont
  {Ema}}, \bibinfo {author} {\bibfnamefont {T.}~\bibnamefont {Gao}},\ and\
  \bibinfo {author} {\bibfnamefont {M.}~\bibnamefont {Pospelov}},\ }\href
  {https://doi.org/10.48550/ARXIV.2202.10524} {\bibinfo {title} {Standard model
  prediction for paramagnetic edms}} (\bibinfo {year} {2022})\BibitemShut
  {NoStop}%
\bibitem [{\citenamefont {Chupp}\ \emph
  {et~al.}(2019{\natexlab{a}})\citenamefont {Chupp}, \citenamefont
  {Fierlinger}, \citenamefont {Ramsey-Musolf},\ and\ \citenamefont
  {Singh}}]{Chupp:2017rkp}%
  \BibitemOpen
  \bibfield  {author} {\bibinfo {author} {\bibfnamefont {T.}~\bibnamefont
  {Chupp}}, \bibinfo {author} {\bibfnamefont {P.}~\bibnamefont {Fierlinger}},
  \bibinfo {author} {\bibfnamefont {M.}~\bibnamefont {Ramsey-Musolf}},\ and\
  \bibinfo {author} {\bibfnamefont {J.}~\bibnamefont {Singh}},\ }\href
  {https://doi.org/10.1103/RevModPhys.91.015001} {\bibfield  {journal}
  {\bibinfo  {journal} {Rev. Mod. Phys.}\ }\textbf {\bibinfo {volume} {91}},\
  \bibinfo {pages} {015001} (\bibinfo {year} {2019}{\natexlab{a}})},\ \Eprint
  {https://arxiv.org/abs/1710.02504} {arXiv:1710.02504 [physics.atom-ph]}
  \BibitemShut {NoStop}%
\bibitem [{\citenamefont {Bennett}\ \emph {et~al.}(2003)\citenamefont {Bennett}
  \emph {et~al.}}]{Bennett:2003bz}%
  \BibitemOpen
  \bibfield  {author} {\bibinfo {author} {\bibfnamefont {C.}~\bibnamefont
  {Bennett}} \emph {et~al.} (\bibinfo {collaboration} {WMAP Collaboration}),\
  }\href {https://doi.org/10.1086/377253} {\bibfield  {journal} {\bibinfo
  {journal} {Astrophys.J.Suppl.}\ }\textbf {\bibinfo {volume} {148}},\ \bibinfo
  {pages} {1} (\bibinfo {year} {2003})},\ \Eprint
  {https://arxiv.org/abs/astro-ph/0302207} {arXiv:astro-ph/0302207 [astro-ph]}
  \BibitemShut {NoStop}%
\bibitem [{\citenamefont {Ade}\ \emph {et~al.}(2016)\citenamefont {Ade} \emph
  {et~al.}}]{Planck:2015fie}%
  \BibitemOpen
  \bibfield  {author} {\bibinfo {author} {\bibfnamefont {P.~A.~R.}\
  \bibnamefont {Ade}} \emph {et~al.} (\bibinfo {collaboration} {Planck}),\
  }\href {https://doi.org/10.1051/0004-6361/201525830} {\bibfield  {journal}
  {\bibinfo  {journal} {Astron. Astrophys.}\ }\textbf {\bibinfo {volume}
  {594}},\ \bibinfo {pages} {A13} (\bibinfo {year} {2016})},\ \Eprint
  {https://arxiv.org/abs/1502.01589} {arXiv:1502.01589 [astro-ph.CO]}
  \BibitemShut {NoStop}%
\bibitem [{\citenamefont {Aghanim}\ \emph {et~al.}(2020)\citenamefont {Aghanim}
  \emph {et~al.}}]{Planck:2018vyg}%
  \BibitemOpen
  \bibfield  {author} {\bibinfo {author} {\bibfnamefont {N.}~\bibnamefont
  {Aghanim}} \emph {et~al.} (\bibinfo {collaboration} {Planck}),\ }\href
  {https://doi.org/10.1051/0004-6361/201833910} {\bibfield  {journal} {\bibinfo
   {journal} {Astron. Astrophys.}\ }\textbf {\bibinfo {volume} {641}},\
  \bibinfo {pages} {A6} (\bibinfo {year} {2020})},\ \bibinfo {note} {[Erratum:
  Astron.Astrophys. 652, C4 (2021)]},\ \Eprint
  {https://arxiv.org/abs/1807.06209} {arXiv:1807.06209 [astro-ph.CO]}
  \BibitemShut {NoStop}%
\bibitem [{\citenamefont {Fields}\ \emph {et~al.}(2020)\citenamefont {Fields},
  \citenamefont {Olive}, \citenamefont {Yeh},\ and\ \citenamefont
  {Young}}]{Fields:2019pfx}%
  \BibitemOpen
  \bibfield  {author} {\bibinfo {author} {\bibfnamefont {B.~D.}\ \bibnamefont
  {Fields}}, \bibinfo {author} {\bibfnamefont {K.~A.}\ \bibnamefont {Olive}},
  \bibinfo {author} {\bibfnamefont {T.-H.}\ \bibnamefont {Yeh}},\ and\ \bibinfo
  {author} {\bibfnamefont {C.}~\bibnamefont {Young}},\ }\href
  {https://doi.org/10.1088/1475-7516/2020/03/010} {\bibfield  {journal}
  {\bibinfo  {journal} {JCAP}\ }\textbf {\bibinfo {volume} {2020}}\bibfield
  {number} {\bibinfo  {number} { (03)},\ \bibinfo {pages} {010}},\ }\bibinfo
  {note} {[Erratum: JCAP 11, E02 (2020)]},\ \Eprint
  {https://arxiv.org/abs/1912.01132} {arXiv:1912.01132 [astro-ph.CO]}
  \BibitemShut {NoStop}%
\bibitem [{\citenamefont {Coppi}(2004)}]{Coppi:2004za}%
  \BibitemOpen
  \bibfield  {author} {\bibinfo {author} {\bibfnamefont {P.}~\bibnamefont
  {Coppi}},\ }\href@noop {} {\bibfield  {journal} {\bibinfo  {journal} {eConf}\
  }\textbf {\bibinfo {volume} {C040802}},\ \bibinfo {pages} {L017} (\bibinfo
  {year} {2004})}\BibitemShut {NoStop}%
\bibitem [{\citenamefont {Sakharov}(1967)}]{Sakharov:1967dj}%
  \BibitemOpen
  \bibfield  {author} {\bibinfo {author} {\bibfnamefont {A.}~\bibnamefont
  {Sakharov}},\ }\href {https://doi.org/10.1070/PU1991v034n05ABEH002497}
  {\bibfield  {journal} {\bibinfo  {journal} {Pisma Zh.Eksp.Teor.Fiz.}\
  }\textbf {\bibinfo {volume} {5}},\ \bibinfo {pages} {32} (\bibinfo {year}
  {1967})}\BibitemShut {NoStop}%
\bibitem [{\citenamefont {Shaposhnikov}(1987)}]{Shaposhnikov:1987tw}%
  \BibitemOpen
  \bibfield  {author} {\bibinfo {author} {\bibfnamefont {M.~E.}\ \bibnamefont
  {Shaposhnikov}},\ }\href {https://doi.org/10.1016/0550-3213(87)90127-1}
  {\bibfield  {journal} {\bibinfo  {journal} {Nucl.\ Phys.}\ }\textbf {\bibinfo
  {volume} {B287}},\ \bibinfo {pages} {757} (\bibinfo {year}
  {1987})}\BibitemShut {NoStop}%
\bibitem [{\citenamefont {Farrar}\ and\ \citenamefont
  {Shaposhnikov}(1993)}]{Farrar:1993sp}%
  \BibitemOpen
  \bibfield  {author} {\bibinfo {author} {\bibfnamefont {G.~R.}\ \bibnamefont
  {Farrar}}\ and\ \bibinfo {author} {\bibfnamefont {M.}~\bibnamefont
  {Shaposhnikov}},\ }\href {https://doi.org/10.1103/PhysRevLett.70.2833}
  {\bibfield  {journal} {\bibinfo  {journal} {Phys.\ Rev.\ Lett.}\ }\textbf
  {\bibinfo {volume} {70}},\ \bibinfo {pages} {2833} (\bibinfo {year}
  {1993})},\ \Eprint {https://arxiv.org/abs/hep-ph/9305274}
  {arXiv:hep-ph/9305274 [hep-ph]} \BibitemShut {NoStop}%
\bibitem [{\citenamefont {Gavela}\ \emph {et~al.}(1994)\citenamefont {Gavela},
  \citenamefont {Hernandez}, \citenamefont {Orloff},\ and\ \citenamefont
  {Pene}}]{Gavela:1993ts}%
  \BibitemOpen
  \bibfield  {author} {\bibinfo {author} {\bibfnamefont {M.~B.}\ \bibnamefont
  {Gavela}}, \bibinfo {author} {\bibfnamefont {P.}~\bibnamefont {Hernandez}},
  \bibinfo {author} {\bibfnamefont {J.}~\bibnamefont {Orloff}},\ and\ \bibinfo
  {author} {\bibfnamefont {O.}~\bibnamefont {Pene}},\ }\href
  {https://doi.org/10.1142/S0217732394000629} {\bibfield  {journal} {\bibinfo
  {journal} {Mod. Phys. Lett.}\ }\textbf {\bibinfo {volume} {A9}},\ \bibinfo
  {pages} {795} (\bibinfo {year} {1994})},\ \Eprint
  {https://arxiv.org/abs/hep-ph/9312215} {arXiv:hep-ph/9312215 [hep-ph]}
  \BibitemShut {NoStop}%
\bibitem [{\citenamefont {Huet}\ and\ \citenamefont
  {Sather}(1995)}]{Huet:1994jb}%
  \BibitemOpen
  \bibfield  {author} {\bibinfo {author} {\bibfnamefont {P.}~\bibnamefont
  {Huet}}\ and\ \bibinfo {author} {\bibfnamefont {E.}~\bibnamefont {Sather}},\
  }\href {https://doi.org/10.1103/PhysRevD.51.379} {\bibfield  {journal}
  {\bibinfo  {journal} {Phys. Rev.}\ }\textbf {\bibinfo {volume} {D51}},\
  \bibinfo {pages} {379} (\bibinfo {year} {1995})},\ \Eprint
  {https://arxiv.org/abs/hep-ph/9404302} {arXiv:hep-ph/9404302 [hep-ph]}
  \BibitemShut {NoStop}%
\bibitem [{\citenamefont {Dolgov}(1992)}]{Dolgov:1991fr}%
  \BibitemOpen
  \bibfield  {author} {\bibinfo {author} {\bibfnamefont {A.}~\bibnamefont
  {Dolgov}},\ }\href {https://doi.org/10.1016/0370-1573(92)90107-B} {\bibfield
  {journal} {\bibinfo  {journal} {Phys.\ Rept.}\ }\textbf {\bibinfo {volume}
  {222}},\ \bibinfo {pages} {309} (\bibinfo {year} {1992})}\BibitemShut
  {NoStop}%
\bibitem [{\citenamefont {Gross}\ \emph {et~al.}(1981)\citenamefont {Gross},
  \citenamefont {Pisarski},\ and\ \citenamefont {Yaffe}}]{Gross:1980br}%
  \BibitemOpen
  \bibfield  {author} {\bibinfo {author} {\bibfnamefont {D.~J.}\ \bibnamefont
  {Gross}}, \bibinfo {author} {\bibfnamefont {R.~D.}\ \bibnamefont
  {Pisarski}},\ and\ \bibinfo {author} {\bibfnamefont {L.~G.}\ \bibnamefont
  {Yaffe}},\ }\href {https://doi.org/10.1103/RevModPhys.53.43} {\bibfield
  {journal} {\bibinfo  {journal} {Rev.\ Mod.\ Phys.}\ }\textbf {\bibinfo
  {volume} {53}},\ \bibinfo {pages} {43} (\bibinfo {year} {1981})}\BibitemShut
  {NoStop}%
\bibitem [{\citenamefont {Morrissey}\ and\ \citenamefont
  {Ramsey-Musolf}(2012)}]{Morrissey:2012db}%
  \BibitemOpen
  \bibfield  {author} {\bibinfo {author} {\bibfnamefont {D.~E.}\ \bibnamefont
  {Morrissey}}\ and\ \bibinfo {author} {\bibfnamefont {M.~J.}\ \bibnamefont
  {Ramsey-Musolf}},\ }\href {https://doi.org/10.1088/1367-2630/14/12/125003}
  {\bibfield  {journal} {\bibinfo  {journal} {New J. Phys.}\ }\textbf {\bibinfo
  {volume} {14}},\ \bibinfo {pages} {125003} (\bibinfo {year} {2012})},\
  \Eprint {https://arxiv.org/abs/1206.2942} {arXiv:1206.2942 [hep-ph]}
  \BibitemShut {NoStop}%
\bibitem [{\citenamefont {Nir}\ and\ \citenamefont
  {Rattazzi}(1996)}]{Nir:1996am}%
  \BibitemOpen
  \bibfield  {author} {\bibinfo {author} {\bibfnamefont {Y.}~\bibnamefont
  {Nir}}\ and\ \bibinfo {author} {\bibfnamefont {R.}~\bibnamefont {Rattazzi}},\
  }\href {https://doi.org/10.1016/0370-2693(96)00571-0} {\bibfield  {journal}
  {\bibinfo  {journal} {Phys. Lett. B}\ }\textbf {\bibinfo {volume} {382}},\
  \bibinfo {pages} {363} (\bibinfo {year} {1996})},\ \Eprint
  {https://arxiv.org/abs/hep-ph/9603233} {arXiv:hep-ph/9603233} \BibitemShut
  {NoStop}%
\bibitem [{\citenamefont {Barr}\ and\ \citenamefont {Zee}(1990)}]{Barr:1990vd}%
  \BibitemOpen
  \bibfield  {author} {\bibinfo {author} {\bibfnamefont {S.~M.}\ \bibnamefont
  {Barr}}\ and\ \bibinfo {author} {\bibfnamefont {A.}~\bibnamefont {Zee}},\
  }\href {https://doi.org/10.1103/PhysRevLett.65.21} {\bibfield  {journal}
  {\bibinfo  {journal} {Phys. Rev. Lett.}\ }\textbf {\bibinfo {volume} {65}},\
  \bibinfo {pages} {21} (\bibinfo {year} {1990})},\ \bibinfo {note} {[Erratum:
  Phys.Rev.Lett. 65, 2920 (1990)]}\BibitemShut {NoStop}%
\bibitem [{\citenamefont {Ellis}\ \emph {et~al.}(1982)\citenamefont {Ellis},
  \citenamefont {Ferrara},\ and\ \citenamefont {Nanopoulos}}]{Ellis:1982tk}%
  \BibitemOpen
  \bibfield  {author} {\bibinfo {author} {\bibfnamefont {J.~R.}\ \bibnamefont
  {Ellis}}, \bibinfo {author} {\bibfnamefont {S.}~\bibnamefont {Ferrara}},\
  and\ \bibinfo {author} {\bibfnamefont {D.~V.}\ \bibnamefont {Nanopoulos}},\
  }\href {https://doi.org/10.1016/0370-2693(82)90484-1} {\bibfield  {journal}
  {\bibinfo  {journal} {Phys. Lett.}\ }\textbf {\bibinfo {volume} {114B}},\
  \bibinfo {pages} {231} (\bibinfo {year} {1982})}\BibitemShut {NoStop}%
\bibitem [{\citenamefont {Chia}\ and\ \citenamefont
  {Nandi}(1982)}]{Chia:1982gp}%
  \BibitemOpen
  \bibfield  {author} {\bibinfo {author} {\bibfnamefont {S.~P.}\ \bibnamefont
  {Chia}}\ and\ \bibinfo {author} {\bibfnamefont {S.}~\bibnamefont {Nandi}},\
  }\href {https://doi.org/10.1016/0370-2693(82)90870-X} {\bibfield  {journal}
  {\bibinfo  {journal} {Phys. Lett.}\ }\textbf {\bibinfo {volume} {117B}},\
  \bibinfo {pages} {45} (\bibinfo {year} {1982})}\BibitemShut {NoStop}%
\bibitem [{\citenamefont {Polchinski}\ and\ \citenamefont
  {Wise}(1983)}]{Polchinski:1983zd}%
  \BibitemOpen
  \bibfield  {author} {\bibinfo {author} {\bibfnamefont {J.}~\bibnamefont
  {Polchinski}}\ and\ \bibinfo {author} {\bibfnamefont {M.~B.}\ \bibnamefont
  {Wise}},\ }\href {https://doi.org/10.1016/0370-2693(83)91310-2} {\bibfield
  {journal} {\bibinfo  {journal} {Phys. Lett.}\ }\textbf {\bibinfo {volume}
  {125B}},\ \bibinfo {pages} {393} (\bibinfo {year} {1983})}\BibitemShut
  {NoStop}%
\bibitem [{\citenamefont {Gavela}\ and\ \citenamefont
  {Georgi}(1982)}]{GavelaLegazpi:1982ud}%
  \BibitemOpen
  \bibfield  {author} {\bibinfo {author} {\bibfnamefont {M.~B.}\ \bibnamefont
  {Gavela}}\ and\ \bibinfo {author} {\bibfnamefont {H.}~\bibnamefont
  {Georgi}},\ }\href {https://doi.org/10.1016/0370-2693(82)90263-5} {\bibfield
  {journal} {\bibinfo  {journal} {Phys. Lett.}\ }\textbf {\bibinfo {volume}
  {119B}},\ \bibinfo {pages} {141} (\bibinfo {year} {1982})}\BibitemShut
  {NoStop}%
\bibitem [{\citenamefont {del Aguila}\ \emph {et~al.}(1983)\citenamefont {del
  Aguila}, \citenamefont {Gavela}, \citenamefont {Grifols},\ and\ \citenamefont
  {Mendez}}]{delAguila:1983dfr}%
  \BibitemOpen
  \bibfield  {author} {\bibinfo {author} {\bibfnamefont {F.}~\bibnamefont {del
  Aguila}}, \bibinfo {author} {\bibfnamefont {M.~B.}\ \bibnamefont {Gavela}},
  \bibinfo {author} {\bibfnamefont {J.~A.}\ \bibnamefont {Grifols}},\ and\
  \bibinfo {author} {\bibfnamefont {A.}~\bibnamefont {Mendez}},\ }\href
  {https://doi.org/10.1016/0370-2693(83)90018-7} {\bibfield  {journal}
  {\bibinfo  {journal} {Phys. Lett.}\ }\textbf {\bibinfo {volume} {126B}},\
  \bibinfo {pages} {71} (\bibinfo {year} {1983})},\ \bibinfo {note} {[Erratum:
  Phys. Lett.129B,473(1983)]}\BibitemShut {NoStop}%
\bibitem [{\citenamefont {Leigh}\ \emph {et~al.}(1991)\citenamefont {Leigh},
  \citenamefont {Paban},\ and\ \citenamefont {Xu}}]{Leigh:1990kf}%
  \BibitemOpen
  \bibfield  {author} {\bibinfo {author} {\bibfnamefont {R.~G.}\ \bibnamefont
  {Leigh}}, \bibinfo {author} {\bibfnamefont {S.}~\bibnamefont {Paban}},\ and\
  \bibinfo {author} {\bibfnamefont {R.~M.}\ \bibnamefont {Xu}},\ }\href
  {https://doi.org/10.1016/0550-3213(91)90128-K} {\bibfield  {journal}
  {\bibinfo  {journal} {Nucl. Phys.}\ }\textbf {\bibinfo {volume} {B352}},\
  \bibinfo {pages} {45} (\bibinfo {year} {1991})}\BibitemShut {NoStop}%
\bibitem [{\citenamefont {Barr}(1992)}]{Barr:1992cm}%
  \BibitemOpen
  \bibfield  {author} {\bibinfo {author} {\bibfnamefont {S.~M.}\ \bibnamefont
  {Barr}},\ }\href {https://doi.org/10.1103/PhysRevD.45.4148} {\bibfield
  {journal} {\bibinfo  {journal} {Phys. Rev. D}\ }\textbf {\bibinfo {volume}
  {45}},\ \bibinfo {pages} {4148} (\bibinfo {year} {1992})}\BibitemShut
  {NoStop}%
\bibitem [{\citenamefont {Fuyuto}\ \emph {et~al.}(2019)\citenamefont {Fuyuto},
  \citenamefont {Ramsey-Musolf},\ and\ \citenamefont {Shen}}]{Fuyuto:2018scm}%
  \BibitemOpen
  \bibfield  {author} {\bibinfo {author} {\bibfnamefont {K.}~\bibnamefont
  {Fuyuto}}, \bibinfo {author} {\bibfnamefont {M.}~\bibnamefont
  {Ramsey-Musolf}},\ and\ \bibinfo {author} {\bibfnamefont {T.}~\bibnamefont
  {Shen}},\ }\href {https://doi.org/10.1016/j.physletb.2018.11.016} {\bibfield
  {journal} {\bibinfo  {journal} {Phys. Lett. B}\ }\textbf {\bibinfo {volume}
  {788}},\ \bibinfo {pages} {52} (\bibinfo {year} {2019})},\ \Eprint
  {https://arxiv.org/abs/1804.01137} {arXiv:1804.01137 [hep-ph]} \BibitemShut
  {NoStop}%
\bibitem [{\citenamefont {Dekens}\ \emph {et~al.}(2019)\citenamefont {Dekens},
  \citenamefont {de~Vries}, \citenamefont {Jung},\ and\ \citenamefont
  {Vos}}]{Dekens:2018bci}%
  \BibitemOpen
  \bibfield  {author} {\bibinfo {author} {\bibfnamefont {W.}~\bibnamefont
  {Dekens}}, \bibinfo {author} {\bibfnamefont {J.}~\bibnamefont {de~Vries}},
  \bibinfo {author} {\bibfnamefont {M.}~\bibnamefont {Jung}},\ and\ \bibinfo
  {author} {\bibfnamefont {K.~K.}\ \bibnamefont {Vos}},\ }\href
  {https://doi.org/10.1007/JHEP01(2019)069} {\bibfield  {journal} {\bibinfo
  {journal} {JHEP}\ }\textbf {\bibinfo {volume} {2019}}\bibfield  {number}
  {\bibinfo  {number} { (01)},\ \bibinfo {pages} {069}},\ }\Eprint
  {https://arxiv.org/abs/1809.09114} {arXiv:1809.09114 [hep-ph]} \BibitemShut
  {NoStop}%
\bibitem [{\citenamefont {Mohapatra}\ and\ \citenamefont
  {Pati}(1975)}]{Mohapatra:1974hk}%
  \BibitemOpen
  \bibfield  {author} {\bibinfo {author} {\bibfnamefont {R.~N.}\ \bibnamefont
  {Mohapatra}}\ and\ \bibinfo {author} {\bibfnamefont {J.~C.}\ \bibnamefont
  {Pati}},\ }\href {https://doi.org/10.1103/PhysRevD.11.566} {\bibfield
  {journal} {\bibinfo  {journal} {Phys. Rev. D}\ }\textbf {\bibinfo {volume}
  {11}},\ \bibinfo {pages} {566} (\bibinfo {year} {1975})}\BibitemShut
  {NoStop}%
\bibitem [{\citenamefont {Senjanovic}\ and\ \citenamefont
  {Mohapatra}(1975)}]{Senjanovic:1975rk}%
  \BibitemOpen
  \bibfield  {author} {\bibinfo {author} {\bibfnamefont {G.}~\bibnamefont
  {Senjanovic}}\ and\ \bibinfo {author} {\bibfnamefont {R.~N.}\ \bibnamefont
  {Mohapatra}},\ }\href {https://doi.org/10.1103/PhysRevD.12.1502} {\bibfield
  {journal} {\bibinfo  {journal} {Phys. Rev. D}\ }\textbf {\bibinfo {volume}
  {12}},\ \bibinfo {pages} {1502} (\bibinfo {year} {1975})}\BibitemShut
  {NoStop}%
\bibitem [{\citenamefont {Maiezza}\ and\ \citenamefont
  {Nemev\v{s}ek}(2014)}]{Maiezza:2014ala}%
  \BibitemOpen
  \bibfield  {author} {\bibinfo {author} {\bibfnamefont {A.}~\bibnamefont
  {Maiezza}}\ and\ \bibinfo {author} {\bibfnamefont {M.}~\bibnamefont
  {Nemev\v{s}ek}},\ }\href {https://doi.org/10.1103/PhysRevD.90.095002}
  {\bibfield  {journal} {\bibinfo  {journal} {Phys. Rev. D}\ }\textbf {\bibinfo
  {volume} {90}},\ \bibinfo {pages} {095002} (\bibinfo {year} {2014})},\
  \Eprint {https://arxiv.org/abs/1407.3678} {arXiv:1407.3678 [hep-ph]}
  \BibitemShut {NoStop}%
\bibitem [{\citenamefont {Peccei}\ and\ \citenamefont
  {Quinn}(1977{\natexlab{a}})}]{Peccei:1977ur}%
  \BibitemOpen
  \bibfield  {author} {\bibinfo {author} {\bibfnamefont {R.~D.}\ \bibnamefont
  {Peccei}}\ and\ \bibinfo {author} {\bibfnamefont {H.~R.}\ \bibnamefont
  {Quinn}},\ }\href {https://doi.org/10.1103/PhysRevD.16.1791} {\bibfield
  {journal} {\bibinfo  {journal} {Phys. Rev.}\ }\textbf {\bibinfo {volume}
  {D16}},\ \bibinfo {pages} {1791} (\bibinfo {year}
  {1977}{\natexlab{a}})}\BibitemShut {NoStop}%
\bibitem [{\citenamefont {Peccei}\ and\ \citenamefont
  {Quinn}(1977{\natexlab{b}})}]{Peccei:1977hh}%
  \BibitemOpen
  \bibfield  {author} {\bibinfo {author} {\bibfnamefont {R.~D.}\ \bibnamefont
  {Peccei}}\ and\ \bibinfo {author} {\bibfnamefont {H.~R.}\ \bibnamefont
  {Quinn}},\ }\href {https://doi.org/10.1103/PhysRevLett.38.1440} {\bibfield
  {journal} {\bibinfo  {journal} {Phys. Rev. Lett.}\ }\textbf {\bibinfo
  {volume} {38}},\ \bibinfo {pages} {1440} (\bibinfo {year}
  {1977}{\natexlab{b}})}\BibitemShut {NoStop}%
\bibitem [{\citenamefont {Weinberg}(1978)}]{Weinberg:1977ma}%
  \BibitemOpen
  \bibfield  {author} {\bibinfo {author} {\bibfnamefont {S.}~\bibnamefont
  {Weinberg}},\ }\href {https://doi.org/10.1103/PhysRevLett.40.223} {\bibfield
  {journal} {\bibinfo  {journal} {Phys. Rev. Lett.}\ }\textbf {\bibinfo
  {volume} {40}},\ \bibinfo {pages} {223} (\bibinfo {year} {1978})}\BibitemShut
  {NoStop}%
\bibitem [{\citenamefont {Wilczek}(1978)}]{Wilczek:1977pj}%
  \BibitemOpen
  \bibfield  {author} {\bibinfo {author} {\bibfnamefont {F.}~\bibnamefont
  {Wilczek}},\ }\href {https://doi.org/10.1103/PhysRevLett.40.279} {\bibfield
  {journal} {\bibinfo  {journal} {Phys. Rev. Lett.}\ }\textbf {\bibinfo
  {volume} {40}},\ \bibinfo {pages} {279} (\bibinfo {year} {1978})}\BibitemShut
  {NoStop}%
\bibitem [{\citenamefont {Barr}\ and\ \citenamefont
  {Seckel}(1992)}]{Barr:1992qq}%
  \BibitemOpen
  \bibfield  {author} {\bibinfo {author} {\bibfnamefont {S.~M.}\ \bibnamefont
  {Barr}}\ and\ \bibinfo {author} {\bibfnamefont {D.}~\bibnamefont {Seckel}},\
  }\href {https://doi.org/10.1103/PhysRevD.46.539} {\bibfield  {journal}
  {\bibinfo  {journal} {Phys. Rev.}\ }\textbf {\bibinfo {volume} {D46}},\
  \bibinfo {pages} {539} (\bibinfo {year} {1992})}\BibitemShut {NoStop}%
\bibitem [{\citenamefont {Kamionkowski}\ and\ \citenamefont
  {March-Russell}(1992)}]{kamionkowski:1992mf}%
  \BibitemOpen
  \bibfield  {author} {\bibinfo {author} {\bibfnamefont {M.}~\bibnamefont
  {Kamionkowski}}\ and\ \bibinfo {author} {\bibfnamefont {J.}~\bibnamefont
  {March-Russell}},\ }\href {https://doi.org/10.1016/0370-2693(92)90492-M}
  {\bibfield  {journal} {\bibinfo  {journal} {Phys.Lett.}\ }\textbf {\bibinfo
  {volume} {B282}},\ \bibinfo {pages} {137} (\bibinfo {year} {1992})},\ \Eprint
  {https://arxiv.org/abs/hep-th/9202003} {arXiv:hep-th/9202003 [hep-th]}
  \BibitemShut {NoStop}%
\bibitem [{\citenamefont {Holman}\ \emph {et~al.}(1992)\citenamefont {Holman},
  \citenamefont {Hsu}, \citenamefont {Kephart}, \citenamefont {Kolb},
  \citenamefont {Watkins},\ and\ \citenamefont {Widrow}}]{holman:1992us}%
  \BibitemOpen
  \bibfield  {author} {\bibinfo {author} {\bibfnamefont {R.}~\bibnamefont
  {Holman}}, \bibinfo {author} {\bibfnamefont {S.~D.}\ \bibnamefont {Hsu}},
  \bibinfo {author} {\bibfnamefont {T.~W.}\ \bibnamefont {Kephart}}, \bibinfo
  {author} {\bibfnamefont {E.~W.}\ \bibnamefont {Kolb}}, \bibinfo {author}
  {\bibfnamefont {R.}~\bibnamefont {Watkins}},\ and\ \bibinfo {author}
  {\bibfnamefont {L.~M.}\ \bibnamefont {Widrow}},\ }\href
  {https://doi.org/10.1016/0370-2693(92)90491-L} {\bibfield  {journal}
  {\bibinfo  {journal} {Phys.Lett.}\ }\textbf {\bibinfo {volume} {B282}},\
  \bibinfo {pages} {132} (\bibinfo {year} {1992})},\ \Eprint
  {https://arxiv.org/abs/hep-ph/9203206} {arXiv:hep-ph/9203206 [hep-ph]}
  \BibitemShut {NoStop}%
\bibitem [{\citenamefont {Ghigna}\ \emph {et~al.}(1992)\citenamefont {Ghigna},
  \citenamefont {Lusignoli},\ and\ \citenamefont {Roncadelli}}]{Ghigna:1992iv}%
  \BibitemOpen
  \bibfield  {author} {\bibinfo {author} {\bibfnamefont {S.}~\bibnamefont
  {Ghigna}}, \bibinfo {author} {\bibfnamefont {M.}~\bibnamefont {Lusignoli}},\
  and\ \bibinfo {author} {\bibfnamefont {M.}~\bibnamefont {Roncadelli}},\
  }\href {https://doi.org/10.1016/0370-2693(92)90019-Z} {\bibfield  {journal}
  {\bibinfo  {journal} {Phys. Lett.}\ }\textbf {\bibinfo {volume} {B283}},\
  \bibinfo {pages} {278} (\bibinfo {year} {1992})}\BibitemShut {NoStop}%
\bibitem [{\citenamefont {Arvanitaki}\ and\ \citenamefont
  {Geraci}(2014)}]{Arvanitaki:2014dfa}%
  \BibitemOpen
  \bibfield  {author} {\bibinfo {author} {\bibfnamefont {A.}~\bibnamefont
  {Arvanitaki}}\ and\ \bibinfo {author} {\bibfnamefont {A.~A.}\ \bibnamefont
  {Geraci}},\ }\href {https://doi.org/10.1103/PhysRevLett.113.161801}
  {\bibfield  {journal} {\bibinfo  {journal} {Phys. Rev. Lett.}\ }\textbf
  {\bibinfo {volume} {113}},\ \bibinfo {pages} {161801} (\bibinfo {year}
  {2014})},\ \Eprint {https://arxiv.org/abs/1403.1290} {arXiv:1403.1290
  [hep-ph]} \BibitemShut {NoStop}%
\bibitem [{\citenamefont {Le~Dall}\ \emph {et~al.}(2015)\citenamefont
  {Le~Dall}, \citenamefont {Pospelov},\ and\ \citenamefont
  {Ritz}}]{LeDall:2015ptt}%
  \BibitemOpen
  \bibfield  {author} {\bibinfo {author} {\bibfnamefont {M.}~\bibnamefont
  {Le~Dall}}, \bibinfo {author} {\bibfnamefont {M.}~\bibnamefont {Pospelov}},\
  and\ \bibinfo {author} {\bibfnamefont {A.}~\bibnamefont {Ritz}},\ }\href
  {https://doi.org/10.1103/PhysRevD.92.016010} {\bibfield  {journal} {\bibinfo
  {journal} {Phys. Rev. D}\ }\textbf {\bibinfo {volume} {92}},\ \bibinfo
  {pages} {016010} (\bibinfo {year} {2015})},\ \Eprint
  {https://arxiv.org/abs/1505.01865} {arXiv:1505.01865 [hep-ph]} \BibitemShut
  {NoStop}%
\bibitem [{\citenamefont {Fuyuto}\ \emph {et~al.}(2020)\citenamefont {Fuyuto},
  \citenamefont {He}, \citenamefont {Li},\ and\ \citenamefont
  {Ramsey-Musolf}}]{Fuyuto:2019vfe}%
  \BibitemOpen
  \bibfield  {author} {\bibinfo {author} {\bibfnamefont {K.}~\bibnamefont
  {Fuyuto}}, \bibinfo {author} {\bibfnamefont {X.-G.}\ \bibnamefont {He}},
  \bibinfo {author} {\bibfnamefont {G.}~\bibnamefont {Li}},\ and\ \bibinfo
  {author} {\bibfnamefont {M.}~\bibnamefont {Ramsey-Musolf}},\ }\href
  {https://doi.org/10.1103/PhysRevD.101.075016} {\bibfield  {journal} {\bibinfo
   {journal} {Phys. Rev. D}\ }\textbf {\bibinfo {volume} {101}},\ \bibinfo
  {pages} {075016} (\bibinfo {year} {2020})},\ \Eprint
  {https://arxiv.org/abs/1902.10340} {arXiv:1902.10340 [hep-ph]} \BibitemShut
  {NoStop}%
\bibitem [{\citenamefont {Engel}\ \emph
  {et~al.}(2013{\natexlab{a}})\citenamefont {Engel}, \citenamefont
  {Ramsey-Musolf},\ and\ \citenamefont {van Kolck}}]{Engel:2013lsa}%
  \BibitemOpen
  \bibfield  {author} {\bibinfo {author} {\bibfnamefont {J.}~\bibnamefont
  {Engel}}, \bibinfo {author} {\bibfnamefont {M.~J.}\ \bibnamefont
  {Ramsey-Musolf}},\ and\ \bibinfo {author} {\bibfnamefont {U.}~\bibnamefont
  {van Kolck}},\ }\href {https://doi.org/10.1016/j.ppnp.2013.03.003} {\bibfield
   {journal} {\bibinfo  {journal} {Prog. Part. Nucl. Phys.}\ }\textbf {\bibinfo
  {volume} {71}},\ \bibinfo {pages} {21} (\bibinfo {year}
  {2013}{\natexlab{a}})},\ \Eprint {https://arxiv.org/abs/1303.2371}
  {arXiv:1303.2371 [nucl-th]} \BibitemShut {NoStop}%
\bibitem [{\citenamefont {Weinberg}(1989)}]{Weinberg:1989dx}%
  \BibitemOpen
  \bibfield  {author} {\bibinfo {author} {\bibfnamefont {S.}~\bibnamefont
  {Weinberg}},\ }\href {https://doi.org/10.1103/PhysRevLett.63.2333} {\bibfield
   {journal} {\bibinfo  {journal} {Phys. Rev. Lett.}\ }\textbf {\bibinfo
  {volume} {63}},\ \bibinfo {pages} {2333} (\bibinfo {year}
  {1989})}\BibitemShut {NoStop}%
\bibitem [{\citenamefont {Mereghetti}\ \emph {et~al.}(2010)\citenamefont
  {Mereghetti}, \citenamefont {Hockings},\ and\ \citenamefont {van
  Kolck}}]{Mereghetti:2010tp}%
  \BibitemOpen
  \bibfield  {author} {\bibinfo {author} {\bibfnamefont {E.}~\bibnamefont
  {Mereghetti}}, \bibinfo {author} {\bibfnamefont {W.~H.}\ \bibnamefont
  {Hockings}},\ and\ \bibinfo {author} {\bibfnamefont {U.}~\bibnamefont {van
  Kolck}},\ }\href {https://doi.org/10.1016/j.aop.2010.03.005} {\bibfield
  {journal} {\bibinfo  {journal} {Annals Phys.}\ }\textbf {\bibinfo {volume}
  {325}},\ \bibinfo {pages} {2363} (\bibinfo {year} {2010})},\ \Eprint
  {https://arxiv.org/abs/1002.2391} {arXiv:1002.2391 [hep-ph]} \BibitemShut
  {NoStop}%
\bibitem [{\citenamefont {de~Vries}\ \emph {et~al.}(2013)\citenamefont
  {de~Vries}, \citenamefont {Mereghetti}, \citenamefont {Timmermans},\ and\
  \citenamefont {van Kolck}}]{deVries:2012ab}%
  \BibitemOpen
  \bibfield  {author} {\bibinfo {author} {\bibfnamefont {J.}~\bibnamefont
  {de~Vries}}, \bibinfo {author} {\bibfnamefont {E.}~\bibnamefont
  {Mereghetti}}, \bibinfo {author} {\bibfnamefont {R.~G.~E.}\ \bibnamefont
  {Timmermans}},\ and\ \bibinfo {author} {\bibfnamefont {U.}~\bibnamefont {van
  Kolck}},\ }\href {https://doi.org/10.1016/j.aop.2013.05.022} {\bibfield
  {journal} {\bibinfo  {journal} {Annals Phys.}\ }\textbf {\bibinfo {volume}
  {338}},\ \bibinfo {pages} {50} (\bibinfo {year} {2013})},\ \Eprint
  {https://arxiv.org/abs/1212.0990} {arXiv:1212.0990 [hep-ph]} \BibitemShut
  {NoStop}%
\bibitem [{\citenamefont {de~Vries}\ \emph {et~al.}(2020)\citenamefont
  {de~Vries}, \citenamefont {Epelbaum}, \citenamefont {Girlanda}, \citenamefont
  {Gnech}, \citenamefont {Mereghetti},\ and\ \citenamefont
  {Viviani}}]{deVries:2020iea}%
  \BibitemOpen
  \bibfield  {author} {\bibinfo {author} {\bibfnamefont {J.}~\bibnamefont
  {de~Vries}}, \bibinfo {author} {\bibfnamefont {E.}~\bibnamefont {Epelbaum}},
  \bibinfo {author} {\bibfnamefont {L.}~\bibnamefont {Girlanda}}, \bibinfo
  {author} {\bibfnamefont {A.}~\bibnamefont {Gnech}}, \bibinfo {author}
  {\bibfnamefont {E.}~\bibnamefont {Mereghetti}},\ and\ \bibinfo {author}
  {\bibfnamefont {M.}~\bibnamefont {Viviani}},\ }\href
  {https://doi.org/10.3389/fphy.2020.00218} {\bibfield  {journal} {\bibinfo
  {journal} {Front. in Phys.}\ }\textbf {\bibinfo {volume} {8}},\ \bibinfo
  {pages} {218} (\bibinfo {year} {2020})},\ \Eprint
  {https://arxiv.org/abs/2001.09050} {arXiv:2001.09050 [nucl-th]} \BibitemShut
  {NoStop}%
\bibitem [{\citenamefont {Shindler}(2021)}]{Shindler:2021bcx}%
  \BibitemOpen
  \bibfield  {author} {\bibinfo {author} {\bibfnamefont {A.}~\bibnamefont
  {Shindler}},\ }\href {https://doi.org/10.1140/epja/s10050-021-00421-y}
  {\bibfield  {journal} {\bibinfo  {journal} {Eur. Phys. J. A}\ }\textbf
  {\bibinfo {volume} {57}},\ \bibinfo {pages} {128} (\bibinfo {year}
  {2021})}\BibitemShut {NoStop}%
\bibitem [{\citenamefont {Dragos}\ \emph {et~al.}(2021)\citenamefont {Dragos},
  \citenamefont {Luu}, \citenamefont {Shindler}, \citenamefont {de~Vries},\
  and\ \citenamefont {Yousif}}]{Dragos:2019oxn}%
  \BibitemOpen
  \bibfield  {author} {\bibinfo {author} {\bibfnamefont {J.}~\bibnamefont
  {Dragos}}, \bibinfo {author} {\bibfnamefont {T.}~\bibnamefont {Luu}},
  \bibinfo {author} {\bibfnamefont {A.}~\bibnamefont {Shindler}}, \bibinfo
  {author} {\bibfnamefont {J.}~\bibnamefont {de~Vries}},\ and\ \bibinfo
  {author} {\bibfnamefont {A.}~\bibnamefont {Yousif}},\ }\href
  {https://doi.org/10.1103/PhysRevC.103.015202} {\bibfield  {journal} {\bibinfo
   {journal} {Phys. Rev. C}\ }\textbf {\bibinfo {volume} {103}},\ \bibinfo
  {pages} {015202} (\bibinfo {year} {2021})},\ \Eprint
  {https://arxiv.org/abs/1902.03254} {arXiv:1902.03254 [hep-lat]} \BibitemShut
  {NoStop}%
\bibitem [{\citenamefont {Bhattacharya}\ \emph {et~al.}(2021)\citenamefont
  {Bhattacharya}, \citenamefont {Cirigliano}, \citenamefont {Gupta},
  \citenamefont {Mereghetti},\ and\ \citenamefont
  {Yoon}}]{Bhattacharya:2021lol}%
  \BibitemOpen
  \bibfield  {author} {\bibinfo {author} {\bibfnamefont {T.}~\bibnamefont
  {Bhattacharya}}, \bibinfo {author} {\bibfnamefont {V.}~\bibnamefont
  {Cirigliano}}, \bibinfo {author} {\bibfnamefont {R.}~\bibnamefont {Gupta}},
  \bibinfo {author} {\bibfnamefont {E.}~\bibnamefont {Mereghetti}},\ and\
  \bibinfo {author} {\bibfnamefont {B.}~\bibnamefont {Yoon}},\ }\href
  {https://doi.org/10.1103/PhysRevD.103.114507} {\bibfield  {journal} {\bibinfo
   {journal} {Phys. Rev. D}\ }\textbf {\bibinfo {volume} {103}},\ \bibinfo
  {pages} {114507} (\bibinfo {year} {2021})},\ \Eprint
  {https://arxiv.org/abs/2101.07230} {arXiv:2101.07230 [hep-lat]} \BibitemShut
  {NoStop}%
\bibitem [{\citenamefont {Gupta}\ \emph {et~al.}(2018)\citenamefont {Gupta},
  \citenamefont {Yoon}, \citenamefont {Bhattacharya}, \citenamefont
  {Cirigliano}, \citenamefont {Jang},\ and\ \citenamefont
  {Lin}}]{Gupta:2018lvp}%
  \BibitemOpen
  \bibfield  {author} {\bibinfo {author} {\bibfnamefont {R.}~\bibnamefont
  {Gupta}}, \bibinfo {author} {\bibfnamefont {B.}~\bibnamefont {Yoon}},
  \bibinfo {author} {\bibfnamefont {T.}~\bibnamefont {Bhattacharya}}, \bibinfo
  {author} {\bibfnamefont {V.}~\bibnamefont {Cirigliano}}, \bibinfo {author}
  {\bibfnamefont {Y.-C.}\ \bibnamefont {Jang}},\ and\ \bibinfo {author}
  {\bibfnamefont {H.-W.}\ \bibnamefont {Lin}},\ }\href
  {https://doi.org/10.1103/PhysRevD.98.091501} {\bibfield  {journal} {\bibinfo
  {journal} {Phys. Rev. D}\ }\textbf {\bibinfo {volume} {98}},\ \bibinfo
  {pages} {091501} (\bibinfo {year} {2018})},\ \Eprint
  {https://arxiv.org/abs/1808.07597} {arXiv:1808.07597 [hep-lat]} \BibitemShut
  {NoStop}%
\bibitem [{\citenamefont {Bhattacharya}\ \emph
  {et~al.}(2015{\natexlab{a}})\citenamefont {Bhattacharya}, \citenamefont
  {Cirigliano}, \citenamefont {Gupta}, \citenamefont {Lin},\ and\ \citenamefont
  {Yoon}}]{Bhattacharya:2015esa}%
  \BibitemOpen
  \bibfield  {author} {\bibinfo {author} {\bibfnamefont {T.}~\bibnamefont
  {Bhattacharya}}, \bibinfo {author} {\bibfnamefont {V.}~\bibnamefont
  {Cirigliano}}, \bibinfo {author} {\bibfnamefont {R.}~\bibnamefont {Gupta}},
  \bibinfo {author} {\bibfnamefont {H.-W.}\ \bibnamefont {Lin}},\ and\ \bibinfo
  {author} {\bibfnamefont {B.}~\bibnamefont {Yoon}},\ }\href
  {https://doi.org/10.1103/PhysRevLett.115.212002} {\bibfield  {journal}
  {\bibinfo  {journal} {Phys. Rev. Lett.}\ }\textbf {\bibinfo {volume} {115}},\
  \bibinfo {pages} {212002} (\bibinfo {year} {2015}{\natexlab{a}})},\ \Eprint
  {https://arxiv.org/abs/1506.04196} {arXiv:1506.04196 [hep-lat]} \BibitemShut
  {NoStop}%
\bibitem [{\citenamefont {Pospelov}\ and\ \citenamefont
  {Ritz}(2001)}]{Pospelov_qCEDM}%
  \BibitemOpen
  \bibfield  {author} {\bibinfo {author} {\bibfnamefont {M.}~\bibnamefont
  {Pospelov}}\ and\ \bibinfo {author} {\bibfnamefont {A.}~\bibnamefont
  {Ritz}},\ }\href {https://doi.org/10.1103/PhysRevD.63.073015} {\bibfield
  {journal} {\bibinfo  {journal} {Phys. Rev. D}\ }\textbf {\bibinfo {volume}
  {63}},\ \bibinfo {pages} {073015} (\bibinfo {year} {2001})},\ \Eprint
  {https://arxiv.org/abs/hep-ph/0010037} {arXiv:hep-ph/0010037 [hep-ph]}
  \BibitemShut {NoStop}%
\bibitem [{\citenamefont {Lebedev}\ \emph
  {et~al.}(2004{\natexlab{a}})\citenamefont {Lebedev}, \citenamefont {Olive},
  \citenamefont {Pospelov},\ and\ \citenamefont {Ritz}}]{Pospelov_deuteron}%
  \BibitemOpen
  \bibfield  {author} {\bibinfo {author} {\bibfnamefont {O.}~\bibnamefont
  {Lebedev}}, \bibinfo {author} {\bibfnamefont {K.~A.}\ \bibnamefont {Olive}},
  \bibinfo {author} {\bibfnamefont {M.}~\bibnamefont {Pospelov}},\ and\
  \bibinfo {author} {\bibfnamefont {A.}~\bibnamefont {Ritz}},\ }\href
  {https://doi.org/10.1103/PhysRevD.70.016003} {\bibfield  {journal} {\bibinfo
  {journal} {Phys. Rev. D}\ }\textbf {\bibinfo {volume} {70}},\ \bibinfo
  {pages} {016003} (\bibinfo {year} {2004}{\natexlab{a}})},\ \Eprint
  {https://arxiv.org/abs/hep-ph/0402023} {arXiv:hep-ph/0402023 [hep-ph]}
  \BibitemShut {NoStop}%
\bibitem [{\citenamefont {Hisano}\ \emph {et~al.}(2012)\citenamefont {Hisano},
  \citenamefont {Lee}, \citenamefont {Nagata},\ and\ \citenamefont
  {Shimizu}}]{Hisano1}%
  \BibitemOpen
  \bibfield  {author} {\bibinfo {author} {\bibfnamefont {J.}~\bibnamefont
  {Hisano}}, \bibinfo {author} {\bibfnamefont {J.~Y.}\ \bibnamefont {Lee}},
  \bibinfo {author} {\bibfnamefont {N.}~\bibnamefont {Nagata}},\ and\ \bibinfo
  {author} {\bibfnamefont {Y.}~\bibnamefont {Shimizu}},\ }\href
  {https://doi.org/10.1103/PhysRevD.85.114044} {\bibfield  {journal} {\bibinfo
  {journal} {Phys. Rev. D}\ }\textbf {\bibinfo {volume} {85}},\ \bibinfo
  {pages} {114044} (\bibinfo {year} {2012})},\ \Eprint
  {https://arxiv.org/abs/1204.2653} {arXiv:1204.2653 [hep-ph]} \BibitemShut
  {NoStop}%
\bibitem [{\citenamefont {Demir}\ \emph {et~al.}(2003)\citenamefont {Demir},
  \citenamefont {Pospelov},\ and\ \citenamefont {Ritz}}]{Pospelov_Weinberg}%
  \BibitemOpen
  \bibfield  {author} {\bibinfo {author} {\bibfnamefont {D.~A.}\ \bibnamefont
  {Demir}}, \bibinfo {author} {\bibfnamefont {M.}~\bibnamefont {Pospelov}},\
  and\ \bibinfo {author} {\bibfnamefont {A.}~\bibnamefont {Ritz}},\ }\href
  {https://doi.org/10.1103/PhysRevD.67.015007} {\bibfield  {journal} {\bibinfo
  {journal} {Phys. Rev. D}\ }\textbf {\bibinfo {volume} {67}},\ \bibinfo
  {pages} {015007} (\bibinfo {year} {2003})},\ \Eprint
  {https://arxiv.org/abs/hep-ph/0208257} {arXiv:hep-ph/0208257 [hep-ph]}
  \BibitemShut {NoStop}%
\bibitem [{\citenamefont {Haisch}\ and\ \citenamefont
  {Hala}(2019)}]{Haisch:2019bml}%
  \BibitemOpen
  \bibfield  {author} {\bibinfo {author} {\bibfnamefont {U.}~\bibnamefont
  {Haisch}}\ and\ \bibinfo {author} {\bibfnamefont {A.}~\bibnamefont {Hala}},\
  }\href {https://doi.org/10.1007/JHEP11(2019)154} {\bibfield  {journal}
  {\bibinfo  {journal} {JHEP}\ }\textbf {\bibinfo {volume} {2019}}\bibfield
  {number} {\bibinfo  {number} { (11)},\ \bibinfo {pages} {154}},\ }\Eprint
  {https://arxiv.org/abs/1909.08955} {arXiv:1909.08955 [hep-ph]} \BibitemShut
  {NoStop}%
\bibitem [{\citenamefont {Chien}\ \emph {et~al.}(2016)\citenamefont {Chien},
  \citenamefont {Cirigliano}, \citenamefont {Dekens}, \citenamefont
  {de~Vries},\ and\ \citenamefont {Mereghetti}}]{Chien:2015xha}%
  \BibitemOpen
  \bibfield  {author} {\bibinfo {author} {\bibfnamefont {Y.~T.}\ \bibnamefont
  {Chien}}, \bibinfo {author} {\bibfnamefont {V.}~\bibnamefont {Cirigliano}},
  \bibinfo {author} {\bibfnamefont {W.}~\bibnamefont {Dekens}}, \bibinfo
  {author} {\bibfnamefont {J.}~\bibnamefont {de~Vries}},\ and\ \bibinfo
  {author} {\bibfnamefont {E.}~\bibnamefont {Mereghetti}},\ }\href
  {https://doi.org/10.1007/JHEP02(2016)011} {\bibfield  {journal} {\bibinfo
  {journal} {JHEP}\ }\textbf {\bibinfo {volume} {2016}}\bibfield  {number}
  {\bibinfo  {number} { (02)},\ \bibinfo {pages} {011}},\ }\Eprint
  {https://arxiv.org/abs/1510.00725} {arXiv:1510.00725 [hep-ph]} \BibitemShut
  {NoStop}%
\bibitem [{\citenamefont {Hasenfratz}\ \emph {et~al.}(2022)\citenamefont
  {Hasenfratz}, \citenamefont {Monahan}, \citenamefont {Rizik}, \citenamefont
  {Shindler},\ and\ \citenamefont {Witzel}}]{Hasenfratz:2022wll}%
  \BibitemOpen
  \bibfield  {author} {\bibinfo {author} {\bibfnamefont {A.}~\bibnamefont
  {Hasenfratz}}, \bibinfo {author} {\bibfnamefont {C.~J.}\ \bibnamefont
  {Monahan}}, \bibinfo {author} {\bibfnamefont {M.~D.}\ \bibnamefont {Rizik}},
  \bibinfo {author} {\bibfnamefont {A.}~\bibnamefont {Shindler}},\ and\
  \bibinfo {author} {\bibfnamefont {O.}~\bibnamefont {Witzel}},\ }in\
  \href@noop {} {\emph {\bibinfo {booktitle} {{38th International Symposium on
  Lattice Field Theory}}}}\ (\bibinfo {year} {2022})\ \Eprint
  {https://arxiv.org/abs/2201.09740} {arXiv:2201.09740 [hep-lat]} \BibitemShut
  {NoStop}%
\bibitem [{\citenamefont {Mereghetti}\ \emph {et~al.}(2021)\citenamefont
  {Mereghetti}, \citenamefont {Monahan}, \citenamefont {Rizik}, \citenamefont
  {Shindler},\ and\ \citenamefont {Stoffer}}]{Mereghetti:2021nkt}%
  \BibitemOpen
  \bibfield  {author} {\bibinfo {author} {\bibfnamefont {E.}~\bibnamefont
  {Mereghetti}}, \bibinfo {author} {\bibfnamefont {C.~J.}\ \bibnamefont
  {Monahan}}, \bibinfo {author} {\bibfnamefont {M.~D.}\ \bibnamefont {Rizik}},
  \bibinfo {author} {\bibfnamefont {A.}~\bibnamefont {Shindler}},\ and\
  \bibinfo {author} {\bibfnamefont {P.}~\bibnamefont {Stoffer}},\ }\href@noop
  {} {\bibinfo {title} {{One-loop matching for quark dipole operators in a
  gradient-flow scheme}}} (\bibinfo {year} {2021}),\ \Eprint
  {https://arxiv.org/abs/2111.11449} {arXiv:2111.11449 [hep-lat]} \BibitemShut
  {NoStop}%
\bibitem [{\citenamefont {Kim}\ \emph {et~al.}(2021)\citenamefont {Kim},
  \citenamefont {Luu}, \citenamefont {Rizik},\ and\ \citenamefont
  {Shindler}}]{Kim:2021qae}%
  \BibitemOpen
  \bibfield  {author} {\bibinfo {author} {\bibfnamefont {J.}~\bibnamefont
  {Kim}}, \bibinfo {author} {\bibfnamefont {T.}~\bibnamefont {Luu}}, \bibinfo
  {author} {\bibfnamefont {M.~D.}\ \bibnamefont {Rizik}},\ and\ \bibinfo
  {author} {\bibfnamefont {A.}~\bibnamefont {Shindler}} (\bibinfo
  {collaboration} {SymLat}),\ }\href
  {https://doi.org/10.1103/PhysRevD.104.074516} {\bibfield  {journal} {\bibinfo
   {journal} {Phys. Rev. D}\ }\textbf {\bibinfo {volume} {104}},\ \bibinfo
  {pages} {074516} (\bibinfo {year} {2021})},\ \Eprint
  {https://arxiv.org/abs/2106.07633} {arXiv:2106.07633 [hep-lat]} \BibitemShut
  {NoStop}%
\bibitem [{\citenamefont {Cirigliano}\ \emph {et~al.}(2020)\citenamefont
  {Cirigliano}, \citenamefont {Mereghetti},\ and\ \citenamefont
  {Stoffer}}]{Cirigliano:2020msr}%
  \BibitemOpen
  \bibfield  {author} {\bibinfo {author} {\bibfnamefont {V.}~\bibnamefont
  {Cirigliano}}, \bibinfo {author} {\bibfnamefont {E.}~\bibnamefont
  {Mereghetti}},\ and\ \bibinfo {author} {\bibfnamefont {P.}~\bibnamefont
  {Stoffer}},\ }\href {https://doi.org/10.1007/JHEP09(2020)094} {\bibfield
  {journal} {\bibinfo  {journal} {JHEP}\ }\textbf {\bibinfo {volume}
  {2020}}\bibfield  {number} {\bibinfo  {number} { (09)},\ \bibinfo {pages}
  {094}},\ }\Eprint {https://arxiv.org/abs/2004.03576} {arXiv:2004.03576
  [hep-ph]} \BibitemShut {NoStop}%
\bibitem [{\citenamefont {Bhattacharya}\ \emph
  {et~al.}(2015{\natexlab{b}})\citenamefont {Bhattacharya}, \citenamefont
  {Cirigliano}, \citenamefont {Gupta}, \citenamefont {Mereghetti},\ and\
  \citenamefont {Yoon}}]{Bhattacharya:2015rsa}%
  \BibitemOpen
  \bibfield  {author} {\bibinfo {author} {\bibfnamefont {T.}~\bibnamefont
  {Bhattacharya}}, \bibinfo {author} {\bibfnamefont {V.}~\bibnamefont
  {Cirigliano}}, \bibinfo {author} {\bibfnamefont {R.}~\bibnamefont {Gupta}},
  \bibinfo {author} {\bibfnamefont {E.}~\bibnamefont {Mereghetti}},\ and\
  \bibinfo {author} {\bibfnamefont {B.}~\bibnamefont {Yoon}},\ }\href
  {https://doi.org/10.1103/PhysRevD.92.114026} {\bibfield  {journal} {\bibinfo
  {journal} {Phys. Rev. D}\ }\textbf {\bibinfo {volume} {92}},\ \bibinfo
  {pages} {114026} (\bibinfo {year} {2015}{\natexlab{b}})},\ \Eprint
  {https://arxiv.org/abs/1502.07325} {arXiv:1502.07325 [hep-ph]} \BibitemShut
  {NoStop}%
\bibitem [{\citenamefont {Kniehl}\ and\ \citenamefont
  {Veretin}(2020)}]{Kniehl:2020sgo}%
  \BibitemOpen
  \bibfield  {author} {\bibinfo {author} {\bibfnamefont {B.~A.}\ \bibnamefont
  {Kniehl}}\ and\ \bibinfo {author} {\bibfnamefont {O.~L.}\ \bibnamefont
  {Veretin}},\ }\href {https://doi.org/10.1016/j.physletb.2020.135398}
  {\bibfield  {journal} {\bibinfo  {journal} {Phys. Lett. B}\ }\textbf
  {\bibinfo {volume} {804}},\ \bibinfo {pages} {135398} (\bibinfo {year}
  {2020})},\ \Eprint {https://arxiv.org/abs/2002.10894} {arXiv:2002.10894
  [hep-ph]} \BibitemShut {NoStop}%
\bibitem [{\citenamefont {Rizik}\ \emph {et~al.}(2020)\citenamefont {Rizik},
  \citenamefont {Monahan},\ and\ \citenamefont {Shindler}}]{Rizik:2020naq}%
  \BibitemOpen
  \bibfield  {author} {\bibinfo {author} {\bibfnamefont {M.~D.}\ \bibnamefont
  {Rizik}}, \bibinfo {author} {\bibfnamefont {C.~J.}\ \bibnamefont {Monahan}},\
  and\ \bibinfo {author} {\bibfnamefont {A.}~\bibnamefont {Shindler}} (\bibinfo
  {collaboration} {SymLat}),\ }\href
  {https://doi.org/10.1103/PhysRevD.102.034509} {\bibfield  {journal} {\bibinfo
   {journal} {Phys. Rev. D}\ }\textbf {\bibinfo {volume} {102}},\ \bibinfo
  {pages} {034509} (\bibinfo {year} {2020})},\ \Eprint
  {https://arxiv.org/abs/2005.04199} {arXiv:2005.04199 [hep-lat]} \BibitemShut
  {NoStop}%
\bibitem [{\citenamefont {Aoki}\ \emph {et~al.}(2021)\citenamefont {Aoki} \emph
  {et~al.}}]{Aoki:2021kgd}%
  \BibitemOpen
  \bibfield  {author} {\bibinfo {author} {\bibfnamefont {Y.}~\bibnamefont
  {Aoki}} \emph {et~al.},\ }\href@noop {} {\bibinfo {title} {{FLAG Review
  2021}}} (\bibinfo {year} {2021}),\ \Eprint {https://arxiv.org/abs/2111.09849}
  {arXiv:2111.09849 [hep-lat]} \BibitemShut {NoStop}%
\bibitem [{\citenamefont {Shintani}\ \emph {et~al.}(2005)\citenamefont
  {Shintani}, \citenamefont {Aoki}, \citenamefont {Ishizuka}, \citenamefont
  {Kanaya}, \citenamefont {Kikukawa}, \citenamefont {Kuramashi}, \citenamefont
  {Okawa}, \citenamefont {Taniguchi}, \citenamefont {Ukawa},\ and\
  \citenamefont {Yoshi\'e}}]{Shintani:2005xg}%
  \BibitemOpen
  \bibfield  {author} {\bibinfo {author} {\bibfnamefont {E.}~\bibnamefont
  {Shintani}}, \bibinfo {author} {\bibfnamefont {S.}~\bibnamefont {Aoki}},
  \bibinfo {author} {\bibfnamefont {N.}~\bibnamefont {Ishizuka}}, \bibinfo
  {author} {\bibfnamefont {K.}~\bibnamefont {Kanaya}}, \bibinfo {author}
  {\bibfnamefont {Y.}~\bibnamefont {Kikukawa}}, \bibinfo {author}
  {\bibfnamefont {Y.}~\bibnamefont {Kuramashi}}, \bibinfo {author}
  {\bibfnamefont {M.}~\bibnamefont {Okawa}}, \bibinfo {author} {\bibfnamefont
  {Y.}~\bibnamefont {Taniguchi}}, \bibinfo {author} {\bibfnamefont
  {A.}~\bibnamefont {Ukawa}},\ and\ \bibinfo {author} {\bibfnamefont
  {T.}~\bibnamefont {Yoshi\'e}},\ }\href
  {https://doi.org/10.1103/PhysRevD.72.014504} {\bibfield  {journal} {\bibinfo
  {journal} {Phys. Rev. D}\ }\textbf {\bibinfo {volume} {72}},\ \bibinfo
  {pages} {014504} (\bibinfo {year} {2005})},\ \Eprint
  {https://arxiv.org/abs/hep-lat/0505022} {arXiv:hep-lat/0505022} \BibitemShut
  {NoStop}%
\bibitem [{\citenamefont {Berruto}\ \emph {et~al.}(2006)\citenamefont
  {Berruto}, \citenamefont {Blum}, \citenamefont {Orginos},\ and\ \citenamefont
  {Soni}}]{Berruto:2005hg}%
  \BibitemOpen
  \bibfield  {author} {\bibinfo {author} {\bibfnamefont {F.}~\bibnamefont
  {Berruto}}, \bibinfo {author} {\bibfnamefont {T.}~\bibnamefont {Blum}},
  \bibinfo {author} {\bibfnamefont {K.}~\bibnamefont {Orginos}},\ and\ \bibinfo
  {author} {\bibfnamefont {A.}~\bibnamefont {Soni}},\ }\href
  {https://doi.org/10.1103/PhysRevD.73.054509} {\bibfield  {journal} {\bibinfo
  {journal} {Phys. Rev. D}\ }\textbf {\bibinfo {volume} {73}},\ \bibinfo
  {pages} {054509} (\bibinfo {year} {2006})},\ \Eprint
  {https://arxiv.org/abs/hep-lat/0512004} {arXiv:hep-lat/0512004} \BibitemShut
  {NoStop}%
\bibitem [{\citenamefont {Shindler}\ \emph {et~al.}(2014)\citenamefont
  {Shindler}, \citenamefont {de~Vries},\ and\ \citenamefont
  {Luu}}]{Shindler:2014oha}%
  \BibitemOpen
  \bibfield  {author} {\bibinfo {author} {\bibfnamefont {A.}~\bibnamefont
  {Shindler}}, \bibinfo {author} {\bibfnamefont {J.}~\bibnamefont {de~Vries}},\
  and\ \bibinfo {author} {\bibfnamefont {T.}~\bibnamefont {Luu}},\ }\href
  {https://doi.org/10.22323/1.214.0251} {\bibfield  {journal} {\bibinfo
  {journal} {PoS}\ }\textbf {\bibinfo {volume} {LATTICE2014}},\ \bibinfo
  {pages} {251} (\bibinfo {year} {2014})},\ \Eprint
  {https://arxiv.org/abs/1409.2735} {arXiv:1409.2735 [hep-lat]} \BibitemShut
  {NoStop}%
\bibitem [{\citenamefont {Guo}\ \emph {et~al.}(2015)\citenamefont {Guo},
  \citenamefont {Horsley}, \citenamefont {Mei\ss{}ner}, \citenamefont
  {Nakamura}, \citenamefont {Perlt}, \citenamefont {Rakow}, \citenamefont
  {Schierholz}, \citenamefont {Schiller},\ and\ \citenamefont
  {Zanotti}}]{Guo:2015tla}%
  \BibitemOpen
  \bibfield  {author} {\bibinfo {author} {\bibfnamefont {F.-K.}\ \bibnamefont
  {Guo}}, \bibinfo {author} {\bibfnamefont {R.}~\bibnamefont {Horsley}},
  \bibinfo {author} {\bibfnamefont {U.-G.}\ \bibnamefont {Mei\ss{}ner}},
  \bibinfo {author} {\bibfnamefont {Y.}~\bibnamefont {Nakamura}}, \bibinfo
  {author} {\bibfnamefont {H.}~\bibnamefont {Perlt}}, \bibinfo {author}
  {\bibfnamefont {P.~E.~L.}\ \bibnamefont {Rakow}}, \bibinfo {author}
  {\bibfnamefont {G.}~\bibnamefont {Schierholz}}, \bibinfo {author}
  {\bibfnamefont {A.}~\bibnamefont {Schiller}},\ and\ \bibinfo {author}
  {\bibfnamefont {J.~M.}\ \bibnamefont {Zanotti}},\ }\href
  {https://doi.org/10.1103/PhysRevLett.115.062001} {\bibfield  {journal}
  {\bibinfo  {journal} {Phys. Rev. Lett.}\ }\textbf {\bibinfo {volume} {115}},\
  \bibinfo {pages} {062001} (\bibinfo {year} {2015})},\ \Eprint
  {https://arxiv.org/abs/1502.02295} {arXiv:1502.02295 [hep-lat]} \BibitemShut
  {NoStop}%
\bibitem [{\citenamefont {Shindler}\ \emph {et~al.}(2015)\citenamefont
  {Shindler}, \citenamefont {Luu},\ and\ \citenamefont
  {de~Vries}}]{Shindler:2015aqa}%
  \BibitemOpen
  \bibfield  {author} {\bibinfo {author} {\bibfnamefont {A.}~\bibnamefont
  {Shindler}}, \bibinfo {author} {\bibfnamefont {T.}~\bibnamefont {Luu}},\ and\
  \bibinfo {author} {\bibfnamefont {J.}~\bibnamefont {de~Vries}},\ }\href
  {https://doi.org/10.1103/PhysRevD.92.094518} {\bibfield  {journal} {\bibinfo
  {journal} {Phys. Rev. D}\ }\textbf {\bibinfo {volume} {92}},\ \bibinfo
  {pages} {094518} (\bibinfo {year} {2015})},\ \Eprint
  {https://arxiv.org/abs/1507.02343} {arXiv:1507.02343 [hep-lat]} \BibitemShut
  {NoStop}%
\bibitem [{\citenamefont {Alexandrou}\ \emph {et~al.}(2016)\citenamefont
  {Alexandrou}, \citenamefont {Athenodorou}, \citenamefont {Constantinou},
  \citenamefont {Hadjiyiannakou}, \citenamefont {Jansen}, \citenamefont
  {Koutsou}, \citenamefont {Ottnad},\ and\ \citenamefont
  {Petschlies}}]{Alexandrou:2015spa}%
  \BibitemOpen
  \bibfield  {author} {\bibinfo {author} {\bibfnamefont {C.}~\bibnamefont
  {Alexandrou}}, \bibinfo {author} {\bibfnamefont {A.}~\bibnamefont
  {Athenodorou}}, \bibinfo {author} {\bibfnamefont {M.}~\bibnamefont
  {Constantinou}}, \bibinfo {author} {\bibfnamefont {K.}~\bibnamefont
  {Hadjiyiannakou}}, \bibinfo {author} {\bibfnamefont {K.}~\bibnamefont
  {Jansen}}, \bibinfo {author} {\bibfnamefont {G.}~\bibnamefont {Koutsou}},
  \bibinfo {author} {\bibfnamefont {K.}~\bibnamefont {Ottnad}},\ and\ \bibinfo
  {author} {\bibfnamefont {M.}~\bibnamefont {Petschlies}},\ }\href
  {https://doi.org/10.1103/PhysRevD.93.074503} {\bibfield  {journal} {\bibinfo
  {journal} {Phys. Rev. D}\ }\textbf {\bibinfo {volume} {93}},\ \bibinfo
  {pages} {074503} (\bibinfo {year} {2016})},\ \Eprint
  {https://arxiv.org/abs/1510.05823} {arXiv:1510.05823 [hep-lat]} \BibitemShut
  {NoStop}%
\bibitem [{\citenamefont {Shintani}\ \emph {et~al.}(2016)\citenamefont
  {Shintani}, \citenamefont {Blum}, \citenamefont {Izubuchi},\ and\
  \citenamefont {Soni}}]{Shintani:2015vsx}%
  \BibitemOpen
  \bibfield  {author} {\bibinfo {author} {\bibfnamefont {E.}~\bibnamefont
  {Shintani}}, \bibinfo {author} {\bibfnamefont {T.}~\bibnamefont {Blum}},
  \bibinfo {author} {\bibfnamefont {T.}~\bibnamefont {Izubuchi}},\ and\
  \bibinfo {author} {\bibfnamefont {A.}~\bibnamefont {Soni}},\ }\href
  {https://doi.org/10.1103/PhysRevD.93.094503} {\bibfield  {journal} {\bibinfo
  {journal} {Phys. Rev. D}\ }\textbf {\bibinfo {volume} {93}},\ \bibinfo
  {pages} {094503} (\bibinfo {year} {2016})},\ \Eprint
  {https://arxiv.org/abs/1512.00566} {arXiv:1512.00566 [hep-lat]} \BibitemShut
  {NoStop}%
\bibitem [{\citenamefont {Alexandrou}\ \emph {et~al.}(2021)\citenamefont
  {Alexandrou}, \citenamefont {Athenodorou}, \citenamefont {Hadjiyiannakou},\
  and\ \citenamefont {Todaro}}]{Alexandrou:2020mds}%
  \BibitemOpen
  \bibfield  {author} {\bibinfo {author} {\bibfnamefont {C.}~\bibnamefont
  {Alexandrou}}, \bibinfo {author} {\bibfnamefont {A.}~\bibnamefont
  {Athenodorou}}, \bibinfo {author} {\bibfnamefont {K.}~\bibnamefont
  {Hadjiyiannakou}},\ and\ \bibinfo {author} {\bibfnamefont {A.}~\bibnamefont
  {Todaro}},\ }\href {https://doi.org/10.1103/PhysRevD.103.054501} {\bibfield
  {journal} {\bibinfo  {journal} {Phys. Rev. D}\ }\textbf {\bibinfo {volume}
  {103}},\ \bibinfo {pages} {054501} (\bibinfo {year} {2021})},\ \Eprint
  {https://arxiv.org/abs/2011.01084} {arXiv:2011.01084 [hep-lat]} \BibitemShut
  {NoStop}%
\bibitem [{\citenamefont {Abramczyk}\ \emph {et~al.}(2017)\citenamefont
  {Abramczyk}, \citenamefont {Aoki}, \citenamefont {Blum}, \citenamefont
  {Izubuchi}, \citenamefont {Ohki},\ and\ \citenamefont
  {Syritsyn}}]{Abramczyk:2017oxr}%
  \BibitemOpen
  \bibfield  {author} {\bibinfo {author} {\bibfnamefont {M.}~\bibnamefont
  {Abramczyk}}, \bibinfo {author} {\bibfnamefont {S.}~\bibnamefont {Aoki}},
  \bibinfo {author} {\bibfnamefont {T.}~\bibnamefont {Blum}}, \bibinfo {author}
  {\bibfnamefont {T.}~\bibnamefont {Izubuchi}}, \bibinfo {author}
  {\bibfnamefont {H.}~\bibnamefont {Ohki}},\ and\ \bibinfo {author}
  {\bibfnamefont {S.}~\bibnamefont {Syritsyn}},\ }\href
  {https://doi.org/10.1103/PhysRevD.96.014501} {\bibfield  {journal} {\bibinfo
  {journal} {Phys. Rev.}\ }\textbf {\bibinfo {volume} {D96}},\ \bibinfo {pages}
  {014501} (\bibinfo {year} {2017})},\ \Eprint
  {https://arxiv.org/abs/1701.07792} {arXiv:1701.07792 [hep-lat]} \BibitemShut
  {NoStop}%
\bibitem [{\citenamefont {Kim}\ \emph {et~al.}(2019)\citenamefont {Kim},
  \citenamefont {Dragos}, \citenamefont {Shindler}, \citenamefont {Luu},\ and\
  \citenamefont {de~Vries}}]{Kim:2018rce}%
  \BibitemOpen
  \bibfield  {author} {\bibinfo {author} {\bibfnamefont {J.}~\bibnamefont
  {Kim}}, \bibinfo {author} {\bibfnamefont {J.}~\bibnamefont {Dragos}},
  \bibinfo {author} {\bibfnamefont {A.}~\bibnamefont {Shindler}}, \bibinfo
  {author} {\bibfnamefont {T.}~\bibnamefont {Luu}},\ and\ \bibinfo {author}
  {\bibfnamefont {J.}~\bibnamefont {de~Vries}},\ }\href
  {https://doi.org/10.22323/1.334.0260} {\bibfield  {journal} {\bibinfo
  {journal} {PoS}\ }\textbf {\bibinfo {volume} {LATTICE2018}},\ \bibinfo
  {pages} {260} (\bibinfo {year} {2019})},\ \Eprint
  {https://arxiv.org/abs/1810.10301} {arXiv:1810.10301 [hep-lat]} \BibitemShut
  {NoStop}%
\bibitem [{\citenamefont {Bhattacharya}\ \emph {et~al.}(2022)\citenamefont
  {Bhattacharya}, \citenamefont {Cirigliano}, \citenamefont {Gupta},
  \citenamefont {Mereghetti},\ and\ \citenamefont {Yoon}}]{TanmoyUnpublished}%
  \BibitemOpen
  \bibfield  {author} {\bibinfo {author} {\bibfnamefont {T.}~\bibnamefont
  {Bhattacharya}}, \bibinfo {author} {\bibfnamefont {V.}~\bibnamefont
  {Cirigliano}}, \bibinfo {author} {\bibfnamefont {R.}~\bibnamefont {Gupta}},
  \bibinfo {author} {\bibfnamefont {E.}~\bibnamefont {Mereghetti}},\ and\
  \bibinfo {author} {\bibfnamefont {B.}~\bibnamefont {Yoon}},\ }\href@noop {}
  {\bibinfo {title} {Calculation of neutron electric dipole moment due to the
  {QCD} topological term, {W}einberg three-gluon operator and the quark
  chromoelectric moment}} (\bibinfo {year} {2022}),\ \Eprint
  {https://arxiv.org/abs/2203.03746} {arXiv:2203.03746 [hep-lat]} \BibitemShut
  {NoStop}%
\bibitem [{\citenamefont {Dragos}\ \emph {et~al.}(2018)\citenamefont {Dragos},
  \citenamefont {Luu}, \citenamefont {Shindler},\ and\ \citenamefont
  {de~Vries}}]{Dragos:2017wms}%
  \BibitemOpen
  \bibfield  {author} {\bibinfo {author} {\bibfnamefont {J.}~\bibnamefont
  {Dragos}}, \bibinfo {author} {\bibfnamefont {T.}~\bibnamefont {Luu}},
  \bibinfo {author} {\bibfnamefont {A.}~\bibnamefont {Shindler}},\ and\
  \bibinfo {author} {\bibfnamefont {J.}~\bibnamefont {de~Vries}},\ }\href
  {https://doi.org/10.1051/epjconf/201817506018} {\bibfield  {journal}
  {\bibinfo  {journal} {EPJ Web Conf.}\ }\textbf {\bibinfo {volume} {175}},\
  \bibinfo {pages} {06018} (\bibinfo {year} {2018})},\ \Eprint
  {https://arxiv.org/abs/1711.04730} {arXiv:1711.04730 [hep-lat]} \BibitemShut
  {NoStop}%
\bibitem [{\citenamefont {Maiani}\ \emph {et~al.}(1992)\citenamefont {Maiani},
  \citenamefont {Martinelli},\ and\ \citenamefont {Sachrajda}}]{Maiani:1991az}%
  \BibitemOpen
  \bibfield  {author} {\bibinfo {author} {\bibfnamefont {L.}~\bibnamefont
  {Maiani}}, \bibinfo {author} {\bibfnamefont {G.}~\bibnamefont {Martinelli}},\
  and\ \bibinfo {author} {\bibfnamefont {C.~T.}\ \bibnamefont {Sachrajda}},\
  }\href {https://doi.org/10.1016/0550-3213(92)90528-J} {\bibfield  {journal}
  {\bibinfo  {journal} {Nucl. Phys. B}\ }\textbf {\bibinfo {volume} {368}},\
  \bibinfo {pages} {281} (\bibinfo {year} {1992})}\BibitemShut {NoStop}%
\bibitem [{\citenamefont {Jang}\ \emph {et~al.}(2020)\citenamefont {Jang},
  \citenamefont {Gupta}, \citenamefont {Yoon},\ and\ \citenamefont
  {Bhattacharya}}]{Jang:2019vkm}%
  \BibitemOpen
  \bibfield  {author} {\bibinfo {author} {\bibfnamefont {Y.-C.}\ \bibnamefont
  {Jang}}, \bibinfo {author} {\bibfnamefont {R.}~\bibnamefont {Gupta}},
  \bibinfo {author} {\bibfnamefont {B.}~\bibnamefont {Yoon}},\ and\ \bibinfo
  {author} {\bibfnamefont {T.}~\bibnamefont {Bhattacharya}},\ }\href
  {https://doi.org/10.1103/PhysRevLett.124.072002} {\bibfield  {journal}
  {\bibinfo  {journal} {Phys. Rev. Lett.}\ }\textbf {\bibinfo {volume} {124}},\
  \bibinfo {pages} {072002} (\bibinfo {year} {2020})},\ \Eprint
  {https://arxiv.org/abs/1905.06470} {arXiv:1905.06470 [hep-lat]} \BibitemShut
  {NoStop}%
\bibitem [{\citenamefont {Detmold}\ \emph {et~al.}(2021)\citenamefont
  {Detmold}, \citenamefont {Kanwar}, \citenamefont {Lamm}, \citenamefont
  {Wagman},\ and\ \citenamefont {Warrington}}]{Detmold:2021ulb}%
  \BibitemOpen
  \bibfield  {author} {\bibinfo {author} {\bibfnamefont {W.}~\bibnamefont
  {Detmold}}, \bibinfo {author} {\bibfnamefont {G.}~\bibnamefont {Kanwar}},
  \bibinfo {author} {\bibfnamefont {H.}~\bibnamefont {Lamm}}, \bibinfo {author}
  {\bibfnamefont {M.~L.}\ \bibnamefont {Wagman}},\ and\ \bibinfo {author}
  {\bibfnamefont {N.~C.}\ \bibnamefont {Warrington}},\ }\href
  {https://doi.org/10.1103/PhysRevD.103.094517} {\bibfield  {journal} {\bibinfo
   {journal} {Phys. Rev. D}\ }\textbf {\bibinfo {volume} {103}},\ \bibinfo
  {pages} {094517} (\bibinfo {year} {2021})},\ \Eprint
  {https://arxiv.org/abs/2101.12668} {arXiv:2101.12668 [hep-lat]} \BibitemShut
  {NoStop}%
\bibitem [{\citenamefont {Detmold}\ \emph {et~al.}(2020)\citenamefont
  {Detmold}, \citenamefont {Kanwar}, \citenamefont {Wagman},\ and\
  \citenamefont {Warrington}}]{Detmold:2020ncp}%
  \BibitemOpen
  \bibfield  {author} {\bibinfo {author} {\bibfnamefont {W.}~\bibnamefont
  {Detmold}}, \bibinfo {author} {\bibfnamefont {G.}~\bibnamefont {Kanwar}},
  \bibinfo {author} {\bibfnamefont {M.~L.}\ \bibnamefont {Wagman}},\ and\
  \bibinfo {author} {\bibfnamefont {N.~C.}\ \bibnamefont {Warrington}},\ }\href
  {https://doi.org/10.1103/PhysRevD.102.014514} {\bibfield  {journal} {\bibinfo
   {journal} {Phys. Rev. D}\ }\textbf {\bibinfo {volume} {102}},\ \bibinfo
  {pages} {014514} (\bibinfo {year} {2020})},\ \Eprint
  {https://arxiv.org/abs/2003.05914} {arXiv:2003.05914 [hep-lat]} \BibitemShut
  {NoStop}%
\bibitem [{\citenamefont {Anastassopoulos}\ \emph {et~al.}(1 10)\citenamefont
  {Anastassopoulos} \emph {et~al.}}]{edm_proposal}%
  \BibitemOpen
  \bibfield  {author} {\bibinfo {author} {\bibfnamefont {V.}~\bibnamefont
  {Anastassopoulos}} \emph {et~al.},\ }\href
  {https://www.bnl.gov/edm/files/pdf/Proton_EDM_proposal_20111027_final.pdf}
  {\bibinfo {title} {A proposal to measure the proton electric dipole moment
  with $10^{-29}e \cdot$ cm sensitivity, by the {S}torage ring {EDM}
  collaboration}} (\bibinfo {year} {2011-10}),\ \bibinfo {note} {access at:
  \url{https://inspirehep.net/files/fedd912e77ee5f1defd288d2ea8f8aeb}}\BibitemShut
  {NoStop}%
\bibitem [{\citenamefont {Abusaif}\ \emph {et~al.}(2021)\citenamefont {Abusaif}
  \emph {et~al.}}]{CPEDM:2019nwp}%
  \BibitemOpen
  \bibfield  {author} {\bibinfo {author} {\bibfnamefont {F.}~\bibnamefont
  {Abusaif}} \emph {et~al.} (\bibinfo {collaboration} {CPEDM}),\ }\href
  {https://doi.org/10.23731/CYRM-2021-003} {\emph {\bibinfo {title} {{Storage
  Ring to Search for Electric Dipole Moments of Charged Particles --
  Feasibility Study}}}}\ (\bibinfo  {publisher} {CERN},\ \bibinfo {address}
  {Geneva},\ \bibinfo {year} {2021})\ \Eprint
  {https://arxiv.org/abs/1912.07881} {arXiv:1912.07881 [hep-ex]} \BibitemShut
  {NoStop}%
\bibitem [{\citenamefont {de~Vries}\ \emph {et~al.}(2011)\citenamefont
  {de~Vries}, \citenamefont {Higa}, \citenamefont {Liu}, \citenamefont
  {Mereghetti}, \citenamefont {Stetcu}, \citenamefont {Timmermans},\ and\
  \citenamefont {van Kolck}}]{deVries:2011an}%
  \BibitemOpen
  \bibfield  {author} {\bibinfo {author} {\bibfnamefont {J.}~\bibnamefont
  {de~Vries}}, \bibinfo {author} {\bibfnamefont {R.}~\bibnamefont {Higa}},
  \bibinfo {author} {\bibfnamefont {C.~P.}\ \bibnamefont {Liu}}, \bibinfo
  {author} {\bibfnamefont {E.}~\bibnamefont {Mereghetti}}, \bibinfo {author}
  {\bibfnamefont {I.}~\bibnamefont {Stetcu}}, \bibinfo {author} {\bibfnamefont
  {R.~G.~E.}\ \bibnamefont {Timmermans}},\ and\ \bibinfo {author}
  {\bibfnamefont {U.}~\bibnamefont {van Kolck}},\ }\href
  {https://doi.org/10.1103/PhysRevC.84.065501} {\bibfield  {journal} {\bibinfo
  {journal} {Phys. Rev. C}\ }\textbf {\bibinfo {volume} {84}},\ \bibinfo
  {pages} {065501} (\bibinfo {year} {2011})},\ \Eprint
  {https://arxiv.org/abs/1109.3604} {arXiv:1109.3604 [hep-ph]} \BibitemShut
  {NoStop}%
\bibitem [{\citenamefont {Bsaisou}\ \emph {et~al.}(2015)\citenamefont
  {Bsaisou}, \citenamefont {de~Vries}, \citenamefont {Hanhart}, \citenamefont
  {Liebig}, \citenamefont {Meissner}, \citenamefont {Minossi}, \citenamefont
  {Nogga},\ and\ \citenamefont {Wirzba}}]{Bsaisou:2014zwa}%
  \BibitemOpen
  \bibfield  {author} {\bibinfo {author} {\bibfnamefont {J.}~\bibnamefont
  {Bsaisou}}, \bibinfo {author} {\bibfnamefont {J.}~\bibnamefont {de~Vries}},
  \bibinfo {author} {\bibfnamefont {C.}~\bibnamefont {Hanhart}}, \bibinfo
  {author} {\bibfnamefont {S.}~\bibnamefont {Liebig}}, \bibinfo {author}
  {\bibfnamefont {U.-G.}\ \bibnamefont {Meissner}}, \bibinfo {author}
  {\bibfnamefont {D.}~\bibnamefont {Minossi}}, \bibinfo {author} {\bibfnamefont
  {A.}~\bibnamefont {Nogga}},\ and\ \bibinfo {author} {\bibfnamefont
  {A.}~\bibnamefont {Wirzba}},\ }\href
  {https://doi.org/10.1007/JHEP03(2015)104} {\bibfield  {journal} {\bibinfo
  {journal} {JHEP}\ }\textbf {\bibinfo {volume} {2015}}\bibfield  {number}
  {\bibinfo  {number} { (03)},\ \bibinfo {pages} {104}},\ }\bibinfo {note}
  {[Erratum: JHEP 05, 083 (2015)]},\ \Eprint {https://arxiv.org/abs/1411.5804}
  {arXiv:1411.5804 [hep-ph]} \BibitemShut {NoStop}%
\bibitem [{\citenamefont {Yang}\ \emph {et~al.}(2021)\citenamefont {Yang},
  \citenamefont {Mereghetti}, \citenamefont {Platter}, \citenamefont
  {Schindler},\ and\ \citenamefont {Vanasse}}]{Yang:2020ges}%
  \BibitemOpen
  \bibfield  {author} {\bibinfo {author} {\bibfnamefont {Z.}~\bibnamefont
  {Yang}}, \bibinfo {author} {\bibfnamefont {E.}~\bibnamefont {Mereghetti}},
  \bibinfo {author} {\bibfnamefont {L.}~\bibnamefont {Platter}}, \bibinfo
  {author} {\bibfnamefont {M.~R.}\ \bibnamefont {Schindler}},\ and\ \bibinfo
  {author} {\bibfnamefont {J.}~\bibnamefont {Vanasse}},\ }\href
  {https://doi.org/10.1103/PhysRevC.104.024002} {\bibfield  {journal} {\bibinfo
   {journal} {Phys. Rev. C}\ }\textbf {\bibinfo {volume} {104}},\ \bibinfo
  {pages} {024002} (\bibinfo {year} {2021})},\ \Eprint
  {https://arxiv.org/abs/2011.01885} {arXiv:2011.01885 [nucl-th]} \BibitemShut
  {NoStop}%
\bibitem [{\citenamefont {Froese}\ and\ \citenamefont
  {Navratil}(2021)}]{Froese:2021civ}%
  \BibitemOpen
  \bibfield  {author} {\bibinfo {author} {\bibfnamefont {P.}~\bibnamefont
  {Froese}}\ and\ \bibinfo {author} {\bibfnamefont {P.}~\bibnamefont
  {Navratil}},\ }\href {https://doi.org/10.1103/PhysRevC.104.025502} {\bibfield
   {journal} {\bibinfo  {journal} {Phys. Rev. C}\ }\textbf {\bibinfo {volume}
  {104}},\ \bibinfo {pages} {025502} (\bibinfo {year} {2021})},\ \Eprint
  {https://arxiv.org/abs/2103.06365} {arXiv:2103.06365 [nucl-th]} \BibitemShut
  {NoStop}%
\bibitem [{\citenamefont {Lebedev}\ \emph
  {et~al.}(2004{\natexlab{b}})\citenamefont {Lebedev}, \citenamefont {Olive},
  \citenamefont {Pospelov},\ and\ \citenamefont {Ritz}}]{Lebedev:2004va}%
  \BibitemOpen
  \bibfield  {author} {\bibinfo {author} {\bibfnamefont {O.}~\bibnamefont
  {Lebedev}}, \bibinfo {author} {\bibfnamefont {K.~A.}\ \bibnamefont {Olive}},
  \bibinfo {author} {\bibfnamefont {M.}~\bibnamefont {Pospelov}},\ and\
  \bibinfo {author} {\bibfnamefont {A.}~\bibnamefont {Ritz}},\ }\href
  {https://doi.org/10.1103/PhysRevD.70.016003} {\bibfield  {journal} {\bibinfo
  {journal} {Phys. Rev. D}\ }\textbf {\bibinfo {volume} {70}},\ \bibinfo
  {pages} {016003} (\bibinfo {year} {2004}{\natexlab{b}})},\ \Eprint
  {https://arxiv.org/abs/hep-ph/0402023} {arXiv:hep-ph/0402023} \BibitemShut
  {NoStop}%
\bibitem [{\citenamefont {Dekens}\ \emph {et~al.}(2014)\citenamefont {Dekens},
  \citenamefont {de~Vries}, \citenamefont {Bsaisou}, \citenamefont
  {Bernreuther}, \citenamefont {Hanhart}, \citenamefont {Mei\ss{}ner},
  \citenamefont {Nogga},\ and\ \citenamefont {Wirzba}}]{Dekens:2014jka}%
  \BibitemOpen
  \bibfield  {author} {\bibinfo {author} {\bibfnamefont {W.}~\bibnamefont
  {Dekens}}, \bibinfo {author} {\bibfnamefont {J.}~\bibnamefont {de~Vries}},
  \bibinfo {author} {\bibfnamefont {J.}~\bibnamefont {Bsaisou}}, \bibinfo
  {author} {\bibfnamefont {W.}~\bibnamefont {Bernreuther}}, \bibinfo {author}
  {\bibfnamefont {C.}~\bibnamefont {Hanhart}}, \bibinfo {author} {\bibfnamefont
  {U.-G.}\ \bibnamefont {Mei\ss{}ner}}, \bibinfo {author} {\bibfnamefont
  {A.}~\bibnamefont {Nogga}},\ and\ \bibinfo {author} {\bibfnamefont
  {A.}~\bibnamefont {Wirzba}},\ }\href
  {https://doi.org/10.1007/JHEP07(2014)069} {\bibfield  {journal} {\bibinfo
  {journal} {JHEP}\ }\textbf {\bibinfo {volume} {2014}}\bibfield  {number}
  {\bibinfo  {number} { (07)},\ \bibinfo {pages} {069}},\ }\Eprint
  {https://arxiv.org/abs/1404.6082} {arXiv:1404.6082 [hep-ph]} \BibitemShut
  {NoStop}%
\bibitem [{\citenamefont {Flambaum}\ \emph
  {et~al.}(2020{\natexlab{a}})\citenamefont {Flambaum}, \citenamefont
  {Pospelov}, \citenamefont {Ritz},\ and\ \citenamefont
  {Stadnik}}]{Flambaum:2019ejc}%
  \BibitemOpen
  \bibfield  {author} {\bibinfo {author} {\bibfnamefont {V.~V.}\ \bibnamefont
  {Flambaum}}, \bibinfo {author} {\bibfnamefont {M.}~\bibnamefont {Pospelov}},
  \bibinfo {author} {\bibfnamefont {A.}~\bibnamefont {Ritz}},\ and\ \bibinfo
  {author} {\bibfnamefont {Y.~V.}\ \bibnamefont {Stadnik}},\ }\href
  {https://doi.org/10.1103/PhysRevD.102.035001} {\bibfield  {journal} {\bibinfo
   {journal} {Phys. Rev. D}\ }\textbf {\bibinfo {volume} {102}},\ \bibinfo
  {pages} {035001} (\bibinfo {year} {2020}{\natexlab{a}})},\ \Eprint
  {https://arxiv.org/abs/1912.13129} {arXiv:1912.13129 [hep-ph]} \BibitemShut
  {NoStop}%
\bibitem [{\citenamefont {Flambaum}\ \emph
  {et~al.}(2020{\natexlab{b}})\citenamefont {Flambaum}, \citenamefont
  {Samsonov},\ and\ \citenamefont {{Tran Tan}}}]{Flambaum2020Hadronic2}%
  \BibitemOpen
  \bibfield  {author} {\bibinfo {author} {\bibfnamefont {V.~V.}\ \bibnamefont
  {Flambaum}}, \bibinfo {author} {\bibfnamefont {I.~B.}\ \bibnamefont
  {Samsonov}},\ and\ \bibinfo {author} {\bibfnamefont {H.~B.}\ \bibnamefont
  {{Tran Tan}}},\ }\href {https://doi.org/10.1007/JHEP10(2020)077} {\bibfield
  {journal} {\bibinfo  {journal} {Journal of High Energy Physics}\ }\textbf
  {\bibinfo {volume} {2020}},\ \bibinfo {pages} {77} (\bibinfo {year}
  {2020}{\natexlab{b}})}\BibitemShut {NoStop}%
\bibitem [{\citenamefont {Flambaum}\ \emph
  {et~al.}(2020{\natexlab{c}})\citenamefont {Flambaum}, \citenamefont
  {Samsonov},\ and\ \citenamefont {Tran~Tan}}]{Flambaum2020Internucleon}%
  \BibitemOpen
  \bibfield  {author} {\bibinfo {author} {\bibfnamefont {V.~V.}\ \bibnamefont
  {Flambaum}}, \bibinfo {author} {\bibfnamefont {I.~B.}\ \bibnamefont
  {Samsonov}},\ and\ \bibinfo {author} {\bibfnamefont {H.~B.}\ \bibnamefont
  {Tran~Tan}},\ }\href {https://doi.org/10.1103/PhysRevD.102.115036} {\bibfield
   {journal} {\bibinfo  {journal} {Phys. Rev. D}\ }\textbf {\bibinfo {volume}
  {102}},\ \bibinfo {pages} {115036} (\bibinfo {year}
  {2020}{\natexlab{c}})}\BibitemShut {NoStop}%
\bibitem [{\citenamefont {Caurier}\ \emph {et~al.}(2005)\citenamefont
  {Caurier}, \citenamefont {Mart\'{\i}nez-Pinedo}, \citenamefont {Nowacki},
  \citenamefont {Poves},\ and\ \citenamefont {Zuker}}]{Caurier05}%
  \BibitemOpen
  \bibfield  {author} {\bibinfo {author} {\bibfnamefont {E.}~\bibnamefont
  {Caurier}}, \bibinfo {author} {\bibfnamefont {G.}~\bibnamefont
  {Mart\'{\i}nez-Pinedo}}, \bibinfo {author} {\bibfnamefont {F.}~\bibnamefont
  {Nowacki}}, \bibinfo {author} {\bibfnamefont {A.}~\bibnamefont {Poves}},\
  and\ \bibinfo {author} {\bibfnamefont {A.~P.}\ \bibnamefont {Zuker}},\ }\href
  {https://doi.org/10.1103/RevModPhys.77.427} {\bibfield  {journal} {\bibinfo
  {journal} {Rev. Mod. Phys.}\ }\textbf {\bibinfo {volume} {77}},\ \bibinfo
  {pages} {427} (\bibinfo {year} {2005})}\BibitemShut {NoStop}%
\bibitem [{\citenamefont {Schunck}(2019)}]{Schunck19}%
  \BibitemOpen
  \bibinfo {editor} {\bibfnamefont {N.}~\bibnamefont {Schunck}},\ ed.,\ \href
  {https://doi.org/10.1088/2053-2563/aae0ed} {\emph {\bibinfo {title} {Energy
  Density Functional Methods for Atomic Nuclei}}},\ 2053-2563\ (\bibinfo
  {publisher} {IOP Publishing},\ \bibinfo {year} {2019})\BibitemShut {NoStop}%
\bibitem [{\citenamefont {Yanase}\ and\ \citenamefont
  {Shimizu}(2020)}]{Yanase20}%
  \BibitemOpen
  \bibfield  {author} {\bibinfo {author} {\bibfnamefont {K.}~\bibnamefont
  {Yanase}}\ and\ \bibinfo {author} {\bibfnamefont {N.}~\bibnamefont
  {Shimizu}},\ }\href {https://doi.org/10.1103/PhysRevC.102.065502} {\bibfield
  {journal} {\bibinfo  {journal} {Phys. Rev. C}\ }\textbf {\bibinfo {volume}
  {102}},\ \bibinfo {pages} {065502} (\bibinfo {year} {2020})}\BibitemShut
  {NoStop}%
\bibitem [{\citenamefont {Dobaczewski}\ and\ \citenamefont
  {Engel}(2005)}]{Dobaczewski2005}%
  \BibitemOpen
  \bibfield  {author} {\bibinfo {author} {\bibfnamefont {J.}~\bibnamefont
  {Dobaczewski}}\ and\ \bibinfo {author} {\bibfnamefont {J.}~\bibnamefont
  {Engel}},\ }\href {https://doi.org/10.1103/PhysRevLett.94.232502} {\bibfield
  {journal} {\bibinfo  {journal} {Physical Review Letters}\ }\textbf {\bibinfo
  {volume} {94}},\ \bibinfo {pages} {232502} (\bibinfo {year}
  {2005})}\BibitemShut {NoStop}%
\bibitem [{\citenamefont {Dobaczewski}\ \emph {et~al.}(2018)\citenamefont
  {Dobaczewski}, \citenamefont {Engel}, \citenamefont {Kortelainen},\ and\
  \citenamefont {Becker}}]{Dobaczewski2018}%
  \BibitemOpen
  \bibfield  {author} {\bibinfo {author} {\bibfnamefont {J.}~\bibnamefont
  {Dobaczewski}}, \bibinfo {author} {\bibfnamefont {J.}~\bibnamefont {Engel}},
  \bibinfo {author} {\bibfnamefont {M.}~\bibnamefont {Kortelainen}},\ and\
  \bibinfo {author} {\bibfnamefont {P.}~\bibnamefont {Becker}},\ }\href
  {https://doi.org/10.1103/PhysRevLett.121.232501} {\bibfield  {journal}
  {\bibinfo  {journal} {Physical Review Letters}\ }\textbf {\bibinfo {volume}
  {121}},\ \bibinfo {pages} {232501} (\bibinfo {year} {2018})}\BibitemShut
  {NoStop}%
\bibitem [{\citenamefont {Hagen}\ \emph {et~al.}(2014)\citenamefont {Hagen},
  \citenamefont {Papenbrock}, \citenamefont {Hjorth-Jensen},\ and\
  \citenamefont {Dean}}]{Hagen14}%
  \BibitemOpen
  \bibfield  {author} {\bibinfo {author} {\bibfnamefont {G.}~\bibnamefont
  {Hagen}}, \bibinfo {author} {\bibfnamefont {T.}~\bibnamefont {Papenbrock}},
  \bibinfo {author} {\bibfnamefont {M.}~\bibnamefont {Hjorth-Jensen}},\ and\
  \bibinfo {author} {\bibfnamefont {D.~J.}\ \bibnamefont {Dean}},\ }\href
  {https://doi.org/10.1088/0034-4885/77/9/096302} {\bibfield  {journal}
  {\bibinfo  {journal} {Reports on Progress in Physics}\ }\textbf {\bibinfo
  {volume} {77}},\ \bibinfo {pages} {096302} (\bibinfo {year}
  {2014})}\BibitemShut {NoStop}%
\bibitem [{\citenamefont {Bonaiti}\ \emph {et~al.}(2022)\citenamefont
  {Bonaiti}, \citenamefont {Bacca},\ and\ \citenamefont {Hagen}}]{Bonaiti21}%
  \BibitemOpen
  \bibfield  {author} {\bibinfo {author} {\bibfnamefont {F.}~\bibnamefont
  {Bonaiti}}, \bibinfo {author} {\bibfnamefont {S.}~\bibnamefont {Bacca}},\
  and\ \bibinfo {author} {\bibfnamefont {G.}~\bibnamefont {Hagen}},\ }\href
  {https://doi.org/10.1103/PhysRevC.105.034313} {\bibfield  {journal} {\bibinfo
   {journal} {Phys. Rev. C}\ }\textbf {\bibinfo {volume} {105}},\ \bibinfo
  {pages} {034313} (\bibinfo {year} {2022})}\BibitemShut {NoStop}%
\bibitem [{\citenamefont {Novario}\ \emph {et~al.}(2021)\citenamefont
  {Novario}, \citenamefont {Gysbers}, \citenamefont {Engel}, \citenamefont
  {Hagen}, \citenamefont {Jansen}, \citenamefont {Morris}, \citenamefont
  {Navr\'atil}, \citenamefont {Papenbrock},\ and\ \citenamefont
  {Quaglioni}}]{Novario21}%
  \BibitemOpen
  \bibfield  {author} {\bibinfo {author} {\bibfnamefont {S.}~\bibnamefont
  {Novario}}, \bibinfo {author} {\bibfnamefont {P.}~\bibnamefont {Gysbers}},
  \bibinfo {author} {\bibfnamefont {J.}~\bibnamefont {Engel}}, \bibinfo
  {author} {\bibfnamefont {G.}~\bibnamefont {Hagen}}, \bibinfo {author}
  {\bibfnamefont {G.~R.}\ \bibnamefont {Jansen}}, \bibinfo {author}
  {\bibfnamefont {T.~D.}\ \bibnamefont {Morris}}, \bibinfo {author}
  {\bibfnamefont {P.}~\bibnamefont {Navr\'atil}}, \bibinfo {author}
  {\bibfnamefont {T.}~\bibnamefont {Papenbrock}},\ and\ \bibinfo {author}
  {\bibfnamefont {S.}~\bibnamefont {Quaglioni}},\ }\href
  {https://doi.org/10.1103/PhysRevLett.126.182502} {\bibfield  {journal}
  {\bibinfo  {journal} {Phys. Rev. Lett.}\ }\textbf {\bibinfo {volume} {126}},\
  \bibinfo {pages} {182502} (\bibinfo {year} {2021})}\BibitemShut {NoStop}%
\bibitem [{\citenamefont {Hergert}\ \emph {et~al.}(2016)\citenamefont
  {Hergert}, \citenamefont {Bogner}, \citenamefont {Morris}, \citenamefont
  {Schwenk},\ and\ \citenamefont {Tsukiyama}}]{Hergert16}%
  \BibitemOpen
  \bibfield  {author} {\bibinfo {author} {\bibfnamefont {H.}~\bibnamefont
  {Hergert}}, \bibinfo {author} {\bibfnamefont {S.~K.}\ \bibnamefont {Bogner}},
  \bibinfo {author} {\bibfnamefont {T.~D.}\ \bibnamefont {Morris}}, \bibinfo
  {author} {\bibfnamefont {A.}~\bibnamefont {Schwenk}},\ and\ \bibinfo {author}
  {\bibfnamefont {K.}~\bibnamefont {Tsukiyama}},\ }\bibfield  {booktitle}
  {\emph {\bibinfo {booktitle} {Memorial Volume in Honor of Gerald E. Brown}},\
  }\href {https://doi.org/10.1016/j.physrep.2015.12.007} {\bibfield  {journal}
  {\bibinfo  {journal} {Physics Reports}\ }\textbf {\bibinfo {volume} {621}},\
  \bibinfo {pages} {165} (\bibinfo {year} {2016})}\BibitemShut {NoStop}%
\bibitem [{\citenamefont {Hergert}(2017)}]{Hergert17}%
  \BibitemOpen
  \bibfield  {author} {\bibinfo {author} {\bibfnamefont {H.}~\bibnamefont
  {Hergert}},\ }\href {http://stacks.iop.org/1402-4896/92/i=2/a=023002}
  {\bibfield  {journal} {\bibinfo  {journal} {Phys. Scripta}\ }\textbf
  {\bibinfo {volume} {92}},\ \bibinfo {pages} {023002} (\bibinfo {year}
  {2017})}\BibitemShut {NoStop}%
\bibitem [{\citenamefont {Stroberg}\ \emph {et~al.}(2019)\citenamefont
  {Stroberg}, \citenamefont {Hergert}, \citenamefont {Bogner},\ and\
  \citenamefont {Holt}}]{Stroberg19}%
  \BibitemOpen
  \bibfield  {author} {\bibinfo {author} {\bibfnamefont {S.~R.}\ \bibnamefont
  {Stroberg}}, \bibinfo {author} {\bibfnamefont {H.}~\bibnamefont {Hergert}},
  \bibinfo {author} {\bibfnamefont {S.~K.}\ \bibnamefont {Bogner}},\ and\
  \bibinfo {author} {\bibfnamefont {J.~D.}\ \bibnamefont {Holt}},\ }\bibfield
  {booktitle} {\emph {\bibinfo {booktitle} {Annual Review of Nuclear and
  Particle Science}},\ }\href
  {https://doi.org/10.1146/annurev-nucl-101917-021120} {\bibfield  {journal}
  {\bibinfo  {journal} {Annual Review of Nuclear and Particle Science}\
  }\textbf {\bibinfo {volume} {69}},\ \bibinfo {pages} {307} (\bibinfo {year}
  {2019})}\BibitemShut {NoStop}%
\bibitem [{\citenamefont {Yao}\ \emph {et~al.}(2020)\citenamefont {Yao},
  \citenamefont {Bally}, \citenamefont {Engel}, \citenamefont {Wirth},
  \citenamefont {Rodr\'{\i}guez},\ and\ \citenamefont {Hergert}}]{Yao20}%
  \BibitemOpen
  \bibfield  {author} {\bibinfo {author} {\bibfnamefont {J.~M.}\ \bibnamefont
  {Yao}}, \bibinfo {author} {\bibfnamefont {B.}~\bibnamefont {Bally}}, \bibinfo
  {author} {\bibfnamefont {J.}~\bibnamefont {Engel}}, \bibinfo {author}
  {\bibfnamefont {R.}~\bibnamefont {Wirth}}, \bibinfo {author} {\bibfnamefont
  {T.~R.}\ \bibnamefont {Rodr\'{\i}guez}},\ and\ \bibinfo {author}
  {\bibfnamefont {H.}~\bibnamefont {Hergert}},\ }\href
  {https://doi.org/10.1103/PhysRevLett.124.232501} {\bibfield  {journal}
  {\bibinfo  {journal} {Phys. Rev. Lett.}\ }\textbf {\bibinfo {volume} {124}},\
  \bibinfo {pages} {232501} (\bibinfo {year} {2020})}\BibitemShut {NoStop}%
\bibitem [{\citenamefont {Belley}\ \emph {et~al.}(2021)\citenamefont {Belley},
  \citenamefont {Payne}, \citenamefont {Stroberg}, \citenamefont {Miyagi},\
  and\ \citenamefont {Holt}}]{Belley21}%
  \BibitemOpen
  \bibfield  {author} {\bibinfo {author} {\bibfnamefont {A.}~\bibnamefont
  {Belley}}, \bibinfo {author} {\bibfnamefont {C.~G.}\ \bibnamefont {Payne}},
  \bibinfo {author} {\bibfnamefont {S.~R.}\ \bibnamefont {Stroberg}}, \bibinfo
  {author} {\bibfnamefont {T.}~\bibnamefont {Miyagi}},\ and\ \bibinfo {author}
  {\bibfnamefont {J.~D.}\ \bibnamefont {Holt}},\ }\href
  {https://doi.org/10.1103/PhysRevLett.126.042502} {\bibfield  {journal}
  {\bibinfo  {journal} {Phys. Rev. Lett.}\ }\textbf {\bibinfo {volume} {126}},\
  \bibinfo {pages} {042502} (\bibinfo {year} {2021})}\BibitemShut {NoStop}%
\bibitem [{\citenamefont {de~Vries}\ \emph {et~al.}(2021)\citenamefont
  {de~Vries}, \citenamefont {Draper}, \citenamefont {Fuyuto}, \citenamefont
  {Kozaczuk},\ and\ \citenamefont {Lillard}}]{deVries:2021sxz}%
  \BibitemOpen
  \bibfield  {author} {\bibinfo {author} {\bibfnamefont {J.}~\bibnamefont
  {de~Vries}}, \bibinfo {author} {\bibfnamefont {P.}~\bibnamefont {Draper}},
  \bibinfo {author} {\bibfnamefont {K.}~\bibnamefont {Fuyuto}}, \bibinfo
  {author} {\bibfnamefont {J.}~\bibnamefont {Kozaczuk}},\ and\ \bibinfo
  {author} {\bibfnamefont {B.}~\bibnamefont {Lillard}},\ }\href
  {https://doi.org/10.1103/PhysRevD.104.055039} {\bibfield  {journal} {\bibinfo
   {journal} {Phys. Rev. D}\ }\textbf {\bibinfo {volume} {104}},\ \bibinfo
  {pages} {055039} (\bibinfo {year} {2021})},\ \Eprint
  {https://arxiv.org/abs/2107.04046} {arXiv:2107.04046 [hep-ph]} \BibitemShut
  {NoStop}%
\bibitem [{\citenamefont {Purcell}\ and\ \citenamefont
  {Ramsey}(1950)}]{EPurcell1950}%
  \BibitemOpen
  \bibfield  {author} {\bibinfo {author} {\bibfnamefont {E.~M.}\ \bibnamefont
  {Purcell}}\ and\ \bibinfo {author} {\bibfnamefont {N.~F.}\ \bibnamefont
  {Ramsey}},\ }\href {https://doi.org/10.1103/PhysRev.78.807} {\bibfield
  {journal} {\bibinfo  {journal} {Phys. Rev.}\ }\textbf {\bibinfo {volume}
  {78}},\ \bibinfo {pages} {807} (\bibinfo {year} {1950})}\BibitemShut
  {NoStop}%
\bibitem [{\citenamefont {Smith}\ \emph {et~al.}(1957)\citenamefont {Smith},
  \citenamefont {Purcell},\ and\ \citenamefont {Ramsey}}]{Smith1957}%
  \BibitemOpen
  \bibfield  {author} {\bibinfo {author} {\bibfnamefont {J.~H.}\ \bibnamefont
  {Smith}}, \bibinfo {author} {\bibfnamefont {E.~M.}\ \bibnamefont {Purcell}},\
  and\ \bibinfo {author} {\bibfnamefont {N.~F.}\ \bibnamefont {Ramsey}},\
  }\href {https://doi.org/10.1103/PhysRev.108.120} {\bibfield  {journal}
  {\bibinfo  {journal} {Phys. Rev.}\ }\textbf {\bibinfo {volume} {108}},\
  \bibinfo {pages} {120} (\bibinfo {year} {1957})}\BibitemShut {NoStop}%
\bibitem [{\citenamefont {Ramsey}(1950)}]{Ramsey1950}%
  \BibitemOpen
  \bibfield  {author} {\bibinfo {author} {\bibfnamefont {N.~F.}\ \bibnamefont
  {Ramsey}},\ }\href {https://doi.org/10.1103/PhysRev.78.695} {\bibfield
  {journal} {\bibinfo  {journal} {Physical Review}\ }\textbf {\bibinfo {volume}
  {78}},\ \bibinfo {pages} {695} (\bibinfo {year} {1950})}\BibitemShut
  {NoStop}%
\bibitem [{\citenamefont {Cohen}\ \emph {et~al.}(1969)\citenamefont {Cohen},
  \citenamefont {Nathans}, \citenamefont {Silsbee}, \citenamefont {Lipworth},\
  and\ \citenamefont {Ramsey}}]{Cohen69}%
  \BibitemOpen
  \bibfield  {author} {\bibinfo {author} {\bibfnamefont {V.~W.}\ \bibnamefont
  {Cohen}}, \bibinfo {author} {\bibfnamefont {R.}~\bibnamefont {Nathans}},
  \bibinfo {author} {\bibfnamefont {H.~B.}\ \bibnamefont {Silsbee}}, \bibinfo
  {author} {\bibfnamefont {E.}~\bibnamefont {Lipworth}},\ and\ \bibinfo
  {author} {\bibfnamefont {N.~F.}\ \bibnamefont {Ramsey}},\ }\href
  {https://doi.org/10.1103/PhysRev.177.1942} {\bibfield  {journal} {\bibinfo
  {journal} {Phys. Rev.}\ }\textbf {\bibinfo {volume} {177}},\ \bibinfo {pages}
  {1942} (\bibinfo {year} {1969})}\BibitemShut {NoStop}%
\bibitem [{\citenamefont {Baird}\ \emph {et~al.}(1969)\citenamefont {Baird},
  \citenamefont {Miller}, \citenamefont {Dress},\ and\ \citenamefont
  {Ramsey}}]{Baird69}%
  \BibitemOpen
  \bibfield  {author} {\bibinfo {author} {\bibfnamefont {J.~K.}\ \bibnamefont
  {Baird}}, \bibinfo {author} {\bibfnamefont {P.~D.}\ \bibnamefont {Miller}},
  \bibinfo {author} {\bibfnamefont {W.~B.}\ \bibnamefont {Dress}},\ and\
  \bibinfo {author} {\bibfnamefont {N.~F.}\ \bibnamefont {Ramsey}},\ }\href
  {https://doi.org/10.1103/PhysRev.179.1285} {\bibfield  {journal} {\bibinfo
  {journal} {Phys. Rev.}\ }\textbf {\bibinfo {volume} {179}},\ \bibinfo {pages}
  {1285} (\bibinfo {year} {1969})}\BibitemShut {NoStop}%
\bibitem [{\citenamefont {Sandars}\ and\ \citenamefont
  {Lipworth}(1964{\natexlab{a}})}]{Sandars1964}%
  \BibitemOpen
  \bibfield  {author} {\bibinfo {author} {\bibfnamefont {P.~G.~H.}\
  \bibnamefont {Sandars}}\ and\ \bibinfo {author} {\bibfnamefont
  {E.}~\bibnamefont {Lipworth}},\ }\href
  {https://doi.org/10.1103/PhysRevLett.13.718} {\bibfield  {journal} {\bibinfo
  {journal} {Phys. Rev. Lett.}\ }\textbf {\bibinfo {volume} {13}},\ \bibinfo
  {pages} {718} (\bibinfo {year} {1964}{\natexlab{a}})}\BibitemShut {NoStop}%
\bibitem [{\citenamefont {Dress}\ \emph {et~al.}(1977)\citenamefont {Dress},
  \citenamefont {Miller}, \citenamefont {Pendlebury}, \citenamefont {Perrin},\
  and\ \citenamefont {Ramsey}}]{Dress77}%
  \BibitemOpen
  \bibfield  {author} {\bibinfo {author} {\bibfnamefont {W.~B.}\ \bibnamefont
  {Dress}}, \bibinfo {author} {\bibfnamefont {P.~D.}\ \bibnamefont {Miller}},
  \bibinfo {author} {\bibfnamefont {J.~M.}\ \bibnamefont {Pendlebury}},
  \bibinfo {author} {\bibfnamefont {P.}~\bibnamefont {Perrin}},\ and\ \bibinfo
  {author} {\bibfnamefont {N.~F.}\ \bibnamefont {Ramsey}},\ }\href
  {https://doi.org/10.1103/PhysRevD.15.9} {\bibfield  {journal} {\bibinfo
  {journal} {Phys. Rev. D}\ }\textbf {\bibinfo {volume} {15}},\ \bibinfo
  {pages} {9} (\bibinfo {year} {1977})}\BibitemShut {NoStop}%
\bibitem [{\citenamefont {Smith}\ \emph {et~al.}(1990)\citenamefont {Smith},
  \citenamefont {Crampin}, \citenamefont {Pendlebury}, \citenamefont
  {Richardson}, \citenamefont {Shiers}, \citenamefont {Green}, \citenamefont
  {Kilvington}, \citenamefont {Moir}, \citenamefont {Prosper}, \citenamefont
  {Thompson}, \citenamefont {Ramsey}, \citenamefont {Heckel}, \citenamefont
  {Lamoreaux}, \citenamefont {Ageron}, \citenamefont {Mampe},\ and\
  \citenamefont {Steyerl}}]{SMITH1990}%
  \BibitemOpen
  \bibfield  {author} {\bibinfo {author} {\bibfnamefont {K.}~\bibnamefont
  {Smith}}, \bibinfo {author} {\bibfnamefont {N.}~\bibnamefont {Crampin}},
  \bibinfo {author} {\bibfnamefont {J.}~\bibnamefont {Pendlebury}}, \bibinfo
  {author} {\bibfnamefont {D.}~\bibnamefont {Richardson}}, \bibinfo {author}
  {\bibfnamefont {D.}~\bibnamefont {Shiers}}, \bibinfo {author} {\bibfnamefont
  {K.}~\bibnamefont {Green}}, \bibinfo {author} {\bibfnamefont
  {A.}~\bibnamefont {Kilvington}}, \bibinfo {author} {\bibfnamefont
  {J.}~\bibnamefont {Moir}}, \bibinfo {author} {\bibfnamefont {H.}~\bibnamefont
  {Prosper}}, \bibinfo {author} {\bibfnamefont {D.}~\bibnamefont {Thompson}},
  \bibinfo {author} {\bibfnamefont {N.}~\bibnamefont {Ramsey}}, \bibinfo
  {author} {\bibfnamefont {B.}~\bibnamefont {Heckel}}, \bibinfo {author}
  {\bibfnamefont {S.}~\bibnamefont {Lamoreaux}}, \bibinfo {author}
  {\bibfnamefont {P.}~\bibnamefont {Ageron}}, \bibinfo {author} {\bibfnamefont
  {W.}~\bibnamefont {Mampe}},\ and\ \bibinfo {author} {\bibfnamefont
  {A.}~\bibnamefont {Steyerl}},\ }\href
  {https://doi.org/10.1016/0370-2693(90)92027-G} {\bibfield  {journal}
  {\bibinfo  {journal} {Physics Letters B}\ }\textbf {\bibinfo {volume}
  {234}},\ \bibinfo {pages} {191} (\bibinfo {year} {1990})}\BibitemShut
  {NoStop}%
\bibitem [{\citenamefont {Baker}\ \emph {et~al.}(2006)\citenamefont {Baker},
  \citenamefont {Doyle}, \citenamefont {Geltenbort}, \citenamefont {Green},
  \citenamefont {van~der Grinten}, \citenamefont {Harris}, \citenamefont
  {Iaydjiev}, \citenamefont {Ivanov}, \citenamefont {May}, \citenamefont
  {Pendlebury}, \citenamefont {Richardson}, \citenamefont {Shiers},\ and\
  \citenamefont {Smith}}]{Baker2006}%
  \BibitemOpen
  \bibfield  {author} {\bibinfo {author} {\bibfnamefont {C.}~\bibnamefont
  {Baker}}, \bibinfo {author} {\bibfnamefont {D.~D.}\ \bibnamefont {Doyle}},
  \bibinfo {author} {\bibfnamefont {P.}~\bibnamefont {Geltenbort}}, \bibinfo
  {author} {\bibfnamefont {K.}~\bibnamefont {Green}}, \bibinfo {author}
  {\bibfnamefont {M.~G.~D.}\ \bibnamefont {van~der Grinten}}, \bibinfo {author}
  {\bibfnamefont {P.~G.}\ \bibnamefont {Harris}}, \bibinfo {author}
  {\bibfnamefont {P.}~\bibnamefont {Iaydjiev}}, \bibinfo {author}
  {\bibfnamefont {S.~N.}\ \bibnamefont {Ivanov}}, \bibinfo {author}
  {\bibfnamefont {D.~J.~R.}\ \bibnamefont {May}}, \bibinfo {author}
  {\bibfnamefont {J.~M.}\ \bibnamefont {Pendlebury}}, \bibinfo {author}
  {\bibfnamefont {J.~D.}\ \bibnamefont {Richardson}}, \bibinfo {author}
  {\bibfnamefont {D.}~\bibnamefont {Shiers}},\ and\ \bibinfo {author}
  {\bibfnamefont {K.~F.}\ \bibnamefont {Smith}},\ }\href
  {https://doi.org/10.1103/PhysRevLett.97.131801} {\bibfield  {journal}
  {\bibinfo  {journal} {Physical Review Letters}\ }\textbf {\bibinfo {volume}
  {97}},\ \bibinfo {pages} {131801} (\bibinfo {year} {2006})}\BibitemShut
  {NoStop}%
\bibitem [{\citenamefont {Serebrov}\ \emph {et~al.}(2014)\citenamefont
  {Serebrov}, \citenamefont {Kolomenskiy}, \citenamefont {Pirozhkov},
  \citenamefont {Krasnoschekova}, \citenamefont {Vassiljev}, \citenamefont
  {Polushkin}, \citenamefont {Lasakov}, \citenamefont {Fomin}, \citenamefont
  {Shoka}, \citenamefont {Solovey}, \citenamefont {Zherebtsov}, \citenamefont
  {Geltenbort}, \citenamefont {Ivanov}, \citenamefont {Zimmer}, \citenamefont
  {Alexandrov}, \citenamefont {Dmitriev},\ and\ \citenamefont
  {Dovator}}]{Serebrov2014}%
  \BibitemOpen
  \bibfield  {author} {\bibinfo {author} {\bibfnamefont {A.~P.}\ \bibnamefont
  {Serebrov}}, \bibinfo {author} {\bibfnamefont {E.~A.}\ \bibnamefont
  {Kolomenskiy}}, \bibinfo {author} {\bibfnamefont {A.~N.}\ \bibnamefont
  {Pirozhkov}}, \bibinfo {author} {\bibfnamefont {I.~A.}\ \bibnamefont
  {Krasnoschekova}}, \bibinfo {author} {\bibfnamefont {A.~V.}\ \bibnamefont
  {Vassiljev}}, \bibinfo {author} {\bibfnamefont {A.~O.}\ \bibnamefont
  {Polushkin}}, \bibinfo {author} {\bibfnamefont {M.~S.}\ \bibnamefont
  {Lasakov}}, \bibinfo {author} {\bibfnamefont {A.~K.}\ \bibnamefont {Fomin}},
  \bibinfo {author} {\bibfnamefont {I.~V.}\ \bibnamefont {Shoka}}, \bibinfo
  {author} {\bibfnamefont {V.~A.}\ \bibnamefont {Solovey}}, \bibinfo {author}
  {\bibfnamefont {O.~M.}\ \bibnamefont {Zherebtsov}}, \bibinfo {author}
  {\bibfnamefont {P.}~\bibnamefont {Geltenbort}}, \bibinfo {author}
  {\bibfnamefont {S.~N.}\ \bibnamefont {Ivanov}}, \bibinfo {author}
  {\bibfnamefont {O.}~\bibnamefont {Zimmer}}, \bibinfo {author} {\bibfnamefont
  {E.~B.}\ \bibnamefont {Alexandrov}}, \bibinfo {author} {\bibfnamefont
  {S.~P.}\ \bibnamefont {Dmitriev}},\ and\ \bibinfo {author} {\bibfnamefont
  {N.~A.}\ \bibnamefont {Dovator}},\ }\href
  {https://doi.org/10.1134/S0021364014010111} {\bibfield  {journal} {\bibinfo
  {journal} {JETP Letters}\ }\textbf {\bibinfo {volume} {99}},\ \bibinfo
  {pages} {4} (\bibinfo {year} {2014})}\BibitemShut {NoStop}%
\bibitem [{\citenamefont {Pendlebury}\ \emph {et~al.}(2004)\citenamefont
  {Pendlebury}, \citenamefont {Heil}, \citenamefont {Sobolev}, \citenamefont
  {Harris}, \citenamefont {Richardson}, \citenamefont {Baskin}, \citenamefont
  {Doyle}, \citenamefont {Geltenbort}, \citenamefont {Green}, \citenamefont
  {van~der Grinten}, \citenamefont {Iaydjiev}, \citenamefont {Ivanov},
  \citenamefont {May},\ and\ \citenamefont {Smith}}]{Pendlebury2004}%
  \BibitemOpen
  \bibfield  {author} {\bibinfo {author} {\bibfnamefont {J.~M.}\ \bibnamefont
  {Pendlebury}}, \bibinfo {author} {\bibfnamefont {W.}~\bibnamefont {Heil}},
  \bibinfo {author} {\bibfnamefont {Y.}~\bibnamefont {Sobolev}}, \bibinfo
  {author} {\bibfnamefont {P.~G.}\ \bibnamefont {Harris}}, \bibinfo {author}
  {\bibfnamefont {J.~D.}\ \bibnamefont {Richardson}}, \bibinfo {author}
  {\bibfnamefont {R.~J.}\ \bibnamefont {Baskin}}, \bibinfo {author}
  {\bibfnamefont {D.~D.}\ \bibnamefont {Doyle}}, \bibinfo {author}
  {\bibfnamefont {P.}~\bibnamefont {Geltenbort}}, \bibinfo {author}
  {\bibfnamefont {K.}~\bibnamefont {Green}}, \bibinfo {author} {\bibfnamefont
  {M.~G.~D.}\ \bibnamefont {van~der Grinten}}, \bibinfo {author} {\bibfnamefont
  {P.~S.}\ \bibnamefont {Iaydjiev}}, \bibinfo {author} {\bibfnamefont {S.~N.}\
  \bibnamefont {Ivanov}}, \bibinfo {author} {\bibfnamefont {D.~J.~R.}\
  \bibnamefont {May}},\ and\ \bibinfo {author} {\bibfnamefont {K.~F.}\
  \bibnamefont {Smith}},\ }\href {https://doi.org/10.1103/PhysRevA.70.032102}
  {\bibfield  {journal} {\bibinfo  {journal} {Phys. Rev. A}\ }\textbf {\bibinfo
  {volume} {70}},\ \bibinfo {pages} {032102} (\bibinfo {year}
  {2004})}\BibitemShut {NoStop}%
\bibitem [{\citenamefont {Abel}\ \emph {et~al.}(2020)\citenamefont {Abel},
  \citenamefont {Afach}, \citenamefont {Ayres}, \citenamefont {Baker},
  \citenamefont {Ban}, \citenamefont {Bison}, \citenamefont {Bodek},
  \citenamefont {Bondar}, \citenamefont {Burghoff}, \citenamefont {Chanel},
  \citenamefont {Chowdhuri}, \citenamefont {Chiu}, \citenamefont {Clement},
  \citenamefont {Crawford}, \citenamefont {Daum}, \citenamefont {Emmenegger},
  \citenamefont {Ferraris-Bouchez}, \citenamefont {Fertl}, \citenamefont
  {Flaux}, \citenamefont {Franke}, \citenamefont {Fratangelo}, \citenamefont
  {Geltenbort}, \citenamefont {Green}, \citenamefont {Griffith}, \citenamefont
  {van~der Grinten}, \citenamefont {Gruji{\'{c}}}, \citenamefont {Harris},
  \citenamefont {Hayen}, \citenamefont {Heil}, \citenamefont {Henneck},
  \citenamefont {H{\'{e}}laine}, \citenamefont {Hild}, \citenamefont {Hodge},
  \citenamefont {Horras}, \citenamefont {Iaydjiev}, \citenamefont {Ivanov},
  \citenamefont {Kasprzak}, \citenamefont {Kermaidic}, \citenamefont {Kirch},
  \citenamefont {Knecht}, \citenamefont {Knowles}, \citenamefont {Koch},
  \citenamefont {Koss}, \citenamefont {Komposch}, \citenamefont {Kozela},
  \citenamefont {Kraft}, \citenamefont {Krempel}, \citenamefont
  {Ku{\'{z}}niak}, \citenamefont {Lauss}, \citenamefont {Lefort}, \citenamefont
  {Lemi{\`{e}}re}, \citenamefont {Leredde}, \citenamefont {Mohanmurthy},
  \citenamefont {Mtchedlishvili}, \citenamefont {Musgrave}, \citenamefont
  {Naviliat-Cuncic}, \citenamefont {Pais}, \citenamefont {Piegsa},
  \citenamefont {Pierre}, \citenamefont {Pignol}, \citenamefont {Plonka-Spehr},
  \citenamefont {Prashanth}, \citenamefont {Qu{\'{e}}m{\'{e}}ner},
  \citenamefont {Rawlik}, \citenamefont {Rebreyend}, \citenamefont
  {Rien{\"{a}}cker}, \citenamefont {Ries}, \citenamefont {Roccia},
  \citenamefont {Rogel}, \citenamefont {Rozpedzik}, \citenamefont {Schnabel},
  \citenamefont {Schmidt-Wellenburg}, \citenamefont {Severijns}, \citenamefont
  {Shiers}, \citenamefont {{Tavakoli Dinani}}, \citenamefont {Thorne},
  \citenamefont {Virot}, \citenamefont {Voigt}, \citenamefont {Weis},
  \citenamefont {Wursten}, \citenamefont {Wyszynski}, \citenamefont {Zejma},
  \citenamefont {Zenner},\ and\ \citenamefont {Zsigmond}}]{Abel2020}%
  \BibitemOpen
  \bibfield  {author} {\bibinfo {author} {\bibfnamefont {C.}~\bibnamefont
  {Abel}}, \bibinfo {author} {\bibfnamefont {S.}~\bibnamefont {Afach}},
  \bibinfo {author} {\bibfnamefont {N.~J.}\ \bibnamefont {Ayres}}, \bibinfo
  {author} {\bibfnamefont {C.~A.}\ \bibnamefont {Baker}}, \bibinfo {author}
  {\bibfnamefont {G.}~\bibnamefont {Ban}}, \bibinfo {author} {\bibfnamefont
  {G.}~\bibnamefont {Bison}}, \bibinfo {author} {\bibfnamefont
  {K.}~\bibnamefont {Bodek}}, \bibinfo {author} {\bibfnamefont
  {V.}~\bibnamefont {Bondar}}, \bibinfo {author} {\bibfnamefont
  {M.}~\bibnamefont {Burghoff}}, \bibinfo {author} {\bibfnamefont
  {E.}~\bibnamefont {Chanel}}, \bibinfo {author} {\bibfnamefont
  {Z.}~\bibnamefont {Chowdhuri}}, \bibinfo {author} {\bibfnamefont {P.-J.}\
  \bibnamefont {Chiu}}, \bibinfo {author} {\bibfnamefont {B.}~\bibnamefont
  {Clement}}, \bibinfo {author} {\bibfnamefont {C.~B.}\ \bibnamefont
  {Crawford}}, \bibinfo {author} {\bibfnamefont {M.}~\bibnamefont {Daum}},
  \bibinfo {author} {\bibfnamefont {S.}~\bibnamefont {Emmenegger}}, \bibinfo
  {author} {\bibfnamefont {L.}~\bibnamefont {Ferraris-Bouchez}}, \bibinfo
  {author} {\bibfnamefont {M.}~\bibnamefont {Fertl}}, \bibinfo {author}
  {\bibfnamefont {P.}~\bibnamefont {Flaux}}, \bibinfo {author} {\bibfnamefont
  {B.}~\bibnamefont {Franke}}, \bibinfo {author} {\bibfnamefont
  {A.}~\bibnamefont {Fratangelo}}, \bibinfo {author} {\bibfnamefont
  {P.}~\bibnamefont {Geltenbort}}, \bibinfo {author} {\bibfnamefont
  {K.}~\bibnamefont {Green}}, \bibinfo {author} {\bibfnamefont {W.~C.}\
  \bibnamefont {Griffith}}, \bibinfo {author} {\bibfnamefont {M.}~\bibnamefont
  {van~der Grinten}}, \bibinfo {author} {\bibfnamefont {Z.~D.}\ \bibnamefont
  {Gruji{\'{c}}}}, \bibinfo {author} {\bibfnamefont {P.~G.}\ \bibnamefont
  {Harris}}, \bibinfo {author} {\bibfnamefont {L.}~\bibnamefont {Hayen}},
  \bibinfo {author} {\bibfnamefont {W.}~\bibnamefont {Heil}}, \bibinfo {author}
  {\bibfnamefont {R.}~\bibnamefont {Henneck}}, \bibinfo {author} {\bibfnamefont
  {V.}~\bibnamefont {H{\'{e}}laine}}, \bibinfo {author} {\bibfnamefont
  {N.}~\bibnamefont {Hild}}, \bibinfo {author} {\bibfnamefont {Z.}~\bibnamefont
  {Hodge}}, \bibinfo {author} {\bibfnamefont {M.}~\bibnamefont {Horras}},
  \bibinfo {author} {\bibfnamefont {P.}~\bibnamefont {Iaydjiev}}, \bibinfo
  {author} {\bibfnamefont {S.~N.}\ \bibnamefont {Ivanov}}, \bibinfo {author}
  {\bibfnamefont {M.}~\bibnamefont {Kasprzak}}, \bibinfo {author}
  {\bibfnamefont {Y.}~\bibnamefont {Kermaidic}}, \bibinfo {author}
  {\bibfnamefont {K.}~\bibnamefont {Kirch}}, \bibinfo {author} {\bibfnamefont
  {A.}~\bibnamefont {Knecht}}, \bibinfo {author} {\bibfnamefont
  {P.}~\bibnamefont {Knowles}}, \bibinfo {author} {\bibfnamefont {H.-C.}\
  \bibnamefont {Koch}}, \bibinfo {author} {\bibfnamefont {P.~A.}\ \bibnamefont
  {Koss}}, \bibinfo {author} {\bibfnamefont {S.}~\bibnamefont {Komposch}},
  \bibinfo {author} {\bibfnamefont {A.}~\bibnamefont {Kozela}}, \bibinfo
  {author} {\bibfnamefont {A.}~\bibnamefont {Kraft}}, \bibinfo {author}
  {\bibfnamefont {J.}~\bibnamefont {Krempel}}, \bibinfo {author} {\bibfnamefont
  {M.}~\bibnamefont {Ku{\'{z}}niak}}, \bibinfo {author} {\bibfnamefont
  {B.}~\bibnamefont {Lauss}}, \bibinfo {author} {\bibfnamefont
  {T.}~\bibnamefont {Lefort}}, \bibinfo {author} {\bibfnamefont
  {Y.}~\bibnamefont {Lemi{\`{e}}re}}, \bibinfo {author} {\bibfnamefont
  {A.}~\bibnamefont {Leredde}}, \bibinfo {author} {\bibfnamefont
  {P.}~\bibnamefont {Mohanmurthy}}, \bibinfo {author} {\bibfnamefont
  {A.}~\bibnamefont {Mtchedlishvili}}, \bibinfo {author} {\bibfnamefont
  {M.}~\bibnamefont {Musgrave}}, \bibinfo {author} {\bibfnamefont
  {O.}~\bibnamefont {Naviliat-Cuncic}}, \bibinfo {author} {\bibfnamefont
  {D.}~\bibnamefont {Pais}}, \bibinfo {author} {\bibfnamefont {F.~M.}\
  \bibnamefont {Piegsa}}, \bibinfo {author} {\bibfnamefont {E.}~\bibnamefont
  {Pierre}}, \bibinfo {author} {\bibfnamefont {G.}~\bibnamefont {Pignol}},
  \bibinfo {author} {\bibfnamefont {C.}~\bibnamefont {Plonka-Spehr}}, \bibinfo
  {author} {\bibfnamefont {P.~N.}\ \bibnamefont {Prashanth}}, \bibinfo {author}
  {\bibfnamefont {G.}~\bibnamefont {Qu{\'{e}}m{\'{e}}ner}}, \bibinfo {author}
  {\bibfnamefont {M.}~\bibnamefont {Rawlik}}, \bibinfo {author} {\bibfnamefont
  {D.}~\bibnamefont {Rebreyend}}, \bibinfo {author} {\bibfnamefont
  {I.}~\bibnamefont {Rien{\"{a}}cker}}, \bibinfo {author} {\bibfnamefont
  {D.}~\bibnamefont {Ries}}, \bibinfo {author} {\bibfnamefont {S.}~\bibnamefont
  {Roccia}}, \bibinfo {author} {\bibfnamefont {G.}~\bibnamefont {Rogel}},
  \bibinfo {author} {\bibfnamefont {D.}~\bibnamefont {Rozpedzik}}, \bibinfo
  {author} {\bibfnamefont {A.}~\bibnamefont {Schnabel}}, \bibinfo {author}
  {\bibfnamefont {P.}~\bibnamefont {Schmidt-Wellenburg}}, \bibinfo {author}
  {\bibfnamefont {N.}~\bibnamefont {Severijns}}, \bibinfo {author}
  {\bibfnamefont {D.}~\bibnamefont {Shiers}}, \bibinfo {author} {\bibfnamefont
  {R.}~\bibnamefont {{Tavakoli Dinani}}}, \bibinfo {author} {\bibfnamefont
  {J.~A.}\ \bibnamefont {Thorne}}, \bibinfo {author} {\bibfnamefont
  {R.}~\bibnamefont {Virot}}, \bibinfo {author} {\bibfnamefont
  {J.}~\bibnamefont {Voigt}}, \bibinfo {author} {\bibfnamefont
  {A.}~\bibnamefont {Weis}}, \bibinfo {author} {\bibfnamefont {E.}~\bibnamefont
  {Wursten}}, \bibinfo {author} {\bibfnamefont {G.}~\bibnamefont {Wyszynski}},
  \bibinfo {author} {\bibfnamefont {J.}~\bibnamefont {Zejma}}, \bibinfo
  {author} {\bibfnamefont {J.}~\bibnamefont {Zenner}},\ and\ \bibinfo {author}
  {\bibfnamefont {G.}~\bibnamefont {Zsigmond}},\ }\href
  {https://doi.org/10.1103/PhysRevLett.124.081803} {\bibfield  {journal}
  {\bibinfo  {journal} {Physical Review Letters}\ }\textbf {\bibinfo {volume}
  {124}},\ \bibinfo {pages} {081803} (\bibinfo {year} {2020})}\BibitemShut
  {NoStop}%
\bibitem [{\citenamefont {Pendlebury}(1992)}]{Pendlebury1992}%
  \BibitemOpen
  \bibfield  {author} {\bibinfo {author} {\bibfnamefont {J.}~\bibnamefont
  {Pendlebury}},\ }\href {https://doi.org/10.1016/0375-9474(92)90520-T}
  {\bibfield  {journal} {\bibinfo  {journal} {Nuclear Physics A}\ }\textbf
  {\bibinfo {volume} {546}},\ \bibinfo {pages} {359} (\bibinfo {year}
  {1992})}\BibitemShut {NoStop}%
\bibitem [{\citenamefont {Graner}\ \emph
  {et~al.}(2016{\natexlab{a}})\citenamefont {Graner}, \citenamefont {Chen},
  \citenamefont {Lindahl},\ and\ \citenamefont {Heckel}}]{BGraner2016}%
  \BibitemOpen
  \bibfield  {author} {\bibinfo {author} {\bibfnamefont {B.}~\bibnamefont
  {Graner}}, \bibinfo {author} {\bibfnamefont {Y.}~\bibnamefont {Chen}},
  \bibinfo {author} {\bibfnamefont {E.~G.}\ \bibnamefont {Lindahl}},\ and\
  \bibinfo {author} {\bibfnamefont {B.~R.}\ \bibnamefont {Heckel}},\ }\href
  {https://doi.org/10.1103/PhysRevLett.116.161601} {\bibfield  {journal}
  {\bibinfo  {journal} {Phys. Rev. Lett.}\ }\textbf {\bibinfo {volume} {116}},\
  \bibinfo {pages} {161601} (\bibinfo {year} {2016}{\natexlab{a}})}\BibitemShut
  {NoStop}%
\bibitem [{\citenamefont {Lamoreaux}\ and\ \citenamefont
  {Golub}(2005)}]{Lamoreaux2005}%
  \BibitemOpen
  \bibfield  {author} {\bibinfo {author} {\bibfnamefont {S.~K.}\ \bibnamefont
  {Lamoreaux}}\ and\ \bibinfo {author} {\bibfnamefont {R.}~\bibnamefont
  {Golub}},\ }\href {https://doi.org/10.1103/PhysRevA.71.032104} {\bibfield
  {journal} {\bibinfo  {journal} {Phys. Rev. A}\ }\textbf {\bibinfo {volume}
  {71}},\ \bibinfo {pages} {032104} (\bibinfo {year} {2005})}\BibitemShut
  {NoStop}%
\bibitem [{\citenamefont {Afach}\ \emph {et~al.}(2015)\citenamefont {Afach},
  \citenamefont {Ayres}, \citenamefont {Ban}, \citenamefont {Bison},
  \citenamefont {Bodek}, \citenamefont {Chowdhuri}, \citenamefont {Daum},
  \citenamefont {Fertl}, \citenamefont {Franke}, \citenamefont {Griffith},
  \citenamefont {Gruji\ifmmode~\acute{c}\else \'{c}\fi{}}, \citenamefont
  {Harris}, \citenamefont {Heil}, \citenamefont {H\'elaine}, \citenamefont
  {Kasprzak}, \citenamefont {Kermaidic}, \citenamefont {Kirch}, \citenamefont
  {Knowles}, \citenamefont {Koch}, \citenamefont {Komposch}, \citenamefont
  {Kozela}, \citenamefont {Krempel}, \citenamefont {Lauss}, \citenamefont
  {Lefort}, \citenamefont {Lemi\`ere}, \citenamefont {Mtchedlishvili},
  \citenamefont {Musgrave}, \citenamefont {Naviliat-Cuncic}, \citenamefont
  {Pendlebury}, \citenamefont {Piegsa}, \citenamefont {Pignol}, \citenamefont
  {Plonka-Spehr}, \citenamefont {Prashanth}, \citenamefont {Qu\'em\'ener},
  \citenamefont {Rawlik}, \citenamefont {Rebreyend}, \citenamefont {Ries},
  \citenamefont {Roccia}, \citenamefont {Rozpedzik}, \citenamefont
  {Schmidt-Wellenburg}, \citenamefont {Severijns}, \citenamefont {Thorne},
  \citenamefont {Weis}, \citenamefont {Wursten}, \citenamefont {Wyszynski},
  \citenamefont {Zejma}, \citenamefont {Zenner},\ and\ \citenamefont
  {Zsigmond}}]{PSI3}%
  \BibitemOpen
  \bibfield  {author} {\bibinfo {author} {\bibfnamefont {S.}~\bibnamefont
  {Afach}}, \bibinfo {author} {\bibfnamefont {N.~J.}\ \bibnamefont {Ayres}},
  \bibinfo {author} {\bibfnamefont {G.}~\bibnamefont {Ban}}, \bibinfo {author}
  {\bibfnamefont {G.}~\bibnamefont {Bison}}, \bibinfo {author} {\bibfnamefont
  {K.}~\bibnamefont {Bodek}}, \bibinfo {author} {\bibfnamefont
  {Z.}~\bibnamefont {Chowdhuri}}, \bibinfo {author} {\bibfnamefont
  {M.}~\bibnamefont {Daum}}, \bibinfo {author} {\bibfnamefont {M.}~\bibnamefont
  {Fertl}}, \bibinfo {author} {\bibfnamefont {B.}~\bibnamefont {Franke}},
  \bibinfo {author} {\bibfnamefont {W.~C.}\ \bibnamefont {Griffith}}, \bibinfo
  {author} {\bibfnamefont {Z.~D.}\ \bibnamefont {Gruji\ifmmode~\acute{c}\else
  \'{c}\fi{}}}, \bibinfo {author} {\bibfnamefont {P.~G.}\ \bibnamefont
  {Harris}}, \bibinfo {author} {\bibfnamefont {W.}~\bibnamefont {Heil}},
  \bibinfo {author} {\bibfnamefont {V.}~\bibnamefont {H\'elaine}}, \bibinfo
  {author} {\bibfnamefont {M.}~\bibnamefont {Kasprzak}}, \bibinfo {author}
  {\bibfnamefont {Y.}~\bibnamefont {Kermaidic}}, \bibinfo {author}
  {\bibfnamefont {K.}~\bibnamefont {Kirch}}, \bibinfo {author} {\bibfnamefont
  {P.}~\bibnamefont {Knowles}}, \bibinfo {author} {\bibfnamefont {H.-C.}\
  \bibnamefont {Koch}}, \bibinfo {author} {\bibfnamefont {S.}~\bibnamefont
  {Komposch}}, \bibinfo {author} {\bibfnamefont {A.}~\bibnamefont {Kozela}},
  \bibinfo {author} {\bibfnamefont {J.}~\bibnamefont {Krempel}}, \bibinfo
  {author} {\bibfnamefont {B.}~\bibnamefont {Lauss}}, \bibinfo {author}
  {\bibfnamefont {T.}~\bibnamefont {Lefort}}, \bibinfo {author} {\bibfnamefont
  {Y.}~\bibnamefont {Lemi\`ere}}, \bibinfo {author} {\bibfnamefont
  {A.}~\bibnamefont {Mtchedlishvili}}, \bibinfo {author} {\bibfnamefont
  {M.}~\bibnamefont {Musgrave}}, \bibinfo {author} {\bibfnamefont
  {O.}~\bibnamefont {Naviliat-Cuncic}}, \bibinfo {author} {\bibfnamefont
  {J.~M.}\ \bibnamefont {Pendlebury}}, \bibinfo {author} {\bibfnamefont
  {F.~M.}\ \bibnamefont {Piegsa}}, \bibinfo {author} {\bibfnamefont
  {G.}~\bibnamefont {Pignol}}, \bibinfo {author} {\bibfnamefont
  {C.}~\bibnamefont {Plonka-Spehr}}, \bibinfo {author} {\bibfnamefont {P.~N.}\
  \bibnamefont {Prashanth}}, \bibinfo {author} {\bibfnamefont {G.}~\bibnamefont
  {Qu\'em\'ener}}, \bibinfo {author} {\bibfnamefont {M.}~\bibnamefont
  {Rawlik}}, \bibinfo {author} {\bibfnamefont {D.}~\bibnamefont {Rebreyend}},
  \bibinfo {author} {\bibfnamefont {D.}~\bibnamefont {Ries}}, \bibinfo {author}
  {\bibfnamefont {S.}~\bibnamefont {Roccia}}, \bibinfo {author} {\bibfnamefont
  {D.}~\bibnamefont {Rozpedzik}}, \bibinfo {author} {\bibfnamefont
  {P.}~\bibnamefont {Schmidt-Wellenburg}}, \bibinfo {author} {\bibfnamefont
  {N.}~\bibnamefont {Severijns}}, \bibinfo {author} {\bibfnamefont {J.~A.}\
  \bibnamefont {Thorne}}, \bibinfo {author} {\bibfnamefont {A.}~\bibnamefont
  {Weis}}, \bibinfo {author} {\bibfnamefont {E.}~\bibnamefont {Wursten}},
  \bibinfo {author} {\bibfnamefont {G.}~\bibnamefont {Wyszynski}}, \bibinfo
  {author} {\bibfnamefont {J.}~\bibnamefont {Zejma}}, \bibinfo {author}
  {\bibfnamefont {J.}~\bibnamefont {Zenner}},\ and\ \bibinfo {author}
  {\bibfnamefont {G.}~\bibnamefont {Zsigmond}},\ }\href
  {https://doi.org/10.1103/PhysRevLett.115.162502} {\bibfield  {journal}
  {\bibinfo  {journal} {Phys. Rev. Lett.}\ }\textbf {\bibinfo {volume} {115}},\
  \bibinfo {pages} {162502} (\bibinfo {year} {2015})}\BibitemShut {NoStop}%
\bibitem [{\citenamefont {Pendlebury}\ \emph {et~al.}(2015)\citenamefont
  {Pendlebury}, \citenamefont {Afach}, \citenamefont {Ayres}, \citenamefont
  {Baker}, \citenamefont {Ban}, \citenamefont {Bison}, \citenamefont {Bodek},
  \citenamefont {Burghoff}, \citenamefont {Geltenbort}, \citenamefont {Green},
  \citenamefont {Griffith}, \citenamefont {{Van Der Grinten}}, \citenamefont
  {Gruji{\'{c}}}, \citenamefont {Harris}, \citenamefont {H{\'{e}}laine},
  \citenamefont {Iaydjiev}, \citenamefont {Ivanov}, \citenamefont {Kasprzak},
  \citenamefont {Kermaidic}, \citenamefont {Kirch}, \citenamefont {Koch},
  \citenamefont {Komposch}, \citenamefont {Kozela}, \citenamefont {Krempel},
  \citenamefont {Lauss}, \citenamefont {Lefort}, \citenamefont {Lemi{\`{e}}re},
  \citenamefont {May}, \citenamefont {Musgrave}, \citenamefont
  {Naviliat-Cuncic}, \citenamefont {Piegsa}, \citenamefont {Pignol},
  \citenamefont {Prashanth}, \citenamefont {Qu{\'{e}}m{\'{e}}ner},
  \citenamefont {Rawlik}, \citenamefont {Rebreyend}, \citenamefont
  {Richardson}, \citenamefont {Ries}, \citenamefont {Roccia}, \citenamefont
  {Rozpedzik}, \citenamefont {Schnabel}, \citenamefont {Schmidt-Wellenburg},
  \citenamefont {Severijns}, \citenamefont {Shiers}, \citenamefont {Thorne},
  \citenamefont {Weis}, \citenamefont {Winston}, \citenamefont {Wursten},
  \citenamefont {Zejma},\ and\ \citenamefont {Zsigmond}}]{Pendlebury2015}%
  \BibitemOpen
  \bibfield  {author} {\bibinfo {author} {\bibfnamefont {J.~M.}\ \bibnamefont
  {Pendlebury}}, \bibinfo {author} {\bibfnamefont {S.}~\bibnamefont {Afach}},
  \bibinfo {author} {\bibfnamefont {N.~J.}\ \bibnamefont {Ayres}}, \bibinfo
  {author} {\bibfnamefont {C.~A.}\ \bibnamefont {Baker}}, \bibinfo {author}
  {\bibfnamefont {G.}~\bibnamefont {Ban}}, \bibinfo {author} {\bibfnamefont
  {G.}~\bibnamefont {Bison}}, \bibinfo {author} {\bibfnamefont
  {K.}~\bibnamefont {Bodek}}, \bibinfo {author} {\bibfnamefont
  {M.}~\bibnamefont {Burghoff}}, \bibinfo {author} {\bibfnamefont
  {P.}~\bibnamefont {Geltenbort}}, \bibinfo {author} {\bibfnamefont
  {K.}~\bibnamefont {Green}}, \bibinfo {author} {\bibfnamefont {W.~C.}\
  \bibnamefont {Griffith}}, \bibinfo {author} {\bibfnamefont {M.}~\bibnamefont
  {{Van Der Grinten}}}, \bibinfo {author} {\bibfnamefont {Z.~D.}\ \bibnamefont
  {Gruji{\'{c}}}}, \bibinfo {author} {\bibfnamefont {P.~G.}\ \bibnamefont
  {Harris}}, \bibinfo {author} {\bibfnamefont {V.}~\bibnamefont
  {H{\'{e}}laine}}, \bibinfo {author} {\bibfnamefont {P.}~\bibnamefont
  {Iaydjiev}}, \bibinfo {author} {\bibfnamefont {S.~N.}\ \bibnamefont
  {Ivanov}}, \bibinfo {author} {\bibfnamefont {M.}~\bibnamefont {Kasprzak}},
  \bibinfo {author} {\bibfnamefont {Y.}~\bibnamefont {Kermaidic}}, \bibinfo
  {author} {\bibfnamefont {K.}~\bibnamefont {Kirch}}, \bibinfo {author}
  {\bibfnamefont {H.~C.}\ \bibnamefont {Koch}}, \bibinfo {author}
  {\bibfnamefont {S.}~\bibnamefont {Komposch}}, \bibinfo {author}
  {\bibfnamefont {A.}~\bibnamefont {Kozela}}, \bibinfo {author} {\bibfnamefont
  {J.}~\bibnamefont {Krempel}}, \bibinfo {author} {\bibfnamefont
  {B.}~\bibnamefont {Lauss}}, \bibinfo {author} {\bibfnamefont
  {T.}~\bibnamefont {Lefort}}, \bibinfo {author} {\bibfnamefont
  {Y.}~\bibnamefont {Lemi{\`{e}}re}}, \bibinfo {author} {\bibfnamefont {D.~J.}\
  \bibnamefont {May}}, \bibinfo {author} {\bibfnamefont {M.}~\bibnamefont
  {Musgrave}}, \bibinfo {author} {\bibfnamefont {O.}~\bibnamefont
  {Naviliat-Cuncic}}, \bibinfo {author} {\bibfnamefont {F.~M.}\ \bibnamefont
  {Piegsa}}, \bibinfo {author} {\bibfnamefont {G.}~\bibnamefont {Pignol}},
  \bibinfo {author} {\bibfnamefont {P.~N.}\ \bibnamefont {Prashanth}}, \bibinfo
  {author} {\bibfnamefont {G.}~\bibnamefont {Qu{\'{e}}m{\'{e}}ner}}, \bibinfo
  {author} {\bibfnamefont {M.}~\bibnamefont {Rawlik}}, \bibinfo {author}
  {\bibfnamefont {D.}~\bibnamefont {Rebreyend}}, \bibinfo {author}
  {\bibfnamefont {J.~D.}\ \bibnamefont {Richardson}}, \bibinfo {author}
  {\bibfnamefont {D.}~\bibnamefont {Ries}}, \bibinfo {author} {\bibfnamefont
  {S.}~\bibnamefont {Roccia}}, \bibinfo {author} {\bibfnamefont
  {D.}~\bibnamefont {Rozpedzik}}, \bibinfo {author} {\bibfnamefont
  {A.}~\bibnamefont {Schnabel}}, \bibinfo {author} {\bibfnamefont
  {P.}~\bibnamefont {Schmidt-Wellenburg}}, \bibinfo {author} {\bibfnamefont
  {N.}~\bibnamefont {Severijns}}, \bibinfo {author} {\bibfnamefont
  {D.}~\bibnamefont {Shiers}}, \bibinfo {author} {\bibfnamefont {J.~A.}\
  \bibnamefont {Thorne}}, \bibinfo {author} {\bibfnamefont {A.}~\bibnamefont
  {Weis}}, \bibinfo {author} {\bibfnamefont {O.~J.}\ \bibnamefont {Winston}},
  \bibinfo {author} {\bibfnamefont {E.}~\bibnamefont {Wursten}}, \bibinfo
  {author} {\bibfnamefont {J.}~\bibnamefont {Zejma}},\ and\ \bibinfo {author}
  {\bibfnamefont {G.}~\bibnamefont {Zsigmond}},\ }\href
  {https://doi.org/10.1103/PhysRevD.92.092003} {\bibfield  {journal} {\bibinfo
  {journal} {Physical Review D}\ }\textbf {\bibinfo {volume} {92}},\ \bibinfo
  {pages} {092003} (\bibinfo {year} {2015})},\ \Eprint
  {https://arxiv.org/abs/1509.04411} {arXiv:1509.04411} \BibitemShut {NoStop}%
\bibitem [{\citenamefont {Piegsa}(2013)}]{Piegsa2013}%
  \BibitemOpen
  \bibfield  {author} {\bibinfo {author} {\bibfnamefont {F.~M.}\ \bibnamefont
  {Piegsa}},\ }\href {https://doi.org/10.1103/PhysRevC.88.045502} {\bibfield
  {journal} {\bibinfo  {journal} {Phys. Rev. C}\ }\textbf {\bibinfo {volume}
  {88}},\ \bibinfo {pages} {045502} (\bibinfo {year} {2013})}\BibitemShut
  {NoStop}%
\bibitem [{\citenamefont {Steyerl}\ \emph {et~al.}(1986)\citenamefont
  {Steyerl}, \citenamefont {Nagel}, \citenamefont {Schreiber}, \citenamefont
  {Steinhauser}, \citenamefont {Gähler}, \citenamefont {Gläser},
  \citenamefont {Ageron}, \citenamefont {Astruc}, \citenamefont {Drexel},
  \citenamefont {Gervais},\ and\ \citenamefont {Mampe}}]{Steyerl1986}%
  \BibitemOpen
  \bibfield  {author} {\bibinfo {author} {\bibfnamefont {A.}~\bibnamefont
  {Steyerl}}, \bibinfo {author} {\bibfnamefont {H.}~\bibnamefont {Nagel}},
  \bibinfo {author} {\bibfnamefont {F.-X.}\ \bibnamefont {Schreiber}}, \bibinfo
  {author} {\bibfnamefont {K.-A.}\ \bibnamefont {Steinhauser}}, \bibinfo
  {author} {\bibfnamefont {R.}~\bibnamefont {Gähler}}, \bibinfo {author}
  {\bibfnamefont {W.}~\bibnamefont {Gläser}}, \bibinfo {author} {\bibfnamefont
  {P.}~\bibnamefont {Ageron}}, \bibinfo {author} {\bibfnamefont
  {J.}~\bibnamefont {Astruc}}, \bibinfo {author} {\bibfnamefont
  {W.}~\bibnamefont {Drexel}}, \bibinfo {author} {\bibfnamefont
  {G.}~\bibnamefont {Gervais}},\ and\ \bibinfo {author} {\bibfnamefont
  {W.}~\bibnamefont {Mampe}},\ }\href
  {https://doi.org/10.1016/0375-9601(86)90587-6} {\bibfield  {journal}
  {\bibinfo  {journal} {Physics Letters A}\ }\textbf {\bibinfo {volume}
  {116}},\ \bibinfo {pages} {347} (\bibinfo {year} {1986})}\BibitemShut
  {NoStop}%
\bibitem [{\citenamefont {Serebrov}\ \emph {et~al.}(2015)\citenamefont
  {Serebrov}, \citenamefont {Kolomenskiy}, \citenamefont {Pirozhkov},
  \citenamefont {Krasnoschekova}, \citenamefont {Vassiljev}, \citenamefont
  {Polyushkin}, \citenamefont {Lasakov}, \citenamefont {Murashkin},
  \citenamefont {Solovey}, \citenamefont {Fomin}, \citenamefont {Shoka},
  \citenamefont {Zherebtsov}, \citenamefont {Geltenbort}, \citenamefont
  {Ivanov}, \citenamefont {Zimmer}, \citenamefont {Alexandrov}, \citenamefont
  {Dmitriev},\ and\ \citenamefont {Dovator}}]{Serebrov2015}%
  \BibitemOpen
  \bibfield  {author} {\bibinfo {author} {\bibfnamefont {A.~P.}\ \bibnamefont
  {Serebrov}}, \bibinfo {author} {\bibfnamefont {E.~A.}\ \bibnamefont
  {Kolomenskiy}}, \bibinfo {author} {\bibfnamefont {A.~N.}\ \bibnamefont
  {Pirozhkov}}, \bibinfo {author} {\bibfnamefont {I.~A.}\ \bibnamefont
  {Krasnoschekova}}, \bibinfo {author} {\bibfnamefont {A.~V.}\ \bibnamefont
  {Vassiljev}}, \bibinfo {author} {\bibfnamefont {A.~O.}\ \bibnamefont
  {Polyushkin}}, \bibinfo {author} {\bibfnamefont {M.~S.}\ \bibnamefont
  {Lasakov}}, \bibinfo {author} {\bibfnamefont {A.~N.}\ \bibnamefont
  {Murashkin}}, \bibinfo {author} {\bibfnamefont {V.~A.}\ \bibnamefont
  {Solovey}}, \bibinfo {author} {\bibfnamefont {A.~K.}\ \bibnamefont {Fomin}},
  \bibinfo {author} {\bibfnamefont {I.~V.}\ \bibnamefont {Shoka}}, \bibinfo
  {author} {\bibfnamefont {O.~M.}\ \bibnamefont {Zherebtsov}}, \bibinfo
  {author} {\bibfnamefont {P.}~\bibnamefont {Geltenbort}}, \bibinfo {author}
  {\bibfnamefont {S.~N.}\ \bibnamefont {Ivanov}}, \bibinfo {author}
  {\bibfnamefont {O.}~\bibnamefont {Zimmer}}, \bibinfo {author} {\bibfnamefont
  {E.~B.}\ \bibnamefont {Alexandrov}}, \bibinfo {author} {\bibfnamefont
  {S.~P.}\ \bibnamefont {Dmitriev}},\ and\ \bibinfo {author} {\bibfnamefont
  {N.~A.}\ \bibnamefont {Dovator}},\ }\href
  {https://doi.org/10.1103/PhysRevC.92.055501} {\bibfield  {journal} {\bibinfo
  {journal} {Phys. Rev. C}\ }\textbf {\bibinfo {volume} {92}},\ \bibinfo
  {pages} {055501} (\bibinfo {year} {2015})}\BibitemShut {NoStop}%
\bibitem [{\citenamefont {Golub}\ and\ \citenamefont
  {Pendlebury}(1975)}]{Golub1975}%
  \BibitemOpen
  \bibfield  {author} {\bibinfo {author} {\bibfnamefont {R.}~\bibnamefont
  {Golub}}\ and\ \bibinfo {author} {\bibfnamefont {J.}~\bibnamefont
  {Pendlebury}},\ }\href {https://doi.org/10.1016/0375-9601(75)90500-9}
  {\bibfield  {journal} {\bibinfo  {journal} {Physics Letters A}\ }\textbf
  {\bibinfo {volume} {53}},\ \bibinfo {pages} {133} (\bibinfo {year}
  {1975})}\BibitemShut {NoStop}%
\bibitem [{\citenamefont {Zimmer}(2016)}]{Zimmer2016}%
  \BibitemOpen
  \bibfield  {author} {\bibinfo {author} {\bibfnamefont {O.}~\bibnamefont
  {Zimmer}}\ }(\bibinfo  {publisher} {Physics of fundamental Symmetries and
  Interactions – PSI2016},\ \bibinfo {year} {2016})\BibitemShut {NoStop}%
\bibitem [{\citenamefont {Anghel}\ \emph {et~al.}(2009)\citenamefont {Anghel},
  \citenamefont {Atchison}, \citenamefont {Blau}, \citenamefont {{van den
  Brandt}}, \citenamefont {Daum}, \citenamefont {Doelling}, \citenamefont
  {Dubs}, \citenamefont {Duperrex}, \citenamefont {Fuchs}, \citenamefont
  {George}, \citenamefont {Gültl}, \citenamefont {Hautle}, \citenamefont
  {Heidenreich}, \citenamefont {Heinrich}, \citenamefont {Henneck},
  \citenamefont {Heule}, \citenamefont {Hofmann}, \citenamefont {Joray},
  \citenamefont {Kasprzak}, \citenamefont {Kirch}, \citenamefont {Knecht},
  \citenamefont {Konter}, \citenamefont {Korhonen}, \citenamefont {Kuzniak},
  \citenamefont {Lauss}, \citenamefont {Mezger}, \citenamefont
  {Mtchedlishvili}, \citenamefont {Petzoldt}, \citenamefont {Pichlmaier},
  \citenamefont {Reggiani}, \citenamefont {Reiser}, \citenamefont {Rohrer},
  \citenamefont {Seidel}, \citenamefont {Spitzer}, \citenamefont {Thomsen},
  \citenamefont {Wagner}, \citenamefont {Wohlmuther}, \citenamefont {Zsigmond},
  \citenamefont {Zuellig}, \citenamefont {Bodek}, \citenamefont {Kistryn},
  \citenamefont {Zejma}, \citenamefont {Geltenbort}, \citenamefont {Plonka},\
  and\ \citenamefont {Grigoriev}}]{Anghel2009}%
  \BibitemOpen
  \bibfield  {author} {\bibinfo {author} {\bibfnamefont {A.}~\bibnamefont
  {Anghel}}, \bibinfo {author} {\bibfnamefont {F.}~\bibnamefont {Atchison}},
  \bibinfo {author} {\bibfnamefont {B.}~\bibnamefont {Blau}}, \bibinfo {author}
  {\bibfnamefont {B.}~\bibnamefont {{van den Brandt}}}, \bibinfo {author}
  {\bibfnamefont {M.}~\bibnamefont {Daum}}, \bibinfo {author} {\bibfnamefont
  {R.}~\bibnamefont {Doelling}}, \bibinfo {author} {\bibfnamefont
  {M.}~\bibnamefont {Dubs}}, \bibinfo {author} {\bibfnamefont {P.-A.}\
  \bibnamefont {Duperrex}}, \bibinfo {author} {\bibfnamefont {A.}~\bibnamefont
  {Fuchs}}, \bibinfo {author} {\bibfnamefont {D.}~\bibnamefont {George}},
  \bibinfo {author} {\bibfnamefont {L.}~\bibnamefont {Gültl}}, \bibinfo
  {author} {\bibfnamefont {P.}~\bibnamefont {Hautle}}, \bibinfo {author}
  {\bibfnamefont {G.}~\bibnamefont {Heidenreich}}, \bibinfo {author}
  {\bibfnamefont {F.}~\bibnamefont {Heinrich}}, \bibinfo {author}
  {\bibfnamefont {R.}~\bibnamefont {Henneck}}, \bibinfo {author} {\bibfnamefont
  {S.}~\bibnamefont {Heule}}, \bibinfo {author} {\bibfnamefont
  {T.}~\bibnamefont {Hofmann}}, \bibinfo {author} {\bibfnamefont
  {S.}~\bibnamefont {Joray}}, \bibinfo {author} {\bibfnamefont
  {M.}~\bibnamefont {Kasprzak}}, \bibinfo {author} {\bibfnamefont
  {K.}~\bibnamefont {Kirch}}, \bibinfo {author} {\bibfnamefont
  {A.}~\bibnamefont {Knecht}}, \bibinfo {author} {\bibfnamefont
  {J.}~\bibnamefont {Konter}}, \bibinfo {author} {\bibfnamefont
  {T.}~\bibnamefont {Korhonen}}, \bibinfo {author} {\bibfnamefont
  {M.}~\bibnamefont {Kuzniak}}, \bibinfo {author} {\bibfnamefont
  {B.}~\bibnamefont {Lauss}}, \bibinfo {author} {\bibfnamefont
  {A.}~\bibnamefont {Mezger}}, \bibinfo {author} {\bibfnamefont
  {A.}~\bibnamefont {Mtchedlishvili}}, \bibinfo {author} {\bibfnamefont
  {G.}~\bibnamefont {Petzoldt}}, \bibinfo {author} {\bibfnamefont
  {A.}~\bibnamefont {Pichlmaier}}, \bibinfo {author} {\bibfnamefont
  {D.}~\bibnamefont {Reggiani}}, \bibinfo {author} {\bibfnamefont
  {R.}~\bibnamefont {Reiser}}, \bibinfo {author} {\bibfnamefont
  {U.}~\bibnamefont {Rohrer}}, \bibinfo {author} {\bibfnamefont
  {M.}~\bibnamefont {Seidel}}, \bibinfo {author} {\bibfnamefont
  {H.}~\bibnamefont {Spitzer}}, \bibinfo {author} {\bibfnamefont
  {K.}~\bibnamefont {Thomsen}}, \bibinfo {author} {\bibfnamefont
  {W.}~\bibnamefont {Wagner}}, \bibinfo {author} {\bibfnamefont
  {M.}~\bibnamefont {Wohlmuther}}, \bibinfo {author} {\bibfnamefont
  {G.}~\bibnamefont {Zsigmond}}, \bibinfo {author} {\bibfnamefont
  {J.}~\bibnamefont {Zuellig}}, \bibinfo {author} {\bibfnamefont
  {K.}~\bibnamefont {Bodek}}, \bibinfo {author} {\bibfnamefont
  {S.}~\bibnamefont {Kistryn}}, \bibinfo {author} {\bibfnamefont
  {J.}~\bibnamefont {Zejma}}, \bibinfo {author} {\bibfnamefont
  {P.}~\bibnamefont {Geltenbort}}, \bibinfo {author} {\bibfnamefont
  {C.}~\bibnamefont {Plonka}},\ and\ \bibinfo {author} {\bibfnamefont
  {S.}~\bibnamefont {Grigoriev}},\ }\href
  {https://doi.org/10.1016/j.nima.2009.07.077} {\bibfield  {journal} {\bibinfo
  {journal} {Nuclear Instruments and Methods in Physics Research Section A:
  Accelerators, Spectrometers, Detectors and Associated Equipment}\ }\textbf
  {\bibinfo {volume} {611}},\ \bibinfo {pages} {272} (\bibinfo {year}
  {2009})},\ \bibinfo {note} {particle Physics with Slow Neutrons}\BibitemShut
  {NoStop}%
\bibitem [{\citenamefont {Ito}\ \emph {et~al.}(2018)\citenamefont {Ito},
  \citenamefont {Adamek}, \citenamefont {Callahan}, \citenamefont {Choi},
  \citenamefont {Clayton}, \citenamefont {Cude-Woods}, \citenamefont {Currie},
  \citenamefont {Ding}, \citenamefont {Fellers}, \citenamefont {Geltenbort},
  \citenamefont {Lamoreaux}, \citenamefont {Liu}, \citenamefont {MacDonald},
  \citenamefont {Makela}, \citenamefont {Morris}, \citenamefont {Pattie},
  \citenamefont {Ramsey}, \citenamefont {Salvat}, \citenamefont {Saunders},
  \citenamefont {Sharapov}, \citenamefont {Sjue}, \citenamefont {Sprow},
  \citenamefont {Tang}, \citenamefont {Weaver}, \citenamefont {Wei},\ and\
  \citenamefont {Young}}]{Ito2018}%
  \BibitemOpen
  \bibfield  {author} {\bibinfo {author} {\bibfnamefont {T.~M.}\ \bibnamefont
  {Ito}}, \bibinfo {author} {\bibfnamefont {E.~R.}\ \bibnamefont {Adamek}},
  \bibinfo {author} {\bibfnamefont {N.~B.}\ \bibnamefont {Callahan}}, \bibinfo
  {author} {\bibfnamefont {J.~H.}\ \bibnamefont {Choi}}, \bibinfo {author}
  {\bibfnamefont {S.~M.}\ \bibnamefont {Clayton}}, \bibinfo {author}
  {\bibfnamefont {C.}~\bibnamefont {Cude-Woods}}, \bibinfo {author}
  {\bibfnamefont {S.}~\bibnamefont {Currie}}, \bibinfo {author} {\bibfnamefont
  {X.}~\bibnamefont {Ding}}, \bibinfo {author} {\bibfnamefont {D.~E.}\
  \bibnamefont {Fellers}}, \bibinfo {author} {\bibfnamefont {P.}~\bibnamefont
  {Geltenbort}}, \bibinfo {author} {\bibfnamefont {S.~K.}\ \bibnamefont
  {Lamoreaux}}, \bibinfo {author} {\bibfnamefont {C.-Y.}\ \bibnamefont {Liu}},
  \bibinfo {author} {\bibfnamefont {S.}~\bibnamefont {MacDonald}}, \bibinfo
  {author} {\bibfnamefont {M.}~\bibnamefont {Makela}}, \bibinfo {author}
  {\bibfnamefont {C.~L.}\ \bibnamefont {Morris}}, \bibinfo {author}
  {\bibfnamefont {R.~W.}\ \bibnamefont {Pattie}}, \bibinfo {author}
  {\bibfnamefont {J.~C.}\ \bibnamefont {Ramsey}}, \bibinfo {author}
  {\bibfnamefont {D.~J.}\ \bibnamefont {Salvat}}, \bibinfo {author}
  {\bibfnamefont {A.}~\bibnamefont {Saunders}}, \bibinfo {author}
  {\bibfnamefont {E.~I.}\ \bibnamefont {Sharapov}}, \bibinfo {author}
  {\bibfnamefont {S.}~\bibnamefont {Sjue}}, \bibinfo {author} {\bibfnamefont
  {A.~P.}\ \bibnamefont {Sprow}}, \bibinfo {author} {\bibfnamefont
  {Z.}~\bibnamefont {Tang}}, \bibinfo {author} {\bibfnamefont {H.~L.}\
  \bibnamefont {Weaver}}, \bibinfo {author} {\bibfnamefont {W.}~\bibnamefont
  {Wei}},\ and\ \bibinfo {author} {\bibfnamefont {A.~R.}\ \bibnamefont
  {Young}},\ }\href {https://doi.org/10.1103/PhysRevC.97.012501} {\bibfield
  {journal} {\bibinfo  {journal} {Phys. Rev. C}\ }\textbf {\bibinfo {volume}
  {97}},\ \bibinfo {pages} {012501} (\bibinfo {year} {2018})}\BibitemShut
  {NoStop}%
\bibitem [{\citenamefont {Ahmed}\ \emph
  {et~al.}(2019{\natexlab{a}})\citenamefont {Ahmed}, \citenamefont {Altiere},
  \citenamefont {Andalib}, \citenamefont {Bell}, \citenamefont {Bidinosti},
  \citenamefont {Cudmore}, \citenamefont {Das}, \citenamefont {Davis},
  \citenamefont {Franke}, \citenamefont {Gericke}, \citenamefont {Giampa},
  \citenamefont {Gnyp}, \citenamefont {Hansen-Romu}, \citenamefont {Hatanaka},
  \citenamefont {Hayamizu}, \citenamefont {Jamieson}, \citenamefont {Jones},
  \citenamefont {Kawasaki}, \citenamefont {Kikawa}, \citenamefont {Kitaguchi},
  \citenamefont {Klassen}, \citenamefont {Konaka}, \citenamefont {Korkmaz},
  \citenamefont {Kuchler}, \citenamefont {Lang}, \citenamefont {Lee},
  \citenamefont {Lindner}, \citenamefont {Madison}, \citenamefont {Makida},
  \citenamefont {Mammei}, \citenamefont {Mammei}, \citenamefont {Martin},
  \citenamefont {Matsumiya}, \citenamefont {Miller}, \citenamefont {Mishima},
  \citenamefont {Momose}, \citenamefont {Okamura}, \citenamefont {Page},
  \citenamefont {Picker}, \citenamefont {Pierre}, \citenamefont {Ramsay},
  \citenamefont {Rebenitsch}, \citenamefont {Rehm}, \citenamefont {Schreyer},
  \citenamefont {Shimizu}, \citenamefont {Sidhu}, \citenamefont {Sikora},
  \citenamefont {Smith}, \citenamefont {Tanihata}, \citenamefont
  {Thorsteinson}, \citenamefont {Vanbergen}, \citenamefont {van Oers},\ and\
  \citenamefont {Watanabe}}]{SAhmed2019}%
  \BibitemOpen
  \bibfield  {author} {\bibinfo {author} {\bibfnamefont {S.}~\bibnamefont
  {Ahmed}}, \bibinfo {author} {\bibfnamefont {E.}~\bibnamefont {Altiere}},
  \bibinfo {author} {\bibfnamefont {T.}~\bibnamefont {Andalib}}, \bibinfo
  {author} {\bibfnamefont {B.}~\bibnamefont {Bell}}, \bibinfo {author}
  {\bibfnamefont {C.~P.}\ \bibnamefont {Bidinosti}}, \bibinfo {author}
  {\bibfnamefont {E.}~\bibnamefont {Cudmore}}, \bibinfo {author} {\bibfnamefont
  {M.}~\bibnamefont {Das}}, \bibinfo {author} {\bibfnamefont {C.~A.}\
  \bibnamefont {Davis}}, \bibinfo {author} {\bibfnamefont {B.}~\bibnamefont
  {Franke}}, \bibinfo {author} {\bibfnamefont {M.}~\bibnamefont {Gericke}},
  \bibinfo {author} {\bibfnamefont {P.}~\bibnamefont {Giampa}}, \bibinfo
  {author} {\bibfnamefont {P.}~\bibnamefont {Gnyp}}, \bibinfo {author}
  {\bibfnamefont {S.}~\bibnamefont {Hansen-Romu}}, \bibinfo {author}
  {\bibfnamefont {K.}~\bibnamefont {Hatanaka}}, \bibinfo {author}
  {\bibfnamefont {T.}~\bibnamefont {Hayamizu}}, \bibinfo {author}
  {\bibfnamefont {B.}~\bibnamefont {Jamieson}}, \bibinfo {author}
  {\bibfnamefont {D.}~\bibnamefont {Jones}}, \bibinfo {author} {\bibfnamefont
  {S.}~\bibnamefont {Kawasaki}}, \bibinfo {author} {\bibfnamefont
  {T.}~\bibnamefont {Kikawa}}, \bibinfo {author} {\bibfnamefont
  {M.}~\bibnamefont {Kitaguchi}}, \bibinfo {author} {\bibfnamefont
  {W.}~\bibnamefont {Klassen}}, \bibinfo {author} {\bibfnamefont
  {A.}~\bibnamefont {Konaka}}, \bibinfo {author} {\bibfnamefont
  {E.}~\bibnamefont {Korkmaz}}, \bibinfo {author} {\bibfnamefont
  {F.}~\bibnamefont {Kuchler}}, \bibinfo {author} {\bibfnamefont
  {M.}~\bibnamefont {Lang}}, \bibinfo {author} {\bibfnamefont {L.}~\bibnamefont
  {Lee}}, \bibinfo {author} {\bibfnamefont {T.}~\bibnamefont {Lindner}},
  \bibinfo {author} {\bibfnamefont {K.~W.}\ \bibnamefont {Madison}}, \bibinfo
  {author} {\bibfnamefont {Y.}~\bibnamefont {Makida}}, \bibinfo {author}
  {\bibfnamefont {J.}~\bibnamefont {Mammei}}, \bibinfo {author} {\bibfnamefont
  {R.}~\bibnamefont {Mammei}}, \bibinfo {author} {\bibfnamefont {J.~W.}\
  \bibnamefont {Martin}}, \bibinfo {author} {\bibfnamefont {R.}~\bibnamefont
  {Matsumiya}}, \bibinfo {author} {\bibfnamefont {E.}~\bibnamefont {Miller}},
  \bibinfo {author} {\bibfnamefont {K.}~\bibnamefont {Mishima}}, \bibinfo
  {author} {\bibfnamefont {T.}~\bibnamefont {Momose}}, \bibinfo {author}
  {\bibfnamefont {T.}~\bibnamefont {Okamura}}, \bibinfo {author} {\bibfnamefont
  {S.}~\bibnamefont {Page}}, \bibinfo {author} {\bibfnamefont {R.}~\bibnamefont
  {Picker}}, \bibinfo {author} {\bibfnamefont {E.}~\bibnamefont {Pierre}},
  \bibinfo {author} {\bibfnamefont {W.~D.}\ \bibnamefont {Ramsay}}, \bibinfo
  {author} {\bibfnamefont {L.}~\bibnamefont {Rebenitsch}}, \bibinfo {author}
  {\bibfnamefont {F.}~\bibnamefont {Rehm}}, \bibinfo {author} {\bibfnamefont
  {W.}~\bibnamefont {Schreyer}}, \bibinfo {author} {\bibfnamefont {H.~M.}\
  \bibnamefont {Shimizu}}, \bibinfo {author} {\bibfnamefont {S.}~\bibnamefont
  {Sidhu}}, \bibinfo {author} {\bibfnamefont {A.}~\bibnamefont {Sikora}},
  \bibinfo {author} {\bibfnamefont {J.}~\bibnamefont {Smith}}, \bibinfo
  {author} {\bibfnamefont {I.}~\bibnamefont {Tanihata}}, \bibinfo {author}
  {\bibfnamefont {B.}~\bibnamefont {Thorsteinson}}, \bibinfo {author}
  {\bibfnamefont {S.}~\bibnamefont {Vanbergen}}, \bibinfo {author}
  {\bibfnamefont {W.~T.~H.}\ \bibnamefont {van Oers}},\ and\ \bibinfo {author}
  {\bibfnamefont {Y.~X.}\ \bibnamefont {Watanabe}} (\bibinfo {collaboration}
  {TUCAN Collaboration9}),\ }\href {https://doi.org/10.1103/PhysRevC.99.025503}
  {\bibfield  {journal} {\bibinfo  {journal} {Phys. Rev. C}\ }\textbf {\bibinfo
  {volume} {99}},\ \bibinfo {pages} {025503} (\bibinfo {year}
  {2019}{\natexlab{a}})}\BibitemShut {NoStop}%
\bibitem [{\citenamefont {Korobkina}\ \emph {et~al.}(2014)\citenamefont
  {Korobkina}, \citenamefont {Medlin}, \citenamefont {Wehring}, \citenamefont
  {Hawari}, \citenamefont {Huffman}, \citenamefont {Young}, \citenamefont
  {Beaumont},\ and\ \citenamefont {Palmquist}}]{Korobkina2014}%
  \BibitemOpen
  \bibfield  {author} {\bibinfo {author} {\bibfnamefont {E.}~\bibnamefont
  {Korobkina}}, \bibinfo {author} {\bibfnamefont {G.}~\bibnamefont {Medlin}},
  \bibinfo {author} {\bibfnamefont {B.}~\bibnamefont {Wehring}}, \bibinfo
  {author} {\bibfnamefont {A.}~\bibnamefont {Hawari}}, \bibinfo {author}
  {\bibfnamefont {P.}~\bibnamefont {Huffman}}, \bibinfo {author} {\bibfnamefont
  {A.}~\bibnamefont {Young}}, \bibinfo {author} {\bibfnamefont
  {B.}~\bibnamefont {Beaumont}},\ and\ \bibinfo {author} {\bibfnamefont
  {G.}~\bibnamefont {Palmquist}},\ }\href
  {https://doi.org/10.1016/j.nima.2014.08.016} {\bibfield  {journal} {\bibinfo
  {journal} {Nuclear Instruments and Methods in Physics Research Section A:
  Accelerators, Spectrometers, Detectors and Associated Equipment}\ }\textbf
  {\bibinfo {volume} {767}},\ \bibinfo {pages} {169} (\bibinfo {year}
  {2014})}\BibitemShut {NoStop}%
\bibitem [{\citenamefont {Kahlenberg}\ \emph {et~al.}(2017)\citenamefont
  {Kahlenberg}, \citenamefont {Ries}, \citenamefont {Ross}, \citenamefont
  {Siemensen}, \citenamefont {Beck}, \citenamefont {Geppert}, \citenamefont
  {Heil}, \citenamefont {Hild}, \citenamefont {Karch}, \citenamefont {Karpuk},
  \citenamefont {Kories}, \citenamefont {Kretschmer}, \citenamefont {Lauss},
  \citenamefont {Reich}, \citenamefont {Sobolev},\ and\ \citenamefont
  {Trautmann}}]{Kahlenberg2017}%
  \BibitemOpen
  \bibfield  {author} {\bibinfo {author} {\bibfnamefont {J.}~\bibnamefont
  {Kahlenberg}}, \bibinfo {author} {\bibfnamefont {D.}~\bibnamefont {Ries}},
  \bibinfo {author} {\bibfnamefont {K.~U.}\ \bibnamefont {Ross}}, \bibinfo
  {author} {\bibfnamefont {C.}~\bibnamefont {Siemensen}}, \bibinfo {author}
  {\bibfnamefont {M.}~\bibnamefont {Beck}}, \bibinfo {author} {\bibfnamefont
  {C.}~\bibnamefont {Geppert}}, \bibinfo {author} {\bibfnamefont
  {W.}~\bibnamefont {Heil}}, \bibinfo {author} {\bibfnamefont {N.}~\bibnamefont
  {Hild}}, \bibinfo {author} {\bibfnamefont {J.}~\bibnamefont {Karch}},
  \bibinfo {author} {\bibfnamefont {S.}~\bibnamefont {Karpuk}}, \bibinfo
  {author} {\bibfnamefont {F.}~\bibnamefont {Kories}}, \bibinfo {author}
  {\bibfnamefont {M.}~\bibnamefont {Kretschmer}}, \bibinfo {author}
  {\bibfnamefont {B.}~\bibnamefont {Lauss}}, \bibinfo {author} {\bibfnamefont
  {T.}~\bibnamefont {Reich}}, \bibinfo {author} {\bibfnamefont
  {Y.}~\bibnamefont {Sobolev}},\ and\ \bibinfo {author} {\bibfnamefont
  {N.}~\bibnamefont {Trautmann}},\ }\href
  {https://doi.org/10.1140/epja/i2017-12428-9} {\bibfield  {journal} {\bibinfo
  {journal} {The European Physical Journal A}\ }\textbf {\bibinfo {volume}
  {53}},\ \bibinfo {pages} {226} (\bibinfo {year} {2017})}\BibitemShut
  {NoStop}%
\bibitem [{FRM(2022)}]{FRMII}%
  \BibitemOpen
  \href@noop {} {\bibinfo {title} {{UCN source at the FRM II}}} (\bibinfo
  {year} {2022}),\ \bibinfo {note}
  {\url{https://www.frm2.tum.de/en/frm2/secondary-sources/ultra-cold-source/},
  Last accessed on 2022-03-08}\BibitemShut {NoStop}%
\bibitem [{\citenamefont {Serebrov}\ \emph {et~al.}(2017)\citenamefont
  {Serebrov}, \citenamefont {Lyamkin}, \citenamefont {Fomin}, \citenamefont
  {Prudnikov}, \citenamefont {Samodurov},\ and\ \citenamefont
  {Kanin}}]{Serebrov_2017}%
  \BibitemOpen
  \bibfield  {author} {\bibinfo {author} {\bibfnamefont {A.~P.}\ \bibnamefont
  {Serebrov}}, \bibinfo {author} {\bibfnamefont {V.~A.}\ \bibnamefont
  {Lyamkin}}, \bibinfo {author} {\bibfnamefont {A.~K.}\ \bibnamefont {Fomin}},
  \bibinfo {author} {\bibfnamefont {D.~V.}\ \bibnamefont {Prudnikov}}, \bibinfo
  {author} {\bibfnamefont {O.~Y.}\ \bibnamefont {Samodurov}},\ and\ \bibinfo
  {author} {\bibfnamefont {A.~S.}\ \bibnamefont {Kanin}},\ }\href
  {https://doi.org/10.1088/1742-6596/798/1/012147} {\bibfield  {journal}
  {\bibinfo  {journal} {Journal of Physics: Conference Series}\ }\textbf
  {\bibinfo {volume} {798}},\ \bibinfo {pages} {012147} (\bibinfo {year}
  {2017})}\BibitemShut {NoStop}%
\bibitem [{\citenamefont {Leung}\ \emph {et~al.}(2019)\citenamefont {Leung},
  \citenamefont {Muhrer}, \citenamefont {Hügle}, \citenamefont {Ito},
  \citenamefont {Lutz}, \citenamefont {Makela}, \citenamefont {Morris},
  \citenamefont {Pattie}, \citenamefont {Saunders},\ and\ \citenamefont
  {Young}}]{Leung2019}%
  \BibitemOpen
  \bibfield  {author} {\bibinfo {author} {\bibfnamefont {K.~K.~H.}\
  \bibnamefont {Leung}}, \bibinfo {author} {\bibfnamefont {G.}~\bibnamefont
  {Muhrer}}, \bibinfo {author} {\bibfnamefont {T.}~\bibnamefont {Hügle}},
  \bibinfo {author} {\bibfnamefont {T.~M.}\ \bibnamefont {Ito}}, \bibinfo
  {author} {\bibfnamefont {E.~M.}\ \bibnamefont {Lutz}}, \bibinfo {author}
  {\bibfnamefont {M.}~\bibnamefont {Makela}}, \bibinfo {author} {\bibfnamefont
  {C.~L.}\ \bibnamefont {Morris}}, \bibinfo {author} {\bibfnamefont {R.~W.}\
  \bibnamefont {Pattie}}, \bibinfo {author} {\bibfnamefont {A.}~\bibnamefont
  {Saunders}},\ and\ \bibinfo {author} {\bibfnamefont {A.~R.}\ \bibnamefont
  {Young}},\ }\href {https://doi.org/10.1063/1.5109879} {\bibfield  {journal}
  {\bibinfo  {journal} {Journal of Applied Physics}\ }\textbf {\bibinfo
  {volume} {126}},\ \bibinfo {pages} {224901} (\bibinfo {year} {2019})},\
  \Eprint {https://arxiv.org/abs/https://doi.org/10.1063/1.5109879}
  {https://doi.org/10.1063/1.5109879} \BibitemShut {NoStop}%
\bibitem [{\citenamefont {Shin}\ \emph {et~al.}(2021)\citenamefont {Shin},
  \citenamefont {Snow}, \citenamefont {Baxter}, \citenamefont {Liu},
  \citenamefont {Kim}, \citenamefont {Kim},\ and\ \citenamefont
  {Semertzidis}}]{Shin2021}%
  \BibitemOpen
  \bibfield  {author} {\bibinfo {author} {\bibfnamefont {Y.~C.}\ \bibnamefont
  {Shin}}, \bibinfo {author} {\bibfnamefont {W.~M.}\ \bibnamefont {Snow}},
  \bibinfo {author} {\bibfnamefont {D.~V.}\ \bibnamefont {Baxter}}, \bibinfo
  {author} {\bibfnamefont {C.-Y.}\ \bibnamefont {Liu}}, \bibinfo {author}
  {\bibfnamefont {D.}~\bibnamefont {Kim}}, \bibinfo {author} {\bibfnamefont
  {Y.}~\bibnamefont {Kim}},\ and\ \bibinfo {author} {\bibfnamefont {Y.~K.}\
  \bibnamefont {Semertzidis}},\ }\href
  {https://doi.org/10.1140/epjp/s13360-021-01740-1} {\bibfield  {journal}
  {\bibinfo  {journal} {The European Physical Journal Plus}\ }\textbf {\bibinfo
  {volume} {136}},\ \bibinfo {pages} {882} (\bibinfo {year}
  {2021})}\BibitemShut {NoStop}%
\bibitem [{\citenamefont {Yoshiki}(2003)}]{Yoshiki_2003}%
  \BibitemOpen
  \bibfield  {author} {\bibinfo {author} {\bibfnamefont {H.}~\bibnamefont
  {Yoshiki}},\ }\href {https://doi.org/10.1016/S0010-4655(02)00819-6}
  {\bibfield  {journal} {\bibinfo  {journal} {Computer Physics Communications}\
  }\textbf {\bibinfo {volume} {151}},\ \bibinfo {pages} {141} (\bibinfo {year}
  {2003})}\BibitemShut {NoStop}%
\bibitem [{\citenamefont {Frei}\ \emph {et~al.}(2010)\citenamefont {Frei},
  \citenamefont {Gutsmiedl}, \citenamefont {Morkel}, \citenamefont {Müller},
  \citenamefont {Paul}, \citenamefont {Rols}, \citenamefont {Schober},\ and\
  \citenamefont {Unruh}}]{Frei2010}%
  \BibitemOpen
  \bibfield  {author} {\bibinfo {author} {\bibfnamefont {A.}~\bibnamefont
  {Frei}}, \bibinfo {author} {\bibfnamefont {E.}~\bibnamefont {Gutsmiedl}},
  \bibinfo {author} {\bibfnamefont {C.}~\bibnamefont {Morkel}}, \bibinfo
  {author} {\bibfnamefont {A.~R.}\ \bibnamefont {Müller}}, \bibinfo {author}
  {\bibfnamefont {S.}~\bibnamefont {Paul}}, \bibinfo {author} {\bibfnamefont
  {S.}~\bibnamefont {Rols}}, \bibinfo {author} {\bibfnamefont {H.}~\bibnamefont
  {Schober}},\ and\ \bibinfo {author} {\bibfnamefont {T.}~\bibnamefont
  {Unruh}},\ }\href {https://doi.org/10.1209/0295-5075/92/62001} {\bibfield
  {journal} {\bibinfo  {journal} {{EPL} (Europhysics Letters)}\ }\textbf
  {\bibinfo {volume} {92}},\ \bibinfo {pages} {62001} (\bibinfo {year}
  {2010})}\BibitemShut {NoStop}%
\bibitem [{\citenamefont {Becker}\ \emph {et~al.}(2015)\citenamefont {Becker},
  \citenamefont {Bison}, \citenamefont {Blau}, \citenamefont {Chowdhuri},
  \citenamefont {Eikenberg}, \citenamefont {Fertl}, \citenamefont {Kirch},
  \citenamefont {Lauss}, \citenamefont {Perret}, \citenamefont {Reggiani},
  \citenamefont {Ries}, \citenamefont {Schmidt-Wellenburg}, \citenamefont
  {Talanov}, \citenamefont {Wohlmuther},\ and\ \citenamefont
  {Zsigmond}}]{Becker2015}%
  \BibitemOpen
  \bibfield  {author} {\bibinfo {author} {\bibfnamefont {H.}~\bibnamefont
  {Becker}}, \bibinfo {author} {\bibfnamefont {G.}~\bibnamefont {Bison}},
  \bibinfo {author} {\bibfnamefont {B.}~\bibnamefont {Blau}}, \bibinfo {author}
  {\bibfnamefont {Z.}~\bibnamefont {Chowdhuri}}, \bibinfo {author}
  {\bibfnamefont {J.}~\bibnamefont {Eikenberg}}, \bibinfo {author}
  {\bibfnamefont {M.}~\bibnamefont {Fertl}}, \bibinfo {author} {\bibfnamefont
  {K.}~\bibnamefont {Kirch}}, \bibinfo {author} {\bibfnamefont
  {B.}~\bibnamefont {Lauss}}, \bibinfo {author} {\bibfnamefont
  {G.}~\bibnamefont {Perret}}, \bibinfo {author} {\bibfnamefont
  {D.}~\bibnamefont {Reggiani}}, \bibinfo {author} {\bibfnamefont
  {D.}~\bibnamefont {Ries}}, \bibinfo {author} {\bibfnamefont {P.}~\bibnamefont
  {Schmidt-Wellenburg}}, \bibinfo {author} {\bibfnamefont {V.}~\bibnamefont
  {Talanov}}, \bibinfo {author} {\bibfnamefont {M.}~\bibnamefont
  {Wohlmuther}},\ and\ \bibinfo {author} {\bibfnamefont {G.}~\bibnamefont
  {Zsigmond}},\ }\href {https://doi.org/10.1016/j.nima.2014.12.091} {\bibfield
  {journal} {\bibinfo  {journal} {Nuclear Instruments and Methods in Physics
  Research Section A: Accelerators, Spectrometers, Detectors and Associated
  Equipment}\ }\textbf {\bibinfo {volume} {777}},\ \bibinfo {pages} {20}
  (\bibinfo {year} {2015})}\BibitemShut {NoStop}%
\bibitem [{\citenamefont {Bison}\ \emph {et~al.}(2020)\citenamefont {Bison},
  \citenamefont {Blau}, \citenamefont {Daum}, \citenamefont {Göltl},
  \citenamefont {Henneck}, \citenamefont {Kirch}, \citenamefont {Lauss},
  \citenamefont {Ries}, \citenamefont {Schmidt-Wellenburg},\ and\ \citenamefont
  {Zsigmond}}]{Bison2020}%
  \BibitemOpen
  \bibfield  {author} {\bibinfo {author} {\bibfnamefont {G.}~\bibnamefont
  {Bison}}, \bibinfo {author} {\bibfnamefont {B.}~\bibnamefont {Blau}},
  \bibinfo {author} {\bibfnamefont {M.}~\bibnamefont {Daum}}, \bibinfo {author}
  {\bibfnamefont {L.}~\bibnamefont {Göltl}}, \bibinfo {author} {\bibfnamefont
  {R.}~\bibnamefont {Henneck}}, \bibinfo {author} {\bibfnamefont
  {K.}~\bibnamefont {Kirch}}, \bibinfo {author} {\bibfnamefont
  {B.}~\bibnamefont {Lauss}}, \bibinfo {author} {\bibfnamefont
  {D.}~\bibnamefont {Ries}}, \bibinfo {author} {\bibfnamefont {P.}~\bibnamefont
  {Schmidt-Wellenburg}},\ and\ \bibinfo {author} {\bibfnamefont
  {G.}~\bibnamefont {Zsigmond}},\ }\href
  {https://doi.org/10.1140/epja/s10050-020-00027-w} {\bibfield  {journal}
  {\bibinfo  {journal} {The European Physical Journal A}\ }\textbf {\bibinfo
  {volume} {56}},\ \bibinfo {pages} {33} (\bibinfo {year} {2020})}\BibitemShut
  {NoStop}%
\bibitem [{\citenamefont {Baker}\ \emph {et~al.}(2010)\citenamefont {Baker},
  \citenamefont {Balashov}, \citenamefont {Francis}, \citenamefont {Green},
  \citenamefont {van~der Grinten}, \citenamefont {Iaydjiev}, \citenamefont
  {Ivanov}, \citenamefont {Khazov}, \citenamefont {Tucker}, \citenamefont
  {Wark}, \citenamefont {Davidson}, \citenamefont {Grozier}, \citenamefont
  {Hardiman}, \citenamefont {Harris}, \citenamefont {Karamath}, \citenamefont
  {Katsika}, \citenamefont {Pendlebury}, \citenamefont {Peeters}, \citenamefont
  {Shiers}, \citenamefont {Smith}, \citenamefont {Townsley}, \citenamefont
  {Wardell}, \citenamefont {Clarke}, \citenamefont {Henry}, \citenamefont
  {Kraus}, \citenamefont {McCann}, \citenamefont {Geltenbort},\ and\
  \citenamefont {Yoshiki}}]{Baker2010}%
  \BibitemOpen
  \bibfield  {author} {\bibinfo {author} {\bibfnamefont {C.~A.}\ \bibnamefont
  {Baker}}, \bibinfo {author} {\bibfnamefont {S.~N.}\ \bibnamefont {Balashov}},
  \bibinfo {author} {\bibfnamefont {V.}~\bibnamefont {Francis}}, \bibinfo
  {author} {\bibfnamefont {K.}~\bibnamefont {Green}}, \bibinfo {author}
  {\bibfnamefont {M.~G.~D.}\ \bibnamefont {van~der Grinten}}, \bibinfo {author}
  {\bibfnamefont {P.~S.}\ \bibnamefont {Iaydjiev}}, \bibinfo {author}
  {\bibfnamefont {S.~N.}\ \bibnamefont {Ivanov}}, \bibinfo {author}
  {\bibfnamefont {A.}~\bibnamefont {Khazov}}, \bibinfo {author} {\bibfnamefont
  {M.~A.~H.}\ \bibnamefont {Tucker}}, \bibinfo {author} {\bibfnamefont {D.~L.}\
  \bibnamefont {Wark}}, \bibinfo {author} {\bibfnamefont {A.}~\bibnamefont
  {Davidson}}, \bibinfo {author} {\bibfnamefont {J.~R.}\ \bibnamefont
  {Grozier}}, \bibinfo {author} {\bibfnamefont {M.}~\bibnamefont {Hardiman}},
  \bibinfo {author} {\bibfnamefont {P.~G.}\ \bibnamefont {Harris}}, \bibinfo
  {author} {\bibfnamefont {J.~R.}\ \bibnamefont {Karamath}}, \bibinfo {author}
  {\bibfnamefont {K.}~\bibnamefont {Katsika}}, \bibinfo {author} {\bibfnamefont
  {J.~M.}\ \bibnamefont {Pendlebury}}, \bibinfo {author} {\bibfnamefont
  {S.~J.~M.}\ \bibnamefont {Peeters}}, \bibinfo {author} {\bibfnamefont
  {D.~B.}\ \bibnamefont {Shiers}}, \bibinfo {author} {\bibfnamefont {P.~N.}\
  \bibnamefont {Smith}}, \bibinfo {author} {\bibfnamefont {C.~M.}\ \bibnamefont
  {Townsley}}, \bibinfo {author} {\bibfnamefont {I.}~\bibnamefont {Wardell}},
  \bibinfo {author} {\bibfnamefont {C.}~\bibnamefont {Clarke}}, \bibinfo
  {author} {\bibfnamefont {S.}~\bibnamefont {Henry}}, \bibinfo {author}
  {\bibfnamefont {H.}~\bibnamefont {Kraus}}, \bibinfo {author} {\bibfnamefont
  {M.}~\bibnamefont {McCann}}, \bibinfo {author} {\bibfnamefont
  {P.}~\bibnamefont {Geltenbort}},\ and\ \bibinfo {author} {\bibfnamefont
  {H.}~\bibnamefont {Yoshiki}},\ }\href
  {https://doi.org/10.1088/1742-6596/251/1/012055} {\bibfield  {journal}
  {\bibinfo  {journal} {Journal of Physics: Conference Series}\ }\textbf
  {\bibinfo {volume} {251}},\ \bibinfo {pages} {012055} (\bibinfo {year}
  {2010})}\BibitemShut {NoStop}%
\bibitem [{\citenamefont {Ahmed}\ \emph
  {et~al.}(2019{\natexlab{b}})\citenamefont {Ahmed}, \citenamefont {Alarcon},
  \citenamefont {Aleksandrova}, \citenamefont {Bae{\ss}ler}, \citenamefont
  {Barron-Palos}, \citenamefont {Bartoszek}, \citenamefont {Beck},
  \citenamefont {Behzadipour}, \citenamefont {Berkutov}, \citenamefont
  {Bessuille}, \citenamefont {Blatnik}, \citenamefont {Broering}, \citenamefont
  {Broussard}, \citenamefont {Busch}, \citenamefont {Carr}, \citenamefont
  {Cianciolo}, \citenamefont {Clayton}, \citenamefont {Cooper}, \citenamefont
  {Crawford}, \citenamefont {Currie}, \citenamefont {Daurer}, \citenamefont
  {Dipert}, \citenamefont {Dow}, \citenamefont {Dutta}, \citenamefont
  {Efremenko}, \citenamefont {Erickson}, \citenamefont {Filippone},
  \citenamefont {Fomin}, \citenamefont {Gao}, \citenamefont {Golub},
  \citenamefont {Gould}, \citenamefont {Greene}, \citenamefont {Haase},
  \citenamefont {Hasell}, \citenamefont {Hawari}, \citenamefont {Hayden},
  \citenamefont {Holley}, \citenamefont {Holt}, \citenamefont {Huffman},
  \citenamefont {Ihloff}, \citenamefont {Imam}, \citenamefont {Ito},
  \citenamefont {Karcz}, \citenamefont {Kelsey}, \citenamefont {Kendellen},
  \citenamefont {Kim}, \citenamefont {Korobkina}, \citenamefont {Korsch},
  \citenamefont {Lamoreaux}, \citenamefont {Leggett}, \citenamefont {Leung},
  \citenamefont {Lipman}, \citenamefont {Liu}, \citenamefont {Long},
  \citenamefont {MacDonald}, \citenamefont {Makela}, \citenamefont {Matlashov},
  \citenamefont {Maxwell}, \citenamefont {Mendenhall}, \citenamefont {Meyer},
  \citenamefont {Milner}, \citenamefont {Mueller}, \citenamefont {Nouri},
  \citenamefont {O{\textquotesingle}Shaughnessy}, \citenamefont {Osthelder},
  \citenamefont {Peng}, \citenamefont {Penttila}, \citenamefont {Phan},
  \citenamefont {Plaster}, \citenamefont {Ramsey}, \citenamefont {Rao},
  \citenamefont {Redwine}, \citenamefont {Reid}, \citenamefont {Saftah},
  \citenamefont {Seidel}, \citenamefont {Silvera}, \citenamefont {Slutsky},
  \citenamefont {Smith}, \citenamefont {Snow}, \citenamefont {Sondheim},
  \citenamefont {Sosothikul}, \citenamefont {Stanislaus}, \citenamefont {Sun},
  \citenamefont {Swank}, \citenamefont {Tang}, \citenamefont {Dinani},
  \citenamefont {Tsentalovich}, \citenamefont {Vidal}, \citenamefont {Wei},
  \citenamefont {White}, \citenamefont {Williamson}, \citenamefont {Yang},
  \citenamefont {Yao},\ and\ \citenamefont {Young}}]{Ahmed2019}%
  \BibitemOpen
  \bibfield  {author} {\bibinfo {author} {\bibfnamefont {M.}~\bibnamefont
  {Ahmed}}, \bibinfo {author} {\bibfnamefont {R.}~\bibnamefont {Alarcon}},
  \bibinfo {author} {\bibfnamefont {A.}~\bibnamefont {Aleksandrova}}, \bibinfo
  {author} {\bibfnamefont {S.}~\bibnamefont {Bae{\ss}ler}}, \bibinfo {author}
  {\bibfnamefont {L.}~\bibnamefont {Barron-Palos}}, \bibinfo {author}
  {\bibfnamefont {L.}~\bibnamefont {Bartoszek}}, \bibinfo {author}
  {\bibfnamefont {D.}~\bibnamefont {Beck}}, \bibinfo {author} {\bibfnamefont
  {M.}~\bibnamefont {Behzadipour}}, \bibinfo {author} {\bibfnamefont
  {I.}~\bibnamefont {Berkutov}}, \bibinfo {author} {\bibfnamefont
  {J.}~\bibnamefont {Bessuille}}, \bibinfo {author} {\bibfnamefont
  {M.}~\bibnamefont {Blatnik}}, \bibinfo {author} {\bibfnamefont
  {M.}~\bibnamefont {Broering}}, \bibinfo {author} {\bibfnamefont
  {L.}~\bibnamefont {Broussard}}, \bibinfo {author} {\bibfnamefont
  {M.}~\bibnamefont {Busch}}, \bibinfo {author} {\bibfnamefont
  {R.}~\bibnamefont {Carr}}, \bibinfo {author} {\bibfnamefont {V.}~\bibnamefont
  {Cianciolo}}, \bibinfo {author} {\bibfnamefont {S.}~\bibnamefont {Clayton}},
  \bibinfo {author} {\bibfnamefont {M.}~\bibnamefont {Cooper}}, \bibinfo
  {author} {\bibfnamefont {C.}~\bibnamefont {Crawford}}, \bibinfo {author}
  {\bibfnamefont {S.}~\bibnamefont {Currie}}, \bibinfo {author} {\bibfnamefont
  {C.}~\bibnamefont {Daurer}}, \bibinfo {author} {\bibfnamefont
  {R.}~\bibnamefont {Dipert}}, \bibinfo {author} {\bibfnamefont
  {K.}~\bibnamefont {Dow}}, \bibinfo {author} {\bibfnamefont {D.}~\bibnamefont
  {Dutta}}, \bibinfo {author} {\bibfnamefont {Y.}~\bibnamefont {Efremenko}},
  \bibinfo {author} {\bibfnamefont {C.}~\bibnamefont {Erickson}}, \bibinfo
  {author} {\bibfnamefont {B.}~\bibnamefont {Filippone}}, \bibinfo {author}
  {\bibfnamefont {N.}~\bibnamefont {Fomin}}, \bibinfo {author} {\bibfnamefont
  {H.}~\bibnamefont {Gao}}, \bibinfo {author} {\bibfnamefont {R.}~\bibnamefont
  {Golub}}, \bibinfo {author} {\bibfnamefont {C.}~\bibnamefont {Gould}},
  \bibinfo {author} {\bibfnamefont {G.}~\bibnamefont {Greene}}, \bibinfo
  {author} {\bibfnamefont {D.}~\bibnamefont {Haase}}, \bibinfo {author}
  {\bibfnamefont {D.}~\bibnamefont {Hasell}}, \bibinfo {author} {\bibfnamefont
  {A.}~\bibnamefont {Hawari}}, \bibinfo {author} {\bibfnamefont
  {M.}~\bibnamefont {Hayden}}, \bibinfo {author} {\bibfnamefont
  {A.}~\bibnamefont {Holley}}, \bibinfo {author} {\bibfnamefont
  {R.}~\bibnamefont {Holt}}, \bibinfo {author} {\bibfnamefont {P.}~\bibnamefont
  {Huffman}}, \bibinfo {author} {\bibfnamefont {E.}~\bibnamefont {Ihloff}},
  \bibinfo {author} {\bibfnamefont {S.}~\bibnamefont {Imam}}, \bibinfo {author}
  {\bibfnamefont {T.}~\bibnamefont {Ito}}, \bibinfo {author} {\bibfnamefont
  {M.}~\bibnamefont {Karcz}}, \bibinfo {author} {\bibfnamefont
  {J.}~\bibnamefont {Kelsey}}, \bibinfo {author} {\bibfnamefont
  {D.}~\bibnamefont {Kendellen}}, \bibinfo {author} {\bibfnamefont
  {Y.}~\bibnamefont {Kim}}, \bibinfo {author} {\bibfnamefont {E.}~\bibnamefont
  {Korobkina}}, \bibinfo {author} {\bibfnamefont {W.}~\bibnamefont {Korsch}},
  \bibinfo {author} {\bibfnamefont {S.}~\bibnamefont {Lamoreaux}}, \bibinfo
  {author} {\bibfnamefont {E.}~\bibnamefont {Leggett}}, \bibinfo {author}
  {\bibfnamefont {K.}~\bibnamefont {Leung}}, \bibinfo {author} {\bibfnamefont
  {A.}~\bibnamefont {Lipman}}, \bibinfo {author} {\bibfnamefont
  {C.}~\bibnamefont {Liu}}, \bibinfo {author} {\bibfnamefont {J.}~\bibnamefont
  {Long}}, \bibinfo {author} {\bibfnamefont {S.}~\bibnamefont {MacDonald}},
  \bibinfo {author} {\bibfnamefont {M.}~\bibnamefont {Makela}}, \bibinfo
  {author} {\bibfnamefont {A.}~\bibnamefont {Matlashov}}, \bibinfo {author}
  {\bibfnamefont {J.}~\bibnamefont {Maxwell}}, \bibinfo {author} {\bibfnamefont
  {M.}~\bibnamefont {Mendenhall}}, \bibinfo {author} {\bibfnamefont
  {H.}~\bibnamefont {Meyer}}, \bibinfo {author} {\bibfnamefont
  {R.}~\bibnamefont {Milner}}, \bibinfo {author} {\bibfnamefont
  {P.}~\bibnamefont {Mueller}}, \bibinfo {author} {\bibfnamefont
  {N.}~\bibnamefont {Nouri}}, \bibinfo {author} {\bibfnamefont
  {C.}~\bibnamefont {O{\textquotesingle}Shaughnessy}}, \bibinfo {author}
  {\bibfnamefont {C.}~\bibnamefont {Osthelder}}, \bibinfo {author}
  {\bibfnamefont {J.}~\bibnamefont {Peng}}, \bibinfo {author} {\bibfnamefont
  {S.}~\bibnamefont {Penttila}}, \bibinfo {author} {\bibfnamefont
  {N.}~\bibnamefont {Phan}}, \bibinfo {author} {\bibfnamefont {B.}~\bibnamefont
  {Plaster}}, \bibinfo {author} {\bibfnamefont {J.}~\bibnamefont {Ramsey}},
  \bibinfo {author} {\bibfnamefont {T.}~\bibnamefont {Rao}}, \bibinfo {author}
  {\bibfnamefont {R.}~\bibnamefont {Redwine}}, \bibinfo {author} {\bibfnamefont
  {A.}~\bibnamefont {Reid}}, \bibinfo {author} {\bibfnamefont {A.}~\bibnamefont
  {Saftah}}, \bibinfo {author} {\bibfnamefont {G.}~\bibnamefont {Seidel}},
  \bibinfo {author} {\bibfnamefont {I.}~\bibnamefont {Silvera}}, \bibinfo
  {author} {\bibfnamefont {S.}~\bibnamefont {Slutsky}}, \bibinfo {author}
  {\bibfnamefont {E.}~\bibnamefont {Smith}}, \bibinfo {author} {\bibfnamefont
  {W.}~\bibnamefont {Snow}}, \bibinfo {author} {\bibfnamefont {W.}~\bibnamefont
  {Sondheim}}, \bibinfo {author} {\bibfnamefont {S.}~\bibnamefont
  {Sosothikul}}, \bibinfo {author} {\bibfnamefont {T.}~\bibnamefont
  {Stanislaus}}, \bibinfo {author} {\bibfnamefont {X.}~\bibnamefont {Sun}},
  \bibinfo {author} {\bibfnamefont {C.}~\bibnamefont {Swank}}, \bibinfo
  {author} {\bibfnamefont {Z.}~\bibnamefont {Tang}}, \bibinfo {author}
  {\bibfnamefont {R.~T.}\ \bibnamefont {Dinani}}, \bibinfo {author}
  {\bibfnamefont {E.}~\bibnamefont {Tsentalovich}}, \bibinfo {author}
  {\bibfnamefont {C.}~\bibnamefont {Vidal}}, \bibinfo {author} {\bibfnamefont
  {W.}~\bibnamefont {Wei}}, \bibinfo {author} {\bibfnamefont {C.}~\bibnamefont
  {White}}, \bibinfo {author} {\bibfnamefont {S.}~\bibnamefont {Williamson}},
  \bibinfo {author} {\bibfnamefont {L.}~\bibnamefont {Yang}}, \bibinfo {author}
  {\bibfnamefont {W.}~\bibnamefont {Yao}},\ and\ \bibinfo {author}
  {\bibfnamefont {A.}~\bibnamefont {Young}},\ }\href
  {https://doi.org/10.1088/1748-0221/14/11/p11017} {\bibfield  {journal}
  {\bibinfo  {journal} {Journal of Instrumentation}\ }\textbf {\bibinfo
  {volume} {14}}\bibinfo  {number} { (11)},\ \bibinfo {pages}
  {P11017}}\BibitemShut {NoStop}%
\bibitem [{\citenamefont {Ito}\ \emph {et~al.}(2016)\citenamefont {Ito},
  \citenamefont {Ramsey}, \citenamefont {Yao}, \citenamefont {Beck},
  \citenamefont {Cianciolo}, \citenamefont {Clayton}, \citenamefont {Crawford},
  \citenamefont {Currie}, \citenamefont {Filippone}, \citenamefont {Griffith},
  \citenamefont {Makela}, \citenamefont {Schmid}, \citenamefont {Seidel},
  \citenamefont {Tang}, \citenamefont {Wagner}, \citenamefont {Wei},\ and\
  \citenamefont {Williamson}}]{Ito2016}%
  \BibitemOpen
\bibfield  {number} {  }\bibfield  {author} {\bibinfo {author} {\bibfnamefont
  {T.~M.}\ \bibnamefont {Ito}}, \bibinfo {author} {\bibfnamefont {J.~C.}\
  \bibnamefont {Ramsey}}, \bibinfo {author} {\bibfnamefont {W.}~\bibnamefont
  {Yao}}, \bibinfo {author} {\bibfnamefont {D.~H.}\ \bibnamefont {Beck}},
  \bibinfo {author} {\bibfnamefont {V.}~\bibnamefont {Cianciolo}}, \bibinfo
  {author} {\bibfnamefont {S.~M.}\ \bibnamefont {Clayton}}, \bibinfo {author}
  {\bibfnamefont {C.}~\bibnamefont {Crawford}}, \bibinfo {author}
  {\bibfnamefont {S.~A.}\ \bibnamefont {Currie}}, \bibinfo {author}
  {\bibfnamefont {B.~W.}\ \bibnamefont {Filippone}}, \bibinfo {author}
  {\bibfnamefont {W.~C.}\ \bibnamefont {Griffith}}, \bibinfo {author}
  {\bibfnamefont {M.}~\bibnamefont {Makela}}, \bibinfo {author} {\bibfnamefont
  {R.}~\bibnamefont {Schmid}}, \bibinfo {author} {\bibfnamefont {G.~M.}\
  \bibnamefont {Seidel}}, \bibinfo {author} {\bibfnamefont {Z.}~\bibnamefont
  {Tang}}, \bibinfo {author} {\bibfnamefont {D.}~\bibnamefont {Wagner}},
  \bibinfo {author} {\bibfnamefont {W.}~\bibnamefont {Wei}},\ and\ \bibinfo
  {author} {\bibfnamefont {S.~E.}\ \bibnamefont {Williamson}},\ }\href
  {https://doi.org/10.1063/1.4946896} {\bibfield  {journal} {\bibinfo
  {journal} {Review of Scientific Instruments}\ }\textbf {\bibinfo {volume}
  {87}},\ \bibinfo {pages} {045113} (\bibinfo {year} {2016})},\ \Eprint
  {https://arxiv.org/abs/https://doi.org/10.1063/1.4946896}
  {https://doi.org/10.1063/1.4946896} \BibitemShut {NoStop}%
\bibitem [{\citenamefont {Phan}\ \emph {et~al.}(2021)\citenamefont {Phan},
  \citenamefont {Wei}, \citenamefont {Beaumont}, \citenamefont {Bouman},
  \citenamefont {Clayton}, \citenamefont {Currie}, \citenamefont {Ito},
  \citenamefont {Ramsey},\ and\ \citenamefont {Seidel}}]{Phan2021}%
  \BibitemOpen
  \bibfield  {author} {\bibinfo {author} {\bibfnamefont {N.~S.}\ \bibnamefont
  {Phan}}, \bibinfo {author} {\bibfnamefont {W.}~\bibnamefont {Wei}}, \bibinfo
  {author} {\bibfnamefont {B.}~\bibnamefont {Beaumont}}, \bibinfo {author}
  {\bibfnamefont {N.}~\bibnamefont {Bouman}}, \bibinfo {author} {\bibfnamefont
  {S.~M.}\ \bibnamefont {Clayton}}, \bibinfo {author} {\bibfnamefont {S.~A.}\
  \bibnamefont {Currie}}, \bibinfo {author} {\bibfnamefont {T.~M.}\
  \bibnamefont {Ito}}, \bibinfo {author} {\bibfnamefont {J.~C.}\ \bibnamefont
  {Ramsey}},\ and\ \bibinfo {author} {\bibfnamefont {G.~M.}\ \bibnamefont
  {Seidel}},\ }\href {https://doi.org/10.1063/5.0037888} {\bibfield  {journal}
  {\bibinfo  {journal} {Journal of Applied Physics}\ }\textbf {\bibinfo
  {volume} {129}},\ \bibinfo {pages} {083301} (\bibinfo {year} {2021})},\
  \Eprint {https://arxiv.org/abs/https://doi.org/10.1063/5.0037888}
  {https://doi.org/10.1063/5.0037888} \BibitemShut {NoStop}%
\bibitem [{\citenamefont {Chanel}\ \emph {et~al.}(2019)\citenamefont {Chanel},
  \citenamefont {Hodge}, \citenamefont {Ries}, \citenamefont {Schulthess},
  \citenamefont {Solar}, \citenamefont {Soldner}, \citenamefont {Stalder},
  \citenamefont {Thorne},\ and\ \citenamefont {Piegsa}}]{Piegsa2019}%
  \BibitemOpen
  \bibfield  {author} {\bibinfo {author} {\bibfnamefont {E.}~\bibnamefont
  {Chanel}}, \bibinfo {author} {\bibfnamefont {Z.}~\bibnamefont {Hodge}},
  \bibinfo {author} {\bibfnamefont {D.}~\bibnamefont {Ries}}, \bibinfo {author}
  {\bibfnamefont {I.}~\bibnamefont {Schulthess}}, \bibinfo {author}
  {\bibfnamefont {M.}~\bibnamefont {Solar}}, \bibinfo {author} {\bibfnamefont
  {T.}~\bibnamefont {Soldner}}, \bibinfo {author} {\bibfnamefont
  {O.}~\bibnamefont {Stalder}}, \bibinfo {author} {\bibfnamefont
  {J.}~\bibnamefont {Thorne}},\ and\ \bibinfo {author} {\bibfnamefont {F.~M.}\
  \bibnamefont {Piegsa}},\ }\href
  {https://doi.org/10.1051/epjconf/201921902004} {\bibfield  {journal}
  {\bibinfo  {journal} {EPJ Web of Conferences}\ }\textbf {\bibinfo {volume}
  {219}},\ \bibinfo {pages} {02004} (\bibinfo {year} {2019})}\BibitemShut
  {NoStop}%
\bibitem [{\citenamefont {Commins}(1991)}]{Commins1991}%
  \BibitemOpen
  \bibfield  {author} {\bibinfo {author} {\bibfnamefont {E.~D.}\ \bibnamefont
  {Commins}},\ }\href {https://doi.org/10.1119/1.16616} {\bibfield  {journal}
  {\bibinfo  {journal} {American Journal of Physics}\ }\textbf {\bibinfo
  {volume} {59}},\ \bibinfo {pages} {1077} (\bibinfo {year}
  {1991})}\BibitemShut {NoStop}%
\bibitem [{\citenamefont {Regan}\ \emph {et~al.}(2002)\citenamefont {Regan},
  \citenamefont {Commins}, \citenamefont {Schmidt},\ and\ \citenamefont
  {DeMille}}]{Regan2002}%
  \BibitemOpen
  \bibfield  {author} {\bibinfo {author} {\bibfnamefont {B.}~\bibnamefont
  {Regan}}, \bibinfo {author} {\bibfnamefont {E.}~\bibnamefont {Commins}},
  \bibinfo {author} {\bibfnamefont {C.}~\bibnamefont {Schmidt}},\ and\ \bibinfo
  {author} {\bibfnamefont {D.}~\bibnamefont {DeMille}},\ }\href
  {https://doi.org/10.1103/PhysRevLett.88.071805} {\bibfield  {journal}
  {\bibinfo  {journal} {Physical Review Letters}\ }\textbf {\bibinfo {volume}
  {88}},\ \bibinfo {pages} {071805} (\bibinfo {year} {2002})}\BibitemShut
  {NoStop}%
\bibitem [{\citenamefont {Barabanov}\ \emph {et~al.}(2006)\citenamefont
  {Barabanov}, \citenamefont {Golub},\ and\ \citenamefont
  {Lamoreaux}}]{Barabanov2006}%
  \BibitemOpen
  \bibfield  {author} {\bibinfo {author} {\bibfnamefont {A.~L.}\ \bibnamefont
  {Barabanov}}, \bibinfo {author} {\bibfnamefont {R.}~\bibnamefont {Golub}},\
  and\ \bibinfo {author} {\bibfnamefont {S.~K.}\ \bibnamefont {Lamoreaux}},\
  }\href {https://doi.org/10.1103/PhysRevA.74.052115} {\bibfield  {journal}
  {\bibinfo  {journal} {Phys. Rev. A}\ }\textbf {\bibinfo {volume} {74}},\
  \bibinfo {pages} {052115} (\bibinfo {year} {2006})}\BibitemShut {NoStop}%
\bibitem [{\citenamefont {Harris}\ and\ \citenamefont
  {Pendlebury}(2006)}]{Harris2006}%
  \BibitemOpen
  \bibfield  {author} {\bibinfo {author} {\bibfnamefont {P.~G.}\ \bibnamefont
  {Harris}}\ and\ \bibinfo {author} {\bibfnamefont {J.~M.}\ \bibnamefont
  {Pendlebury}},\ }\href {https://doi.org/10.1103/PhysRevA.73.014101}
  {\bibfield  {journal} {\bibinfo  {journal} {Phys. Rev. A}\ }\textbf {\bibinfo
  {volume} {73}},\ \bibinfo {pages} {014101} (\bibinfo {year}
  {2006})}\BibitemShut {NoStop}%
\bibitem [{\citenamefont {Clayton}(2011)}]{Clayton2011}%
  \BibitemOpen
  \bibfield  {author} {\bibinfo {author} {\bibfnamefont {S.~M.}\ \bibnamefont
  {Clayton}},\ }\href {https://doi.org/10.1016/j.jmr.2011.04.008} {\bibfield
  {journal} {\bibinfo  {journal} {Journal of Magnetic Resonance}\ }\textbf
  {\bibinfo {volume} {211}},\ \bibinfo {pages} {89} (\bibinfo {year}
  {2011})}\BibitemShut {NoStop}%
\bibitem [{\citenamefont {Pignol}\ and\ \citenamefont
  {Roccia}(2012)}]{Pignol2012}%
  \BibitemOpen
  \bibfield  {author} {\bibinfo {author} {\bibfnamefont {G.}~\bibnamefont
  {Pignol}}\ and\ \bibinfo {author} {\bibfnamefont {S.}~\bibnamefont
  {Roccia}},\ }\href {https://doi.org/10.1103/PhysRevA.85.042105} {\bibfield
  {journal} {\bibinfo  {journal} {Phys. Rev. A}\ }\textbf {\bibinfo {volume}
  {85}},\ \bibinfo {pages} {042105} (\bibinfo {year} {2012})}\BibitemShut
  {NoStop}%
\bibitem [{\citenamefont {Pignol}\ \emph
  {et~al.}(2015{\natexlab{a}})\citenamefont {Pignol}, \citenamefont {Guigue},
  \citenamefont {Petukhov},\ and\ \citenamefont {Golub}}]{Pignol2015}%
  \BibitemOpen
  \bibfield  {author} {\bibinfo {author} {\bibfnamefont {G.}~\bibnamefont
  {Pignol}}, \bibinfo {author} {\bibfnamefont {M.}~\bibnamefont {Guigue}},
  \bibinfo {author} {\bibfnamefont {A.}~\bibnamefont {Petukhov}},\ and\
  \bibinfo {author} {\bibfnamefont {R.}~\bibnamefont {Golub}},\ }\href
  {https://doi.org/10.1103/PhysRevA.92.053407} {\bibfield  {journal} {\bibinfo
  {journal} {Phys. Rev. A}\ }\textbf {\bibinfo {volume} {92}},\ \bibinfo
  {pages} {053407} (\bibinfo {year} {2015}{\natexlab{a}})}\BibitemShut
  {NoStop}%
\bibitem [{\citenamefont {Swank}\ \emph {et~al.}(2016)\citenamefont {Swank},
  \citenamefont {Petukhov},\ and\ \citenamefont {Golub}}]{Swank2016}%
  \BibitemOpen
  \bibfield  {author} {\bibinfo {author} {\bibfnamefont {C.~M.}\ \bibnamefont
  {Swank}}, \bibinfo {author} {\bibfnamefont {A.~K.}\ \bibnamefont
  {Petukhov}},\ and\ \bibinfo {author} {\bibfnamefont {R.}~\bibnamefont
  {Golub}},\ }\href {https://doi.org/10.1103/PhysRevA.93.062703} {\bibfield
  {journal} {\bibinfo  {journal} {Phys. Rev. A}\ }\textbf {\bibinfo {volume}
  {93}},\ \bibinfo {pages} {062703} (\bibinfo {year} {2016})}\BibitemShut
  {NoStop}%
\bibitem [{\citenamefont {Pignol}(2019)}]{Pignol2019}%
  \BibitemOpen
  \bibfield  {author} {\bibinfo {author} {\bibfnamefont {G.}~\bibnamefont
  {Pignol}},\ }\href {https://doi.org/10.1016/j.physletb.2019.05.014}
  {\bibfield  {journal} {\bibinfo  {journal} {Physics Letters B}\ }\textbf
  {\bibinfo {volume} {793}},\ \bibinfo {pages} {440} (\bibinfo {year}
  {2019})}\BibitemShut {NoStop}%
\bibitem [{\citenamefont {Ayres}\ \emph {et~al.}(2021)\citenamefont {Ayres},
  \citenamefont {Ban}, \citenamefont {Bienstman}, \citenamefont {Bison},
  \citenamefont {Bodek}, \citenamefont {Bondar}, \citenamefont {Bouillaud},
  \citenamefont {Chanel}, \citenamefont {Chen}, \citenamefont {Chiu},
  \citenamefont {Clément}, \citenamefont {Crawford}, \citenamefont {Daum},
  \citenamefont {Dechenaux}, \citenamefont {Doorenbos}, \citenamefont
  {Emmenegger}, \citenamefont {Ferraris-Bouchez}, \citenamefont {Fertl},
  \citenamefont {Fratangelo}, \citenamefont {Flaux}, \citenamefont
  {Goupillière}, \citenamefont {Griffith}, \citenamefont {Grujic},
  \citenamefont {Harris}, \citenamefont {Kirch}, \citenamefont {Koss},
  \citenamefont {Krempel}, \citenamefont {Lauss}, \citenamefont {Lefort},
  \citenamefont {Lemière}, \citenamefont {Leredde}, \citenamefont {Meier},
  \citenamefont {Menu}, \citenamefont {Mullins}, \citenamefont
  {Naviliat-Cuncic}, \citenamefont {Pais}, \citenamefont {Piegsa},
  \citenamefont {Pignol}, \citenamefont {Quéméner}, \citenamefont {Rawlik},
  \citenamefont {Rebreyend}, \citenamefont {Rienäcker}, \citenamefont {Ries},
  \citenamefont {Roccia}, \citenamefont {Ross}, \citenamefont {Rozpedzik},
  \citenamefont {Saenz}, \citenamefont {Schmidt-Wellenburg}, \citenamefont
  {Schnabel}, \citenamefont {Severijns}, \citenamefont {Shen}, \citenamefont
  {Stapf}, \citenamefont {Svirina}, \citenamefont {Tavakoli~Dinani},
  \citenamefont {Touati}, \citenamefont {Thorne}, \citenamefont {Virot},
  \citenamefont {Voigt}, \citenamefont {Wursten}, \citenamefont {Yazdandoost},
  \citenamefont {Zejma},\ and\ \citenamefont {Zsigmond}}]{PSI_apparatus_2021}%
  \BibitemOpen
  \bibfield  {author} {\bibinfo {author} {\bibfnamefont {N.~J.}\ \bibnamefont
  {Ayres}}, \bibinfo {author} {\bibfnamefont {G.}~\bibnamefont {Ban}}, \bibinfo
  {author} {\bibfnamefont {L.}~\bibnamefont {Bienstman}}, \bibinfo {author}
  {\bibfnamefont {G.}~\bibnamefont {Bison}}, \bibinfo {author} {\bibfnamefont
  {K.}~\bibnamefont {Bodek}}, \bibinfo {author} {\bibfnamefont
  {V.}~\bibnamefont {Bondar}}, \bibinfo {author} {\bibfnamefont
  {T.}~\bibnamefont {Bouillaud}}, \bibinfo {author} {\bibfnamefont
  {E.}~\bibnamefont {Chanel}}, \bibinfo {author} {\bibfnamefont
  {J.}~\bibnamefont {Chen}}, \bibinfo {author} {\bibfnamefont {P.~J.}\
  \bibnamefont {Chiu}}, \bibinfo {author} {\bibfnamefont {B.}~\bibnamefont
  {Clément}}, \bibinfo {author} {\bibfnamefont {C.~B.}\ \bibnamefont
  {Crawford}}, \bibinfo {author} {\bibfnamefont {M.}~\bibnamefont {Daum}},
  \bibinfo {author} {\bibfnamefont {B.}~\bibnamefont {Dechenaux}}, \bibinfo
  {author} {\bibfnamefont {C.~B.}\ \bibnamefont {Doorenbos}}, \bibinfo {author}
  {\bibfnamefont {S.}~\bibnamefont {Emmenegger}}, \bibinfo {author}
  {\bibfnamefont {L.}~\bibnamefont {Ferraris-Bouchez}}, \bibinfo {author}
  {\bibfnamefont {M.}~\bibnamefont {Fertl}}, \bibinfo {author} {\bibfnamefont
  {A.}~\bibnamefont {Fratangelo}}, \bibinfo {author} {\bibfnamefont
  {P.}~\bibnamefont {Flaux}}, \bibinfo {author} {\bibfnamefont
  {D.}~\bibnamefont {Goupillière}}, \bibinfo {author} {\bibfnamefont {W.~C.}\
  \bibnamefont {Griffith}}, \bibinfo {author} {\bibfnamefont {Z.~D.}\
  \bibnamefont {Grujic}}, \bibinfo {author} {\bibfnamefont {P.~G.}\
  \bibnamefont {Harris}}, \bibinfo {author} {\bibfnamefont {K.}~\bibnamefont
  {Kirch}}, \bibinfo {author} {\bibfnamefont {P.~A.}\ \bibnamefont {Koss}},
  \bibinfo {author} {\bibfnamefont {J.}~\bibnamefont {Krempel}}, \bibinfo
  {author} {\bibfnamefont {B.}~\bibnamefont {Lauss}}, \bibinfo {author}
  {\bibfnamefont {T.}~\bibnamefont {Lefort}}, \bibinfo {author} {\bibfnamefont
  {Y.}~\bibnamefont {Lemière}}, \bibinfo {author} {\bibfnamefont
  {A.}~\bibnamefont {Leredde}}, \bibinfo {author} {\bibfnamefont
  {M.}~\bibnamefont {Meier}}, \bibinfo {author} {\bibfnamefont
  {J.}~\bibnamefont {Menu}}, \bibinfo {author} {\bibfnamefont {D.~A.}\
  \bibnamefont {Mullins}}, \bibinfo {author} {\bibfnamefont {O.}~\bibnamefont
  {Naviliat-Cuncic}}, \bibinfo {author} {\bibfnamefont {D.}~\bibnamefont
  {Pais}}, \bibinfo {author} {\bibfnamefont {F.~M.}\ \bibnamefont {Piegsa}},
  \bibinfo {author} {\bibfnamefont {G.}~\bibnamefont {Pignol}}, \bibinfo
  {author} {\bibfnamefont {G.}~\bibnamefont {Quéméner}}, \bibinfo {author}
  {\bibfnamefont {M.}~\bibnamefont {Rawlik}}, \bibinfo {author} {\bibfnamefont
  {D.}~\bibnamefont {Rebreyend}}, \bibinfo {author} {\bibfnamefont
  {I.}~\bibnamefont {Rienäcker}}, \bibinfo {author} {\bibfnamefont
  {D.}~\bibnamefont {Ries}}, \bibinfo {author} {\bibfnamefont {S.}~\bibnamefont
  {Roccia}}, \bibinfo {author} {\bibfnamefont {K.~U.}\ \bibnamefont {Ross}},
  \bibinfo {author} {\bibfnamefont {D.}~\bibnamefont {Rozpedzik}}, \bibinfo
  {author} {\bibfnamefont {W.}~\bibnamefont {Saenz}}, \bibinfo {author}
  {\bibfnamefont {P.}~\bibnamefont {Schmidt-Wellenburg}}, \bibinfo {author}
  {\bibfnamefont {A.}~\bibnamefont {Schnabel}}, \bibinfo {author}
  {\bibfnamefont {N.}~\bibnamefont {Severijns}}, \bibinfo {author}
  {\bibfnamefont {B.}~\bibnamefont {Shen}}, \bibinfo {author} {\bibfnamefont
  {T.}~\bibnamefont {Stapf}}, \bibinfo {author} {\bibfnamefont
  {K.}~\bibnamefont {Svirina}}, \bibinfo {author} {\bibfnamefont
  {R.}~\bibnamefont {Tavakoli~Dinani}}, \bibinfo {author} {\bibfnamefont
  {S.}~\bibnamefont {Touati}}, \bibinfo {author} {\bibfnamefont
  {J.}~\bibnamefont {Thorne}}, \bibinfo {author} {\bibfnamefont
  {R.}~\bibnamefont {Virot}}, \bibinfo {author} {\bibfnamefont
  {J.}~\bibnamefont {Voigt}}, \bibinfo {author} {\bibfnamefont
  {E.}~\bibnamefont {Wursten}}, \bibinfo {author} {\bibfnamefont
  {N.}~\bibnamefont {Yazdandoost}}, \bibinfo {author} {\bibfnamefont
  {J.}~\bibnamefont {Zejma}},\ and\ \bibinfo {author} {\bibfnamefont
  {G.}~\bibnamefont {Zsigmond}},\ }\href
  {https://doi.org/10.1140/epjc/s10052-021-09298-z} {\bibfield  {journal}
  {\bibinfo  {journal} {The European Physical Journal C}\ }\textbf {\bibinfo
  {volume} {81}},\ \bibinfo {pages} {512} (\bibinfo {year} {2021})}\BibitemShut
  {NoStop}%
\bibitem [{\citenamefont {Wurm}\ \emph {et~al.}(2019)\citenamefont {Wurm} \emph
  {et~al.}}]{Wurm2019}%
  \BibitemOpen
  \bibfield  {author} {\bibinfo {author} {\bibfnamefont {D.}~\bibnamefont
  {Wurm}} \emph {et~al.},\ }\href
  {https://doi.org/10.1051/epjconf/201921902006} {\bibfield  {journal}
  {\bibinfo  {journal} {EPJ Web Conf.}\ }\textbf {\bibinfo {volume} {219}},\
  \bibinfo {pages} {02006} (\bibinfo {year} {2019})},\ \Eprint
  {https://arxiv.org/abs/1911.09161} {arXiv:1911.09161 [physics.ins-det]}
  \BibitemShut {NoStop}%
\bibitem [{\citenamefont {Martin}(2020)}]{Martin2020}%
  \BibitemOpen
  \bibfield  {author} {\bibinfo {author} {\bibfnamefont {J.}~\bibnamefont
  {Martin}},\ }\href {https://doi.org/10.1088/1742-6596/1643/1/012002}
  {\bibfield  {journal} {\bibinfo  {journal} {Journal of Physics: Conference
  Series}\ }\textbf {\bibinfo {volume} {1643}},\ \bibinfo {pages} {012002}
  (\bibinfo {year} {2020})}\BibitemShut {NoStop}%
\bibitem [{\citenamefont {Abel}\ \emph {et~al.}(2019)\citenamefont {Abel},
  \citenamefont {Ayres}, \citenamefont {Baker}, \citenamefont {Ban},
  \citenamefont {Bison}, \citenamefont {Bodek}, \citenamefont {Bondar},
  \citenamefont {Crawford}, \citenamefont {Chiu}, \citenamefont {Chanel},
  \citenamefont {Chowdhuri}, \citenamefont {Daum}, \citenamefont {Dechenaux},
  \citenamefont {Emmenegger}, \citenamefont {Ferraris-Bouchez}, \citenamefont
  {Flaux}, \citenamefont {Geltenbort}, \citenamefont {Green}, \citenamefont
  {Griffith}, \citenamefont {van~der Grinten}, \citenamefont {Harris},
  \citenamefont {Henneck}, \citenamefont {Hild}, \citenamefont {Iaydjiev},
  \citenamefont {Ivanov}, \citenamefont {Kasprzak}, \citenamefont {Kermaidic},
  \citenamefont {Kirch}, \citenamefont {Koch}, \citenamefont {Komposch},
  \citenamefont {Koss}, \citenamefont {Kozela}, \citenamefont {Krempel},
  \citenamefont {Lauss}, \citenamefont {Lefort}, \citenamefont {Lemiere},
  \citenamefont {Leredde}, \citenamefont {Mohanmurthy}, \citenamefont {Pais},
  \citenamefont {Piegsa}, \citenamefont {Pignol}, \citenamefont {Qu\'em\'ener},
  \citenamefont {Rawlik}, \citenamefont {Rebreyend}, \citenamefont {Ries},
  \citenamefont {Roccia}, \citenamefont {Rozpedzik}, \citenamefont
  {Schmidt-Wellenburg}, \citenamefont {Schnabel}, \citenamefont {Severijns},
  \citenamefont {Virot}, \citenamefont {Weis}, \citenamefont {Wursten},
  \citenamefont {Wyszynski}, \citenamefont {Zejma},\ and\ \citenamefont
  {Zsigmond}}]{PSI2}%
  \BibitemOpen
  \bibfield  {author} {\bibinfo {author} {\bibfnamefont {C.}~\bibnamefont
  {Abel}}, \bibinfo {author} {\bibfnamefont {N.~J.}\ \bibnamefont {Ayres}},
  \bibinfo {author} {\bibfnamefont {T.}~\bibnamefont {Baker}}, \bibinfo
  {author} {\bibfnamefont {G.}~\bibnamefont {Ban}}, \bibinfo {author}
  {\bibfnamefont {G.}~\bibnamefont {Bison}}, \bibinfo {author} {\bibfnamefont
  {K.}~\bibnamefont {Bodek}}, \bibinfo {author} {\bibfnamefont
  {V.}~\bibnamefont {Bondar}}, \bibinfo {author} {\bibfnamefont {C.~B.}\
  \bibnamefont {Crawford}}, \bibinfo {author} {\bibfnamefont {P.-J.}\
  \bibnamefont {Chiu}}, \bibinfo {author} {\bibfnamefont {E.}~\bibnamefont
  {Chanel}}, \bibinfo {author} {\bibfnamefont {Z.}~\bibnamefont {Chowdhuri}},
  \bibinfo {author} {\bibfnamefont {M.}~\bibnamefont {Daum}}, \bibinfo {author}
  {\bibfnamefont {B.}~\bibnamefont {Dechenaux}}, \bibinfo {author}
  {\bibfnamefont {S.}~\bibnamefont {Emmenegger}}, \bibinfo {author}
  {\bibfnamefont {L.}~\bibnamefont {Ferraris-Bouchez}}, \bibinfo {author}
  {\bibfnamefont {P.}~\bibnamefont {Flaux}}, \bibinfo {author} {\bibfnamefont
  {P.}~\bibnamefont {Geltenbort}}, \bibinfo {author} {\bibfnamefont
  {K.}~\bibnamefont {Green}}, \bibinfo {author} {\bibfnamefont {W.~C.}\
  \bibnamefont {Griffith}}, \bibinfo {author} {\bibfnamefont {M.}~\bibnamefont
  {van~der Grinten}}, \bibinfo {author} {\bibfnamefont {P.~G.}\ \bibnamefont
  {Harris}}, \bibinfo {author} {\bibfnamefont {R.}~\bibnamefont {Henneck}},
  \bibinfo {author} {\bibfnamefont {N.}~\bibnamefont {Hild}}, \bibinfo {author}
  {\bibfnamefont {P.}~\bibnamefont {Iaydjiev}}, \bibinfo {author}
  {\bibfnamefont {S.~N.}\ \bibnamefont {Ivanov}}, \bibinfo {author}
  {\bibfnamefont {M.}~\bibnamefont {Kasprzak}}, \bibinfo {author}
  {\bibfnamefont {Y.}~\bibnamefont {Kermaidic}}, \bibinfo {author}
  {\bibfnamefont {K.}~\bibnamefont {Kirch}}, \bibinfo {author} {\bibfnamefont
  {H.-C.}\ \bibnamefont {Koch}}, \bibinfo {author} {\bibfnamefont
  {S.}~\bibnamefont {Komposch}}, \bibinfo {author} {\bibfnamefont {P.~A.}\
  \bibnamefont {Koss}}, \bibinfo {author} {\bibfnamefont {A.}~\bibnamefont
  {Kozela}}, \bibinfo {author} {\bibfnamefont {J.}~\bibnamefont {Krempel}},
  \bibinfo {author} {\bibfnamefont {B.}~\bibnamefont {Lauss}}, \bibinfo
  {author} {\bibfnamefont {T.}~\bibnamefont {Lefort}}, \bibinfo {author}
  {\bibfnamefont {Y.}~\bibnamefont {Lemiere}}, \bibinfo {author} {\bibfnamefont
  {A.}~\bibnamefont {Leredde}}, \bibinfo {author} {\bibfnamefont
  {P.}~\bibnamefont {Mohanmurthy}}, \bibinfo {author} {\bibfnamefont
  {D.}~\bibnamefont {Pais}}, \bibinfo {author} {\bibfnamefont {F.~M.}\
  \bibnamefont {Piegsa}}, \bibinfo {author} {\bibfnamefont {G.}~\bibnamefont
  {Pignol}}, \bibinfo {author} {\bibfnamefont {G.}~\bibnamefont
  {Qu\'em\'ener}}, \bibinfo {author} {\bibfnamefont {M.}~\bibnamefont
  {Rawlik}}, \bibinfo {author} {\bibfnamefont {D.}~\bibnamefont {Rebreyend}},
  \bibinfo {author} {\bibfnamefont {D.}~\bibnamefont {Ries}}, \bibinfo {author}
  {\bibfnamefont {S.}~\bibnamefont {Roccia}}, \bibinfo {author} {\bibfnamefont
  {D.}~\bibnamefont {Rozpedzik}}, \bibinfo {author} {\bibfnamefont
  {P.}~\bibnamefont {Schmidt-Wellenburg}}, \bibinfo {author} {\bibfnamefont
  {A.}~\bibnamefont {Schnabel}}, \bibinfo {author} {\bibfnamefont
  {N.}~\bibnamefont {Severijns}}, \bibinfo {author} {\bibfnamefont
  {R.}~\bibnamefont {Virot}}, \bibinfo {author} {\bibfnamefont
  {A.}~\bibnamefont {Weis}}, \bibinfo {author} {\bibfnamefont {E.}~\bibnamefont
  {Wursten}}, \bibinfo {author} {\bibfnamefont {G.}~\bibnamefont {Wyszynski}},
  \bibinfo {author} {\bibfnamefont {J.}~\bibnamefont {Zejma}},\ and\ \bibinfo
  {author} {\bibfnamefont {G.}~\bibnamefont {Zsigmond}},\ }\href
  {https://doi.org/10.1103/PhysRevA.99.042112} {\bibfield  {journal} {\bibinfo
  {journal} {Phys. Rev. A}\ }\textbf {\bibinfo {volume} {99}},\ \bibinfo
  {pages} {042112} (\bibinfo {year} {2019})}\BibitemShut {NoStop}%
\bibitem [{\citenamefont {Altarev}\ \emph {et~al.}(2015)\citenamefont {Altarev}
  \emph {et~al.}}]{Altarev:2015fra}%
  \BibitemOpen
  \bibfield  {author} {\bibinfo {author} {\bibfnamefont {I.}~\bibnamefont
  {Altarev}} \emph {et~al.},\ }\href {https://doi.org/10.1063/1.4919366}
  {\bibfield  {journal} {\bibinfo  {journal} {J. Appl. Phys.}\ }\textbf
  {\bibinfo {volume} {117}},\ \bibinfo {pages} {183903} (\bibinfo {year}
  {2015})},\ \Eprint {https://arxiv.org/abs/1501.07861} {arXiv:1501.07861
  [physics.ins-det]} \BibitemShut {NoStop}%
\bibitem [{\citenamefont {Gonzalez}\ \emph {et~al.}(2021)\citenamefont
  {Gonzalez}, \citenamefont {Fries}, \citenamefont {Cude-Woods}, \citenamefont
  {Bailey}, \citenamefont {Blatnik}, \citenamefont {Broussard}, \citenamefont
  {Callahan}, \citenamefont {Choi}, \citenamefont {Clayton}, \citenamefont
  {Currie}, \citenamefont {Dawid}, \citenamefont {Dees}, \citenamefont
  {Filippone}, \citenamefont {Fox}, \citenamefont {Geltenbort}, \citenamefont
  {George}, \citenamefont {Hayen}, \citenamefont {Hickerson}, \citenamefont
  {Hoffbauer}, \citenamefont {Hoffman}, \citenamefont {Holley}, \citenamefont
  {Ito}, \citenamefont {Komives}, \citenamefont {Liu}, \citenamefont {Makela},
  \citenamefont {Morris}, \citenamefont {Musedinovic}, \citenamefont
  {O'Shaughnessy}, \citenamefont {Pattie}, \citenamefont {Ramsey},
  \citenamefont {Salvat}, \citenamefont {Saunders}, \citenamefont {Sharapov},
  \citenamefont {Slutsky}, \citenamefont {Su}, \citenamefont {Sun},
  \citenamefont {Swank}, \citenamefont {Tang}, \citenamefont {Uhrich},
  \citenamefont {Vanderwerp}, \citenamefont {Walstrom}, \citenamefont {Wang},
  \citenamefont {Wei},\ and\ \citenamefont {Young}}]{Gonzalez2021}%
  \BibitemOpen
  \bibfield  {author} {\bibinfo {author} {\bibfnamefont {F.~M.}\ \bibnamefont
  {Gonzalez}}, \bibinfo {author} {\bibfnamefont {E.~M.}\ \bibnamefont {Fries}},
  \bibinfo {author} {\bibfnamefont {C.}~\bibnamefont {Cude-Woods}}, \bibinfo
  {author} {\bibfnamefont {T.}~\bibnamefont {Bailey}}, \bibinfo {author}
  {\bibfnamefont {M.}~\bibnamefont {Blatnik}}, \bibinfo {author} {\bibfnamefont
  {L.~J.}\ \bibnamefont {Broussard}}, \bibinfo {author} {\bibfnamefont {N.~B.}\
  \bibnamefont {Callahan}}, \bibinfo {author} {\bibfnamefont {J.~H.}\
  \bibnamefont {Choi}}, \bibinfo {author} {\bibfnamefont {S.~M.}\ \bibnamefont
  {Clayton}}, \bibinfo {author} {\bibfnamefont {S.~A.}\ \bibnamefont {Currie}},
  \bibinfo {author} {\bibfnamefont {M.}~\bibnamefont {Dawid}}, \bibinfo
  {author} {\bibfnamefont {E.~B.}\ \bibnamefont {Dees}}, \bibinfo {author}
  {\bibfnamefont {B.~W.}\ \bibnamefont {Filippone}}, \bibinfo {author}
  {\bibfnamefont {W.}~\bibnamefont {Fox}}, \bibinfo {author} {\bibfnamefont
  {P.}~\bibnamefont {Geltenbort}}, \bibinfo {author} {\bibfnamefont
  {E.}~\bibnamefont {George}}, \bibinfo {author} {\bibfnamefont
  {L.}~\bibnamefont {Hayen}}, \bibinfo {author} {\bibfnamefont {K.~P.}\
  \bibnamefont {Hickerson}}, \bibinfo {author} {\bibfnamefont {M.~A.}\
  \bibnamefont {Hoffbauer}}, \bibinfo {author} {\bibfnamefont {K.}~\bibnamefont
  {Hoffman}}, \bibinfo {author} {\bibfnamefont {A.~T.}\ \bibnamefont {Holley}},
  \bibinfo {author} {\bibfnamefont {T.~M.}\ \bibnamefont {Ito}}, \bibinfo
  {author} {\bibfnamefont {A.}~\bibnamefont {Komives}}, \bibinfo {author}
  {\bibfnamefont {C.-Y.}\ \bibnamefont {Liu}}, \bibinfo {author} {\bibfnamefont
  {M.}~\bibnamefont {Makela}}, \bibinfo {author} {\bibfnamefont {C.~L.}\
  \bibnamefont {Morris}}, \bibinfo {author} {\bibfnamefont {R.}~\bibnamefont
  {Musedinovic}}, \bibinfo {author} {\bibfnamefont {C.}~\bibnamefont
  {O'Shaughnessy}}, \bibinfo {author} {\bibfnamefont {R.~W.}\ \bibnamefont
  {Pattie}}, \bibinfo {author} {\bibfnamefont {J.}~\bibnamefont {Ramsey}},
  \bibinfo {author} {\bibfnamefont {D.~J.}\ \bibnamefont {Salvat}}, \bibinfo
  {author} {\bibfnamefont {A.}~\bibnamefont {Saunders}}, \bibinfo {author}
  {\bibfnamefont {E.~I.}\ \bibnamefont {Sharapov}}, \bibinfo {author}
  {\bibfnamefont {S.}~\bibnamefont {Slutsky}}, \bibinfo {author} {\bibfnamefont
  {V.}~\bibnamefont {Su}}, \bibinfo {author} {\bibfnamefont {X.}~\bibnamefont
  {Sun}}, \bibinfo {author} {\bibfnamefont {C.}~\bibnamefont {Swank}}, \bibinfo
  {author} {\bibfnamefont {Z.}~\bibnamefont {Tang}}, \bibinfo {author}
  {\bibfnamefont {W.}~\bibnamefont {Uhrich}}, \bibinfo {author} {\bibfnamefont
  {J.}~\bibnamefont {Vanderwerp}}, \bibinfo {author} {\bibfnamefont
  {P.}~\bibnamefont {Walstrom}}, \bibinfo {author} {\bibfnamefont
  {Z.}~\bibnamefont {Wang}}, \bibinfo {author} {\bibfnamefont {W.}~\bibnamefont
  {Wei}},\ and\ \bibinfo {author} {\bibfnamefont {A.~R.}\ \bibnamefont {Young}}
  (\bibinfo {collaboration} {$\mathrm{UCN}\ensuremath{\tau}$ Collaboration}),\
  }\href {https://doi.org/10.1103/PhysRevLett.127.162501} {\bibfield  {journal}
  {\bibinfo  {journal} {Phys. Rev. Lett.}\ }\textbf {\bibinfo {volume} {127}},\
  \bibinfo {pages} {162501} (\bibinfo {year} {2021})}\BibitemShut {NoStop}%
\bibitem [{\citenamefont {Pendlebury}\ and\ \citenamefont
  {Hinds}(2000)}]{Pendlebury2000}%
  \BibitemOpen
  \bibfield  {author} {\bibinfo {author} {\bibfnamefont {J.}~\bibnamefont
  {Pendlebury}}\ and\ \bibinfo {author} {\bibfnamefont {E.}~\bibnamefont
  {Hinds}},\ }\href {https://doi.org/10.1016/S0168-9002(99)01023-2} {\bibfield
  {journal} {\bibinfo  {journal} {Nuclear Instruments and Methods in Physics
  Research Section A: Accelerators, Spectrometers, Detectors and Associated
  Equipment}\ }\textbf {\bibinfo {volume} {440}},\ \bibinfo {pages} {471}
  (\bibinfo {year} {2000})}\BibitemShut {NoStop}%
\bibitem [{\citenamefont {Golub}\ and\ \citenamefont
  {Lamoreaux}(1994)}]{GOLUB1994}%
  \BibitemOpen
  \bibfield  {author} {\bibinfo {author} {\bibfnamefont {R.}~\bibnamefont
  {Golub}}\ and\ \bibinfo {author} {\bibfnamefont {S.~K.}\ \bibnamefont
  {Lamoreaux}},\ }\href {https://doi.org/10.1016/0370-1573(94)90084-1}
  {\bibfield  {journal} {\bibinfo  {journal} {Physics Reports}\ }\textbf
  {\bibinfo {volume} {237}},\ \bibinfo {pages} {1} (\bibinfo {year}
  {1994})}\BibitemShut {NoStop}%
\bibitem [{\citenamefont {Lamoreaux}\ \emph {et~al.}(2002)\citenamefont
  {Lamoreaux}, \citenamefont {Archibald}, \citenamefont {Barnes}, \citenamefont
  {Buttler}, \citenamefont {Clark}, \citenamefont {Cooper}, \citenamefont
  {Espy}, \citenamefont {Greene}, \citenamefont {Golub}, \citenamefont
  {Hayden}, \citenamefont {Lei}, \citenamefont {Marek}, \citenamefont {Peng},\
  and\ \citenamefont {Penttila}}]{Lamoreaux2002}%
  \BibitemOpen
  \bibfield  {author} {\bibinfo {author} {\bibfnamefont {S.~K.}\ \bibnamefont
  {Lamoreaux}}, \bibinfo {author} {\bibfnamefont {G.}~\bibnamefont
  {Archibald}}, \bibinfo {author} {\bibfnamefont {P.~D.}\ \bibnamefont
  {Barnes}}, \bibinfo {author} {\bibfnamefont {W.~T.}\ \bibnamefont {Buttler}},
  \bibinfo {author} {\bibfnamefont {D.~J.}\ \bibnamefont {Clark}}, \bibinfo
  {author} {\bibfnamefont {M.~D.}\ \bibnamefont {Cooper}}, \bibinfo {author}
  {\bibfnamefont {M.}~\bibnamefont {Espy}}, \bibinfo {author} {\bibfnamefont
  {G.~L.}\ \bibnamefont {Greene}}, \bibinfo {author} {\bibfnamefont
  {R.}~\bibnamefont {Golub}}, \bibinfo {author} {\bibfnamefont {M.~E.}\
  \bibnamefont {Hayden}}, \bibinfo {author} {\bibfnamefont {C.}~\bibnamefont
  {Lei}}, \bibinfo {author} {\bibfnamefont {L.~J.}\ \bibnamefont {Marek}},
  \bibinfo {author} {\bibfnamefont {J.-C.}\ \bibnamefont {Peng}},\ and\
  \bibinfo {author} {\bibfnamefont {S.}~\bibnamefont {Penttila}},\ }\href
  {https://doi.org/10.1209/epl/i2002-00408-4} {\bibfield  {journal} {\bibinfo
  {journal} {Europhysics Letters ({EPL})}\ }\textbf {\bibinfo {volume} {58}},\
  \bibinfo {pages} {718} (\bibinfo {year} {2002})}\BibitemShut {NoStop}%
\bibitem [{\citenamefont {Ye}\ \emph {et~al.}(2008)\citenamefont {Ye},
  \citenamefont {Dutta}, \citenamefont {Gao}, \citenamefont {Kramer},
  \citenamefont {Qian}, \citenamefont {Zong}, \citenamefont {Hannelius},
  \citenamefont {McKeown}, \citenamefont {Heyburn}, \citenamefont {Singer},
  \citenamefont {Golub},\ and\ \citenamefont {Korobkina}}]{Ye2008}%
  \BibitemOpen
  \bibfield  {author} {\bibinfo {author} {\bibfnamefont {Q.}~\bibnamefont
  {Ye}}, \bibinfo {author} {\bibfnamefont {D.}~\bibnamefont {Dutta}}, \bibinfo
  {author} {\bibfnamefont {H.}~\bibnamefont {Gao}}, \bibinfo {author}
  {\bibfnamefont {K.}~\bibnamefont {Kramer}}, \bibinfo {author} {\bibfnamefont
  {X.}~\bibnamefont {Qian}}, \bibinfo {author} {\bibfnamefont {X.}~\bibnamefont
  {Zong}}, \bibinfo {author} {\bibfnamefont {L.}~\bibnamefont {Hannelius}},
  \bibinfo {author} {\bibfnamefont {R.~D.}\ \bibnamefont {McKeown}}, \bibinfo
  {author} {\bibfnamefont {B.}~\bibnamefont {Heyburn}}, \bibinfo {author}
  {\bibfnamefont {S.}~\bibnamefont {Singer}}, \bibinfo {author} {\bibfnamefont
  {R.}~\bibnamefont {Golub}},\ and\ \bibinfo {author} {\bibfnamefont
  {E.}~\bibnamefont {Korobkina}},\ }\href
  {https://doi.org/10.1103/PhysRevA.77.053408} {\bibfield  {journal} {\bibinfo
  {journal} {Phys. Rev. A}\ }\textbf {\bibinfo {volume} {77}},\ \bibinfo
  {pages} {053408} (\bibinfo {year} {2008})}\BibitemShut {NoStop}%
\bibitem [{\citenamefont {Ye}\ \emph {et~al.}(2009)\citenamefont {Ye},
  \citenamefont {Gao}, \citenamefont {Zheng}, \citenamefont {Dutta},
  \citenamefont {Dubose}, \citenamefont {Golub}, \citenamefont {Huffman},
  \citenamefont {Swank},\ and\ \citenamefont {Korobkina}}]{Ye2009}%
  \BibitemOpen
  \bibfield  {author} {\bibinfo {author} {\bibfnamefont {Q.}~\bibnamefont
  {Ye}}, \bibinfo {author} {\bibfnamefont {H.}~\bibnamefont {Gao}}, \bibinfo
  {author} {\bibfnamefont {W.}~\bibnamefont {Zheng}}, \bibinfo {author}
  {\bibfnamefont {D.}~\bibnamefont {Dutta}}, \bibinfo {author} {\bibfnamefont
  {F.}~\bibnamefont {Dubose}}, \bibinfo {author} {\bibfnamefont
  {R.}~\bibnamefont {Golub}}, \bibinfo {author} {\bibfnamefont
  {P.}~\bibnamefont {Huffman}}, \bibinfo {author} {\bibfnamefont
  {C.}~\bibnamefont {Swank}},\ and\ \bibinfo {author} {\bibfnamefont
  {E.}~\bibnamefont {Korobkina}},\ }\href
  {https://doi.org/10.1103/PhysRevA.80.023403} {\bibfield  {journal} {\bibinfo
  {journal} {Phys. Rev. A}\ }\textbf {\bibinfo {volume} {80}},\ \bibinfo
  {pages} {023403} (\bibinfo {year} {2009})}\BibitemShut {NoStop}%
\bibitem [{\citenamefont {Eckel}\ \emph
  {et~al.}(2012{\natexlab{a}})\citenamefont {Eckel}, \citenamefont {Lamoreaux},
  \citenamefont {Hayden},\ and\ \citenamefont {Ito}}]{SEckel2012}%
  \BibitemOpen
  \bibfield  {author} {\bibinfo {author} {\bibfnamefont {S.}~\bibnamefont
  {Eckel}}, \bibinfo {author} {\bibfnamefont {S.~K.}\ \bibnamefont
  {Lamoreaux}}, \bibinfo {author} {\bibfnamefont {M.~E.}\ \bibnamefont
  {Hayden}},\ and\ \bibinfo {author} {\bibfnamefont {T.~M.}\ \bibnamefont
  {Ito}},\ }\href {https://doi.org/10.1103/PhysRevA.85.032124} {\bibfield
  {journal} {\bibinfo  {journal} {Phys. Rev. A}\ }\textbf {\bibinfo {volume}
  {85}},\ \bibinfo {pages} {032124} (\bibinfo {year}
  {2012}{\natexlab{a}})}\BibitemShut {NoStop}%
\bibitem [{\citenamefont {Baym}\ \emph {et~al.}(2013)\citenamefont {Baym},
  \citenamefont {Beck},\ and\ \citenamefont {Pethick}}]{Baym2013}%
  \BibitemOpen
  \bibfield  {author} {\bibinfo {author} {\bibfnamefont {G.}~\bibnamefont
  {Baym}}, \bibinfo {author} {\bibfnamefont {D.~H.}\ \bibnamefont {Beck}},\
  and\ \bibinfo {author} {\bibfnamefont {C.~J.}\ \bibnamefont {Pethick}},\
  }\href {https://doi.org/10.1103/PhysRevB.88.014512} {\bibfield  {journal}
  {\bibinfo  {journal} {Phys. Rev. B}\ }\textbf {\bibinfo {volume} {88}},\
  \bibinfo {pages} {014512} (\bibinfo {year} {2013})}\BibitemShut {NoStop}%
\bibitem [{\citenamefont {Baym}\ \emph
  {et~al.}(2015{\natexlab{a}})\citenamefont {Baym}, \citenamefont {Beck},\ and\
  \citenamefont {Pethick}}]{Baym2015}%
  \BibitemOpen
  \bibfield  {author} {\bibinfo {author} {\bibfnamefont {G.}~\bibnamefont
  {Baym}}, \bibinfo {author} {\bibfnamefont {D.~H.}\ \bibnamefont {Beck}},\
  and\ \bibinfo {author} {\bibfnamefont {C.~J.}\ \bibnamefont {Pethick}},\
  }\href {https://doi.org/10.1103/PhysRevB.92.024504} {\bibfield  {journal}
  {\bibinfo  {journal} {Phys. Rev. B}\ }\textbf {\bibinfo {volume} {92}},\
  \bibinfo {pages} {024504} (\bibinfo {year} {2015}{\natexlab{a}})}\BibitemShut
  {NoStop}%
\bibitem [{\citenamefont {Baym}\ \emph
  {et~al.}(2015{\natexlab{b}})\citenamefont {Baym}, \citenamefont {Beck},\ and\
  \citenamefont {Pethick}}]{Baym2015b}%
  \BibitemOpen
  \bibfield  {author} {\bibinfo {author} {\bibfnamefont {G.}~\bibnamefont
  {Baym}}, \bibinfo {author} {\bibfnamefont {D.~H.}\ \bibnamefont {Beck}},\
  and\ \bibinfo {author} {\bibfnamefont {C.~J.}\ \bibnamefont {Pethick}},\
  }\href {https://doi.org/10.1007/s10909-014-1235-0} {\bibfield  {journal}
  {\bibinfo  {journal} {Journal of Low Temperature Physics}\ }\textbf {\bibinfo
  {volume} {178}},\ \bibinfo {pages} {200} (\bibinfo {year}
  {2015}{\natexlab{b}})}\BibitemShut {NoStop}%
\bibitem [{\citenamefont {Esler}\ \emph {et~al.}(2007)\citenamefont {Esler},
  \citenamefont {Peng}, \citenamefont {Chandler}, \citenamefont {Howell},
  \citenamefont {Lamoreaux}, \citenamefont {Liu},\ and\ \citenamefont
  {Torgerson}}]{Esler2007}%
  \BibitemOpen
  \bibfield  {author} {\bibinfo {author} {\bibfnamefont {A.}~\bibnamefont
  {Esler}}, \bibinfo {author} {\bibfnamefont {J.~C.}\ \bibnamefont {Peng}},
  \bibinfo {author} {\bibfnamefont {D.}~\bibnamefont {Chandler}}, \bibinfo
  {author} {\bibfnamefont {D.}~\bibnamefont {Howell}}, \bibinfo {author}
  {\bibfnamefont {S.~K.}\ \bibnamefont {Lamoreaux}}, \bibinfo {author}
  {\bibfnamefont {C.~Y.}\ \bibnamefont {Liu}},\ and\ \bibinfo {author}
  {\bibfnamefont {J.~R.}\ \bibnamefont {Torgerson}},\ }\href
  {https://doi.org/10.1103/PhysRevC.76.051302} {\bibfield  {journal} {\bibinfo
  {journal} {Phys. Rev. C}\ }\textbf {\bibinfo {volume} {76}},\ \bibinfo
  {pages} {051302} (\bibinfo {year} {2007})}\BibitemShut {NoStop}%
\bibitem [{\citenamefont {Chu}\ \emph {et~al.}(2011)\citenamefont {Chu},
  \citenamefont {Esler}, \citenamefont {Peng}, \citenamefont {Beck},
  \citenamefont {Chandler}, \citenamefont {Clayton}, \citenamefont {Hu},
  \citenamefont {Ngan}, \citenamefont {Sham}, \citenamefont {So}, \citenamefont
  {Williamson},\ and\ \citenamefont {Yoder}}]{Chu2011}%
  \BibitemOpen
  \bibfield  {author} {\bibinfo {author} {\bibfnamefont {P.-H.}\ \bibnamefont
  {Chu}}, \bibinfo {author} {\bibfnamefont {A.~M.}\ \bibnamefont {Esler}},
  \bibinfo {author} {\bibfnamefont {J.~C.}\ \bibnamefont {Peng}}, \bibinfo
  {author} {\bibfnamefont {D.~H.}\ \bibnamefont {Beck}}, \bibinfo {author}
  {\bibfnamefont {D.~E.}\ \bibnamefont {Chandler}}, \bibinfo {author}
  {\bibfnamefont {S.}~\bibnamefont {Clayton}}, \bibinfo {author} {\bibfnamefont
  {B.-Z.}\ \bibnamefont {Hu}}, \bibinfo {author} {\bibfnamefont {S.~Y.}\
  \bibnamefont {Ngan}}, \bibinfo {author} {\bibfnamefont {C.~H.}\ \bibnamefont
  {Sham}}, \bibinfo {author} {\bibfnamefont {L.~H.}\ \bibnamefont {So}},
  \bibinfo {author} {\bibfnamefont {S.}~\bibnamefont {Williamson}},\ and\
  \bibinfo {author} {\bibfnamefont {J.}~\bibnamefont {Yoder}},\ }\href
  {https://doi.org/10.1103/PhysRevC.84.022501} {\bibfield  {journal} {\bibinfo
  {journal} {Phys. Rev. C}\ }\textbf {\bibinfo {volume} {84}},\ \bibinfo
  {pages} {022501} (\bibinfo {year} {2011})}\BibitemShut {NoStop}%
\bibitem [{\citenamefont {Chu}\ and\ \citenamefont {Peng}(2015)}]{Chu2015}%
  \BibitemOpen
  \bibfield  {author} {\bibinfo {author} {\bibfnamefont {P.-H.}\ \bibnamefont
  {Chu}}\ and\ \bibinfo {author} {\bibfnamefont {J.-C.}\ \bibnamefont {Peng}},\
  }\href {https://doi.org/10.1016/j.nima.2015.05.062} {\bibfield  {journal}
  {\bibinfo  {journal} {Nuclear Instruments and Methods in Physics Research
  Section A: Accelerators, Spectrometers, Detectors and Associated Equipment}\
  }\textbf {\bibinfo {volume} {795}},\ \bibinfo {pages} {128} (\bibinfo {year}
  {2015})}\BibitemShut {NoStop}%
\bibitem [{\citenamefont {Nouri}\ \emph {et~al.}(2015)\citenamefont {Nouri},
  \citenamefont {Biswas}, \citenamefont {Brown}, \citenamefont {Carr},
  \citenamefont {Filippone}, \citenamefont {Osthelder}, \citenamefont
  {Plaster}, \citenamefont {Slutsky},\ and\ \citenamefont {Swank}}]{Nouri2015}%
  \BibitemOpen
  \bibfield  {author} {\bibinfo {author} {\bibfnamefont {N.}~\bibnamefont
  {Nouri}}, \bibinfo {author} {\bibfnamefont {A.}~\bibnamefont {Biswas}},
  \bibinfo {author} {\bibfnamefont {M.}~\bibnamefont {Brown}}, \bibinfo
  {author} {\bibfnamefont {R.}~\bibnamefont {Carr}}, \bibinfo {author}
  {\bibfnamefont {B.}~\bibnamefont {Filippone}}, \bibinfo {author}
  {\bibfnamefont {C.}~\bibnamefont {Osthelder}}, \bibinfo {author}
  {\bibfnamefont {B.}~\bibnamefont {Plaster}}, \bibinfo {author} {\bibfnamefont
  {S.}~\bibnamefont {Slutsky}},\ and\ \bibinfo {author} {\bibfnamefont
  {C.}~\bibnamefont {Swank}},\ }\href
  {https://doi.org/10.1088/1748-0221/10/12/p12003} {\bibfield  {journal}
  {\bibinfo  {journal} {Journal of Instrumentation}\ }\textbf {\bibinfo
  {volume} {10}}\bibinfo  {number} { (12)},\ \bibinfo {pages}
  {P12003}}\BibitemShut {NoStop}%
\bibitem [{\citenamefont {Cianciolo}\ \emph {et~al.}(2018)\citenamefont
  {Cianciolo}, \citenamefont {Ramsey},\ and\ \citenamefont
  {Fabris}}]{Cianciolo2018}%
  \BibitemOpen
\bibfield  {number} {  }\bibfield  {author} {\bibinfo {author} {\bibfnamefont
  {V.}~\bibnamefont {Cianciolo}}, \bibinfo {author} {\bibfnamefont
  {J.}~\bibnamefont {Ramsey}},\ and\ \bibinfo {author} {\bibfnamefont
  {L.}~\bibnamefont {Fabris}},\ }\href
  {https://doi.org/10.1088/1748-0221/13/09/p09010} {\bibfield  {journal}
  {\bibinfo  {journal} {Journal of Instrumentation}\ }\textbf {\bibinfo
  {volume} {13}}\bibinfo  {number} { (09)},\ \bibinfo {pages}
  {P09010}}\BibitemShut {NoStop}%
\bibitem [{\citenamefont {Kim}\ and\ \citenamefont
  {Clayton}(2015)}]{YJKim2015}%
  \BibitemOpen
\bibfield  {number} {  }\bibfield  {author} {\bibinfo {author} {\bibfnamefont
  {Y.~J.}\ \bibnamefont {Kim}}\ and\ \bibinfo {author} {\bibfnamefont {S.~M.}\
  \bibnamefont {Clayton}},\ }\href {https://doi.org/10.1109/TASC.2014.2359336}
  {\bibfield  {journal} {\bibinfo  {journal} {IEEE Transactions on Applied
  Superconductivity}\ }\textbf {\bibinfo {volume} {25}},\ \bibinfo {pages} {1}
  (\bibinfo {year} {2015})}\BibitemShut {NoStop}%
\bibitem [{\citenamefont {Ito}\ \emph {et~al.}(2012)\citenamefont {Ito},
  \citenamefont {Clayton}, \citenamefont {Ramsey}, \citenamefont {Karcz},
  \citenamefont {Liu}, \citenamefont {Long}, \citenamefont {Reddy},\ and\
  \citenamefont {Seidel}}]{Ito2012}%
  \BibitemOpen
  \bibfield  {author} {\bibinfo {author} {\bibfnamefont {T.~M.}\ \bibnamefont
  {Ito}}, \bibinfo {author} {\bibfnamefont {S.~M.}\ \bibnamefont {Clayton}},
  \bibinfo {author} {\bibfnamefont {J.}~\bibnamefont {Ramsey}}, \bibinfo
  {author} {\bibfnamefont {M.}~\bibnamefont {Karcz}}, \bibinfo {author}
  {\bibfnamefont {C.-Y.}\ \bibnamefont {Liu}}, \bibinfo {author} {\bibfnamefont
  {J.~C.}\ \bibnamefont {Long}}, \bibinfo {author} {\bibfnamefont {T.~G.}\
  \bibnamefont {Reddy}},\ and\ \bibinfo {author} {\bibfnamefont {G.~M.}\
  \bibnamefont {Seidel}},\ }\href {https://doi.org/10.1103/PhysRevA.85.042718}
  {\bibfield  {journal} {\bibinfo  {journal} {Phys. Rev. A}\ }\textbf {\bibinfo
  {volume} {85}},\ \bibinfo {pages} {042718} (\bibinfo {year}
  {2012})}\BibitemShut {NoStop}%
\bibitem [{\citenamefont {Ito}\ and\ \citenamefont {Seidel}(2013)}]{Ito2013}%
  \BibitemOpen
  \bibfield  {author} {\bibinfo {author} {\bibfnamefont {T.~M.}\ \bibnamefont
  {Ito}}\ and\ \bibinfo {author} {\bibfnamefont {G.~M.}\ \bibnamefont
  {Seidel}},\ }\href {https://doi.org/10.1103/PhysRevC.88.025805} {\bibfield
  {journal} {\bibinfo  {journal} {Phys. Rev. C}\ }\textbf {\bibinfo {volume}
  {88}},\ \bibinfo {pages} {025805} (\bibinfo {year} {2013})}\BibitemShut
  {NoStop}%
\bibitem [{\citenamefont {Gehman}\ \emph {et~al.}(2013)\citenamefont {Gehman},
  \citenamefont {Ito}, \citenamefont {Griffith},\ and\ \citenamefont
  {Seibert}}]{Gehman2013}%
  \BibitemOpen
  \bibfield  {author} {\bibinfo {author} {\bibfnamefont {V.~M.}\ \bibnamefont
  {Gehman}}, \bibinfo {author} {\bibfnamefont {T.~M.}\ \bibnamefont {Ito}},
  \bibinfo {author} {\bibfnamefont {W.~C.}\ \bibnamefont {Griffith}},\ and\
  \bibinfo {author} {\bibfnamefont {S.~R.}\ \bibnamefont {Seibert}},\ }\href
  {https://doi.org/10.1088/1748-0221/8/04/p04024} {\bibfield  {journal}
  {\bibinfo  {journal} {Journal of Instrumentation}\ }\textbf {\bibinfo
  {volume} {8}}\bibinfo  {number} { (04)},\ \bibinfo {pages}
  {P04024}}\BibitemShut {NoStop}%
\bibitem [{\citenamefont {Seidel}\ \emph {et~al.}(2014)\citenamefont {Seidel},
  \citenamefont {Ito}, \citenamefont {Ghosh},\ and\ \citenamefont
  {Sethumadhavan}}]{Seidel2014}%
  \BibitemOpen
\bibfield  {number} {  }\bibfield  {author} {\bibinfo {author} {\bibfnamefont
  {G.~M.}\ \bibnamefont {Seidel}}, \bibinfo {author} {\bibfnamefont {T.~M.}\
  \bibnamefont {Ito}}, \bibinfo {author} {\bibfnamefont {A.}~\bibnamefont
  {Ghosh}},\ and\ \bibinfo {author} {\bibfnamefont {B.}~\bibnamefont
  {Sethumadhavan}},\ }\href {https://doi.org/10.1103/PhysRevC.89.025808}
  {\bibfield  {journal} {\bibinfo  {journal} {Phys. Rev. C}\ }\textbf {\bibinfo
  {volume} {89}},\ \bibinfo {pages} {025808} (\bibinfo {year}
  {2014})}\BibitemShut {NoStop}%
\bibitem [{\citenamefont {Phan}\ \emph {et~al.}(2020)\citenamefont {Phan},
  \citenamefont {Cianciolo}, \citenamefont {Clayton}, \citenamefont {Currie},
  \citenamefont {Dipert}, \citenamefont {Ito}, \citenamefont {MacDonald},
  \citenamefont {O'Shaughnessy}, \citenamefont {Ramsey}, \citenamefont
  {Seidel}, \citenamefont {Smith}, \citenamefont {Tang}, \citenamefont {Tang},\
  and\ \citenamefont {Yao}}]{Phan2020}%
  \BibitemOpen
  \bibfield  {author} {\bibinfo {author} {\bibfnamefont {N.~S.}\ \bibnamefont
  {Phan}}, \bibinfo {author} {\bibfnamefont {V.}~\bibnamefont {Cianciolo}},
  \bibinfo {author} {\bibfnamefont {S.~M.}\ \bibnamefont {Clayton}}, \bibinfo
  {author} {\bibfnamefont {S.~A.}\ \bibnamefont {Currie}}, \bibinfo {author}
  {\bibfnamefont {R.}~\bibnamefont {Dipert}}, \bibinfo {author} {\bibfnamefont
  {T.~M.}\ \bibnamefont {Ito}}, \bibinfo {author} {\bibfnamefont {S.~W.~T.}\
  \bibnamefont {MacDonald}}, \bibinfo {author} {\bibfnamefont {C.~M.}\
  \bibnamefont {O'Shaughnessy}}, \bibinfo {author} {\bibfnamefont {J.~C.}\
  \bibnamefont {Ramsey}}, \bibinfo {author} {\bibfnamefont {G.~M.}\
  \bibnamefont {Seidel}}, \bibinfo {author} {\bibfnamefont {E.}~\bibnamefont
  {Smith}}, \bibinfo {author} {\bibfnamefont {E.}~\bibnamefont {Tang}},
  \bibinfo {author} {\bibfnamefont {Z.}~\bibnamefont {Tang}},\ and\ \bibinfo
  {author} {\bibfnamefont {W.}~\bibnamefont {Yao}},\ }\href
  {https://doi.org/10.1103/PhysRevC.102.035503} {\bibfield  {journal} {\bibinfo
   {journal} {Phys. Rev. C}\ }\textbf {\bibinfo {volume} {102}},\ \bibinfo
  {pages} {035503} (\bibinfo {year} {2020})}\BibitemShut {NoStop}%
\bibitem [{\citenamefont {Loomis}\ \emph {et~al.}(2021)\citenamefont {Loomis},
  \citenamefont {Cianciolo},\ and\ \citenamefont {Leggett}}]{Loomis2021}%
  \BibitemOpen
  \bibfield  {author} {\bibinfo {author} {\bibfnamefont {D.~A.}\ \bibnamefont
  {Loomis}}, \bibinfo {author} {\bibfnamefont {V.}~\bibnamefont {Cianciolo}},\
  and\ \bibinfo {author} {\bibfnamefont {E.}~\bibnamefont {Leggett}},\
  }\href@noop {} {\bibfield  {journal} {\bibinfo  {journal} {ArXiv e-prints}\ }
  (\bibinfo {year} {2021})},\ \Eprint {https://arxiv.org/abs/2110.03930}
  {arXiv:2110.03930 [nucl-ex]} \BibitemShut {NoStop}%
\bibitem [{\citenamefont {Clayton}\ \emph {et~al.}(2018)\citenamefont
  {Clayton}, \citenamefont {Ito}, \citenamefont {Ramsey}, \citenamefont {Wei},
  \citenamefont {Blatnik}, \citenamefont {Filippone},\ and\ \citenamefont
  {Seidel}}]{Clayton2018}%
  \BibitemOpen
  \bibfield  {author} {\bibinfo {author} {\bibfnamefont {S.}~\bibnamefont
  {Clayton}}, \bibinfo {author} {\bibfnamefont {T.}~\bibnamefont {Ito}},
  \bibinfo {author} {\bibfnamefont {J.}~\bibnamefont {Ramsey}}, \bibinfo
  {author} {\bibfnamefont {W.}~\bibnamefont {Wei}}, \bibinfo {author}
  {\bibfnamefont {M.}~\bibnamefont {Blatnik}}, \bibinfo {author} {\bibfnamefont
  {B.}~\bibnamefont {Filippone}},\ and\ \bibinfo {author} {\bibfnamefont
  {G.}~\bibnamefont {Seidel}},\ }\href
  {https://doi.org/10.1088/1748-0221/13/05/p05017} {\bibfield  {journal}
  {\bibinfo  {journal} {Journal of Instrumentation}\ }\textbf {\bibinfo
  {volume} {13}}\bibinfo  {number} { (05)},\ \bibinfo {pages}
  {P05017}}\BibitemShut {NoStop}%
\bibitem [{\citenamefont {Slutsky}\ \emph {et~al.}(2017)\citenamefont
  {Slutsky}, \citenamefont {Swank}, \citenamefont {Biswas}, \citenamefont
  {Carr}, \citenamefont {Escribano}, \citenamefont {Filippone}, \citenamefont
  {Griffith}, \citenamefont {Mendenhall}, \citenamefont {Nouri}, \citenamefont
  {Osthelder}, \citenamefont {{Pérez Galván}}, \citenamefont {Picker},\ and\
  \citenamefont {Plaster}}]{Slutsky2017}%
  \BibitemOpen
\bibfield  {number} {  }\bibfield  {author} {\bibinfo {author} {\bibfnamefont
  {S.}~\bibnamefont {Slutsky}}, \bibinfo {author} {\bibfnamefont
  {C.}~\bibnamefont {Swank}}, \bibinfo {author} {\bibfnamefont
  {A.}~\bibnamefont {Biswas}}, \bibinfo {author} {\bibfnamefont
  {R.}~\bibnamefont {Carr}}, \bibinfo {author} {\bibfnamefont {J.}~\bibnamefont
  {Escribano}}, \bibinfo {author} {\bibfnamefont {B.}~\bibnamefont
  {Filippone}}, \bibinfo {author} {\bibfnamefont {W.}~\bibnamefont {Griffith}},
  \bibinfo {author} {\bibfnamefont {M.}~\bibnamefont {Mendenhall}}, \bibinfo
  {author} {\bibfnamefont {N.}~\bibnamefont {Nouri}}, \bibinfo {author}
  {\bibfnamefont {C.}~\bibnamefont {Osthelder}}, \bibinfo {author}
  {\bibfnamefont {A.}~\bibnamefont {{Pérez Galván}}}, \bibinfo {author}
  {\bibfnamefont {R.}~\bibnamefont {Picker}},\ and\ \bibinfo {author}
  {\bibfnamefont {B.}~\bibnamefont {Plaster}},\ }\href
  {https://doi.org/10.1016/j.nima.2017.05.005} {\bibfield  {journal} {\bibinfo
  {journal} {Nuclear Instruments and Methods in Physics Research Section A:
  Accelerators, Spectrometers, Detectors and Associated Equipment}\ }\textbf
  {\bibinfo {volume} {862}},\ \bibinfo {pages} {36} (\bibinfo {year}
  {2017})}\BibitemShut {NoStop}%
\bibitem [{\citenamefont {Schmid}\ \emph {et~al.}(2008)\citenamefont {Schmid},
  \citenamefont {Plaster},\ and\ \citenamefont {Filippone}}]{Schmid2008}%
  \BibitemOpen
  \bibfield  {author} {\bibinfo {author} {\bibfnamefont {R.}~\bibnamefont
  {Schmid}}, \bibinfo {author} {\bibfnamefont {B.}~\bibnamefont {Plaster}},\
  and\ \bibinfo {author} {\bibfnamefont {B.~W.}\ \bibnamefont {Filippone}},\
  }\href {https://doi.org/10.1103/PhysRevA.78.023401} {\bibfield  {journal}
  {\bibinfo  {journal} {Phys. Rev. A}\ }\textbf {\bibinfo {volume} {78}},\
  \bibinfo {pages} {023401} (\bibinfo {year} {2008})}\BibitemShut {NoStop}%
\bibitem [{\citenamefont {Steyerl}\ \emph {et~al.}(2014)\citenamefont
  {Steyerl}, \citenamefont {Kaufman}, \citenamefont {M\"uller}, \citenamefont
  {Malik}, \citenamefont {Desai},\ and\ \citenamefont {Golub}}]{Steyerl2014}%
  \BibitemOpen
  \bibfield  {author} {\bibinfo {author} {\bibfnamefont {A.}~\bibnamefont
  {Steyerl}}, \bibinfo {author} {\bibfnamefont {C.}~\bibnamefont {Kaufman}},
  \bibinfo {author} {\bibfnamefont {G.}~\bibnamefont {M\"uller}}, \bibinfo
  {author} {\bibfnamefont {S.~S.}\ \bibnamefont {Malik}}, \bibinfo {author}
  {\bibfnamefont {A.~M.}\ \bibnamefont {Desai}},\ and\ \bibinfo {author}
  {\bibfnamefont {R.}~\bibnamefont {Golub}},\ }\href
  {https://doi.org/10.1103/PhysRevA.89.052129} {\bibfield  {journal} {\bibinfo
  {journal} {Phys. Rev. A}\ }\textbf {\bibinfo {volume} {89}},\ \bibinfo
  {pages} {052129} (\bibinfo {year} {2014})}\BibitemShut {NoStop}%
\bibitem [{\citenamefont {Pignol}\ \emph
  {et~al.}(2015{\natexlab{b}})\citenamefont {Pignol}, \citenamefont {Guigue},
  \citenamefont {Petukhov},\ and\ \citenamefont {Golub}}]{Golub2015}%
  \BibitemOpen
  \bibfield  {author} {\bibinfo {author} {\bibfnamefont {G.}~\bibnamefont
  {Pignol}}, \bibinfo {author} {\bibfnamefont {M.}~\bibnamefont {Guigue}},
  \bibinfo {author} {\bibfnamefont {A.}~\bibnamefont {Petukhov}},\ and\
  \bibinfo {author} {\bibfnamefont {R.}~\bibnamefont {Golub}},\ }\href
  {https://doi.org/10.1103/PhysRevA.92.053407} {\bibfield  {journal} {\bibinfo
  {journal} {Phys. Rev. A}\ }\textbf {\bibinfo {volume} {92}},\ \bibinfo
  {pages} {053407} (\bibinfo {year} {2015}{\natexlab{b}})}\BibitemShut
  {NoStop}%
\bibitem [{\citenamefont {Golub}\ \emph {et~al.}(2015)\citenamefont {Golub},
  \citenamefont {Kaufman}, \citenamefont {M\"uller},\ and\ \citenamefont
  {Steyerl}}]{Golub2015b}%
  \BibitemOpen
  \bibfield  {author} {\bibinfo {author} {\bibfnamefont {R.}~\bibnamefont
  {Golub}}, \bibinfo {author} {\bibfnamefont {C.}~\bibnamefont {Kaufman}},
  \bibinfo {author} {\bibfnamefont {G.}~\bibnamefont {M\"uller}},\ and\
  \bibinfo {author} {\bibfnamefont {A.}~\bibnamefont {Steyerl}},\ }\href
  {https://doi.org/10.1103/PhysRevA.92.062123} {\bibfield  {journal} {\bibinfo
  {journal} {Phys. Rev. A}\ }\textbf {\bibinfo {volume} {92}},\ \bibinfo
  {pages} {062123} (\bibinfo {year} {2015})}\BibitemShut {NoStop}%
\bibitem [{\citenamefont {Safronova}\ \emph {et~al.}(2018)\citenamefont
  {Safronova}, \citenamefont {Budker}, \citenamefont {DeMille}, \citenamefont
  {Kimball}, \citenamefont {Derevianko},\ and\ \citenamefont
  {Clark}}]{Safronova2018}%
  \BibitemOpen
  \bibfield  {author} {\bibinfo {author} {\bibfnamefont {M.~S.}\ \bibnamefont
  {Safronova}}, \bibinfo {author} {\bibfnamefont {D.}~\bibnamefont {Budker}},
  \bibinfo {author} {\bibfnamefont {D.}~\bibnamefont {DeMille}}, \bibinfo
  {author} {\bibfnamefont {D.~F.~J.}\ \bibnamefont {Kimball}}, \bibinfo
  {author} {\bibfnamefont {A.}~\bibnamefont {Derevianko}},\ and\ \bibinfo
  {author} {\bibfnamefont {C.~W.}\ \bibnamefont {Clark}},\ }\href
  {https://doi.org/10.1103/RevModPhys.90.025008} {\bibfield  {journal}
  {\bibinfo  {journal} {Reviews of Modern Physics}\ }\textbf {\bibinfo {volume}
  {90}},\ \bibinfo {pages} {025008} (\bibinfo {year} {2018})}\BibitemShut
  {NoStop}%
\bibitem [{\citenamefont {{V. Andreev}}\ \emph {et~al.}(2018)\citenamefont {{V.
  Andreev}}, \citenamefont {Ang}, \citenamefont {DeMille}, \citenamefont
  {Doyle}, \citenamefont {Gabrielse}, \citenamefont {Haefner}, \citenamefont
  {Hutzler}, \citenamefont {Lasner}, \citenamefont {Meisenhelder},
  \citenamefont {O'Leary}, \citenamefont {Panda}, \citenamefont {West},
  \citenamefont {West},\ and\ \citenamefont {Wu}}]{ACME2018}%
  \BibitemOpen
  \bibfield  {author} {\bibinfo {author} {\bibnamefont {{V. Andreev}}},
  \bibinfo {author} {\bibfnamefont {D.~G.}\ \bibnamefont {Ang}}, \bibinfo
  {author} {\bibfnamefont {D.}~\bibnamefont {DeMille}}, \bibinfo {author}
  {\bibfnamefont {J.~M.}\ \bibnamefont {Doyle}}, \bibinfo {author}
  {\bibfnamefont {G.}~\bibnamefont {Gabrielse}}, \bibinfo {author}
  {\bibfnamefont {J.}~\bibnamefont {Haefner}}, \bibinfo {author} {\bibfnamefont
  {N.~R.}\ \bibnamefont {Hutzler}}, \bibinfo {author} {\bibfnamefont
  {Z.}~\bibnamefont {Lasner}}, \bibinfo {author} {\bibfnamefont
  {C.}~\bibnamefont {Meisenhelder}}, \bibinfo {author} {\bibfnamefont {B.~R.}\
  \bibnamefont {O'Leary}}, \bibinfo {author} {\bibfnamefont {C.~D.}\
  \bibnamefont {Panda}}, \bibinfo {author} {\bibfnamefont {A.~D.}\ \bibnamefont
  {West}}, \bibinfo {author} {\bibfnamefont {E.~P.}\ \bibnamefont {West}},\
  and\ \bibinfo {author} {\bibfnamefont {X.}~\bibnamefont {Wu}},\ }\href
  {https://doi.org/10.1038/s41586-018-0599-8} {\bibfield  {journal} {\bibinfo
  {journal} {Nature}\ }\textbf {\bibinfo {volume} {562}},\ \bibinfo {pages}
  {355} (\bibinfo {year} {2018})}\BibitemShut {NoStop}%
\bibitem [{\citenamefont {Graner}\ \emph
  {et~al.}(2016{\natexlab{b}})\citenamefont {Graner}, \citenamefont {Chen},
  \citenamefont {Lindahl},\ and\ \citenamefont {Heckel}}]{Graner2016}%
  \BibitemOpen
  \bibfield  {author} {\bibinfo {author} {\bibfnamefont {B.}~\bibnamefont
  {Graner}}, \bibinfo {author} {\bibfnamefont {Y.}~\bibnamefont {Chen}},
  \bibinfo {author} {\bibfnamefont {E.~G.}\ \bibnamefont {Lindahl}},\ and\
  \bibinfo {author} {\bibfnamefont {B.~R.}\ \bibnamefont {Heckel}},\ }\href
  {https://doi.org/10.1103/PhysRevLett.116.161601} {\bibfield  {journal}
  {\bibinfo  {journal} {Physical Review Letters}\ }\textbf {\bibinfo {volume}
  {116}},\ \bibinfo {pages} {161601} (\bibinfo {year}
  {2016}{\natexlab{b}})}\BibitemShut {NoStop}%
\bibitem [{\citenamefont {Schiff}(1963)}]{Schiff1963}%
  \BibitemOpen
  \bibfield  {author} {\bibinfo {author} {\bibfnamefont {L.~I.}\ \bibnamefont
  {Schiff}},\ }\href@noop {} {\bibfield  {journal} {\bibinfo  {journal}
  {Physical Review}\ }\textbf {\bibinfo {volume} {132}},\ \bibinfo {pages}
  {2194} (\bibinfo {year} {1963})}\BibitemShut {NoStop}%
\bibitem [{\citenamefont {Sandars}\ and\ \citenamefont
  {Lipworth}(1964{\natexlab{b}})}]{Sandars1964a}%
  \BibitemOpen
  \bibfield  {author} {\bibinfo {author} {\bibfnamefont {P.~G.~H.}\
  \bibnamefont {Sandars}}\ and\ \bibinfo {author} {\bibfnamefont
  {E.}~\bibnamefont {Lipworth}},\ }\href@noop {} {\bibfield  {journal}
  {\bibinfo  {journal} {Physical Review Letters}\ }\textbf {\bibinfo {volume}
  {13}},\ \bibinfo {pages} {718} (\bibinfo {year}
  {1964}{\natexlab{b}})}\BibitemShut {NoStop}%
\bibitem [{\citenamefont {Sandars}(1965)}]{Sandars1965}%
  \BibitemOpen
  \bibfield  {author} {\bibinfo {author} {\bibfnamefont {P.}~\bibnamefont
  {Sandars}},\ }\href {https://doi.org/10.1016/0031-9163(65)90583-4} {\bibfield
   {journal} {\bibinfo  {journal} {Physics Letters}\ }\textbf {\bibinfo
  {volume} {14}},\ \bibinfo {pages} {194} (\bibinfo {year} {1965})}\BibitemShut
  {NoStop}%
\bibitem [{\citenamefont {Flambaum}\ and\ \citenamefont
  {Ginges}(2002)}]{Flambaum2002}%
  \BibitemOpen
  \bibfield  {author} {\bibinfo {author} {\bibfnamefont {V.~V.}\ \bibnamefont
  {Flambaum}}\ and\ \bibinfo {author} {\bibfnamefont {J.~S.~M.}\ \bibnamefont
  {Ginges}},\ }\href {https://doi.org/10.1103/PhysRevA.65.032113} {\bibfield
  {journal} {\bibinfo  {journal} {Physical Review A}\ }\textbf {\bibinfo
  {volume} {65}},\ \bibinfo {pages} {032113} (\bibinfo {year}
  {2002})}\BibitemShut {NoStop}%
\bibitem [{\citenamefont {Ginges}\ and\ \citenamefont
  {Flambaum}(2004)}]{Ginges2004}%
  \BibitemOpen
  \bibfield  {author} {\bibinfo {author} {\bibfnamefont {J.}~\bibnamefont
  {Ginges}}\ and\ \bibinfo {author} {\bibfnamefont {V.}~\bibnamefont
  {Flambaum}},\ }\href {https://doi.org/10.1016/j.physrep.2004.03.005}
  {\bibfield  {journal} {\bibinfo  {journal} {Physics Reports}\ }\textbf
  {\bibinfo {volume} {397}},\ \bibinfo {pages} {63} (\bibinfo {year} {2004})},\
  \Eprint {https://arxiv.org/abs/0309054} {arXiv:0309054 [physics]}
  \BibitemShut {NoStop}%
\bibitem [{\citenamefont {Porsev}\ \emph {et~al.}(2011)\citenamefont {Porsev},
  \citenamefont {Ginges},\ and\ \citenamefont {Flambaum}}]{Porsev2011}%
  \BibitemOpen
  \bibfield  {author} {\bibinfo {author} {\bibfnamefont {S.}~\bibnamefont
  {Porsev}}, \bibinfo {author} {\bibfnamefont {J.}~\bibnamefont {Ginges}},\
  and\ \bibinfo {author} {\bibfnamefont {V.}~\bibnamefont {Flambaum}},\ }\href
  {https://doi.org/10.1103/PhysRevA.83.042507} {\bibfield  {journal} {\bibinfo
  {journal} {Physical Review A}\ }\textbf {\bibinfo {volume} {83}},\ \bibinfo
  {pages} {1} (\bibinfo {year} {2011})}\BibitemShut {NoStop}%
\bibitem [{\citenamefont {Flambaum}(2018)}]{Flambaum2018}%
  \BibitemOpen
  \bibfield  {author} {\bibinfo {author} {\bibfnamefont {V.~V.}\ \bibnamefont
  {Flambaum}},\ }\href {https://doi.org/10.1103/PhysRevA.98.043408} {\bibfield
  {journal} {\bibinfo  {journal} {Phys. Rev. A}\ }\textbf {\bibinfo {volume}
  {98}},\ \bibinfo {pages} {043408} (\bibinfo {year} {2018})}\BibitemShut
  {NoStop}%
\bibitem [{\citenamefont {Tan}\ \emph {et~al.}(2019)\citenamefont {Tan},
  \citenamefont {Flambaum},\ and\ \citenamefont {Samsonov}}]{Tan2019}%
  \BibitemOpen
  \bibfield  {author} {\bibinfo {author} {\bibfnamefont {H.~B.~T.}\
  \bibnamefont {Tan}}, \bibinfo {author} {\bibfnamefont {V.~V.}\ \bibnamefont
  {Flambaum}},\ and\ \bibinfo {author} {\bibfnamefont {I.~B.}\ \bibnamefont
  {Samsonov}},\ }\bibfield  {journal} {\bibinfo  {journal} {Physical Review A}\
  }\textbf {\bibinfo {volume} {99}},\ \href
  {https://doi.org/10.1103/physreva.99.013430} {10.1103/physreva.99.013430}
  (\bibinfo {year} {2019})\BibitemShut {NoStop}%
\bibitem [{\citenamefont {Flambaum}\ and\ \citenamefont
  {Samsonov}(2020)}]{Flambaum2020Solids}%
  \BibitemOpen
  \bibfield  {author} {\bibinfo {author} {\bibfnamefont {V.~V.}\ \bibnamefont
  {Flambaum}}\ and\ \bibinfo {author} {\bibfnamefont {I.~B.}\ \bibnamefont
  {Samsonov}},\ }\bibfield  {journal} {\bibinfo  {journal} {Physical Review
  Research}\ }\textbf {\bibinfo {volume} {2}},\ \href
  {https://doi.org/10.1103/physrevresearch.2.023042}
  {10.1103/physrevresearch.2.023042} (\bibinfo {year} {2020})\BibitemShut
  {NoStop}%
\bibitem [{\citenamefont {Sushkov}\ \emph {et~al.}(1985)\citenamefont
  {Sushkov}, \citenamefont {Flambaum},\ and\ \citenamefont
  {Khriplovich}}]{Sushkov1985}%
  \BibitemOpen
  \bibfield  {author} {\bibinfo {author} {\bibfnamefont {O.~P.}\ \bibnamefont
  {Sushkov}}, \bibinfo {author} {\bibfnamefont {V.~V.}\ \bibnamefont
  {Flambaum}},\ and\ \bibinfo {author} {\bibfnamefont {I.~B.}\ \bibnamefont
  {Khriplovich}},\ }\href
  {http://www.jetp.ac.ru/cgi-bin/dn/e{\_}060{\_}05{\_}0873.pdf} {\bibfield
  {journal} {\bibinfo  {journal} {JETP}\ }\textbf {\bibinfo {volume} {60}},\
  \bibinfo {pages} {873} (\bibinfo {year} {1985})}\BibitemShut {NoStop}%
\bibitem [{\citenamefont {Khriplovich}\ and\ \citenamefont
  {Lamoreaux}(1997)}]{Khriplovich1997}%
  \BibitemOpen
  \bibfield  {author} {\bibinfo {author} {\bibfnamefont {I.~B.}\ \bibnamefont
  {Khriplovich}}\ and\ \bibinfo {author} {\bibfnamefont {S.~K.}\ \bibnamefont
  {Lamoreaux}},\ }\href {https://doi.org/10.1007/978-3-642-60838-4} {\emph
  {\bibinfo {title} {{CP Violation Without Strangeness}}}}\ (\bibinfo
  {publisher} {Springer-Verlag Berlin Heidelberg},\ \bibinfo {year}
  {1997})\BibitemShut {NoStop}%
\bibitem [{\citenamefont {Petrov}\ \emph {et~al.}(2018)\citenamefont {Petrov},
  \citenamefont {Skripnikov}, \citenamefont {Titov},\ and\ \citenamefont
  {Flambaum}}]{Petrov2018}%
  \BibitemOpen
  \bibfield  {author} {\bibinfo {author} {\bibfnamefont {A.~N.}\ \bibnamefont
  {Petrov}}, \bibinfo {author} {\bibfnamefont {L.~V.}\ \bibnamefont
  {Skripnikov}}, \bibinfo {author} {\bibfnamefont {A.~V.}\ \bibnamefont
  {Titov}},\ and\ \bibinfo {author} {\bibfnamefont {V.~V.}\ \bibnamefont
  {Flambaum}},\ }\bibfield  {journal} {\bibinfo  {journal} {Physical Review A}\
  }\textbf {\bibinfo {volume} {98}},\ \href
  {https://doi.org/10.1103/physreva.98.042502} {10.1103/physreva.98.042502}
  (\bibinfo {year} {2018})\BibitemShut {NoStop}%
\bibitem [{\citenamefont {Kurchavov}\ and\ \citenamefont
  {Petrov}(2020)}]{Kurchavov2020}%
  \BibitemOpen
  \bibfield  {author} {\bibinfo {author} {\bibfnamefont {I.~P.}\ \bibnamefont
  {Kurchavov}}\ and\ \bibinfo {author} {\bibfnamefont {A.~N.}\ \bibnamefont
  {Petrov}},\ }\bibfield  {journal} {\bibinfo  {journal} {Physical Review A}\
  }\textbf {\bibinfo {volume} {102}},\ \href
  {https://doi.org/10.1103/physreva.102.032805} {10.1103/physreva.102.032805}
  (\bibinfo {year} {2020})\BibitemShut {NoStop}%
\bibitem [{\citenamefont {Baturo}\ \emph {et~al.}(2021)\citenamefont {Baturo},
  \citenamefont {Rupasinghe}, \citenamefont {Sears}, \citenamefont {Mawhorter},
  \citenamefont {Grabow},\ and\ \citenamefont {Petrov}}]{Baturo2021}%
  \BibitemOpen
  \bibfield  {author} {\bibinfo {author} {\bibfnamefont {V.~V.}\ \bibnamefont
  {Baturo}}, \bibinfo {author} {\bibfnamefont {P.~M.}\ \bibnamefont
  {Rupasinghe}}, \bibinfo {author} {\bibfnamefont {T.~J.}\ \bibnamefont
  {Sears}}, \bibinfo {author} {\bibfnamefont {R.~J.}\ \bibnamefont
  {Mawhorter}}, \bibinfo {author} {\bibfnamefont {J.-U.}\ \bibnamefont
  {Grabow}},\ and\ \bibinfo {author} {\bibfnamefont {A.~N.}\ \bibnamefont
  {Petrov}},\ }\bibfield  {journal} {\bibinfo  {journal} {Physical Review A}\
  }\textbf {\bibinfo {volume} {104}},\ \href
  {https://doi.org/10.1103/physreva.104.012811} {10.1103/physreva.104.012811}
  (\bibinfo {year} {2021})\BibitemShut {NoStop}%
\bibitem [{\citenamefont {Petrov}\ and\ \citenamefont
  {Zakharova}(2021)}]{Petrov2021}%
  \BibitemOpen
  \bibfield  {author} {\bibinfo {author} {\bibfnamefont {A.}~\bibnamefont
  {Petrov}}\ and\ \bibinfo {author} {\bibfnamefont {A.}~\bibnamefont
  {Zakharova}},\ }\href {http://arxiv.org/abs/2111.02772} {\bibfield  {journal}
  {\bibinfo  {journal} {arXiv:2111.02772}\ } (\bibinfo {year} {2021})},\
  \Eprint {https://arxiv.org/abs/2111.02772} {arXiv:2111.02772} \BibitemShut
  {NoStop}%
\bibitem [{\citenamefont {Kurchavov}\ and\ \citenamefont
  {Petrov}(2021)}]{Kurchavov2021}%
  \BibitemOpen
  \bibfield  {author} {\bibinfo {author} {\bibfnamefont {I.~P.}\ \bibnamefont
  {Kurchavov}}\ and\ \bibinfo {author} {\bibfnamefont {A.~N.}\ \bibnamefont
  {Petrov}},\ }\bibfield  {journal} {\bibinfo  {journal} {Optics and
  Spectroscopy}\ }\href {https://doi.org/10.1134/s0030400x21070109}
  {10.1134/s0030400x21070109} (\bibinfo {year} {2021})\BibitemShut {NoStop}%
\bibitem [{\citenamefont {Cairncross}\ \emph {et~al.}(2017)\citenamefont
  {Cairncross}, \citenamefont {Gresh}, \citenamefont {Grau}, \citenamefont
  {Cossel}, \citenamefont {Roussy}, \citenamefont {Ni}, \citenamefont {Zhou},
  \citenamefont {Ye},\ and\ \citenamefont {Cornell}}]{Cairncross2017}%
  \BibitemOpen
  \bibfield  {author} {\bibinfo {author} {\bibfnamefont {W.~B.}\ \bibnamefont
  {Cairncross}}, \bibinfo {author} {\bibfnamefont {D.~N.}\ \bibnamefont
  {Gresh}}, \bibinfo {author} {\bibfnamefont {M.}~\bibnamefont {Grau}},
  \bibinfo {author} {\bibfnamefont {K.~C.}\ \bibnamefont {Cossel}}, \bibinfo
  {author} {\bibfnamefont {T.~S.}\ \bibnamefont {Roussy}}, \bibinfo {author}
  {\bibfnamefont {Y.}~\bibnamefont {Ni}}, \bibinfo {author} {\bibfnamefont
  {Y.}~\bibnamefont {Zhou}}, \bibinfo {author} {\bibfnamefont {J.}~\bibnamefont
  {Ye}},\ and\ \bibinfo {author} {\bibfnamefont {E.~A.}\ \bibnamefont
  {Cornell}},\ }\href {https://doi.org/10.1103/PhysRevLett.119.153001}
  {\bibfield  {journal} {\bibinfo  {journal} {Physical Review Letters}\
  }\textbf {\bibinfo {volume} {119}},\ \bibinfo {pages} {153001} (\bibinfo
  {year} {2017})}\BibitemShut {NoStop}%
\bibitem [{\citenamefont {Petrov}(2018)}]{Petrov2018b}%
  \BibitemOpen
  \bibfield  {author} {\bibinfo {author} {\bibfnamefont {A.~N.}\ \bibnamefont
  {Petrov}},\ }\bibfield  {journal} {\bibinfo  {journal} {Physical Review A}\
  }\textbf {\bibinfo {volume} {97}},\ \href
  {https://doi.org/10.1103/physreva.97.052504} {10.1103/physreva.97.052504}
  (\bibinfo {year} {2018})\BibitemShut {NoStop}%
\bibitem [{\citenamefont {Flambaum}(1994)}]{Flambaum1994}%
  \BibitemOpen
  \bibfield  {author} {\bibinfo {author} {\bibfnamefont {V.~V.}\ \bibnamefont
  {Flambaum}},\ }\href {https://doi.org/10.1016/0370-2693(94)90646-7}
  {\bibfield  {journal} {\bibinfo  {journal} {Physics Letters B}\ }\textbf
  {\bibinfo {volume} {320}},\ \bibinfo {pages} {211} (\bibinfo {year}
  {1994})}\BibitemShut {NoStop}%
\bibitem [{\citenamefont {Flambaum}\ \emph {et~al.}(2014)\citenamefont
  {Flambaum}, \citenamefont {DeMille},\ and\ \citenamefont
  {Kozlov}}]{Flambaum2014}%
  \BibitemOpen
  \bibfield  {author} {\bibinfo {author} {\bibfnamefont {V.~V.}\ \bibnamefont
  {Flambaum}}, \bibinfo {author} {\bibfnamefont {D.}~\bibnamefont {DeMille}},\
  and\ \bibinfo {author} {\bibfnamefont {M.~G.}\ \bibnamefont {Kozlov}},\
  }\href {https://doi.org/10.1103/PhysRevLett.113.103003} {\bibfield  {journal}
  {\bibinfo  {journal} {Physical Review Letters}\ }\textbf {\bibinfo {volume}
  {113}},\ \bibinfo {pages} {103003} (\bibinfo {year} {2014})}\BibitemShut
  {NoStop}%
\bibitem [{\citenamefont {Lackenby}\ and\ \citenamefont
  {Flambaum}(2018)}]{Lackenby2018}%
  \BibitemOpen
  \bibfield  {author} {\bibinfo {author} {\bibfnamefont {B.~G.}\ \bibnamefont
  {Lackenby}}\ and\ \bibinfo {author} {\bibfnamefont {V.~V.}\ \bibnamefont
  {Flambaum}},\ }\href {https://doi.org/10.1103/PhysRevD.98.115019} {\bibfield
  {journal} {\bibinfo  {journal} {Physical Review D}\ }\textbf {\bibinfo
  {volume} {98}},\ \bibinfo {pages} {115019} (\bibinfo {year} {2018})},\
  \Eprint {https://arxiv.org/abs/1810.02477} {arXiv:1810.02477} \BibitemShut
  {NoStop}%
\bibitem [{\citenamefont {Kudashov}\ \emph {et~al.}(2014)\citenamefont
  {Kudashov}, \citenamefont {Petrov}, \citenamefont {Skripnikov}, \citenamefont
  {Mosyagin}, \citenamefont {Isaev}, \citenamefont {Berger},\ and\
  \citenamefont {Titov}}]{Kudashov2014}%
  \BibitemOpen
  \bibfield  {author} {\bibinfo {author} {\bibfnamefont {A.~D.}\ \bibnamefont
  {Kudashov}}, \bibinfo {author} {\bibfnamefont {A.~N.}\ \bibnamefont
  {Petrov}}, \bibinfo {author} {\bibfnamefont {L.~V.}\ \bibnamefont
  {Skripnikov}}, \bibinfo {author} {\bibfnamefont {N.~S.}\ \bibnamefont
  {Mosyagin}}, \bibinfo {author} {\bibfnamefont {T.~A.}\ \bibnamefont {Isaev}},
  \bibinfo {author} {\bibfnamefont {R.}~\bibnamefont {Berger}},\ and\ \bibinfo
  {author} {\bibfnamefont {A.~V.}\ \bibnamefont {Titov}},\ }\href
  {https://doi.org/10.1103/PhysRevA.90.052513} {\bibfield  {journal} {\bibinfo
  {journal} {Physical Review A}\ }\textbf {\bibinfo {volume} {90}},\ \bibinfo
  {pages} {052513} (\bibinfo {year} {2014})}\BibitemShut {NoStop}%
\bibitem [{\citenamefont {Gaul}\ and\ \citenamefont {Berger}(2017)}]{Gaul2017}%
  \BibitemOpen
  \bibfield  {author} {\bibinfo {author} {\bibfnamefont {K.}~\bibnamefont
  {Gaul}}\ and\ \bibinfo {author} {\bibfnamefont {R.}~\bibnamefont {Berger}},\
  }\href {https://doi.org/10.1063/1.4985567} {\bibfield  {journal} {\bibinfo
  {journal} {The Journal of Chemical Physics}\ }\textbf {\bibinfo {volume}
  {147}},\ \bibinfo {pages} {014109} (\bibinfo {year} {2017})}\BibitemShut
  {NoStop}%
\bibitem [{\citenamefont {Gaul}\ and\ \citenamefont {Berger}(2020)}]{Gaul2020}%
  \BibitemOpen
  \bibfield  {author} {\bibinfo {author} {\bibfnamefont {K.}~\bibnamefont
  {Gaul}}\ and\ \bibinfo {author} {\bibfnamefont {R.}~\bibnamefont {Berger}},\
  }\href {https://doi.org/10.1103/PhysRevA.101.012508} {\bibfield  {journal}
  {\bibinfo  {journal} {Physical Review A}\ }\textbf {\bibinfo {volume}
  {101}},\ \bibinfo {pages} {012508} (\bibinfo {year} {2020})}\BibitemShut
  {NoStop}%
\bibitem [{\citenamefont {Skripnikov}(2016)}]{Skripnikov2016}%
  \BibitemOpen
  \bibfield  {author} {\bibinfo {author} {\bibfnamefont {L.~V.}\ \bibnamefont
  {Skripnikov}},\ }\href {https://doi.org/10.1063/1.4968229} {\bibfield
  {journal} {\bibinfo  {journal} {The Journal of Chemical Physics}\ }\textbf
  {\bibinfo {volume} {145}},\ \bibinfo {pages} {214301} (\bibinfo {year}
  {2016})}\BibitemShut {NoStop}%
\bibitem [{\citenamefont {Sasmal}\ \emph {et~al.}(2016)\citenamefont {Sasmal},
  \citenamefont {Pathak}, \citenamefont {Nayak}, \citenamefont {Vaval},\ and\
  \citenamefont {Pal}}]{Sudip:2016b}%
  \BibitemOpen
  \bibfield  {author} {\bibinfo {author} {\bibfnamefont {S.}~\bibnamefont
  {Sasmal}}, \bibinfo {author} {\bibfnamefont {H.}~\bibnamefont {Pathak}},
  \bibinfo {author} {\bibfnamefont {M.~K.}\ \bibnamefont {Nayak}}, \bibinfo
  {author} {\bibfnamefont {N.}~\bibnamefont {Vaval}},\ and\ \bibinfo {author}
  {\bibfnamefont {S.}~\bibnamefont {Pal}},\ }\href
  {https://doi.org/10.1103/PhysRevA.93.062506} {\bibfield  {journal} {\bibinfo
  {journal} {Phys. Rev. A}\ }\textbf {\bibinfo {volume} {93}},\ \bibinfo
  {pages} {062506} (\bibinfo {year} {2016})}\BibitemShut {NoStop}%
\bibitem [{\citenamefont {Abe}\ \emph {et~al.}(2018)\citenamefont {Abe},
  \citenamefont {Prasannaa},\ and\ \citenamefont {Das}}]{Abe:2018}%
  \BibitemOpen
  \bibfield  {author} {\bibinfo {author} {\bibfnamefont {M.}~\bibnamefont
  {Abe}}, \bibinfo {author} {\bibfnamefont {V.~S.}\ \bibnamefont {Prasannaa}},\
  and\ \bibinfo {author} {\bibfnamefont {B.~P.}\ \bibnamefont {Das}},\ }\href
  {https://doi.org/10.1103/PhysRevA.97.032515} {\bibfield  {journal} {\bibinfo
  {journal} {Phys. Rev. A}\ }\textbf {\bibinfo {volume} {97}},\ \bibinfo
  {pages} {032515} (\bibinfo {year} {2018})}\BibitemShut {NoStop}%
\bibitem [{\citenamefont {Zhang}\ \emph {et~al.}(2021)\citenamefont {Zhang},
  \citenamefont {Zheng},\ and\ \citenamefont {Cheng}}]{Zhang:2021}%
  \BibitemOpen
  \bibfield  {author} {\bibinfo {author} {\bibfnamefont {C.}~\bibnamefont
  {Zhang}}, \bibinfo {author} {\bibfnamefont {X.}~\bibnamefont {Zheng}},\ and\
  \bibinfo {author} {\bibfnamefont {L.}~\bibnamefont {Cheng}},\ }\href
  {https://doi.org/10.1103/PhysRevA.104.012814} {\bibfield  {journal} {\bibinfo
   {journal} {Phys. Rev. A}\ }\textbf {\bibinfo {volume} {104}},\ \bibinfo
  {pages} {012814} (\bibinfo {year} {2021})}\BibitemShut {NoStop}%
\bibitem [{\citenamefont {Skripnikov}\ and\ \citenamefont
  {Titov}(2015)}]{Skripnikov2015ThO}%
  \BibitemOpen
  \bibfield  {author} {\bibinfo {author} {\bibfnamefont {L.~V.}\ \bibnamefont
  {Skripnikov}}\ and\ \bibinfo {author} {\bibfnamefont {A.~V.}\ \bibnamefont
  {Titov}},\ }\href {https://doi.org/10.1063/1.4904877} {\bibfield  {journal}
  {\bibinfo  {journal} {The Journal of Chemical Physics}\ }\textbf {\bibinfo
  {volume} {142}},\ \bibinfo {pages} {024301} (\bibinfo {year}
  {2015})}\BibitemShut {NoStop}%
\bibitem [{\citenamefont {Haase}\ \emph {et~al.}(2020)\citenamefont {Haase},
  \citenamefont {Eliav}, \citenamefont {Iliaš},\ and\ \citenamefont
  {Borschevsky}}]{HaaEliIli20}%
  \BibitemOpen
  \bibfield  {author} {\bibinfo {author} {\bibfnamefont {P.~A.~B.}\
  \bibnamefont {Haase}}, \bibinfo {author} {\bibfnamefont {E.}~\bibnamefont
  {Eliav}}, \bibinfo {author} {\bibfnamefont {M.}~\bibnamefont {Iliaš}},\ and\
  \bibinfo {author} {\bibfnamefont {A.}~\bibnamefont {Borschevsky}},\ }\href
  {https://doi.org/10.1021/acs.jpca.0c00877} {\bibfield  {journal} {\bibinfo
  {journal} {The Journal of Physical Chemistry A}\ }\textbf {\bibinfo {volume}
  {124}},\ \bibinfo {pages} {3157} (\bibinfo {year} {2020})}\BibitemShut
  {NoStop}%
\bibitem [{\citenamefont {Haase}\ \emph {et~al.}(2021)\citenamefont {Haase},
  \citenamefont {Doeglas}, \citenamefont {Boeschoten}, \citenamefont {Eliav},
  \citenamefont {Iliaš}, \citenamefont {Aggarwal}, \citenamefont {Bethlem},
  \citenamefont {Borschevsky}, \citenamefont {Esajas}, \citenamefont {Hao},
  \citenamefont {Hoekstra}, \citenamefont {Marshall}, \citenamefont
  {Meijknecht}, \citenamefont {Mooij}, \citenamefont {Steinebach},
  \citenamefont {Timmermans}, \citenamefont {Touwen}, \citenamefont {Ubachs},
  \citenamefont {Willmann},\ and\ \citenamefont {Yin}}]{HaaDoeBoe21}%
  \BibitemOpen
  \bibfield  {author} {\bibinfo {author} {\bibfnamefont {P.~A.~B.}\
  \bibnamefont {Haase}}, \bibinfo {author} {\bibfnamefont {D.~J.}\ \bibnamefont
  {Doeglas}}, \bibinfo {author} {\bibfnamefont {A.}~\bibnamefont {Boeschoten}},
  \bibinfo {author} {\bibfnamefont {E.}~\bibnamefont {Eliav}}, \bibinfo
  {author} {\bibfnamefont {M.}~\bibnamefont {Iliaš}}, \bibinfo {author}
  {\bibfnamefont {P.}~\bibnamefont {Aggarwal}}, \bibinfo {author}
  {\bibfnamefont {H.~L.}\ \bibnamefont {Bethlem}}, \bibinfo {author}
  {\bibfnamefont {A.}~\bibnamefont {Borschevsky}}, \bibinfo {author}
  {\bibfnamefont {K.}~\bibnamefont {Esajas}}, \bibinfo {author} {\bibfnamefont
  {Y.}~\bibnamefont {Hao}}, \bibinfo {author} {\bibfnamefont {S.}~\bibnamefont
  {Hoekstra}}, \bibinfo {author} {\bibfnamefont {V.~R.}\ \bibnamefont
  {Marshall}}, \bibinfo {author} {\bibfnamefont {T.~B.}\ \bibnamefont
  {Meijknecht}}, \bibinfo {author} {\bibfnamefont {M.~C.}\ \bibnamefont
  {Mooij}}, \bibinfo {author} {\bibfnamefont {K.}~\bibnamefont {Steinebach}},
  \bibinfo {author} {\bibfnamefont {R.~G.~E.}\ \bibnamefont {Timmermans}},
  \bibinfo {author} {\bibfnamefont {A.~P.}\ \bibnamefont {Touwen}}, \bibinfo
  {author} {\bibfnamefont {W.}~\bibnamefont {Ubachs}}, \bibinfo {author}
  {\bibfnamefont {L.}~\bibnamefont {Willmann}},\ and\ \bibinfo {author}
  {\bibfnamefont {Y.}~\bibnamefont {Yin}},\ }\href
  {https://doi.org/10.1063/5.0047344} {\bibfield  {journal} {\bibinfo
  {journal} {The Journal of Chemical Physics}\ }\textbf {\bibinfo {volume}
  {155}},\ \bibinfo {pages} {034309} (\bibinfo {year} {2021})}\BibitemShut
  {NoStop}%
\bibitem [{\citenamefont {Abe}\ \emph {et~al.}(2014)\citenamefont {Abe},
  \citenamefont {Gopakumar}, \citenamefont {Hada}, \citenamefont {Das},
  \citenamefont {Tatewaki},\ and\ \citenamefont {Mukherjee}}]{Abe2014}%
  \BibitemOpen
  \bibfield  {author} {\bibinfo {author} {\bibfnamefont {M.}~\bibnamefont
  {Abe}}, \bibinfo {author} {\bibfnamefont {G.}~\bibnamefont {Gopakumar}},
  \bibinfo {author} {\bibfnamefont {M.}~\bibnamefont {Hada}}, \bibinfo {author}
  {\bibfnamefont {B.~P.}\ \bibnamefont {Das}}, \bibinfo {author} {\bibfnamefont
  {H.}~\bibnamefont {Tatewaki}},\ and\ \bibinfo {author} {\bibfnamefont
  {D.}~\bibnamefont {Mukherjee}},\ }\href
  {https://doi.org/10.1103/PhysRevA.90.022501} {\bibfield  {journal} {\bibinfo
  {journal} {Physical Review A}\ }\textbf {\bibinfo {volume} {90}},\ \bibinfo
  {pages} {022501} (\bibinfo {year} {2014})}\BibitemShut {NoStop}%
\bibitem [{\citenamefont {Prasannaa}\ \emph {et~al.}(2020)\citenamefont
  {Prasannaa}, \citenamefont {Sahoo}, \citenamefont {Abe},\ and\ \citenamefont
  {Das}}]{Prasannaa2020}%
  \BibitemOpen
  \bibfield  {author} {\bibinfo {author} {\bibfnamefont {V.~S.}\ \bibnamefont
  {Prasannaa}}, \bibinfo {author} {\bibfnamefont {B.~K.}\ \bibnamefont
  {Sahoo}}, \bibinfo {author} {\bibfnamefont {M.}~\bibnamefont {Abe}},\ and\
  \bibinfo {author} {\bibfnamefont {B.~P.}\ \bibnamefont {Das}},\ }\href
  {https://doi.org/10.3390/sym12050811} {\bibfield  {journal} {\bibinfo
  {journal} {Symmetry}\ }\textbf {\bibinfo {volume} {12}},\ \bibinfo {pages}
  {811} (\bibinfo {year} {2020})}\BibitemShut {NoStop}%
\bibitem [{\citenamefont {Zakharova}\ and\ \citenamefont
  {Petrov}(2021)}]{Zakharova2021RaOH}%
  \BibitemOpen
  \bibfield  {author} {\bibinfo {author} {\bibfnamefont {A.}~\bibnamefont
  {Zakharova}}\ and\ \bibinfo {author} {\bibfnamefont {A.}~\bibnamefont
  {Petrov}},\ }\bibfield  {journal} {\bibinfo  {journal} {Physical Review A}\
  }\textbf {\bibinfo {volume} {103}},\ \href
  {https://doi.org/10.1103/physreva.103.032819} {10.1103/physreva.103.032819}
  (\bibinfo {year} {2021})\BibitemShut {NoStop}%
\bibitem [{\citenamefont {Zakharova}\ \emph {et~al.}(2021)\citenamefont
  {Zakharova}, \citenamefont {Kurchavov},\ and\ \citenamefont
  {Petrov}}]{Zakharova2021YbOH}%
  \BibitemOpen
  \bibfield  {author} {\bibinfo {author} {\bibfnamefont {A.}~\bibnamefont
  {Zakharova}}, \bibinfo {author} {\bibfnamefont {I.}~\bibnamefont
  {Kurchavov}},\ and\ \bibinfo {author} {\bibfnamefont {A.}~\bibnamefont
  {Petrov}},\ }\href {https://doi.org/10.1063/5.0069281} {\bibfield  {journal}
  {\bibinfo  {journal} {The Journal of Chemical Physics}\ }\textbf {\bibinfo
  {volume} {155}},\ \bibinfo {pages} {164301} (\bibinfo {year}
  {2021})}\BibitemShut {NoStop}%
\bibitem [{\citenamefont {Oleynichenko}\ \emph {et~al.}(2022)\citenamefont
  {Oleynichenko}, \citenamefont {Skripnikov}, \citenamefont {Zaitsevskii},\
  and\ \citenamefont {Flambaum}}]{Oleynichenko2022}%
  \BibitemOpen
  \bibfield  {author} {\bibinfo {author} {\bibfnamefont {A.~V.}\ \bibnamefont
  {Oleynichenko}}, \bibinfo {author} {\bibfnamefont {L.~V.}\ \bibnamefont
  {Skripnikov}}, \bibinfo {author} {\bibfnamefont {A.~V.}\ \bibnamefont
  {Zaitsevskii}},\ and\ \bibinfo {author} {\bibfnamefont {V.~V.}\ \bibnamefont
  {Flambaum}},\ }\bibfield  {journal} {\bibinfo  {journal} {Physical Review A}\
  }\textbf {\bibinfo {volume} {105}},\ \href
  {https://doi.org/10.1103/physreva.105.022825} {10.1103/physreva.105.022825}
  (\bibinfo {year} {2022})\BibitemShut {NoStop}%
\bibitem [{\citenamefont {Fleig}(2017{\natexlab{a}})}]{Fleig2017HfF}%
  \BibitemOpen
  \bibfield  {author} {\bibinfo {author} {\bibfnamefont {T.}~\bibnamefont
  {Fleig}},\ }\bibfield  {journal} {\bibinfo  {journal} {Physical Review A}\
  }\textbf {\bibinfo {volume} {96}},\ \href
  {https://doi.org/10.1103/physreva.96.040502} {10.1103/physreva.96.040502}
  (\bibinfo {year} {2017}{\natexlab{a}})\BibitemShut {NoStop}%
\bibitem [{\citenamefont {Skripnikov}(2017)}]{Skripnikov2017}%
  \BibitemOpen
  \bibfield  {author} {\bibinfo {author} {\bibfnamefont {L.~V.}\ \bibnamefont
  {Skripnikov}},\ }\href {https://doi.org/10.1063/1.4993622} {\bibfield
  {journal} {\bibinfo  {journal} {The Journal of Chemical Physics}\ }\textbf
  {\bibinfo {volume} {147}},\ \bibinfo {pages} {021101} (\bibinfo {year}
  {2017})}\BibitemShut {NoStop}%
\bibitem [{\citenamefont {Kozlov}(1997)}]{Kozlov1997}%
  \BibitemOpen
  \bibfield  {author} {\bibinfo {author} {\bibfnamefont {M.~G.}\ \bibnamefont
  {Kozlov}},\ }\href {https://doi.org/10.1088/0953-4075/30/18/003} {\bibfield
  {journal} {\bibinfo  {journal} {Journal of Physics B: Atomic, Molecular and
  Optical Physics}\ }\textbf {\bibinfo {volume} {30}},\ \bibinfo {pages} {L607}
  (\bibinfo {year} {1997})}\BibitemShut {NoStop}%
\bibitem [{\citenamefont {Skripnikov}(2020)}]{Skripnikov2020BW}%
  \BibitemOpen
  \bibfield  {author} {\bibinfo {author} {\bibfnamefont {L.~V.}\ \bibnamefont
  {Skripnikov}},\ }\href {https://doi.org/10.1063/5.0024103} {\bibfield
  {journal} {\bibinfo  {journal} {The Journal of Chemical Physics}\ }\textbf
  {\bibinfo {volume} {153}},\ \bibinfo {pages} {114114} (\bibinfo {year}
  {2020})}\BibitemShut {NoStop}%
\bibitem [{\citenamefont {Peruzzo}\ \emph {et~al.}(2014)\citenamefont
  {Peruzzo}, \citenamefont {McClean}, \citenamefont {Shadbolt}, \citenamefont
  {Yung}, \citenamefont {Zhou}, \citenamefont {Love}, \citenamefont
  {Aspuru-Guzik},\ and\ \citenamefont {O'Brien}}]{Peruzzo2014}%
  \BibitemOpen
  \bibfield  {author} {\bibinfo {author} {\bibfnamefont {A.}~\bibnamefont
  {Peruzzo}}, \bibinfo {author} {\bibfnamefont {J.}~\bibnamefont {McClean}},
  \bibinfo {author} {\bibfnamefont {P.}~\bibnamefont {Shadbolt}}, \bibinfo
  {author} {\bibfnamefont {M.-H.}\ \bibnamefont {Yung}}, \bibinfo {author}
  {\bibfnamefont {X.-Q.}\ \bibnamefont {Zhou}}, \bibinfo {author}
  {\bibfnamefont {P.~J.}\ \bibnamefont {Love}}, \bibinfo {author}
  {\bibfnamefont {A.}~\bibnamefont {Aspuru-Guzik}},\ and\ \bibinfo {author}
  {\bibfnamefont {J.~L.}\ \bibnamefont {O'Brien}},\ }\bibfield  {journal}
  {\bibinfo  {journal} {Nature Communications}\ }\textbf {\bibinfo {volume}
  {5}},\ \href {https://doi.org/10.1038/ncomms5213} {10.1038/ncomms5213}
  (\bibinfo {year} {2014})\BibitemShut {NoStop}%
\bibitem [{\citenamefont {Villela}\ \emph {et~al.}(2021)\citenamefont
  {Villela}, \citenamefont {Prasannaa},\ and\ \citenamefont
  {Das}}]{Villela2021}%
  \BibitemOpen
  \bibfield  {author} {\bibinfo {author} {\bibfnamefont {R.}~\bibnamefont
  {Villela}}, \bibinfo {author} {\bibfnamefont {V.~S.}\ \bibnamefont
  {Prasannaa}},\ and\ \bibinfo {author} {\bibfnamefont {B.~P.}\ \bibnamefont
  {Das}},\ }\href {https://doi.org/10.48550/ARXIV.2109.12583} {\bibinfo {title}
  {Ionization energies in lithium and boron atoms using the variational quantum
  eigensolver algorithm}} (\bibinfo {year} {2021})\BibitemShut {NoStop}%
\bibitem [{\citenamefont {Engel}\ \emph
  {et~al.}(2013{\natexlab{b}})\citenamefont {Engel}, \citenamefont
  {Ramsey-Musolf},\ and\ \citenamefont {van Kolck}}]{Engel2013}%
  \BibitemOpen
  \bibfield  {author} {\bibinfo {author} {\bibfnamefont {J.}~\bibnamefont
  {Engel}}, \bibinfo {author} {\bibfnamefont {M.~J.}\ \bibnamefont
  {Ramsey-Musolf}},\ and\ \bibinfo {author} {\bibfnamefont {U.}~\bibnamefont
  {van Kolck}},\ }\href {https://doi.org/10.1016/j.ppnp.2013.03.003} {\bibfield
   {journal} {\bibinfo  {journal} {Progress in Particle and Nuclear Physics}\
  }\textbf {\bibinfo {volume} {71}},\ \bibinfo {pages} {21} (\bibinfo {year}
  {2013}{\natexlab{b}})}\BibitemShut {NoStop}%
\bibitem [{\citenamefont {Auerbach}\ \emph {et~al.}(1996)\citenamefont
  {Auerbach}, \citenamefont {Flambaum},\ and\ \citenamefont
  {Spevak}}]{Auerbach1996}%
  \BibitemOpen
  \bibfield  {author} {\bibinfo {author} {\bibfnamefont {N.}~\bibnamefont
  {Auerbach}}, \bibinfo {author} {\bibfnamefont {V.~V.}\ \bibnamefont
  {Flambaum}},\ and\ \bibinfo {author} {\bibfnamefont {V.}~\bibnamefont
  {Spevak}},\ }\href {https://doi.org/10.1103/PhysRevLett.76.4316} {\bibfield
  {journal} {\bibinfo  {journal} {Physical Review Letters}\ }\textbf {\bibinfo
  {volume} {76}},\ \bibinfo {pages} {4316} (\bibinfo {year}
  {1996})}\BibitemShut {NoStop}%
\bibitem [{\citenamefont {Moses}\ \emph {et~al.}(2017)\citenamefont {Moses},
  \citenamefont {Covey}, \citenamefont {Miecnikowski}, \citenamefont {Jin},\
  and\ \citenamefont {Ye}}]{Moses2017}%
  \BibitemOpen
  \bibfield  {author} {\bibinfo {author} {\bibfnamefont {S.~A.}\ \bibnamefont
  {Moses}}, \bibinfo {author} {\bibfnamefont {J.~P.}\ \bibnamefont {Covey}},
  \bibinfo {author} {\bibfnamefont {M.~T.}\ \bibnamefont {Miecnikowski}},
  \bibinfo {author} {\bibfnamefont {D.~S.}\ \bibnamefont {Jin}},\ and\ \bibinfo
  {author} {\bibfnamefont {J.}~\bibnamefont {Ye}},\ }\href@noop {} {\bibfield
  {journal} {\bibinfo  {journal} {Nature Physics}\ }\textbf {\bibinfo {volume}
  {13}},\ \bibinfo {pages} {13} (\bibinfo {year} {2017})}\BibitemShut {NoStop}%
\bibitem [{\citenamefont {Isaev}\ and\ \citenamefont
  {Berger}(2018)}]{Isaev2018}%
  \BibitemOpen
  \bibfield  {author} {\bibinfo {author} {\bibfnamefont {T.~A.}\ \bibnamefont
  {Isaev}}\ and\ \bibinfo {author} {\bibfnamefont {R.}~\bibnamefont {Berger}},\
  }\href {https://doi.org/10.2533/chimia.2018.375} {\bibfield  {journal}
  {\bibinfo  {journal} {CHIMIA International Journal for Chemistry}\ }\textbf
  {\bibinfo {volume} {72}},\ \bibinfo {pages} {375} (\bibinfo {year}
  {2018})}\BibitemShut {NoStop}%
\bibitem [{\citenamefont {Hutzler}(2020{\natexlab{a}})}]{Hutzler2020Review}%
  \BibitemOpen
  \bibfield  {author} {\bibinfo {author} {\bibfnamefont {N.~R.}\ \bibnamefont
  {Hutzler}},\ }\href {https://doi.org/10.1088/2058-9565/abb9c5} {\bibfield
  {journal} {\bibinfo  {journal} {Quantum Science and Technology}\ }\textbf
  {\bibinfo {volume} {5}},\ \bibinfo {pages} {044011} (\bibinfo {year}
  {2020}{\natexlab{a}})}\BibitemShut {NoStop}%
\bibitem [{\citenamefont {Fitch}\ and\ \citenamefont
  {Tarbutt}(2021)}]{Fitch2021Review}%
  \BibitemOpen
  \bibfield  {author} {\bibinfo {author} {\bibfnamefont {N.}~\bibnamefont
  {Fitch}}\ and\ \bibinfo {author} {\bibfnamefont {M.}~\bibnamefont
  {Tarbutt}},\ }\href {https://doi.org/10.1016/bs.aamop.2021.04.003} {\bibfield
   {journal} {\bibinfo  {journal} {Advances in Atomic, Molecular and Optical
  Physics}\ }\textbf {\bibinfo {volume} {70}},\ \bibinfo {pages} {157}
  (\bibinfo {year} {2021})}\BibitemShut {NoStop}%
\bibitem [{\citenamefont {Vutha}\ \emph
  {et~al.}(2018{\natexlab{a}})\citenamefont {Vutha}, \citenamefont
  {Horbatsch},\ and\ \citenamefont {Hessels}}]{Vutha2018Atoms}%
  \BibitemOpen
  \bibfield  {author} {\bibinfo {author} {\bibfnamefont {A.}~\bibnamefont
  {Vutha}}, \bibinfo {author} {\bibfnamefont {M.}~\bibnamefont {Horbatsch}},\
  and\ \bibinfo {author} {\bibfnamefont {E.}~\bibnamefont {Hessels}},\ }\href
  {https://doi.org/10.3390/atoms6010003} {\bibfield  {journal} {\bibinfo
  {journal} {Atoms}\ }\textbf {\bibinfo {volume} {6}},\ \bibinfo {pages} {3}
  (\bibinfo {year} {2018}{\natexlab{a}})}\BibitemShut {NoStop}%
\bibitem [{\citenamefont {Vutha}\ \emph
  {et~al.}(2018{\natexlab{b}})\citenamefont {Vutha}, \citenamefont
  {Horbatsch},\ and\ \citenamefont {Hessels}}]{Vutha2018PRA}%
  \BibitemOpen
  \bibfield  {author} {\bibinfo {author} {\bibfnamefont {A.~C.}\ \bibnamefont
  {Vutha}}, \bibinfo {author} {\bibfnamefont {M.}~\bibnamefont {Horbatsch}},\
  and\ \bibinfo {author} {\bibfnamefont {E.~A.}\ \bibnamefont {Hessels}},\
  }\href {https://doi.org/10.1103/PhysRevA.98.032513} {\bibfield  {journal}
  {\bibinfo  {journal} {Phys. Rev. A}\ }\textbf {\bibinfo {volume} {98}},\
  \bibinfo {pages} {032513} (\bibinfo {year} {2018}{\natexlab{b}})}\BibitemShut
  {NoStop}%
\bibitem [{\citenamefont {Singh}(2019)}]{Singh2019}%
  \BibitemOpen
  \bibfield  {author} {\bibinfo {author} {\bibfnamefont {J.~T.}\ \bibnamefont
  {Singh}},\ }\href {https://doi.org/10.1007/s10751-019-1573-z} {\bibfield
  {journal} {\bibinfo  {journal} {Hyperfine Interactions}\ }\textbf {\bibinfo
  {volume} {240}},\ \bibinfo {pages} {29} (\bibinfo {year} {2019})}\BibitemShut
  {NoStop}%
\bibitem [{\citenamefont {Upadhyay}\ \emph {et~al.}(2020)\citenamefont
  {Upadhyay}, \citenamefont {Dargyte}, \citenamefont {Patterson},\ and\
  \citenamefont {Weinstein}}]{Upadhyay2020}%
  \BibitemOpen
  \bibfield  {author} {\bibinfo {author} {\bibfnamefont {S.}~\bibnamefont
  {Upadhyay}}, \bibinfo {author} {\bibfnamefont {U.}~\bibnamefont {Dargyte}},
  \bibinfo {author} {\bibfnamefont {D.}~\bibnamefont {Patterson}},\ and\
  \bibinfo {author} {\bibfnamefont {J.~D.}\ \bibnamefont {Weinstein}},\ }\href
  {https://doi.org/10.1103/PhysRevLett.125.043601} {\bibfield  {journal}
  {\bibinfo  {journal} {Phys. Rev. Lett.}\ }\textbf {\bibinfo {volume} {125}},\
  \bibinfo {pages} {043601} (\bibinfo {year} {2020})}\BibitemShut {NoStop}%
\bibitem [{\citenamefont {Zhou}\ \emph {et~al.}(2020)\citenamefont {Zhou},
  \citenamefont {Shagam}, \citenamefont {Cairncross}, \citenamefont {Ng},
  \citenamefont {Roussy}, \citenamefont {Grogan}, \citenamefont {Boyce},
  \citenamefont {Vigil}, \citenamefont {Pettine}, \citenamefont {Zelevinsky},
  \citenamefont {Ye},\ and\ \citenamefont {Cornell}}]{Zhou2020}%
  \BibitemOpen
  \bibfield  {author} {\bibinfo {author} {\bibfnamefont {Y.}~\bibnamefont
  {Zhou}}, \bibinfo {author} {\bibfnamefont {Y.}~\bibnamefont {Shagam}},
  \bibinfo {author} {\bibfnamefont {W.~B.}\ \bibnamefont {Cairncross}},
  \bibinfo {author} {\bibfnamefont {K.~B.}\ \bibnamefont {Ng}}, \bibinfo
  {author} {\bibfnamefont {T.~S.}\ \bibnamefont {Roussy}}, \bibinfo {author}
  {\bibfnamefont {T.}~\bibnamefont {Grogan}}, \bibinfo {author} {\bibfnamefont
  {K.}~\bibnamefont {Boyce}}, \bibinfo {author} {\bibfnamefont
  {A.}~\bibnamefont {Vigil}}, \bibinfo {author} {\bibfnamefont
  {M.}~\bibnamefont {Pettine}}, \bibinfo {author} {\bibfnamefont
  {T.}~\bibnamefont {Zelevinsky}}, \bibinfo {author} {\bibfnamefont
  {J.}~\bibnamefont {Ye}},\ and\ \bibinfo {author} {\bibfnamefont {E.~A.}\
  \bibnamefont {Cornell}},\ }\href
  {https://doi.org/10.1103/PhysRevLett.124.053201} {\bibfield  {journal}
  {\bibinfo  {journal} {Physical Review Letters}\ }\textbf {\bibinfo {volume}
  {124}},\ \bibinfo {pages} {053201} (\bibinfo {year} {2020})}\BibitemShut
  {NoStop}%
\bibitem [{\citenamefont {Fan}\ \emph {et~al.}(2021)\citenamefont {Fan},
  \citenamefont {Holliman}, \citenamefont {Shi}, \citenamefont {Zhang},
  \citenamefont {Straus}, \citenamefont {Li}, \citenamefont {Buechele},\ and\
  \citenamefont {Jayich}}]{Fan2021}%
  \BibitemOpen
  \bibfield  {author} {\bibinfo {author} {\bibfnamefont {M.}~\bibnamefont
  {Fan}}, \bibinfo {author} {\bibfnamefont {C.~A.}\ \bibnamefont {Holliman}},
  \bibinfo {author} {\bibfnamefont {X.}~\bibnamefont {Shi}}, \bibinfo {author}
  {\bibfnamefont {H.}~\bibnamefont {Zhang}}, \bibinfo {author} {\bibfnamefont
  {M.~W.}\ \bibnamefont {Straus}}, \bibinfo {author} {\bibfnamefont
  {X.}~\bibnamefont {Li}}, \bibinfo {author} {\bibfnamefont {S.~W.}\
  \bibnamefont {Buechele}},\ and\ \bibinfo {author} {\bibfnamefont {A.~M.}\
  \bibnamefont {Jayich}},\ }\href
  {https://doi.org/10.1103/PhysRevLett.126.023002} {\bibfield  {journal}
  {\bibinfo  {journal} {Physical Review Letters}\ }\textbf {\bibinfo {volume}
  {126}},\ \bibinfo {pages} {023002} (\bibinfo {year} {2021})}\BibitemShut
  {NoStop}%
\bibitem [{\citenamefont {Parker}\ \emph {et~al.}(2015)\citenamefont {Parker},
  \citenamefont {Dietrich}, \citenamefont {Kalita}, \citenamefont {Lemke},
  \citenamefont {Bailey}, \citenamefont {Bishof}, \citenamefont {Greene},
  \citenamefont {Holt}, \citenamefont {Korsch}, \citenamefont {Lu},
  \citenamefont {Mueller}, \citenamefont {O'Connor},\ and\ \citenamefont
  {Singh}}]{Parker2015}%
  \BibitemOpen
  \bibfield  {author} {\bibinfo {author} {\bibfnamefont {R.~H.}\ \bibnamefont
  {Parker}}, \bibinfo {author} {\bibfnamefont {M.~R.}\ \bibnamefont
  {Dietrich}}, \bibinfo {author} {\bibfnamefont {M.~R.}\ \bibnamefont
  {Kalita}}, \bibinfo {author} {\bibfnamefont {N.~D.}\ \bibnamefont {Lemke}},
  \bibinfo {author} {\bibfnamefont {K.~G.}\ \bibnamefont {Bailey}}, \bibinfo
  {author} {\bibfnamefont {M.}~\bibnamefont {Bishof}}, \bibinfo {author}
  {\bibfnamefont {J.~P.}\ \bibnamefont {Greene}}, \bibinfo {author}
  {\bibfnamefont {R.~J.}\ \bibnamefont {Holt}}, \bibinfo {author}
  {\bibfnamefont {W.}~\bibnamefont {Korsch}}, \bibinfo {author} {\bibfnamefont
  {Z.-T.}\ \bibnamefont {Lu}}, \bibinfo {author} {\bibfnamefont
  {P.}~\bibnamefont {Mueller}}, \bibinfo {author} {\bibfnamefont {T.~P.}\
  \bibnamefont {O'Connor}},\ and\ \bibinfo {author} {\bibfnamefont {J.~T.}\
  \bibnamefont {Singh}},\ }\href
  {https://doi.org/10.1103/PhysRevLett.114.233002} {\bibfield  {journal}
  {\bibinfo  {journal} {Physical Review Letters}\ }\textbf {\bibinfo {volume}
  {114}},\ \bibinfo {pages} {233002} (\bibinfo {year} {2015})}\BibitemShut
  {NoStop}%
\bibitem [{\citenamefont {Flambaum}(2019)}]{Flambaum2019Schiff}%
  \BibitemOpen
  \bibfield  {author} {\bibinfo {author} {\bibfnamefont {V.~V.}\ \bibnamefont
  {Flambaum}},\ }\href {https://doi.org/10.1103/PhysRevC.99.035501} {\bibfield
  {journal} {\bibinfo  {journal} {Physical Review C}\ }\textbf {\bibinfo
  {volume} {99}},\ \bibinfo {pages} {035501} (\bibinfo {year}
  {2019})}\BibitemShut {NoStop}%
\bibitem [{\citenamefont {Swallows}\ \emph {et~al.}(2013)\citenamefont
  {Swallows}, \citenamefont {Loftus}, \citenamefont {Griffith}, \citenamefont
  {Heckel}, \citenamefont {Fortson},\ and\ \citenamefont
  {Romalis}}]{Swallows2013}%
  \BibitemOpen
  \bibfield  {author} {\bibinfo {author} {\bibfnamefont {M.~D.}\ \bibnamefont
  {Swallows}}, \bibinfo {author} {\bibfnamefont {T.~H.}\ \bibnamefont
  {Loftus}}, \bibinfo {author} {\bibfnamefont {W.~C.}\ \bibnamefont
  {Griffith}}, \bibinfo {author} {\bibfnamefont {B.~R.}\ \bibnamefont
  {Heckel}}, \bibinfo {author} {\bibfnamefont {E.~N.}\ \bibnamefont
  {Fortson}},\ and\ \bibinfo {author} {\bibfnamefont {M.~V.}\ \bibnamefont
  {Romalis}},\ }\href {https://doi.org/10.1103/PhysRevA.87.012102} {\bibfield
  {journal} {\bibinfo  {journal} {Physical Review A}\ }\textbf {\bibinfo
  {volume} {87}},\ \bibinfo {pages} {012102} (\bibinfo {year}
  {2013})}\BibitemShut {NoStop}%
\bibitem [{\citenamefont {Sahoo}\ and\ \citenamefont {Das}(2018)}]{Sahoo2018}%
  \BibitemOpen
  \bibfield  {author} {\bibinfo {author} {\bibfnamefont {B.}~\bibnamefont
  {Sahoo}}\ and\ \bibinfo {author} {\bibfnamefont {B.}~\bibnamefont {Das}},\
  }\bibfield  {journal} {\bibinfo  {journal} {Physical Review Letters}\
  }\textbf {\bibinfo {volume} {120}},\ \href
  {https://doi.org/10.1103/physrevlett.120.203001}
  {10.1103/physrevlett.120.203001} (\bibinfo {year} {2018})\BibitemShut
  {NoStop}%
\bibitem [{\citenamefont {Allmendinger}\ \emph {et~al.}(2019)\citenamefont
  {Allmendinger}, \citenamefont {Engin}, \citenamefont {Heil}, \citenamefont
  {Karpuk}, \citenamefont {Krause}, \citenamefont {Niederl{\"{a}}nder},
  \citenamefont {Offenh{\"{a}}usser}, \citenamefont {Repetto}, \citenamefont
  {Schmidt},\ and\ \citenamefont {Zimmer}}]{Allmendinger2019}%
  \BibitemOpen
  \bibfield  {author} {\bibinfo {author} {\bibfnamefont {F.}~\bibnamefont
  {Allmendinger}}, \bibinfo {author} {\bibfnamefont {I.}~\bibnamefont {Engin}},
  \bibinfo {author} {\bibfnamefont {W.}~\bibnamefont {Heil}}, \bibinfo {author}
  {\bibfnamefont {S.}~\bibnamefont {Karpuk}}, \bibinfo {author} {\bibfnamefont
  {H.-J.}\ \bibnamefont {Krause}}, \bibinfo {author} {\bibfnamefont
  {B.}~\bibnamefont {Niederl{\"{a}}nder}}, \bibinfo {author} {\bibfnamefont
  {A.}~\bibnamefont {Offenh{\"{a}}usser}}, \bibinfo {author} {\bibfnamefont
  {M.}~\bibnamefont {Repetto}}, \bibinfo {author} {\bibfnamefont
  {U.}~\bibnamefont {Schmidt}},\ and\ \bibinfo {author} {\bibfnamefont
  {S.}~\bibnamefont {Zimmer}},\ }\href
  {https://doi.org/10.1103/PhysRevA.100.022505} {\bibfield  {journal} {\bibinfo
   {journal} {Physical Review A}\ }\textbf {\bibinfo {volume} {100}},\ \bibinfo
  {pages} {022505} (\bibinfo {year} {2019})}\BibitemShut {NoStop}%
\bibitem [{\citenamefont {Sachdeva}\ \emph {et~al.}(2019)\citenamefont
  {Sachdeva}, \citenamefont {Fan}, \citenamefont {Babcock}, \citenamefont
  {Burghoff}, \citenamefont {Chupp}, \citenamefont {Degenkolb}, \citenamefont
  {Fierlinger}, \citenamefont {Haude}, \citenamefont {Kraegeloh}, \citenamefont
  {Kilian}, \citenamefont {Knappe-Gr{\"{u}}neberg}, \citenamefont {Kuchler},
  \citenamefont {Liu}, \citenamefont {Marino}, \citenamefont {Meinel},
  \citenamefont {Rolfs}, \citenamefont {Salhi}, \citenamefont {Schnabel},
  \citenamefont {Singh}, \citenamefont {Stuiber}, \citenamefont {Terrano},
  \citenamefont {Trahms},\ and\ \citenamefont {Voigt}}]{Sachdeva2019}%
  \BibitemOpen
  \bibfield  {author} {\bibinfo {author} {\bibfnamefont {N.}~\bibnamefont
  {Sachdeva}}, \bibinfo {author} {\bibfnamefont {I.}~\bibnamefont {Fan}},
  \bibinfo {author} {\bibfnamefont {E.}~\bibnamefont {Babcock}}, \bibinfo
  {author} {\bibfnamefont {M.}~\bibnamefont {Burghoff}}, \bibinfo {author}
  {\bibfnamefont {T.~E.}\ \bibnamefont {Chupp}}, \bibinfo {author}
  {\bibfnamefont {S.}~\bibnamefont {Degenkolb}}, \bibinfo {author}
  {\bibfnamefont {P.}~\bibnamefont {Fierlinger}}, \bibinfo {author}
  {\bibfnamefont {S.}~\bibnamefont {Haude}}, \bibinfo {author} {\bibfnamefont
  {E.}~\bibnamefont {Kraegeloh}}, \bibinfo {author} {\bibfnamefont
  {W.}~\bibnamefont {Kilian}}, \bibinfo {author} {\bibfnamefont
  {S.}~\bibnamefont {Knappe-Gr{\"{u}}neberg}}, \bibinfo {author} {\bibfnamefont
  {F.}~\bibnamefont {Kuchler}}, \bibinfo {author} {\bibfnamefont
  {T.}~\bibnamefont {Liu}}, \bibinfo {author} {\bibfnamefont {M.}~\bibnamefont
  {Marino}}, \bibinfo {author} {\bibfnamefont {J.}~\bibnamefont {Meinel}},
  \bibinfo {author} {\bibfnamefont {K.}~\bibnamefont {Rolfs}}, \bibinfo
  {author} {\bibfnamefont {Z.}~\bibnamefont {Salhi}}, \bibinfo {author}
  {\bibfnamefont {A.}~\bibnamefont {Schnabel}}, \bibinfo {author}
  {\bibfnamefont {J.~T.}\ \bibnamefont {Singh}}, \bibinfo {author}
  {\bibfnamefont {S.}~\bibnamefont {Stuiber}}, \bibinfo {author} {\bibfnamefont
  {W.~A.}\ \bibnamefont {Terrano}}, \bibinfo {author} {\bibfnamefont
  {L.}~\bibnamefont {Trahms}},\ and\ \bibinfo {author} {\bibfnamefont
  {J.}~\bibnamefont {Voigt}},\ }\href
  {https://doi.org/10.1103/PhysRevLett.123.143003} {\bibfield  {journal}
  {\bibinfo  {journal} {Physical Review Letters}\ }\textbf {\bibinfo {volume}
  {123}},\ \bibinfo {pages} {143003} (\bibinfo {year} {2019})}\BibitemShut
  {NoStop}%
\bibitem [{\citenamefont {Sakurai}\ \emph {et~al.}(2019)\citenamefont
  {Sakurai}, \citenamefont {Sahoo}, \citenamefont {Asahi},\ and\ \citenamefont
  {Das}}]{Sakurai2019}%
  \BibitemOpen
  \bibfield  {author} {\bibinfo {author} {\bibfnamefont {A.}~\bibnamefont
  {Sakurai}}, \bibinfo {author} {\bibfnamefont {B.~K.}\ \bibnamefont {Sahoo}},
  \bibinfo {author} {\bibfnamefont {K.}~\bibnamefont {Asahi}},\ and\ \bibinfo
  {author} {\bibfnamefont {B.~P.}\ \bibnamefont {Das}},\ }\bibfield  {journal}
  {\bibinfo  {journal} {Physical Review A}\ }\textbf {\bibinfo {volume}
  {100}},\ \href {https://doi.org/10.1103/physreva.100.020502}
  {10.1103/physreva.100.020502} (\bibinfo {year} {2019})\BibitemShut {NoStop}%
\bibitem [{\citenamefont {Chupp}\ and\ \citenamefont
  {Ramsey-Musolf}(2015)}]{Chupp2015}%
  \BibitemOpen
  \bibfield  {author} {\bibinfo {author} {\bibfnamefont {T.}~\bibnamefont
  {Chupp}}\ and\ \bibinfo {author} {\bibfnamefont {M.}~\bibnamefont
  {Ramsey-Musolf}},\ }\href {https://doi.org/10.1103/PhysRevC.91.035502}
  {\bibfield  {journal} {\bibinfo  {journal} {Physical Review C}\ }\textbf
  {\bibinfo {volume} {91}},\ \bibinfo {pages} {035502} (\bibinfo {year}
  {2015})}\BibitemShut {NoStop}%
\bibitem [{\citenamefont {Chupp}\ \emph
  {et~al.}(2019{\natexlab{b}})\citenamefont {Chupp}, \citenamefont
  {Fierlinger}, \citenamefont {Ramsey-Musolf},\ and\ \citenamefont
  {Singh}}]{Chupp2019}%
  \BibitemOpen
  \bibfield  {author} {\bibinfo {author} {\bibfnamefont {T.~E.}\ \bibnamefont
  {Chupp}}, \bibinfo {author} {\bibfnamefont {P.}~\bibnamefont {Fierlinger}},
  \bibinfo {author} {\bibfnamefont {M.~J.}\ \bibnamefont {Ramsey-Musolf}},\
  and\ \bibinfo {author} {\bibfnamefont {J.~T.}\ \bibnamefont {Singh}},\ }\href
  {https://doi.org/10.1103/RevModPhys.91.015001} {\bibfield  {journal}
  {\bibinfo  {journal} {Reviews of Modern Physics}\ }\textbf {\bibinfo {volume}
  {91}},\ \bibinfo {pages} {015001} (\bibinfo {year}
  {2019}{\natexlab{b}})}\BibitemShut {NoStop}%
\bibitem [{\citenamefont {Zhu}\ \emph {et~al.}(2013)\citenamefont {Zhu},
  \citenamefont {Solmeyer}, \citenamefont {Tang},\ and\ \citenamefont
  {Weiss}}]{Zhu2013}%
  \BibitemOpen
  \bibfield  {author} {\bibinfo {author} {\bibfnamefont {K.}~\bibnamefont
  {Zhu}}, \bibinfo {author} {\bibfnamefont {N.}~\bibnamefont {Solmeyer}},
  \bibinfo {author} {\bibfnamefont {C.}~\bibnamefont {Tang}},\ and\ \bibinfo
  {author} {\bibfnamefont {D.~S.}\ \bibnamefont {Weiss}},\ }\href
  {https://doi.org/10.1103/PhysRevLett.111.243006} {\bibfield  {journal}
  {\bibinfo  {journal} {Physical Review Letters}\ }\textbf {\bibinfo {volume}
  {111}},\ \bibinfo {pages} {243006} (\bibinfo {year} {2013})}\BibitemShut
  {NoStop}%
\bibitem [{\citenamefont {Tang}\ \emph {et~al.}(2018)\citenamefont {Tang},
  \citenamefont {Zhang},\ and\ \citenamefont {Weiss}}]{Tang2018}%
  \BibitemOpen
  \bibfield  {author} {\bibinfo {author} {\bibfnamefont {C.}~\bibnamefont
  {Tang}}, \bibinfo {author} {\bibfnamefont {T.}~\bibnamefont {Zhang}},\ and\
  \bibinfo {author} {\bibfnamefont {D.~S.}\ \bibnamefont {Weiss}},\ }\href
  {https://doi.org/10.1103/PhysRevA.97.033404} {\bibfield  {journal} {\bibinfo
  {journal} {Physical Review A}\ }\textbf {\bibinfo {volume} {97}},\ \bibinfo
  {pages} {033404} (\bibinfo {year} {2018})}\BibitemShut {NoStop}%
\bibitem [{\citenamefont {Wundt}\ \emph {et~al.}(2012)\citenamefont {Wundt},
  \citenamefont {Munger},\ and\ \citenamefont {Jentschura}}]{Wundt2012}%
  \BibitemOpen
  \bibfield  {author} {\bibinfo {author} {\bibfnamefont {B.~J.}\ \bibnamefont
  {Wundt}}, \bibinfo {author} {\bibfnamefont {C.~T.}\ \bibnamefont {Munger}},\
  and\ \bibinfo {author} {\bibfnamefont {U.~D.}\ \bibnamefont {Jentschura}},\
  }\href {https://doi.org/10.1103/PhysRevX.2.041009} {\bibfield  {journal}
  {\bibinfo  {journal} {Physical Review X}\ }\textbf {\bibinfo {volume} {2}},\
  \bibinfo {pages} {041009} (\bibinfo {year} {2012})},\ \Eprint
  {https://arxiv.org/abs/1211.4057} {arXiv:1211.4057} \BibitemShut {NoStop}%
\bibitem [{\citenamefont {Inoue}\ \emph {et~al.}(2014)\citenamefont {Inoue},
  \citenamefont {Ando}, \citenamefont {Aoki}, \citenamefont {Arikawa},
  \citenamefont {Ezure}, \citenamefont {Harada}, \citenamefont {Hayamizu},
  \citenamefont {Ishikawa}, \citenamefont {Itoh}, \citenamefont {Kato},
  \citenamefont {Kawamura}, \citenamefont {Uchiyama}, \citenamefont {Aoki},
  \citenamefont {Asahi}, \citenamefont {Furukawa}, \citenamefont {Hatakeyama},
  \citenamefont {Hatanaka}, \citenamefont {Imai}, \citenamefont {Murakami},
  \citenamefont {Nataraj}, \citenamefont {Sato}, \citenamefont {Shimizu},
  \citenamefont {Wakasa}, \citenamefont {Yoshida}, \citenamefont {Yoshimi},\
  and\ \citenamefont {Sakemi}}]{Inoue2014}%
  \BibitemOpen
  \bibfield  {author} {\bibinfo {author} {\bibfnamefont {T.}~\bibnamefont
  {Inoue}}, \bibinfo {author} {\bibfnamefont {S.}~\bibnamefont {Ando}},
  \bibinfo {author} {\bibfnamefont {T.}~\bibnamefont {Aoki}}, \bibinfo {author}
  {\bibfnamefont {H.}~\bibnamefont {Arikawa}}, \bibinfo {author} {\bibfnamefont
  {S.}~\bibnamefont {Ezure}}, \bibinfo {author} {\bibfnamefont
  {K.}~\bibnamefont {Harada}}, \bibinfo {author} {\bibfnamefont
  {T.}~\bibnamefont {Hayamizu}}, \bibinfo {author} {\bibfnamefont
  {T.}~\bibnamefont {Ishikawa}}, \bibinfo {author} {\bibfnamefont
  {M.}~\bibnamefont {Itoh}}, \bibinfo {author} {\bibfnamefont {K.}~\bibnamefont
  {Kato}}, \bibinfo {author} {\bibfnamefont {H.}~\bibnamefont {Kawamura}},
  \bibinfo {author} {\bibfnamefont {A.}~\bibnamefont {Uchiyama}}, \bibinfo
  {author} {\bibfnamefont {T.}~\bibnamefont {Aoki}}, \bibinfo {author}
  {\bibfnamefont {K.}~\bibnamefont {Asahi}}, \bibinfo {author} {\bibfnamefont
  {T.}~\bibnamefont {Furukawa}}, \bibinfo {author} {\bibfnamefont
  {A.}~\bibnamefont {Hatakeyama}}, \bibinfo {author} {\bibfnamefont
  {K.}~\bibnamefont {Hatanaka}}, \bibinfo {author} {\bibfnamefont
  {K.}~\bibnamefont {Imai}}, \bibinfo {author} {\bibfnamefont {T.}~\bibnamefont
  {Murakami}}, \bibinfo {author} {\bibfnamefont {H.~S.}\ \bibnamefont
  {Nataraj}}, \bibinfo {author} {\bibfnamefont {T.}~\bibnamefont {Sato}},
  \bibinfo {author} {\bibfnamefont {Y.}~\bibnamefont {Shimizu}}, \bibinfo
  {author} {\bibfnamefont {T.}~\bibnamefont {Wakasa}}, \bibinfo {author}
  {\bibfnamefont {H.~P.}\ \bibnamefont {Yoshida}}, \bibinfo {author}
  {\bibfnamefont {A.}~\bibnamefont {Yoshimi}},\ and\ \bibinfo {author}
  {\bibfnamefont {Y.}~\bibnamefont {Sakemi}},\ }\href
  {https://doi.org/10.1007/s10751-014-1100-1} {\bibfield  {journal} {\bibinfo
  {journal} {Hyperfine Interactions}\ }\textbf {\bibinfo {volume} {231}},\
  \bibinfo {pages} {157} (\bibinfo {year} {2014})}\BibitemShut {NoStop}%
\bibitem [{\citenamefont {Feinberg}\ and\ \citenamefont
  {Gould}(2018)}]{Feinberg2018}%
  \BibitemOpen
  \bibfield  {author} {\bibinfo {author} {\bibfnamefont {B.}~\bibnamefont
  {Feinberg}}\ and\ \bibinfo {author} {\bibfnamefont {H.}~\bibnamefont
  {Gould}},\ }\href {https://doi.org/10.1063/1.5009926} {\bibfield  {journal}
  {\bibinfo  {journal} {{AIP} Advances}\ }\textbf {\bibinfo {volume} {8}},\
  \bibinfo {pages} {035303} (\bibinfo {year} {2018})}\BibitemShut {NoStop}%
\bibitem [{\citenamefont {{Aoki}}\ \emph {et~al.}(2021)\citenamefont {{Aoki}},
  \citenamefont {{Sreekantham}}, \citenamefont {{Sahoo}}, \citenamefont
  {{Arora}}, \citenamefont {{Kastberg}}, \citenamefont {{Sato}}, \citenamefont
  {{Ikeda}}, \citenamefont {{Okamoto}}, \citenamefont {{Torii}}, \citenamefont
  {{Hayamizu}}, \citenamefont {{Nakamura}}, \citenamefont {{Nagase}},
  \citenamefont {{Ohtsuka}}, \citenamefont {{Nagahama}}, \citenamefont
  {{Ozawa}}, \citenamefont {{Sato}}, \citenamefont {{Nakashita}}, \citenamefont
  {{Yamane}}, \citenamefont {{Tanaka}}, \citenamefont {{Harada}}, \citenamefont
  {{Kawamura}}, \citenamefont {{Inoue}}, \citenamefont {{Uchiyama}},
  \citenamefont {{Hatakeyama}}, \citenamefont {{Takamine}}, \citenamefont
  {{Ueno}}, \citenamefont {{Ichikawa}}, \citenamefont {{Matsuda}},
  \citenamefont {{Haba}},\ and\ \citenamefont {{Sakemi}}}]{Aoki2021}%
  \BibitemOpen
  \bibfield  {author} {\bibinfo {author} {\bibfnamefont {T.}~\bibnamefont
  {{Aoki}}}, \bibinfo {author} {\bibfnamefont {R.}~\bibnamefont
  {{Sreekantham}}}, \bibinfo {author} {\bibfnamefont {B.~K.}\ \bibnamefont
  {{Sahoo}}}, \bibinfo {author} {\bibfnamefont {B.}~\bibnamefont {{Arora}}},
  \bibinfo {author} {\bibfnamefont {A.}~\bibnamefont {{Kastberg}}}, \bibinfo
  {author} {\bibfnamefont {T.}~\bibnamefont {{Sato}}}, \bibinfo {author}
  {\bibfnamefont {H.}~\bibnamefont {{Ikeda}}}, \bibinfo {author} {\bibfnamefont
  {N.}~\bibnamefont {{Okamoto}}}, \bibinfo {author} {\bibfnamefont
  {Y.}~\bibnamefont {{Torii}}}, \bibinfo {author} {\bibfnamefont
  {T.}~\bibnamefont {{Hayamizu}}}, \bibinfo {author} {\bibfnamefont
  {K.}~\bibnamefont {{Nakamura}}}, \bibinfo {author} {\bibfnamefont
  {S.}~\bibnamefont {{Nagase}}}, \bibinfo {author} {\bibfnamefont
  {M.}~\bibnamefont {{Ohtsuka}}}, \bibinfo {author} {\bibfnamefont
  {H.}~\bibnamefont {{Nagahama}}}, \bibinfo {author} {\bibfnamefont
  {N.}~\bibnamefont {{Ozawa}}}, \bibinfo {author} {\bibfnamefont
  {M.}~\bibnamefont {{Sato}}}, \bibinfo {author} {\bibfnamefont
  {T.}~\bibnamefont {{Nakashita}}}, \bibinfo {author} {\bibfnamefont
  {K.}~\bibnamefont {{Yamane}}}, \bibinfo {author} {\bibfnamefont {K.~S.}\
  \bibnamefont {{Tanaka}}}, \bibinfo {author} {\bibfnamefont {K.}~\bibnamefont
  {{Harada}}}, \bibinfo {author} {\bibfnamefont {H.}~\bibnamefont
  {{Kawamura}}}, \bibinfo {author} {\bibfnamefont {T.}~\bibnamefont {{Inoue}}},
  \bibinfo {author} {\bibfnamefont {A.}~\bibnamefont {{Uchiyama}}}, \bibinfo
  {author} {\bibfnamefont {A.}~\bibnamefont {{Hatakeyama}}}, \bibinfo {author}
  {\bibfnamefont {A.}~\bibnamefont {{Takamine}}}, \bibinfo {author}
  {\bibfnamefont {H.}~\bibnamefont {{Ueno}}}, \bibinfo {author} {\bibfnamefont
  {Y.}~\bibnamefont {{Ichikawa}}}, \bibinfo {author} {\bibfnamefont
  {Y.}~\bibnamefont {{Matsuda}}}, \bibinfo {author} {\bibfnamefont
  {H.}~\bibnamefont {{Haba}}},\ and\ \bibinfo {author} {\bibfnamefont
  {Y.}~\bibnamefont {{Sakemi}}},\ }\href
  {https://doi.org/10.1088/2058-9565/ac1b6a} {\bibfield  {journal} {\bibinfo
  {journal} {Quantum Science and Technology}\ }\textbf {\bibinfo {volume}
  {6}},\ \bibinfo {eid} {044008} (\bibinfo {year} {2021})}\BibitemShut
  {NoStop}%
\bibitem [{\citenamefont {Shitara}\ \emph {et~al.}(2021)\citenamefont
  {Shitara}, \citenamefont {Yamanaka}, \citenamefont {Sahoo}, \citenamefont
  {Watanabe},\ and\ \citenamefont {Das}}]{Shitara2021}%
  \BibitemOpen
  \bibfield  {author} {\bibinfo {author} {\bibfnamefont {N.}~\bibnamefont
  {Shitara}}, \bibinfo {author} {\bibfnamefont {N.}~\bibnamefont {Yamanaka}},
  \bibinfo {author} {\bibfnamefont {B.~K.}\ \bibnamefont {Sahoo}}, \bibinfo
  {author} {\bibfnamefont {T.}~\bibnamefont {Watanabe}},\ and\ \bibinfo
  {author} {\bibfnamefont {B.~P.}\ \bibnamefont {Das}},\ }\bibfield  {journal}
  {\bibinfo  {journal} {Journal of High Energy Physics}\ }\textbf {\bibinfo
  {volume} {2021}},\ \href {https://doi.org/10.1007/jhep02(2021)124}
  {10.1007/jhep02(2021)124} (\bibinfo {year} {2021})\BibitemShut {NoStop}%
\bibitem [{\citenamefont {Feinberg}\ \emph {et~al.}(2020)\citenamefont
  {Feinberg}, \citenamefont {Gould},\ and\ \citenamefont {Munger~Jr.}}]{FrLOI}%
  \BibitemOpen
  \bibfield  {author} {\bibinfo {author} {\bibfnamefont {B.}~\bibnamefont
  {Feinberg}}, \bibinfo {author} {\bibfnamefont {H.}~\bibnamefont {Gould}},\
  and\ \bibinfo {author} {\bibfnamefont {C.~T.}\ \bibnamefont {Munger~Jr.}},\
  }\href {https://mis.triumf.ca/science/experiment/view/S1324LOI} {\bibinfo
  {title} {{TRIUMF EEC Letter of Intent S1324: Electron electric dipole moment
  using francium}}} (\bibinfo {year} {2020})\BibitemShut {NoStop}%
\bibitem [{\citenamefont {Bishof}\ \emph {et~al.}(2016)\citenamefont {Bishof},
  \citenamefont {Parker}, \citenamefont {Bailey}, \citenamefont {Greene},
  \citenamefont {Holt}, \citenamefont {Kalita}, \citenamefont {Korsch},
  \citenamefont {Lemke}, \citenamefont {Lu}, \citenamefont {Mueller},
  \citenamefont {O'Connor}, \citenamefont {Singh},\ and\ \citenamefont
  {Dietrich}}]{Bishof2016}%
  \BibitemOpen
  \bibfield  {author} {\bibinfo {author} {\bibfnamefont {M.}~\bibnamefont
  {Bishof}}, \bibinfo {author} {\bibfnamefont {R.~H.}\ \bibnamefont {Parker}},
  \bibinfo {author} {\bibfnamefont {K.~G.}\ \bibnamefont {Bailey}}, \bibinfo
  {author} {\bibfnamefont {J.~P.}\ \bibnamefont {Greene}}, \bibinfo {author}
  {\bibfnamefont {R.~J.}\ \bibnamefont {Holt}}, \bibinfo {author}
  {\bibfnamefont {M.~R.}\ \bibnamefont {Kalita}}, \bibinfo {author}
  {\bibfnamefont {W.}~\bibnamefont {Korsch}}, \bibinfo {author} {\bibfnamefont
  {N.~D.}\ \bibnamefont {Lemke}}, \bibinfo {author} {\bibfnamefont {Z.-T.}\
  \bibnamefont {Lu}}, \bibinfo {author} {\bibfnamefont {P.}~\bibnamefont
  {Mueller}}, \bibinfo {author} {\bibfnamefont {T.~P.}\ \bibnamefont
  {O'Connor}}, \bibinfo {author} {\bibfnamefont {J.~T.}\ \bibnamefont
  {Singh}},\ and\ \bibinfo {author} {\bibfnamefont {M.~R.}\ \bibnamefont
  {Dietrich}},\ }\href {https://doi.org/10.1103/PhysRevC.94.025501} {\bibfield
  {journal} {\bibinfo  {journal} {Physical Review C}\ }\textbf {\bibinfo
  {volume} {94}},\ \bibinfo {pages} {025501} (\bibinfo {year}
  {2016})}\BibitemShut {NoStop}%
\bibitem [{\citenamefont {Spevak}\ \emph {et~al.}(1997)\citenamefont {Spevak},
  \citenamefont {Auerbach},\ and\ \citenamefont {Flambaum}}]{Spevak1997}%
  \BibitemOpen
  \bibfield  {author} {\bibinfo {author} {\bibfnamefont {V.}~\bibnamefont
  {Spevak}}, \bibinfo {author} {\bibfnamefont {N.}~\bibnamefont {Auerbach}},\
  and\ \bibinfo {author} {\bibfnamefont {V.~V.}\ \bibnamefont {Flambaum}},\
  }\href {https://doi.org/10.1103/PhysRevC.56.1357} {\bibfield  {journal}
  {\bibinfo  {journal} {Physical Review C - Nuclear Physics}\ }\textbf
  {\bibinfo {volume} {56}},\ \bibinfo {pages} {1357} (\bibinfo {year}
  {1997})},\ \Eprint {https://arxiv.org/abs/9612044} {arXiv:9612044 [nucl-th]}
  \BibitemShut {NoStop}%
\bibitem [{\citenamefont {Ready}\ \emph {et~al.}(2021)\citenamefont {Ready},
  \citenamefont {Arrowsmith-Kron}, \citenamefont {Bailey}, \citenamefont
  {Battaglia}, \citenamefont {Bishof}, \citenamefont {Coulter}, \citenamefont
  {Dietrich}, \citenamefont {Fang}, \citenamefont {Hanley}, \citenamefont
  {Huneau}, \citenamefont {Kennedy}, \citenamefont {Lalain}, \citenamefont
  {Loseth}, \citenamefont {McGee}, \citenamefont {Mueller}, \citenamefont
  {O'Connor}, \citenamefont {O'Kronley}, \citenamefont {Powers}, \citenamefont
  {Rabga}, \citenamefont {Sanchez}, \citenamefont {Schalk}, \citenamefont
  {Waldo}, \citenamefont {Wescott},\ and\ \citenamefont {Singh}}]{Ready2021}%
  \BibitemOpen
  \bibfield  {author} {\bibinfo {author} {\bibfnamefont {R.~A.}\ \bibnamefont
  {Ready}}, \bibinfo {author} {\bibfnamefont {G.}~\bibnamefont
  {Arrowsmith-Kron}}, \bibinfo {author} {\bibfnamefont {K.~G.}\ \bibnamefont
  {Bailey}}, \bibinfo {author} {\bibfnamefont {D.}~\bibnamefont {Battaglia}},
  \bibinfo {author} {\bibfnamefont {M.}~\bibnamefont {Bishof}}, \bibinfo
  {author} {\bibfnamefont {D.}~\bibnamefont {Coulter}}, \bibinfo {author}
  {\bibfnamefont {M.~R.}\ \bibnamefont {Dietrich}}, \bibinfo {author}
  {\bibfnamefont {R.}~\bibnamefont {Fang}}, \bibinfo {author} {\bibfnamefont
  {B.}~\bibnamefont {Hanley}}, \bibinfo {author} {\bibfnamefont
  {J.}~\bibnamefont {Huneau}}, \bibinfo {author} {\bibfnamefont
  {S.}~\bibnamefont {Kennedy}}, \bibinfo {author} {\bibfnamefont
  {P.}~\bibnamefont {Lalain}}, \bibinfo {author} {\bibfnamefont
  {B.}~\bibnamefont {Loseth}}, \bibinfo {author} {\bibfnamefont
  {K.}~\bibnamefont {McGee}}, \bibinfo {author} {\bibfnamefont
  {P.}~\bibnamefont {Mueller}}, \bibinfo {author} {\bibfnamefont {T.~P.}\
  \bibnamefont {O'Connor}}, \bibinfo {author} {\bibfnamefont {J.}~\bibnamefont
  {O'Kronley}}, \bibinfo {author} {\bibfnamefont {A.}~\bibnamefont {Powers}},
  \bibinfo {author} {\bibfnamefont {T.}~\bibnamefont {Rabga}}, \bibinfo
  {author} {\bibfnamefont {A.}~\bibnamefont {Sanchez}}, \bibinfo {author}
  {\bibfnamefont {E.}~\bibnamefont {Schalk}}, \bibinfo {author} {\bibfnamefont
  {D.}~\bibnamefont {Waldo}}, \bibinfo {author} {\bibfnamefont
  {J.}~\bibnamefont {Wescott}},\ and\ \bibinfo {author} {\bibfnamefont {J.~T.}\
  \bibnamefont {Singh}},\ }\href {https://doi.org/10.1016/j.nima.2021.165738}
  {\bibfield  {journal} {\bibinfo  {journal} {Nuclear Instruments and Methods
  in Physics Research Section A: Accelerators, Spectrometers, Detectors and
  Associated Equipment}\ }\textbf {\bibinfo {volume} {1014}},\ \bibinfo {pages}
  {165738} (\bibinfo {year} {2021})}\BibitemShut {NoStop}%
\bibitem [{\citenamefont {Booth}\ \emph {et~al.}(2020)\citenamefont {Booth},
  \citenamefont {Rabga}, \citenamefont {Ready}, \citenamefont {Bailey},
  \citenamefont {Bishof}, \citenamefont {Dietrich}, \citenamefont {Greene},
  \citenamefont {Mueller}, \citenamefont {O{\textquotesingle}Connor},\ and\
  \citenamefont {Singh}}]{Booth2020}%
  \BibitemOpen
  \bibfield  {author} {\bibinfo {author} {\bibfnamefont {D.}~\bibnamefont
  {Booth}}, \bibinfo {author} {\bibfnamefont {T.}~\bibnamefont {Rabga}},
  \bibinfo {author} {\bibfnamefont {R.}~\bibnamefont {Ready}}, \bibinfo
  {author} {\bibfnamefont {K.}~\bibnamefont {Bailey}}, \bibinfo {author}
  {\bibfnamefont {M.}~\bibnamefont {Bishof}}, \bibinfo {author} {\bibfnamefont
  {M.}~\bibnamefont {Dietrich}}, \bibinfo {author} {\bibfnamefont
  {J.}~\bibnamefont {Greene}}, \bibinfo {author} {\bibfnamefont
  {P.}~\bibnamefont {Mueller}}, \bibinfo {author} {\bibfnamefont
  {T.}~\bibnamefont {O{\textquotesingle}Connor}},\ and\ \bibinfo {author}
  {\bibfnamefont {J.}~\bibnamefont {Singh}},\ }\href
  {https://doi.org/10.1016/j.sab.2020.105967} {\bibfield  {journal} {\bibinfo
  {journal} {Spectrochimica Acta Part B: Atomic Spectroscopy}\ }\textbf
  {\bibinfo {volume} {172}},\ \bibinfo {pages} {105967} (\bibinfo {year}
  {2020})}\BibitemShut {NoStop}%
\bibitem [{\citenamefont {{Rabga, Tenzin}}(2020)}]{Rabga2020}%
  \BibitemOpen
  \bibfield  {author} {\bibinfo {author} {\bibnamefont {{Rabga, Tenzin}}},\
  }\emph {\bibinfo {title} {UPGRADES FOR AN IMPROVED MEASUREMENT OF THE
  PERMANENT ELECTRIC DIPOLE MOMENT OF RADIUM}},\ \href
  {https://doi.org/10.25335/MJN9-HZ76} {Ph.D. thesis},\ \bibinfo  {school}
  {Michigan State University} (\bibinfo {year} {2020})\BibitemShut {NoStop}%
\bibitem [{\citenamefont {DeMille}\ \emph {et~al.}(2017)\citenamefont
  {DeMille}, \citenamefont {Doyle},\ and\ \citenamefont
  {Sushkov}}]{DeMille2017}%
  \BibitemOpen
  \bibfield  {author} {\bibinfo {author} {\bibfnamefont {D.}~\bibnamefont
  {DeMille}}, \bibinfo {author} {\bibfnamefont {J.~M.}\ \bibnamefont {Doyle}},\
  and\ \bibinfo {author} {\bibfnamefont {A.~O.}\ \bibnamefont {Sushkov}},\
  }\href {https://doi.org/10.1126/science.aal3003} {\bibfield  {journal}
  {\bibinfo  {journal} {Science}\ }\textbf {\bibinfo {volume} {357}},\ \bibinfo
  {pages} {990} (\bibinfo {year} {2017})}\BibitemShut {NoStop}%
\bibitem [{\citenamefont {Cairncross}\ and\ \citenamefont
  {Ye}(2019)}]{Cairncross2019}%
  \BibitemOpen
  \bibfield  {author} {\bibinfo {author} {\bibfnamefont {W.~B.}\ \bibnamefont
  {Cairncross}}\ and\ \bibinfo {author} {\bibfnamefont {J.}~\bibnamefont
  {Ye}},\ }\href {https://doi.org/10.1038/s42254-019-0080-0} {\bibfield
  {journal} {\bibinfo  {journal} {Nature Reviews Physics}\ }\textbf {\bibinfo
  {volume} {1}},\ \bibinfo {pages} {510} (\bibinfo {year} {2019})}\BibitemShut
  {NoStop}%
\bibitem [{\citenamefont {Hudson}\ \emph {et~al.}(2011)\citenamefont {Hudson},
  \citenamefont {Kara}, \citenamefont {Smallman}, \citenamefont {Sauer},
  \citenamefont {Tarbutt},\ and\ \citenamefont {Hinds}}]{Hudson2011}%
  \BibitemOpen
  \bibfield  {author} {\bibinfo {author} {\bibfnamefont {J.~J.}\ \bibnamefont
  {Hudson}}, \bibinfo {author} {\bibfnamefont {D.~M.}\ \bibnamefont {Kara}},
  \bibinfo {author} {\bibfnamefont {I.~J.}\ \bibnamefont {Smallman}}, \bibinfo
  {author} {\bibfnamefont {B.~E.}\ \bibnamefont {Sauer}}, \bibinfo {author}
  {\bibfnamefont {M.~R.}\ \bibnamefont {Tarbutt}},\ and\ \bibinfo {author}
  {\bibfnamefont {E.~A.}\ \bibnamefont {Hinds}},\ }\href
  {https://doi.org/10.1038/nature10104} {\bibfield  {journal} {\bibinfo
  {journal} {Nature}\ }\textbf {\bibinfo {volume} {473}},\ \bibinfo {pages}
  {493} (\bibinfo {year} {2011})}\BibitemShut {NoStop}%
\bibitem [{\citenamefont {Baron}\ \emph {et~al.}(2014)\citenamefont {Baron},
  \citenamefont {Campbell}, \citenamefont {DeMille}, \citenamefont {Doyle},
  \citenamefont {Gabrielse}, \citenamefont {Gurevich}, \citenamefont {Hess},
  \citenamefont {Hutzler}, \citenamefont {Kirilov}, \citenamefont {Kozyryev},
  \citenamefont {O'Leary}, \citenamefont {Panda}, \citenamefont {Parsons},
  \citenamefont {Petrik}, \citenamefont {Spaun}, \citenamefont {Vutha},\ and\
  \citenamefont {West}}]{Baron2014}%
  \BibitemOpen
  \bibfield  {author} {\bibinfo {author} {\bibfnamefont {J.}~\bibnamefont
  {Baron}}, \bibinfo {author} {\bibfnamefont {W.~C.}\ \bibnamefont {Campbell}},
  \bibinfo {author} {\bibfnamefont {D.}~\bibnamefont {DeMille}}, \bibinfo
  {author} {\bibfnamefont {J.~M.}\ \bibnamefont {Doyle}}, \bibinfo {author}
  {\bibfnamefont {G.}~\bibnamefont {Gabrielse}}, \bibinfo {author}
  {\bibfnamefont {Y.~V.}\ \bibnamefont {Gurevich}}, \bibinfo {author}
  {\bibfnamefont {P.~W.}\ \bibnamefont {Hess}}, \bibinfo {author}
  {\bibfnamefont {N.~R.}\ \bibnamefont {Hutzler}}, \bibinfo {author}
  {\bibfnamefont {E.}~\bibnamefont {Kirilov}}, \bibinfo {author} {\bibfnamefont
  {I.}~\bibnamefont {Kozyryev}}, \bibinfo {author} {\bibfnamefont {B.~R.}\
  \bibnamefont {O'Leary}}, \bibinfo {author} {\bibfnamefont {C.~D.}\
  \bibnamefont {Panda}}, \bibinfo {author} {\bibfnamefont {M.~F.}\ \bibnamefont
  {Parsons}}, \bibinfo {author} {\bibfnamefont {E.~S.}\ \bibnamefont {Petrik}},
  \bibinfo {author} {\bibfnamefont {B.}~\bibnamefont {Spaun}}, \bibinfo
  {author} {\bibfnamefont {A.~C.}\ \bibnamefont {Vutha}},\ and\ \bibinfo
  {author} {\bibfnamefont {A.~D.}\ \bibnamefont {West}},\ }\href
  {https://doi.org/10.1126/science.1248213} {\bibfield  {journal} {\bibinfo
  {journal} {Science}\ }\textbf {\bibinfo {volume} {343}},\ \bibinfo {pages}
  {269} (\bibinfo {year} {2014})}\BibitemShut {NoStop}%
\bibitem [{\citenamefont {Fleig}\ and\ \citenamefont {Jung}(2018)}]{Fleig2018}%
  \BibitemOpen
  \bibfield  {author} {\bibinfo {author} {\bibfnamefont {T.}~\bibnamefont
  {Fleig}}\ and\ \bibinfo {author} {\bibfnamefont {M.}~\bibnamefont {Jung}},\
  }\bibfield  {journal} {\bibinfo  {journal} {Journal of High Energy Physics}\
  }\textbf {\bibinfo {volume} {2018}},\ \href
  {https://doi.org/10.1007/jhep07(2018)012} {10.1007/jhep07(2018)012} (\bibinfo
  {year} {2018})\BibitemShut {NoStop}%
\bibitem [{\citenamefont {Gaul}\ \emph {et~al.}(2019)\citenamefont {Gaul},
  \citenamefont {Marquardt}, \citenamefont {Isaev},\ and\ \citenamefont
  {Berger}}]{Gaul2019}%
  \BibitemOpen
  \bibfield  {author} {\bibinfo {author} {\bibfnamefont {K.}~\bibnamefont
  {Gaul}}, \bibinfo {author} {\bibfnamefont {S.}~\bibnamefont {Marquardt}},
  \bibinfo {author} {\bibfnamefont {T.}~\bibnamefont {Isaev}},\ and\ \bibinfo
  {author} {\bibfnamefont {R.}~\bibnamefont {Berger}},\ }\bibfield  {journal}
  {\bibinfo  {journal} {Physical Review A}\ }\textbf {\bibinfo {volume} {99}},\
  \href {https://doi.org/10.1103/physreva.99.032509}
  {10.1103/physreva.99.032509} (\bibinfo {year} {2019})\BibitemShut {NoStop}%
\bibitem [{\citenamefont {Panda}\ \emph {et~al.}(2019)\citenamefont {Panda},
  \citenamefont {Meisenhelder}, \citenamefont {Verma}, \citenamefont {Ang},
  \citenamefont {Chow}, \citenamefont {Lasner}, \citenamefont {Wu},
  \citenamefont {DeMille}, \citenamefont {Doyle},\ and\ \citenamefont
  {Gabrielse}}]{Panda2019}%
  \BibitemOpen
  \bibfield  {author} {\bibinfo {author} {\bibfnamefont {C.~D.}\ \bibnamefont
  {Panda}}, \bibinfo {author} {\bibfnamefont {C.}~\bibnamefont {Meisenhelder}},
  \bibinfo {author} {\bibfnamefont {M.}~\bibnamefont {Verma}}, \bibinfo
  {author} {\bibfnamefont {D.~G.}\ \bibnamefont {Ang}}, \bibinfo {author}
  {\bibfnamefont {J.}~\bibnamefont {Chow}}, \bibinfo {author} {\bibfnamefont
  {Z.}~\bibnamefont {Lasner}}, \bibinfo {author} {\bibfnamefont
  {X.}~\bibnamefont {Wu}}, \bibinfo {author} {\bibfnamefont {D.}~\bibnamefont
  {DeMille}}, \bibinfo {author} {\bibfnamefont {J.~M.}\ \bibnamefont {Doyle}},\
  and\ \bibinfo {author} {\bibfnamefont {G.}~\bibnamefont {Gabrielse}},\ }\href
  {https://doi.org/10.1088/1361-6455/ab4a61} {\bibfield  {journal} {\bibinfo
  {journal} {Journal of Physics B: Atomic, Molecular and Optical Physics}\
  }\textbf {\bibinfo {volume} {52}},\ \bibinfo {pages} {235003} (\bibinfo
  {year} {2019})}\BibitemShut {NoStop}%
\bibitem [{\citenamefont {Wu}\ \emph {et~al.}(2020)\citenamefont {Wu},
  \citenamefont {Han}, \citenamefont {Chow}, \citenamefont {Ang}, \citenamefont
  {Meisenhelder}, \citenamefont {Panda}, \citenamefont {West}, \citenamefont
  {Gabrielse}, \citenamefont {Doyle},\ and\ \citenamefont {DeMille}}]{Wu2020}%
  \BibitemOpen
  \bibfield  {author} {\bibinfo {author} {\bibfnamefont {X.}~\bibnamefont
  {Wu}}, \bibinfo {author} {\bibfnamefont {Z.}~\bibnamefont {Han}}, \bibinfo
  {author} {\bibfnamefont {J.}~\bibnamefont {Chow}}, \bibinfo {author}
  {\bibfnamefont {D.~G.}\ \bibnamefont {Ang}}, \bibinfo {author} {\bibfnamefont
  {C.}~\bibnamefont {Meisenhelder}}, \bibinfo {author} {\bibfnamefont {C.~D.}\
  \bibnamefont {Panda}}, \bibinfo {author} {\bibfnamefont {E.~P.}\ \bibnamefont
  {West}}, \bibinfo {author} {\bibfnamefont {G.}~\bibnamefont {Gabrielse}},
  \bibinfo {author} {\bibfnamefont {J.~M.}\ \bibnamefont {Doyle}},\ and\
  \bibinfo {author} {\bibfnamefont {D.}~\bibnamefont {DeMille}},\ }\href
  {https://doi.org/10.1088/1367-2630/ab6a3a} {\bibfield  {journal} {\bibinfo
  {journal} {New Journal of Physics}\ }\textbf {\bibinfo {volume} {22}},\
  \bibinfo {pages} {023013} (\bibinfo {year} {2020})}\BibitemShut {NoStop}%
\bibitem [{\citenamefont {Masuda}\ \emph {et~al.}(2021)\citenamefont {Masuda},
  \citenamefont {Ang}, \citenamefont {Hutzler}, \citenamefont {Meisenhelder},
  \citenamefont {Sasao}, \citenamefont {Uetake}, \citenamefont {Wu},
  \citenamefont {DeMille}, \citenamefont {Gabrielse}, \citenamefont {Doyle},\
  and\ \citenamefont {Yoshimura}}]{Masuda2021}%
  \BibitemOpen
  \bibfield  {author} {\bibinfo {author} {\bibfnamefont {T.}~\bibnamefont
  {Masuda}}, \bibinfo {author} {\bibfnamefont {D.~G.}\ \bibnamefont {Ang}},
  \bibinfo {author} {\bibfnamefont {N.~R.}\ \bibnamefont {Hutzler}}, \bibinfo
  {author} {\bibfnamefont {C.}~\bibnamefont {Meisenhelder}}, \bibinfo {author}
  {\bibfnamefont {N.}~\bibnamefont {Sasao}}, \bibinfo {author} {\bibfnamefont
  {S.}~\bibnamefont {Uetake}}, \bibinfo {author} {\bibfnamefont
  {X.}~\bibnamefont {Wu}}, \bibinfo {author} {\bibfnamefont {D.}~\bibnamefont
  {DeMille}}, \bibinfo {author} {\bibfnamefont {G.}~\bibnamefont {Gabrielse}},
  \bibinfo {author} {\bibfnamefont {J.~M.}\ \bibnamefont {Doyle}},\ and\
  \bibinfo {author} {\bibfnamefont {K.}~\bibnamefont {Yoshimura}},\ }\href
  {https://doi.org/10.1364/OE.424460} {\bibfield  {journal} {\bibinfo
  {journal} {Optics Express}\ }\textbf {\bibinfo {volume} {29}},\ \bibinfo
  {pages} {16914} (\bibinfo {year} {2021})}\BibitemShut {NoStop}%
\bibitem [{\citenamefont {Gresh}\ \emph {et~al.}(2016)\citenamefont {Gresh},
  \citenamefont {Cossel}, \citenamefont {Zhou}, \citenamefont {Ye},\ and\
  \citenamefont {Cornell}}]{Gresh2016}%
  \BibitemOpen
  \bibfield  {author} {\bibinfo {author} {\bibfnamefont {D.~N.}\ \bibnamefont
  {Gresh}}, \bibinfo {author} {\bibfnamefont {K.~C.}\ \bibnamefont {Cossel}},
  \bibinfo {author} {\bibfnamefont {Y.}~\bibnamefont {Zhou}}, \bibinfo {author}
  {\bibfnamefont {J.}~\bibnamefont {Ye}},\ and\ \bibinfo {author}
  {\bibfnamefont {E.~A.}\ \bibnamefont {Cornell}},\ }\href
  {https://doi.org/10.1016/j.jms.2015.11.001} {\bibfield  {journal} {\bibinfo
  {journal} {Journal of Molecular Spectroscopy}\ }\textbf {\bibinfo {volume}
  {319}},\ \bibinfo {pages} {1} (\bibinfo {year} {2016})}\BibitemShut {NoStop}%
\bibitem [{\citenamefont {Ng}\ \emph {et~al.}(2022)\citenamefont {Ng},
  \citenamefont {Zhou}, \citenamefont {Cheng}, \citenamefont {Schlossberger},
  \citenamefont {Park}, \citenamefont {Roussy}, \citenamefont {Caldwell},
  \citenamefont {Shagam}, \citenamefont {Vigil}, \citenamefont {Cornell},\ and\
  \citenamefont {Ye}}]{Ng2022}%
  \BibitemOpen
  \bibfield  {author} {\bibinfo {author} {\bibfnamefont {K.~B.}\ \bibnamefont
  {Ng}}, \bibinfo {author} {\bibfnamefont {Y.}~\bibnamefont {Zhou}}, \bibinfo
  {author} {\bibfnamefont {L.}~\bibnamefont {Cheng}}, \bibinfo {author}
  {\bibfnamefont {N.}~\bibnamefont {Schlossberger}}, \bibinfo {author}
  {\bibfnamefont {S.~Y.}\ \bibnamefont {Park}}, \bibinfo {author}
  {\bibfnamefont {T.~S.}\ \bibnamefont {Roussy}}, \bibinfo {author}
  {\bibfnamefont {L.}~\bibnamefont {Caldwell}}, \bibinfo {author}
  {\bibfnamefont {Y.}~\bibnamefont {Shagam}}, \bibinfo {author} {\bibfnamefont
  {A.~J.}\ \bibnamefont {Vigil}}, \bibinfo {author} {\bibfnamefont {E.~A.}\
  \bibnamefont {Cornell}},\ and\ \bibinfo {author} {\bibfnamefont
  {J.}~\bibnamefont {Ye}},\ }\bibfield  {journal} {\bibinfo  {journal}
  {Physical Review A}\ }\textbf {\bibinfo {volume} {105}},\ \href
  {https://doi.org/10.1103/physreva.105.022823} {10.1103/physreva.105.022823}
  (\bibinfo {year} {2022})\BibitemShut {NoStop}%
\bibitem [{\citenamefont {Ho}\ \emph {et~al.}(2020)\citenamefont {Ho},
  \citenamefont {Devlin}, \citenamefont {Rabey}, \citenamefont {Yzombard},
  \citenamefont {Lim}, \citenamefont {Wright}, \citenamefont {Fitch},
  \citenamefont {Hinds}, \citenamefont {Tarbutt},\ and\ \citenamefont
  {Sauer}}]{Ho2020}%
  \BibitemOpen
  \bibfield  {author} {\bibinfo {author} {\bibfnamefont {C.~J.}\ \bibnamefont
  {Ho}}, \bibinfo {author} {\bibfnamefont {J.~A.}\ \bibnamefont {Devlin}},
  \bibinfo {author} {\bibfnamefont {I.~M.}\ \bibnamefont {Rabey}}, \bibinfo
  {author} {\bibfnamefont {P.}~\bibnamefont {Yzombard}}, \bibinfo {author}
  {\bibfnamefont {J.}~\bibnamefont {Lim}}, \bibinfo {author} {\bibfnamefont
  {S.~C.}\ \bibnamefont {Wright}}, \bibinfo {author} {\bibfnamefont {N.~J.}\
  \bibnamefont {Fitch}}, \bibinfo {author} {\bibfnamefont {E.~A.}\ \bibnamefont
  {Hinds}}, \bibinfo {author} {\bibfnamefont {M.~R.}\ \bibnamefont {Tarbutt}},\
  and\ \bibinfo {author} {\bibfnamefont {B.~E.}\ \bibnamefont {Sauer}},\ }\href
  {https://doi.org/10.1088/1367-2630/ab83d2} {\bibfield  {journal} {\bibinfo
  {journal} {New J. Phys.}\ }\textbf {\bibinfo {volume} {22}},\ \bibinfo
  {pages} {053031} (\bibinfo {year} {2020})}\BibitemShut {NoStop}%
\bibitem [{\citenamefont {Alauze}\ \emph {et~al.}(2021)\citenamefont {Alauze},
  \citenamefont {Lim}, \citenamefont {Trigatzis}, \citenamefont {Swarbrick},
  \citenamefont {Collings}, \citenamefont {Fitch}, \citenamefont {Sauer},\ and\
  \citenamefont {Tarbutt}}]{Alauze2021}%
  \BibitemOpen
  \bibfield  {author} {\bibinfo {author} {\bibfnamefont {X.}~\bibnamefont
  {Alauze}}, \bibinfo {author} {\bibfnamefont {J.}~\bibnamefont {Lim}},
  \bibinfo {author} {\bibfnamefont {M.~A.}\ \bibnamefont {Trigatzis}}, \bibinfo
  {author} {\bibfnamefont {S.}~\bibnamefont {Swarbrick}}, \bibinfo {author}
  {\bibfnamefont {F.~J.}\ \bibnamefont {Collings}}, \bibinfo {author}
  {\bibfnamefont {N.~J.}\ \bibnamefont {Fitch}}, \bibinfo {author}
  {\bibfnamefont {B.~E.}\ \bibnamefont {Sauer}},\ and\ \bibinfo {author}
  {\bibfnamefont {M.~R.}\ \bibnamefont {Tarbutt}},\ }\href
  {https://doi.org/10.1088/2058-9565/ac107e} {\bibfield  {journal} {\bibinfo
  {journal} {Quantum Science and Technology}\ }\textbf {\bibinfo {volume}
  {6}},\ \bibinfo {pages} {044005} (\bibinfo {year} {2021})}\BibitemShut
  {NoStop}%
\bibitem [{\citenamefont {Hunter}\ \emph {et~al.}(2012)\citenamefont {Hunter},
  \citenamefont {Peck}, \citenamefont {Greenspon}, \citenamefont {Alam},\ and\
  \citenamefont {DeMille}}]{Hunter2012}%
  \BibitemOpen
  \bibfield  {author} {\bibinfo {author} {\bibfnamefont {L.~R.}\ \bibnamefont
  {Hunter}}, \bibinfo {author} {\bibfnamefont {S.~K.}\ \bibnamefont {Peck}},
  \bibinfo {author} {\bibfnamefont {A.~S.}\ \bibnamefont {Greenspon}}, \bibinfo
  {author} {\bibfnamefont {S.~S.}\ \bibnamefont {Alam}},\ and\ \bibinfo
  {author} {\bibfnamefont {D.}~\bibnamefont {DeMille}},\ }\href
  {https://doi.org/10.1103/PhysRevA.85.012511} {\bibfield  {journal} {\bibinfo
  {journal} {Physical Review A}\ }\textbf {\bibinfo {volume} {85}},\ \bibinfo
  {pages} {012511} (\bibinfo {year} {2012})}\BibitemShut {NoStop}%
\bibitem [{\citenamefont {Grasdijk}\ \emph {et~al.}(2021)\citenamefont
  {Grasdijk}, \citenamefont {Timgren}, \citenamefont {Kastelic}, \citenamefont
  {Wright}, \citenamefont {Lamoreaux}, \citenamefont {DeMille}, \citenamefont
  {Wenz}, \citenamefont {Aitken}, \citenamefont {Zelevinsky}, \citenamefont
  {Winick},\ and\ \citenamefont {Kawall}}]{Grasdijk2021}%
  \BibitemOpen
  \bibfield  {author} {\bibinfo {author} {\bibfnamefont {O.}~\bibnamefont
  {Grasdijk}}, \bibinfo {author} {\bibfnamefont {O.}~\bibnamefont {Timgren}},
  \bibinfo {author} {\bibfnamefont {J.}~\bibnamefont {Kastelic}}, \bibinfo
  {author} {\bibfnamefont {T.}~\bibnamefont {Wright}}, \bibinfo {author}
  {\bibfnamefont {S.}~\bibnamefont {Lamoreaux}}, \bibinfo {author}
  {\bibfnamefont {D.}~\bibnamefont {DeMille}}, \bibinfo {author} {\bibfnamefont
  {K.}~\bibnamefont {Wenz}}, \bibinfo {author} {\bibfnamefont {M.}~\bibnamefont
  {Aitken}}, \bibinfo {author} {\bibfnamefont {T.}~\bibnamefont {Zelevinsky}},
  \bibinfo {author} {\bibfnamefont {T.}~\bibnamefont {Winick}},\ and\ \bibinfo
  {author} {\bibfnamefont {D.}~\bibnamefont {Kawall}},\ }\href
  {https://doi.org/10.1088/2058-9565/abdca3} {\bibfield  {journal} {\bibinfo
  {journal} {Quantum Science and Technology}\ }\textbf {\bibinfo {volume}
  {6}},\ \bibinfo {pages} {044007} (\bibinfo {year} {2021})}\BibitemShut
  {NoStop}%
\bibitem [{\citenamefont {Kozyryev}\ and\ \citenamefont
  {Hutzler}(2017)}]{Kozyryev2017PolyEDM}%
  \BibitemOpen
  \bibfield  {author} {\bibinfo {author} {\bibfnamefont {I.}~\bibnamefont
  {Kozyryev}}\ and\ \bibinfo {author} {\bibfnamefont {N.~R.}\ \bibnamefont
  {Hutzler}},\ }\href {https://doi.org/10.1103/PhysRevLett.119.133002}
  {\bibfield  {journal} {\bibinfo  {journal} {Physical Review Letters}\
  }\textbf {\bibinfo {volume} {119}},\ \bibinfo {pages} {133002} (\bibinfo
  {year} {2017})}\BibitemShut {NoStop}%
\bibitem [{\citenamefont {Maison}\ \emph {et~al.}(2019)\citenamefont {Maison},
  \citenamefont {Skripnikov},\ and\ \citenamefont {Flambaum}}]{Maison2019}%
  \BibitemOpen
  \bibfield  {author} {\bibinfo {author} {\bibfnamefont {D.~E.}\ \bibnamefont
  {Maison}}, \bibinfo {author} {\bibfnamefont {L.~V.}\ \bibnamefont
  {Skripnikov}},\ and\ \bibinfo {author} {\bibfnamefont {V.~V.}\ \bibnamefont
  {Flambaum}},\ }\href {https://doi.org/10.1103/PhysRevA.100.032514} {\bibfield
   {journal} {\bibinfo  {journal} {Physical Review A}\ }\textbf {\bibinfo
  {volume} {100}},\ \bibinfo {pages} {032514} (\bibinfo {year}
  {2019})}\BibitemShut {NoStop}%
\bibitem [{\citenamefont {Denis}\ \emph {et~al.}(2020)\citenamefont {Denis},
  \citenamefont {Hao}, \citenamefont {Eliav}, \citenamefont {Hutzler},
  \citenamefont {Nayak}, \citenamefont {Timmermans},\ and\ \citenamefont
  {Borschesvky}}]{Denis2020}%
  \BibitemOpen
  \bibfield  {author} {\bibinfo {author} {\bibfnamefont {M.}~\bibnamefont
  {Denis}}, \bibinfo {author} {\bibfnamefont {Y.}~\bibnamefont {Hao}}, \bibinfo
  {author} {\bibfnamefont {E.}~\bibnamefont {Eliav}}, \bibinfo {author}
  {\bibfnamefont {N.~R.}\ \bibnamefont {Hutzler}}, \bibinfo {author}
  {\bibfnamefont {M.~K.}\ \bibnamefont {Nayak}}, \bibinfo {author}
  {\bibfnamefont {R.~G.~E.}\ \bibnamefont {Timmermans}},\ and\ \bibinfo
  {author} {\bibfnamefont {A.}~\bibnamefont {Borschesvky}},\ }\href
  {https://doi.org/10.1063/1.5141065} {\bibfield  {journal} {\bibinfo
  {journal} {The Journal of Chemical Physics}\ }\textbf {\bibinfo {volume}
  {152}},\ \bibinfo {pages} {084303} (\bibinfo {year} {2020})}\BibitemShut
  {NoStop}%
\bibitem [{\citenamefont {Pilgram}\ \emph {et~al.}(2021)\citenamefont
  {Pilgram}, \citenamefont {Jadbabaie}, \citenamefont {Zeng}, \citenamefont
  {Hutzler},\ and\ \citenamefont {Steimle}}]{Pilgram2021}%
  \BibitemOpen
  \bibfield  {author} {\bibinfo {author} {\bibfnamefont {N.~H.}\ \bibnamefont
  {Pilgram}}, \bibinfo {author} {\bibfnamefont {A.}~\bibnamefont {Jadbabaie}},
  \bibinfo {author} {\bibfnamefont {Y.}~\bibnamefont {Zeng}}, \bibinfo {author}
  {\bibfnamefont {N.~R.}\ \bibnamefont {Hutzler}},\ and\ \bibinfo {author}
  {\bibfnamefont {T.~C.}\ \bibnamefont {Steimle}},\ }\href
  {https://doi.org/10.1063/5.0055293} {\bibfield  {journal} {\bibinfo
  {journal} {The Journal of Chemical Physics}\ }\textbf {\bibinfo {volume}
  {154}},\ \bibinfo {pages} {244309} (\bibinfo {year} {2021})}\BibitemShut
  {NoStop}%
\bibitem [{\citenamefont {Fleig}(2017{\natexlab{b}})}]{Fleig2017TaO}%
  \BibitemOpen
  \bibfield  {author} {\bibinfo {author} {\bibfnamefont {T.}~\bibnamefont
  {Fleig}},\ }\href {https://doi.org/10.1103/PhysRevA.95.022504} {\bibfield
  {journal} {\bibinfo  {journal} {Phys. Rev. A}\ }\textbf {\bibinfo {volume}
  {95}},\ \bibinfo {pages} {022504} (\bibinfo {year}
  {2017}{\natexlab{b}})}\BibitemShut {NoStop}%
\bibitem [{\citenamefont {Chung}\ \emph {et~al.}(2021)\citenamefont {Chung},
  \citenamefont {Cooper},\ and\ \citenamefont {Zhou}}]{Chung2021}%
  \BibitemOpen
  \bibfield  {author} {\bibinfo {author} {\bibfnamefont {T.}~\bibnamefont
  {Chung}}, \bibinfo {author} {\bibfnamefont {M.~C.}\ \bibnamefont {Cooper}},\
  and\ \bibinfo {author} {\bibfnamefont {Y.}~\bibnamefont {Zhou}},\ }in\
  \href@noop {} {\emph {\bibinfo {booktitle} {Int. Symp. Mol. Spectrosc.}}}\
  (\bibinfo {year} {2021})\ p.\ \bibinfo {pages} {RB06}\BibitemShut {NoStop}%
\bibitem [{\citenamefont {{Garcia Ruiz}}\ \emph {et~al.}(2020)\citenamefont
  {{Garcia Ruiz}}, \citenamefont {Berger}, \citenamefont {Billowes},
  \citenamefont {Binnersley}, \citenamefont {Bissell}, \citenamefont {Breier},
  \citenamefont {Brinson}, \citenamefont {Chrysalidis}, \citenamefont
  {Cocolios}, \citenamefont {Cooper}, \citenamefont {Flanagan}, \citenamefont
  {Giesen}, \citenamefont {de~Groote}, \citenamefont {Franchoo}, \citenamefont
  {Gustafsson}, \citenamefont {Isaev}, \citenamefont {Koszorus}, \citenamefont
  {Neyens}, \citenamefont {Perrett}, \citenamefont {Ricketts}, \citenamefont
  {Rothe}, \citenamefont {Schweikhard}, \citenamefont {Vernon}, \citenamefont
  {Wendt}, \citenamefont {Wienholtz}, \citenamefont {Wilkins},\ and\
  \citenamefont {Yang}}]{GarciaRuiz2020}%
  \BibitemOpen
  \bibfield  {author} {\bibinfo {author} {\bibfnamefont {R.~F.}\ \bibnamefont
  {{Garcia Ruiz}}}, \bibinfo {author} {\bibfnamefont {R.}~\bibnamefont
  {Berger}}, \bibinfo {author} {\bibfnamefont {J.}~\bibnamefont {Billowes}},
  \bibinfo {author} {\bibfnamefont {C.~L.}\ \bibnamefont {Binnersley}},
  \bibinfo {author} {\bibfnamefont {M.~L.}\ \bibnamefont {Bissell}}, \bibinfo
  {author} {\bibfnamefont {A.~A.}\ \bibnamefont {Breier}}, \bibinfo {author}
  {\bibfnamefont {A.~J.}\ \bibnamefont {Brinson}}, \bibinfo {author}
  {\bibfnamefont {K.}~\bibnamefont {Chrysalidis}}, \bibinfo {author}
  {\bibfnamefont {T.~E.}\ \bibnamefont {Cocolios}}, \bibinfo {author}
  {\bibfnamefont {B.~S.}\ \bibnamefont {Cooper}}, \bibinfo {author}
  {\bibfnamefont {K.~T.}\ \bibnamefont {Flanagan}}, \bibinfo {author}
  {\bibfnamefont {T.~F.}\ \bibnamefont {Giesen}}, \bibinfo {author}
  {\bibfnamefont {R.~P.}\ \bibnamefont {de~Groote}}, \bibinfo {author}
  {\bibfnamefont {S.}~\bibnamefont {Franchoo}}, \bibinfo {author}
  {\bibfnamefont {F.~P.}\ \bibnamefont {Gustafsson}}, \bibinfo {author}
  {\bibfnamefont {T.~A.}\ \bibnamefont {Isaev}}, \bibinfo {author}
  {\bibfnamefont {A.}~\bibnamefont {Koszorus}}, \bibinfo {author}
  {\bibfnamefont {G.}~\bibnamefont {Neyens}}, \bibinfo {author} {\bibfnamefont
  {H.~A.}\ \bibnamefont {Perrett}}, \bibinfo {author} {\bibfnamefont {C.~M.}\
  \bibnamefont {Ricketts}}, \bibinfo {author} {\bibfnamefont {S.}~\bibnamefont
  {Rothe}}, \bibinfo {author} {\bibfnamefont {L.}~\bibnamefont {Schweikhard}},
  \bibinfo {author} {\bibfnamefont {A.~R.}\ \bibnamefont {Vernon}}, \bibinfo
  {author} {\bibfnamefont {K.~D.~A.}\ \bibnamefont {Wendt}}, \bibinfo {author}
  {\bibfnamefont {F.}~\bibnamefont {Wienholtz}}, \bibinfo {author}
  {\bibfnamefont {S.~G.}\ \bibnamefont {Wilkins}},\ and\ \bibinfo {author}
  {\bibfnamefont {X.~F.}\ \bibnamefont {Yang}},\ }\href
  {https://doi.org/10.1038/s41586-020-2299-4} {\bibfield  {journal} {\bibinfo
  {journal} {Nature}\ }\textbf {\bibinfo {volume} {581}},\ \bibinfo {pages}
  {396} (\bibinfo {year} {2020})}\BibitemShut {NoStop}%
\bibitem [{\citenamefont {Udrescu}\ \emph {et~al.}(2021)\citenamefont
  {Udrescu}, \citenamefont {Brinson}, \citenamefont {Ruiz}, \citenamefont
  {Gaul}, \citenamefont {Berger}, \citenamefont {Billowes}, \citenamefont
  {Binnersley}, \citenamefont {Bissell}, \citenamefont {Breier}, \citenamefont
  {Chrysalidis}, \citenamefont {Cocolios}, \citenamefont {Cooper},
  \citenamefont {Flanagan}, \citenamefont {Giesen}, \citenamefont {de~Groote},
  \citenamefont {Franchoo}, \citenamefont {Gustafsson}, \citenamefont {Isaev},
  \citenamefont {Koszor{\'{u}}s}, \citenamefont {Neyens}, \citenamefont
  {Perrett}, \citenamefont {Ricketts}, \citenamefont {Rothe}, \citenamefont
  {Vernon}, \citenamefont {Wendt}, \citenamefont {Wienholtz}, \citenamefont
  {Wilkins},\ and\ \citenamefont {Yang}}]{Udrescu2021}%
  \BibitemOpen
  \bibfield  {author} {\bibinfo {author} {\bibfnamefont {S.~M.}\ \bibnamefont
  {Udrescu}}, \bibinfo {author} {\bibfnamefont {A.~J.}\ \bibnamefont
  {Brinson}}, \bibinfo {author} {\bibfnamefont {R.~F.~G.}\ \bibnamefont
  {Ruiz}}, \bibinfo {author} {\bibfnamefont {K.}~\bibnamefont {Gaul}}, \bibinfo
  {author} {\bibfnamefont {R.}~\bibnamefont {Berger}}, \bibinfo {author}
  {\bibfnamefont {J.}~\bibnamefont {Billowes}}, \bibinfo {author}
  {\bibfnamefont {C.~L.}\ \bibnamefont {Binnersley}}, \bibinfo {author}
  {\bibfnamefont {M.~L.}\ \bibnamefont {Bissell}}, \bibinfo {author}
  {\bibfnamefont {A.~A.}\ \bibnamefont {Breier}}, \bibinfo {author}
  {\bibfnamefont {K.}~\bibnamefont {Chrysalidis}}, \bibinfo {author}
  {\bibfnamefont {T.~E.}\ \bibnamefont {Cocolios}}, \bibinfo {author}
  {\bibfnamefont {B.~S.}\ \bibnamefont {Cooper}}, \bibinfo {author}
  {\bibfnamefont {K.~T.}\ \bibnamefont {Flanagan}}, \bibinfo {author}
  {\bibfnamefont {T.~F.}\ \bibnamefont {Giesen}}, \bibinfo {author}
  {\bibfnamefont {R.~P.}\ \bibnamefont {de~Groote}}, \bibinfo {author}
  {\bibfnamefont {S.}~\bibnamefont {Franchoo}}, \bibinfo {author}
  {\bibfnamefont {F.~P.}\ \bibnamefont {Gustafsson}}, \bibinfo {author}
  {\bibfnamefont {T.~A.}\ \bibnamefont {Isaev}}, \bibinfo {author}
  {\bibfnamefont {{\'{A}}.}~\bibnamefont {Koszor{\'{u}}s}}, \bibinfo {author}
  {\bibfnamefont {G.}~\bibnamefont {Neyens}}, \bibinfo {author} {\bibfnamefont
  {H.~A.}\ \bibnamefont {Perrett}}, \bibinfo {author} {\bibfnamefont {C.~M.}\
  \bibnamefont {Ricketts}}, \bibinfo {author} {\bibfnamefont {S.}~\bibnamefont
  {Rothe}}, \bibinfo {author} {\bibfnamefont {A.~R.}\ \bibnamefont {Vernon}},
  \bibinfo {author} {\bibfnamefont {K.~D.~A.}\ \bibnamefont {Wendt}}, \bibinfo
  {author} {\bibfnamefont {F.}~\bibnamefont {Wienholtz}}, \bibinfo {author}
  {\bibfnamefont {S.~G.}\ \bibnamefont {Wilkins}},\ and\ \bibinfo {author}
  {\bibfnamefont {X.~F.}\ \bibnamefont {Yang}},\ }\href
  {https://doi.org/10.1103/PhysRevLett.127.033001} {\bibfield  {journal}
  {\bibinfo  {journal} {Phys. Rev. Lett.}\ }\textbf {\bibinfo {volume} {127}},\
  \bibinfo {pages} {033001} (\bibinfo {year} {2021})}\BibitemShut {NoStop}%
\bibitem [{\citenamefont {Yu}\ and\ \citenamefont {Hutzler}(2021)}]{Yu2021}%
  \BibitemOpen
  \bibfield  {author} {\bibinfo {author} {\bibfnamefont {P.}~\bibnamefont
  {Yu}}\ and\ \bibinfo {author} {\bibfnamefont {N.~R.}\ \bibnamefont
  {Hutzler}},\ }\href {https://doi.org/10.1103/PhysRevLett.126.023003}
  {\bibfield  {journal} {\bibinfo  {journal} {Physical Review Letters}\
  }\textbf {\bibinfo {volume} {126}},\ \bibinfo {pages} {023003} (\bibinfo
  {year} {2021})}\BibitemShut {NoStop}%
\bibitem [{\citenamefont {Graham}\ and\ \citenamefont
  {Rajendran}(2011)}]{Graham2011}%
  \BibitemOpen
  \bibfield  {author} {\bibinfo {author} {\bibfnamefont {P.~W.}\ \bibnamefont
  {Graham}}\ and\ \bibinfo {author} {\bibfnamefont {S.}~\bibnamefont
  {Rajendran}},\ }\href {https://doi.org/10.1103/PhysRevD.84.055013} {\bibfield
   {journal} {\bibinfo  {journal} {Phys. Rev. D}\ ,\ \bibinfo {pages} {055013}}
  (\bibinfo {year} {2011})}\BibitemShut {NoStop}%
\bibitem [{\citenamefont {Graham}\ and\ \citenamefont
  {Rajendran}(2013)}]{Graham2013}%
  \BibitemOpen
  \bibfield  {author} {\bibinfo {author} {\bibfnamefont {P.~W.}\ \bibnamefont
  {Graham}}\ and\ \bibinfo {author} {\bibfnamefont {S.}~\bibnamefont
  {Rajendran}},\ }\href {https://doi.org/10.1103/PhysRevD.88.035023} {\bibfield
   {journal} {\bibinfo  {journal} {Physical Review D}\ }\textbf {\bibinfo
  {volume} {88}},\ \bibinfo {pages} {1} (\bibinfo {year} {2013})}\BibitemShut
  {NoStop}%
\bibitem [{\citenamefont {Stadnik}\ and\ \citenamefont
  {Flambaum}(2014)}]{Stadnik2014}%
  \BibitemOpen
  \bibfield  {author} {\bibinfo {author} {\bibfnamefont {Y.~V.}\ \bibnamefont
  {Stadnik}}\ and\ \bibinfo {author} {\bibfnamefont {V.~V.}\ \bibnamefont
  {Flambaum}},\ }\href {https://doi.org/10.1103/PhysRevD.89.043522} {\bibfield
  {journal} {\bibinfo  {journal} {Physical Review D}\ }\textbf {\bibinfo
  {volume} {89}},\ \bibinfo {pages} {043522} (\bibinfo {year}
  {2014})}\BibitemShut {NoStop}%
\bibitem [{\citenamefont {Budker}\ \emph {et~al.}(2014)\citenamefont {Budker},
  \citenamefont {Graham}, \citenamefont {Ledbetter}, \citenamefont
  {Rajendran},\ and\ \citenamefont {Sushkov}}]{Budker2014}%
  \BibitemOpen
  \bibfield  {author} {\bibinfo {author} {\bibfnamefont {D.}~\bibnamefont
  {Budker}}, \bibinfo {author} {\bibfnamefont {P.~W.}\ \bibnamefont {Graham}},
  \bibinfo {author} {\bibfnamefont {M.}~\bibnamefont {Ledbetter}}, \bibinfo
  {author} {\bibfnamefont {S.}~\bibnamefont {Rajendran}},\ and\ \bibinfo
  {author} {\bibfnamefont {A.~O.}\ \bibnamefont {Sushkov}},\ }\href
  {https://doi.org/10.1103/PhysRevX.4.021030} {\bibfield  {journal} {\bibinfo
  {journal} {Physical Review X}\ }\textbf {\bibinfo {volume} {4}},\ \bibinfo
  {pages} {021030} (\bibinfo {year} {2014})}\BibitemShut {NoStop}%
\bibitem [{\citenamefont {Flambaum}\ and\ \citenamefont
  {Tan}(2019)}]{Flambaum2020SpinRotation}%
  \BibitemOpen
  \bibfield  {author} {\bibinfo {author} {\bibfnamefont {V.~V.}\ \bibnamefont
  {Flambaum}}\ and\ \bibinfo {author} {\bibfnamefont {H.~B.~T.}\ \bibnamefont
  {Tan}},\ }\href {https://doi.org/10.1103/PhysRevD.100.111301} {\bibfield
  {journal} {\bibinfo  {journal} {Physical Review D}\ }\textbf {\bibinfo
  {volume} {100}},\ \bibinfo {pages} {111301} (\bibinfo {year}
  {2019})}\BibitemShut {NoStop}%
\bibitem [{\citenamefont {A.~Arvanitaki}(2021)}]{Arvanitaki2021}%
  \BibitemOpen
  \bibfield  {author} {\bibinfo {author} {\bibfnamefont {K.~V.~T.}\
  \bibnamefont {A.~Arvanitaki}, \bibfnamefont {A.~Madden}},\ }\href@noop {}
  {\bibfield  {journal} {\bibinfo  {journal} {arXiv:2112.11466}\ } (\bibinfo
  {year} {2021})}\BibitemShut {NoStop}%
\bibitem [{\citenamefont {Stadnik}\ \emph {et~al.}(2018)\citenamefont
  {Stadnik}, \citenamefont {Dzuba},\ and\ \citenamefont
  {Flambaum}}]{Stadnik2018}%
  \BibitemOpen
  \bibfield  {author} {\bibinfo {author} {\bibfnamefont {Y.~V.}\ \bibnamefont
  {Stadnik}}, \bibinfo {author} {\bibfnamefont {V.~A.}\ \bibnamefont {Dzuba}},\
  and\ \bibinfo {author} {\bibfnamefont {V.~V.}\ \bibnamefont {Flambaum}},\
  }\href {https://doi.org/10.1103/PhysRevLett.120.013202} {\bibfield  {journal}
  {\bibinfo  {journal} {Physical Review Letters}\ }\textbf {\bibinfo {volume}
  {120}},\ \bibinfo {pages} {013202} (\bibinfo {year} {2018})}\BibitemShut
  {NoStop}%
\bibitem [{\citenamefont {Maison}\ \emph
  {et~al.}(2021{\natexlab{a}})\citenamefont {Maison}, \citenamefont {Flambaum},
  \citenamefont {Hutzler},\ and\ \citenamefont {Skripnikov}}]{Maison2021}%
  \BibitemOpen
  \bibfield  {author} {\bibinfo {author} {\bibfnamefont {D.~E.}\ \bibnamefont
  {Maison}}, \bibinfo {author} {\bibfnamefont {V.~V.}\ \bibnamefont
  {Flambaum}}, \bibinfo {author} {\bibfnamefont {N.~R.}\ \bibnamefont
  {Hutzler}},\ and\ \bibinfo {author} {\bibfnamefont {L.~V.}\ \bibnamefont
  {Skripnikov}},\ }\href {https://doi.org/10.1103/PhysRevA.103.022813}
  {\bibfield  {journal} {\bibinfo  {journal} {Physical Review A}\ }\textbf
  {\bibinfo {volume} {103}},\ \bibinfo {pages} {022813} (\bibinfo {year}
  {2021}{\natexlab{a}})}\BibitemShut {NoStop}%
\bibitem [{\citenamefont {Maison}\ \emph
  {et~al.}(2021{\natexlab{b}})\citenamefont {Maison}, \citenamefont
  {Skripnikov}, \citenamefont {Oleynichenko},\ and\ \citenamefont
  {Zaitsevskii}}]{Maison2021Axion}%
  \BibitemOpen
  \bibfield  {author} {\bibinfo {author} {\bibfnamefont {D.~E.}\ \bibnamefont
  {Maison}}, \bibinfo {author} {\bibfnamefont {L.~V.}\ \bibnamefont
  {Skripnikov}}, \bibinfo {author} {\bibfnamefont {A.~V.}\ \bibnamefont
  {Oleynichenko}},\ and\ \bibinfo {author} {\bibfnamefont {A.~V.}\ \bibnamefont
  {Zaitsevskii}},\ }\href {https://doi.org/10.1063/5.0051590} {\bibfield
  {journal} {\bibinfo  {journal} {The Journal of Chemical Physics}\ }\textbf
  {\bibinfo {volume} {154}},\ \bibinfo {pages} {224303} (\bibinfo {year}
  {2021}{\natexlab{b}})}\BibitemShut {NoStop}%
\bibitem [{\citenamefont {Bothwell}\ \emph {et~al.}(2022)\citenamefont
  {Bothwell}, \citenamefont {Kennedy}, \citenamefont {Aeppli}, \citenamefont
  {Kedar}, \citenamefont {Robinson}, \citenamefont {Oelker}, \citenamefont
  {Staron},\ and\ \citenamefont {Ye}}]{Bothwell2022}%
  \BibitemOpen
  \bibfield  {author} {\bibinfo {author} {\bibfnamefont {T.}~\bibnamefont
  {Bothwell}}, \bibinfo {author} {\bibfnamefont {C.~J.}\ \bibnamefont
  {Kennedy}}, \bibinfo {author} {\bibfnamefont {A.}~\bibnamefont {Aeppli}},
  \bibinfo {author} {\bibfnamefont {D.}~\bibnamefont {Kedar}}, \bibinfo
  {author} {\bibfnamefont {J.~M.}\ \bibnamefont {Robinson}}, \bibinfo {author}
  {\bibfnamefont {E.}~\bibnamefont {Oelker}}, \bibinfo {author} {\bibfnamefont
  {A.}~\bibnamefont {Staron}},\ and\ \bibinfo {author} {\bibfnamefont
  {J.}~\bibnamefont {Ye}},\ }\href {https://doi.org/10.1038/s41586-021-04349-7}
  {\bibfield  {journal} {\bibinfo  {journal} {Nature}\ }\textbf {\bibinfo
  {volume} {602}},\ \bibinfo {pages} {420} (\bibinfo {year}
  {2022})}\BibitemShut {NoStop}%
\bibitem [{\citenamefont {{Di Rosa}}(2004)}]{DiRosa2004}%
  \BibitemOpen
  \bibfield  {author} {\bibinfo {author} {\bibfnamefont {M.~D.}\ \bibnamefont
  {{Di Rosa}}},\ }\href {https://doi.org/10.1140/epjd/e2004-00167-2} {\bibfield
   {journal} {\bibinfo  {journal} {The European Physical Journal D}\ }\textbf
  {\bibinfo {volume} {31}},\ \bibinfo {pages} {395} (\bibinfo {year}
  {2004})}\BibitemShut {NoStop}%
\bibitem [{\citenamefont {Isaev}\ and\ \citenamefont
  {Berger}(2016)}]{Isaev2016Poly}%
  \BibitemOpen
  \bibfield  {author} {\bibinfo {author} {\bibfnamefont {T.~A.}\ \bibnamefont
  {Isaev}}\ and\ \bibinfo {author} {\bibfnamefont {R.}~\bibnamefont {Berger}},\
  }\href {https://doi.org/10.1103/PhysRevLett.116.063006} {\bibfield  {journal}
  {\bibinfo  {journal} {Physical Review Letters}\ }\textbf {\bibinfo {volume}
  {116}},\ \bibinfo {pages} {063006} (\bibinfo {year} {2016})}\BibitemShut
  {NoStop}%
\bibitem [{\citenamefont {Shuman}\ \emph {et~al.}(2010)\citenamefont {Shuman},
  \citenamefont {Barry},\ and\ \citenamefont {DeMille}}]{Shuman2010}%
  \BibitemOpen
  \bibfield  {author} {\bibinfo {author} {\bibfnamefont {E.~S.}\ \bibnamefont
  {Shuman}}, \bibinfo {author} {\bibfnamefont {J.~F.}\ \bibnamefont {Barry}},\
  and\ \bibinfo {author} {\bibfnamefont {D.}~\bibnamefont {DeMille}},\ }\href
  {https://doi.org/10.1038/nature09443} {\bibfield  {journal} {\bibinfo
  {journal} {Nature}\ }\textbf {\bibinfo {volume} {467}},\ \bibinfo {pages}
  {820} (\bibinfo {year} {2010})}\BibitemShut {NoStop}%
\bibitem [{\citenamefont {McCarron}\ \emph {et~al.}(2018)\citenamefont
  {McCarron}, \citenamefont {Steinecker}, \citenamefont {Zhu},\ and\
  \citenamefont {DeMille}}]{McCarron2018SrF}%
  \BibitemOpen
  \bibfield  {author} {\bibinfo {author} {\bibfnamefont {D.~J.}\ \bibnamefont
  {McCarron}}, \bibinfo {author} {\bibfnamefont {M.~H.}\ \bibnamefont
  {Steinecker}}, \bibinfo {author} {\bibfnamefont {Y.}~\bibnamefont {Zhu}},\
  and\ \bibinfo {author} {\bibfnamefont {D.}~\bibnamefont {DeMille}},\ }\href
  {https://doi.org/10.1103/PhysRevLett.121.013202} {\bibfield  {journal}
  {\bibinfo  {journal} {Physical Review Letters}\ }\textbf {\bibinfo {volume}
  {121}},\ \bibinfo {pages} {013202} (\bibinfo {year} {2018})}\BibitemShut
  {NoStop}%
\bibitem [{\citenamefont {Caldwell}\ \emph {et~al.}(2019)\citenamefont
  {Caldwell}, \citenamefont {Devlin}, \citenamefont {Williams}, \citenamefont
  {Fitch}, \citenamefont {Hinds}, \citenamefont {Sauer},\ and\ \citenamefont
  {Tarbutt}}]{Caldwell2019}%
  \BibitemOpen
  \bibfield  {author} {\bibinfo {author} {\bibfnamefont {L.}~\bibnamefont
  {Caldwell}}, \bibinfo {author} {\bibfnamefont {J.~A.}\ \bibnamefont
  {Devlin}}, \bibinfo {author} {\bibfnamefont {H.~J.}\ \bibnamefont
  {Williams}}, \bibinfo {author} {\bibfnamefont {N.~J.}\ \bibnamefont {Fitch}},
  \bibinfo {author} {\bibfnamefont {E.~A.}\ \bibnamefont {Hinds}}, \bibinfo
  {author} {\bibfnamefont {B.~E.}\ \bibnamefont {Sauer}},\ and\ \bibinfo
  {author} {\bibfnamefont {M.~R.}\ \bibnamefont {Tarbutt}},\ }\href
  {https://doi.org/10.1103/PhysRevLett.123.033202} {\bibfield  {journal}
  {\bibinfo  {journal} {Physical Review Letters}\ }\textbf {\bibinfo {volume}
  {123}},\ \bibinfo {pages} {033202} (\bibinfo {year} {2019})}\BibitemShut
  {NoStop}%
\bibitem [{\citenamefont {Anderegg}\ \emph {et~al.}(2019)\citenamefont
  {Anderegg}, \citenamefont {Cheuk}, \citenamefont {Bao}, \citenamefont
  {Burchesky}, \citenamefont {Ketterle}, \citenamefont {Ni},\ and\
  \citenamefont {Doyle}}]{Anderegg2019}%
  \BibitemOpen
  \bibfield  {author} {\bibinfo {author} {\bibfnamefont {L.}~\bibnamefont
  {Anderegg}}, \bibinfo {author} {\bibfnamefont {L.~W.}\ \bibnamefont {Cheuk}},
  \bibinfo {author} {\bibfnamefont {Y.}~\bibnamefont {Bao}}, \bibinfo {author}
  {\bibfnamefont {S.}~\bibnamefont {Burchesky}}, \bibinfo {author}
  {\bibfnamefont {W.}~\bibnamefont {Ketterle}}, \bibinfo {author}
  {\bibfnamefont {K.-K.}\ \bibnamefont {Ni}},\ and\ \bibinfo {author}
  {\bibfnamefont {J.~M.}\ \bibnamefont {Doyle}},\ }\href
  {https://doi.org/10.1126/science.aax1265} {\bibfield  {journal} {\bibinfo
  {journal} {Science}\ }\textbf {\bibinfo {volume} {365}},\ \bibinfo {pages}
  {1156} (\bibinfo {year} {2019})}\BibitemShut {NoStop}%
\bibitem [{\citenamefont {Ding}\ \emph {et~al.}(2020)\citenamefont {Ding},
  \citenamefont {Wu}, \citenamefont {Finneran}, \citenamefont {Burau},\ and\
  \citenamefont {Ye}}]{Ding2020}%
  \BibitemOpen
  \bibfield  {author} {\bibinfo {author} {\bibfnamefont {S.}~\bibnamefont
  {Ding}}, \bibinfo {author} {\bibfnamefont {Y.}~\bibnamefont {Wu}}, \bibinfo
  {author} {\bibfnamefont {I.~A.}\ \bibnamefont {Finneran}}, \bibinfo {author}
  {\bibfnamefont {J.~J.}\ \bibnamefont {Burau}},\ and\ \bibinfo {author}
  {\bibfnamefont {J.}~\bibnamefont {Ye}},\ }\href
  {https://doi.org/10.1103/PhysRevX.10.021049} {\bibfield  {journal} {\bibinfo
  {journal} {Physical Review X}\ }\textbf {\bibinfo {volume} {10}},\ \bibinfo
  {pages} {021049} (\bibinfo {year} {2020})}\BibitemShut {NoStop}%
\bibitem [{\citenamefont {Wu}\ \emph {et~al.}(2021)\citenamefont {Wu},
  \citenamefont {Burau}, \citenamefont {Mehling}, \citenamefont {Ye},\ and\
  \citenamefont {Ding}}]{Wu2021}%
  \BibitemOpen
  \bibfield  {author} {\bibinfo {author} {\bibfnamefont {Y.}~\bibnamefont
  {Wu}}, \bibinfo {author} {\bibfnamefont {J.~J.}\ \bibnamefont {Burau}},
  \bibinfo {author} {\bibfnamefont {K.}~\bibnamefont {Mehling}}, \bibinfo
  {author} {\bibfnamefont {J.}~\bibnamefont {Ye}},\ and\ \bibinfo {author}
  {\bibfnamefont {S.}~\bibnamefont {Ding}},\ }\href
  {https://doi.org/10.1103/PhysRevLett.127.263201} {\bibfield  {journal}
  {\bibinfo  {journal} {Physical Review Letters}\ }\textbf {\bibinfo {volume}
  {127}},\ \bibinfo {pages} {263201} (\bibinfo {year} {2021})}\BibitemShut
  {NoStop}%
\bibitem [{\citenamefont {Tarbutt}\ \emph {et~al.}(2013)\citenamefont
  {Tarbutt}, \citenamefont {Sauer}, \citenamefont {Hudson},\ and\ \citenamefont
  {Hinds}}]{Tarbutt2013}%
  \BibitemOpen
  \bibfield  {author} {\bibinfo {author} {\bibfnamefont {M.~R.}\ \bibnamefont
  {Tarbutt}}, \bibinfo {author} {\bibfnamefont {B.~E.}\ \bibnamefont {Sauer}},
  \bibinfo {author} {\bibfnamefont {J.~J.}\ \bibnamefont {Hudson}},\ and\
  \bibinfo {author} {\bibfnamefont {E.~A.}\ \bibnamefont {Hinds}},\ }\href
  {https://doi.org/10.1088/1367-2630/15/5/053034} {\bibfield  {journal}
  {\bibinfo  {journal} {New Journal of Physics}\ }\textbf {\bibinfo {volume}
  {15}},\ \bibinfo {pages} {053034} (\bibinfo {year} {2013})}\BibitemShut
  {NoStop}%
\bibitem [{\citenamefont {Lim}\ \emph {et~al.}(2018)\citenamefont {Lim},
  \citenamefont {Almond}, \citenamefont {Trigatzis}, \citenamefont {Devlin},
  \citenamefont {Fitch}, \citenamefont {Sauer}, \citenamefont {Tarbutt},\ and\
  \citenamefont {Hinds}}]{Lim2018}%
  \BibitemOpen
  \bibfield  {author} {\bibinfo {author} {\bibfnamefont {J.}~\bibnamefont
  {Lim}}, \bibinfo {author} {\bibfnamefont {J.~R.}\ \bibnamefont {Almond}},
  \bibinfo {author} {\bibfnamefont {M.~A.}\ \bibnamefont {Trigatzis}}, \bibinfo
  {author} {\bibfnamefont {J.~A.}\ \bibnamefont {Devlin}}, \bibinfo {author}
  {\bibfnamefont {N.~J.}\ \bibnamefont {Fitch}}, \bibinfo {author}
  {\bibfnamefont {B.~E.}\ \bibnamefont {Sauer}}, \bibinfo {author}
  {\bibfnamefont {M.~R.}\ \bibnamefont {Tarbutt}},\ and\ \bibinfo {author}
  {\bibfnamefont {E.~A.}\ \bibnamefont {Hinds}},\ }\href
  {https://doi.org/10.1103/PhysRevLett.120.123201} {\bibfield  {journal}
  {\bibinfo  {journal} {Physical Review Letters}\ }\textbf {\bibinfo {volume}
  {120}},\ \bibinfo {pages} {123201} (\bibinfo {year} {2018})}\BibitemShut
  {NoStop}%
\bibitem [{\citenamefont {Fitch}\ \emph {et~al.}(2021)\citenamefont {Fitch},
  \citenamefont {Lim}, \citenamefont {Hinds}, \citenamefont {Sauer},\ and\
  \citenamefont {Tarbutt}}]{Fitch2021YbF}%
  \BibitemOpen
  \bibfield  {author} {\bibinfo {author} {\bibfnamefont {N.~J.}\ \bibnamefont
  {Fitch}}, \bibinfo {author} {\bibfnamefont {J.}~\bibnamefont {Lim}}, \bibinfo
  {author} {\bibfnamefont {E.~A.}\ \bibnamefont {Hinds}}, \bibinfo {author}
  {\bibfnamefont {B.~E.}\ \bibnamefont {Sauer}},\ and\ \bibinfo {author}
  {\bibfnamefont {M.~R.}\ \bibnamefont {Tarbutt}},\ }\href
  {https://doi.org/10.1088/2058-9565/abc931} {\bibfield  {journal} {\bibinfo
  {journal} {Quantum Science and Technology}\ }\textbf {\bibinfo {volume}
  {6}},\ \bibinfo {pages} {014006} (\bibinfo {year} {2021})}\BibitemShut
  {NoStop}%
\bibitem [{\citenamefont {Aggarwal}\ \emph {et~al.}(2018)\citenamefont
  {Aggarwal}, \citenamefont {Bethlem}, \citenamefont {Borschevsky},
  \citenamefont {Denis}, \citenamefont {Esajas}, \citenamefont {Haase},
  \citenamefont {Hao}, \citenamefont {Hoekstra}, \citenamefont {Jungmann},
  \citenamefont {Meijknecht}, \citenamefont {Mooij}, \citenamefont
  {Timmermans}, \citenamefont {Ubachs}, \citenamefont {Willmann},\ and\
  \citenamefont {Zapara}}]{Aggarwal2018}%
  \BibitemOpen
  \bibfield  {author} {\bibinfo {author} {\bibfnamefont {P.}~\bibnamefont
  {Aggarwal}}, \bibinfo {author} {\bibfnamefont {H.~L.}\ \bibnamefont
  {Bethlem}}, \bibinfo {author} {\bibfnamefont {A.}~\bibnamefont
  {Borschevsky}}, \bibinfo {author} {\bibfnamefont {M.}~\bibnamefont {Denis}},
  \bibinfo {author} {\bibfnamefont {K.}~\bibnamefont {Esajas}}, \bibinfo
  {author} {\bibfnamefont {P.~A.~B.}\ \bibnamefont {Haase}}, \bibinfo {author}
  {\bibfnamefont {Y.}~\bibnamefont {Hao}}, \bibinfo {author} {\bibfnamefont
  {S.}~\bibnamefont {Hoekstra}}, \bibinfo {author} {\bibfnamefont
  {K.}~\bibnamefont {Jungmann}}, \bibinfo {author} {\bibfnamefont {T.~B.}\
  \bibnamefont {Meijknecht}}, \bibinfo {author} {\bibfnamefont {M.~C.}\
  \bibnamefont {Mooij}}, \bibinfo {author} {\bibfnamefont {R.~G.~E.}\
  \bibnamefont {Timmermans}}, \bibinfo {author} {\bibfnamefont
  {W.}~\bibnamefont {Ubachs}}, \bibinfo {author} {\bibfnamefont
  {L.}~\bibnamefont {Willmann}},\ and\ \bibinfo {author} {\bibfnamefont
  {A.}~\bibnamefont {Zapara}},\ }\href
  {https://doi.org/10.1140/epjd/e2018-90192-9} {\bibfield  {journal} {\bibinfo
  {journal} {The European Physical Journal D}\ }\textbf {\bibinfo {volume}
  {72}},\ \bibinfo {pages} {197} (\bibinfo {year} {2018})}\BibitemShut
  {NoStop}%
\bibitem [{\citenamefont {Isaev}\ \emph {et~al.}(2010)\citenamefont {Isaev},
  \citenamefont {Hoekstra},\ and\ \citenamefont {Berger}}]{Isaev2010}%
  \BibitemOpen
  \bibfield  {author} {\bibinfo {author} {\bibfnamefont {T.}~\bibnamefont
  {Isaev}}, \bibinfo {author} {\bibfnamefont {S.}~\bibnamefont {Hoekstra}},\
  and\ \bibinfo {author} {\bibfnamefont {R.}~\bibnamefont {Berger}},\ }\href
  {https://doi.org/10.1103/PhysRevA.82.052521} {\bibfield  {journal} {\bibinfo
  {journal} {Physical Review A}\ }\textbf {\bibinfo {volume} {82}},\ \bibinfo
  {pages} {052521} (\bibinfo {year} {2010})}\BibitemShut {NoStop}%
\bibitem [{\citenamefont {Petrov}\ and\ \citenamefont
  {Skripnikov}(2020)}]{Petrov2020}%
  \BibitemOpen
  \bibfield  {author} {\bibinfo {author} {\bibfnamefont {A.~N.}\ \bibnamefont
  {Petrov}}\ and\ \bibinfo {author} {\bibfnamefont {L.~V.}\ \bibnamefont
  {Skripnikov}},\ }\bibfield  {journal} {\bibinfo  {journal} {Physical Review
  A}\ }\textbf {\bibinfo {volume} {102}},\ \href
  {https://doi.org/10.1103/physreva.102.062801} {10.1103/physreva.102.062801}
  (\bibinfo {year} {2020})\BibitemShut {NoStop}%
\bibitem [{\citenamefont {Prasannaa}\ \emph {et~al.}(2015)\citenamefont
  {Prasannaa}, \citenamefont {Vutha}, \citenamefont {Abe},\ and\ \citenamefont
  {Das}}]{Prasannaa2015}%
  \BibitemOpen
  \bibfield  {author} {\bibinfo {author} {\bibfnamefont {V.~S.}\ \bibnamefont
  {Prasannaa}}, \bibinfo {author} {\bibfnamefont {A.~C.}\ \bibnamefont
  {Vutha}}, \bibinfo {author} {\bibfnamefont {M.}~\bibnamefont {Abe}},\ and\
  \bibinfo {author} {\bibfnamefont {B.~P.}\ \bibnamefont {Das}},\ }\href
  {https://doi.org/10.1103/PhysRevLett.114.183001} {\bibfield  {journal}
  {\bibinfo  {journal} {Physical Review Letters}\ }\textbf {\bibinfo {volume}
  {114}},\ \bibinfo {pages} {183001} (\bibinfo {year} {2015})},\ \Eprint
  {https://arxiv.org/abs/1410.5138} {arXiv:1410.5138} \BibitemShut {NoStop}%
\bibitem [{\citenamefont {Yang}\ \emph {et~al.}(2019)\citenamefont {Yang},
  \citenamefont {Li}, \citenamefont {Lin}, \citenamefont {Xu}, \citenamefont
  {Wang}, \citenamefont {Yang},\ and\ \citenamefont {Yin}}]{Yang2019}%
  \BibitemOpen
  \bibfield  {author} {\bibinfo {author} {\bibfnamefont {Z.}~\bibnamefont
  {Yang}}, \bibinfo {author} {\bibfnamefont {J.}~\bibnamefont {Li}}, \bibinfo
  {author} {\bibfnamefont {Q.}~\bibnamefont {Lin}}, \bibinfo {author}
  {\bibfnamefont {L.}~\bibnamefont {Xu}}, \bibinfo {author} {\bibfnamefont
  {H.}~\bibnamefont {Wang}}, \bibinfo {author} {\bibfnamefont {T.}~\bibnamefont
  {Yang}},\ and\ \bibinfo {author} {\bibfnamefont {J.}~\bibnamefont {Yin}},\
  }\href {https://doi.org/10.1103/PhysRevA.99.032502} {\bibfield  {journal}
  {\bibinfo  {journal} {Physical Review A}\ }\textbf {\bibinfo {volume} {99}},\
  \bibinfo {pages} {032502} (\bibinfo {year} {2019})}\BibitemShut {NoStop}%
\bibitem [{\citenamefont {Cho}\ \emph {et~al.}(1991)\citenamefont {Cho},
  \citenamefont {Sangster},\ and\ \citenamefont {Hinds}}]{Cho1991}%
  \BibitemOpen
  \bibfield  {author} {\bibinfo {author} {\bibfnamefont {D.}~\bibnamefont
  {Cho}}, \bibinfo {author} {\bibfnamefont {K.}~\bibnamefont {Sangster}},\ and\
  \bibinfo {author} {\bibfnamefont {E.~A.}\ \bibnamefont {Hinds}},\ }\href
  {https://doi.org/10.1103/PhysRevA.44.2783} {\bibfield  {journal} {\bibinfo
  {journal} {Physical Review A}\ }\textbf {\bibinfo {volume} {44}},\ \bibinfo
  {pages} {2783} (\bibinfo {year} {1991})}\BibitemShut {NoStop}%
\bibitem [{\citenamefont {Denis}\ \emph {et~al.}(2019)\citenamefont {Denis},
  \citenamefont {Haase}, \citenamefont {Timmermans}, \citenamefont {Eliav},
  \citenamefont {Hutzler},\ and\ \citenamefont {Borschevsky}}]{Denis2019}%
  \BibitemOpen
  \bibfield  {author} {\bibinfo {author} {\bibfnamefont {M.}~\bibnamefont
  {Denis}}, \bibinfo {author} {\bibfnamefont {P.~A.~B.}\ \bibnamefont {Haase}},
  \bibinfo {author} {\bibfnamefont {R.~G.~E.}\ \bibnamefont {Timmermans}},
  \bibinfo {author} {\bibfnamefont {E.}~\bibnamefont {Eliav}}, \bibinfo
  {author} {\bibfnamefont {N.~R.}\ \bibnamefont {Hutzler}},\ and\ \bibinfo
  {author} {\bibfnamefont {A.}~\bibnamefont {Borschevsky}},\ }\href
  {https://doi.org/10.1103/PhysRevA.99.042512} {\bibfield  {journal} {\bibinfo
  {journal} {Physical Review A}\ }\textbf {\bibinfo {volume} {99}},\ \bibinfo
  {pages} {042512} (\bibinfo {year} {2019})}\BibitemShut {NoStop}%
\bibitem [{\citenamefont {Prasannaa}\ \emph {et~al.}(2019)\citenamefont
  {Prasannaa}, \citenamefont {Shitara}, \citenamefont {Sakurai}, \citenamefont
  {Abe},\ and\ \citenamefont {Das}}]{Prasannaa2019}%
  \BibitemOpen
  \bibfield  {author} {\bibinfo {author} {\bibfnamefont {V.~S.}\ \bibnamefont
  {Prasannaa}}, \bibinfo {author} {\bibfnamefont {N.}~\bibnamefont {Shitara}},
  \bibinfo {author} {\bibfnamefont {A.}~\bibnamefont {Sakurai}}, \bibinfo
  {author} {\bibfnamefont {M.}~\bibnamefont {Abe}},\ and\ \bibinfo {author}
  {\bibfnamefont {B.~P.}\ \bibnamefont {Das}},\ }\href
  {https://doi.org/10.1103/PhysRevA.99.062502} {\bibfield  {journal} {\bibinfo
  {journal} {Physical Review A}\ }\textbf {\bibinfo {volume} {99}},\ \bibinfo
  {pages} {062502} (\bibinfo {year} {2019})}\BibitemShut {NoStop}%
\bibitem [{\citenamefont {Augenbraun}\ \emph {et~al.}(2020)\citenamefont
  {Augenbraun}, \citenamefont {Lasner}, \citenamefont {Frenett}, \citenamefont
  {Sawaoka}, \citenamefont {Miller}, \citenamefont {Steimle},\ and\
  \citenamefont {Doyle}}]{Augenbraun2020YbOH}%
  \BibitemOpen
  \bibfield  {author} {\bibinfo {author} {\bibfnamefont {B.~L.}\ \bibnamefont
  {Augenbraun}}, \bibinfo {author} {\bibfnamefont {Z.~D.}\ \bibnamefont
  {Lasner}}, \bibinfo {author} {\bibfnamefont {A.}~\bibnamefont {Frenett}},
  \bibinfo {author} {\bibfnamefont {H.}~\bibnamefont {Sawaoka}}, \bibinfo
  {author} {\bibfnamefont {C.}~\bibnamefont {Miller}}, \bibinfo {author}
  {\bibfnamefont {T.~C.}\ \bibnamefont {Steimle}},\ and\ \bibinfo {author}
  {\bibfnamefont {J.~M.}\ \bibnamefont {Doyle}},\ }\href
  {https://doi.org/10.1088/1367-2630/ab687b} {\bibfield  {journal} {\bibinfo
  {journal} {New Journal of Physics}\ }\textbf {\bibinfo {volume} {22}},\
  \bibinfo {pages} {022003} (\bibinfo {year} {2020})}\BibitemShut {NoStop}%
\bibitem [{\citenamefont {Isaev}\ \emph {et~al.}(2017)\citenamefont {Isaev},
  \citenamefont {Zaitsevskii},\ and\ \citenamefont {Eliav}}]{Isaev2017RaOH}%
  \BibitemOpen
  \bibfield  {author} {\bibinfo {author} {\bibfnamefont {T.~A.}\ \bibnamefont
  {Isaev}}, \bibinfo {author} {\bibfnamefont {A.~V.}\ \bibnamefont
  {Zaitsevskii}},\ and\ \bibinfo {author} {\bibfnamefont {E.}~\bibnamefont
  {Eliav}},\ }\href {https://doi.org/10.1088/1361-6455/aa8f34} {\bibfield
  {journal} {\bibinfo  {journal} {Journal of Physics B}\ }\textbf {\bibinfo
  {volume} {50}},\ \bibinfo {pages} {225101} (\bibinfo {year}
  {2017})}\BibitemShut {NoStop}%
\bibitem [{\citenamefont {Kozyryev}\ \emph {et~al.}(2017)\citenamefont
  {Kozyryev}, \citenamefont {Baum}, \citenamefont {Matsuda}, \citenamefont
  {Augenbraun}, \citenamefont {Anderegg}, \citenamefont {Sedlack},\ and\
  \citenamefont {Doyle}}]{Kozyryev2017SrOH}%
  \BibitemOpen
  \bibfield  {author} {\bibinfo {author} {\bibfnamefont {I.}~\bibnamefont
  {Kozyryev}}, \bibinfo {author} {\bibfnamefont {L.}~\bibnamefont {Baum}},
  \bibinfo {author} {\bibfnamefont {K.}~\bibnamefont {Matsuda}}, \bibinfo
  {author} {\bibfnamefont {B.~L.}\ \bibnamefont {Augenbraun}}, \bibinfo
  {author} {\bibfnamefont {L.}~\bibnamefont {Anderegg}}, \bibinfo {author}
  {\bibfnamefont {A.~P.}\ \bibnamefont {Sedlack}},\ and\ \bibinfo {author}
  {\bibfnamefont {J.~M.}\ \bibnamefont {Doyle}},\ }\href
  {https://doi.org/10.1103/PhysRevLett.118.173201} {\bibfield  {journal}
  {\bibinfo  {journal} {Physical Review Letters}\ }\textbf {\bibinfo {volume}
  {118}},\ \bibinfo {pages} {173201} (\bibinfo {year} {2017})}\BibitemShut
  {NoStop}%
\bibitem [{\citenamefont {Cheng}\ \emph {et~al.}(2016)\citenamefont {Cheng},
  \citenamefont {van~der Poel}, \citenamefont {Jansen}, \citenamefont
  {Quintero-P{\'{e}}rez}, \citenamefont {Wall}, \citenamefont {Ubachs},\ and\
  \citenamefont {Bethlem}}]{Cheng2016}%
  \BibitemOpen
  \bibfield  {author} {\bibinfo {author} {\bibfnamefont {C.}~\bibnamefont
  {Cheng}}, \bibinfo {author} {\bibfnamefont {A.~P.}\ \bibnamefont {van~der
  Poel}}, \bibinfo {author} {\bibfnamefont {P.}~\bibnamefont {Jansen}},
  \bibinfo {author} {\bibfnamefont {M.}~\bibnamefont {Quintero-P{\'{e}}rez}},
  \bibinfo {author} {\bibfnamefont {T.~E.}\ \bibnamefont {Wall}}, \bibinfo
  {author} {\bibfnamefont {W.}~\bibnamefont {Ubachs}},\ and\ \bibinfo {author}
  {\bibfnamefont {H.~L.}\ \bibnamefont {Bethlem}},\ }\bibfield  {journal}
  {\bibinfo  {journal} {Physical Review Letters}\ }\textbf {\bibinfo {volume}
  {117}},\ \href {https://doi.org/10.1103/physrevlett.117.253201}
  {10.1103/physrevlett.117.253201} (\bibinfo {year} {2016})\BibitemShut
  {NoStop}%
\bibitem [{\citenamefont {Aggarwal}\ \emph {et~al.}(2021)\citenamefont
  {Aggarwal}, \citenamefont {Yin}, \citenamefont {Esajas}, \citenamefont
  {Bethlem}, \citenamefont {Boeschoten}, \citenamefont {Borschevsky},
  \citenamefont {Hoekstra}, \citenamefont {Jungmann}, \citenamefont {Marshall},
  \citenamefont {Meijknecht}, \citenamefont {Mooij}, \citenamefont
  {Timmermans}, \citenamefont {Touwen}, \citenamefont {Ubachs},\ and\
  \citenamefont {and}}]{Aggarwal2021}%
  \BibitemOpen
  \bibfield  {author} {\bibinfo {author} {\bibfnamefont {P.}~\bibnamefont
  {Aggarwal}}, \bibinfo {author} {\bibfnamefont {Y.}~\bibnamefont {Yin}},
  \bibinfo {author} {\bibfnamefont {K.}~\bibnamefont {Esajas}}, \bibinfo
  {author} {\bibfnamefont {H.}~\bibnamefont {Bethlem}}, \bibinfo {author}
  {\bibfnamefont {A.}~\bibnamefont {Boeschoten}}, \bibinfo {author}
  {\bibfnamefont {A.}~\bibnamefont {Borschevsky}}, \bibinfo {author}
  {\bibfnamefont {S.}~\bibnamefont {Hoekstra}}, \bibinfo {author}
  {\bibfnamefont {K.}~\bibnamefont {Jungmann}}, \bibinfo {author}
  {\bibfnamefont {V.}~\bibnamefont {Marshall}}, \bibinfo {author}
  {\bibfnamefont {T.}~\bibnamefont {Meijknecht}}, \bibinfo {author}
  {\bibfnamefont {M.}~\bibnamefont {Mooij}}, \bibinfo {author} {\bibfnamefont
  {R.}~\bibnamefont {Timmermans}}, \bibinfo {author} {\bibfnamefont
  {A.}~\bibnamefont {Touwen}}, \bibinfo {author} {\bibfnamefont
  {W.}~\bibnamefont {Ubachs}},\ and\ \bibinfo {author} {\bibfnamefont {L.~W.}\
  \bibnamefont {and}},\ }\bibfield  {journal} {\bibinfo  {journal} {Physical
  Review Letters}\ }\textbf {\bibinfo {volume} {127}},\ \href
  {https://doi.org/10.1103/physrevlett.127.173201}
  {10.1103/physrevlett.127.173201} (\bibinfo {year} {2021})\BibitemShut
  {NoStop}%
\bibitem [{\citenamefont {Augenbraun}\ \emph {et~al.}(2021)\citenamefont
  {Augenbraun}, \citenamefont {Frenett}, \citenamefont {Sawaoka}, \citenamefont
  {Hallas}, \citenamefont {Vilas}, \citenamefont {Nasir}, \citenamefont
  {Lasner},\ and\ \citenamefont {Doyle}}]{Augenbraun2021ZS}%
  \BibitemOpen
  \bibfield  {author} {\bibinfo {author} {\bibfnamefont {B.~L.}\ \bibnamefont
  {Augenbraun}}, \bibinfo {author} {\bibfnamefont {A.}~\bibnamefont {Frenett}},
  \bibinfo {author} {\bibfnamefont {H.}~\bibnamefont {Sawaoka}}, \bibinfo
  {author} {\bibfnamefont {C.}~\bibnamefont {Hallas}}, \bibinfo {author}
  {\bibfnamefont {N.~B.}\ \bibnamefont {Vilas}}, \bibinfo {author}
  {\bibfnamefont {A.}~\bibnamefont {Nasir}}, \bibinfo {author} {\bibfnamefont
  {Z.~D.}\ \bibnamefont {Lasner}},\ and\ \bibinfo {author} {\bibfnamefont
  {J.~M.}\ \bibnamefont {Doyle}},\ }\href
  {https://doi.org/10.1103/PhysRevLett.127.263002} {\bibfield  {journal}
  {\bibinfo  {journal} {Physical Review Letters}\ }\textbf {\bibinfo {volume}
  {127}},\ \bibinfo {pages} {263002} (\bibinfo {year} {2021})}\BibitemShut
  {NoStop}%
\bibitem [{\citenamefont {Ni}\ \emph {et~al.}(2008)\citenamefont {Ni},
  \citenamefont {Ospelkaus}, \citenamefont {de~Miranda}, \citenamefont {Pe'er},
  \citenamefont {Neyenhuis}, \citenamefont {Zirbel}, \citenamefont
  {Kotochigova}, \citenamefont {Julienne}, \citenamefont {Jin},\ and\
  \citenamefont {Ye}}]{Ni2008}%
  \BibitemOpen
  \bibfield  {author} {\bibinfo {author} {\bibfnamefont {K.-K.}\ \bibnamefont
  {Ni}}, \bibinfo {author} {\bibfnamefont {S.}~\bibnamefont {Ospelkaus}},
  \bibinfo {author} {\bibfnamefont {M.~H.~G.}\ \bibnamefont {de~Miranda}},
  \bibinfo {author} {\bibfnamefont {A.}~\bibnamefont {Pe'er}}, \bibinfo
  {author} {\bibfnamefont {B.}~\bibnamefont {Neyenhuis}}, \bibinfo {author}
  {\bibfnamefont {J.~J.}\ \bibnamefont {Zirbel}}, \bibinfo {author}
  {\bibfnamefont {S.}~\bibnamefont {Kotochigova}}, \bibinfo {author}
  {\bibfnamefont {P.~S.}\ \bibnamefont {Julienne}}, \bibinfo {author}
  {\bibfnamefont {D.~S.}\ \bibnamefont {Jin}},\ and\ \bibinfo {author}
  {\bibfnamefont {J.}~\bibnamefont {Ye}},\ }\href
  {https://doi.org/10.1126/science.1163861} {\bibfield  {journal} {\bibinfo
  {journal} {Science}\ }\textbf {\bibinfo {volume} {322}},\ \bibinfo {pages}
  {231} (\bibinfo {year} {2008})}\BibitemShut {NoStop}%
\bibitem [{\citenamefont {Meyer}\ and\ \citenamefont {Bohn}(2009)}]{Meyer2009}%
  \BibitemOpen
  \bibfield  {author} {\bibinfo {author} {\bibfnamefont {E.~R.}\ \bibnamefont
  {Meyer}}\ and\ \bibinfo {author} {\bibfnamefont {J.~L.}\ \bibnamefont
  {Bohn}},\ }\href {https://doi.org/10.1103/PhysRevA.80.042508} {\bibfield
  {journal} {\bibinfo  {journal} {Physical Review A}\ }\textbf {\bibinfo
  {volume} {80}},\ \bibinfo {pages} {042508} (\bibinfo {year}
  {2009})}\BibitemShut {NoStop}%
\bibitem [{\citenamefont {Fleig}\ and\ \citenamefont
  {DeMille}(2021)}]{Fleig2021}%
  \BibitemOpen
  \bibfield  {author} {\bibinfo {author} {\bibfnamefont {T.}~\bibnamefont
  {Fleig}}\ and\ \bibinfo {author} {\bibfnamefont {D.}~\bibnamefont
  {DeMille}},\ }\href {https://doi.org/10.1088/1367-2630/ac3619} {\bibfield
  {journal} {\bibinfo  {journal} {New Journal of Physics}\ }\textbf {\bibinfo
  {volume} {23}},\ \bibinfo {pages} {113039} (\bibinfo {year}
  {2021})}\BibitemShut {NoStop}%
\bibitem [{\citenamefont {K{\l}os}\ \emph {et~al.}(2022)\citenamefont
  {K{\l}os}, \citenamefont {Li}, \citenamefont {Tiesinga},\ and\ \citenamefont
  {Kotochigova}}]{Klos2022}%
  \BibitemOpen
  \bibfield  {author} {\bibinfo {author} {\bibfnamefont {J.}~\bibnamefont
  {K{\l}os}}, \bibinfo {author} {\bibfnamefont {H.}~\bibnamefont {Li}},
  \bibinfo {author} {\bibfnamefont {E.}~\bibnamefont {Tiesinga}},\ and\
  \bibinfo {author} {\bibfnamefont {S.}~\bibnamefont {Kotochigova}},\ }\href
  {https://doi.org/10.1088/1367-2630/ac50ea} {\bibfield  {journal} {\bibinfo
  {journal} {New J. Phys.}\ }\textbf {\bibinfo {volume} {24}},\ \bibinfo
  {pages} {025005} (\bibinfo {year} {2022})}\BibitemShut {NoStop}%
\bibitem [{\citenamefont {Sunaga}\ \emph
  {et~al.}(2019{\natexlab{a}})\citenamefont {Sunaga}, \citenamefont {Abe},
  \citenamefont {Hada},\ and\ \citenamefont {Das}}]{Sunaga2019Heavy}%
  \BibitemOpen
  \bibfield  {author} {\bibinfo {author} {\bibfnamefont {A.}~\bibnamefont
  {Sunaga}}, \bibinfo {author} {\bibfnamefont {M.}~\bibnamefont {Abe}},
  \bibinfo {author} {\bibfnamefont {M.}~\bibnamefont {Hada}},\ and\ \bibinfo
  {author} {\bibfnamefont {B.~P.}\ \bibnamefont {Das}},\ }\href
  {https://doi.org/10.1103/PhysRevA.99.062506} {\bibfield  {journal} {\bibinfo
  {journal} {Physical Review A}\ }\textbf {\bibinfo {volume} {99}},\ \bibinfo
  {pages} {062506} (\bibinfo {year} {2019}{\natexlab{a}})}\BibitemShut
  {NoStop}%
\bibitem [{\citenamefont {Sunaga}\ \emph
  {et~al.}(2019{\natexlab{b}})\citenamefont {Sunaga}, \citenamefont {Abe},
  \citenamefont {Prasannaa}, \citenamefont {Aoki},\ and\ \citenamefont
  {Hada}}]{Sunaga2019AMD}%
  \BibitemOpen
  \bibfield  {author} {\bibinfo {author} {\bibfnamefont {A.}~\bibnamefont
  {Sunaga}}, \bibinfo {author} {\bibfnamefont {M.}~\bibnamefont {Abe}},
  \bibinfo {author} {\bibfnamefont {V.~S.}\ \bibnamefont {Prasannaa}}, \bibinfo
  {author} {\bibfnamefont {T.}~\bibnamefont {Aoki}},\ and\ \bibinfo {author}
  {\bibfnamefont {M.}~\bibnamefont {Hada}},\ }\href
  {https://doi.org/10.1088/1361-6455/ab5255} {\bibfield  {journal} {\bibinfo
  {journal} {Journal of Physics B: Atomic, Molecular and Optical Physics}\
  }\textbf {\bibinfo {volume} {53}},\ \bibinfo {pages} {015102} (\bibinfo
  {year} {2019}{\natexlab{b}})}\BibitemShut {NoStop}%
\bibitem [{\citenamefont {Pryor}\ and\ \citenamefont
  {Wilczek}(1987)}]{Pryor1987}%
  \BibitemOpen
  \bibfield  {author} {\bibinfo {author} {\bibfnamefont {C.}~\bibnamefont
  {Pryor}}\ and\ \bibinfo {author} {\bibfnamefont {F.}~\bibnamefont
  {Wilczek}},\ }\href {https://doi.org/10.1016/0370-2693(87)90783-0} {\bibfield
   {journal} {\bibinfo  {journal} {Phys. Lett. B}\ }\textbf {\bibinfo {volume}
  {194}},\ \bibinfo {pages} {137} (\bibinfo {year} {1987})}\BibitemShut
  {NoStop}%
\bibitem [{\citenamefont {Arndt}\ \emph {et~al.}(1993)\citenamefont {Arndt},
  \citenamefont {Kanorsky}, \citenamefont {Weis},\ and\ \citenamefont
  {H{\"{a}}nsch}}]{Arndt1993}%
  \BibitemOpen
  \bibfield  {author} {\bibinfo {author} {\bibfnamefont {M.}~\bibnamefont
  {Arndt}}, \bibinfo {author} {\bibfnamefont {S.}~\bibnamefont {Kanorsky}},
  \bibinfo {author} {\bibfnamefont {A.}~\bibnamefont {Weis}},\ and\ \bibinfo
  {author} {\bibfnamefont {T.}~\bibnamefont {H{\"{a}}nsch}},\ }\href
  {https://doi.org/10.1016/0375-9601(93)90142-M} {\bibfield  {journal}
  {\bibinfo  {journal} {Phys. Lett. A}\ }\textbf {\bibinfo {volume} {174}},\
  \bibinfo {pages} {298} (\bibinfo {year} {1993})}\BibitemShut {NoStop}%
\bibitem [{\citenamefont {Kozlov}\ and\ \citenamefont
  {Derevianko}(2006)}]{Kozlov2006EDM}%
  \BibitemOpen
  \bibfield  {author} {\bibinfo {author} {\bibfnamefont {M.~G.}\ \bibnamefont
  {Kozlov}}\ and\ \bibinfo {author} {\bibfnamefont {A.}~\bibnamefont
  {Derevianko}},\ }\href {https://doi.org/10.1103/PhysRevLett.97.063001}
  {\bibfield  {journal} {\bibinfo  {journal} {Phys. Rev. Lett.}\ }\textbf
  {\bibinfo {volume} {97}},\ \bibinfo {pages} {063001} (\bibinfo {year}
  {2006})}\BibitemShut {NoStop}%
\bibitem [{\citenamefont {Vasil'ev}\ and\ \citenamefont
  {Kolycheva}(1978)}]{VK78}%
  \BibitemOpen
  \bibfield  {author} {\bibinfo {author} {\bibfnamefont {B.~V.}\ \bibnamefont
  {Vasil'ev}}\ and\ \bibinfo {author} {\bibfnamefont {E.~V.}\ \bibnamefont
  {Kolycheva}},\ }\href {http://jetp.ras.ru/cgi-bin/e/index/e/47/2/p243?a=list}
  {\bibfield  {journal} {\bibinfo  {journal} {Sov. Phys. - JETP (Engl.
  Transl.)}\ }\textbf {\bibinfo {volume} {74}} (\bibinfo {year}
  {1978})}\BibitemShut {NoStop}%
\bibitem [{\citenamefont {Heidenreich}\ \emph {et~al.}(2005)\citenamefont
  {Heidenreich}, \citenamefont {Elliott}, \citenamefont {Charney},
  \citenamefont {Virgien}, \citenamefont {Bridges}, \citenamefont {McKeon},
  \citenamefont {Peck}, \citenamefont {Krause}, \citenamefont {Gordon},
  \citenamefont {Hunter},\ and\ \citenamefont {Lamoreaux}}]{Hei2005}%
  \BibitemOpen
  \bibfield  {author} {\bibinfo {author} {\bibfnamefont {B.~J.}\ \bibnamefont
  {Heidenreich}}, \bibinfo {author} {\bibfnamefont {O.~T.}\ \bibnamefont
  {Elliott}}, \bibinfo {author} {\bibfnamefont {N.~D.}\ \bibnamefont
  {Charney}}, \bibinfo {author} {\bibfnamefont {K.~A.}\ \bibnamefont
  {Virgien}}, \bibinfo {author} {\bibfnamefont {A.~W.}\ \bibnamefont
  {Bridges}}, \bibinfo {author} {\bibfnamefont {M.~A.}\ \bibnamefont {McKeon}},
  \bibinfo {author} {\bibfnamefont {S.~K.}\ \bibnamefont {Peck}}, \bibinfo
  {author} {\bibfnamefont {D.}~\bibnamefont {Krause}}, \bibinfo {author}
  {\bibfnamefont {J.~E.}\ \bibnamefont {Gordon}}, \bibinfo {author}
  {\bibfnamefont {L.~R.}\ \bibnamefont {Hunter}},\ and\ \bibinfo {author}
  {\bibfnamefont {S.~K.}\ \bibnamefont {Lamoreaux}},\ }\bibfield  {journal}
  {\bibinfo  {journal} {Physical Review Letters}\ }\textbf {\bibinfo {volume}
  {95}},\ \href {https://doi.org/10.1103/physrevlett.95.253004}
  {10.1103/physrevlett.95.253004} (\bibinfo {year} {2005})\BibitemShut
  {NoStop}%
\bibitem [{\citenamefont {Eckel}\ \emph
  {et~al.}(2012{\natexlab{b}})\citenamefont {Eckel}, \citenamefont {Sushkov},\
  and\ \citenamefont {Lamoreaux}}]{Eckel2012}%
  \BibitemOpen
  \bibfield  {author} {\bibinfo {author} {\bibfnamefont {S.}~\bibnamefont
  {Eckel}}, \bibinfo {author} {\bibfnamefont {a.~O.}\ \bibnamefont {Sushkov}},\
  and\ \bibinfo {author} {\bibfnamefont {S.~K.}\ \bibnamefont {Lamoreaux}},\
  }\href {https://doi.org/10.1103/PhysRevLett.109.193003} {\bibfield  {journal}
  {\bibinfo  {journal} {Physical Review Letters}\ }\textbf {\bibinfo {volume}
  {109}},\ \bibinfo {pages} {193003} (\bibinfo {year}
  {2012}{\natexlab{b}})}\BibitemShut {NoStop}%
\bibitem [{\citenamefont {Kim}\ \emph {et~al.}(2015)\citenamefont {Kim},
  \citenamefont {Liu}, \citenamefont {Lamoreaux}, \citenamefont {Visser},
  \citenamefont {Kunkler}, \citenamefont {Matlashov}, \citenamefont {Long},\
  and\ \citenamefont {Reddy}}]{Kim2015}%
  \BibitemOpen
  \bibfield  {author} {\bibinfo {author} {\bibfnamefont {Y.}~\bibnamefont
  {Kim}}, \bibinfo {author} {\bibfnamefont {C.-Y.}\ \bibnamefont {Liu}},
  \bibinfo {author} {\bibfnamefont {S.}~\bibnamefont {Lamoreaux}}, \bibinfo
  {author} {\bibfnamefont {G.}~\bibnamefont {Visser}}, \bibinfo {author}
  {\bibfnamefont {B.}~\bibnamefont {Kunkler}}, \bibinfo {author} {\bibfnamefont
  {A.}~\bibnamefont {Matlashov}}, \bibinfo {author} {\bibfnamefont
  {J.}~\bibnamefont {Long}},\ and\ \bibinfo {author} {\bibfnamefont
  {T.}~\bibnamefont {Reddy}},\ }\bibfield  {journal} {\bibinfo  {journal}
  {Physical Review D}\ }\textbf {\bibinfo {volume} {91}},\ \href
  {https://doi.org/10.1103/physrevd.91.102004} {10.1103/physrevd.91.102004}
  (\bibinfo {year} {2015})\BibitemShut {NoStop}%
\bibitem [{\citenamefont {Shapiro}(1968)}]{Shapiro68}%
  \BibitemOpen
  \bibfield  {author} {\bibinfo {author} {\bibfnamefont {F.~L.}\ \bibnamefont
  {Shapiro}},\ }\href {https://doi.org/10.1070/pu1968v011n03abeh003840}
  {\bibfield  {journal} {\bibinfo  {journal} {Soviet Physics Uspekhi}\ }\textbf
  {\bibinfo {volume} {11}},\ \bibinfo {pages} {345} (\bibinfo {year}
  {1968})}\BibitemShut {NoStop}%
\bibitem [{\citenamefont {Ignatovich}(1969)}]{Ignat69}%
  \BibitemOpen
  \bibfield  {author} {\bibinfo {author} {\bibfnamefont {V.~K.}\ \bibnamefont
  {Ignatovich}},\ }\href
  {http://jetp.ras.ru/cgi-bin/e/index/e/29/6/p1084?a=list} {\bibfield
  {journal} {\bibinfo  {journal} {Sov. Phys. - JETP (Engl. Transl.)}\ }\textbf
  {\bibinfo {volume} {56}} (\bibinfo {year} {1969})}\BibitemShut {NoStop}%
\bibitem [{\citenamefont {Flambaum}\ and\ \citenamefont
  {Dzuba}(2020)}]{Flambaum2020Schiff}%
  \BibitemOpen
  \bibfield  {author} {\bibinfo {author} {\bibfnamefont {V.~V.}\ \bibnamefont
  {Flambaum}}\ and\ \bibinfo {author} {\bibfnamefont {V.~A.}\ \bibnamefont
  {Dzuba}},\ }\href {https://doi.org/10.1103/PhysRevA.101.042504} {\bibfield
  {journal} {\bibinfo  {journal} {Physical Review A}\ }\textbf {\bibinfo
  {volume} {101}},\ \bibinfo {pages} {042504} (\bibinfo {year}
  {2020})}\BibitemShut {NoStop}%
\bibitem [{\citenamefont {Flambaum}\ and\ \citenamefont
  {Feldmeier}(2020)}]{Flambaum2020SchiffStable}%
  \BibitemOpen
  \bibfield  {author} {\bibinfo {author} {\bibfnamefont {V.~V.}\ \bibnamefont
  {Flambaum}}\ and\ \bibinfo {author} {\bibfnamefont {H.}~\bibnamefont
  {Feldmeier}},\ }\bibfield  {journal} {\bibinfo  {journal} {Physical Review
  C}\ }\textbf {\bibinfo {volume} {101}},\ \href
  {https://doi.org/10.1103/physrevc.101.015502} {10.1103/physrevc.101.015502}
  (\bibinfo {year} {2020})\BibitemShut {NoStop}%
\bibitem [{\citenamefont {Gaffney}\ \emph {et~al.}(2013)\citenamefont
  {Gaffney}, \citenamefont {Butler}, \citenamefont {Scheck}, \citenamefont
  {Hayes}, \citenamefont {Wenander}, \citenamefont {Albers}, \citenamefont
  {Bastin}, \citenamefont {Bauer}, \citenamefont {Blazhev}, \citenamefont
  {B{\"{o}}nig}, \citenamefont {Bree}, \citenamefont {Cederk{\"{a}}ll},
  \citenamefont {Chupp}, \citenamefont {Cline}, \citenamefont {Cocolios},
  \citenamefont {Davinson}, \citenamefont {{De Witte}}, \citenamefont
  {Diriken}, \citenamefont {Grahn}, \citenamefont {Herzan}, \citenamefont
  {Huyse}, \citenamefont {Jenkins}, \citenamefont {Joss}, \citenamefont
  {Kesteloot}, \citenamefont {Konki}, \citenamefont {Kowalczyk}, \citenamefont
  {Kr{\"{o}}ll}, \citenamefont {Kwan}, \citenamefont {Lutter}, \citenamefont
  {Moschner}, \citenamefont {Napiorkowski}, \citenamefont {Pakarinen},
  \citenamefont {Pfeiffer}, \citenamefont {Radeck}, \citenamefont {Reiter},
  \citenamefont {Reynders}, \citenamefont {Rigby}, \citenamefont {Robledo},
  \citenamefont {Rudigier}, \citenamefont {Sambi}, \citenamefont {Seidlitz},
  \citenamefont {Siebeck}, \citenamefont {Stora}, \citenamefont {Thoele},
  \citenamefont {{Van Duppen}}, \citenamefont {Vermeulen}, \citenamefont {von
  Schmid}, \citenamefont {Voulot}, \citenamefont {Warr}, \citenamefont
  {Wimmer}, \citenamefont {Wrzosek-Lipska}, \citenamefont {Wu},\ and\
  \citenamefont {Zielinska}}]{Gaffney2013}%
  \BibitemOpen
  \bibfield  {author} {\bibinfo {author} {\bibfnamefont {L.~P.}\ \bibnamefont
  {Gaffney}}, \bibinfo {author} {\bibfnamefont {P.~A.}\ \bibnamefont {Butler}},
  \bibinfo {author} {\bibfnamefont {M.}~\bibnamefont {Scheck}}, \bibinfo
  {author} {\bibfnamefont {A.~B.}\ \bibnamefont {Hayes}}, \bibinfo {author}
  {\bibfnamefont {F.}~\bibnamefont {Wenander}}, \bibinfo {author}
  {\bibfnamefont {M.}~\bibnamefont {Albers}}, \bibinfo {author} {\bibfnamefont
  {B.}~\bibnamefont {Bastin}}, \bibinfo {author} {\bibfnamefont
  {C.}~\bibnamefont {Bauer}}, \bibinfo {author} {\bibfnamefont
  {A.}~\bibnamefont {Blazhev}}, \bibinfo {author} {\bibfnamefont
  {S.}~\bibnamefont {B{\"{o}}nig}}, \bibinfo {author} {\bibfnamefont
  {N.}~\bibnamefont {Bree}}, \bibinfo {author} {\bibfnamefont {J.}~\bibnamefont
  {Cederk{\"{a}}ll}}, \bibinfo {author} {\bibfnamefont {T.}~\bibnamefont
  {Chupp}}, \bibinfo {author} {\bibfnamefont {D.}~\bibnamefont {Cline}},
  \bibinfo {author} {\bibfnamefont {T.~E.}\ \bibnamefont {Cocolios}}, \bibinfo
  {author} {\bibfnamefont {T.}~\bibnamefont {Davinson}}, \bibinfo {author}
  {\bibfnamefont {H.}~\bibnamefont {{De Witte}}}, \bibinfo {author}
  {\bibfnamefont {J.}~\bibnamefont {Diriken}}, \bibinfo {author} {\bibfnamefont
  {T.}~\bibnamefont {Grahn}}, \bibinfo {author} {\bibfnamefont
  {A.}~\bibnamefont {Herzan}}, \bibinfo {author} {\bibfnamefont
  {M.}~\bibnamefont {Huyse}}, \bibinfo {author} {\bibfnamefont {D.~G.}\
  \bibnamefont {Jenkins}}, \bibinfo {author} {\bibfnamefont {D.~T.}\
  \bibnamefont {Joss}}, \bibinfo {author} {\bibfnamefont {N.}~\bibnamefont
  {Kesteloot}}, \bibinfo {author} {\bibfnamefont {J.}~\bibnamefont {Konki}},
  \bibinfo {author} {\bibfnamefont {M.}~\bibnamefont {Kowalczyk}}, \bibinfo
  {author} {\bibfnamefont {T.}~\bibnamefont {Kr{\"{o}}ll}}, \bibinfo {author}
  {\bibfnamefont {E.}~\bibnamefont {Kwan}}, \bibinfo {author} {\bibfnamefont
  {R.}~\bibnamefont {Lutter}}, \bibinfo {author} {\bibfnamefont
  {K.}~\bibnamefont {Moschner}}, \bibinfo {author} {\bibfnamefont
  {P.}~\bibnamefont {Napiorkowski}}, \bibinfo {author} {\bibfnamefont
  {J.}~\bibnamefont {Pakarinen}}, \bibinfo {author} {\bibfnamefont
  {M.}~\bibnamefont {Pfeiffer}}, \bibinfo {author} {\bibfnamefont
  {D.}~\bibnamefont {Radeck}}, \bibinfo {author} {\bibfnamefont
  {P.}~\bibnamefont {Reiter}}, \bibinfo {author} {\bibfnamefont
  {K.}~\bibnamefont {Reynders}}, \bibinfo {author} {\bibfnamefont {S.~V.}\
  \bibnamefont {Rigby}}, \bibinfo {author} {\bibfnamefont {L.~M.}\ \bibnamefont
  {Robledo}}, \bibinfo {author} {\bibfnamefont {M.}~\bibnamefont {Rudigier}},
  \bibinfo {author} {\bibfnamefont {S.}~\bibnamefont {Sambi}}, \bibinfo
  {author} {\bibfnamefont {M.}~\bibnamefont {Seidlitz}}, \bibinfo {author}
  {\bibfnamefont {B.}~\bibnamefont {Siebeck}}, \bibinfo {author} {\bibfnamefont
  {T.}~\bibnamefont {Stora}}, \bibinfo {author} {\bibfnamefont
  {P.}~\bibnamefont {Thoele}}, \bibinfo {author} {\bibfnamefont
  {P.}~\bibnamefont {{Van Duppen}}}, \bibinfo {author} {\bibfnamefont {M.~J.}\
  \bibnamefont {Vermeulen}}, \bibinfo {author} {\bibfnamefont {M.}~\bibnamefont
  {von Schmid}}, \bibinfo {author} {\bibfnamefont {D.}~\bibnamefont {Voulot}},
  \bibinfo {author} {\bibfnamefont {N.}~\bibnamefont {Warr}}, \bibinfo {author}
  {\bibfnamefont {K.}~\bibnamefont {Wimmer}}, \bibinfo {author} {\bibfnamefont
  {K.}~\bibnamefont {Wrzosek-Lipska}}, \bibinfo {author} {\bibfnamefont
  {C.~Y.}\ \bibnamefont {Wu}},\ and\ \bibinfo {author} {\bibfnamefont
  {M.}~\bibnamefont {Zielinska}},\ }\href {https://doi.org/10.1038/nature12073}
  {\bibfield  {journal} {\bibinfo  {journal} {Nature}\ }\textbf {\bibinfo
  {volume} {497}},\ \bibinfo {pages} {199} (\bibinfo {year}
  {2013})}\BibitemShut {NoStop}%
\bibitem [{\citenamefont {Butler}\ \emph {et~al.}(2020)\citenamefont {Butler},
  \citenamefont {Gaffney}, \citenamefont {Spagnoletti}, \citenamefont
  {Abrahams}, \citenamefont {Bowry}, \citenamefont {Cederk{\"{a}}ll},
  \citenamefont {de~Angelis}, \citenamefont {{De Witte}}, \citenamefont
  {Garrett}, \citenamefont {Goldkuhle}, \citenamefont {Henrich}, \citenamefont
  {Illana}, \citenamefont {Johnston}, \citenamefont {Joss}, \citenamefont
  {Keatings}, \citenamefont {Kelly}, \citenamefont {Komorowska}, \citenamefont
  {Konki}, \citenamefont {Kr{\"{o}}ll}, \citenamefont {Lozano}, \citenamefont
  {{Nara Singh}}, \citenamefont {O'Donnell}, \citenamefont {Ojala},
  \citenamefont {Page}, \citenamefont {Pedersen}, \citenamefont {Raison},
  \citenamefont {Reiter}, \citenamefont {Rodriguez}, \citenamefont {Rosiak},
  \citenamefont {Rothe}, \citenamefont {Scheck}, \citenamefont {Seidlitz},
  \citenamefont {Shneidman}, \citenamefont {Siebeck}, \citenamefont {Sinclair},
  \citenamefont {Smith}, \citenamefont {Stryjczyk}, \citenamefont {{Van
  Duppen}}, \citenamefont {Vinals}, \citenamefont {Virtanen}, \citenamefont
  {Warr}, \citenamefont {Wrzosek-Lipska},\ and\ \citenamefont
  {Zieli{\'{n}}ska}}]{Butler2020}%
  \BibitemOpen
  \bibfield  {author} {\bibinfo {author} {\bibfnamefont {P.~A.}\ \bibnamefont
  {Butler}}, \bibinfo {author} {\bibfnamefont {L.~P.}\ \bibnamefont {Gaffney}},
  \bibinfo {author} {\bibfnamefont {P.}~\bibnamefont {Spagnoletti}}, \bibinfo
  {author} {\bibfnamefont {K.}~\bibnamefont {Abrahams}}, \bibinfo {author}
  {\bibfnamefont {M.}~\bibnamefont {Bowry}}, \bibinfo {author} {\bibfnamefont
  {J.}~\bibnamefont {Cederk{\"{a}}ll}}, \bibinfo {author} {\bibfnamefont
  {G.}~\bibnamefont {de~Angelis}}, \bibinfo {author} {\bibfnamefont
  {H.}~\bibnamefont {{De Witte}}}, \bibinfo {author} {\bibfnamefont {P.~E.}\
  \bibnamefont {Garrett}}, \bibinfo {author} {\bibfnamefont {A.}~\bibnamefont
  {Goldkuhle}}, \bibinfo {author} {\bibfnamefont {C.}~\bibnamefont {Henrich}},
  \bibinfo {author} {\bibfnamefont {A.}~\bibnamefont {Illana}}, \bibinfo
  {author} {\bibfnamefont {K.}~\bibnamefont {Johnston}}, \bibinfo {author}
  {\bibfnamefont {D.~T.}\ \bibnamefont {Joss}}, \bibinfo {author}
  {\bibfnamefont {J.~M.}\ \bibnamefont {Keatings}}, \bibinfo {author}
  {\bibfnamefont {N.~A.}\ \bibnamefont {Kelly}}, \bibinfo {author}
  {\bibfnamefont {M.}~\bibnamefont {Komorowska}}, \bibinfo {author}
  {\bibfnamefont {J.}~\bibnamefont {Konki}}, \bibinfo {author} {\bibfnamefont
  {T.}~\bibnamefont {Kr{\"{o}}ll}}, \bibinfo {author} {\bibfnamefont
  {M.}~\bibnamefont {Lozano}}, \bibinfo {author} {\bibfnamefont {B.~S.}\
  \bibnamefont {{Nara Singh}}}, \bibinfo {author} {\bibfnamefont
  {D.}~\bibnamefont {O'Donnell}}, \bibinfo {author} {\bibfnamefont
  {J.}~\bibnamefont {Ojala}}, \bibinfo {author} {\bibfnamefont {R.~D.}\
  \bibnamefont {Page}}, \bibinfo {author} {\bibfnamefont {L.~G.}\ \bibnamefont
  {Pedersen}}, \bibinfo {author} {\bibfnamefont {C.}~\bibnamefont {Raison}},
  \bibinfo {author} {\bibfnamefont {P.}~\bibnamefont {Reiter}}, \bibinfo
  {author} {\bibfnamefont {J.~A.}\ \bibnamefont {Rodriguez}}, \bibinfo {author}
  {\bibfnamefont {D.}~\bibnamefont {Rosiak}}, \bibinfo {author} {\bibfnamefont
  {S.}~\bibnamefont {Rothe}}, \bibinfo {author} {\bibfnamefont
  {M.}~\bibnamefont {Scheck}}, \bibinfo {author} {\bibfnamefont
  {M.}~\bibnamefont {Seidlitz}}, \bibinfo {author} {\bibfnamefont {T.~M.}\
  \bibnamefont {Shneidman}}, \bibinfo {author} {\bibfnamefont {B.}~\bibnamefont
  {Siebeck}}, \bibinfo {author} {\bibfnamefont {J.}~\bibnamefont {Sinclair}},
  \bibinfo {author} {\bibfnamefont {J.~F.}\ \bibnamefont {Smith}}, \bibinfo
  {author} {\bibfnamefont {M.}~\bibnamefont {Stryjczyk}}, \bibinfo {author}
  {\bibfnamefont {P.}~\bibnamefont {{Van Duppen}}}, \bibinfo {author}
  {\bibfnamefont {S.}~\bibnamefont {Vinals}}, \bibinfo {author} {\bibfnamefont
  {V.}~\bibnamefont {Virtanen}}, \bibinfo {author} {\bibfnamefont
  {N.}~\bibnamefont {Warr}}, \bibinfo {author} {\bibfnamefont {K.}~\bibnamefont
  {Wrzosek-Lipska}},\ and\ \bibinfo {author} {\bibfnamefont {M.}~\bibnamefont
  {Zieli{\'{n}}ska}},\ }\href {https://doi.org/10.1103/PhysRevLett.124.042503}
  {\bibfield  {journal} {\bibinfo  {journal} {Physical Review Letters}\
  }\textbf {\bibinfo {volume} {124}},\ \bibinfo {pages} {042503} (\bibinfo
  {year} {2020})}\BibitemShut {NoStop}%
\bibitem [{\citenamefont {Fan}\ \emph {et~al.}(2019)\citenamefont {Fan},
  \citenamefont {Holliman}, \citenamefont {Wang},\ and\ \citenamefont
  {Jayich}}]{Fan2019}%
  \BibitemOpen
  \bibfield  {author} {\bibinfo {author} {\bibfnamefont {M.}~\bibnamefont
  {Fan}}, \bibinfo {author} {\bibfnamefont {C.~A.}\ \bibnamefont {Holliman}},
  \bibinfo {author} {\bibfnamefont {A.~L.}\ \bibnamefont {Wang}},\ and\
  \bibinfo {author} {\bibfnamefont {A.~M.}\ \bibnamefont {Jayich}},\ }\href
  {https://doi.org/10.1103/PhysRevLett.122.223001} {\bibfield  {journal}
  {\bibinfo  {journal} {Phys. Rev. Lett.}\ }\textbf {\bibinfo {volume} {122}},\
  \bibinfo {pages} {223001} (\bibinfo {year} {2019})}\BibitemShut {NoStop}%
\bibitem [{\citenamefont {Skripnikov}\ \emph {et~al.}(2020)\citenamefont
  {Skripnikov}, \citenamefont {Mosyagin}, \citenamefont {Titov},\ and\
  \citenamefont {Flambaum}}]{Skripnikov2020Act}%
  \BibitemOpen
  \bibfield  {author} {\bibinfo {author} {\bibfnamefont {L.~V.}\ \bibnamefont
  {Skripnikov}}, \bibinfo {author} {\bibfnamefont {N.~S.}\ \bibnamefont
  {Mosyagin}}, \bibinfo {author} {\bibfnamefont {A.~V.}\ \bibnamefont
  {Titov}},\ and\ \bibinfo {author} {\bibfnamefont {V.~V.}\ \bibnamefont
  {Flambaum}},\ }\href {https://doi.org/10.1039/D0CP01989E} {\bibfield
  {journal} {\bibinfo  {journal} {Physical Chemistry Chemical Physics}\
  }\textbf {\bibinfo {volume} {22}},\ \bibinfo {pages} {18374} (\bibinfo {year}
  {2020})}\BibitemShut {NoStop}%
\bibitem [{\citenamefont {Ahmad}\ \emph {et~al.}(1982)\citenamefont {Ahmad},
  \citenamefont {Gindler}, \citenamefont {Betts}, \citenamefont {Chasman},\
  and\ \citenamefont {Friedman}}]{Ahmad1982}%
  \BibitemOpen
  \bibfield  {author} {\bibinfo {author} {\bibfnamefont {I.}~\bibnamefont
  {Ahmad}}, \bibinfo {author} {\bibfnamefont {J.~E.}\ \bibnamefont {Gindler}},
  \bibinfo {author} {\bibfnamefont {R.~R.}\ \bibnamefont {Betts}}, \bibinfo
  {author} {\bibfnamefont {R.~R.}\ \bibnamefont {Chasman}},\ and\ \bibinfo
  {author} {\bibfnamefont {A.~M.}\ \bibnamefont {Friedman}},\ }\href
  {https://doi.org/10.1103/PhysRevLett.49.1758} {\bibfield  {journal} {\bibinfo
   {journal} {Physical Review Letters}\ }\textbf {\bibinfo {volume} {49}},\
  \bibinfo {pages} {1758} (\bibinfo {year} {1982})}\BibitemShut {NoStop}%
\bibitem [{\citenamefont {Ahmad}\ \emph {et~al.}(2015)\citenamefont {Ahmad},
  \citenamefont {Chasman}, \citenamefont {Greene}, \citenamefont {Kondev},\
  and\ \citenamefont {Zhu}}]{Ahmad2015}%
  \BibitemOpen
  \bibfield  {author} {\bibinfo {author} {\bibfnamefont {I.}~\bibnamefont
  {Ahmad}}, \bibinfo {author} {\bibfnamefont {R.~R.}\ \bibnamefont {Chasman}},
  \bibinfo {author} {\bibfnamefont {J.~P.}\ \bibnamefont {Greene}}, \bibinfo
  {author} {\bibfnamefont {F.~G.}\ \bibnamefont {Kondev}},\ and\ \bibinfo
  {author} {\bibfnamefont {S.}~\bibnamefont {Zhu}},\ }\href
  {https://doi.org/10.1103/PhysRevC.92.024313} {\bibfield  {journal} {\bibinfo
  {journal} {Phys. Rev. C}\ }\textbf {\bibinfo {volume} {92}},\ \bibinfo
  {pages} {024313} (\bibinfo {year} {2015})}\BibitemShut {NoStop}%
\bibitem [{\citenamefont {Flambaum}(2008)}]{Flambaum2008}%
  \BibitemOpen
  \bibfield  {author} {\bibinfo {author} {\bibfnamefont {V.~V.}\ \bibnamefont
  {Flambaum}},\ }\href {https://doi.org/10.1103/PhysRevA.77.024501} {\bibfield
  {journal} {\bibinfo  {journal} {Physical Review A}\ }\textbf {\bibinfo
  {volume} {77}},\ \bibinfo {pages} {024501} (\bibinfo {year}
  {2008})}\BibitemShut {NoStop}%
\bibitem [{\citenamefont {Mitra}\ \emph {et~al.}(2021)\citenamefont {Mitra},
  \citenamefont {Prasannaa}, \citenamefont {Ruiz}, \citenamefont {Sato},
  \citenamefont {Abe}, \citenamefont {Sakemi}, \citenamefont {Das},\ and\
  \citenamefont {Sahoo}}]{Mitra2021Superheavy}%
  \BibitemOpen
  \bibfield  {author} {\bibinfo {author} {\bibfnamefont {R.}~\bibnamefont
  {Mitra}}, \bibinfo {author} {\bibfnamefont {V.~S.}\ \bibnamefont
  {Prasannaa}}, \bibinfo {author} {\bibfnamefont {R.~F.~G.}\ \bibnamefont
  {Ruiz}}, \bibinfo {author} {\bibfnamefont {T.~K.}\ \bibnamefont {Sato}},
  \bibinfo {author} {\bibfnamefont {M.}~\bibnamefont {Abe}}, \bibinfo {author}
  {\bibfnamefont {Y.}~\bibnamefont {Sakemi}}, \bibinfo {author} {\bibfnamefont
  {B.~P.}\ \bibnamefont {Das}},\ and\ \bibinfo {author} {\bibfnamefont {B.~K.}\
  \bibnamefont {Sahoo}},\ }\bibfield  {journal} {\bibinfo  {journal} {Physical
  Review A}\ }\textbf {\bibinfo {volume} {104}},\ \href
  {https://doi.org/10.1103/physreva.104.062801} {10.1103/physreva.104.062801}
  (\bibinfo {year} {2021})\BibitemShut {NoStop}%
\bibitem [{\citenamefont {Clo{\"{e}}t}\ \emph {et~al.}(2019)\citenamefont
  {Clo{\"{e}}t}, \citenamefont {Dietrich}, \citenamefont {Arrington},
  \citenamefont {Bazavov}, \citenamefont {Bishof}, \citenamefont {Freese},
  \citenamefont {Gorshkov}, \citenamefont {Grassellino}, \citenamefont
  {Hafidi}, \citenamefont {Jacob}, \citenamefont {McGuigan}, \citenamefont
  {Meurice}, \citenamefont {Meziani}, \citenamefont {Mueller}, \citenamefont
  {Muschik}, \citenamefont {Osborn}, \citenamefont {Otten}, \citenamefont
  {Petreczky}, \citenamefont {Polakovic}, \citenamefont {Poon}, \citenamefont
  {Pooser}, \citenamefont {Roggero}, \citenamefont {Saffman}, \citenamefont
  {VanDevender}, \citenamefont {Zhang},\ and\ \citenamefont
  {Zohar}}]{Cloet2019}%
  \BibitemOpen
  \bibfield  {author} {\bibinfo {author} {\bibfnamefont {I.~C.}\ \bibnamefont
  {Clo{\"{e}}t}}, \bibinfo {author} {\bibfnamefont {M.~R.}\ \bibnamefont
  {Dietrich}}, \bibinfo {author} {\bibfnamefont {J.}~\bibnamefont {Arrington}},
  \bibinfo {author} {\bibfnamefont {A.}~\bibnamefont {Bazavov}}, \bibinfo
  {author} {\bibfnamefont {M.}~\bibnamefont {Bishof}}, \bibinfo {author}
  {\bibfnamefont {A.}~\bibnamefont {Freese}}, \bibinfo {author} {\bibfnamefont
  {A.~V.}\ \bibnamefont {Gorshkov}}, \bibinfo {author} {\bibfnamefont
  {A.}~\bibnamefont {Grassellino}}, \bibinfo {author} {\bibfnamefont
  {K.}~\bibnamefont {Hafidi}}, \bibinfo {author} {\bibfnamefont
  {Z.}~\bibnamefont {Jacob}}, \bibinfo {author} {\bibfnamefont
  {M.}~\bibnamefont {McGuigan}}, \bibinfo {author} {\bibfnamefont
  {Y.}~\bibnamefont {Meurice}}, \bibinfo {author} {\bibfnamefont {Z.-E.}\
  \bibnamefont {Meziani}}, \bibinfo {author} {\bibfnamefont {P.}~\bibnamefont
  {Mueller}}, \bibinfo {author} {\bibfnamefont {C.}~\bibnamefont {Muschik}},
  \bibinfo {author} {\bibfnamefont {J.}~\bibnamefont {Osborn}}, \bibinfo
  {author} {\bibfnamefont {M.}~\bibnamefont {Otten}}, \bibinfo {author}
  {\bibfnamefont {P.}~\bibnamefont {Petreczky}}, \bibinfo {author}
  {\bibfnamefont {T.}~\bibnamefont {Polakovic}}, \bibinfo {author}
  {\bibfnamefont {A.}~\bibnamefont {Poon}}, \bibinfo {author} {\bibfnamefont
  {R.}~\bibnamefont {Pooser}}, \bibinfo {author} {\bibfnamefont
  {A.}~\bibnamefont {Roggero}}, \bibinfo {author} {\bibfnamefont
  {M.}~\bibnamefont {Saffman}}, \bibinfo {author} {\bibfnamefont
  {B.}~\bibnamefont {VanDevender}}, \bibinfo {author} {\bibfnamefont
  {J.}~\bibnamefont {Zhang}},\ and\ \bibinfo {author} {\bibfnamefont
  {E.}~\bibnamefont {Zohar}},\ }\href@noop {} {\bibfield  {journal} {\bibinfo
  {journal} {arXiv:1903.05453}\ } (\bibinfo {year} {2019})},\ \Eprint
  {https://arxiv.org/abs/1903.05453} {arXiv:1903.05453} \BibitemShut {NoStop}%
\bibitem [{\citenamefont {Hosten}\ \emph {et~al.}(2016)\citenamefont {Hosten},
  \citenamefont {Engelsen}, \citenamefont {Krishnakumar},\ and\ \citenamefont
  {Kasevich}}]{Hosten2016}%
  \BibitemOpen
  \bibfield  {author} {\bibinfo {author} {\bibfnamefont {O.}~\bibnamefont
  {Hosten}}, \bibinfo {author} {\bibfnamefont {N.~J.}\ \bibnamefont
  {Engelsen}}, \bibinfo {author} {\bibfnamefont {R.}~\bibnamefont
  {Krishnakumar}},\ and\ \bibinfo {author} {\bibfnamefont {M.~A.}\ \bibnamefont
  {Kasevich}},\ }\href {https://doi.org/10.1038/nature16176} {\bibfield
  {journal} {\bibinfo  {journal} {Nature}\ }\textbf {\bibinfo {volume} {529}},\
  \bibinfo {pages} {505} (\bibinfo {year} {2016})}\BibitemShut {NoStop}%
\bibitem [{\citenamefont {Chang}\ \emph {et~al.}(2019)\citenamefont {Chang},
  \citenamefont {Hac\ifmmode \imath \else \i
  \fi{}\"omero\ifmmode~\breve{g}\else \u{g}\fi{}lu}, \citenamefont {Kim},
  \citenamefont {Lee}, \citenamefont {Park},\ and\ \citenamefont
  {Semertzidis}}]{SeungPyo19}%
  \BibitemOpen
  \bibfield  {author} {\bibinfo {author} {\bibfnamefont {S.~P.}\ \bibnamefont
  {Chang}}, \bibinfo {author} {\bibfnamefont {S.~m.~c.}\ \bibnamefont
  {Hac\ifmmode \imath \else \i \fi{}\"omero\ifmmode~\breve{g}\else
  \u{g}\fi{}lu}}, \bibinfo {author} {\bibfnamefont {O.}~\bibnamefont {Kim}},
  \bibinfo {author} {\bibfnamefont {S.}~\bibnamefont {Lee}}, \bibinfo {author}
  {\bibfnamefont {S.}~\bibnamefont {Park}},\ and\ \bibinfo {author}
  {\bibfnamefont {Y.~K.}\ \bibnamefont {Semertzidis}},\ }\href
  {https://doi.org/10.1103/PhysRevD.99.083002} {\bibfield  {journal} {\bibinfo
  {journal} {Phys. Rev. D}\ }\textbf {\bibinfo {volume} {99}},\ \bibinfo
  {pages} {083002} (\bibinfo {year} {2019})}\BibitemShut {NoStop}%
\bibitem [{\citenamefont {Pretz}\ \emph {et~al.}(2020)\citenamefont {Pretz},
  \citenamefont {Chang}, \citenamefont {Hejny}, \citenamefont {Karanth},
  \citenamefont {Park}, \citenamefont {Semertzidis}, \citenamefont
  {Stephenson},\ and\ \citenamefont {Str{\"o}her}}]{pretz2020statistical}%
  \BibitemOpen
  \bibfield  {author} {\bibinfo {author} {\bibfnamefont {J.}~\bibnamefont
  {Pretz}}, \bibinfo {author} {\bibfnamefont {S.~P.}\ \bibnamefont {Chang}},
  \bibinfo {author} {\bibfnamefont {V.}~\bibnamefont {Hejny}}, \bibinfo
  {author} {\bibfnamefont {S.}~\bibnamefont {Karanth}}, \bibinfo {author}
  {\bibfnamefont {S.}~\bibnamefont {Park}}, \bibinfo {author} {\bibfnamefont
  {Y.}~\bibnamefont {Semertzidis}}, \bibinfo {author} {\bibfnamefont
  {E.}~\bibnamefont {Stephenson}},\ and\ \bibinfo {author} {\bibfnamefont
  {H.}~\bibnamefont {Str{\"o}her}},\ }\href
  {https://doi.org/10.1140/epjc/s10052-020-7664-9} {\bibfield  {journal}
  {\bibinfo  {journal} {The European Physical Journal C}\ }\textbf {\bibinfo
  {volume} {80}},\ \bibinfo {pages} {1} (\bibinfo {year} {2020})}\BibitemShut
  {NoStop}%
\bibitem [{\citenamefont {Kim}\ and\ \citenamefont
  {Semertzidis}(2021)}]{kim_new_2021}%
  \BibitemOpen
  \bibfield  {author} {\bibinfo {author} {\bibfnamefont {O.}~\bibnamefont
  {Kim}}\ and\ \bibinfo {author} {\bibfnamefont {Y.~K.}\ \bibnamefont
  {Semertzidis}},\ }\href {https://doi.org/10.1103/PhysRevD.104.096006}
  {\bibfield  {journal} {\bibinfo  {journal} {Phys. Rev. D}\ }\textbf {\bibinfo
  {volume} {104}},\ \bibinfo {pages} {096006} (\bibinfo {year}
  {2021})}\BibitemShut {NoStop}%
\bibitem [{\citenamefont {Suleiman}\ \emph {et~al.}(2021)\citenamefont
  {Suleiman}, \citenamefont {Morozov},\ and\ \citenamefont
  {Derbenev}}]{Suleiman:2021whz}%
  \BibitemOpen
  \bibfield  {author} {\bibinfo {author} {\bibfnamefont {R.}~\bibnamefont
  {Suleiman}}, \bibinfo {author} {\bibfnamefont {V.~S.}\ \bibnamefont
  {Morozov}},\ and\ \bibinfo {author} {\bibfnamefont {Y.~S.}\ \bibnamefont
  {Derbenev}},\ }\href {https://doi.org/10.48550/ARXIV.2105.11575} {\bibinfo
  {title} {On possibilities of high precision experiments in fundamental
  physics in storage rings of low energy polarized electron beams}} (\bibinfo
  {year} {2021})\BibitemShut {NoStop}%
\bibitem [{\citenamefont {Filatov}\ \emph {et~al.}(2020)\citenamefont
  {Filatov}, \citenamefont {Kondratenko}, \citenamefont {Kondratenko},
  \citenamefont {Derbenev},\ and\ \citenamefont
  {Morozov}}]{PhysRevLett.124.194801}%
  \BibitemOpen
  \bibfield  {author} {\bibinfo {author} {\bibfnamefont {Y.~N.}\ \bibnamefont
  {Filatov}}, \bibinfo {author} {\bibfnamefont {A.~M.}\ \bibnamefont
  {Kondratenko}}, \bibinfo {author} {\bibfnamefont {M.~A.}\ \bibnamefont
  {Kondratenko}}, \bibinfo {author} {\bibfnamefont {Y.~S.}\ \bibnamefont
  {Derbenev}},\ and\ \bibinfo {author} {\bibfnamefont {V.~S.}\ \bibnamefont
  {Morozov}},\ }\href {https://doi.org/10.1103/PhysRevLett.124.194801}
  {\bibfield  {journal} {\bibinfo  {journal} {Phys. Rev. Lett.}\ }\textbf
  {\bibinfo {volume} {124}},\ \bibinfo {pages} {194801} (\bibinfo {year}
  {2020})}\BibitemShut {NoStop}%
\bibitem [{\citenamefont {Semertzidis}\ \emph {et~al.}(1999)\citenamefont
  {Semertzidis} \emph {et~al.}}]{Semertzidis:1999kv}%
  \BibitemOpen
  \bibfield  {author} {\bibinfo {author} {\bibfnamefont {Y.~K.}\ \bibnamefont
  {Semertzidis}} \emph {et~al.},\ }in\ \href
  {https://doi.org/10.1142/9789812791849_0007} {\emph {\bibinfo {booktitle}
  {{KEK International Workshop on High Intensity Muon Sources (HIMUS 99)}}}}\
  (\bibinfo {year} {1999})\ pp.\ \bibinfo {pages} {81--96},\ \Eprint
  {https://arxiv.org/abs/hep-ph/0012087} {arXiv:hep-ph/0012087} \BibitemShut
  {NoStop}%
\bibitem [{\citenamefont {Silenko}\ \emph {et~al.}(2003)\citenamefont {Silenko}
  \emph {et~al.}}]{JPARCmedmLOI}%
  \BibitemOpen
  \bibfield  {author} {\bibinfo {author} {\bibfnamefont {A.}~\bibnamefont
  {Silenko}} \emph {et~al.},\ }\href
  {https://www.bnl.gov/edm/papers/jparc_loi_030109.pdf} {\bibinfo {title}
  {{J-PARC} {L}etter of {I}ntent: {S}earch for a {P}ermanent {M}uon {E}lectric
  {D}ipole {M}oment at the $10^{-24} \, e \cdot {\rm cm}$ {L}evel}} (\bibinfo
  {year} {2003})\BibitemShut {NoStop}%
\bibitem [{\citenamefont {Semertzidis}\ \emph {et~al.}(2004)\citenamefont
  {Semertzidis} \emph {et~al.}}]{EDM:2003olr}%
  \BibitemOpen
  \bibfield  {author} {\bibinfo {author} {\bibfnamefont {Y.~K.}\ \bibnamefont
  {Semertzidis}} \emph {et~al.} (\bibinfo {collaboration} {EDM}),\ }\href
  {https://doi.org/10.1063/1.1664226} {\bibfield  {journal} {\bibinfo
  {journal} {AIP Conf. Proc.}\ }\textbf {\bibinfo {volume} {698}},\ \bibinfo
  {pages} {200} (\bibinfo {year} {2004})},\ \Eprint
  {https://arxiv.org/abs/hep-ex/0308063} {arXiv:hep-ex/0308063} \BibitemShut
  {NoStop}%
\bibitem [{\citenamefont {Miller}\ \emph {et~al.}(2004)\citenamefont {Miller},
  \citenamefont {Carey}, \citenamefont {Logashenko}, \citenamefont {Lynch},
  \citenamefont {Roberts}, \citenamefont {Silenko}, \citenamefont {Bennett},
  \citenamefont {Lazarus}, \citenamefont {Leipuner}, \citenamefont {Marciano},
  \citenamefont {Meng}, \citenamefont {Morse}, \citenamefont {Prigl},
  \citenamefont {Semertzidis}, \citenamefont {Balakin}, \citenamefont {Bazhan},
  \citenamefont {Dunikov}, \citenamefont {Khazin}, \citenamefont {Khriplovich},
  \citenamefont {Sylvestrov}, \citenamefont {Orlov}, \citenamefont {Jungmann},
  \citenamefont {Debevec}, \citenamefont {Hertzog}, \citenamefont {Onderwater},
  \citenamefont {Özben}, \citenamefont {Stephenson}, \citenamefont {Auzinsh},
  \citenamefont {Cushman}, \citenamefont {McNabb}, \citenamefont
  {Shafer‐Ray}, \citenamefont {Yoshimura}, \citenamefont {Aoki},
  \citenamefont {Kuno}, \citenamefont {Sato}, \citenamefont {Iwasaki},\ and\
  \citenamefont {Farley}}]{Miller2004}%
  \BibitemOpen
  \bibfield  {author} {\bibinfo {author} {\bibfnamefont {J.~P.}\ \bibnamefont
  {Miller}}, \bibinfo {author} {\bibfnamefont {R.~M.}\ \bibnamefont {Carey}},
  \bibinfo {author} {\bibfnamefont {V.}~\bibnamefont {Logashenko}}, \bibinfo
  {author} {\bibfnamefont {K.~R.}\ \bibnamefont {Lynch}}, \bibinfo {author}
  {\bibfnamefont {B.~L.}\ \bibnamefont {Roberts}}, \bibinfo {author}
  {\bibfnamefont {A.}~\bibnamefont {Silenko}}, \bibinfo {author} {\bibfnamefont
  {G.}~\bibnamefont {Bennett}}, \bibinfo {author} {\bibfnamefont {D.~M.}\
  \bibnamefont {Lazarus}}, \bibinfo {author} {\bibfnamefont {L.~B.}\
  \bibnamefont {Leipuner}}, \bibinfo {author} {\bibfnamefont {W.}~\bibnamefont
  {Marciano}}, \bibinfo {author} {\bibfnamefont {W.}~\bibnamefont {Meng}},
  \bibinfo {author} {\bibfnamefont {W.~M.}\ \bibnamefont {Morse}}, \bibinfo
  {author} {\bibfnamefont {R.}~\bibnamefont {Prigl}}, \bibinfo {author}
  {\bibfnamefont {Y.~K.}\ \bibnamefont {Semertzidis}}, \bibinfo {author}
  {\bibfnamefont {V.}~\bibnamefont {Balakin}}, \bibinfo {author} {\bibfnamefont
  {A.}~\bibnamefont {Bazhan}}, \bibinfo {author} {\bibfnamefont
  {A.}~\bibnamefont {Dunikov}}, \bibinfo {author} {\bibfnamefont
  {B.}~\bibnamefont {Khazin}}, \bibinfo {author} {\bibfnamefont {I.~B.}\
  \bibnamefont {Khriplovich}}, \bibinfo {author} {\bibfnamefont
  {G.}~\bibnamefont {Sylvestrov}}, \bibinfo {author} {\bibfnamefont
  {Y.}~\bibnamefont {Orlov}}, \bibinfo {author} {\bibfnamefont
  {K.}~\bibnamefont {Jungmann}}, \bibinfo {author} {\bibfnamefont {P.~T.}\
  \bibnamefont {Debevec}}, \bibinfo {author} {\bibfnamefont {D.~W.}\
  \bibnamefont {Hertzog}}, \bibinfo {author} {\bibfnamefont {C.~J.~G.}\
  \bibnamefont {Onderwater}}, \bibinfo {author} {\bibfnamefont {C.~S.}\
  \bibnamefont {Özben}}, \bibinfo {author} {\bibfnamefont {E.}~\bibnamefont
  {Stephenson}}, \bibinfo {author} {\bibfnamefont {M.}~\bibnamefont {Auzinsh}},
  \bibinfo {author} {\bibfnamefont {P.}~\bibnamefont {Cushman}}, \bibinfo
  {author} {\bibfnamefont {R.}~\bibnamefont {McNabb}}, \bibinfo {author}
  {\bibfnamefont {N.}~\bibnamefont {Shafer‐Ray}}, \bibinfo {author}
  {\bibfnamefont {K.}~\bibnamefont {Yoshimura}}, \bibinfo {author}
  {\bibfnamefont {A.}~\bibnamefont {Aoki}}, \bibinfo {author} {\bibfnamefont
  {Y.}~\bibnamefont {Kuno}}, \bibinfo {author} {\bibfnamefont {A.}~\bibnamefont
  {Sato}}, \bibinfo {author} {\bibfnamefont {M.}~\bibnamefont {Iwasaki}},\ and\
  \bibinfo {author} {\bibfnamefont {F.~J.~M.}\ \bibnamefont {Farley}},\ }in\
  \href {https://doi.org/10.1063/1.1664225} {\emph {\bibinfo {booktitle} {{AIP}
  Conference Proceedings}}}\ (\bibinfo  {publisher} {{AIP}},\ \bibinfo {year}
  {2004})\BibitemShut {NoStop}%
\bibitem [{\citenamefont {Adelmann}\ \emph {et~al.}(2010)\citenamefont
  {Adelmann}, \citenamefont {Kirch}, \citenamefont {Onderwater},\ and\
  \citenamefont {Schietinger}}]{Adelmann:2010zz}%
  \BibitemOpen
  \bibfield  {author} {\bibinfo {author} {\bibfnamefont {A.}~\bibnamefont
  {Adelmann}}, \bibinfo {author} {\bibfnamefont {K.}~\bibnamefont {Kirch}},
  \bibinfo {author} {\bibfnamefont {C.~J.~G.}\ \bibnamefont {Onderwater}},\
  and\ \bibinfo {author} {\bibfnamefont {T.}~\bibnamefont {Schietinger}},\
  }\href {https://doi.org/10.1088/0954-3899/37/8/085001} {\bibfield  {journal}
  {\bibinfo  {journal} {J. Phys. G}\ }\textbf {\bibinfo {volume} {37}},\
  \bibinfo {pages} {085001} (\bibinfo {year} {2010})}\BibitemShut {NoStop}%
\bibitem [{\citenamefont {Adelmann}\ \emph {et~al.}(2021)\citenamefont
  {Adelmann}, \citenamefont {Backhaus}, \citenamefont {Barajas}, \citenamefont
  {Berger}, \citenamefont {Bowcock}, \citenamefont {Calzolaio}, \citenamefont
  {Cavoto}, \citenamefont {Chislett}, \citenamefont {Crivellin}, \citenamefont
  {Daum}, \citenamefont {Fertl}, \citenamefont {Giovannozzi}, \citenamefont
  {Hesketh}, \citenamefont {Hildebrandt}, \citenamefont {Keshelashvili},
  \citenamefont {Keshavarzi}, \citenamefont {Khaw}, \citenamefont {Kirch},
  \citenamefont {Kozlinskiy}, \citenamefont {Knecht}, \citenamefont
  {Lancaster}, \citenamefont {M\"{a}rkisch}, \citenamefont {Aeschbacher},
  \citenamefont {Méot}, \citenamefont {Nass}, \citenamefont {Papa},
  \citenamefont {Pretz}, \citenamefont {Price}, \citenamefont {Rathmann},
  \citenamefont {Renga}, \citenamefont {Sakurai}, \citenamefont
  {Schmidt-Wellenburg}, \citenamefont {Sch\"{o}ning}, \citenamefont {Schott},
  \citenamefont {Voena}, \citenamefont {Vossebeld}, \citenamefont {Wauters},\
  and\ \citenamefont {Winter}}]{Adelmann:2021udj}%
  \BibitemOpen
  \bibfield  {author} {\bibinfo {author} {\bibfnamefont {A.}~\bibnamefont
  {Adelmann}}, \bibinfo {author} {\bibfnamefont {M.}~\bibnamefont {Backhaus}},
  \bibinfo {author} {\bibfnamefont {C.~C.}\ \bibnamefont {Barajas}}, \bibinfo
  {author} {\bibfnamefont {N.}~\bibnamefont {Berger}}, \bibinfo {author}
  {\bibfnamefont {T.}~\bibnamefont {Bowcock}}, \bibinfo {author} {\bibfnamefont
  {C.}~\bibnamefont {Calzolaio}}, \bibinfo {author} {\bibfnamefont
  {G.}~\bibnamefont {Cavoto}}, \bibinfo {author} {\bibfnamefont
  {R.}~\bibnamefont {Chislett}}, \bibinfo {author} {\bibfnamefont
  {A.}~\bibnamefont {Crivellin}}, \bibinfo {author} {\bibfnamefont
  {M.}~\bibnamefont {Daum}}, \bibinfo {author} {\bibfnamefont {M.}~\bibnamefont
  {Fertl}}, \bibinfo {author} {\bibfnamefont {M.}~\bibnamefont {Giovannozzi}},
  \bibinfo {author} {\bibfnamefont {G.}~\bibnamefont {Hesketh}}, \bibinfo
  {author} {\bibfnamefont {M.}~\bibnamefont {Hildebrandt}}, \bibinfo {author}
  {\bibfnamefont {I.}~\bibnamefont {Keshelashvili}}, \bibinfo {author}
  {\bibfnamefont {A.}~\bibnamefont {Keshavarzi}}, \bibinfo {author}
  {\bibfnamefont {K.~S.}\ \bibnamefont {Khaw}}, \bibinfo {author}
  {\bibfnamefont {K.}~\bibnamefont {Kirch}}, \bibinfo {author} {\bibfnamefont
  {A.}~\bibnamefont {Kozlinskiy}}, \bibinfo {author} {\bibfnamefont
  {A.}~\bibnamefont {Knecht}}, \bibinfo {author} {\bibfnamefont
  {M.}~\bibnamefont {Lancaster}}, \bibinfo {author} {\bibfnamefont
  {B.}~\bibnamefont {M\"{a}rkisch}}, \bibinfo {author} {\bibfnamefont {F.~M.}\
  \bibnamefont {Aeschbacher}}, \bibinfo {author} {\bibfnamefont
  {F.}~\bibnamefont {Méot}}, \bibinfo {author} {\bibfnamefont
  {A.}~\bibnamefont {Nass}}, \bibinfo {author} {\bibfnamefont {A.}~\bibnamefont
  {Papa}}, \bibinfo {author} {\bibfnamefont {J.}~\bibnamefont {Pretz}},
  \bibinfo {author} {\bibfnamefont {J.}~\bibnamefont {Price}}, \bibinfo
  {author} {\bibfnamefont {F.}~\bibnamefont {Rathmann}}, \bibinfo {author}
  {\bibfnamefont {F.}~\bibnamefont {Renga}}, \bibinfo {author} {\bibfnamefont
  {M.}~\bibnamefont {Sakurai}}, \bibinfo {author} {\bibfnamefont
  {P.}~\bibnamefont {Schmidt-Wellenburg}}, \bibinfo {author} {\bibfnamefont
  {A.}~\bibnamefont {Sch\"{o}ning}}, \bibinfo {author} {\bibfnamefont
  {M.}~\bibnamefont {Schott}}, \bibinfo {author} {\bibfnamefont
  {C.}~\bibnamefont {Voena}}, \bibinfo {author} {\bibfnamefont
  {J.}~\bibnamefont {Vossebeld}}, \bibinfo {author} {\bibfnamefont
  {F.}~\bibnamefont {Wauters}},\ and\ \bibinfo {author} {\bibfnamefont
  {P.}~\bibnamefont {Winter}},\ }\href
  {https://doi.org/10.48550/ARXIV.2102.08838} {\bibinfo {title} {Search for a
  muon edm using the frozen-spin technique}} (\bibinfo {year}
  {2021})\BibitemShut {NoStop}%
\bibitem [{\citenamefont {Crivellin}\ \emph {et~al.}(2018)\citenamefont
  {Crivellin}, \citenamefont {Hoferichter},\ and\ \citenamefont
  {Schmidt-Wellenburg}}]{Crivellin:2018qmi}%
  \BibitemOpen
  \bibfield  {author} {\bibinfo {author} {\bibfnamefont {A.}~\bibnamefont
  {Crivellin}}, \bibinfo {author} {\bibfnamefont {M.}~\bibnamefont
  {Hoferichter}},\ and\ \bibinfo {author} {\bibfnamefont {P.}~\bibnamefont
  {Schmidt-Wellenburg}},\ }\href {https://doi.org/10.1103/PhysRevD.98.113002}
  {\bibfield  {journal} {\bibinfo  {journal} {Phys. Rev. D}\ }\textbf {\bibinfo
  {volume} {98}},\ \bibinfo {pages} {113002} (\bibinfo {year} {2018})},\
  \Eprint {https://arxiv.org/abs/1807.11484} {arXiv:1807.11484 [hep-ph]}
  \BibitemShut {NoStop}%
\bibitem [{\citenamefont {Aiba}\ \emph {et~al.}(2021)\citenamefont {Aiba},
  \citenamefont {Amato}, \citenamefont {Antognini}, \citenamefont {Ban},
  \citenamefont {Berger}, \citenamefont {Caminada}, \citenamefont {Chislett},
  \citenamefont {Crivelli}, \citenamefont {Crivellin}, \citenamefont {Maso},
  \citenamefont {Davidson}, \citenamefont {Hoferichter}, \citenamefont {Iwai},
  \citenamefont {Iwamoto}, \citenamefont {Kirch}, \citenamefont {Knecht},
  \citenamefont {Langenegger}, \citenamefont {Lombardi}, \citenamefont
  {Luetkens}, \citenamefont {Aeschbacher}, \citenamefont {Mori}, \citenamefont
  {Nuber}, \citenamefont {Ootani}, \citenamefont {Papa}, \citenamefont
  {Prokscha}, \citenamefont {Renga}, \citenamefont {Ritt}, \citenamefont
  {Sakurai}, \citenamefont {Salman}, \citenamefont {Schmidt-Wellenburg},
  \citenamefont {Sch\"{o}ning}, \citenamefont {Signer}, \citenamefont {Soter},
  \citenamefont {Stingelin}, \citenamefont {Uchiyama},\ and\ \citenamefont
  {Wauters}}]{Aiba:2021bxe}%
  \BibitemOpen
  \bibfield  {author} {\bibinfo {author} {\bibfnamefont {M.}~\bibnamefont
  {Aiba}}, \bibinfo {author} {\bibfnamefont {A.}~\bibnamefont {Amato}},
  \bibinfo {author} {\bibfnamefont {A.}~\bibnamefont {Antognini}}, \bibinfo
  {author} {\bibfnamefont {S.}~\bibnamefont {Ban}}, \bibinfo {author}
  {\bibfnamefont {N.}~\bibnamefont {Berger}}, \bibinfo {author} {\bibfnamefont
  {L.}~\bibnamefont {Caminada}}, \bibinfo {author} {\bibfnamefont
  {R.}~\bibnamefont {Chislett}}, \bibinfo {author} {\bibfnamefont
  {P.}~\bibnamefont {Crivelli}}, \bibinfo {author} {\bibfnamefont
  {A.}~\bibnamefont {Crivellin}}, \bibinfo {author} {\bibfnamefont {G.~D.}\
  \bibnamefont {Maso}}, \bibinfo {author} {\bibfnamefont {S.}~\bibnamefont
  {Davidson}}, \bibinfo {author} {\bibfnamefont {M.}~\bibnamefont
  {Hoferichter}}, \bibinfo {author} {\bibfnamefont {R.}~\bibnamefont {Iwai}},
  \bibinfo {author} {\bibfnamefont {T.}~\bibnamefont {Iwamoto}}, \bibinfo
  {author} {\bibfnamefont {K.}~\bibnamefont {Kirch}}, \bibinfo {author}
  {\bibfnamefont {A.}~\bibnamefont {Knecht}}, \bibinfo {author} {\bibfnamefont
  {U.}~\bibnamefont {Langenegger}}, \bibinfo {author} {\bibfnamefont {A.~M.}\
  \bibnamefont {Lombardi}}, \bibinfo {author} {\bibfnamefont {H.}~\bibnamefont
  {Luetkens}}, \bibinfo {author} {\bibfnamefont {F.~M.}\ \bibnamefont
  {Aeschbacher}}, \bibinfo {author} {\bibfnamefont {T.}~\bibnamefont {Mori}},
  \bibinfo {author} {\bibfnamefont {J.}~\bibnamefont {Nuber}}, \bibinfo
  {author} {\bibfnamefont {W.}~\bibnamefont {Ootani}}, \bibinfo {author}
  {\bibfnamefont {A.}~\bibnamefont {Papa}}, \bibinfo {author} {\bibfnamefont
  {T.}~\bibnamefont {Prokscha}}, \bibinfo {author} {\bibfnamefont
  {F.}~\bibnamefont {Renga}}, \bibinfo {author} {\bibfnamefont
  {S.}~\bibnamefont {Ritt}}, \bibinfo {author} {\bibfnamefont {M.}~\bibnamefont
  {Sakurai}}, \bibinfo {author} {\bibfnamefont {Z.}~\bibnamefont {Salman}},
  \bibinfo {author} {\bibfnamefont {P.}~\bibnamefont {Schmidt-Wellenburg}},
  \bibinfo {author} {\bibfnamefont {A.}~\bibnamefont {Sch\"{o}ning}}, \bibinfo
  {author} {\bibfnamefont {A.}~\bibnamefont {Signer}}, \bibinfo {author}
  {\bibfnamefont {A.}~\bibnamefont {Soter}}, \bibinfo {author} {\bibfnamefont
  {L.}~\bibnamefont {Stingelin}}, \bibinfo {author} {\bibfnamefont
  {Y.}~\bibnamefont {Uchiyama}},\ and\ \bibinfo {author} {\bibfnamefont
  {F.}~\bibnamefont {Wauters}},\ }\href
  {https://doi.org/10.48550/ARXIV.2111.05788} {\bibinfo {title} {Science case
  for the new high-intensity muon beams himb at psi}} (\bibinfo {year}
  {2021})\BibitemShut {NoStop}%
\bibitem [{\citenamefont {Marciano}(2020)}]{edmtheory}%
  \BibitemOpen
  \bibfield  {author} {\bibinfo {author} {\bibfnamefont {W.}~\bibnamefont
  {Marciano}},\ }\href@noop {} {\bibinfo {title} {Overview {EDM} theory}}
  (\bibinfo {year} {2020}),\ \bibinfo {note}
  {\url{https://indico.fnal.gov/event/44782/timetable/?view=nicecompact}}\BibitemShut
  {NoStop}%
\bibitem [{\citenamefont {Graham}\ \emph {et~al.}(2021)\citenamefont {Graham},
  \citenamefont {Hac{\i}{\"o}mero{\u{g}}lu}, \citenamefont {Kaplan},
  \citenamefont {Omarov}, \citenamefont {Rajendran},\ and\ \citenamefont
  {Semertzidis}}]{graham_paper}%
  \BibitemOpen
  \bibfield  {author} {\bibinfo {author} {\bibfnamefont {P.~W.}\ \bibnamefont
  {Graham}}, \bibinfo {author} {\bibfnamefont {S.}~\bibnamefont
  {Hac{\i}{\"o}mero{\u{g}}lu}}, \bibinfo {author} {\bibfnamefont {D.~E.}\
  \bibnamefont {Kaplan}}, \bibinfo {author} {\bibfnamefont {Z.}~\bibnamefont
  {Omarov}}, \bibinfo {author} {\bibfnamefont {S.}~\bibnamefont {Rajendran}},\
  and\ \bibinfo {author} {\bibfnamefont {Y.~K.}\ \bibnamefont {Semertzidis}},\
  }\href {https://doi.org/10.1103/PhysRevD.103.055010} {\bibfield  {journal}
  {\bibinfo  {journal} {Phys. Rev. D}\ }\textbf {\bibinfo {volume} {103}},\
  \bibinfo {pages} {055010} (\bibinfo {year} {2021})}\BibitemShut {NoStop}%
\bibitem [{\citenamefont {Hutzler}(2020{\natexlab{b}})}]{edmtheory2}%
  \BibitemOpen
  \bibfield  {author} {\bibinfo {author} {\bibfnamefont {N.}~\bibnamefont
  {Hutzler}},\ }\href@noop {} {\bibinfo {title} {Developing new directions in
  fundamental physics 2020}} (\bibinfo {year} {2020}{\natexlab{b}}),\ \bibinfo
  {note}
  {\url{https://meetings.triumf.ca/event/89/contributions/2707}}\BibitemShut
  {NoStop}%
\bibitem [{\citenamefont {Combley}\ \emph {et~al.}(1981)\citenamefont
  {Combley}, \citenamefont {Farley},\ and\ \citenamefont
  {Picasso}}]{cern_report}%
  \BibitemOpen
  \bibfield  {author} {\bibinfo {author} {\bibfnamefont {F.}~\bibnamefont
  {Combley}}, \bibinfo {author} {\bibfnamefont {F.}~\bibnamefont {Farley}},\
  and\ \bibinfo {author} {\bibfnamefont {E.}~\bibnamefont {Picasso}},\ }\href
  {https://doi.org/10.1016/0370-1573(81)90028-4} {\bibfield  {journal}
  {\bibinfo  {journal} {Physics Reports}\ }\textbf {\bibinfo {volume} {68}},\
  \bibinfo {pages} {93} (\bibinfo {year} {1981})}\BibitemShut {NoStop}%
\bibitem [{\citenamefont {Bailey}\ \emph {et~al.}(1979)\citenamefont {Bailey},
  \citenamefont {Borer}, \citenamefont {Combley}, \citenamefont {Drumm},
  \citenamefont {Eck}, \citenamefont {Farley}, \citenamefont {Field},
  \citenamefont {Flegel}, \citenamefont {Hattersley}, \citenamefont {Krienen}
  \emph {et~al.}}]{cern3}%
  \BibitemOpen
  \bibfield  {author} {\bibinfo {author} {\bibfnamefont {J.}~\bibnamefont
  {Bailey}}, \bibinfo {author} {\bibfnamefont {K.}~\bibnamefont {Borer}},
  \bibinfo {author} {\bibfnamefont {F.}~\bibnamefont {Combley}}, \bibinfo
  {author} {\bibfnamefont {H.}~\bibnamefont {Drumm}}, \bibinfo {author}
  {\bibfnamefont {C.}~\bibnamefont {Eck}}, \bibinfo {author} {\bibfnamefont
  {F.}~\bibnamefont {Farley}}, \bibinfo {author} {\bibfnamefont
  {J.}~\bibnamefont {Field}}, \bibinfo {author} {\bibfnamefont
  {W.}~\bibnamefont {Flegel}}, \bibinfo {author} {\bibfnamefont
  {P.}~\bibnamefont {Hattersley}}, \bibinfo {author} {\bibfnamefont
  {F.}~\bibnamefont {Krienen}}, \emph {et~al.},\ }\href
  {https://doi.org/10.1016/0550-3213(79)90292-X} {\bibfield  {journal}
  {\bibinfo  {journal} {Nuclear Physics B}\ }\textbf {\bibinfo {volume}
  {150}},\ \bibinfo {pages} {1} (\bibinfo {year} {1979})}\BibitemShut {NoStop}%
\bibitem [{\citenamefont {Bennett}\ \emph {et~al.}(4 07)\citenamefont
  {Bennett}, \citenamefont {Bousquet}, \citenamefont {Brown}, \citenamefont
  {Bunce}, \citenamefont {Carey}, \citenamefont {Cushman}, \citenamefont
  {Danby}, \citenamefont {Debevec}, \citenamefont {Deile}, \citenamefont
  {Deng}, \citenamefont {Deninger}, \citenamefont {Dhawan}, \citenamefont
  {Druzhinin}, \citenamefont {Duong}, \citenamefont {Efstathiadis},
  \citenamefont {Farley}, \citenamefont {Fedotovich}, \citenamefont {Giron},
  \citenamefont {Gray}, \citenamefont {Grigoriev}, \citenamefont
  {Grosse-Perdekamp}, \citenamefont {Grossmann}, \citenamefont {Hare},
  \citenamefont {Hertzog}, \citenamefont {Huang}, \citenamefont {Hughes},
  \citenamefont {Iwasaki}, \citenamefont {Jungmann}, \citenamefont {Kawall},
  \citenamefont {Kawamura}, \citenamefont {Khazin}, \citenamefont {Kindem},
  \citenamefont {Krienen}, \citenamefont {Kronkvist}, \citenamefont {Lam},
  \citenamefont {Larsen}, \citenamefont {Lee}, \citenamefont {Logashenko},
  \citenamefont {{McNabb}}, \citenamefont {Meng}, \citenamefont {Mi},
  \citenamefont {Miller}, \citenamefont {Mizumachi}, \citenamefont {Morse},
  \citenamefont {Nikas}, \citenamefont {Onderwater}, \citenamefont {Orlov},
  \citenamefont {Özben}, \citenamefont {Paley}, \citenamefont {Peng},
  \citenamefont {Polly}, \citenamefont {Pretz}, \citenamefont {Prigl},
  \citenamefont {zu~Putlitz}, \citenamefont {Qian}, \citenamefont {Redin},
  \citenamefont {Rind}, \citenamefont {Roberts}, \citenamefont {Ryskulov},
  \citenamefont {Sedykh}, \citenamefont {Semertzidis}, \citenamefont {Shagin},
  \citenamefont {Shatunov}, \citenamefont {Sichtermann}, \citenamefont
  {Solodov}, \citenamefont {Sossong}, \citenamefont {Steinmetz}, \citenamefont
  {Sulak}, \citenamefont {Timmermans}, \citenamefont {Trofimov}, \citenamefont
  {Urner}, \citenamefont {von Walter}, \citenamefont {Warburton}, \citenamefont
  {Winn}, \citenamefont {Yamamoto},\ and\ \citenamefont
  {Zimmerman}}]{bennett_final_2006}%
  \BibitemOpen
  \bibfield  {author} {\bibinfo {author} {\bibfnamefont {G.~W.}\ \bibnamefont
  {Bennett}}, \bibinfo {author} {\bibfnamefont {B.}~\bibnamefont {Bousquet}},
  \bibinfo {author} {\bibfnamefont {H.~N.}\ \bibnamefont {Brown}}, \bibinfo
  {author} {\bibfnamefont {G.}~\bibnamefont {Bunce}}, \bibinfo {author}
  {\bibfnamefont {R.~M.}\ \bibnamefont {Carey}}, \bibinfo {author}
  {\bibfnamefont {P.}~\bibnamefont {Cushman}}, \bibinfo {author} {\bibfnamefont
  {G.~T.}\ \bibnamefont {Danby}}, \bibinfo {author} {\bibfnamefont {P.~T.}\
  \bibnamefont {Debevec}}, \bibinfo {author} {\bibfnamefont {M.}~\bibnamefont
  {Deile}}, \bibinfo {author} {\bibfnamefont {H.}~\bibnamefont {Deng}},
  \bibinfo {author} {\bibfnamefont {W.}~\bibnamefont {Deninger}}, \bibinfo
  {author} {\bibfnamefont {S.~K.}\ \bibnamefont {Dhawan}}, \bibinfo {author}
  {\bibfnamefont {V.~P.}\ \bibnamefont {Druzhinin}}, \bibinfo {author}
  {\bibfnamefont {L.}~\bibnamefont {Duong}}, \bibinfo {author} {\bibfnamefont
  {E.}~\bibnamefont {Efstathiadis}}, \bibinfo {author} {\bibfnamefont
  {F.~J.~M.}\ \bibnamefont {Farley}}, \bibinfo {author} {\bibfnamefont {G.~V.}\
  \bibnamefont {Fedotovich}}, \bibinfo {author} {\bibfnamefont
  {S.}~\bibnamefont {Giron}}, \bibinfo {author} {\bibfnamefont {F.~E.}\
  \bibnamefont {Gray}}, \bibinfo {author} {\bibfnamefont {D.}~\bibnamefont
  {Grigoriev}}, \bibinfo {author} {\bibfnamefont {M.}~\bibnamefont
  {Grosse-Perdekamp}}, \bibinfo {author} {\bibfnamefont {A.}~\bibnamefont
  {Grossmann}}, \bibinfo {author} {\bibfnamefont {M.~F.}\ \bibnamefont {Hare}},
  \bibinfo {author} {\bibfnamefont {D.~W.}\ \bibnamefont {Hertzog}}, \bibinfo
  {author} {\bibfnamefont {X.}~\bibnamefont {Huang}}, \bibinfo {author}
  {\bibfnamefont {V.~W.}\ \bibnamefont {Hughes}}, \bibinfo {author}
  {\bibfnamefont {M.}~\bibnamefont {Iwasaki}}, \bibinfo {author} {\bibfnamefont
  {K.}~\bibnamefont {Jungmann}}, \bibinfo {author} {\bibfnamefont
  {D.}~\bibnamefont {Kawall}}, \bibinfo {author} {\bibfnamefont
  {M.}~\bibnamefont {Kawamura}}, \bibinfo {author} {\bibfnamefont {B.~I.}\
  \bibnamefont {Khazin}}, \bibinfo {author} {\bibfnamefont {J.}~\bibnamefont
  {Kindem}}, \bibinfo {author} {\bibfnamefont {F.}~\bibnamefont {Krienen}},
  \bibinfo {author} {\bibfnamefont {I.}~\bibnamefont {Kronkvist}}, \bibinfo
  {author} {\bibfnamefont {A.}~\bibnamefont {Lam}}, \bibinfo {author}
  {\bibfnamefont {R.}~\bibnamefont {Larsen}}, \bibinfo {author} {\bibfnamefont
  {Y.~Y.}\ \bibnamefont {Lee}}, \bibinfo {author} {\bibfnamefont
  {I.}~\bibnamefont {Logashenko}}, \bibinfo {author} {\bibfnamefont
  {R.}~\bibnamefont {{McNabb}}}, \bibinfo {author} {\bibfnamefont
  {W.}~\bibnamefont {Meng}}, \bibinfo {author} {\bibfnamefont {J.}~\bibnamefont
  {Mi}}, \bibinfo {author} {\bibfnamefont {J.~P.}\ \bibnamefont {Miller}},
  \bibinfo {author} {\bibfnamefont {Y.}~\bibnamefont {Mizumachi}}, \bibinfo
  {author} {\bibfnamefont {W.~M.}\ \bibnamefont {Morse}}, \bibinfo {author}
  {\bibfnamefont {D.}~\bibnamefont {Nikas}}, \bibinfo {author} {\bibfnamefont
  {C.~J.~G.}\ \bibnamefont {Onderwater}}, \bibinfo {author} {\bibfnamefont
  {Y.}~\bibnamefont {Orlov}}, \bibinfo {author} {\bibfnamefont {C.~S.}\
  \bibnamefont {Özben}}, \bibinfo {author} {\bibfnamefont {J.~M.}\
  \bibnamefont {Paley}}, \bibinfo {author} {\bibfnamefont {Q.}~\bibnamefont
  {Peng}}, \bibinfo {author} {\bibfnamefont {C.~C.}\ \bibnamefont {Polly}},
  \bibinfo {author} {\bibfnamefont {J.}~\bibnamefont {Pretz}}, \bibinfo
  {author} {\bibfnamefont {R.}~\bibnamefont {Prigl}}, \bibinfo {author}
  {\bibfnamefont {G.}~\bibnamefont {zu~Putlitz}}, \bibinfo {author}
  {\bibfnamefont {T.}~\bibnamefont {Qian}}, \bibinfo {author} {\bibfnamefont
  {S.~I.}\ \bibnamefont {Redin}}, \bibinfo {author} {\bibfnamefont
  {O.}~\bibnamefont {Rind}}, \bibinfo {author} {\bibfnamefont {B.~L.}\
  \bibnamefont {Roberts}}, \bibinfo {author} {\bibfnamefont {N.}~\bibnamefont
  {Ryskulov}}, \bibinfo {author} {\bibfnamefont {S.}~\bibnamefont {Sedykh}},
  \bibinfo {author} {\bibfnamefont {Y.~K.}\ \bibnamefont {Semertzidis}},
  \bibinfo {author} {\bibfnamefont {P.}~\bibnamefont {Shagin}}, \bibinfo
  {author} {\bibfnamefont {Y.~M.}\ \bibnamefont {Shatunov}}, \bibinfo {author}
  {\bibfnamefont {E.~P.}\ \bibnamefont {Sichtermann}}, \bibinfo {author}
  {\bibfnamefont {E.}~\bibnamefont {Solodov}}, \bibinfo {author} {\bibfnamefont
  {M.}~\bibnamefont {Sossong}}, \bibinfo {author} {\bibfnamefont
  {A.}~\bibnamefont {Steinmetz}}, \bibinfo {author} {\bibfnamefont {L.~R.}\
  \bibnamefont {Sulak}}, \bibinfo {author} {\bibfnamefont {C.}~\bibnamefont
  {Timmermans}}, \bibinfo {author} {\bibfnamefont {A.}~\bibnamefont
  {Trofimov}}, \bibinfo {author} {\bibfnamefont {D.}~\bibnamefont {Urner}},
  \bibinfo {author} {\bibfnamefont {P.}~\bibnamefont {von Walter}}, \bibinfo
  {author} {\bibfnamefont {D.}~\bibnamefont {Warburton}}, \bibinfo {author}
  {\bibfnamefont {D.}~\bibnamefont {Winn}}, \bibinfo {author} {\bibfnamefont
  {A.}~\bibnamefont {Yamamoto}},\ and\ \bibinfo {author} {\bibfnamefont
  {D.}~\bibnamefont {Zimmerman}},\ }\bibfield  {journal} {\bibinfo  {journal}
  {Physical Review D}\ }\textbf {\bibinfo {volume} {73}},\ \href
  {https://doi.org/10.1103/PhysRevD.73.072003} {10.1103/PhysRevD.73.072003}
  (\bibinfo {year} {2006-04-07})\BibitemShut {NoStop}%
\bibitem [{\citenamefont {Abi}\ \emph {et~al.}(2021)\citenamefont {Abi},
  \citenamefont {Albahri}, \citenamefont {Al-Kilani}, \citenamefont {Allspach},
  \citenamefont {Alonzi}, \citenamefont {Anastasi}, \citenamefont {Anisenkov},
  \citenamefont {Azfar}, \citenamefont {Badgley}, \citenamefont {Bae\ss{}ler},
  \citenamefont {Bailey}, \citenamefont {Baranov}, \citenamefont
  {Barlas-Yucel}, \citenamefont {Barrett}, \citenamefont {Barzi}, \citenamefont
  {Basti}, \citenamefont {Bedeschi}, \citenamefont {Behnke}, \citenamefont
  {Berz}, \citenamefont {Bhattacharya}, \citenamefont {Binney}, \citenamefont
  {Bjorkquist}, \citenamefont {Bloom}, \citenamefont {Bono}, \citenamefont
  {Bottalico}, \citenamefont {Bowcock}, \citenamefont {Boyden}, \citenamefont
  {Cantatore}, \citenamefont {Carey}, \citenamefont {Carroll}, \citenamefont
  {Casey}, \citenamefont {Cauz}, \citenamefont {Ceravolo}, \citenamefont
  {Chakraborty}, \citenamefont {Chang}, \citenamefont {Chapelain},
  \citenamefont {Chappa}, \citenamefont {Charity}, \citenamefont {Chislett},
  \citenamefont {Choi}, \citenamefont {Chu}, \citenamefont {Chupp},
  \citenamefont {Convery}, \citenamefont {Conway}, \citenamefont {Corradi},
  \citenamefont {Corrodi}, \citenamefont {Cotrozzi}, \citenamefont {Crnkovic},
  \citenamefont {Dabagov}, \citenamefont {De~Lurgio}, \citenamefont {Debevec},
  \citenamefont {Di~Falco}, \citenamefont {Di~Meo}, \citenamefont
  {Di~Sciascio}, \citenamefont {Di~Stefano}, \citenamefont {Drendel},
  \citenamefont {Driutti}, \citenamefont {Duginov}, \citenamefont {Eads},
  \citenamefont {Eggert}, \citenamefont {Epps}, \citenamefont {Esquivel},
  \citenamefont {Farooq}, \citenamefont {Fatemi}, \citenamefont {Ferrari},
  \citenamefont {Fertl}, \citenamefont {Fiedler}, \citenamefont {Fienberg},
  \citenamefont {Fioretti}, \citenamefont {Flay}, \citenamefont {Foster},
  \citenamefont {Friedsam}, \citenamefont {Frle\ifmmode~\check{z}\else
  \v{z}\fi{}}, \citenamefont {Froemming}, \citenamefont {Fry}, \citenamefont
  {Fu}, \citenamefont {Gabbanini}, \citenamefont {Galati}, \citenamefont
  {Ganguly}, \citenamefont {Garcia}, \citenamefont {Gastler}, \citenamefont
  {George}, \citenamefont {Gibbons}, \citenamefont {Gioiosa}, \citenamefont
  {Giovanetti}, \citenamefont {Girotti}, \citenamefont {Gohn}, \citenamefont
  {Gorringe}, \citenamefont {Grange}, \citenamefont {Grant}, \citenamefont
  {Gray}, \citenamefont {Haciomeroglu}, \citenamefont {Hahn}, \citenamefont
  {Halewood-Leagas}, \citenamefont {Hampai}, \citenamefont {Han}, \citenamefont
  {Hazen}, \citenamefont {Hempstead}, \citenamefont {Henry}, \citenamefont
  {Herrod}, \citenamefont {Hertzog}, \citenamefont {Hesketh}, \citenamefont
  {Hibbert}, \citenamefont {Hodge}, \citenamefont {Holzbauer}, \citenamefont
  {Hong}, \citenamefont {Hong}, \citenamefont {Iacovacci}, \citenamefont
  {Incagli}, \citenamefont {Johnstone}, \citenamefont {Johnstone},
  \citenamefont {Kammel}, \citenamefont {Kargiantoulakis}, \citenamefont
  {Karuza}, \citenamefont {Kaspar}, \citenamefont {Kawall}, \citenamefont
  {Kelton}, \citenamefont {Keshavarzi}, \citenamefont {Kessler}, \citenamefont
  {Khaw}, \citenamefont {Khechadoorian}, \citenamefont {Khomutov},
  \citenamefont {Kiburg}, \citenamefont {Kiburg}, \citenamefont {Kim},
  \citenamefont {Kim}, \citenamefont {Kim}, \citenamefont {King}, \citenamefont
  {Kinnaird}, \citenamefont {Korostelev}, \citenamefont {Kourbanis},
  \citenamefont {Kraegeloh}, \citenamefont {Krylov}, \citenamefont
  {Kuchibhotla}, \citenamefont {Kuchinskiy}, \citenamefont {Labe},
  \citenamefont {LaBounty}, \citenamefont {Lancaster}, \citenamefont {Lee},
  \citenamefont {Lee}, \citenamefont {Leo}, \citenamefont {Li}, \citenamefont
  {Li}, \citenamefont {Li}, \citenamefont {Logashenko}, \citenamefont
  {Lorente~Campos}, \citenamefont {Luc\`a}, \citenamefont {Lukicov},
  \citenamefont {Luo}, \citenamefont {Lusiani}, \citenamefont {Lyon},
  \citenamefont {MacCoy}, \citenamefont {Madrak}, \citenamefont {Makino},
  \citenamefont {Marignetti}, \citenamefont {Mastroianni}, \citenamefont
  {Maxfield}, \citenamefont {McEvoy}, \citenamefont {Merritt}, \citenamefont
  {Mikhailichenko}, \citenamefont {Miller}, \citenamefont {Miozzi},
  \citenamefont {Morgan}, \citenamefont {Morse}, \citenamefont {Mott},
  \citenamefont {Motuk}, \citenamefont {Nath}, \citenamefont {Newton},
  \citenamefont {Nguyen}, \citenamefont {Oberling}, \citenamefont {Osofsky},
  \citenamefont {Ostiguy}, \citenamefont {Park}, \citenamefont {Pauletta},
  \citenamefont {Piacentino}, \citenamefont {Pilato}, \citenamefont {Pitts},
  \citenamefont {Plaster}, \citenamefont {Po\ifmmode \check{c}\else
  \v{c}\fi{}ani\ifmmode~\acute{c}\else \'{c}\fi{}}, \citenamefont {Pohlman},
  \citenamefont {Polly}, \citenamefont {Popovic}, \citenamefont {Price},
  \citenamefont {Quinn}, \citenamefont {Raha}, \citenamefont {Ramachandran},
  \citenamefont {Ramberg}, \citenamefont {Rider}, \citenamefont {Ritchie},
  \citenamefont {Roberts}, \citenamefont {Rubin}, \citenamefont {Santi},
  \citenamefont {Sathyan}, \citenamefont {Schellman}, \citenamefont
  {Schlesier}, \citenamefont {Schreckenberger}, \citenamefont {Semertzidis},
  \citenamefont {Shatunov}, \citenamefont {Shemyakin}, \citenamefont {Shenk},
  \citenamefont {Sim}, \citenamefont {Smith}, \citenamefont {Smith},
  \citenamefont {Soha}, \citenamefont {Sorbara}, \citenamefont {St\"ockinger},
  \citenamefont {Stapleton}, \citenamefont {Still}, \citenamefont {Stoughton},
  \citenamefont {Stratakis}, \citenamefont {Strohman}, \citenamefont
  {Stuttard}, \citenamefont {Swanson}, \citenamefont {Sweetmore}, \citenamefont
  {Sweigart}, \citenamefont {Syphers}, \citenamefont {Tarazona}, \citenamefont
  {Teubner}, \citenamefont {Tewsley-Booth}, \citenamefont {Thomson},
  \citenamefont {Tishchenko}, \citenamefont {Tran}, \citenamefont {Turner},
  \citenamefont {Valetov}, \citenamefont {Vasilkova}, \citenamefont
  {Venanzoni}, \citenamefont {Volnykh}, \citenamefont {Walton}, \citenamefont
  {Warren}, \citenamefont {Weisskopf}, \citenamefont {Welty-Rieger},
  \citenamefont {Whitley}, \citenamefont {Winter}, \citenamefont {Wolski},
  \citenamefont {Wormald}, \citenamefont {Wu},\ and\ \citenamefont
  {Yoshikawa}}]{fnal1}%
  \BibitemOpen
  \bibfield  {author} {\bibinfo {author} {\bibfnamefont {B.}~\bibnamefont
  {Abi}}, \bibinfo {author} {\bibfnamefont {T.}~\bibnamefont {Albahri}},
  \bibinfo {author} {\bibfnamefont {S.}~\bibnamefont {Al-Kilani}}, \bibinfo
  {author} {\bibfnamefont {D.}~\bibnamefont {Allspach}}, \bibinfo {author}
  {\bibfnamefont {L.~P.}\ \bibnamefont {Alonzi}}, \bibinfo {author}
  {\bibfnamefont {A.}~\bibnamefont {Anastasi}}, \bibinfo {author}
  {\bibfnamefont {A.}~\bibnamefont {Anisenkov}}, \bibinfo {author}
  {\bibfnamefont {F.}~\bibnamefont {Azfar}}, \bibinfo {author} {\bibfnamefont
  {K.}~\bibnamefont {Badgley}}, \bibinfo {author} {\bibfnamefont
  {S.}~\bibnamefont {Bae\ss{}ler}}, \bibinfo {author} {\bibfnamefont
  {I.}~\bibnamefont {Bailey}}, \bibinfo {author} {\bibfnamefont {V.~A.}\
  \bibnamefont {Baranov}}, \bibinfo {author} {\bibfnamefont {E.}~\bibnamefont
  {Barlas-Yucel}}, \bibinfo {author} {\bibfnamefont {T.}~\bibnamefont
  {Barrett}}, \bibinfo {author} {\bibfnamefont {E.}~\bibnamefont {Barzi}},
  \bibinfo {author} {\bibfnamefont {A.}~\bibnamefont {Basti}}, \bibinfo
  {author} {\bibfnamefont {F.}~\bibnamefont {Bedeschi}}, \bibinfo {author}
  {\bibfnamefont {A.}~\bibnamefont {Behnke}}, \bibinfo {author} {\bibfnamefont
  {M.}~\bibnamefont {Berz}}, \bibinfo {author} {\bibfnamefont {M.}~\bibnamefont
  {Bhattacharya}}, \bibinfo {author} {\bibfnamefont {H.~P.}\ \bibnamefont
  {Binney}}, \bibinfo {author} {\bibfnamefont {R.}~\bibnamefont {Bjorkquist}},
  \bibinfo {author} {\bibfnamefont {P.}~\bibnamefont {Bloom}}, \bibinfo
  {author} {\bibfnamefont {J.}~\bibnamefont {Bono}}, \bibinfo {author}
  {\bibfnamefont {E.}~\bibnamefont {Bottalico}}, \bibinfo {author}
  {\bibfnamefont {T.}~\bibnamefont {Bowcock}}, \bibinfo {author} {\bibfnamefont
  {D.}~\bibnamefont {Boyden}}, \bibinfo {author} {\bibfnamefont
  {G.}~\bibnamefont {Cantatore}}, \bibinfo {author} {\bibfnamefont {R.~M.}\
  \bibnamefont {Carey}}, \bibinfo {author} {\bibfnamefont {J.}~\bibnamefont
  {Carroll}}, \bibinfo {author} {\bibfnamefont {B.~C.~K.}\ \bibnamefont
  {Casey}}, \bibinfo {author} {\bibfnamefont {D.}~\bibnamefont {Cauz}},
  \bibinfo {author} {\bibfnamefont {S.}~\bibnamefont {Ceravolo}}, \bibinfo
  {author} {\bibfnamefont {R.}~\bibnamefont {Chakraborty}}, \bibinfo {author}
  {\bibfnamefont {S.~P.}\ \bibnamefont {Chang}}, \bibinfo {author}
  {\bibfnamefont {A.}~\bibnamefont {Chapelain}}, \bibinfo {author}
  {\bibfnamefont {S.}~\bibnamefont {Chappa}}, \bibinfo {author} {\bibfnamefont
  {S.}~\bibnamefont {Charity}}, \bibinfo {author} {\bibfnamefont
  {R.}~\bibnamefont {Chislett}}, \bibinfo {author} {\bibfnamefont
  {J.}~\bibnamefont {Choi}}, \bibinfo {author} {\bibfnamefont {Z.}~\bibnamefont
  {Chu}}, \bibinfo {author} {\bibfnamefont {T.~E.}\ \bibnamefont {Chupp}},
  \bibinfo {author} {\bibfnamefont {M.~E.}\ \bibnamefont {Convery}}, \bibinfo
  {author} {\bibfnamefont {A.}~\bibnamefont {Conway}}, \bibinfo {author}
  {\bibfnamefont {G.}~\bibnamefont {Corradi}}, \bibinfo {author} {\bibfnamefont
  {S.}~\bibnamefont {Corrodi}}, \bibinfo {author} {\bibfnamefont
  {L.}~\bibnamefont {Cotrozzi}}, \bibinfo {author} {\bibfnamefont {J.~D.}\
  \bibnamefont {Crnkovic}}, \bibinfo {author} {\bibfnamefont {S.}~\bibnamefont
  {Dabagov}}, \bibinfo {author} {\bibfnamefont {P.~M.}\ \bibnamefont
  {De~Lurgio}}, \bibinfo {author} {\bibfnamefont {P.~T.}\ \bibnamefont
  {Debevec}}, \bibinfo {author} {\bibfnamefont {S.}~\bibnamefont {Di~Falco}},
  \bibinfo {author} {\bibfnamefont {P.}~\bibnamefont {Di~Meo}}, \bibinfo
  {author} {\bibfnamefont {G.}~\bibnamefont {Di~Sciascio}}, \bibinfo {author}
  {\bibfnamefont {R.}~\bibnamefont {Di~Stefano}}, \bibinfo {author}
  {\bibfnamefont {B.}~\bibnamefont {Drendel}}, \bibinfo {author} {\bibfnamefont
  {A.}~\bibnamefont {Driutti}}, \bibinfo {author} {\bibfnamefont {V.~N.}\
  \bibnamefont {Duginov}}, \bibinfo {author} {\bibfnamefont {M.}~\bibnamefont
  {Eads}}, \bibinfo {author} {\bibfnamefont {N.}~\bibnamefont {Eggert}},
  \bibinfo {author} {\bibfnamefont {A.}~\bibnamefont {Epps}}, \bibinfo {author}
  {\bibfnamefont {J.}~\bibnamefont {Esquivel}}, \bibinfo {author}
  {\bibfnamefont {M.}~\bibnamefont {Farooq}}, \bibinfo {author} {\bibfnamefont
  {R.}~\bibnamefont {Fatemi}}, \bibinfo {author} {\bibfnamefont
  {C.}~\bibnamefont {Ferrari}}, \bibinfo {author} {\bibfnamefont
  {M.}~\bibnamefont {Fertl}}, \bibinfo {author} {\bibfnamefont
  {A.}~\bibnamefont {Fiedler}}, \bibinfo {author} {\bibfnamefont {A.~T.}\
  \bibnamefont {Fienberg}}, \bibinfo {author} {\bibfnamefont {A.}~\bibnamefont
  {Fioretti}}, \bibinfo {author} {\bibfnamefont {D.}~\bibnamefont {Flay}},
  \bibinfo {author} {\bibfnamefont {S.~B.}\ \bibnamefont {Foster}}, \bibinfo
  {author} {\bibfnamefont {H.}~\bibnamefont {Friedsam}}, \bibinfo {author}
  {\bibfnamefont {E.}~\bibnamefont {Frle\ifmmode~\check{z}\else \v{z}\fi{}}},
  \bibinfo {author} {\bibfnamefont {N.~S.}\ \bibnamefont {Froemming}}, \bibinfo
  {author} {\bibfnamefont {J.}~\bibnamefont {Fry}}, \bibinfo {author}
  {\bibfnamefont {C.}~\bibnamefont {Fu}}, \bibinfo {author} {\bibfnamefont
  {C.}~\bibnamefont {Gabbanini}}, \bibinfo {author} {\bibfnamefont {M.~D.}\
  \bibnamefont {Galati}}, \bibinfo {author} {\bibfnamefont {S.}~\bibnamefont
  {Ganguly}}, \bibinfo {author} {\bibfnamefont {A.}~\bibnamefont {Garcia}},
  \bibinfo {author} {\bibfnamefont {D.~E.}\ \bibnamefont {Gastler}}, \bibinfo
  {author} {\bibfnamefont {J.}~\bibnamefont {George}}, \bibinfo {author}
  {\bibfnamefont {L.~K.}\ \bibnamefont {Gibbons}}, \bibinfo {author}
  {\bibfnamefont {A.}~\bibnamefont {Gioiosa}}, \bibinfo {author} {\bibfnamefont
  {K.~L.}\ \bibnamefont {Giovanetti}}, \bibinfo {author} {\bibfnamefont
  {P.}~\bibnamefont {Girotti}}, \bibinfo {author} {\bibfnamefont
  {W.}~\bibnamefont {Gohn}}, \bibinfo {author} {\bibfnamefont {T.}~\bibnamefont
  {Gorringe}}, \bibinfo {author} {\bibfnamefont {J.}~\bibnamefont {Grange}},
  \bibinfo {author} {\bibfnamefont {S.}~\bibnamefont {Grant}}, \bibinfo
  {author} {\bibfnamefont {F.}~\bibnamefont {Gray}}, \bibinfo {author}
  {\bibfnamefont {S.}~\bibnamefont {Haciomeroglu}}, \bibinfo {author}
  {\bibfnamefont {D.}~\bibnamefont {Hahn}}, \bibinfo {author} {\bibfnamefont
  {T.}~\bibnamefont {Halewood-Leagas}}, \bibinfo {author} {\bibfnamefont
  {D.}~\bibnamefont {Hampai}}, \bibinfo {author} {\bibfnamefont
  {F.}~\bibnamefont {Han}}, \bibinfo {author} {\bibfnamefont {E.}~\bibnamefont
  {Hazen}}, \bibinfo {author} {\bibfnamefont {J.}~\bibnamefont {Hempstead}},
  \bibinfo {author} {\bibfnamefont {S.}~\bibnamefont {Henry}}, \bibinfo
  {author} {\bibfnamefont {A.~T.}\ \bibnamefont {Herrod}}, \bibinfo {author}
  {\bibfnamefont {D.~W.}\ \bibnamefont {Hertzog}}, \bibinfo {author}
  {\bibfnamefont {G.}~\bibnamefont {Hesketh}}, \bibinfo {author} {\bibfnamefont
  {A.}~\bibnamefont {Hibbert}}, \bibinfo {author} {\bibfnamefont
  {Z.}~\bibnamefont {Hodge}}, \bibinfo {author} {\bibfnamefont {J.~L.}\
  \bibnamefont {Holzbauer}}, \bibinfo {author} {\bibfnamefont {K.~W.}\
  \bibnamefont {Hong}}, \bibinfo {author} {\bibfnamefont {R.}~\bibnamefont
  {Hong}}, \bibinfo {author} {\bibfnamefont {M.}~\bibnamefont {Iacovacci}},
  \bibinfo {author} {\bibfnamefont {M.}~\bibnamefont {Incagli}}, \bibinfo
  {author} {\bibfnamefont {C.}~\bibnamefont {Johnstone}}, \bibinfo {author}
  {\bibfnamefont {J.~A.}\ \bibnamefont {Johnstone}}, \bibinfo {author}
  {\bibfnamefont {P.}~\bibnamefont {Kammel}}, \bibinfo {author} {\bibfnamefont
  {M.}~\bibnamefont {Kargiantoulakis}}, \bibinfo {author} {\bibfnamefont
  {M.}~\bibnamefont {Karuza}}, \bibinfo {author} {\bibfnamefont
  {J.}~\bibnamefont {Kaspar}}, \bibinfo {author} {\bibfnamefont
  {D.}~\bibnamefont {Kawall}}, \bibinfo {author} {\bibfnamefont
  {L.}~\bibnamefont {Kelton}}, \bibinfo {author} {\bibfnamefont
  {A.}~\bibnamefont {Keshavarzi}}, \bibinfo {author} {\bibfnamefont
  {D.}~\bibnamefont {Kessler}}, \bibinfo {author} {\bibfnamefont {K.~S.}\
  \bibnamefont {Khaw}}, \bibinfo {author} {\bibfnamefont {Z.}~\bibnamefont
  {Khechadoorian}}, \bibinfo {author} {\bibfnamefont {N.~V.}\ \bibnamefont
  {Khomutov}}, \bibinfo {author} {\bibfnamefont {B.}~\bibnamefont {Kiburg}},
  \bibinfo {author} {\bibfnamefont {M.}~\bibnamefont {Kiburg}}, \bibinfo
  {author} {\bibfnamefont {O.}~\bibnamefont {Kim}}, \bibinfo {author}
  {\bibfnamefont {S.~C.}\ \bibnamefont {Kim}}, \bibinfo {author} {\bibfnamefont
  {Y.~I.}\ \bibnamefont {Kim}}, \bibinfo {author} {\bibfnamefont
  {B.}~\bibnamefont {King}}, \bibinfo {author} {\bibfnamefont {N.}~\bibnamefont
  {Kinnaird}}, \bibinfo {author} {\bibfnamefont {M.}~\bibnamefont
  {Korostelev}}, \bibinfo {author} {\bibfnamefont {I.}~\bibnamefont
  {Kourbanis}}, \bibinfo {author} {\bibfnamefont {E.}~\bibnamefont
  {Kraegeloh}}, \bibinfo {author} {\bibfnamefont {V.~A.}\ \bibnamefont
  {Krylov}}, \bibinfo {author} {\bibfnamefont {A.}~\bibnamefont {Kuchibhotla}},
  \bibinfo {author} {\bibfnamefont {N.~A.}\ \bibnamefont {Kuchinskiy}},
  \bibinfo {author} {\bibfnamefont {K.~R.}\ \bibnamefont {Labe}}, \bibinfo
  {author} {\bibfnamefont {J.}~\bibnamefont {LaBounty}}, \bibinfo {author}
  {\bibfnamefont {M.}~\bibnamefont {Lancaster}}, \bibinfo {author}
  {\bibfnamefont {M.~J.}\ \bibnamefont {Lee}}, \bibinfo {author} {\bibfnamefont
  {S.}~\bibnamefont {Lee}}, \bibinfo {author} {\bibfnamefont {S.}~\bibnamefont
  {Leo}}, \bibinfo {author} {\bibfnamefont {B.}~\bibnamefont {Li}}, \bibinfo
  {author} {\bibfnamefont {D.}~\bibnamefont {Li}}, \bibinfo {author}
  {\bibfnamefont {L.}~\bibnamefont {Li}}, \bibinfo {author} {\bibfnamefont
  {I.}~\bibnamefont {Logashenko}}, \bibinfo {author} {\bibfnamefont
  {A.}~\bibnamefont {Lorente~Campos}}, \bibinfo {author} {\bibfnamefont
  {A.}~\bibnamefont {Luc\`a}}, \bibinfo {author} {\bibfnamefont
  {G.}~\bibnamefont {Lukicov}}, \bibinfo {author} {\bibfnamefont
  {G.}~\bibnamefont {Luo}}, \bibinfo {author} {\bibfnamefont {A.}~\bibnamefont
  {Lusiani}}, \bibinfo {author} {\bibfnamefont {A.~L.}\ \bibnamefont {Lyon}},
  \bibinfo {author} {\bibfnamefont {B.}~\bibnamefont {MacCoy}}, \bibinfo
  {author} {\bibfnamefont {R.}~\bibnamefont {Madrak}}, \bibinfo {author}
  {\bibfnamefont {K.}~\bibnamefont {Makino}}, \bibinfo {author} {\bibfnamefont
  {F.}~\bibnamefont {Marignetti}}, \bibinfo {author} {\bibfnamefont
  {S.}~\bibnamefont {Mastroianni}}, \bibinfo {author} {\bibfnamefont
  {S.}~\bibnamefont {Maxfield}}, \bibinfo {author} {\bibfnamefont
  {M.}~\bibnamefont {McEvoy}}, \bibinfo {author} {\bibfnamefont
  {W.}~\bibnamefont {Merritt}}, \bibinfo {author} {\bibfnamefont {A.~A.}\
  \bibnamefont {Mikhailichenko}}, \bibinfo {author} {\bibfnamefont {J.~P.}\
  \bibnamefont {Miller}}, \bibinfo {author} {\bibfnamefont {S.}~\bibnamefont
  {Miozzi}}, \bibinfo {author} {\bibfnamefont {J.~P.}\ \bibnamefont {Morgan}},
  \bibinfo {author} {\bibfnamefont {W.~M.}\ \bibnamefont {Morse}}, \bibinfo
  {author} {\bibfnamefont {J.}~\bibnamefont {Mott}}, \bibinfo {author}
  {\bibfnamefont {E.}~\bibnamefont {Motuk}}, \bibinfo {author} {\bibfnamefont
  {A.}~\bibnamefont {Nath}}, \bibinfo {author} {\bibfnamefont {D.}~\bibnamefont
  {Newton}}, \bibinfo {author} {\bibfnamefont {H.}~\bibnamefont {Nguyen}},
  \bibinfo {author} {\bibfnamefont {M.}~\bibnamefont {Oberling}}, \bibinfo
  {author} {\bibfnamefont {R.}~\bibnamefont {Osofsky}}, \bibinfo {author}
  {\bibfnamefont {J.-F.}\ \bibnamefont {Ostiguy}}, \bibinfo {author}
  {\bibfnamefont {S.}~\bibnamefont {Park}}, \bibinfo {author} {\bibfnamefont
  {G.}~\bibnamefont {Pauletta}}, \bibinfo {author} {\bibfnamefont {G.~M.}\
  \bibnamefont {Piacentino}}, \bibinfo {author} {\bibfnamefont {R.~N.}\
  \bibnamefont {Pilato}}, \bibinfo {author} {\bibfnamefont {K.~T.}\
  \bibnamefont {Pitts}}, \bibinfo {author} {\bibfnamefont {B.}~\bibnamefont
  {Plaster}}, \bibinfo {author} {\bibfnamefont {D.}~\bibnamefont {Po\ifmmode
  \check{c}\else \v{c}\fi{}ani\ifmmode~\acute{c}\else \'{c}\fi{}}}, \bibinfo
  {author} {\bibfnamefont {N.}~\bibnamefont {Pohlman}}, \bibinfo {author}
  {\bibfnamefont {C.~C.}\ \bibnamefont {Polly}}, \bibinfo {author}
  {\bibfnamefont {M.}~\bibnamefont {Popovic}}, \bibinfo {author} {\bibfnamefont
  {J.}~\bibnamefont {Price}}, \bibinfo {author} {\bibfnamefont
  {B.}~\bibnamefont {Quinn}}, \bibinfo {author} {\bibfnamefont
  {N.}~\bibnamefont {Raha}}, \bibinfo {author} {\bibfnamefont {S.}~\bibnamefont
  {Ramachandran}}, \bibinfo {author} {\bibfnamefont {E.}~\bibnamefont
  {Ramberg}}, \bibinfo {author} {\bibfnamefont {N.~T.}\ \bibnamefont {Rider}},
  \bibinfo {author} {\bibfnamefont {J.~L.}\ \bibnamefont {Ritchie}}, \bibinfo
  {author} {\bibfnamefont {B.~L.}\ \bibnamefont {Roberts}}, \bibinfo {author}
  {\bibfnamefont {D.~L.}\ \bibnamefont {Rubin}}, \bibinfo {author}
  {\bibfnamefont {L.}~\bibnamefont {Santi}}, \bibinfo {author} {\bibfnamefont
  {D.}~\bibnamefont {Sathyan}}, \bibinfo {author} {\bibfnamefont
  {H.}~\bibnamefont {Schellman}}, \bibinfo {author} {\bibfnamefont
  {C.}~\bibnamefont {Schlesier}}, \bibinfo {author} {\bibfnamefont
  {A.}~\bibnamefont {Schreckenberger}}, \bibinfo {author} {\bibfnamefont
  {Y.~K.}\ \bibnamefont {Semertzidis}}, \bibinfo {author} {\bibfnamefont
  {Y.~M.}\ \bibnamefont {Shatunov}}, \bibinfo {author} {\bibfnamefont
  {D.}~\bibnamefont {Shemyakin}}, \bibinfo {author} {\bibfnamefont
  {M.}~\bibnamefont {Shenk}}, \bibinfo {author} {\bibfnamefont
  {D.}~\bibnamefont {Sim}}, \bibinfo {author} {\bibfnamefont {M.~W.}\
  \bibnamefont {Smith}}, \bibinfo {author} {\bibfnamefont {A.}~\bibnamefont
  {Smith}}, \bibinfo {author} {\bibfnamefont {A.~K.}\ \bibnamefont {Soha}},
  \bibinfo {author} {\bibfnamefont {M.}~\bibnamefont {Sorbara}}, \bibinfo
  {author} {\bibfnamefont {D.}~\bibnamefont {St\"ockinger}}, \bibinfo {author}
  {\bibfnamefont {J.}~\bibnamefont {Stapleton}}, \bibinfo {author}
  {\bibfnamefont {D.}~\bibnamefont {Still}}, \bibinfo {author} {\bibfnamefont
  {C.}~\bibnamefont {Stoughton}}, \bibinfo {author} {\bibfnamefont
  {D.}~\bibnamefont {Stratakis}}, \bibinfo {author} {\bibfnamefont
  {C.}~\bibnamefont {Strohman}}, \bibinfo {author} {\bibfnamefont
  {T.}~\bibnamefont {Stuttard}}, \bibinfo {author} {\bibfnamefont {H.~E.}\
  \bibnamefont {Swanson}}, \bibinfo {author} {\bibfnamefont {G.}~\bibnamefont
  {Sweetmore}}, \bibinfo {author} {\bibfnamefont {D.~A.}\ \bibnamefont
  {Sweigart}}, \bibinfo {author} {\bibfnamefont {M.~J.}\ \bibnamefont
  {Syphers}}, \bibinfo {author} {\bibfnamefont {D.~A.}\ \bibnamefont
  {Tarazona}}, \bibinfo {author} {\bibfnamefont {T.}~\bibnamefont {Teubner}},
  \bibinfo {author} {\bibfnamefont {A.~E.}\ \bibnamefont {Tewsley-Booth}},
  \bibinfo {author} {\bibfnamefont {K.}~\bibnamefont {Thomson}}, \bibinfo
  {author} {\bibfnamefont {V.}~\bibnamefont {Tishchenko}}, \bibinfo {author}
  {\bibfnamefont {N.~H.}\ \bibnamefont {Tran}}, \bibinfo {author}
  {\bibfnamefont {W.}~\bibnamefont {Turner}}, \bibinfo {author} {\bibfnamefont
  {E.}~\bibnamefont {Valetov}}, \bibinfo {author} {\bibfnamefont
  {D.}~\bibnamefont {Vasilkova}}, \bibinfo {author} {\bibfnamefont
  {G.}~\bibnamefont {Venanzoni}}, \bibinfo {author} {\bibfnamefont {V.~P.}\
  \bibnamefont {Volnykh}}, \bibinfo {author} {\bibfnamefont {T.}~\bibnamefont
  {Walton}}, \bibinfo {author} {\bibfnamefont {M.}~\bibnamefont {Warren}},
  \bibinfo {author} {\bibfnamefont {A.}~\bibnamefont {Weisskopf}}, \bibinfo
  {author} {\bibfnamefont {L.}~\bibnamefont {Welty-Rieger}}, \bibinfo {author}
  {\bibfnamefont {M.}~\bibnamefont {Whitley}}, \bibinfo {author} {\bibfnamefont
  {P.}~\bibnamefont {Winter}}, \bibinfo {author} {\bibfnamefont
  {A.}~\bibnamefont {Wolski}}, \bibinfo {author} {\bibfnamefont
  {M.}~\bibnamefont {Wormald}}, \bibinfo {author} {\bibfnamefont
  {W.}~\bibnamefont {Wu}},\ and\ \bibinfo {author} {\bibfnamefont
  {C.}~\bibnamefont {Yoshikawa}} (\bibinfo {collaboration} {Muon
  $g\ensuremath{-}2$ Collaboration}),\ }\href
  {https://doi.org/10.1103/PhysRevLett.126.141801} {\bibfield  {journal}
  {\bibinfo  {journal} {Phys. Rev. Lett.}\ }\textbf {\bibinfo {volume} {126}},\
  \bibinfo {pages} {141801} (\bibinfo {year} {2021})}\BibitemShut {NoStop}%
\bibitem [{\citenamefont {Bennett}\ \emph {et~al.}(2009)\citenamefont
  {Bennett}, \citenamefont {Bousquet}, \citenamefont {Brown}, \citenamefont
  {Bunce}, \citenamefont {Carey}, \citenamefont {Cushman}, \citenamefont
  {Danby}, \citenamefont {Debevec}, \citenamefont {Deile}, \citenamefont
  {Deng}, \citenamefont {Deninger}, \citenamefont {Dhawan}, \citenamefont
  {Druzhinin}, \citenamefont {Duong}, \citenamefont {Efstathiadis},
  \citenamefont {Farley}, \citenamefont {Fedotovich}, \citenamefont {Giron},
  \citenamefont {Gray}, \citenamefont {Grigoriev}, \citenamefont
  {Grosse-Perdekamp}, \citenamefont {Grossmann}, \citenamefont {Hare},
  \citenamefont {Hertzog}, \citenamefont {Huang}, \citenamefont {Hughes},
  \citenamefont {Iwasaki}, \citenamefont {Jungmann}, \citenamefont {Kawall},
  \citenamefont {Kawamura}, \citenamefont {Khazin}, \citenamefont {Kindem},
  \citenamefont {Krienen}, \citenamefont {Kronkvist}, \citenamefont {Lam},
  \citenamefont {Larsen}, \citenamefont {Lee}, \citenamefont {Logashenko},
  \citenamefont {McNabb}, \citenamefont {Meng}, \citenamefont {Mi},
  \citenamefont {Miller}, \citenamefont {Mizumachi}, \citenamefont {Morse},
  \citenamefont {Nikas}, \citenamefont {Onderwater}, \citenamefont {Orlov},
  \citenamefont {\"Ozben}, \citenamefont {Paley}, \citenamefont {Peng},
  \citenamefont {Polly}, \citenamefont {Pretz}, \citenamefont {Prigl},
  \citenamefont {zu~Putlitz}, \citenamefont {Qian}, \citenamefont {Redin},
  \citenamefont {Rind}, \citenamefont {Roberts}, \citenamefont {Ryskulov},
  \citenamefont {Sedykh}, \citenamefont {Semertzidis}, \citenamefont {Shagin},
  \citenamefont {Shatunov}, \citenamefont {Sichtermann}, \citenamefont
  {Solodov}, \citenamefont {Sossong}, \citenamefont {Steinmetz}, \citenamefont
  {Sulak}, \citenamefont {Timmermans}, \citenamefont {Trofimov}, \citenamefont
  {Urner}, \citenamefont {von Walter}, \citenamefont {Warburton}, \citenamefont
  {Winn}, \citenamefont {Yamamoto},\ and\ \citenamefont {Zimmerman}}]{bnl_edm}%
  \BibitemOpen
  \bibfield  {author} {\bibinfo {author} {\bibfnamefont {G.~W.}\ \bibnamefont
  {Bennett}}, \bibinfo {author} {\bibfnamefont {B.}~\bibnamefont {Bousquet}},
  \bibinfo {author} {\bibfnamefont {H.~N.}\ \bibnamefont {Brown}}, \bibinfo
  {author} {\bibfnamefont {G.}~\bibnamefont {Bunce}}, \bibinfo {author}
  {\bibfnamefont {R.~M.}\ \bibnamefont {Carey}}, \bibinfo {author}
  {\bibfnamefont {P.}~\bibnamefont {Cushman}}, \bibinfo {author} {\bibfnamefont
  {G.~T.}\ \bibnamefont {Danby}}, \bibinfo {author} {\bibfnamefont {P.~T.}\
  \bibnamefont {Debevec}}, \bibinfo {author} {\bibfnamefont {M.}~\bibnamefont
  {Deile}}, \bibinfo {author} {\bibfnamefont {H.}~\bibnamefont {Deng}},
  \bibinfo {author} {\bibfnamefont {W.}~\bibnamefont {Deninger}}, \bibinfo
  {author} {\bibfnamefont {S.~K.}\ \bibnamefont {Dhawan}}, \bibinfo {author}
  {\bibfnamefont {V.~P.}\ \bibnamefont {Druzhinin}}, \bibinfo {author}
  {\bibfnamefont {L.}~\bibnamefont {Duong}}, \bibinfo {author} {\bibfnamefont
  {E.}~\bibnamefont {Efstathiadis}}, \bibinfo {author} {\bibfnamefont
  {F.~J.~M.}\ \bibnamefont {Farley}}, \bibinfo {author} {\bibfnamefont {G.~V.}\
  \bibnamefont {Fedotovich}}, \bibinfo {author} {\bibfnamefont
  {S.}~\bibnamefont {Giron}}, \bibinfo {author} {\bibfnamefont {F.~E.}\
  \bibnamefont {Gray}}, \bibinfo {author} {\bibfnamefont {D.}~\bibnamefont
  {Grigoriev}}, \bibinfo {author} {\bibfnamefont {M.}~\bibnamefont
  {Grosse-Perdekamp}}, \bibinfo {author} {\bibfnamefont {A.}~\bibnamefont
  {Grossmann}}, \bibinfo {author} {\bibfnamefont {M.~F.}\ \bibnamefont {Hare}},
  \bibinfo {author} {\bibfnamefont {D.~W.}\ \bibnamefont {Hertzog}}, \bibinfo
  {author} {\bibfnamefont {X.}~\bibnamefont {Huang}}, \bibinfo {author}
  {\bibfnamefont {V.~W.}\ \bibnamefont {Hughes}}, \bibinfo {author}
  {\bibfnamefont {M.}~\bibnamefont {Iwasaki}}, \bibinfo {author} {\bibfnamefont
  {K.}~\bibnamefont {Jungmann}}, \bibinfo {author} {\bibfnamefont
  {D.}~\bibnamefont {Kawall}}, \bibinfo {author} {\bibfnamefont
  {M.}~\bibnamefont {Kawamura}}, \bibinfo {author} {\bibfnamefont {B.~I.}\
  \bibnamefont {Khazin}}, \bibinfo {author} {\bibfnamefont {J.}~\bibnamefont
  {Kindem}}, \bibinfo {author} {\bibfnamefont {F.}~\bibnamefont {Krienen}},
  \bibinfo {author} {\bibfnamefont {I.}~\bibnamefont {Kronkvist}}, \bibinfo
  {author} {\bibfnamefont {A.}~\bibnamefont {Lam}}, \bibinfo {author}
  {\bibfnamefont {R.}~\bibnamefont {Larsen}}, \bibinfo {author} {\bibfnamefont
  {Y.~Y.}\ \bibnamefont {Lee}}, \bibinfo {author} {\bibfnamefont
  {I.}~\bibnamefont {Logashenko}}, \bibinfo {author} {\bibfnamefont
  {R.}~\bibnamefont {McNabb}}, \bibinfo {author} {\bibfnamefont
  {W.}~\bibnamefont {Meng}}, \bibinfo {author} {\bibfnamefont {J.}~\bibnamefont
  {Mi}}, \bibinfo {author} {\bibfnamefont {J.~P.}\ \bibnamefont {Miller}},
  \bibinfo {author} {\bibfnamefont {Y.}~\bibnamefont {Mizumachi}}, \bibinfo
  {author} {\bibfnamefont {W.~M.}\ \bibnamefont {Morse}}, \bibinfo {author}
  {\bibfnamefont {D.}~\bibnamefont {Nikas}}, \bibinfo {author} {\bibfnamefont
  {C.~J.~G.}\ \bibnamefont {Onderwater}}, \bibinfo {author} {\bibfnamefont
  {Y.}~\bibnamefont {Orlov}}, \bibinfo {author} {\bibfnamefont {C.~S.}\
  \bibnamefont {\"Ozben}}, \bibinfo {author} {\bibfnamefont {J.~M.}\
  \bibnamefont {Paley}}, \bibinfo {author} {\bibfnamefont {Q.}~\bibnamefont
  {Peng}}, \bibinfo {author} {\bibfnamefont {C.~C.}\ \bibnamefont {Polly}},
  \bibinfo {author} {\bibfnamefont {J.}~\bibnamefont {Pretz}}, \bibinfo
  {author} {\bibfnamefont {R.}~\bibnamefont {Prigl}}, \bibinfo {author}
  {\bibfnamefont {G.}~\bibnamefont {zu~Putlitz}}, \bibinfo {author}
  {\bibfnamefont {T.}~\bibnamefont {Qian}}, \bibinfo {author} {\bibfnamefont
  {S.~I.}\ \bibnamefont {Redin}}, \bibinfo {author} {\bibfnamefont
  {O.}~\bibnamefont {Rind}}, \bibinfo {author} {\bibfnamefont {B.~L.}\
  \bibnamefont {Roberts}}, \bibinfo {author} {\bibfnamefont {N.}~\bibnamefont
  {Ryskulov}}, \bibinfo {author} {\bibfnamefont {S.}~\bibnamefont {Sedykh}},
  \bibinfo {author} {\bibfnamefont {Y.~K.}\ \bibnamefont {Semertzidis}},
  \bibinfo {author} {\bibfnamefont {P.}~\bibnamefont {Shagin}}, \bibinfo
  {author} {\bibfnamefont {Y.~M.}\ \bibnamefont {Shatunov}}, \bibinfo {author}
  {\bibfnamefont {E.~P.}\ \bibnamefont {Sichtermann}}, \bibinfo {author}
  {\bibfnamefont {E.}~\bibnamefont {Solodov}}, \bibinfo {author} {\bibfnamefont
  {M.}~\bibnamefont {Sossong}}, \bibinfo {author} {\bibfnamefont
  {A.}~\bibnamefont {Steinmetz}}, \bibinfo {author} {\bibfnamefont {L.~R.}\
  \bibnamefont {Sulak}}, \bibinfo {author} {\bibfnamefont {C.}~\bibnamefont
  {Timmermans}}, \bibinfo {author} {\bibfnamefont {A.}~\bibnamefont
  {Trofimov}}, \bibinfo {author} {\bibfnamefont {D.}~\bibnamefont {Urner}},
  \bibinfo {author} {\bibfnamefont {P.}~\bibnamefont {von Walter}}, \bibinfo
  {author} {\bibfnamefont {D.}~\bibnamefont {Warburton}}, \bibinfo {author}
  {\bibfnamefont {D.}~\bibnamefont {Winn}}, \bibinfo {author} {\bibfnamefont
  {A.}~\bibnamefont {Yamamoto}},\ and\ \bibinfo {author} {\bibfnamefont
  {D.}~\bibnamefont {Zimmerman}} (\bibinfo {collaboration} {Muon (g-2)
  Collaboration}),\ }\href {https://doi.org/10.1103/PhysRevD.80.052008}
  {\bibfield  {journal} {\bibinfo  {journal} {Phys. Rev. D}\ }\textbf {\bibinfo
  {volume} {80}},\ \bibinfo {pages} {052008} (\bibinfo {year}
  {2009})}\BibitemShut {NoStop}%
\bibitem [{\citenamefont {Farley}\ \emph {et~al.}(7 27)\citenamefont {Farley},
  \citenamefont {Jungmann}, \citenamefont {Miller}, \citenamefont {Morse},
  \citenamefont {Orlov}, \citenamefont {Roberts}, \citenamefont {Semertzidis},
  \citenamefont {Silenko},\ and\ \citenamefont {Stephenson}}]{farley_new_2004}%
  \BibitemOpen
  \bibfield  {author} {\bibinfo {author} {\bibfnamefont {F.~J.~M.}\
  \bibnamefont {Farley}}, \bibinfo {author} {\bibfnamefont {K.}~\bibnamefont
  {Jungmann}}, \bibinfo {author} {\bibfnamefont {J.~P.}\ \bibnamefont
  {Miller}}, \bibinfo {author} {\bibfnamefont {W.~M.}\ \bibnamefont {Morse}},
  \bibinfo {author} {\bibfnamefont {Y.~F.}\ \bibnamefont {Orlov}}, \bibinfo
  {author} {\bibfnamefont {B.~L.}\ \bibnamefont {Roberts}}, \bibinfo {author}
  {\bibfnamefont {Y.~K.}\ \bibnamefont {Semertzidis}}, \bibinfo {author}
  {\bibfnamefont {A.}~\bibnamefont {Silenko}},\ and\ \bibinfo {author}
  {\bibfnamefont {E.~J.}\ \bibnamefont {Stephenson}},\ }\href
  {https://doi.org/10.1103/PhysRevLett.93.052001} {\bibfield  {journal}
  {\bibinfo  {journal} {Physical Review Letters}\ }\textbf {\bibinfo {volume}
  {93}},\ \bibinfo {pages} {052001} (\bibinfo {year} {2004-07-27})}\BibitemShut
  {NoStop}%
\bibitem [{\citenamefont {Omarov}\ \emph {et~al.}(2022)\citenamefont {Omarov},
  \citenamefont {Davoudiasl}, \citenamefont {Hac{\i}{\"o}mero{\u{g}}lu},
  \citenamefont {Lebedev}, \citenamefont {Morse}, \citenamefont {Semertzidis},
  \citenamefont {Silenko}, \citenamefont {Stephenson},\ and\ \citenamefont
  {Suleiman}}]{symmetric}%
  \BibitemOpen
  \bibfield  {author} {\bibinfo {author} {\bibfnamefont {Z.}~\bibnamefont
  {Omarov}}, \bibinfo {author} {\bibfnamefont {H.}~\bibnamefont {Davoudiasl}},
  \bibinfo {author} {\bibfnamefont {S.}~\bibnamefont
  {Hac{\i}{\"o}mero{\u{g}}lu}}, \bibinfo {author} {\bibfnamefont
  {V.}~\bibnamefont {Lebedev}}, \bibinfo {author} {\bibfnamefont {W.~M.}\
  \bibnamefont {Morse}}, \bibinfo {author} {\bibfnamefont {Y.~K.}\ \bibnamefont
  {Semertzidis}}, \bibinfo {author} {\bibfnamefont {A.~J.}\ \bibnamefont
  {Silenko}}, \bibinfo {author} {\bibfnamefont {E.~J.}\ \bibnamefont
  {Stephenson}},\ and\ \bibinfo {author} {\bibfnamefont {R.}~\bibnamefont
  {Suleiman}},\ }\href {https://doi.org/10.1103/PhysRevD.105.032001} {\bibfield
   {journal} {\bibinfo  {journal} {Phys. Rev. D}\ }\textbf {\bibinfo {volume}
  {105}},\ \bibinfo {pages} {032001} (\bibinfo {year} {2022})}\BibitemShut
  {NoStop}%
\bibitem [{\citenamefont {BastaniNejad}\ \emph {et~al.}(2012)\citenamefont
  {BastaniNejad}, \citenamefont {Mohamed}, \citenamefont {Elmustafa},
  \citenamefont {Adderley}, \citenamefont {Clark}, \citenamefont {Covert},
  \citenamefont {Hansknecht}, \citenamefont {Hernandez-Garcia}, \citenamefont
  {Poelker}, \citenamefont {Mammei}, \citenamefont {Surles-Law},\ and\
  \citenamefont {Williams}}]{electrode1}%
  \BibitemOpen
  \bibfield  {author} {\bibinfo {author} {\bibfnamefont {M.}~\bibnamefont
  {BastaniNejad}}, \bibinfo {author} {\bibfnamefont {M.~A.}\ \bibnamefont
  {Mohamed}}, \bibinfo {author} {\bibfnamefont {A.~A.}\ \bibnamefont
  {Elmustafa}}, \bibinfo {author} {\bibfnamefont {P.}~\bibnamefont {Adderley}},
  \bibinfo {author} {\bibfnamefont {J.}~\bibnamefont {Clark}}, \bibinfo
  {author} {\bibfnamefont {S.}~\bibnamefont {Covert}}, \bibinfo {author}
  {\bibfnamefont {J.}~\bibnamefont {Hansknecht}}, \bibinfo {author}
  {\bibfnamefont {C.}~\bibnamefont {Hernandez-Garcia}}, \bibinfo {author}
  {\bibfnamefont {M.}~\bibnamefont {Poelker}}, \bibinfo {author} {\bibfnamefont
  {R.}~\bibnamefont {Mammei}}, \bibinfo {author} {\bibfnamefont
  {K.}~\bibnamefont {Surles-Law}},\ and\ \bibinfo {author} {\bibfnamefont
  {P.}~\bibnamefont {Williams}},\ }\href
  {https://doi.org/10.1103/PhysRevSTAB.15.083502} {\bibfield  {journal}
  {\bibinfo  {journal} {Phys. Rev. ST Accel. Beams}\ }\textbf {\bibinfo
  {volume} {15}},\ \bibinfo {pages} {083502} (\bibinfo {year}
  {2012})}\BibitemShut {NoStop}%
\bibitem [{\citenamefont {Mamun}\ \emph {et~al.}(2015)\citenamefont {Mamun},
  \citenamefont {Elmustafa}, \citenamefont {Taus}, \citenamefont {Forman},\
  and\ \citenamefont {Poelker}}]{electrode2}%
  \BibitemOpen
  \bibfield  {author} {\bibinfo {author} {\bibfnamefont {M.~A.~A.}\
  \bibnamefont {Mamun}}, \bibinfo {author} {\bibfnamefont {A.~A.}\ \bibnamefont
  {Elmustafa}}, \bibinfo {author} {\bibfnamefont {R.}~\bibnamefont {Taus}},
  \bibinfo {author} {\bibfnamefont {E.}~\bibnamefont {Forman}},\ and\ \bibinfo
  {author} {\bibfnamefont {M.}~\bibnamefont {Poelker}},\ }\href
  {https://doi.org/10.1116/1.4916574} {\bibfield  {journal} {\bibinfo
  {journal} {Journal of Vacuum Science \& Technology A: Vacuum, Surfaces, and
  Films}\ }\textbf {\bibinfo {volume} {33}},\ \bibinfo {pages} {031604}
  (\bibinfo {year} {2015})}\BibitemShut {NoStop}%
\bibitem [{\citenamefont {Palacios-Serrano}\ \emph {et~al.}(2018)\citenamefont
  {Palacios-Serrano}, \citenamefont {Hannon}, \citenamefont {Hernandez-Garcia},
  \citenamefont {Poelker},\ and\ \citenamefont {Baumgart}}]{electrode3}%
  \BibitemOpen
  \bibfield  {author} {\bibinfo {author} {\bibfnamefont {G.}~\bibnamefont
  {Palacios-Serrano}}, \bibinfo {author} {\bibfnamefont {F.}~\bibnamefont
  {Hannon}}, \bibinfo {author} {\bibfnamefont {C.}~\bibnamefont
  {Hernandez-Garcia}}, \bibinfo {author} {\bibfnamefont {M.}~\bibnamefont
  {Poelker}},\ and\ \bibinfo {author} {\bibfnamefont {H.}~\bibnamefont
  {Baumgart}},\ }\href {https://doi.org/10.1063/1.5048700} {\bibfield
  {journal} {\bibinfo  {journal} {Review of Scientific Instruments}\ }\textbf
  {\bibinfo {volume} {89}},\ \bibinfo {pages} {104703} (\bibinfo {year}
  {2018})}\BibitemShut {NoStop}%
\bibitem [{\citenamefont {Mooser}\ \emph {et~al.}(4 05)\citenamefont {Mooser},
  \citenamefont {Ulmer}, \citenamefont {Blaum}, \citenamefont {Franke},
  \citenamefont {Kracke}, \citenamefont {Leiteritz}, \citenamefont {Quint},
  \citenamefont {Rodegheri}, \citenamefont {Smorra},\ and\ \citenamefont
  {Walz}}]{mooser_direct_2014}%
  \BibitemOpen
  \bibfield  {author} {\bibinfo {author} {\bibfnamefont {A.}~\bibnamefont
  {Mooser}}, \bibinfo {author} {\bibfnamefont {S.}~\bibnamefont {Ulmer}},
  \bibinfo {author} {\bibfnamefont {K.}~\bibnamefont {Blaum}}, \bibinfo
  {author} {\bibfnamefont {K.}~\bibnamefont {Franke}}, \bibinfo {author}
  {\bibfnamefont {H.}~\bibnamefont {Kracke}}, \bibinfo {author} {\bibfnamefont
  {C.}~\bibnamefont {Leiteritz}}, \bibinfo {author} {\bibfnamefont
  {W.}~\bibnamefont {Quint}}, \bibinfo {author} {\bibfnamefont {C.~C.}\
  \bibnamefont {Rodegheri}}, \bibinfo {author} {\bibfnamefont {C.}~\bibnamefont
  {Smorra}},\ and\ \bibinfo {author} {\bibfnamefont {J.}~\bibnamefont {Walz}},\
  }\href {https://doi.org/10.1038/nature13388} {\bibfield  {journal} {\bibinfo
  {journal} {Nature}\ }\textbf {\bibinfo {volume} {509}},\ \bibinfo {pages}
  {596} (\bibinfo {year} {2014-05})}\BibitemShut {NoStop}%
\bibitem [{\citenamefont {Brantjes}\ \emph {et~al.}(2012)\citenamefont
  {Brantjes}, \citenamefont {Dzordzhadze}, \citenamefont {Gebel}, \citenamefont
  {Gonnella}, \citenamefont {Gray}, \citenamefont {Van Der~Hoek}, \citenamefont
  {Imig}, \citenamefont {Kruithof}, \citenamefont {Lazarus}, \citenamefont
  {Lehrach} \emph {et~al.}}]{brantjes2012correcting}%
  \BibitemOpen
  \bibfield  {author} {\bibinfo {author} {\bibfnamefont {N.}~\bibnamefont
  {Brantjes}}, \bibinfo {author} {\bibfnamefont {V.}~\bibnamefont
  {Dzordzhadze}}, \bibinfo {author} {\bibfnamefont {R.}~\bibnamefont {Gebel}},
  \bibinfo {author} {\bibfnamefont {F.}~\bibnamefont {Gonnella}}, \bibinfo
  {author} {\bibfnamefont {F.}~\bibnamefont {Gray}}, \bibinfo {author}
  {\bibfnamefont {D.}~\bibnamefont {Van Der~Hoek}}, \bibinfo {author}
  {\bibfnamefont {A.}~\bibnamefont {Imig}}, \bibinfo {author} {\bibfnamefont
  {W.}~\bibnamefont {Kruithof}}, \bibinfo {author} {\bibfnamefont
  {D.}~\bibnamefont {Lazarus}}, \bibinfo {author} {\bibfnamefont
  {A.}~\bibnamefont {Lehrach}}, \emph {et~al.},\ }\href
  {https://doi.org/10.1016/j.nima.2011.09.055} {\bibfield  {journal} {\bibinfo
  {journal} {Nuclear Instruments and Methods in Physics Research Section A:
  Accelerators, Spectrometers, Detectors and Associated Equipment}\ }\textbf
  {\bibinfo {volume} {664}},\ \bibinfo {pages} {49} (\bibinfo {year}
  {2012})}\BibitemShut {NoStop}%
\bibitem [{\citenamefont {Hempelmann}\ \emph {et~al.}(2018)\citenamefont
  {Hempelmann}, \citenamefont {Hejny}, \citenamefont {Pretz}, \citenamefont
  {Soltner}, \citenamefont {Augustyniak}, \citenamefont {Bagdasarian},
  \citenamefont {Bai}, \citenamefont {Barion}, \citenamefont {Berz},
  \citenamefont {Chekmenev}, \citenamefont {Ciullo}, \citenamefont {Dymov},
  \citenamefont {Eversmann}, \citenamefont {Gaisser}, \citenamefont {Gebel},
  \citenamefont {Grigoryev}, \citenamefont {Grzonka}, \citenamefont
  {Guidoboni}, \citenamefont {Heberling}, \citenamefont {Hetzel}, \citenamefont
  {Hinder}, \citenamefont {Kacharava}, \citenamefont {Kamerdzhiev},
  \citenamefont {Keshelashvili}, \citenamefont {Koop}, \citenamefont {Kulikov},
  \citenamefont {Lehrach}, \citenamefont {Lenisa}, \citenamefont {Lomidze},
  \citenamefont {Lorentz}, \citenamefont {Maanen}, \citenamefont
  {Macharashvili}, \citenamefont {Magiera}, \citenamefont {Mchedlishvili},
  \citenamefont {Mey}, \citenamefont {M\"uller}, \citenamefont {Nass},
  \citenamefont {Nikolaev}, \citenamefont {Nioradze}, \citenamefont {Pesce},
  \citenamefont {Prasuhn}, \citenamefont {Rathmann}, \citenamefont {Rosenthal},
  \citenamefont {Saleev}, \citenamefont {Schmidt}, \citenamefont {Semertzidis},
  \citenamefont {Senichev}, \citenamefont {Shmakova}, \citenamefont {Silenko},
  \citenamefont {Slim}, \citenamefont {Stahl}, \citenamefont {Stassen},
  \citenamefont {Stephenson}, \citenamefont {Stockhorst}, \citenamefont
  {Str\"oher}, \citenamefont {Tabidze}, \citenamefont {Tagliente},
  \citenamefont {Talman}, \citenamefont {Th\"orngren~Engblom}, \citenamefont
  {Trinkel}, \citenamefont {Uzikov}, \citenamefont {Valdau}, \citenamefont
  {Valetov}, \citenamefont {Vassiliev}, \citenamefont {Weidemann},
  \citenamefont {Wro\ifmmode~\acute{n}\else \'{n}\fi{}ska}, \citenamefont
  {W\"ustner}, \citenamefont {Zupra\ifmmode~\acute{n}\else \'{n}\fi{}ski},\
  and\ \citenamefont {\ifmmode~\dot{Z}\else \.{Z}\fi{}urek}}]{ref1}%
  \BibitemOpen
  \bibfield  {author} {\bibinfo {author} {\bibfnamefont {N.}~\bibnamefont
  {Hempelmann}}, \bibinfo {author} {\bibfnamefont {V.}~\bibnamefont {Hejny}},
  \bibinfo {author} {\bibfnamefont {J.}~\bibnamefont {Pretz}}, \bibinfo
  {author} {\bibfnamefont {H.}~\bibnamefont {Soltner}}, \bibinfo {author}
  {\bibfnamefont {W.}~\bibnamefont {Augustyniak}}, \bibinfo {author}
  {\bibfnamefont {Z.}~\bibnamefont {Bagdasarian}}, \bibinfo {author}
  {\bibfnamefont {M.}~\bibnamefont {Bai}}, \bibinfo {author} {\bibfnamefont
  {L.}~\bibnamefont {Barion}}, \bibinfo {author} {\bibfnamefont
  {M.}~\bibnamefont {Berz}}, \bibinfo {author} {\bibfnamefont {S.}~\bibnamefont
  {Chekmenev}}, \bibinfo {author} {\bibfnamefont {G.}~\bibnamefont {Ciullo}},
  \bibinfo {author} {\bibfnamefont {S.}~\bibnamefont {Dymov}}, \bibinfo
  {author} {\bibfnamefont {D.}~\bibnamefont {Eversmann}}, \bibinfo {author}
  {\bibfnamefont {M.}~\bibnamefont {Gaisser}}, \bibinfo {author} {\bibfnamefont
  {R.}~\bibnamefont {Gebel}}, \bibinfo {author} {\bibfnamefont
  {K.}~\bibnamefont {Grigoryev}}, \bibinfo {author} {\bibfnamefont
  {D.}~\bibnamefont {Grzonka}}, \bibinfo {author} {\bibfnamefont
  {G.}~\bibnamefont {Guidoboni}}, \bibinfo {author} {\bibfnamefont
  {D.}~\bibnamefont {Heberling}}, \bibinfo {author} {\bibfnamefont
  {J.}~\bibnamefont {Hetzel}}, \bibinfo {author} {\bibfnamefont
  {F.}~\bibnamefont {Hinder}}, \bibinfo {author} {\bibfnamefont
  {A.}~\bibnamefont {Kacharava}}, \bibinfo {author} {\bibfnamefont
  {V.}~\bibnamefont {Kamerdzhiev}}, \bibinfo {author} {\bibfnamefont
  {I.}~\bibnamefont {Keshelashvili}}, \bibinfo {author} {\bibfnamefont
  {I.}~\bibnamefont {Koop}}, \bibinfo {author} {\bibfnamefont {A.}~\bibnamefont
  {Kulikov}}, \bibinfo {author} {\bibfnamefont {A.}~\bibnamefont {Lehrach}},
  \bibinfo {author} {\bibfnamefont {P.}~\bibnamefont {Lenisa}}, \bibinfo
  {author} {\bibfnamefont {N.}~\bibnamefont {Lomidze}}, \bibinfo {author}
  {\bibfnamefont {B.}~\bibnamefont {Lorentz}}, \bibinfo {author} {\bibfnamefont
  {P.}~\bibnamefont {Maanen}}, \bibinfo {author} {\bibfnamefont
  {G.}~\bibnamefont {Macharashvili}}, \bibinfo {author} {\bibfnamefont
  {A.}~\bibnamefont {Magiera}}, \bibinfo {author} {\bibfnamefont
  {D.}~\bibnamefont {Mchedlishvili}}, \bibinfo {author} {\bibfnamefont
  {S.}~\bibnamefont {Mey}}, \bibinfo {author} {\bibfnamefont {F.}~\bibnamefont
  {M\"uller}}, \bibinfo {author} {\bibfnamefont {A.}~\bibnamefont {Nass}},
  \bibinfo {author} {\bibfnamefont {N.~N.}\ \bibnamefont {Nikolaev}}, \bibinfo
  {author} {\bibfnamefont {M.}~\bibnamefont {Nioradze}}, \bibinfo {author}
  {\bibfnamefont {A.}~\bibnamefont {Pesce}}, \bibinfo {author} {\bibfnamefont
  {D.}~\bibnamefont {Prasuhn}}, \bibinfo {author} {\bibfnamefont
  {F.}~\bibnamefont {Rathmann}}, \bibinfo {author} {\bibfnamefont
  {M.}~\bibnamefont {Rosenthal}}, \bibinfo {author} {\bibfnamefont
  {A.}~\bibnamefont {Saleev}}, \bibinfo {author} {\bibfnamefont
  {V.}~\bibnamefont {Schmidt}}, \bibinfo {author} {\bibfnamefont
  {Y.}~\bibnamefont {Semertzidis}}, \bibinfo {author} {\bibfnamefont
  {Y.}~\bibnamefont {Senichev}}, \bibinfo {author} {\bibfnamefont
  {V.}~\bibnamefont {Shmakova}}, \bibinfo {author} {\bibfnamefont
  {A.}~\bibnamefont {Silenko}}, \bibinfo {author} {\bibfnamefont
  {J.}~\bibnamefont {Slim}}, \bibinfo {author} {\bibfnamefont {A.}~\bibnamefont
  {Stahl}}, \bibinfo {author} {\bibfnamefont {R.}~\bibnamefont {Stassen}},
  \bibinfo {author} {\bibfnamefont {E.}~\bibnamefont {Stephenson}}, \bibinfo
  {author} {\bibfnamefont {H.}~\bibnamefont {Stockhorst}}, \bibinfo {author}
  {\bibfnamefont {H.}~\bibnamefont {Str\"oher}}, \bibinfo {author}
  {\bibfnamefont {M.}~\bibnamefont {Tabidze}}, \bibinfo {author} {\bibfnamefont
  {G.}~\bibnamefont {Tagliente}}, \bibinfo {author} {\bibfnamefont
  {R.}~\bibnamefont {Talman}}, \bibinfo {author} {\bibfnamefont
  {P.}~\bibnamefont {Th\"orngren~Engblom}}, \bibinfo {author} {\bibfnamefont
  {F.}~\bibnamefont {Trinkel}}, \bibinfo {author} {\bibfnamefont
  {Y.}~\bibnamefont {Uzikov}}, \bibinfo {author} {\bibfnamefont
  {Y.}~\bibnamefont {Valdau}}, \bibinfo {author} {\bibfnamefont
  {E.}~\bibnamefont {Valetov}}, \bibinfo {author} {\bibfnamefont
  {A.}~\bibnamefont {Vassiliev}}, \bibinfo {author} {\bibfnamefont
  {C.}~\bibnamefont {Weidemann}}, \bibinfo {author} {\bibfnamefont
  {A.}~\bibnamefont {Wro\ifmmode~\acute{n}\else \'{n}\fi{}ska}}, \bibinfo
  {author} {\bibfnamefont {P.}~\bibnamefont {W\"ustner}}, \bibinfo {author}
  {\bibfnamefont {P.}~\bibnamefont {Zupra\ifmmode~\acute{n}\else
  \'{n}\fi{}ski}},\ and\ \bibinfo {author} {\bibfnamefont {M.}~\bibnamefont
  {\ifmmode~\dot{Z}\else \.{Z}\fi{}urek}} (\bibinfo {collaboration} {JEDI
  Collaboration}),\ }\href
  {https://doi.org/10.1103/PhysRevAccelBeams.21.042002} {\bibfield  {journal}
  {\bibinfo  {journal} {Phys. Rev. Accel. Beams}\ }\textbf {\bibinfo {volume}
  {21}},\ \bibinfo {pages} {042002} (\bibinfo {year} {2018})}\BibitemShut
  {NoStop}%
\bibitem [{\citenamefont {Guidoboni}\ \emph {et~al.}(2018)\citenamefont
  {Guidoboni}, \citenamefont {Stephenson}, \citenamefont
  {Wro\ifmmode~\acute{n}\else \'{n}\fi{}ska}, \citenamefont {Bagdasarian},
  \citenamefont {Bsaisou}, \citenamefont {Chekmenev}, \citenamefont {Ciullo},
  \citenamefont {Dymov}, \citenamefont {Eversmann}, \citenamefont {Gaisser},
  \citenamefont {Gebel}, \citenamefont {Hejny}, \citenamefont {Hempelmann},
  \citenamefont {Hinder}, \citenamefont {Kacharava}, \citenamefont
  {Keshelashvili}, \citenamefont {Kulessa}, \citenamefont {Lenisa},
  \citenamefont {Lehrach}, \citenamefont {Lorentz}, \citenamefont {Maanen},
  \citenamefont {Maier}, \citenamefont {Mchedlishvili}, \citenamefont {Mey},
  \citenamefont {Nass}, \citenamefont {Pesce}, \citenamefont {Orlov},
  \citenamefont {Pretz}, \citenamefont {Prasuhn}, \citenamefont {Rathmann},
  \citenamefont {Rosenthal}, \citenamefont {Saleev}, \citenamefont
  {Semertzidis}, \citenamefont {Senichev}, \citenamefont {Shmakova},
  \citenamefont {Stockhorst}, \citenamefont {Str\"oher}, \citenamefont
  {Talman}, \citenamefont {Th\"orngren~Engblom}, \citenamefont {Trinkel},
  \citenamefont {Valdau}, \citenamefont {Weidemann}, \citenamefont {W\"ustner},
  \citenamefont {\ifmmode~\dot{Z}\else \.{Z}\fi{}urek},\ and\ \citenamefont
  {Zyuzin}}]{ref2}%
  \BibitemOpen
  \bibfield  {author} {\bibinfo {author} {\bibfnamefont {G.}~\bibnamefont
  {Guidoboni}}, \bibinfo {author} {\bibfnamefont {E.~J.}\ \bibnamefont
  {Stephenson}}, \bibinfo {author} {\bibfnamefont {A.}~\bibnamefont
  {Wro\ifmmode~\acute{n}\else \'{n}\fi{}ska}}, \bibinfo {author} {\bibfnamefont
  {Z.}~\bibnamefont {Bagdasarian}}, \bibinfo {author} {\bibfnamefont
  {J.}~\bibnamefont {Bsaisou}}, \bibinfo {author} {\bibfnamefont
  {S.}~\bibnamefont {Chekmenev}}, \bibinfo {author} {\bibfnamefont
  {G.}~\bibnamefont {Ciullo}}, \bibinfo {author} {\bibfnamefont
  {S.}~\bibnamefont {Dymov}}, \bibinfo {author} {\bibfnamefont
  {D.}~\bibnamefont {Eversmann}}, \bibinfo {author} {\bibfnamefont
  {M.}~\bibnamefont {Gaisser}}, \bibinfo {author} {\bibfnamefont
  {R.}~\bibnamefont {Gebel}}, \bibinfo {author} {\bibfnamefont
  {V.}~\bibnamefont {Hejny}}, \bibinfo {author} {\bibfnamefont
  {N.}~\bibnamefont {Hempelmann}}, \bibinfo {author} {\bibfnamefont
  {F.}~\bibnamefont {Hinder}}, \bibinfo {author} {\bibfnamefont
  {A.}~\bibnamefont {Kacharava}}, \bibinfo {author} {\bibfnamefont
  {I.}~\bibnamefont {Keshelashvili}}, \bibinfo {author} {\bibfnamefont
  {P.}~\bibnamefont {Kulessa}}, \bibinfo {author} {\bibfnamefont
  {P.}~\bibnamefont {Lenisa}}, \bibinfo {author} {\bibfnamefont
  {A.}~\bibnamefont {Lehrach}}, \bibinfo {author} {\bibfnamefont
  {B.}~\bibnamefont {Lorentz}}, \bibinfo {author} {\bibfnamefont
  {P.}~\bibnamefont {Maanen}}, \bibinfo {author} {\bibfnamefont
  {R.}~\bibnamefont {Maier}}, \bibinfo {author} {\bibfnamefont
  {D.}~\bibnamefont {Mchedlishvili}}, \bibinfo {author} {\bibfnamefont
  {S.}~\bibnamefont {Mey}}, \bibinfo {author} {\bibfnamefont {A.}~\bibnamefont
  {Nass}}, \bibinfo {author} {\bibfnamefont {A.}~\bibnamefont {Pesce}},
  \bibinfo {author} {\bibfnamefont {Y.}~\bibnamefont {Orlov}}, \bibinfo
  {author} {\bibfnamefont {J.}~\bibnamefont {Pretz}}, \bibinfo {author}
  {\bibfnamefont {D.}~\bibnamefont {Prasuhn}}, \bibinfo {author} {\bibfnamefont
  {F.}~\bibnamefont {Rathmann}}, \bibinfo {author} {\bibfnamefont
  {M.}~\bibnamefont {Rosenthal}}, \bibinfo {author} {\bibfnamefont
  {A.}~\bibnamefont {Saleev}}, \bibinfo {author} {\bibfnamefont {Y.~K.}\
  \bibnamefont {Semertzidis}}, \bibinfo {author} {\bibfnamefont
  {Y.}~\bibnamefont {Senichev}}, \bibinfo {author} {\bibfnamefont
  {V.}~\bibnamefont {Shmakova}}, \bibinfo {author} {\bibfnamefont
  {H.}~\bibnamefont {Stockhorst}}, \bibinfo {author} {\bibfnamefont
  {H.}~\bibnamefont {Str\"oher}}, \bibinfo {author} {\bibfnamefont
  {R.}~\bibnamefont {Talman}}, \bibinfo {author} {\bibfnamefont
  {P.}~\bibnamefont {Th\"orngren~Engblom}}, \bibinfo {author} {\bibfnamefont
  {F.}~\bibnamefont {Trinkel}}, \bibinfo {author} {\bibfnamefont
  {Y.}~\bibnamefont {Valdau}}, \bibinfo {author} {\bibfnamefont
  {C.}~\bibnamefont {Weidemann}}, \bibinfo {author} {\bibfnamefont
  {P.}~\bibnamefont {W\"ustner}}, \bibinfo {author} {\bibfnamefont
  {M.}~\bibnamefont {\ifmmode~\dot{Z}\else \.{Z}\fi{}urek}},\ and\ \bibinfo
  {author} {\bibfnamefont {D.}~\bibnamefont {Zyuzin}} (\bibinfo {collaboration}
  {JEDI Collaboration}),\ }\href
  {https://doi.org/10.1103/PhysRevAccelBeams.21.024201} {\bibfield  {journal}
  {\bibinfo  {journal} {Phys. Rev. Accel. Beams}\ }\textbf {\bibinfo {volume}
  {21}},\ \bibinfo {pages} {024201} (\bibinfo {year} {2018})}\BibitemShut
  {NoStop}%
\bibitem [{\citenamefont {Saleev}\ \emph {et~al.}(2017)\citenamefont {Saleev},
  \citenamefont {Nikolaev}, \citenamefont {Rathmann}, \citenamefont
  {Augustyniak}, \citenamefont {Bagdasarian}, \citenamefont {Bai},
  \citenamefont {Barion}, \citenamefont {Berz}, \citenamefont {Chekmenev},
  \citenamefont {Ciullo}, \citenamefont {Dymov}, \citenamefont {Eversmann},
  \citenamefont {Gaisser}, \citenamefont {Gebel}, \citenamefont {Grigoryev},
  \citenamefont {Grzonka}, \citenamefont {Guidoboni}, \citenamefont
  {Heberling}, \citenamefont {Hejny}, \citenamefont {Hempelmann}, \citenamefont
  {Hetzel}, \citenamefont {Hinder}, \citenamefont {Kacharava}, \citenamefont
  {Kamerdzhiev}, \citenamefont {Keshelashvili}, \citenamefont {Koop},
  \citenamefont {Kulikov}, \citenamefont {Lehrach}, \citenamefont {Lenisa},
  \citenamefont {Lomidze}, \citenamefont {Lorentz}, \citenamefont {Maanen},
  \citenamefont {Macharashvili}, \citenamefont {Magiera}, \citenamefont
  {Mchedlishvili}, \citenamefont {Mey}, \citenamefont {M\"uller}, \citenamefont
  {Nass}, \citenamefont {Pesce}, \citenamefont {Prasuhn}, \citenamefont
  {Pretz}, \citenamefont {Rosenthal}, \citenamefont {Schmidt}, \citenamefont
  {Semertzidis}, \citenamefont {Senichev}, \citenamefont {Shmakova},
  \citenamefont {Silenko}, \citenamefont {Slim}, \citenamefont {Soltner},
  \citenamefont {Stahl}, \citenamefont {Stassen}, \citenamefont {Stephenson},
  \citenamefont {Stockhorst}, \citenamefont {Str\"oher}, \citenamefont
  {Tabidze}, \citenamefont {Tagliente}, \citenamefont {Talman}, \citenamefont
  {Engblom}, \citenamefont {Trinkel}, \citenamefont {Uzikov}, \citenamefont
  {Valdau}, \citenamefont {Valetov}, \citenamefont {Vassiliev}, \citenamefont
  {Weidemann}, \citenamefont {Wro\ifmmode~\acute{n}\else \'{n}\fi{}ska},
  \citenamefont {W\"ustner}, \citenamefont {Zupra\ifmmode~\acute{n}\else
  \'{n}\fi{}ski},\ and\ \citenamefont {Zurek}}]{ref3}%
  \BibitemOpen
  \bibfield  {author} {\bibinfo {author} {\bibfnamefont {A.}~\bibnamefont
  {Saleev}}, \bibinfo {author} {\bibfnamefont {N.~N.}\ \bibnamefont
  {Nikolaev}}, \bibinfo {author} {\bibfnamefont {F.}~\bibnamefont {Rathmann}},
  \bibinfo {author} {\bibfnamefont {W.}~\bibnamefont {Augustyniak}}, \bibinfo
  {author} {\bibfnamefont {Z.}~\bibnamefont {Bagdasarian}}, \bibinfo {author}
  {\bibfnamefont {M.}~\bibnamefont {Bai}}, \bibinfo {author} {\bibfnamefont
  {L.}~\bibnamefont {Barion}}, \bibinfo {author} {\bibfnamefont
  {M.}~\bibnamefont {Berz}}, \bibinfo {author} {\bibfnamefont {S.}~\bibnamefont
  {Chekmenev}}, \bibinfo {author} {\bibfnamefont {G.}~\bibnamefont {Ciullo}},
  \bibinfo {author} {\bibfnamefont {S.}~\bibnamefont {Dymov}}, \bibinfo
  {author} {\bibfnamefont {D.}~\bibnamefont {Eversmann}}, \bibinfo {author}
  {\bibfnamefont {M.}~\bibnamefont {Gaisser}}, \bibinfo {author} {\bibfnamefont
  {R.}~\bibnamefont {Gebel}}, \bibinfo {author} {\bibfnamefont
  {K.}~\bibnamefont {Grigoryev}}, \bibinfo {author} {\bibfnamefont
  {D.}~\bibnamefont {Grzonka}}, \bibinfo {author} {\bibfnamefont
  {G.}~\bibnamefont {Guidoboni}}, \bibinfo {author} {\bibfnamefont
  {D.}~\bibnamefont {Heberling}}, \bibinfo {author} {\bibfnamefont
  {V.}~\bibnamefont {Hejny}}, \bibinfo {author} {\bibfnamefont
  {N.}~\bibnamefont {Hempelmann}}, \bibinfo {author} {\bibfnamefont
  {J.}~\bibnamefont {Hetzel}}, \bibinfo {author} {\bibfnamefont
  {F.}~\bibnamefont {Hinder}}, \bibinfo {author} {\bibfnamefont
  {A.}~\bibnamefont {Kacharava}}, \bibinfo {author} {\bibfnamefont
  {V.}~\bibnamefont {Kamerdzhiev}}, \bibinfo {author} {\bibfnamefont
  {I.}~\bibnamefont {Keshelashvili}}, \bibinfo {author} {\bibfnamefont
  {I.}~\bibnamefont {Koop}}, \bibinfo {author} {\bibfnamefont {A.}~\bibnamefont
  {Kulikov}}, \bibinfo {author} {\bibfnamefont {A.}~\bibnamefont {Lehrach}},
  \bibinfo {author} {\bibfnamefont {P.}~\bibnamefont {Lenisa}}, \bibinfo
  {author} {\bibfnamefont {N.}~\bibnamefont {Lomidze}}, \bibinfo {author}
  {\bibfnamefont {B.}~\bibnamefont {Lorentz}}, \bibinfo {author} {\bibfnamefont
  {P.}~\bibnamefont {Maanen}}, \bibinfo {author} {\bibfnamefont
  {G.}~\bibnamefont {Macharashvili}}, \bibinfo {author} {\bibfnamefont
  {A.}~\bibnamefont {Magiera}}, \bibinfo {author} {\bibfnamefont
  {D.}~\bibnamefont {Mchedlishvili}}, \bibinfo {author} {\bibfnamefont
  {S.}~\bibnamefont {Mey}}, \bibinfo {author} {\bibfnamefont {F.}~\bibnamefont
  {M\"uller}}, \bibinfo {author} {\bibfnamefont {A.}~\bibnamefont {Nass}},
  \bibinfo {author} {\bibfnamefont {A.}~\bibnamefont {Pesce}}, \bibinfo
  {author} {\bibfnamefont {D.}~\bibnamefont {Prasuhn}}, \bibinfo {author}
  {\bibfnamefont {J.}~\bibnamefont {Pretz}}, \bibinfo {author} {\bibfnamefont
  {M.}~\bibnamefont {Rosenthal}}, \bibinfo {author} {\bibfnamefont
  {V.}~\bibnamefont {Schmidt}}, \bibinfo {author} {\bibfnamefont
  {Y.}~\bibnamefont {Semertzidis}}, \bibinfo {author} {\bibfnamefont
  {Y.}~\bibnamefont {Senichev}}, \bibinfo {author} {\bibfnamefont
  {V.}~\bibnamefont {Shmakova}}, \bibinfo {author} {\bibfnamefont
  {A.}~\bibnamefont {Silenko}}, \bibinfo {author} {\bibfnamefont
  {J.}~\bibnamefont {Slim}}, \bibinfo {author} {\bibfnamefont {H.}~\bibnamefont
  {Soltner}}, \bibinfo {author} {\bibfnamefont {A.}~\bibnamefont {Stahl}},
  \bibinfo {author} {\bibfnamefont {R.}~\bibnamefont {Stassen}}, \bibinfo
  {author} {\bibfnamefont {E.}~\bibnamefont {Stephenson}}, \bibinfo {author}
  {\bibfnamefont {H.}~\bibnamefont {Stockhorst}}, \bibinfo {author}
  {\bibfnamefont {H.}~\bibnamefont {Str\"oher}}, \bibinfo {author}
  {\bibfnamefont {M.}~\bibnamefont {Tabidze}}, \bibinfo {author} {\bibfnamefont
  {G.}~\bibnamefont {Tagliente}}, \bibinfo {author} {\bibfnamefont
  {R.}~\bibnamefont {Talman}}, \bibinfo {author} {\bibfnamefont {P.~T.}\
  \bibnamefont {Engblom}}, \bibinfo {author} {\bibfnamefont {F.}~\bibnamefont
  {Trinkel}}, \bibinfo {author} {\bibfnamefont {Y.}~\bibnamefont {Uzikov}},
  \bibinfo {author} {\bibfnamefont {Y.}~\bibnamefont {Valdau}}, \bibinfo
  {author} {\bibfnamefont {E.}~\bibnamefont {Valetov}}, \bibinfo {author}
  {\bibfnamefont {A.}~\bibnamefont {Vassiliev}}, \bibinfo {author}
  {\bibfnamefont {C.}~\bibnamefont {Weidemann}}, \bibinfo {author}
  {\bibfnamefont {A.}~\bibnamefont {Wro\ifmmode~\acute{n}\else \'{n}\fi{}ska}},
  \bibinfo {author} {\bibfnamefont {P.}~\bibnamefont {W\"ustner}}, \bibinfo
  {author} {\bibfnamefont {P.}~\bibnamefont {Zupra\ifmmode~\acute{n}\else
  \'{n}\fi{}ski}},\ and\ \bibinfo {author} {\bibfnamefont {M.}~\bibnamefont
  {Zurek}} (\bibinfo {collaboration} {JEDI collaboration}),\ }\href
  {https://doi.org/10.1103/PhysRevAccelBeams.20.072801} {\bibfield  {journal}
  {\bibinfo  {journal} {Phys. Rev. Accel. Beams}\ }\textbf {\bibinfo {volume}
  {20}},\ \bibinfo {pages} {072801} (\bibinfo {year} {2017})}\BibitemShut
  {NoStop}%
\bibitem [{\citenamefont {Hempelmann}\ \emph {et~al.}(2017)\citenamefont
  {Hempelmann}, \citenamefont {Hejny}, \citenamefont {Pretz}, \citenamefont
  {Stephenson}, \citenamefont {Augustyniak}, \citenamefont {Bagdasarian},
  \citenamefont {Bai}, \citenamefont {Barion}, \citenamefont {Berz},
  \citenamefont {Chekmenev}, \citenamefont {Ciullo}, \citenamefont {Dymov},
  \citenamefont {Etzkorn}, \citenamefont {Eversmann}, \citenamefont {Gaisser},
  \citenamefont {Gebel}, \citenamefont {Grigoryev}, \citenamefont {Grzonka},
  \citenamefont {Guidoboni}, \citenamefont {Hanraths}, \citenamefont
  {Heberling}, \citenamefont {Hetzel}, \citenamefont {Hinder}, \citenamefont
  {Kacharava}, \citenamefont {Kamerdzhiev}, \citenamefont {Keshelashvili},
  \citenamefont {Koop}, \citenamefont {Kulikov}, \citenamefont {Lehrach},
  \citenamefont {Lenisa}, \citenamefont {Lomidze}, \citenamefont {Lorentz},
  \citenamefont {Maanen}, \citenamefont {Macharashvili}, \citenamefont
  {Magiera}, \citenamefont {Mchedlishvili}, \citenamefont {Mey}, \citenamefont
  {M\"uller}, \citenamefont {Nass}, \citenamefont {Nikolaev}, \citenamefont
  {Pesce}, \citenamefont {Prasuhn}, \citenamefont {Rathmann}, \citenamefont
  {Rosenthal}, \citenamefont {Saleev}, \citenamefont {Schmidt}, \citenamefont
  {Semertzidis}, \citenamefont {Shmakova}, \citenamefont {Silenko},
  \citenamefont {Slim}, \citenamefont {Soltner}, \citenamefont {Stahl},
  \citenamefont {Stassen}, \citenamefont {Stockhorst}, \citenamefont
  {Str\"oher}, \citenamefont {Tabidze}, \citenamefont {Tagliente},
  \citenamefont {Talman}, \citenamefont {Th\"orngren~Engblom}, \citenamefont
  {Trinkel}, \citenamefont {Uzikov}, \citenamefont {Valdau}, \citenamefont
  {Valetov}, \citenamefont {Vassiliev}, \citenamefont {Weidemann},
  \citenamefont {Wro\ifmmode~\acute{n}\else \'{n}\fi{}ska}, \citenamefont
  {W\"ustner}, \citenamefont {Zupra\ifmmode~\acute{n}\else \'{n}\fi{}ski},\
  and\ \citenamefont {\ifmmode~\dot{Z}\else
  \.{Z}\fi{}urek}}]{hempelmann-phase-2017}%
  \BibitemOpen
  \bibfield  {author} {\bibinfo {author} {\bibfnamefont {N.}~\bibnamefont
  {Hempelmann}}, \bibinfo {author} {\bibfnamefont {V.}~\bibnamefont {Hejny}},
  \bibinfo {author} {\bibfnamefont {J.}~\bibnamefont {Pretz}}, \bibinfo
  {author} {\bibfnamefont {E.}~\bibnamefont {Stephenson}}, \bibinfo {author}
  {\bibfnamefont {W.}~\bibnamefont {Augustyniak}}, \bibinfo {author}
  {\bibfnamefont {Z.}~\bibnamefont {Bagdasarian}}, \bibinfo {author}
  {\bibfnamefont {M.}~\bibnamefont {Bai}}, \bibinfo {author} {\bibfnamefont
  {L.}~\bibnamefont {Barion}}, \bibinfo {author} {\bibfnamefont
  {M.}~\bibnamefont {Berz}}, \bibinfo {author} {\bibfnamefont {S.}~\bibnamefont
  {Chekmenev}}, \bibinfo {author} {\bibfnamefont {G.}~\bibnamefont {Ciullo}},
  \bibinfo {author} {\bibfnamefont {S.}~\bibnamefont {Dymov}}, \bibinfo
  {author} {\bibfnamefont {F.-J.}\ \bibnamefont {Etzkorn}}, \bibinfo {author}
  {\bibfnamefont {D.}~\bibnamefont {Eversmann}}, \bibinfo {author}
  {\bibfnamefont {M.}~\bibnamefont {Gaisser}}, \bibinfo {author} {\bibfnamefont
  {R.}~\bibnamefont {Gebel}}, \bibinfo {author} {\bibfnamefont
  {K.}~\bibnamefont {Grigoryev}}, \bibinfo {author} {\bibfnamefont
  {D.}~\bibnamefont {Grzonka}}, \bibinfo {author} {\bibfnamefont
  {G.}~\bibnamefont {Guidoboni}}, \bibinfo {author} {\bibfnamefont
  {T.}~\bibnamefont {Hanraths}}, \bibinfo {author} {\bibfnamefont
  {D.}~\bibnamefont {Heberling}}, \bibinfo {author} {\bibfnamefont
  {J.}~\bibnamefont {Hetzel}}, \bibinfo {author} {\bibfnamefont
  {F.}~\bibnamefont {Hinder}}, \bibinfo {author} {\bibfnamefont
  {A.}~\bibnamefont {Kacharava}}, \bibinfo {author} {\bibfnamefont
  {V.}~\bibnamefont {Kamerdzhiev}}, \bibinfo {author} {\bibfnamefont
  {I.}~\bibnamefont {Keshelashvili}}, \bibinfo {author} {\bibfnamefont
  {I.}~\bibnamefont {Koop}}, \bibinfo {author} {\bibfnamefont {A.}~\bibnamefont
  {Kulikov}}, \bibinfo {author} {\bibfnamefont {A.}~\bibnamefont {Lehrach}},
  \bibinfo {author} {\bibfnamefont {P.}~\bibnamefont {Lenisa}}, \bibinfo
  {author} {\bibfnamefont {N.}~\bibnamefont {Lomidze}}, \bibinfo {author}
  {\bibfnamefont {B.}~\bibnamefont {Lorentz}}, \bibinfo {author} {\bibfnamefont
  {P.}~\bibnamefont {Maanen}}, \bibinfo {author} {\bibfnamefont
  {G.}~\bibnamefont {Macharashvili}}, \bibinfo {author} {\bibfnamefont
  {A.}~\bibnamefont {Magiera}}, \bibinfo {author} {\bibfnamefont
  {D.}~\bibnamefont {Mchedlishvili}}, \bibinfo {author} {\bibfnamefont
  {S.}~\bibnamefont {Mey}}, \bibinfo {author} {\bibfnamefont {F.}~\bibnamefont
  {M\"uller}}, \bibinfo {author} {\bibfnamefont {A.}~\bibnamefont {Nass}},
  \bibinfo {author} {\bibfnamefont {N.~N.}\ \bibnamefont {Nikolaev}}, \bibinfo
  {author} {\bibfnamefont {A.}~\bibnamefont {Pesce}}, \bibinfo {author}
  {\bibfnamefont {D.}~\bibnamefont {Prasuhn}}, \bibinfo {author} {\bibfnamefont
  {F.}~\bibnamefont {Rathmann}}, \bibinfo {author} {\bibfnamefont
  {M.}~\bibnamefont {Rosenthal}}, \bibinfo {author} {\bibfnamefont
  {A.}~\bibnamefont {Saleev}}, \bibinfo {author} {\bibfnamefont
  {V.}~\bibnamefont {Schmidt}}, \bibinfo {author} {\bibfnamefont
  {Y.}~\bibnamefont {Semertzidis}}, \bibinfo {author} {\bibfnamefont
  {V.}~\bibnamefont {Shmakova}}, \bibinfo {author} {\bibfnamefont
  {A.}~\bibnamefont {Silenko}}, \bibinfo {author} {\bibfnamefont
  {J.}~\bibnamefont {Slim}}, \bibinfo {author} {\bibfnamefont {H.}~\bibnamefont
  {Soltner}}, \bibinfo {author} {\bibfnamefont {A.}~\bibnamefont {Stahl}},
  \bibinfo {author} {\bibfnamefont {R.}~\bibnamefont {Stassen}}, \bibinfo
  {author} {\bibfnamefont {H.}~\bibnamefont {Stockhorst}}, \bibinfo {author}
  {\bibfnamefont {H.}~\bibnamefont {Str\"oher}}, \bibinfo {author}
  {\bibfnamefont {M.}~\bibnamefont {Tabidze}}, \bibinfo {author} {\bibfnamefont
  {G.}~\bibnamefont {Tagliente}}, \bibinfo {author} {\bibfnamefont
  {R.}~\bibnamefont {Talman}}, \bibinfo {author} {\bibfnamefont
  {P.}~\bibnamefont {Th\"orngren~Engblom}}, \bibinfo {author} {\bibfnamefont
  {F.}~\bibnamefont {Trinkel}}, \bibinfo {author} {\bibfnamefont
  {Y.}~\bibnamefont {Uzikov}}, \bibinfo {author} {\bibfnamefont
  {Y.}~\bibnamefont {Valdau}}, \bibinfo {author} {\bibfnamefont
  {E.}~\bibnamefont {Valetov}}, \bibinfo {author} {\bibfnamefont
  {A.}~\bibnamefont {Vassiliev}}, \bibinfo {author} {\bibfnamefont
  {C.}~\bibnamefont {Weidemann}}, \bibinfo {author} {\bibfnamefont
  {A.}~\bibnamefont {Wro\ifmmode~\acute{n}\else \'{n}\fi{}ska}}, \bibinfo
  {author} {\bibfnamefont {P.}~\bibnamefont {W\"ustner}}, \bibinfo {author}
  {\bibfnamefont {P.}~\bibnamefont {Zupra\ifmmode~\acute{n}\else
  \'{n}\fi{}ski}},\ and\ \bibinfo {author} {\bibfnamefont {M.}~\bibnamefont
  {\ifmmode~\dot{Z}\else \.{Z}\fi{}urek}} (\bibinfo {collaboration} {JEDI
  Collaboration}),\ }\href {https://doi.org/10.1103/PhysRevLett.119.014801}
  {\bibfield  {journal} {\bibinfo  {journal} {Phys. Rev. Lett.}\ }\textbf
  {\bibinfo {volume} {119}},\ \bibinfo {pages} {014801} (\bibinfo {year}
  {2017})}\BibitemShut {NoStop}%
\bibitem [{\citenamefont {Guidoboni}\ \emph {et~al.}(2016)\citenamefont
  {Guidoboni}, \citenamefont {Stephenson}, \citenamefont {Andrianov},
  \citenamefont {Augustyniak}, \citenamefont {Bagdasarian}, \citenamefont
  {Bai}, \citenamefont {Baylac}, \citenamefont {Bernreuther}, \citenamefont
  {Bertelli}, \citenamefont {Berz}, \citenamefont {B\"oker}, \citenamefont
  {B\"ohme}, \citenamefont {Bsaisou}, \citenamefont {Chekmenev}, \citenamefont
  {Chiladze}, \citenamefont {Ciullo}, \citenamefont {Contalbrigo},
  \citenamefont {de~Conto}, \citenamefont {Dymov}, \citenamefont {Engels},
  \citenamefont {Esser}, \citenamefont {Eversmann}, \citenamefont {Felden},
  \citenamefont {Gaisser}, \citenamefont {Gebel}, \citenamefont {Gl\"uckler},
  \citenamefont {Goldenbaum}, \citenamefont {Grigoryev}, \citenamefont
  {Grzonka}, \citenamefont {Hahnraths}, \citenamefont {Heberling},
  \citenamefont {Hejny}, \citenamefont {Hempelmann}, \citenamefont {Hetzel},
  \citenamefont {Hinder}, \citenamefont {Hipple}, \citenamefont {H\"olscher},
  \citenamefont {Ivanov}, \citenamefont {Kacharava}, \citenamefont
  {Kamerdzhiev}, \citenamefont {Kamys}, \citenamefont {Keshelashvili},
  \citenamefont {Khoukaz}, \citenamefont {Koop}, \citenamefont {Krause},
  \citenamefont {Krewald}, \citenamefont {Kulikov}, \citenamefont {Lehrach},
  \citenamefont {Lenisa}, \citenamefont {Lomidze}, \citenamefont {Lorentz},
  \citenamefont {Maanen}, \citenamefont {Macharashvili}, \citenamefont
  {Magiera}, \citenamefont {Maier}, \citenamefont {Makino}, \citenamefont
  {Maria\ifmmode~\acute{n}\else \'{n}\fi{}ski}, \citenamefont {Mchedlishvili},
  \citenamefont {Mei\ss{}ner}, \citenamefont {Mey}, \citenamefont {Morse},
  \citenamefont {M\"uller}, \citenamefont {Nass}, \citenamefont {Natour},
  \citenamefont {Nikolaev}, \citenamefont {Nioradze}, \citenamefont
  {Nowakowski}, \citenamefont {Orlov}, \citenamefont {Pesce}, \citenamefont
  {Prasuhn}, \citenamefont {Pretz}, \citenamefont {Rathmann}, \citenamefont
  {Ritman}, \citenamefont {Rosenthal}, \citenamefont {Rudy}, \citenamefont
  {Saleev}, \citenamefont {Sefzick}, \citenamefont {Semertzidis}, \citenamefont
  {Senichev}, \citenamefont {Shmakova}, \citenamefont {Silenko}, \citenamefont
  {Simon}, \citenamefont {Slim}, \citenamefont {Soltner}, \citenamefont
  {Stahl}, \citenamefont {Stassen}, \citenamefont {Statera}, \citenamefont
  {Stockhorst}, \citenamefont {Straatmann}, \citenamefont {Str\"oher},
  \citenamefont {Tabidze}, \citenamefont {Talman}, \citenamefont
  {Th\"orngren~Engblom}, \citenamefont {Trinkel}, \citenamefont
  {Trzci\ifmmode~\acute{n}\else \'{n}\fi{}ski}, \citenamefont {Uzikov},
  \citenamefont {Valdau}, \citenamefont {Valetov}, \citenamefont {Vassiliev},
  \citenamefont {Weidemann}, \citenamefont {Wilkin}, \citenamefont
  {Wro\ifmmode~\acute{n}\else \'{n}\fi{}ska}, \citenamefont {W\"ustner},
  \citenamefont {Zakrzewska}, \citenamefont {Zupra\ifmmode~\acute{n}\else
  \'{n}\fi{}ski},\ and\ \citenamefont {Zyuzin}}]{jedi_collaboration_how_2016}%
  \BibitemOpen
  \bibfield  {author} {\bibinfo {author} {\bibfnamefont {G.}~\bibnamefont
  {Guidoboni}}, \bibinfo {author} {\bibfnamefont {E.}~\bibnamefont
  {Stephenson}}, \bibinfo {author} {\bibfnamefont {S.}~\bibnamefont
  {Andrianov}}, \bibinfo {author} {\bibfnamefont {W.}~\bibnamefont
  {Augustyniak}}, \bibinfo {author} {\bibfnamefont {Z.}~\bibnamefont
  {Bagdasarian}}, \bibinfo {author} {\bibfnamefont {M.}~\bibnamefont {Bai}},
  \bibinfo {author} {\bibfnamefont {M.}~\bibnamefont {Baylac}}, \bibinfo
  {author} {\bibfnamefont {W.}~\bibnamefont {Bernreuther}}, \bibinfo {author}
  {\bibfnamefont {S.}~\bibnamefont {Bertelli}}, \bibinfo {author}
  {\bibfnamefont {M.}~\bibnamefont {Berz}}, \bibinfo {author} {\bibfnamefont
  {J.}~\bibnamefont {B\"oker}}, \bibinfo {author} {\bibfnamefont
  {C.}~\bibnamefont {B\"ohme}}, \bibinfo {author} {\bibfnamefont
  {J.}~\bibnamefont {Bsaisou}}, \bibinfo {author} {\bibfnamefont
  {S.}~\bibnamefont {Chekmenev}}, \bibinfo {author} {\bibfnamefont
  {D.}~\bibnamefont {Chiladze}}, \bibinfo {author} {\bibfnamefont
  {G.}~\bibnamefont {Ciullo}}, \bibinfo {author} {\bibfnamefont
  {M.}~\bibnamefont {Contalbrigo}}, \bibinfo {author} {\bibfnamefont {J.-M.}\
  \bibnamefont {de~Conto}}, \bibinfo {author} {\bibfnamefont {S.}~\bibnamefont
  {Dymov}}, \bibinfo {author} {\bibfnamefont {R.}~\bibnamefont {Engels}},
  \bibinfo {author} {\bibfnamefont {F.~M.}\ \bibnamefont {Esser}}, \bibinfo
  {author} {\bibfnamefont {D.}~\bibnamefont {Eversmann}}, \bibinfo {author}
  {\bibfnamefont {O.}~\bibnamefont {Felden}}, \bibinfo {author} {\bibfnamefont
  {M.}~\bibnamefont {Gaisser}}, \bibinfo {author} {\bibfnamefont
  {R.}~\bibnamefont {Gebel}}, \bibinfo {author} {\bibfnamefont
  {H.}~\bibnamefont {Gl\"uckler}}, \bibinfo {author} {\bibfnamefont
  {F.}~\bibnamefont {Goldenbaum}}, \bibinfo {author} {\bibfnamefont
  {K.}~\bibnamefont {Grigoryev}}, \bibinfo {author} {\bibfnamefont
  {D.}~\bibnamefont {Grzonka}}, \bibinfo {author} {\bibfnamefont
  {T.}~\bibnamefont {Hahnraths}}, \bibinfo {author} {\bibfnamefont
  {D.}~\bibnamefont {Heberling}}, \bibinfo {author} {\bibfnamefont
  {V.}~\bibnamefont {Hejny}}, \bibinfo {author} {\bibfnamefont
  {N.}~\bibnamefont {Hempelmann}}, \bibinfo {author} {\bibfnamefont
  {J.}~\bibnamefont {Hetzel}}, \bibinfo {author} {\bibfnamefont
  {F.}~\bibnamefont {Hinder}}, \bibinfo {author} {\bibfnamefont
  {R.}~\bibnamefont {Hipple}}, \bibinfo {author} {\bibfnamefont
  {D.}~\bibnamefont {H\"olscher}}, \bibinfo {author} {\bibfnamefont
  {A.}~\bibnamefont {Ivanov}}, \bibinfo {author} {\bibfnamefont
  {A.}~\bibnamefont {Kacharava}}, \bibinfo {author} {\bibfnamefont
  {V.}~\bibnamefont {Kamerdzhiev}}, \bibinfo {author} {\bibfnamefont
  {B.}~\bibnamefont {Kamys}}, \bibinfo {author} {\bibfnamefont
  {I.}~\bibnamefont {Keshelashvili}}, \bibinfo {author} {\bibfnamefont
  {A.}~\bibnamefont {Khoukaz}}, \bibinfo {author} {\bibfnamefont
  {I.}~\bibnamefont {Koop}}, \bibinfo {author} {\bibfnamefont {H.-J.}\
  \bibnamefont {Krause}}, \bibinfo {author} {\bibfnamefont {S.}~\bibnamefont
  {Krewald}}, \bibinfo {author} {\bibfnamefont {A.}~\bibnamefont {Kulikov}},
  \bibinfo {author} {\bibfnamefont {A.}~\bibnamefont {Lehrach}}, \bibinfo
  {author} {\bibfnamefont {P.}~\bibnamefont {Lenisa}}, \bibinfo {author}
  {\bibfnamefont {N.}~\bibnamefont {Lomidze}}, \bibinfo {author} {\bibfnamefont
  {B.}~\bibnamefont {Lorentz}}, \bibinfo {author} {\bibfnamefont
  {P.}~\bibnamefont {Maanen}}, \bibinfo {author} {\bibfnamefont
  {G.}~\bibnamefont {Macharashvili}}, \bibinfo {author} {\bibfnamefont
  {A.}~\bibnamefont {Magiera}}, \bibinfo {author} {\bibfnamefont
  {R.}~\bibnamefont {Maier}}, \bibinfo {author} {\bibfnamefont
  {K.}~\bibnamefont {Makino}}, \bibinfo {author} {\bibfnamefont
  {B.}~\bibnamefont {Maria\ifmmode~\acute{n}\else \'{n}\fi{}ski}}, \bibinfo
  {author} {\bibfnamefont {D.}~\bibnamefont {Mchedlishvili}}, \bibinfo {author}
  {\bibfnamefont {U.-G.}\ \bibnamefont {Mei\ss{}ner}}, \bibinfo {author}
  {\bibfnamefont {S.}~\bibnamefont {Mey}}, \bibinfo {author} {\bibfnamefont
  {W.}~\bibnamefont {Morse}}, \bibinfo {author} {\bibfnamefont
  {F.}~\bibnamefont {M\"uller}}, \bibinfo {author} {\bibfnamefont
  {A.}~\bibnamefont {Nass}}, \bibinfo {author} {\bibfnamefont {G.}~\bibnamefont
  {Natour}}, \bibinfo {author} {\bibfnamefont {N.}~\bibnamefont {Nikolaev}},
  \bibinfo {author} {\bibfnamefont {M.}~\bibnamefont {Nioradze}}, \bibinfo
  {author} {\bibfnamefont {K.}~\bibnamefont {Nowakowski}}, \bibinfo {author}
  {\bibfnamefont {Y.}~\bibnamefont {Orlov}}, \bibinfo {author} {\bibfnamefont
  {A.}~\bibnamefont {Pesce}}, \bibinfo {author} {\bibfnamefont
  {D.}~\bibnamefont {Prasuhn}}, \bibinfo {author} {\bibfnamefont
  {J.}~\bibnamefont {Pretz}}, \bibinfo {author} {\bibfnamefont
  {F.}~\bibnamefont {Rathmann}}, \bibinfo {author} {\bibfnamefont
  {J.}~\bibnamefont {Ritman}}, \bibinfo {author} {\bibfnamefont
  {M.}~\bibnamefont {Rosenthal}}, \bibinfo {author} {\bibfnamefont
  {Z.}~\bibnamefont {Rudy}}, \bibinfo {author} {\bibfnamefont {A.}~\bibnamefont
  {Saleev}}, \bibinfo {author} {\bibfnamefont {T.}~\bibnamefont {Sefzick}},
  \bibinfo {author} {\bibfnamefont {Y.~K.}\ \bibnamefont {Semertzidis}},
  \bibinfo {author} {\bibfnamefont {Y.}~\bibnamefont {Senichev}}, \bibinfo
  {author} {\bibfnamefont {V.}~\bibnamefont {Shmakova}}, \bibinfo {author}
  {\bibfnamefont {A.}~\bibnamefont {Silenko}}, \bibinfo {author} {\bibfnamefont
  {M.}~\bibnamefont {Simon}}, \bibinfo {author} {\bibfnamefont
  {J.}~\bibnamefont {Slim}}, \bibinfo {author} {\bibfnamefont {H.}~\bibnamefont
  {Soltner}}, \bibinfo {author} {\bibfnamefont {A.}~\bibnamefont {Stahl}},
  \bibinfo {author} {\bibfnamefont {R.}~\bibnamefont {Stassen}}, \bibinfo
  {author} {\bibfnamefont {M.}~\bibnamefont {Statera}}, \bibinfo {author}
  {\bibfnamefont {H.}~\bibnamefont {Stockhorst}}, \bibinfo {author}
  {\bibfnamefont {H.}~\bibnamefont {Straatmann}}, \bibinfo {author}
  {\bibfnamefont {H.}~\bibnamefont {Str\"oher}}, \bibinfo {author}
  {\bibfnamefont {M.}~\bibnamefont {Tabidze}}, \bibinfo {author} {\bibfnamefont
  {R.}~\bibnamefont {Talman}}, \bibinfo {author} {\bibfnamefont
  {P.}~\bibnamefont {Th\"orngren~Engblom}}, \bibinfo {author} {\bibfnamefont
  {F.}~\bibnamefont {Trinkel}}, \bibinfo {author} {\bibfnamefont
  {A.}~\bibnamefont {Trzci\ifmmode~\acute{n}\else \'{n}\fi{}ski}}, \bibinfo
  {author} {\bibfnamefont {Y.}~\bibnamefont {Uzikov}}, \bibinfo {author}
  {\bibfnamefont {Y.}~\bibnamefont {Valdau}}, \bibinfo {author} {\bibfnamefont
  {E.}~\bibnamefont {Valetov}}, \bibinfo {author} {\bibfnamefont
  {A.}~\bibnamefont {Vassiliev}}, \bibinfo {author} {\bibfnamefont
  {C.}~\bibnamefont {Weidemann}}, \bibinfo {author} {\bibfnamefont
  {C.}~\bibnamefont {Wilkin}}, \bibinfo {author} {\bibfnamefont
  {A.}~\bibnamefont {Wro\ifmmode~\acute{n}\else \'{n}\fi{}ska}}, \bibinfo
  {author} {\bibfnamefont {P.}~\bibnamefont {W\"ustner}}, \bibinfo {author}
  {\bibfnamefont {M.}~\bibnamefont {Zakrzewska}}, \bibinfo {author}
  {\bibfnamefont {P.}~\bibnamefont {Zupra\ifmmode~\acute{n}\else
  \'{n}\fi{}ski}},\ and\ \bibinfo {author} {\bibfnamefont {D.}~\bibnamefont
  {Zyuzin}} (\bibinfo {collaboration} {JEDI Collaboration}),\ }\href
  {https://doi.org/10.1103/PhysRevLett.117.054801} {\bibfield  {journal}
  {\bibinfo  {journal} {Phys. Rev. Lett.}\ }\textbf {\bibinfo {volume} {117}},\
  \bibinfo {pages} {054801} (\bibinfo {year} {2016})}\BibitemShut {NoStop}%
\bibitem [{\citenamefont {Eversmann}\ \emph {et~al.}(2015)\citenamefont
  {Eversmann}, \citenamefont {Hejny}, \citenamefont {Hinder}, \citenamefont
  {Kacharava}, \citenamefont {Pretz}, \citenamefont {Rathmann}, \citenamefont
  {Rosenthal}, \citenamefont {Trinkel}, \citenamefont {Andrianov},
  \citenamefont {Augustyniak}, \citenamefont {Bagdasarian}, \citenamefont
  {Bai}, \citenamefont {Bernreuther}, \citenamefont {Bertelli}, \citenamefont
  {Berz}, \citenamefont {Bsaisou}, \citenamefont {Chekmenev}, \citenamefont
  {Chiladze}, \citenamefont {Ciullo}, \citenamefont {Contalbrigo},
  \citenamefont {de~Vries}, \citenamefont {Dymov}, \citenamefont {Engels},
  \citenamefont {Esser}, \citenamefont {Felden}, \citenamefont {Gaisser},
  \citenamefont {Gebel}, \citenamefont {Gl\"uckler}, \citenamefont
  {Goldenbaum}, \citenamefont {Grigoryev}, \citenamefont {Grzonka},
  \citenamefont {Guidoboni}, \citenamefont {Hanhart}, \citenamefont
  {Heberling}, \citenamefont {Hempelmann}, \citenamefont {Hetzel},
  \citenamefont {Hipple}, \citenamefont {H\"olscher}, \citenamefont {Ivanov},
  \citenamefont {Kamerdzhiev}, \citenamefont {Kamys}, \citenamefont
  {Keshelashvili}, \citenamefont {Khoukaz}, \citenamefont {Koop}, \citenamefont
  {Krause}, \citenamefont {Krewald}, \citenamefont {Kulikov}, \citenamefont
  {Lehrach}, \citenamefont {Lenisa}, \citenamefont {Lomidze}, \citenamefont
  {Lorentz}, \citenamefont {Maanen}, \citenamefont {Macharashvili},
  \citenamefont {Magiera}, \citenamefont {Maier}, \citenamefont {Makino},
  \citenamefont {Maria\ifmmode~\acute{n}\else \'{n}\fi{}ski}, \citenamefont
  {Mchedlishvili}, \citenamefont {Mei\ss{}ner}, \citenamefont {Mey},
  \citenamefont {Nass}, \citenamefont {Natour}, \citenamefont {Nikolaev},
  \citenamefont {Nioradze}, \citenamefont {Nogga}, \citenamefont {Nowakowski},
  \citenamefont {Pesce}, \citenamefont {Prasuhn}, \citenamefont {Ritman},
  \citenamefont {Rudy}, \citenamefont {Saleev}, \citenamefont {Semertzidis},
  \citenamefont {Senichev}, \citenamefont {Shmakova}, \citenamefont {Silenko},
  \citenamefont {Slim}, \citenamefont {Soltner}, \citenamefont {Stahl},
  \citenamefont {Stassen}, \citenamefont {Statera}, \citenamefont {Stephenson},
  \citenamefont {Stockhorst}, \citenamefont {Straatmann}, \citenamefont
  {Str\"oher}, \citenamefont {Tabidze}, \citenamefont {Talman}, \citenamefont
  {Th\"orngren~Engblom}, \citenamefont {Trzci\ifmmode~\acute{n}\else
  \'{n}\fi{}ski}, \citenamefont {Uzikov}, \citenamefont {Valdau}, \citenamefont
  {Valetov}, \citenamefont {Vassiliev}, \citenamefont {Weidemann},
  \citenamefont {Wilkin}, \citenamefont {Wirzba}, \citenamefont
  {Wro\ifmmode~\acute{n}\else \'{n}\fi{}ska}, \citenamefont {W\"ustner},
  \citenamefont {Zakrzewska}, \citenamefont {Zupra\ifmmode~\acute{n}\else
  \'{n}\fi{}ski},\ and\ \citenamefont {Zyuzin}}]{ref6}%
  \BibitemOpen
  \bibfield  {author} {\bibinfo {author} {\bibfnamefont {D.}~\bibnamefont
  {Eversmann}}, \bibinfo {author} {\bibfnamefont {V.}~\bibnamefont {Hejny}},
  \bibinfo {author} {\bibfnamefont {F.}~\bibnamefont {Hinder}}, \bibinfo
  {author} {\bibfnamefont {A.}~\bibnamefont {Kacharava}}, \bibinfo {author}
  {\bibfnamefont {J.}~\bibnamefont {Pretz}}, \bibinfo {author} {\bibfnamefont
  {F.}~\bibnamefont {Rathmann}}, \bibinfo {author} {\bibfnamefont
  {M.}~\bibnamefont {Rosenthal}}, \bibinfo {author} {\bibfnamefont
  {F.}~\bibnamefont {Trinkel}}, \bibinfo {author} {\bibfnamefont
  {S.}~\bibnamefont {Andrianov}}, \bibinfo {author} {\bibfnamefont
  {W.}~\bibnamefont {Augustyniak}}, \bibinfo {author} {\bibfnamefont
  {Z.}~\bibnamefont {Bagdasarian}}, \bibinfo {author} {\bibfnamefont
  {M.}~\bibnamefont {Bai}}, \bibinfo {author} {\bibfnamefont {W.}~\bibnamefont
  {Bernreuther}}, \bibinfo {author} {\bibfnamefont {S.}~\bibnamefont
  {Bertelli}}, \bibinfo {author} {\bibfnamefont {M.}~\bibnamefont {Berz}},
  \bibinfo {author} {\bibfnamefont {J.}~\bibnamefont {Bsaisou}}, \bibinfo
  {author} {\bibfnamefont {S.}~\bibnamefont {Chekmenev}}, \bibinfo {author}
  {\bibfnamefont {D.}~\bibnamefont {Chiladze}}, \bibinfo {author}
  {\bibfnamefont {G.}~\bibnamefont {Ciullo}}, \bibinfo {author} {\bibfnamefont
  {M.}~\bibnamefont {Contalbrigo}}, \bibinfo {author} {\bibfnamefont
  {J.}~\bibnamefont {de~Vries}}, \bibinfo {author} {\bibfnamefont
  {S.}~\bibnamefont {Dymov}}, \bibinfo {author} {\bibfnamefont
  {R.}~\bibnamefont {Engels}}, \bibinfo {author} {\bibfnamefont {F.~M.}\
  \bibnamefont {Esser}}, \bibinfo {author} {\bibfnamefont {O.}~\bibnamefont
  {Felden}}, \bibinfo {author} {\bibfnamefont {M.}~\bibnamefont {Gaisser}},
  \bibinfo {author} {\bibfnamefont {R.}~\bibnamefont {Gebel}}, \bibinfo
  {author} {\bibfnamefont {H.}~\bibnamefont {Gl\"uckler}}, \bibinfo {author}
  {\bibfnamefont {F.}~\bibnamefont {Goldenbaum}}, \bibinfo {author}
  {\bibfnamefont {K.}~\bibnamefont {Grigoryev}}, \bibinfo {author}
  {\bibfnamefont {D.}~\bibnamefont {Grzonka}}, \bibinfo {author} {\bibfnamefont
  {G.}~\bibnamefont {Guidoboni}}, \bibinfo {author} {\bibfnamefont
  {C.}~\bibnamefont {Hanhart}}, \bibinfo {author} {\bibfnamefont
  {D.}~\bibnamefont {Heberling}}, \bibinfo {author} {\bibfnamefont
  {N.}~\bibnamefont {Hempelmann}}, \bibinfo {author} {\bibfnamefont
  {J.}~\bibnamefont {Hetzel}}, \bibinfo {author} {\bibfnamefont
  {R.}~\bibnamefont {Hipple}}, \bibinfo {author} {\bibfnamefont
  {D.}~\bibnamefont {H\"olscher}}, \bibinfo {author} {\bibfnamefont
  {A.}~\bibnamefont {Ivanov}}, \bibinfo {author} {\bibfnamefont
  {V.}~\bibnamefont {Kamerdzhiev}}, \bibinfo {author} {\bibfnamefont
  {B.}~\bibnamefont {Kamys}}, \bibinfo {author} {\bibfnamefont
  {I.}~\bibnamefont {Keshelashvili}}, \bibinfo {author} {\bibfnamefont
  {A.}~\bibnamefont {Khoukaz}}, \bibinfo {author} {\bibfnamefont
  {I.}~\bibnamefont {Koop}}, \bibinfo {author} {\bibfnamefont {H.-J.}\
  \bibnamefont {Krause}}, \bibinfo {author} {\bibfnamefont {S.}~\bibnamefont
  {Krewald}}, \bibinfo {author} {\bibfnamefont {A.}~\bibnamefont {Kulikov}},
  \bibinfo {author} {\bibfnamefont {A.}~\bibnamefont {Lehrach}}, \bibinfo
  {author} {\bibfnamefont {P.}~\bibnamefont {Lenisa}}, \bibinfo {author}
  {\bibfnamefont {N.}~\bibnamefont {Lomidze}}, \bibinfo {author} {\bibfnamefont
  {B.}~\bibnamefont {Lorentz}}, \bibinfo {author} {\bibfnamefont
  {P.}~\bibnamefont {Maanen}}, \bibinfo {author} {\bibfnamefont
  {G.}~\bibnamefont {Macharashvili}}, \bibinfo {author} {\bibfnamefont
  {A.}~\bibnamefont {Magiera}}, \bibinfo {author} {\bibfnamefont
  {R.}~\bibnamefont {Maier}}, \bibinfo {author} {\bibfnamefont
  {K.}~\bibnamefont {Makino}}, \bibinfo {author} {\bibfnamefont
  {B.}~\bibnamefont {Maria\ifmmode~\acute{n}\else \'{n}\fi{}ski}}, \bibinfo
  {author} {\bibfnamefont {D.}~\bibnamefont {Mchedlishvili}}, \bibinfo {author}
  {\bibfnamefont {U.-G.}\ \bibnamefont {Mei\ss{}ner}}, \bibinfo {author}
  {\bibfnamefont {S.}~\bibnamefont {Mey}}, \bibinfo {author} {\bibfnamefont
  {A.}~\bibnamefont {Nass}}, \bibinfo {author} {\bibfnamefont {G.}~\bibnamefont
  {Natour}}, \bibinfo {author} {\bibfnamefont {N.}~\bibnamefont {Nikolaev}},
  \bibinfo {author} {\bibfnamefont {M.}~\bibnamefont {Nioradze}}, \bibinfo
  {author} {\bibfnamefont {A.}~\bibnamefont {Nogga}}, \bibinfo {author}
  {\bibfnamefont {K.}~\bibnamefont {Nowakowski}}, \bibinfo {author}
  {\bibfnamefont {A.}~\bibnamefont {Pesce}}, \bibinfo {author} {\bibfnamefont
  {D.}~\bibnamefont {Prasuhn}}, \bibinfo {author} {\bibfnamefont
  {J.}~\bibnamefont {Ritman}}, \bibinfo {author} {\bibfnamefont
  {Z.}~\bibnamefont {Rudy}}, \bibinfo {author} {\bibfnamefont {A.}~\bibnamefont
  {Saleev}}, \bibinfo {author} {\bibfnamefont {Y.}~\bibnamefont {Semertzidis}},
  \bibinfo {author} {\bibfnamefont {Y.}~\bibnamefont {Senichev}}, \bibinfo
  {author} {\bibfnamefont {V.}~\bibnamefont {Shmakova}}, \bibinfo {author}
  {\bibfnamefont {A.}~\bibnamefont {Silenko}}, \bibinfo {author} {\bibfnamefont
  {J.}~\bibnamefont {Slim}}, \bibinfo {author} {\bibfnamefont {H.}~\bibnamefont
  {Soltner}}, \bibinfo {author} {\bibfnamefont {A.}~\bibnamefont {Stahl}},
  \bibinfo {author} {\bibfnamefont {R.}~\bibnamefont {Stassen}}, \bibinfo
  {author} {\bibfnamefont {M.}~\bibnamefont {Statera}}, \bibinfo {author}
  {\bibfnamefont {E.}~\bibnamefont {Stephenson}}, \bibinfo {author}
  {\bibfnamefont {H.}~\bibnamefont {Stockhorst}}, \bibinfo {author}
  {\bibfnamefont {H.}~\bibnamefont {Straatmann}}, \bibinfo {author}
  {\bibfnamefont {H.}~\bibnamefont {Str\"oher}}, \bibinfo {author}
  {\bibfnamefont {M.}~\bibnamefont {Tabidze}}, \bibinfo {author} {\bibfnamefont
  {R.}~\bibnamefont {Talman}}, \bibinfo {author} {\bibfnamefont
  {P.}~\bibnamefont {Th\"orngren~Engblom}}, \bibinfo {author} {\bibfnamefont
  {A.}~\bibnamefont {Trzci\ifmmode~\acute{n}\else \'{n}\fi{}ski}}, \bibinfo
  {author} {\bibfnamefont {Y.}~\bibnamefont {Uzikov}}, \bibinfo {author}
  {\bibfnamefont {Y.}~\bibnamefont {Valdau}}, \bibinfo {author} {\bibfnamefont
  {E.}~\bibnamefont {Valetov}}, \bibinfo {author} {\bibfnamefont
  {A.}~\bibnamefont {Vassiliev}}, \bibinfo {author} {\bibfnamefont
  {C.}~\bibnamefont {Weidemann}}, \bibinfo {author} {\bibfnamefont
  {C.}~\bibnamefont {Wilkin}}, \bibinfo {author} {\bibfnamefont
  {A.}~\bibnamefont {Wirzba}}, \bibinfo {author} {\bibfnamefont
  {A.}~\bibnamefont {Wro\ifmmode~\acute{n}\else \'{n}\fi{}ska}}, \bibinfo
  {author} {\bibfnamefont {P.}~\bibnamefont {W\"ustner}}, \bibinfo {author}
  {\bibfnamefont {M.}~\bibnamefont {Zakrzewska}}, \bibinfo {author}
  {\bibfnamefont {P.}~\bibnamefont {Zupra\ifmmode~\acute{n}\else
  \'{n}\fi{}ski}},\ and\ \bibinfo {author} {\bibfnamefont {D.}~\bibnamefont
  {Zyuzin}} (\bibinfo {collaboration} {JEDI collaboration}),\ }\href
  {https://doi.org/10.1103/PhysRevLett.115.094801} {\bibfield  {journal}
  {\bibinfo  {journal} {Phys. Rev. Lett.}\ }\textbf {\bibinfo {volume} {115}},\
  \bibinfo {pages} {094801} (\bibinfo {year} {2015})}\BibitemShut {NoStop}%
\bibitem [{\citenamefont {Bagdasarian}\ \emph {et~al.}(2014)\citenamefont
  {Bagdasarian}, \citenamefont {Bertelli}, \citenamefont {Chiladze},
  \citenamefont {Ciullo}, \citenamefont {Dietrich}, \citenamefont {Dymov},
  \citenamefont {Eversmann}, \citenamefont {Fanourakis}, \citenamefont
  {Gaisser}, \citenamefont {Gebel}, \citenamefont {Gou}, \citenamefont
  {Guidoboni}, \citenamefont {Hejny}, \citenamefont {Kacharava}, \citenamefont
  {Kamerdzhiev}, \citenamefont {Lehrach}, \citenamefont {Lenisa}, \citenamefont
  {Lorentz}, \citenamefont {Magallanes}, \citenamefont {Maier}, \citenamefont
  {Mchedlishvili}, \citenamefont {Morse}, \citenamefont {Nass}, \citenamefont
  {Oellers}, \citenamefont {Pesce}, \citenamefont {Prasuhn}, \citenamefont
  {Pretz}, \citenamefont {Rathmann}, \citenamefont {Shmakova}, \citenamefont
  {Semertzidis}, \citenamefont {Stephenson}, \citenamefont {Stockhorst},
  \citenamefont {Str\"oher}, \citenamefont {Talman}, \citenamefont
  {Th\"orngren~Engblom}, \citenamefont {Valdau}, \citenamefont {Weidemann},\
  and\ \citenamefont {W\"ustner}}]{ref7}%
  \BibitemOpen
  \bibfield  {author} {\bibinfo {author} {\bibfnamefont {Z.}~\bibnamefont
  {Bagdasarian}}, \bibinfo {author} {\bibfnamefont {S.}~\bibnamefont
  {Bertelli}}, \bibinfo {author} {\bibfnamefont {D.}~\bibnamefont {Chiladze}},
  \bibinfo {author} {\bibfnamefont {G.}~\bibnamefont {Ciullo}}, \bibinfo
  {author} {\bibfnamefont {J.}~\bibnamefont {Dietrich}}, \bibinfo {author}
  {\bibfnamefont {S.}~\bibnamefont {Dymov}}, \bibinfo {author} {\bibfnamefont
  {D.}~\bibnamefont {Eversmann}}, \bibinfo {author} {\bibfnamefont
  {G.}~\bibnamefont {Fanourakis}}, \bibinfo {author} {\bibfnamefont
  {M.}~\bibnamefont {Gaisser}}, \bibinfo {author} {\bibfnamefont
  {R.}~\bibnamefont {Gebel}}, \bibinfo {author} {\bibfnamefont
  {B.}~\bibnamefont {Gou}}, \bibinfo {author} {\bibfnamefont {G.}~\bibnamefont
  {Guidoboni}}, \bibinfo {author} {\bibfnamefont {V.}~\bibnamefont {Hejny}},
  \bibinfo {author} {\bibfnamefont {A.}~\bibnamefont {Kacharava}}, \bibinfo
  {author} {\bibfnamefont {V.}~\bibnamefont {Kamerdzhiev}}, \bibinfo {author}
  {\bibfnamefont {A.}~\bibnamefont {Lehrach}}, \bibinfo {author} {\bibfnamefont
  {P.}~\bibnamefont {Lenisa}}, \bibinfo {author} {\bibfnamefont
  {B.}~\bibnamefont {Lorentz}}, \bibinfo {author} {\bibfnamefont
  {L.}~\bibnamefont {Magallanes}}, \bibinfo {author} {\bibfnamefont
  {R.}~\bibnamefont {Maier}}, \bibinfo {author} {\bibfnamefont
  {D.}~\bibnamefont {Mchedlishvili}}, \bibinfo {author} {\bibfnamefont {W.~M.}\
  \bibnamefont {Morse}}, \bibinfo {author} {\bibfnamefont {A.}~\bibnamefont
  {Nass}}, \bibinfo {author} {\bibfnamefont {D.}~\bibnamefont {Oellers}},
  \bibinfo {author} {\bibfnamefont {A.}~\bibnamefont {Pesce}}, \bibinfo
  {author} {\bibfnamefont {D.}~\bibnamefont {Prasuhn}}, \bibinfo {author}
  {\bibfnamefont {J.}~\bibnamefont {Pretz}}, \bibinfo {author} {\bibfnamefont
  {F.}~\bibnamefont {Rathmann}}, \bibinfo {author} {\bibfnamefont
  {V.}~\bibnamefont {Shmakova}}, \bibinfo {author} {\bibfnamefont {Y.~K.}\
  \bibnamefont {Semertzidis}}, \bibinfo {author} {\bibfnamefont {E.~J.}\
  \bibnamefont {Stephenson}}, \bibinfo {author} {\bibfnamefont
  {H.}~\bibnamefont {Stockhorst}}, \bibinfo {author} {\bibfnamefont
  {H.}~\bibnamefont {Str\"oher}}, \bibinfo {author} {\bibfnamefont
  {R.}~\bibnamefont {Talman}}, \bibinfo {author} {\bibfnamefont
  {P.}~\bibnamefont {Th\"orngren~Engblom}}, \bibinfo {author} {\bibfnamefont
  {Y.}~\bibnamefont {Valdau}}, \bibinfo {author} {\bibfnamefont
  {C.}~\bibnamefont {Weidemann}},\ and\ \bibinfo {author} {\bibfnamefont
  {P.}~\bibnamefont {W\"ustner}},\ }\href
  {https://doi.org/10.1103/PhysRevSTAB.17.052803} {\bibfield  {journal}
  {\bibinfo  {journal} {Phys. Rev. ST Accel. Beams}\ }\textbf {\bibinfo
  {volume} {17}},\ \bibinfo {pages} {052803} (\bibinfo {year}
  {2014})}\BibitemShut {NoStop}%
\bibitem [{\citenamefont {Meyer}\ \emph {et~al.}(1988)\citenamefont {Meyer},
  \citenamefont {Schwandt}, \citenamefont {Abegg}, \citenamefont {Miller},
  \citenamefont {Jackson}, \citenamefont {Yen}, \citenamefont {Gaillard},
  \citenamefont {Hugi}, \citenamefont {Helmer}, \citenamefont {Frekers},\ and\
  \citenamefont {Saxena}}]{hom88}%
  \BibitemOpen
  \bibfield  {author} {\bibinfo {author} {\bibfnamefont {H.~O.}\ \bibnamefont
  {Meyer}}, \bibinfo {author} {\bibfnamefont {P.}~\bibnamefont {Schwandt}},
  \bibinfo {author} {\bibfnamefont {R.}~\bibnamefont {Abegg}}, \bibinfo
  {author} {\bibfnamefont {C.~A.}\ \bibnamefont {Miller}}, \bibinfo {author}
  {\bibfnamefont {K.~P.}\ \bibnamefont {Jackson}}, \bibinfo {author}
  {\bibfnamefont {S.}~\bibnamefont {Yen}}, \bibinfo {author} {\bibfnamefont
  {G.}~\bibnamefont {Gaillard}}, \bibinfo {author} {\bibfnamefont
  {M.}~\bibnamefont {Hugi}}, \bibinfo {author} {\bibfnamefont {R.}~\bibnamefont
  {Helmer}}, \bibinfo {author} {\bibfnamefont {D.}~\bibnamefont {Frekers}},\
  and\ \bibinfo {author} {\bibfnamefont {A.}~\bibnamefont {Saxena}},\ }\href
  {https://doi.org/10.1103/PhysRevC.37.544} {\bibfield  {journal} {\bibinfo
  {journal} {Phys. Rev. C}\ }\textbf {\bibinfo {volume} {37}},\ \bibinfo
  {pages} {544} (\bibinfo {year} {1988})}\BibitemShut {NoStop}%
\bibitem [{\citenamefont {Farley}\ \emph {et~al.}(2004)\citenamefont {Farley},
  \citenamefont {Jungmann}, \citenamefont {Miller}, \citenamefont {Morse},
  \citenamefont {Orlov}, \citenamefont {Roberts}, \citenamefont {Semertzidis},
  \citenamefont {Silenko},\ and\ \citenamefont {Stephenson}}]{Farley2004}%
  \BibitemOpen
  \bibfield  {author} {\bibinfo {author} {\bibfnamefont {F.}~\bibnamefont
  {Farley}}, \bibinfo {author} {\bibfnamefont {K.}~\bibnamefont {Jungmann}},
  \bibinfo {author} {\bibfnamefont {J.}~\bibnamefont {Miller}}, \bibinfo
  {author} {\bibfnamefont {W.}~\bibnamefont {Morse}}, \bibinfo {author}
  {\bibfnamefont {Y.}~\bibnamefont {Orlov}}, \bibinfo {author} {\bibfnamefont
  {B.}~\bibnamefont {Roberts}}, \bibinfo {author} {\bibfnamefont
  {Y.}~\bibnamefont {Semertzidis}}, \bibinfo {author} {\bibfnamefont
  {A.}~\bibnamefont {Silenko}},\ and\ \bibinfo {author} {\bibfnamefont
  {E.}~\bibnamefont {Stephenson}},\ }\href
  {https://doi.org/10.1103/PhysRevLett.93.052001} {\bibfield  {journal}
  {\bibinfo  {journal} {Physical Review Letters}\ }\textbf {\bibinfo {volume}
  {93}},\ \bibinfo {pages} {052001} (\bibinfo {year} {2004})}\BibitemShut
  {NoStop}%
\end{thebibliography}%
\end{document}